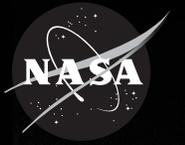

# L U V O I R

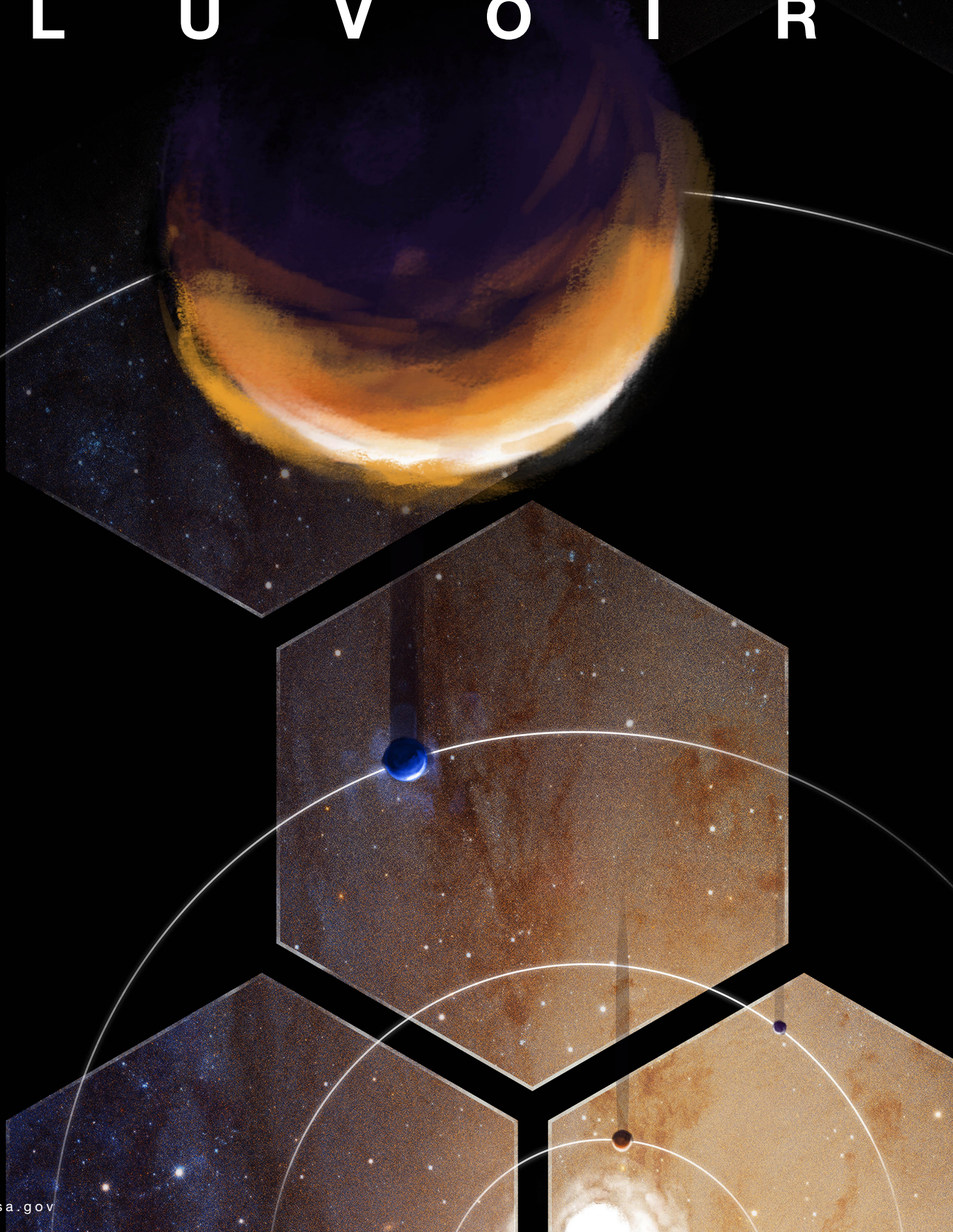

## LUVOIR Study Team

### Science and Technology Definition Team – Voting Members

Debra Fischer (Yale University), **Co-Chair**
Bradley Peterson (Ohio State University / STScI), **Co-Chair**
Jacob Bean (University of Chicago)
Daniela Calzetti (University of Massachusetts – Amherst)
Rebekah Dawson (Penn State)
Courtney Dressing (University of California – Berkeley)
Lee Feinberg (NASA GSFC)
Kevin France (University of Colorado – Boulder), **LUMOS Instrument Lead**
Olivier Guyon (University of Arizona / Subaru Telescope)
Walter Harris (University of Arizona / LPL), **Solar System Working Group Lead**
Mark Marley (NASA Ames), **Exoplanet Working Group Lead**
Victoria Meadows (University of Washington)
Leonidas Moustakas (JPL)
John O'Meara (St. Michael's College), **Cosmic Origins Working Group Lead**
Ilaria Pascucci (University of Arizona / LPL)
Marc Postman (STScI), **HDI Instrument Lead**
Laurent Pueyo (STScI), **ECLIPS Instrument Lead**
David Redding (JPL), **Technology Working Group Lead**
Jane Rigby (NASA GSFC)
Aki Roberge (NASA GSFC), **Study Scientist**
David Schiminovich (Columbia University)
Britney Schmidt (Georgia Institute of Technology)
Karl Stapelfeldt (JPL)
Christopher Stark (STScI)
Jason Tumlinson (STScI), **Simulations Working Group Lead**

### Science and Technology Definition Team – International Non-Voting Members

Martin Barstow (University of Leicester, UK)
Lars Buchhave (DTU Space, National Space Institute, Denmark)
Nicolas Cowan (McGill University, Canada)
José Dias do Nascimento Jr. (Brazilian Federal University, Brazil)
Marc Ferrari (Laboratoire d'Astrophysique de Marseille, France)
Ana Gomez de Castro (Universidad Complutense de Madrid, Spain)
Kevin Heng (University of Bern, Switzerland)
Thomas Henning (Max Planck Institute for Astronomy, Germany)
Michiel Min (Netherlands Institute for Space Research, Netherlands)
Antonella Nota (European Space Agency)
Takahiro Sumi (Osaka University, Japan)

## Science and Technology Definition Team – Ex-Officio Non-Voting Members

Shawn Domagal-Goldman (NASA GSFC), **Deputy Study Scientist**
Mario Perez (NASA HQ), **Program Scientist**
Michael Garcia (NASA HQ), **Deputy Program Scientist**
Susan Neff (NASA GSFC), **COR Program Chief Scientist**
Erin Smith (NASA Ames), **COR Deputy Chief Scientist**

## LUVOIR Study Office & Engineering Team

Julie Crooke (NASA GSFC), **Study Manager**
Matthew Bolcar (NASA GSFC), **Lead Engineer**
Jason Hylan (NASA GSFC), **Deputy Lead Engineer**
Giada Arney (NASA GSFC), **Science Support Analysis Team Lead**
Steve Aloezos (NASA GSFC)
Adrienne Beamer (NASA GSFC)
Carl Blaurock (Night Sky Systems)
Vince Bly (NASA GSFC)
Ginger Bronke (NASA GSFC)
Christine Collins (NASA GSFC)
Knicole Colón (NASA GSFC)
James Corsetti (NASA GSFC)
Don Dichmann (NASA GSFC)
Jean-Etienne Dongmo (NASA GSFC)
Lou Fantano (NASA GSFC)
Thomas Fauchez (NASA GSFC / USRA)
Joseph Generie (NASA GSFC)
Gene Gochar (NASA GSFC)
Qian Gong (NASA GSFC)
Tyler Groff (NASA GSFC)
Kong Ha (NASA GSFC)
Bill Hayden (NASA GSFC)
Andrew Jones (NASA GSFC)
Roser Juanola Parramon (NASA GSFC / USRA)
Ravi Kopparappu (NASA GSFC / University of Maryland – College Park)
Irving Linares (NASA GSFC)
Alice Liu (NASA GSFC)
Eric Lopez (NASA GSFC)
Avi Mandell (NASA GSFC)
Bryan Matonak (NASA GSFC)
Sang Park (Smithsonian Astrophysical Observatory)
Shannon Rodriguez (NASA GSFC)
Lia Sacks (NASA GSFC)
Lisa Smith (NASA MSFC)
Hari Subedi (NASA GSFC, Princeton University)
Steve Tompkins (NASA GSFC)

Geronimo Villanueva (NASA GSFC)
Garrett West (NASA GSFC)
Neil Zimmerman (NASA GSFC)

## POLLUX Team

Coralie Neiner (Observatoire de Meudon, France), **Co-Lead**
Jean-Claude Bouret (Laboratoire d'Astrophysique de Marseille, France), **Co-Lead**
Stéphane Charlot (IAP, France), **Extragalactic Science Working Group Lead**
Jean-Yves Chauffray (LATMOS, France), **Solar System Science Working Group Lead**
Chris Evans (University of Edinburgh, UK), **Stellar Physics Science Working Group Co-Lead**
Luca Fossati (Space Research Institute, Austria), **Exoplanet Science Working Group Lead**
Ana Inès Gomez de Castro (Complutense University of Madrid, Spain), **Stellar Physics Science Working Group Co-Lead**
Cecile Gry (LAM, France), **ISM/IGM Science Working Group Lead**
Pasquier Noterdaeme (IAP, France), **Cosmology Science Working Group Lead**
Eduard Muslimov (LAM, France), **Lead Engineer, Optical Designer**
Arturo Lopez Ariste (IRAP, France), **Polarimeter Designer**
Martin Barstow (University of Leicester, UK), **Detector Working Group Manager**
David Montgomery (UK Astronomy Technology Center, UK), **Mechanical Designer**
Pierre Royer (Catholic University of Leuven, Belgium), **Calibration Unit Manager**
Udo Schuehle (MPS, Germany), **Coatings Working Group Manager**
Louise Lopes (CNES, France), **CNES Manager**
Marc Ferrari (LAM, France), **CNES Representative on LUVOIR STDT**
Dietrich Baade (ESO, Germany)
Richard Desselle (CSL and University of Liège, Belgium)
Boris Gaensicke (University of Warwick, UK)
Miriam Garcia (CAB/CSIC, Madrid, Spain)
Serge Habracken (University of Liège, Belgium)
Jon Lapington (University of Leicester, UK)
Vianney Lebouteiller (AIM/CEA, France)
Maelle Le Gal (LESIA, France)
Frédéric Marin (Strasbourg Observatory, France)
Fabrice Martins (University of Montpelier, France)
Yael Nazé (University of Liège, Belgium)
Hadi Rahmani (GEPI, France)
Hugues Sana (Catholic University of Leuven, Belgium)
Steve Shore (University of Pisa, Italy)
Daphne Stam (TU Delft, The Netherlands)
Luca Teriaca (MPS, Germany)
Jorick Vink (Armagh Observatory, UK)
Huirong Yan (DESY, Germany)

## Community Working Group Members / Additional Contributors

### *Cosmic Origins*

Nate Bastian (Liverpool John Moores University)
Brendan Bowler (University of Texas – Austin)
Charlie Conroy (Harvard University)
Paul Crowther (University of Sheffield)
Selma de Mink (University of Amsterdam)
Ruobing Dong (University of Arizona)
Bruce Elmegreen (IBM)
Steven Finkelstein (University of Texas – Austin)
Brian Fleming (University of Colorado – Boulder)
Michele Fumagalli (Durham University)
Jay Gallagher (University of Wisconsin – Madison)
Melissa Graham (University of California – Berkeley)
Eva Grebel (University of Heidelberg)
Gregory Herczeg (KIAA / University of Peking)
Benne Holwerda (University of Louisville)
Christopher Howk (University of Notre Dame)
Mark Krumholz (Australian National University)
Søren S. Larsen (Radboud University)
Tod Lauer (NOAO)
Janice Lee (IPAC)
Yoshiki Matsuoka (Ehime University)
Stephan McCandliss (Johns Hopkins University)
Stella Offner (University of Texas – Austin)
Ian Roderer (University of Michigan)
Elena Sabbi (STScI)
P. Christian Schneider (University of Hamburg)
Vicky Scowcroft (University of Bath)
Paul Scowen (Arizona State University)
Warren Skidmore (Thirty Meter Telescope)
Linda Smith (ESA / STScI)
Russell Smith (Durham University)
Rachel Somerville (Rutgers University)
Harry Teplitz (IPAC)
Christy Tremonti (University of Wisconsin – Madison)
Kate Whitaker (University of Massachusetts – Amherst)
Gerard Williger (University of Louisville)
Rosemary Wyse (Johns Hopkins University)
Allison Youngblood (NPP / NASA GSFC)

### *Exoplanets*

Eric Agol (University of Washington)
Daniel Apai (University of Arizona)

Natalie Batalha (NASA Ames)
Ruslan Belikov (NASA Ames)
Luca Fossati (Austrian Academy of Sciences)
Yuka Fujii (Tokyo Institute of Technology)
Claire Marie Guimond (McGill University)
Eric Hébrard (University of Exeter)
Tiffany Jansen (Columbia University)
Eliza Kempton (Grinnell College / University of Maryland)
Brianna Lacy (Princeton University)
Andrew Lincowski (University of Washington)
Jacob Lustig-Yeager (University of Washington)
Timothy Lyons (University of California Riverside)
Gijs Mulders (LPL)
Stephanie Olson (University of California Riverside)
Ramses Ramirez (Tokyo Institute of Technology)
Christopher Reinhard (Georgia Tech)
Tyler Robinson (Northern Arizona University)
Edward Schwieterman (University of California – Riverside)
Stoney Simons (National Institutes of Health)
Dillon Teal (NASA GSFC)
Guadalupe Tovar (University of Washington)
Sara Walker (Arizona State University)

## Solar System

Alvaro Alvarez-Candal (Observatorio Nacional, Brazil)
Gerbs Bauer (JPL)
Dennis Bodewits (University of Maryland – College Park)
Valeria Cottini (NASA GSFC)
Lori Glaze (NASA GSFC)
Jeff Morgenthaler (Planetary Science Institute)
Alex Parker (Southwest Research Institute)
Noah Petro (NASA GSFC)
Noemi Pinilla-Alonso (University of Central Florida)
Silvia Protopapa (University of Maryland – College Park)
Andrew Rivkin (Johns Hopkins University – Applied Physics Lab)

## Technology and Design

Lynn Allen (Harris Corporation)
David Allred (Brigham Young University)
Jon Arenberg (Northrop Grumman)
Kunjithapatham Balasubramanian (JPL)
Scott Basinger (JPL)
Ruslan Belikov (NASA Ames)
Ray Bell (Lockheed Martin)
Paul Bierdan (Boston Micromachines)

Ron Broccolo (Harris Corporation)
Eric Cady (JPL)
Kerri Cahoy (Massachusetts Institute of Technology)
Jeff Cavaco (Northrop Grumman)
Alberto Conti (Northrop Grumman)
Jim Contreras (Ball Aerospace)
Laura Coyle (Ball Aerospace)
Brendan Crill (JPL)
Javier Del Hoyo (NASA GSFC)
Larry Dewell (Lockheed Martin)
Jeanette Domber (Ball Aerospace)
Matthew East (Harris Corporation)
Mike Eisenhower (Smithsonian Astrophysical Observatory)
Michael Feinberg (Boston Micromachines)
Greg Fellers (Lockheed Martin)
James R. Fienup (University of Rochester)
Don Figer (Rochester Institute of Technology)
Brian Fleming (University of Colorado – Boulder)
Kevin Fogarty (STScI)
Pascal Hallibert (ESA)
Erika Hamden (California Institute of Technology)
Alex Harwit (Ball Aerospace)
Michael Helmbrecht (Iris AO)
John Hennessy (JPL)
Peter Hill (Zygo / Ametek)
Joe Ho (Ball Aerospace)
Sona Hosseini (JPL)
Tony Hull (University of New Mexico)
Jeffrey Jewell (JPL)
Jeremy Kasdin (Princeton University)
Scott Knight (Ball Aerospace)
Mary Li (NASA GSFC)
Paul Lightsey (Ball Aerospace)
Chris Lindensmith (JPL)
Sarah Lipscy (Ball Aerospace)
John Lou (JPL)
Makenzie Lystrup (Ball Aerospace)
Gary Matthews (ATA Aerospace)
Stephan McCandliss (Johns Hopkins University)
Ted Mooney (Harris Corporation)
Dustin Moore (JPL)
Didier Morancais (Airbus)
Mamadou N'Diaye (Université Côte d'Azur)
Shouleh Nikzad (JPL)
Joel Nissen (JPL)

Alison Nordt (Lockheed Martin)
Jim Oschmann (Ball Aerospace)
Sang Park (Smithsonian Astrophysical Observatory)
Enrico Pinna (Istituto Nazionale di Astrofisica)
Bill Purcell (Ball Aerospace)
Manuel Quijada (NASA GSFC)
Bernie Rauscher (NASA GSFC)
Norman Rioux (NASA GSFC)
Michael Rodgers (Synopsys / JPL)
Garreth Ruane (Caltech)
Derek Sabatke (Ball Aerospace)
John Sadleir (NASA GSFC)
Babak Saif (NASA GSFC)
Eric Schindhelm (Ball Aerospace)
Gene Serabyn (JPL)
Stuart Shaklan (JPL)
Chris Shelton (JPL)
Fang Shi (JPL)
Evgenya Shkolnik (Arizona State University)
Nick Siegler (JPL)
Remi Soummer (STScI)
Joe Sullivan (Ball Aerospace)
Phil Stahl (NASA MSFC)
Kathryn St Laurent (STScI)
Kiarash Tajdaran (Lockheed Martin)
John Trauger (JPL)
Steve Turley (Brigham Young University)
John Vayda (Northrop Grumman)
Michael Werner (JPL)
Scott Will (University of Rochester)
Marco Xompero (National Institute of Astrophysics, Italy)
David Yanatsis (Harris Corporation)
John Ziemer (JPL)

## Acknowledgements

The STDT would like to thank Daniel Apai (University of Arizona), Steven Finkelstein (University of Texas – Austin), Luca Fossati (Austrian Academy of Sciences), and Tyler Robinson (Northern Arizona University) for their major contributions to the science cases that appear in this report. We also thank Brian Fleming (University of Colorado – Boulder), John MacKenty (STScI), and Stephan McCandliss (Johns Hopkins University) for their extensive participation in instrument design. We are very grateful to our industrial cooperative agreement partners: Northrop Grumman, Lockheed Martin, Ball Aerospace, and Harris Corporation. We express sincere appreciation to the LUVOIR Senior Advisors group, Natalie Batalha (NASA Ames), Julianne Dalcanton (University of Washington), Alan Dressler (Carnegie Observatories), Jay Gallagher (University of Wisconsin – Madison), Jonathan Fortney (University of California – Santa Cruz), and Garth Illingworth (University of California – Santa Cruz), for their valuable guidance and feedback. Finally, we thank Pat Tyler, Jay Friedlander, T. Britt Griswold (NASA GSFC), Chad Smith, and Graham Kanarek (STScI) for their hard work laying out this document, creating graphics, and writing simulation tools.

# Contents

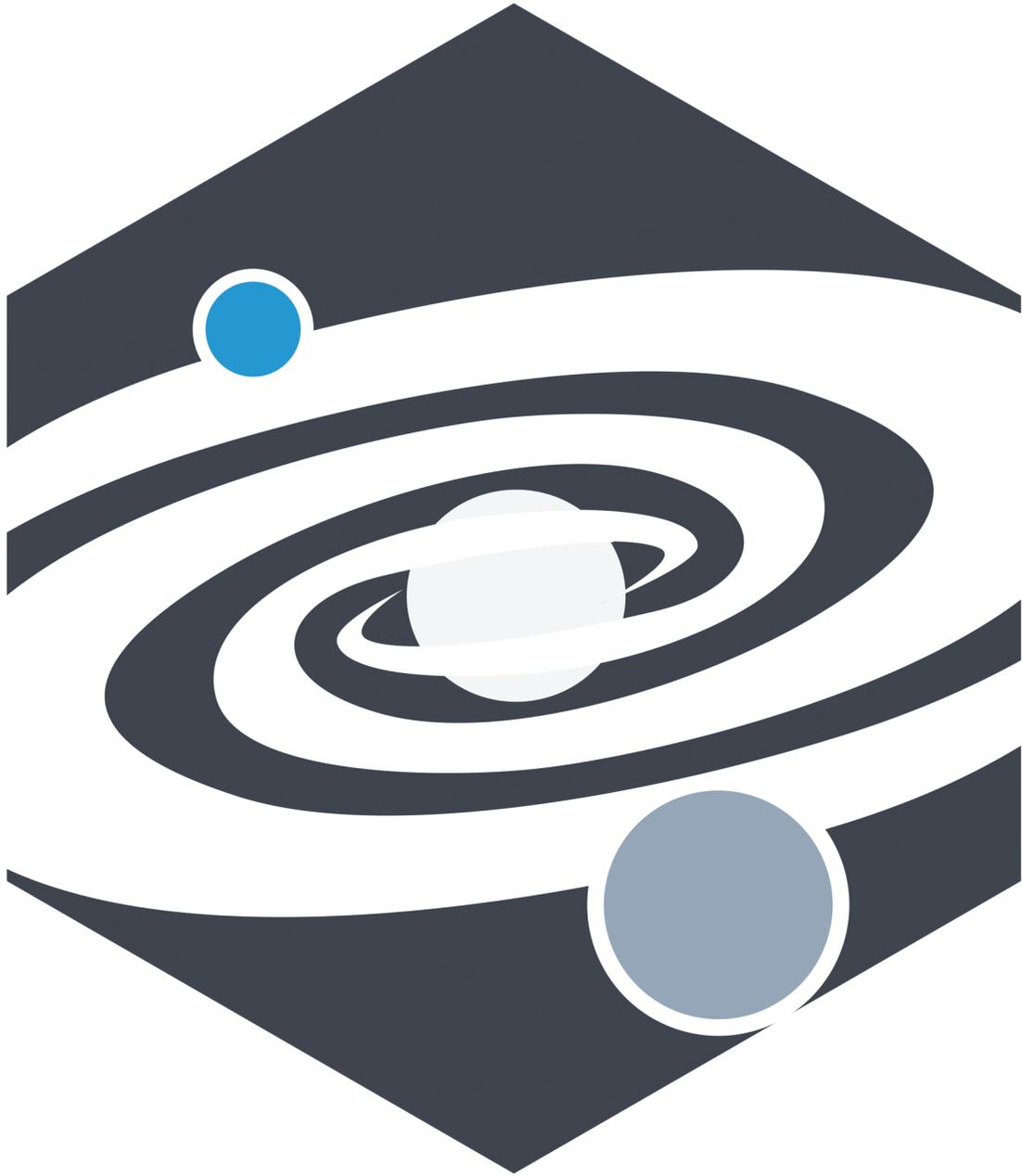

Executive Summary

# 1   LUVOIR: Telling the Story of Life in the Universe

*Interim Report Executive Summary*

Over four hundred years have passed since Galileo's first observations of the heavens fundamentally transformed humanity's understanding of the cosmos. We have learned that we live not in a static eternal universe, but in an expanding one, comprised of matter, radiation, dark matter, and dark energy, with the latter two still poorly understood. We exist not in an island universe, but instead within one of hundreds of billions of galaxies, comprised of tens of trillions of stars. We have seen that our solar system is not alone, but instead one of unknown billions of planetary systems in the Milky Way. These discoveries have emerged from a history of advances in astronomical theory, observation, and technology culminating in telescopes on the ground and in space covering every observable wavelength of light, new frontiers in gravitational wave and neutrino observations, and entire universes residing in simulations on supercomputers.

Despite all these triumphs, many basic, fundamental, and essential questions remain: *Is there life elsewhere? Is our world unique? How do stars form? What are the building blocks of structure? How did our galaxy, solar system, and Earth arise and evolve?* These questions demand an observatory beyond any in existence or in development to answer. They require a large aperture to capture the very faintest objects across cosmic time. They require spatial resolution high enough to separate planets from their host stars and the ability to block out the latter's blinding light. They require access to a range of wavelengths broad enough to read the atomic fingerprints of matter across all temperatures and densities. In this document, we describe such an observatory: The Large Ultraviolet / Optical /

Infrared Surveyor (LUVOIR) mission concept. LUVOIR represents an ambitious yet feasible observatory designed to answer not only the questions astronomy presents today, but also the as of yet unknown questions of tomorrow.

LUVOIR's main features are:

- A large (15-m LUVOIR-A, 8-m LUVOIR-B), segmented aperture designed to fit in a single launch vehicle.
- Scalable, serviceable architecture design for an observatory at Earth-Sun L2.
- A total wavelength range of 100 nm–2.5 μm.
- ECLIPS: An ultra-high contrast corona-graph with an imaging camera and integral field spectrograph spanning 200–2000 nm, capable of directly observing a wide range of exoplanets and obtaining spectra of their atmospheres.
- HDI: A near-UV to near-IR imager covering 200–2500 nm, Nyquist sampled at 400 nm and diffraction limited at 500 nm, with high precision astrometry capability.
- LUMOS: A far-UV imager and far-UV + near-UV multi-resolution, multi-object spectrograph covering 100 nm–400 nm, capable of simultaneous observations of up to hundreds of sources.
- POLLUX: A UV spectropolarimeter covering 100–400 nm with high-resolution point-source spectroscopic capability. A consortium of European institutions, with leadership from the French Space Agency, is contributing this instrument concept study.

LUVOIR continues the vision of the Great Observatories as a community-driven facility with the power and capability to answer scientists' questions across the full NASA

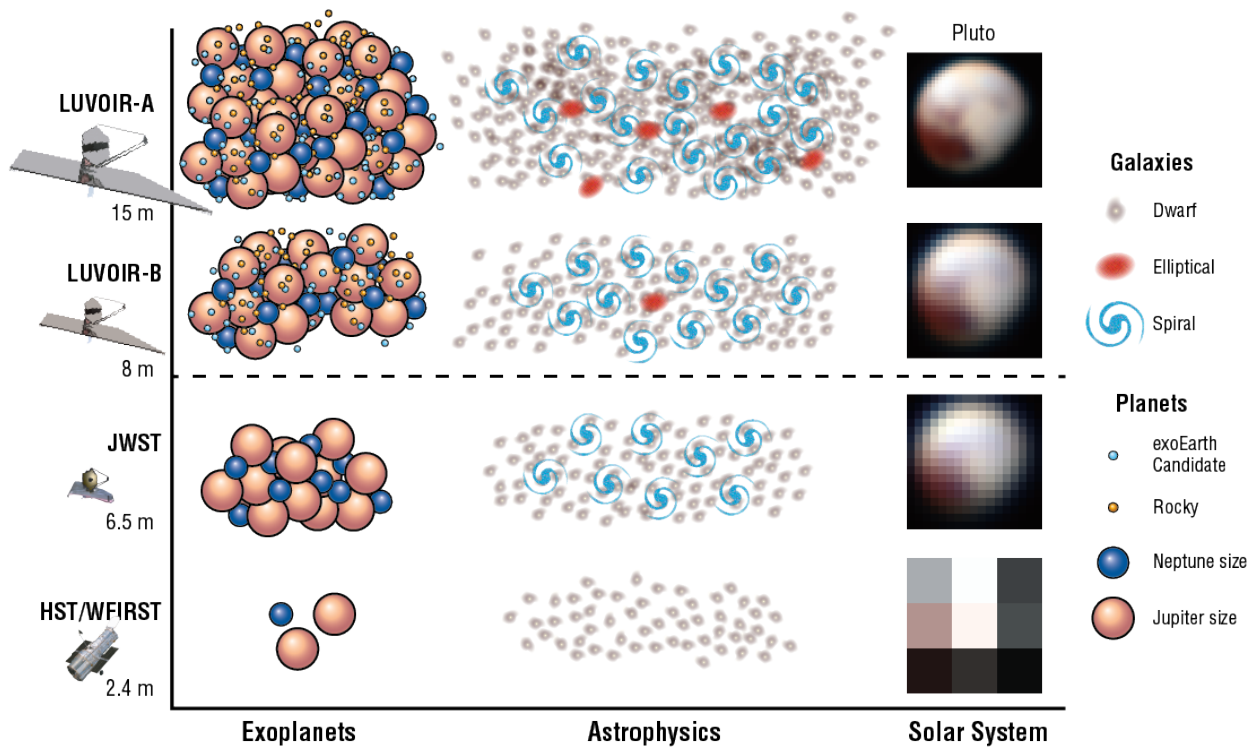

**Figure 1.1.** *Illustration of LUVOIR's range of science and capability as compared to other facilities. LUVOIR dramatically increases the sample size and diversity of exoplanets that can be studied, pushing to dozens of Earth-like planets around Sun-like stars. LUVOIR's sensitivity and spatial resolution opens access to the ultra-faint and ultra-distant regime, enabling observations of individual stars in the full variety of galaxies. Those capabilities also create new opportunities in remote sensing of solar system bodies. Credit: NASA / New Horizons / J. Friedlander & T. B. Griswold (NASA GSFC)*

portfolio of Cosmic Origins, Physics of the Cosmos, Exoplanet, and Solar System Exploration science. These science areas include:

- Exploring the full diversity of exoplanets.
- Discovering and characterizing exoplanets in the habitable zones of Sun-like stars across a range of ages and searching for biosignatures in their atmospheres, in a survey large enough to provide evidence for (or against) the presence of habitable planets and life.
- Remote sensing of the planets, moons, and minor bodies of the solar system.
- Exploring the building blocks of galaxies both in the local universe and at their

emergence in the distant past, and elucidating the nature of dark matter.
- Understanding how galaxies form and evolve from active to passive, both by studying their stars and their gaseous fuel across all temperatures and phases.
- Following the history of stars in the local volume out to tens of megaparsecs to understand how they form and how they depend on their environment.
- Observing the birth of planets and understanding how the diversity of planetary systems arises.

A transformative facility like LUVOIR with its large aperture and need for ultra-high contrast observations will require maturation of some key technologies, including coronagraph design, wavefront sensing and control, and ultra-stable structures. With input from NASA, the aerospace industry, and commercial partners, this study has identified the technological gaps and has defined a path to realize LUVOIR. The observatory is being designed to leverage:

- Segmented mirror technologies from the ground and the James Webb Space Telescope (JWST)
- Coronagraph development from the ground and the Wide Field Infrared Survey Telescope (WFIRST)

- Serviceable design heritage from the Hubble Space Telescope (HST) and other missions
- Detector and optics technologies from multiple missions
- Industry and commercial innovations in vibration isolation and launch vehicles

LUVOIR's power, flexibility, and serviceability ensure that it can perform not only the "signature" science examples in this report, but also the unknown science of the future. It is an observatory to serve the astronomical community not just for the 2030s, but also for many decades to follow.

Chapter 2

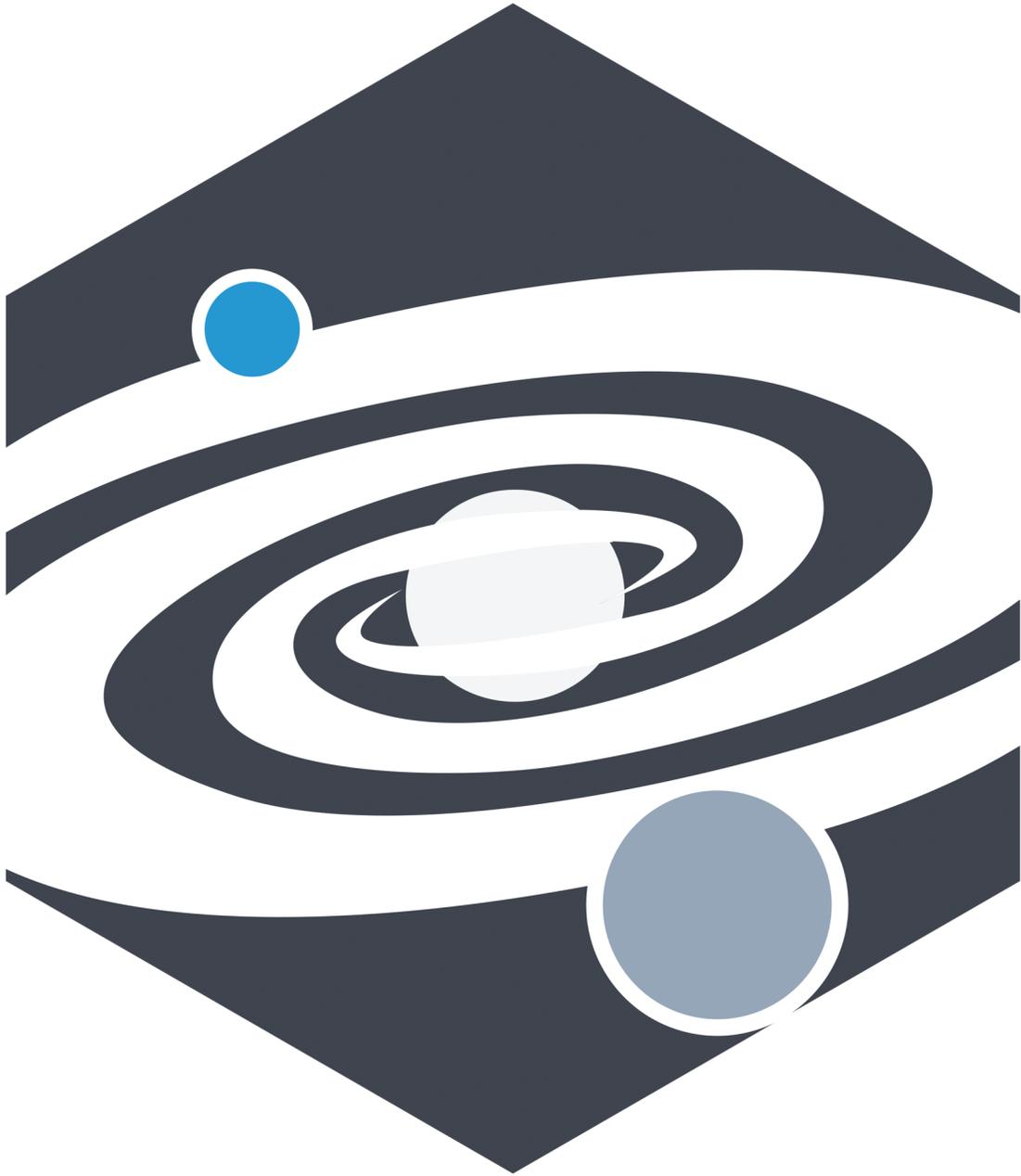

Introduction

## 2   Introduction

Humanity is defined by the drive to know about our place within the world around us. The very name homo sapiens captures what we prize in ourselves, and our never-ending quest for wisdom leads us ever further outward. The value of that quest for the betterment of our species is immeasurable. Ages-old questions and investigations earned us the revelations that the stars are Suns swirling in a vast galaxy, itself one of a myriad of islands in a boundless cosmos. Now we have crossed another threshold of discovery: there are planets around other stars. Tracing a path from the dawn of the universe to life-bearing worlds is within our grasp. The further revelations we will find on that path have long belonged to the realm of speculation—now we begin to find out if our imaginary musings are real.

This monumental objective demands powerful and flexible new tools, and different ways of combining our scientific skills. We have found that the range and diversity of worlds is far greater than we imagined, yet we see planetary systems reminiscent of the solar system as well. At the moment, the vast majority of known exoplanets are "small black shadows" indirectly detected through their effect on their host stars. Our knowledge of exoplanet properties is largely limited to orbits, masses, and sizes. We have just begun to measure the atmospheric constituents of gas giant exoplanets; such studies will greatly expand in the coming years.

The next frontier is to extend these capabilities to rocky planets and find the "pale blue dots" in the solar neighborhood (**Figure 2.1**). We can determine whether habitable, Earth-like conditions are common or rare on nearby worlds and probe them for signs of life. Concurrently, we will nurture a new discipline—comparative exoplanetology—by studying a huge range of exoplanets, thereby gaining invaluable information for placing the solar system in the broader context of planetary systems. A vital part of establishing that context is deeper understanding of the bodies in our own system.

Our drive to know goes beyond asking the question "what exists?" to "why does it exist?" and pushes us to understand the origins of all we see around us. The boundary of what we can see now stretches all the way to the dawn of the universe, but like our first steps in the study of new worlds, our vision lacks both completeness and context. We seek to understand the environments and processes that gave rise to a life-supporting cosmos, from the formation of the earliest structures and the rise of the elements essential for life

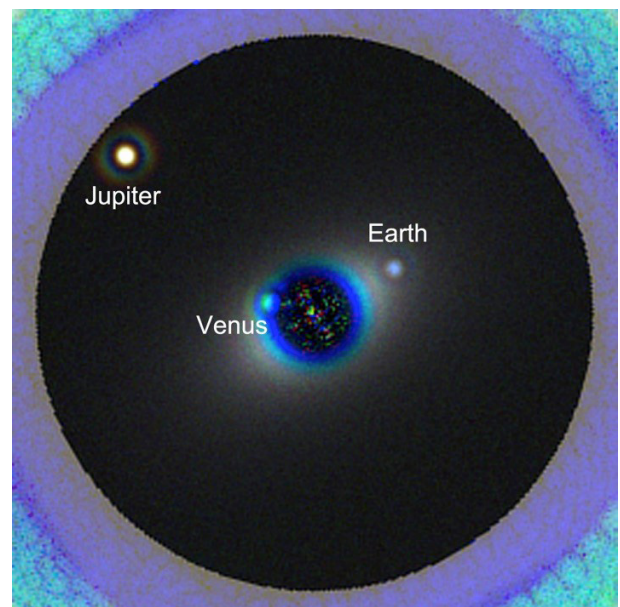

**Figure 2.1.** *Imaging Earth 2.0. Simulation of the inner solar system viewed at visible wavelengths from a distance of 13 parsec with LUVOIR. The enormous glare from the central star has been suppressed with a coronagraph so the faint planets can be seen. The atmosphere of each planet can be probed with spectra to reveal its composition. Credit: L. Pueyo, M. N'Diaye (STScI)/A. Roberge (NASA GSFC)*

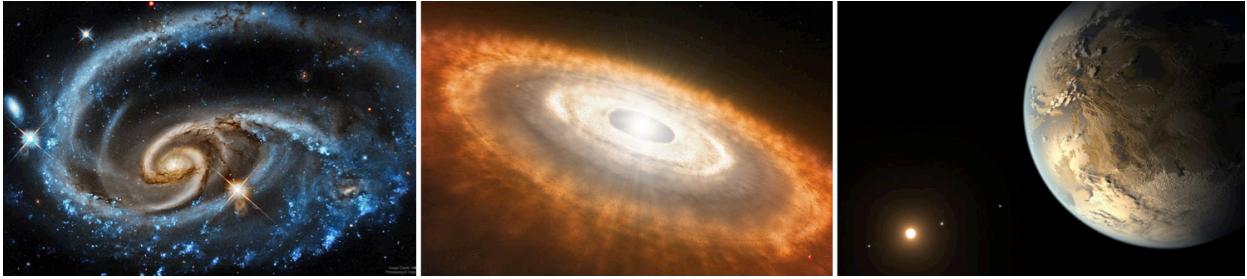

**Figure 2.2.** *From galaxy evolution to star and planet formation to habitable worlds. Credits: NASA/ESA/Hubble; ESO/L. Calcada; NASA Ames/SETI Institute/JPL-Caltech*

as we know it, to the assembly and evolution of galaxies, to the detailed mechanisms of star and planet formation (**Figure 2.2**). The boundaries of physics will be tested and stretched while exploring the birth and evolution of the cosmos.

In subsequent chapters of this report, we will delve into far greater detail on all of these scientific questions and investigations, which define the tool we need to leap forward. That is the Large Ultraviolet/Optical/Infrared Surveyor (LUVOIR), a concept for an ambitious future space telescope in the tradition of NASA's Great Observatories. LUVOIR had its genesis in the 2013 NASA Astrophysics Roadmap "Enduring Quests, Daring Visions." The current LUVOIR study builds upon telescope concepts going back decades (e.g., ATLAST, the High Definition Space Telescope) but goes far beyond them in both scientific and technical detail. LUVOIR will provide the powerful observing capabilities required to execute the revolutionary investigations we will discuss. Furthermore, LUVOIR's power, flexibility, and longevity will allow it to answer the as-yet-unknown questions of the 2030s and beyond.

## 2.1   The worlds around us

We have long speculated about the existence of planets outside the solar system. The names of imagined worlds are familiar to us: Tatooine, Vulcan, Arrakis. Our own planetary system is largely bimodal in nature, with small rocky planets close to the Sun and massive gas giants in the cold outer reaches. Scientists naturally expected that other systems would be similar—if they existed at all. In the mid-1990s, decades of persistence and technological innovation were finally rewarded with the discovery of the first exoplanets. We have been both shocked and delighted to find that exoplanets are both common and diverse—reality far surpassed our scientific predictions. We can now look into the sky and realize that most of the stars we see harbor worlds. Soon, we will more fully reveal the character of giant planets, which will further stretch the bounds of our theories. The next steps in this quest are even more amazing.

### 2.1.1   The census of Earth-like exoplanets

Indirect planet discovery techniques, like the transit method used by NASA's Kepler Mission, have shown that small rocky planets are not rare. Kepler indicates that about 24% of sun-like stars have roughly Earth-size planets in their habitable zones (Kopparapu et al. 2018, and references therein). These zones span the range of distances from the stars where we think rocky planets can have liquid water—an essential material for all life on Earth—on their surfaces. This revelation drives us to proceed to the next step: build the capabilities to reveal the character of

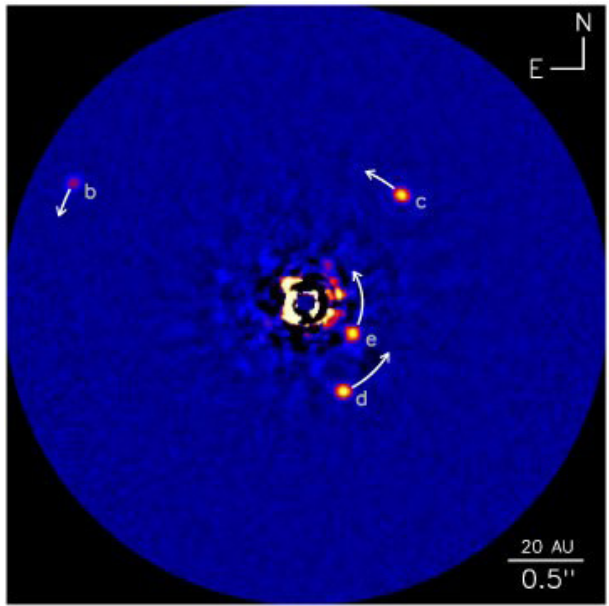

**Figure 2.3.** *Direct image of four young, warm super-Jupiters orbiting the star HR8799. The bright light from the central star has been suppressed with a coronagraph. Credit: NRC-HIA/C. Marois/Keck Observatory.*

The first direct images of exoplanets have been obtained with ground-based observatories, revealing the relatively faint planets by suppressing the bright light from the central stars using coronagraph instruments (**Figure 2.3**). High-performance coronagraphs on future 30-meter-class ground-based telescopes (the Extremely Large Telescopes, or ELTs) may be able to study the atmospheres of rocky planets orbiting in the habitable zones of the nearest cool red dwarf stars (e.g., Wang et al. 2017). However, direct observations of rocky exoplanets around Sun-like stars demand the ultra-high contrast possible only with a space-based telescope coupled to a high-performance starlight suppression system (**Figure 2.1**).

Furthermore, measuring the frequency of Earth-like conditions requires observations of a large number of candidate exoplanets (dozens; **Figure 2.4**). To find and study that many candidates, an even larger number of stars must be surveyed (hundreds). This goal of quantitatively measuring the frequency of habitable conditions on rocky worlds is a strong driver on the required space telescope size. By executing a systematic study of a large number of rocky exoplanets, we can guarantee that whatever we find will be of scientific value. In the absence of positive detections of water vapor, we would know that global-scale surface oceans are relatively rare on rocky worlds in the habitable zone. This null result would still transform our understanding of habitability as a planetary process, and further confirm the special nature of our home. On the other hand, if we find that Earth-like conditions are common on rocky worlds near our Sun, then a stunning vista of hospitable new worlds will be unveiled.

rocky exoplanets and measure the fraction that are truly Earth-like. We need to probe rocky planet atmospheres and determine their thermal/chemical states by measuring their molecular abundances, including water vapor and other greenhouse gases.

Astronomers and planetary scientists have calculated that to achieve this goal for Earth-like planets around Sun-like stars, we must collect and analyze light from the planets themselves, i.e., obtain direct images and spectra (e.g., Kaltenneger & Traub 2009, Snellen et al. 2015). But doing this for habitable planet candidates around Sun-like stars is made extremely challenging by the fact that the Earth is 10 billion times fainter than the Sun and orbits close to its host star. When viewed from an interstellar distance of 10 parsecs, the apparent separation of the Earth from the Sun is only about the width of a human hair at the distance of two football fields.

## LUVOIR Frequently Asked Questions

**How big is LUVOIR?**
We are developing 2 architectures, LUVOIR-A and LUVOIR-B. LUVOIR-A has a primary mirror with a diameter of 15 meters. This size corresponds to the largest observatory that can be launched in a SLS Block 2 vehicle. LUVOIR-B has a primary mirror with a diameter of 8 meters, which is the largest observatory that can be launched in a 5-meter fairing similar to those in use today). [**Chapter 8**]

**What if the SLS Block 2 launch vehicle is not available in the 2030s?**
LUVOIR's segmented design facilitates scalability, as will be demonstrated with LUVOIR-B. LUVOIR can take advantage of future opportunities in heavy lift commercial launch vehicles with large fairings (e.g., Blue Origin's New Glenn rocket and SpaceX's BFR). [**Appendix D**]

**How long will LUVOIR last?**
The prime mission is 5 years, which is the standard prime mission duration for large space missions, with 10 years of on-board consumables. Furthermore, LUVOIR is designed to be serviceable and upgradeable, with a lifetime goal of 25 years for non-serviceable components. [**Chapter 8**]

**What are LUVOIR's key science goals?**
This report describes a number of "signature science" programs that demand LUVOIR. Additionally, we are actively collecting community input on science with LUVOIR and have developed a number of public on-line tools to help create science programs (https://asd.gsfc.nasa.gov/luvoir/tools/). Finally, LUVOIR is being designed to be flexible and powerful enough to enable the as-of-yet unknown science of the 2030s and beyond. [**Chapters 3–7**, **Appendix A**]

**How much of LUVOIR's time will be for community observers?**
LUVOIR is envisioned as a facility in the tradition of NASA's Great Observatories (Hubble, Compton, Chandra, Spitzer), with Guest Observer-driven operations. [**Chapter 12**]

**Why is direct spectroscopy of habitable exoplanet candidates a key goal for LUVOIR?**
Both scientists and the public are strongly motivated to find habitable planets around a range of stars and see if any show signs of life as we know it. Achieving this goal for the exoplanet systems most like the solar system—ones with Earth-like planets around Sun-like stars—requires direct spectroscopy of light coming from the planets after suppression of starlight with a coronagraph or starshade. [**Chapter 3**]

**Why is LUVOIR focused on exoplanet spectroscopy at near-ultraviolet/visible/near-infrared wavelengths?**
This wavelength range contains key molecular markers of habitable surface conditions (e.g., water vapor, $CO_2$), atmospheric biosignatures (e.g., $O_2$, ozone), and major atmospheric constituents for a wide range of exoplanets (e.g., methane). [**Chapters 3** & **4**]

**Can high-performance coronagraphy be done with an obscured, segmented telescope?**
While most coronagraphs perform better with unobscured telescopes, recent advances via ground-based instruments and the WFIRST coronagraph technology demonstration program have shown that high-performance coronagraphs can be designed for obscured apertures. Other studies have shown that telescope segmentation has only a minor impact on coronagraph performance. [**Chapters 9** & **11**]

## LUVOIR Frequently Asked Questions Continued

**Can you do high-performance coronagraphy with a UV-capable telescope?**
Yes. Recent lab studies have shown that UV-compatible coated aluminum mirrors provide better coronagraph performance over broad wavelength ranges than UV-incompatible silver mirrors. [**Appendix D**]

**Why is the telescope warm?**
High-performance coronagraphy demands extreme stability of the wavefront entering the instrument, which is easier to achieve by actively stabilizing the telescope temperature. Maintaining a temperature near or above the freezing point of water also avoids risk of optics contamination that would harm the telescope's ultraviolet sensitivity. [**Chapter 8**, **Appendix D**]

**Why make LUVOIR capable of observing so far into the ultraviolet (< 1150 Å)?**
To understand the history of atoms in the universe, they must be observed over a wide range of physical conditions. Many key atomic and molecular transitions, spanning the full range of temperatures and densities, lie at wavelengths < 1150 Å. [**Chapters 5–7**]

**What unique capabilities will LUVOIR offer compared to ground-based observatories in the 2030s?**
LUVOIR will a) provide continuous wavelength coverage from the far-UV to the near-IR, b) achieve the $10^{-10}$ contrast levels required for direct spectroscopy of Earth-like planets around Sun-like stars, c) provide sensitive observations of faint objects below the sky background, d) provide diffraction limited imaging resolution over wide fields of view. [**Chapters 3–7**]

**Why is there no high-resolution visible/near-IR spectrograph in the first-generation LUVOIR instruments?**
The LUVOIR team developed a preliminary concept for such an instrument (called ONIRS). Given a limited number of instrument bays (four on LUVOIR-A) and some uncertainty about the performance of similar instruments planned for the future ground-based Extremely Large Telescopes (ELTs), ONIRS is currently considered an "alternate or second-generation instrument." We are also studying extending the bandpass of a first-generation UV spectrograph to cover the visible. [**Appendix E**, **Chapter 9**]

**What is LUVOIR's primary technological challenge?**
Achieving the extreme wavefront stability needed for direct observations of Earth-like exoplanets using a large telescope. [**Chapters 8**, **9**, & **11**]

**How much will LUVOIR cost?**
Determination of an accurate cost for LUVOIR is a critical desired result of the study. The LUVOIR team will begin internal cost and risk estimates in mid 2018. Furthermore, NASA will contract for independent cost and risk analysis, to take place in early 2019. The LUVOIR study aims to achieve greater cost fidelity though increased design and technological detail, since cost estimates based on single-parameter scaling relations have been shown to be inadequate for large space missions. A technology development plan, with schedule and cost, will also be presented in the Final Report. Commercial and industry partners will also contribute cost, technology development, and risk analyses. The results of all these analyses will appear in the LUVOIR Final Report.

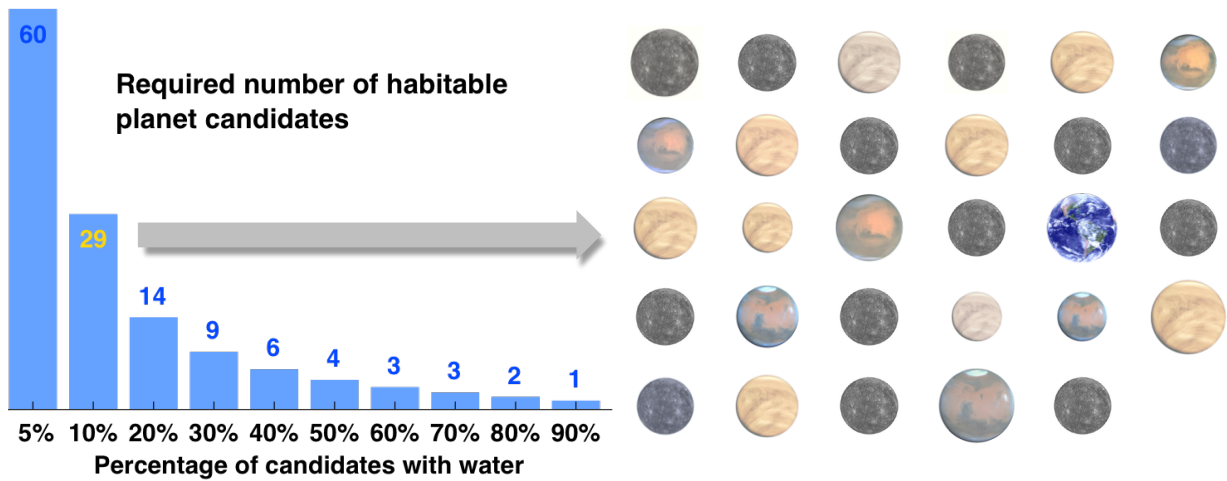

**Figure 2.4.** *How many rocky habitable zone exoplanets need to be observed? The bar chart shows the number of candidates needed to discover one potentially habitable planet (at the 95% confidence level), for different values of the percentage of rocky habitable zone planets that have atmospheric water vapor. If the percentage is 10%, then 29 candidates need to be observed, as illustrated in the planet montage on the right. To put it another way, if 29 candidates are observed and no water vapor is detected, then we learn that < 10% of rocky habitable zone planets are actually Earth-like. Credit: C. Stark (STScI)/A. Roberge (NASA GSFC)*

## 2.1.2   The search for life

The tool to achieve the exoEarth census can also probe the atmospheres of those planets for biosignature gases and find the first signs of life outside our home planet. The spectrum of the Earth contains features arising from gases of biological origin, like $O_2$ and methane, in addition to strong features from water vapor and greenhouse gases (**Figure 2.5**). The Earth is the only planet we know of teeming with surface life that strongly affects the planet's atmosphere. Thus, there are good reasons to focus serious searches for life outside the solar system on exoplanets that are as much like the Earth as possible. By targeting exoplanets with sizes and orbits similar to Earth's and host stars similar to the Sun, we hope to increase our chances of finding—and recognizing—extraterrestrial life.

However, it is important to understand that no one molecule (or pair of molecules) is automatically a biosignature in every context. The key is to calculate the abundances of molecules in the whole atmosphere, and then compare those abundances to the production and destruction rates predicted by physics and chemistry. To calculate the production/destruction rates, a good deal of knowledge about the atmosphere's context will be needed, including the planet's orbit and the characteristics of the host star. If the abundance of a molecule is too high, then we know the production rate is underestimated and an additional biological source is indicated. In essence, we first will try to explain the character of an atmosphere with physics and chemistry. If we fail, then we succeed at finding signs of life.

Therefore, confident life-detection has at least three fundamental requirements: 1) assess a wide range of atmospheric molecules, 2) actually measure their abundances with some precision, and 3) understand the planet's context within its system. The first requirement demands direct spectra with broad wavelength coverage from the near-UV to the near-IR. The LUVOIR telescope and starlight suppression system can span these wavelengths of light and access the important constituents of planetary atmospheres, including water, carbon dioxide, molecular oxygen ($O_2$), ozone, and methane. The second requirement demands good quality spectra to measure abundances, which drives the required telescope and instrument sensitivity. Additional secondary signs of life, such as seasonal variations in a planet's spectral features, will increase confidence in any life-detection claim. The third requirement demands collection of a range of supporting information, including far-UV spectra of stellar activity-driven emission from the host stars. LUVOIR's great sensitivity and flexibility will enable the variety of observations needed to properly place a planet in context.

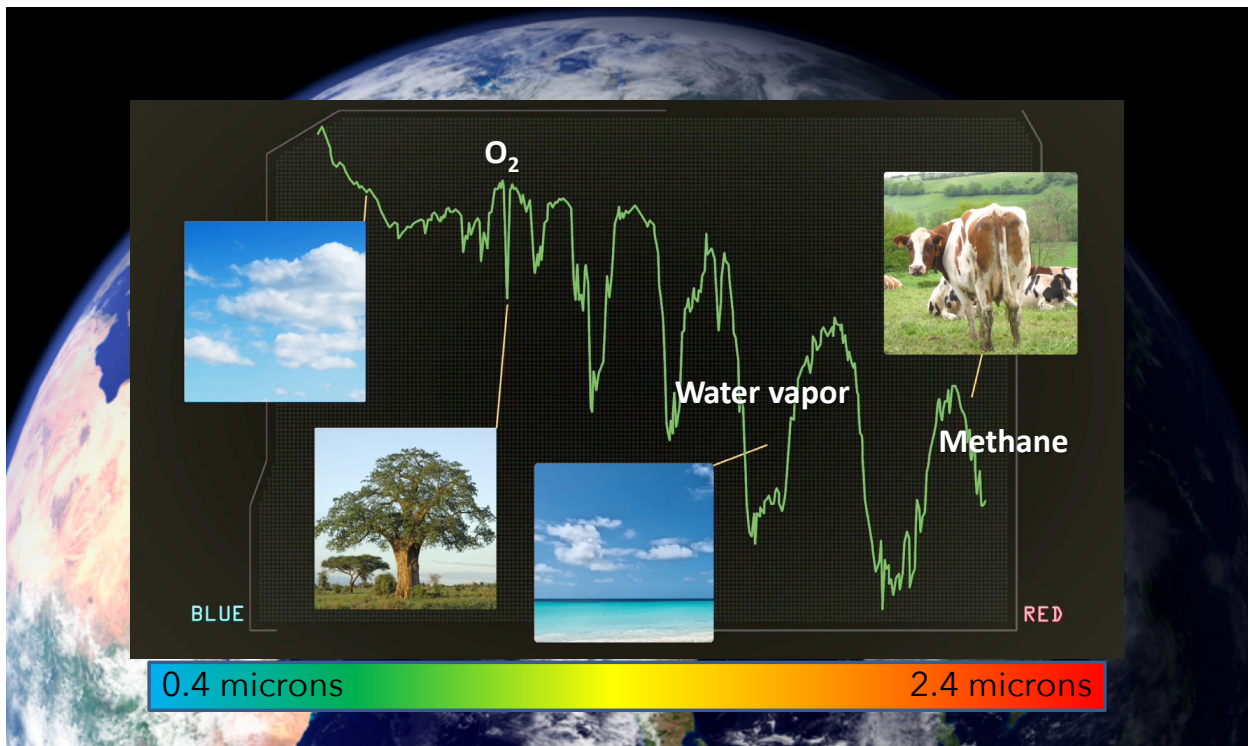

**Figure 2.5.** *The character of a living world. This spectrum of visible and near-infrared light reflected from the Earth reveals the composition of its atmosphere, including gases arising from life. The steep rise in brightness on the left end is the sign of our blue sky. Deep absorption features of water vapor are ubiquitous. A sharp feature comes from molecular oxygen, produced by photosynthesis. And a portion of a broad methane feature is visible on the right. Methane in Earth's atmosphere comes from bacteria in the guts of livestock and in swamps. Credit: NASA GSFC*

The dozens of rocky worlds LUVOIR will explore will inevitably have a range of sizes, orbits, and chemical compositions, and orbit various stars with different ages. Earth has harbored life for most of its history, but in the earlier stages, the atmosphere was very different from that of the modern Earth (e.g., low $O_2$ abundances). Recognizing such a planet as inhabited also demands the careful assessment of atmospheric abundances and planetary context described above. If LUVOIR confirms the habitability of rocky exoplanets, or finds signs of life, we will learn how those traits respond to different planet and star characteristics. This will turn the habitable zone from a theoretical concept based upon a single planet to one that is empirically derived from multiple worlds, and usher in a new era of planetary science: one of comparative astrobiology.

The search for life also takes place closer to home. Over the last decades, we have discovered that several moons of the outer solar system have liquid water beneath their icy surfaces. These sub-surface oceans must be heated from below, which may also provide the energy needed for life. The geysers recently observed from Enceladus and Europa (**Figure 2.6**) may allow glimpses into their deep oceans. LUVOIR has an important role to play by determining the currently unknown strength and frequency of this phenomenon, through high resolution monitoring of icy moons. Such information will be a valuable partner for future spacecraft visiting these other ocean worlds to search for signs of life.

### 2.1.3   The solar system's place in the planet family

When we look at the solar system, it appears delicately balanced to produce a living world, with our largest planet—Jupiter—playing a dominant role. However, our exoplanet discoveries show that the planet formation process is robust and leads to a wide range of outcomes. How do we understand the solar system in this context? By exploring the character of many kinds of exoplanets, including types that do not exist in the solar system, like warm Jupiters, sub-Neptunes, and super-Earths (**Figure 2.7**). Only then can we truly know what is typical about our own system and what is unusual. The capabilities that will allow LUVOIR to investigate dozens of potentially habitable worlds will also allow

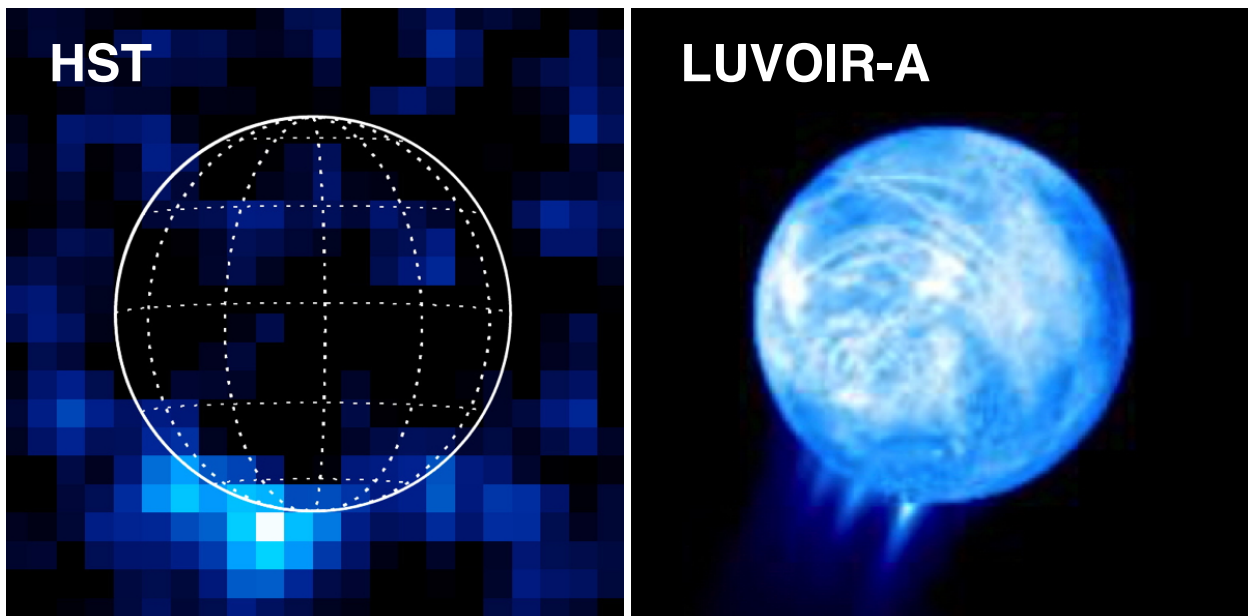

**Figure 2.6.** *Spectroscopic imaging of Europa and its water jets. The left panel shows an aurora on Europa observed with HST (Roth et al. 2014). This UV hydrogen emission (Lyman-alpha) comes from dissociation of water vapor in jets emanating from the surface. The right panel shows a simulation of how hydrogen emission from Europa would look to a 15-m UV telescope. The moon's surface is bright due to reflected solar Ly-alpha emission. With LUVOIR, one could monitor the ocean worlds of the outer solar system for such activity and image the individual jets. Credit: G. Ballester (LPL)*

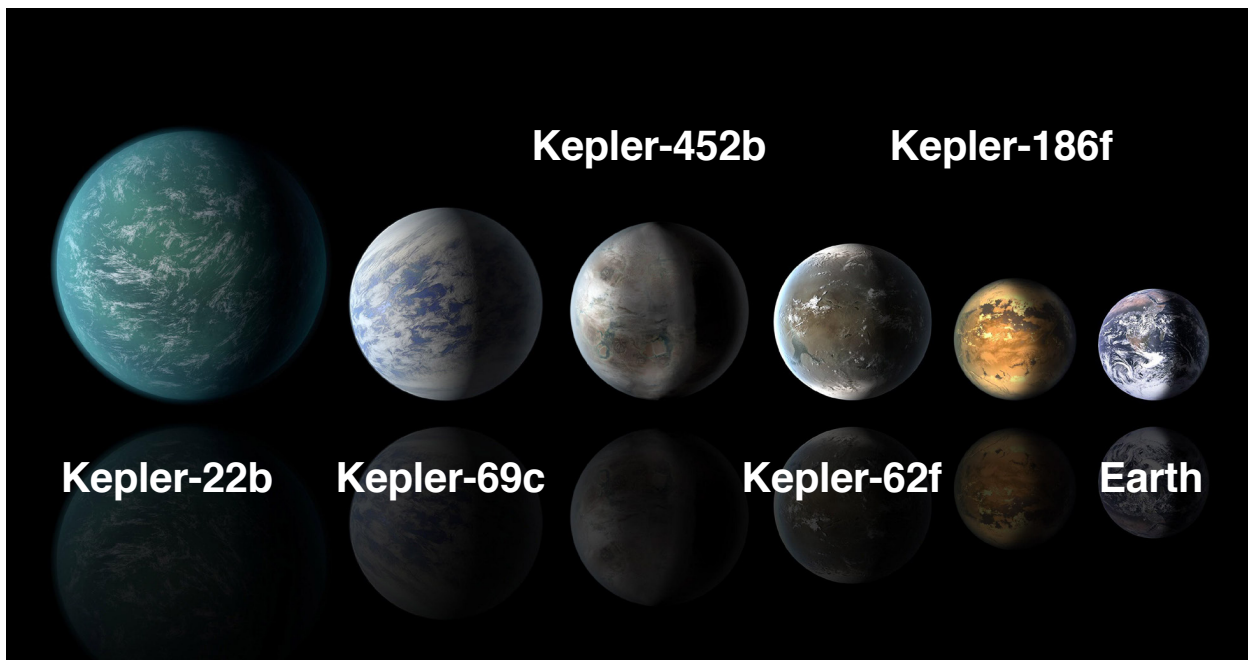

**Figure 2.7.** *Artists conceptions of an assortment of sub-Neptunes and super-Earths discovered with the Kepler Mission. Credit: NASA/Ames/JPL-Caltech*

comprehensive study of a huge range of exotic exoplanets, which by and large will be easier to observe than Earth-like exoplanets.

Beyond individual exoplanets, LUVOIR will study the architectures of planetary systems, which bear witness to formation processes that operated eons ago. The relations and interactions between all the bodies in a system are also vital for understanding their character. For example, the upper atmosphere of Jupiter is polluted by impacts of comets, which continue to this day. We can also study in real time the violent formation of rocky worlds and final sculpting of planetary systems by observing massive belts of colliding asteroids and comets around nearby young stars (aka. debris disks).

There are many things still to be discovered and understood about the bodies within the solar system itself. LUVOIR can provide up to about 25 km imaging resolution at Jupiter in visible light, permitting detailed monitoring of atmospheric dynamics in Jupiter, Saturn, Uranus, and Neptune over long timescales. Sensitive, high resolution imaging and spectroscopy of solar system comets, asteroids, moons, and Kuiper Belt objects that will not be visited by spacecraft in the foreseeable future can provide vital information on the processes that formed the solar system ages ago. These exciting studies only scratch the surface of the solar system remote sensing that LUVOIR can do.

## 2.2   The evolution of the cosmos

The scope of astrophysics investigations that LUVOIR can do is truly vast, covering all the topics addressed with the Hubble Space Telescope (HST) and more. Experiments extremely difficult or impossible to execute with Hubble and its 2.4-meter diameter telescope become routine with LUVOIR. A detailed discussion of every part of LUVOIR's astrophysics portfolio would be far beyond the scope of this report. We have therefore chosen to focus on a few major areas, spread throughout the history and hierarchy of the universe. The LUVOIR team is actively soliciting further contributions from the broader science community, in the areas of astrophysics, comparative exoplanetology, and solar system remote sensing (see **Appendix A**).

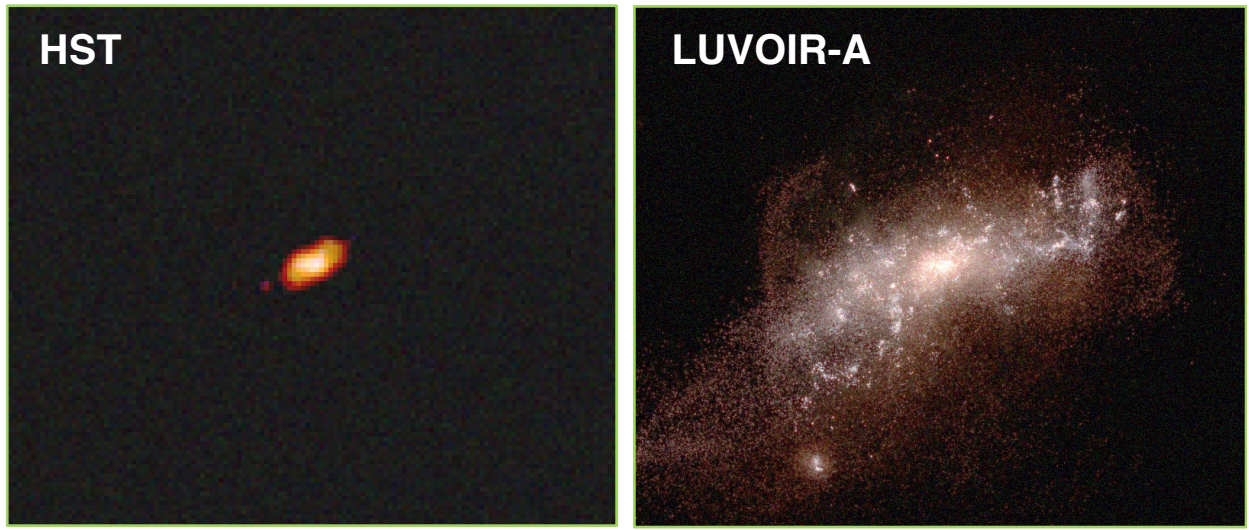

**Figure 2.8.** *Simulation of a z~2 low mass galaxy imaged with the Hubble Space Telescope and LUVOIR. The images are 5.84 arcseconds across. Credit: G. Snyder (STScI)*

### 2.2.1   The lives of galaxies

The destinies of stars and planets are controlled by the formation and evolution of their home galaxies. These are not uniform, static assemblages. The range of galaxy masses and structures is enormous, from massive giant elliptical galaxies, to complex spirals with super-massive black holes at their centers (sometimes seen as active galactic nuclei), to tiny but massive dwarf galaxies housing only dozens of stars. Throughout their lives, galaxies interact with both their neighbors and their neighborhood. Some galaxies are born small but become the building blocks of giant spirals like our own Milky Way. Most mysteriously, we see that some galaxies are actively forming new stars, while others are undergoing a slow death as their stars age and no new ones are born. Our current understanding of this complex interplay is highly incomplete.

Gas flows from the space between the galaxies (intergalactic medium), into the region around a galaxy (circumgalactic medium), then into the galaxy itself where it can become the fuel for new stars. Eventually, massive stars, supernovae, and active galactic nuclei drive material back out. Much of this cycling of matter is currently unobserved, since the gas flows are hot and tenuous. To understand the rates of star formation in galaxies, we need to measure the balance of these inward and outward flows. This demands sensitive ultraviolet spectroscopy, which can only be done from space. Galaxy mergers result in bursts of star formation as well but may also be responsible for the ultimate "death" of merged galaxies. Understanding the role this process plays in giving rise to the massive galaxies of today demands deep, high-resolution visible and near-infrared imaging of all types of galaxies (**Figure 2.8**) at earlier times.

Finally, stars produce essentially all the elements heavier than hydrogen and helium, the materials out of which planets are built. The births and deaths of stars within galaxies enrich the universe in elements needed for life (carbon, oxygen, nitrogen) and our technology (true metals). The global enrichment of the universe is intimately tied to star formation and the cycling of matter into and out of galaxies. LUVOIR will provide the ultraviolet spectroscopic capability and the high-resolution ultraviolet, visible, and near-infrared imaging necessary for major advancement in all these areas.

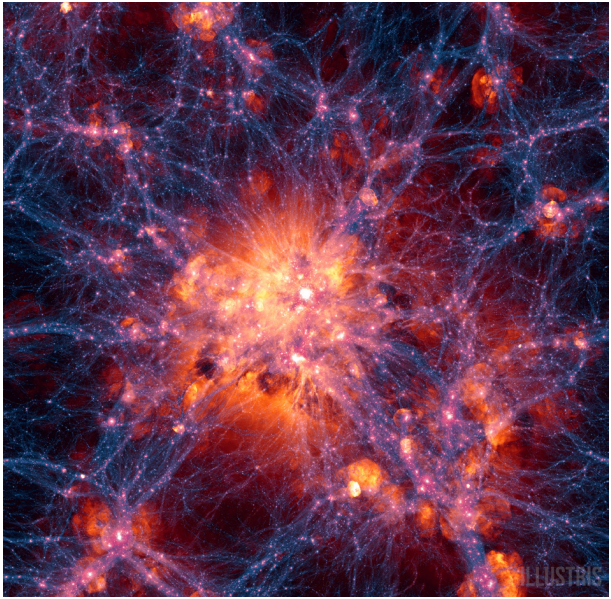

**Figure 2.9.** *Theoretical simulation of galaxy cluster formation in the presence of the dark matter web. Credit: Illustris Collaboration*

### 2.2.2    The far reaches of space and time

How an initially amorphous universe grew to host the diverse structures we see today is a driving question for astrophysicists. We understand that the complex interplay between normal matter and dark matter during the rapid early expansion of the universe was vital for producing a cosmos with galaxies, stars, and planets. Clusters of galaxies are strung out along dark matter webs like jewels in space (**Figure 2.9**). But in some ways, we know very little. What is the nature of dark matter? At the moment, we roughly know what it does on the largest size scales but not what it is. Constraining the nature of dark matter requires probing scales and masses smaller than can be probed by any facility today.

Do our models of galaxy formation still work at smaller size scales? The models were built to explain the largest structures we can see today and they provide constraints on the fundamental physical parameters of the universe. Whether they can explain smaller structures is a key test that will reveal whether we have captured the physics of interactions between normal matter, dark matter, and ionizing light. LUVOIR will facilitate detailed studies of the population of ultra-faint, low mass dwarf galaxies in the furthest reaches of time and space all the way to today. These studies are enabled by visible and near-infrared imaging surveys that take advantage of LUVOIR's sensitivity, high spatial resolution, and high-precision astrometry capabilities. Its UV spectroscopic capabilities will provide a critical tool towards understanding of the emergence of structure and light from the cosmic dark ages.

### 2.2.3    The origins of stars and planets

Descending even further in size scale, stars form through the collapse of dense cores within cold clouds of interstellar gas and dust. Planets then form in rotating disks of remnant gas and dust around the newborn stars (protoplanetary disks). But the fact is we do not have comprehensive theories for either process that can explain what we see, much less predict the currently unseen. We are unable to explain the fundamental outcome of the star formation process—the initial stellar mass function (IMF), which describes how many stars of which masses are born. How the cores of interstellar clouds relate to the stars they form and how the IMF varies with time and environment are complexities that we have only begun to wrestle with. With ultraviolet/visible imaging at exquisite spatial resolution and powerful multi-object spectroscopy, LUVOIR will follow high-mass stars out to unprecedented distances, hunt for binaries down to small separations, and provide access to the lowest-mass ends of the IMF.

Planet formation theories were originally built to describe the birth of the Solar System. As we learned more about protoplanetary disks by observing them in star forming regions (**Figure 2.10**), even the solar system became difficult to explain. When it comes to exoplanets, our planet formation theories did not predict and cannot explain the wide range we have already discovered. We are not able to trace the planet formation process all the way from its beginning in gas and tiny dust grains to any fully formed planet, and it is abundantly obvious that important processes are missing from our theories.

The fundamental problem is that we need better theories to explain what we see but cannot formulate those theories without information that we do not currently have. Stars and protoplanetary disks are small; currently, we can only study them in detail when they are relatively nearby. To increase the numbers we can study in different environments and thereby provide fodder for our theories, observations at higher spatial resolution are needed.

Furthermore, several key observational tests must be done in the ultraviolet. For example, massive stars are rare but play a vital role in star formation. Finding more of them requires extending our ultraviolet observing capabilities to the galaxies farther beyond the Milky Way. Ultraviolet spectroscopy is also needed to observe molecular hydrogen—the most important gas in protoplanetary disks—at the critical time when disk gas is starting to disappear and planet formation is ending. The high spatial resolution and ultraviolet capability provided by LUVOIR will be vital for advancing our theories of star and planet formation.

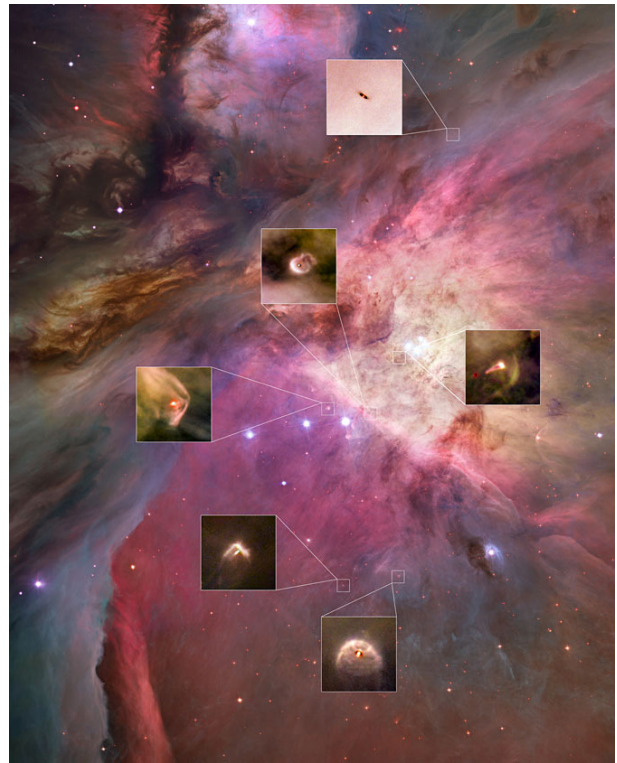

**Figure 2.10.** *Planet-forming systems in the Orion Star Forming Region. Credit: NASA/ESA/Hubble.*

## 2.3    A big observatory for big goals

Each of these revolutionary science investigations demands distinct observations, but they share a common need for a large space telescope that can collect light from the far-ultraviolet to the near-infrared. Further, that observatory must have a diverse and powerful toolkit of instruments. The Hubble Space Telescope, which has been one of the most productive scientific tools ever built, shares these characteristics. But our transformative goals demand a great leap in capability over Hubble and every other current or planned observatory. The design of LUVOIR is driven by our need for greater sensitivity, finer resolution, and higher contrast. At the same time, the design draws on a decades-long wealth of expertise and technology development garnered from several current and near-future space missions.

<div style="background:blue">

### The LUVOIR Mission Characteristics

**Community driven observing program**

**Operating at Earth-Sun L2**

**Prime mission lifetime of 5 years**

**Serviceable and upgradeable**

**Lifetime goal of 25 years for non-serviceable components**

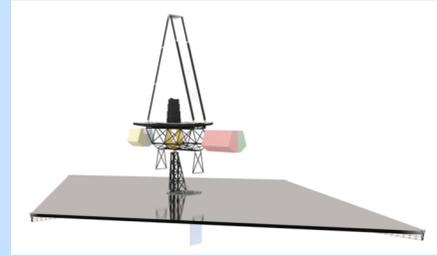

</div>

### 2.3.1 The LUVOIR mission concepts

NASA tasked the LUVOIR team with considering space telescopes in the 8-to-16-meter diameter range. We are studying two distinct observatory architectures, both designed to operate at the Sun-Earth Lagrange 2 point (L2) with 5-year prime missions. By developing two observatory concepts, we gain better understanding of a complex trade space, reveal how science return scales with different technical choices, and establish robustness to uncertainties such as future launch vehicle capabilities and budget constraints. LUVOIR Architecture A (LUVOIR-A) features a 15-m diameter primary telescope aperture and four serviceable instruments (**Figure 2.11**). It was designed for launch on NASA's planned Space Launch System (SLS) in a Block 2 vehicle. The team has completed the bulk of the study-phase design work for LUVOIR-A; refinements are in progress.

Architecture B (LUVOIR-B) has an 8-m telescope aperture and 3 instruments; design work on this concept began in Sept 2017. LUVOIR-B is being designed to fit within a 5-m diameter fairing, similar to those in use today. Call-out boxes throughout this chapter present the high-level characteristics of LUVOIR-A and its instruments. Full details of the Architecture A design and plans for Architecture B appear in **Chapters 8–10**.

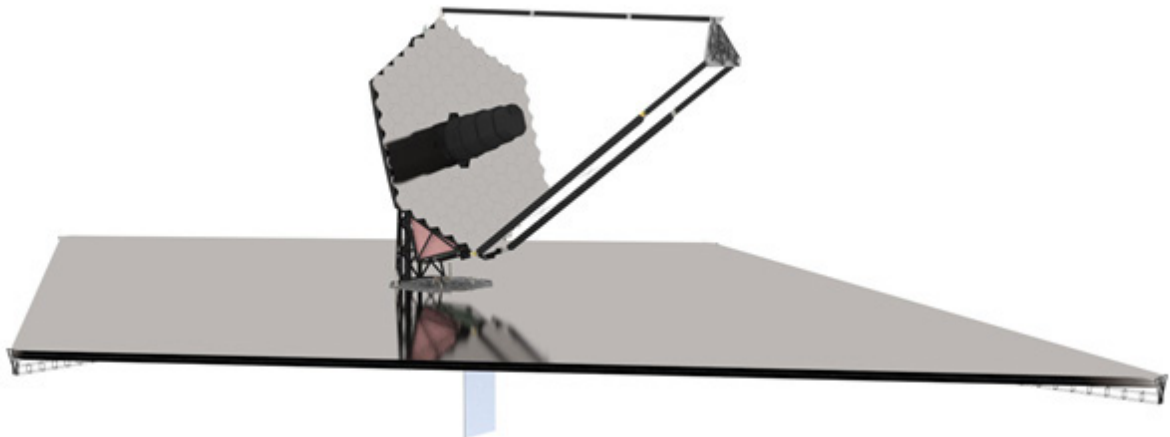

**Figure 2.11.** *Preliminary rendering of the LUVOIR Architecture A observatory, which has a 15-meter diameter primary mirror and four instrument bays. An animation of the observatory deployment may be viewed at* https://asd.gsfc.nasa.gov/luvoir/design/. *Credit: A. Jones (NASA GSFC)*

## LUVOIR's Heritage

**HST for serviceability**

**JWST for deployable telescopes, segmented wavefront control**

**WFIRST for high-performance coronagraphy**

**Sounding rockets and cubesats for UV technology development**

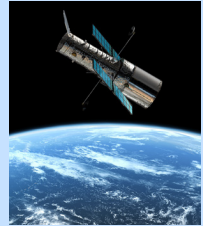
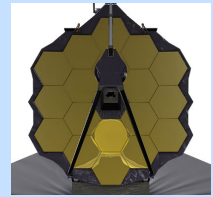
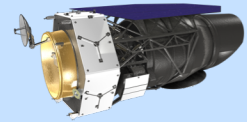

### 2.3.2   Designing for serviceability

The LUVOIR-A science program described in this report can be achieved within a 5-year prime mission using the as-designed instruments. LUVOIR should be judged on the strength of that science, without relying on an extended mission. However, Congress has mandated that all future large space telescopes be serviceable to the extent practicable. The LUVOIR Science and Technology Definition Team (STDT) has embraced this requirement, since gaining the greatest scientific return from a strategic mission with an extended development time and significant cost demands long operational lifetimes. LUVOIR's serviceability is a key part of enabling a series of "Greater Observatories" in the tradition of Hubble, Compton, Chandra, and Spitzer.

Our preliminary observatory designs incorporate standardized mechanical components (e.g., valves, latches, rails) and allow replacement of many spacecraft elements (e.g., reaction wheels, gyros, batteries, solar arrays, computers). Consumables (e.g., propellant, cryogen) can be replenished and science instruments easily removed for replacement. Incorporating these features in the LUVOIR design from the outset will also result in faster and easier Integration and Test procedures on the ground before launch, advantages that will be realized regardless of whether LUVOIR is actually serviced.

We have considered the question "how will LUVOIR be serviced?" We first note that a design commitment to serviceability is not a commitment to servicing. If LUVOIR were to launch in the late 2030s, presumably the first servicing mission would be desired in the 2040s, so a decision to service would not have to be made in the near future. Whether or not LUVOIR would be serviced will depend on political and budgetary realities that are unpredictable. If history is a guide, it is doubtful that NASA's Science Mission Directorate on its own will have the

## The LUVOIR Telescopes

**Architecture A: 15-m diameter, 4 instrument bays**

**Architecture B: 8-m diameter, 3 instrument bays**

**Segmented, deployable, ultra-stable telescopes**

**Diffraction limited at 500 nm**

**Far-UV to near-IR bandpass (100 nm to 2500 nm)**

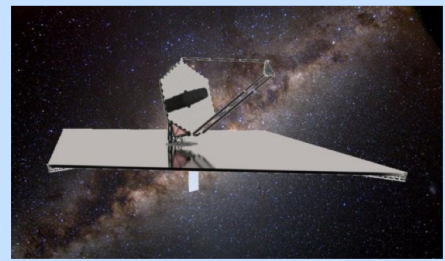

## Instrument : ECLIPS-A

**Ultra-high contrast coronagraph (10$^{-10}$)**

**Total bandpass: 200 nm – 2000 nm**

**Inner working angle ~ 3.5 λ / D**

**Outer working angle ~ 64 λ / D**

**Imaging and imaging spectroscopy**
**(Vis R = 140, NIR R = 70, 200)**

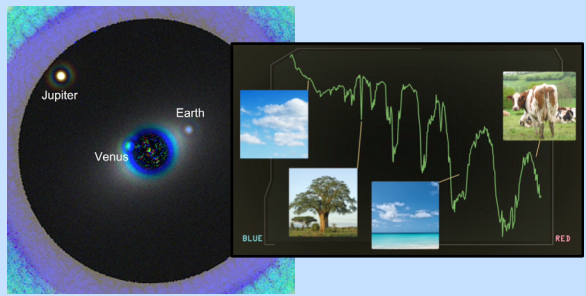

resources to create the infrastructure for servicing LUVOIR.

Servicing LUVOIR will therefore have to be approached opportunistically. As an illustrative example, the Human Exploration and Operations Mission Directorate is considering a Deep Space Gateway in cislunar space which, in principle, could be used for servicing scientific assets on-orbit. The energy required to move LUVOIR from Sun-Earth L2 to Earth-Moon L1 or L2 and back is small compared to that required for launch of the observatory. Returning the fully deployed LUVOIR to cis-lunar space for servicing by some combination of astronauts and/or robots is a technical challenge, but not an impossible task (**Chapter 8**).

### 2.3.3 Enabling LUVOIR

Revolutionary science goals require a technologically challenging observatory. The large LUVOIR apertures demanded by our sensitivity and spatial resolution goals must be segmented and deployable to fit within a launch vehicle. Fortunately, the James Webb Space Telescope (JWST) to be launched in 2021 has paved the way in this area. Our most technically challenging observational goal is the extreme starlight suppression needed to directly observe exoplanets. There are two basic instruments to do this: coronagraphs and starshades. The latter are deemed unfeasible for telescopes as large as LUVOIR-A (details appear in **Appendix E**). Current coronagraphs cannot reach the contrast values needed to detect and study small exoplanets. For LUVOIR to achieve its goal of directly studying rocky exoplanets in the habitable zones of Sun-like stars, coronagraph performance must be improved in several ways.

As part of addressing this challenge, NASA has embarked on several development programs for starlight suppression technology. The Segmented Coronagraph Design Study, under the leadership of NASA's Exoplanet Exploration Program, has demonstrated that coronagraphs can be designed for

## Instrument : LUMOS-A

**UV multi-object spectrograph + imager**

**Total bandpass: 100 – 400 nm**

**Resolution 500 < R < 63,000**

**Microshutter array: 0.14" x 0.07" shutters, 3' x 1.6' field of view**

**Far-UV imaging channel**

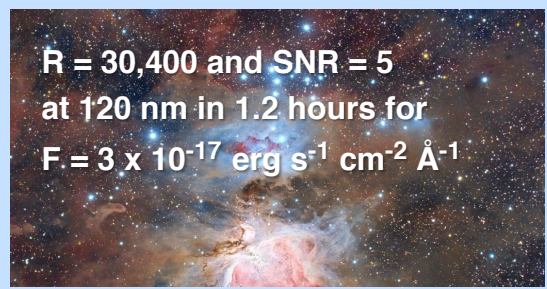

R = 30,400 and SNR = 5 at 120 nm in 1.2 hours for F = 3 x 10$^{-17}$ erg s$^{-1}$ cm$^{-2}$ Å$^{-1}$

### Instrument : High-Definition Imager-A

**2' x 3' imaging camera**

**Total bandpass: 200 nm – 2500 nm**

**Nyquist sampled at 400 nm**

**Simultaneous UV/Vis and NIR channels**

**~ 70 filter slots**

**High-precision astrometry mode**

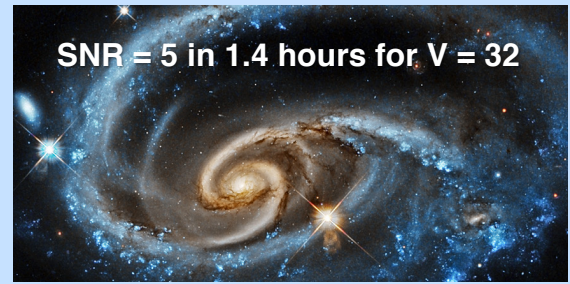

SNR = 5 in 1.4 hours for V = 32

segmented, obscured apertures (a point of doubt a decade ago). The most important current effort is the Wide-Field Infrared Survey Telescope (WFIRST) Coronagraph Instrument (CGI) technology demonstration program currently working to design a high-performance coronagraph coupled to a complex, obscured telescope aperture. Further, a coronagraph's performance is directly linked to stability of the wavefront of light entering it. The CGI program has made great strides in developing technologies and algorithms to stabilize the wavefront coming from the telescope (low-order wavefront sensors, deformable mirrors). The highest contrast measured to date (~$10^{-9}$) was demonstrated in testbeds at JPL through the CGI program (details appear in **Chapter 11**).

The LUVOIR coronagraph must achieve about another 1–2 orders of magnitude improvement in contrast beyond the current record. To do this, the LUVOIR observatory was designed with wavefront stability, sensing, and control in mind at all times.

Mechanical and thermal disturbances have been minimized and several layers of wavefront control implemented. As part of this mission study, we have assessed the technological maturity of each piece of the starlight suppression system; all are currently at Technology Readiness Level 3 or higher. The remaining major challenge is to show that this whole design works together as a system to provide the needed performance. In **Chapter 11**, we present our technology assessments and preliminary development plan.

## 2.4 What is the difference between LUVOIR and HabEx?

The Habitable Exoplanet Imaging Mission (HabEx) is another concept currently being studied in preparation for the 2020 Astrophysics Decadal Survey. Here we answer a question frequently asked about the relationship between the LUVOIR and

### Instrument : POLLUX

**Instrument study from European consortium, under leadership of CNES**

**Point-source UV spectropolarimeter**

**Total bandpass: 100 nm – 400 nm**

**Resolution R = 120,000**

**Circular + linear polarization capability**

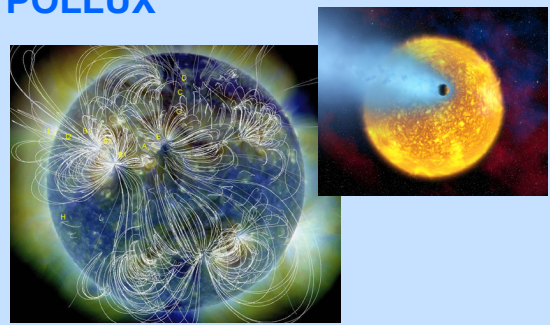

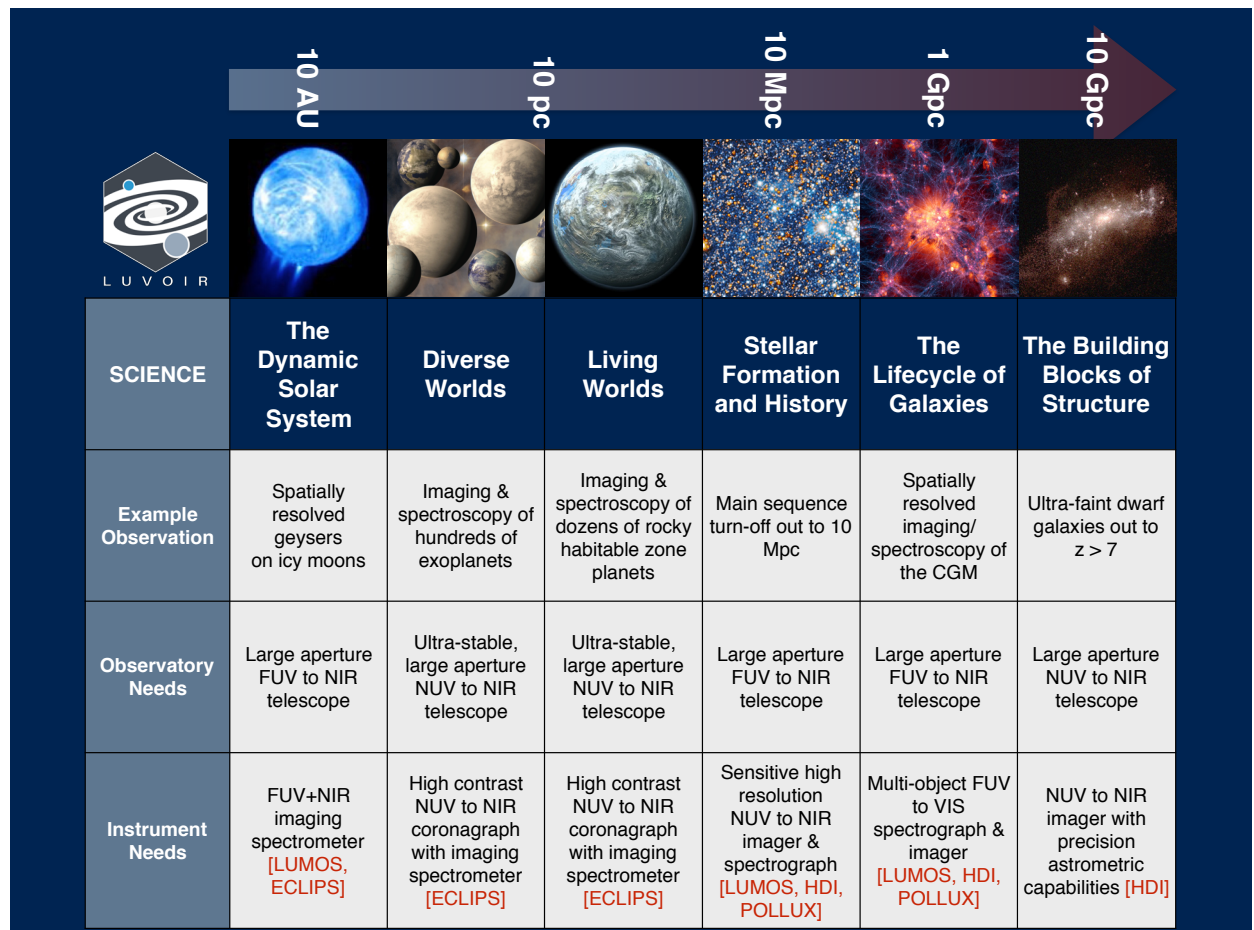

| | 10 AU | 10 pc | 10 Mpc | 1 Gpc | 10 Gpc |
|---|---|---|---|---|---|
| **SCIENCE** | **The Dynamic Solar System** | **Diverse Worlds** | **Living Worlds** | **Stellar Formation and History** | **The Lifecycle of Galaxies** | **The Building Blocks of Structure** |
| **Example Observation** | Spatially resolved geysers on icy moons | Imaging & spectroscopy of hundreds of exoplanets | Imaging & spectroscopy of dozens of rocky habitable zone planets | Main sequence turn-off out to 10 Mpc | Spatially resolved imaging/ spectroscopy of the CGM | Ultra-faint dwarf galaxies out to z > 7 |
| **Observatory Needs** | Large aperture FUV to NIR telescope | Ultra-stable, large aperture NUV to NIR telescope | Ultra-stable, large aperture NUV to NIR telescope | Large aperture FUV to NIR telescope | Large aperture FUV to NIR telescope | Large aperture NUV to NIR telescope |
| **Instrument Needs** | FUV+NIR imaging spectrometer [LUMOS, ECLIPS] | High contrast NUV to NIR coronagraph with imaging spectrometer [ECLIPS] | High contrast NUV to NIR coronagraph with imaging spectrometer [ECLIPS] | Sensitive high resolution NUV to NIR imager & spectrograph [LUMOS, HDI, POLLUX] | Multi-object FUV to VIS spectrograph & imager [LUMOS, HDI, POLLUX] | NUV to NIR imager with precision astrometric capabilities [HDI] |

**Figure 2.12.** *Example observations with LUVOIR as a function of increasing distance and how they drive telescope and instrument needs.*

HabEx concepts. The two teams developed the text of this section together.

LUVOIR and HabEx share two primary science goals: 1) studying habitability and biosignatures in the atmospheres of exoplanets around Sun-like stars and 2) executing a broad range of general astrophysics studies. The two mission concepts are distinguished by a difference in focus. For LUVOIR, Goals 1 and 2 are on equal footing. It will be a general purpose "Great Observatory," a successor to Hubble and JWST in the 8–16 m class. HabEx will be optimized for exoplanet direct observations (Goal 1) but will also enable a range of general astrophysics. It is a more focused mission in the ~ 4 m class.

The two missions do have similar goals for exoplanet science but differ in their quantitative levels of ambition. HabEx will *explore* the nearest stars to "search for" signs of habitability & biosignatures via direct detection of reflected light. LUVOIR will *survey* more stars to "constrain the frequency" of habitability and biosignatures and produce a statistically meaningful sample of temperate terrestrial planets. In sum, the two studies will provide a continuum of options for a range of futures.

## 2.5   The LUVOIR Interim Report

This document begins by describing the Signature Science Cases of the LUVOIR mission identified by the STDT with input from the broader scientific community. In each science chapter, we explain the motivations for choosing these science cases, identify the key measurements that must be made, and describe the observations needed to obtain those measurements. The telescope and instrument characteristics required for the observations are identified. A high-level flow down of some example science cases appears in **Figure 2.12**. Simplified summary science traceability matrices appear at the end of each science chapter and a complete, detailed science traceability matrix appears in **Appendix C**.

The Signature Science Cases discussed in this report represent some of the most compelling types of observing programs that scientists might do with LUVOIR at the limits of its performance. As compelling as they are, they should not be taken as a complete specification LUVOIR's future scientific potential. We have developed concrete examples to ensure that the point designs can execute this compelling science. We fully expect that the creativity of the community, empowered by the revolutionary capabilities of the observatory, will ask questions, acquire data, and solve problems beyond those discussed here—including those that we cannot envision today.

**Chapter 3** addresses the question "Is There Life Elsewhere?" by motivating and describing a strategy to first find habitable exoplanets then search them for signs of life. **Chapter 4**—"How Do We Fit In?"—delves into the richness of planetary systems and comparative exoplanetology. In both chapters, key contributions from LUVOIR observations of solar system bodies are also discussed. The dynamic lives of galaxies are discussed in **Chapter 5**—"How Do Galaxies Evolve?" The far reaches of the universe and its earliest times are the topic of **Chapter 6**—"What Are the Building Blocks of Cosmic Structure?" In the final science chapter, **Chapter 7**—"How Do Stars and Planets

Form?"—the complex and mysterious births of stars and planetary systems are investigated.

An overview of the LUVOIR observatories and mission-level considerations are given in **Chapter 8**—"The LUVOIR Architectures." Details on the LUVOIR-A telescope and each of the US-studied instruments (ECLIPS, LUMOS, and HDI) appear in **Chapter 9**—"The LUVOIR Telescope and Instruments." **Chapter 10** describes both the science and design of POLLUX, an instrument being studied by a consortium of European institutions, with support from the French Space Agency (CNES). Chapter 11—"LUVOIR Technology Development" presents LUVOIR's key technologies, their current status, and initial plans for advancing them. Finally, **Chapter 12**—"LUVOIR Cycle 1"—contains an exciting vision of what the scientific community might chose to do with LUVOIR in its first year of operations in the late 2030s. LUVOIR is an observatory not just for the science of the 2030s, a selected sample of which appears in this report, but also for decades after. Just as Hubble today is doing science never envisioned at the time of its design or launch, LUVOIR is designed to be both flexible and powerful enough to do the as-of-yet unknown science of the 2040s and beyond.

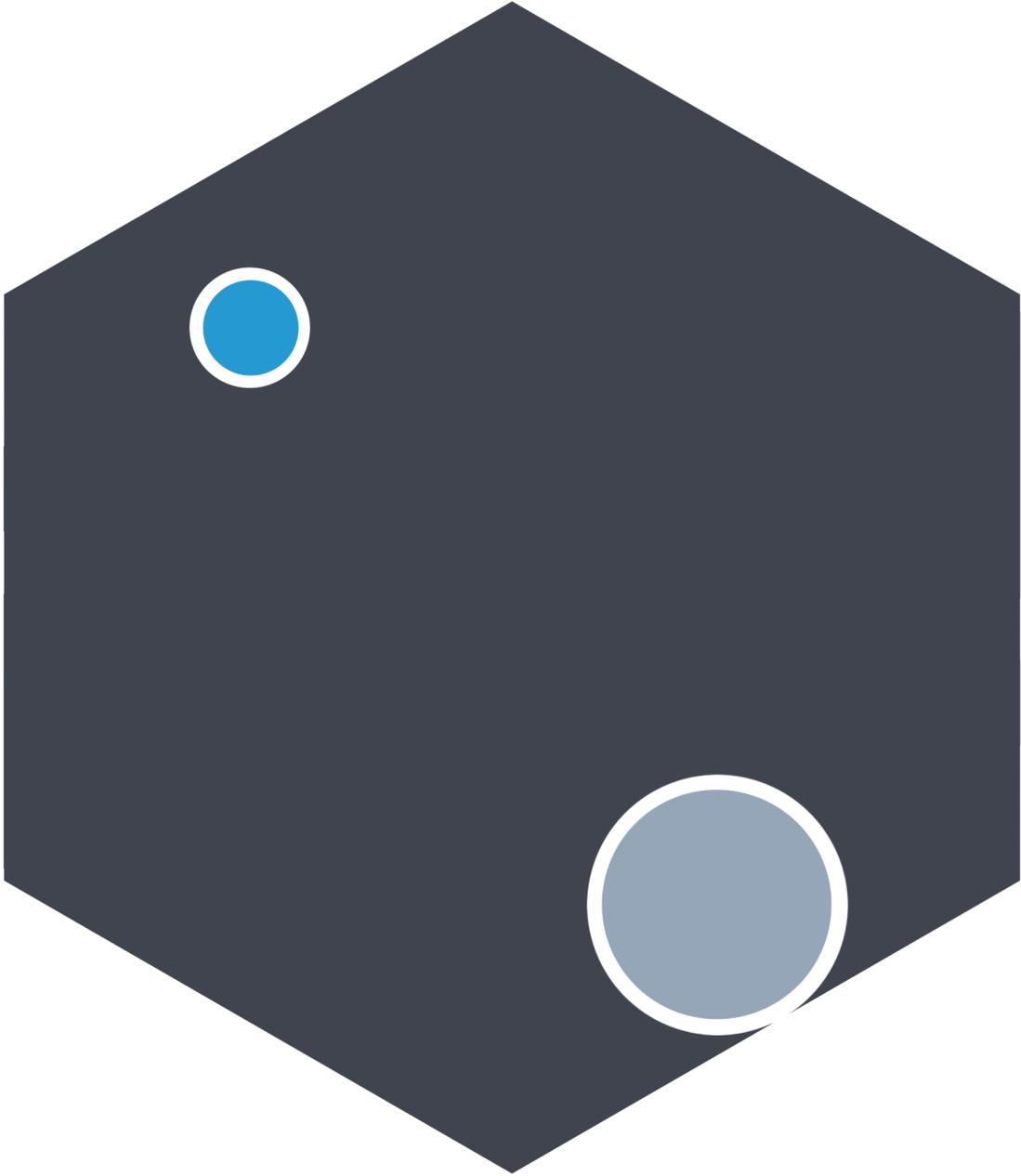

Is there life elsewhere?

# 3 Is there life elsewhere? Habitable exoplanets and solar system ocean worlds

After millennia of wondering whether humanity is alone in the universe, LUVOIR's large aperture and compelling instrument suite will enable the search for habitability and life on dozens of nearby worlds. LUVOIR's exoplanet survey will be both broad and comprehensive, which will revolutionize our understanding of planetary habitability, and will provide the first estimates for the frequency of habitable conditions and life in the local Solar neighborhood (**Figure 3.1**). LUVOIR's detailed study of the orbital parameters, atmospheric compositions, surface properties, and temporal variability of many rocky planets in the habitable zones (HZs) of different stars will reveal environmental conditions for a diversity of worlds at different stages of evolution, and potentially usher in a new era of comparative astrobiology.

LUVOIR will address three profound scientific questions for exoplanets:

- How common are habitable environments on worlds around other stars?
- How common is life beyond the solar system?
- How does life co-evolve with its exoplanetary environment?

LUVOIR's ability to measure fundamental environmental properties (e.g., atmospheric composition) for many exoplanets will help us answer the first two questions, and answering the last question will require detailed, in depth study of LUVOIR's most promising worlds to understand the interactions between the biosphere and its environment.

LUVOIR will also revolutionize the search for habitability and life closer to home, in our outer solar system. With spatial resolution comparable to flyby missions, UV

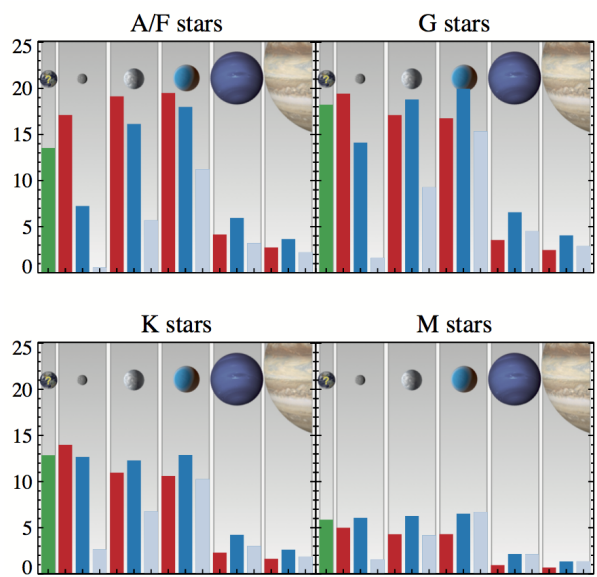

**Figure 3.1.** *Expected exoplanet yields for different types of stars. We anticipate observing ~50 exoEarth planet candidates (green bars) with the LUVOIR-A 15-m architecture during an initial 2-year habitable planet survey. Other planets shown here are detected "for free" during the 2-year search. Color photometry and orbits are obtained for all planets. Two partial spectra are obtained for all planets in systems with exoEarth candidates. Red, blue, and ice blue bars indicate hot, warm, and cold planets, respectively. Planet class types from left to right are: exoEarth candidates, rocky planets, super-Earths, sub-Neptunes, Neptunes, and Jupiters. Credit: C. Stark (STScI)*

capability, and its unprecedented space-based collecting area, LUVOIR will provide sensitive global (i.e., planet-wide) imaging and spectroscopy of a diverse population of potentially habitable solar system worlds, particularly Europa and Enceladus. These observations, which could be made available on a long temporal baseline with an extended mission lifetime, will complement NASA's *in situ* exploration by robotic flyby, orbiter, and

## State of the Field in the 2030s

We anticipate significant advances in understanding solar system habitability and in small exoplanet detection and characterization between now and the mid 2030s. Exoplanet advances will necessarily focus almost exclusively on the more readily detectable planets orbiting M dwarf stars, whose worlds undergo early evolution and star-planet interactions that are not experienced by planets orbiting Sun-like stars. A mission like LUVOIR is essential for the study of planetary systems more like our own—i.e., around Sun-like stars.

**The Europa Clipper:** The planned Europa Clipper mission (2022–2025) will deliver in-situ (mass spectroscopy) and directed spectroscopic measurements of this interesting potentially habitable icy world in the outer solar system.

**Frequency of habitable planets:** On average, early M dwarfs harbor ~0.25 potentially habitable planets per star (Dressing & Charbonneau 2013, 2015). Recent detections of Proxima Cen b (Anglada-Escudé et al. 2016), LHS 1140 b (Dittmann et al. 2017), the TRAPPIST-1 planets (Gillon et al. 2017), and Ross 128b (Bonfils et al. 2017) indicate that many terrestrial M dwarf habitable zone planets are nearby, with further discoveries expected.

**Transiting Exoplanet Survey Satellite (TESS):** NASA's TESS (Ricker et al. 2014) will launch in 2018 and survey the entire sky for transiting habitable zone planets around nearby M dwarfs, with an anticipated yield of 10–20 potentially habitable M dwarf planets (Sullivan et al. 2015), which will be prime targets for transit spectroscopy and possibly direct imaging and transit spectroscopy with LUVOIR.

**James Webb Space Telescope:** NASA's JWST is a 6.5-m infrared telescope that will launch in 2019. Its early exoplanet observations will likely be dominated by hot Jupiters and sub-Neptunes, but JWST may also provide our first opportunity to probe the atmospheres of a few transiting M dwarf HZ planets and search for biosignatures. These transit transmission measurements, while an important step forward, may be stymied by a lack of observable features (e.g., Kreidberg et al. 2014) and the instrumental noise floor (Greene et al. 2016).

**Planetary Transits and Oscillations of stars (PLATO):** ESA's PLATO mission is scheduled to launch in 2026 and will use planetary transits to search for exoplanets around stars ranging from M to G dwarfs (Rauer et al. 2014). PLATO's closest targets may be observable with LUVOIR.

**Ground-based observations:** Ground-based Extremely Large Telescopes (ELTs; 30–40-m diameter) will complement and enhance space-based observatories (Brogi et al. 2016), including LUVOIR. Ground-based observations will likely focus efforts on M dwarf stars, which have a more favorable planet-star contrast ratio ($10^{-8}$ for an M5 star with an Earth), and because adaptive optics systems perform best in the NIR (e.g., Bouchez et al. 2014). Upgrades to instrumentation (e.g., Lovis et al. 2017) may allow observations of the nearest M dwarf planets over 3–4 years with high-resolution transit spectroscopy and/or direct spectroscopy (e.g., Snellen et al. 2013; Crossfield et al. 2011). ELT direct imaging at 3–10 µm of terrestrial planets orbiting Sun-like stars may provide information on thermal emission that is complementary to LUVOIR's direct imaging in reflected light (Quanz et al. 2015).

lander spacecraft. LUVOIR will therefore also address the scientific question:

- Are there habitable environments and life elsewhere in the solar system?

To answer these four scientific questions concerning habitability and life beyond the Earth, we have identified two Signature Science Cases to address the overarching questions: #1 Which Worlds are Habitable? and #2 Which Worlds are Inhabited?

In the sections that follow we will briefly summarize the anticipated state of the field of exoplanets and solar system habitability during LUVOIR's operation lifetime, and we describe the science motivation and required measurements that will enable LUVOIR to discover small habitable zone (HZ) planets. We then describe example observing sequences that illustrate how LUVOIR will conduct the Signature Science Cases and enable the search for habitability and life on planets in and beyond the solar system.

The "Signature Science" cases discussed in this chapter represent some of the most compelling types of observing programs on habitable exoplanets, potentially habitable solar system worlds, and the search for life that scientists might do with LUVOIR at the limits of its performance. As compelling as they are, they should not be taken as a complete specification LUVOIR's future potential in these areas. We have developed concrete examples to ensure that the nominal design can do this compelling science. We fully expect that the creativity of the community, empowered by the revolutionary capabilities of the observatory, will ask questions, acquire data, and solve problems beyond those discussed here—including those that we cannot envision today.

## 3.1 A habitable and inhabited planet: The Earth through time

What does an exo-biosphere look like? LUVOIR's search for habitable worlds and life beyond the solar system will sample habitable zone planets orbiting stars of spectral type from F to M (the habitable zone is defined as that region around a star in which a rocky planet can support surface liquid water; Kasting et al. 1993; Kopparapu et al. 2013). These planetary systems will be at different ages, potentially providing a window into the different stages of habitable planet evolution and possibly with dramatically different biospheres from our own planet. This rich diversity of worlds requires a mission that is versatile and capable of exploring and characterizing environments unlike the modern Earth's, that may nonetheless be habitable. To motivate and enhance the scientific footing of this search, we consider the different habitable environments known to exist throughout Earth's 4.6 billion-year history, and the significant impact that our evolving biosphere had on these environments. These different environmental conditions and dominant biospheres generated significantly different spectral features. The observable features created by life acting on its environment are called "biosignatures." By understanding the likely observable characteristics of these varied environments (e.g., Kaltenegger et al. 2007; Rugheimer & Kaltenegger 2017), we can identify measurements needed to understand habitable exoplanet evolution. While modern Earth offers detailed, high fidelity information on a currently inhabited world, the diverse environments and biospheres of early Earth provide valuable data on "alien" globally habitable or inhabited environments.

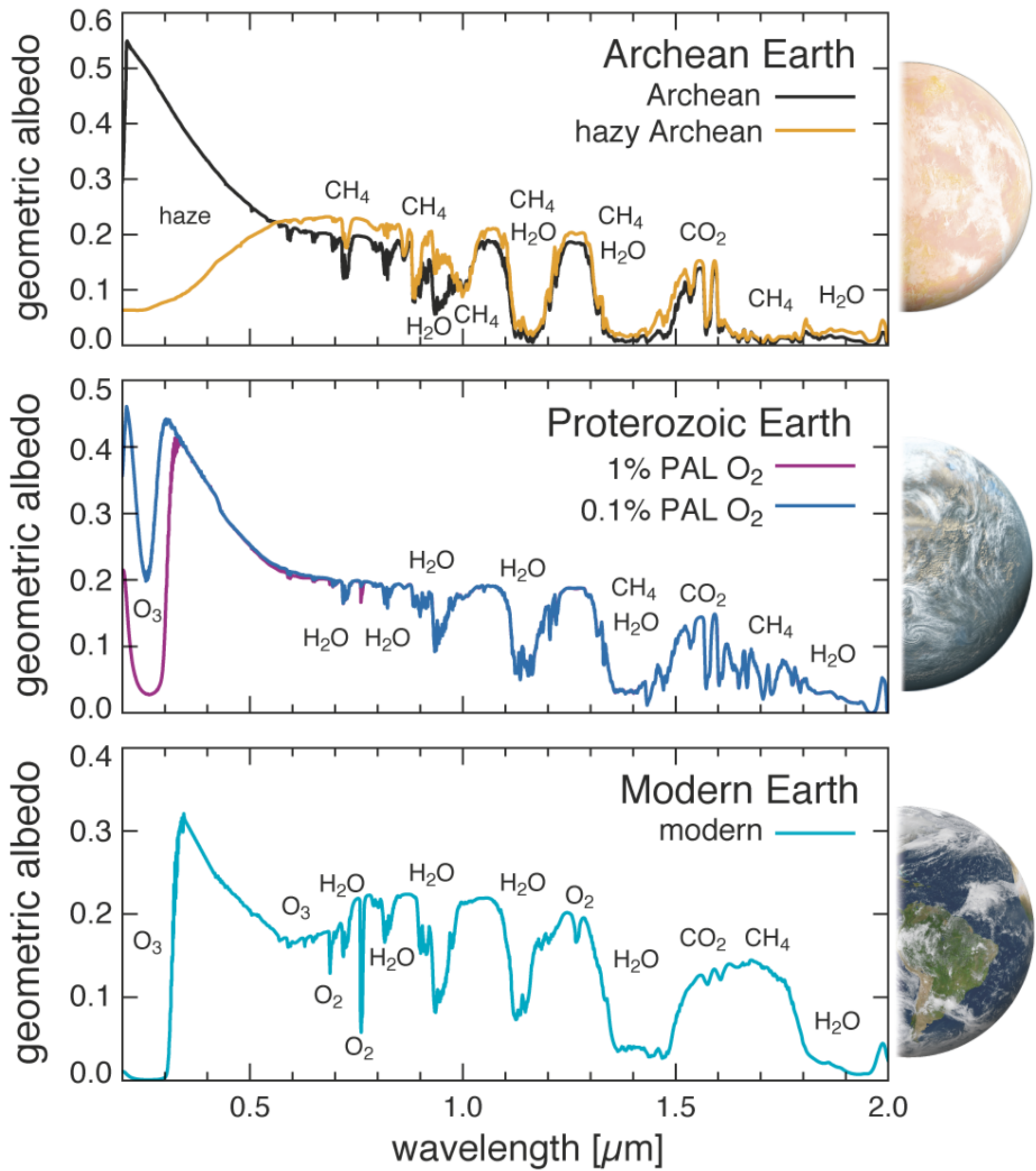

**Figure 3.2.** *The atmospheric composition of Earth has changed significantly over its history, as has its spectral features. For instance, the canonical biosignatures of modern Earth (e.g., $O_2$, $O_3$) cannot be observed in the Archean eon. A wide wavelength range enables higher fidelity spectral characterization by providing observations of many atmospheric species, and often multiple absorption bands of a given species. This breaks degeneracies between overlapping bands (e.g., overlapping $CH_4$ and $H_2O$ absorption features, as can be seen in the near-infrared in the Archean Earth spectra) and enables the search for non-traditional biosignatures. "PAL" is present atmospheric level. Credit: G. Arney (NASA GSFC).*

**Figure 3.2** shows globally-averaged spectra of Earth throughout its history. The environments of the Hadean, Archean, Proterozoic and Phanerozoic eons of Earth history are outlined below, and their distinctive remote observable properties are discussed briefly, with **Sections 3.2** and **3.3** providing a more in-depth analysis of the spectral properties of Earthlike worlds. These spectra provide context and motivation for our search strategies for a diversity of habitable worlds.

The environmental conditions of the **Hadean** eon (before 4 billion years ago) are poorly constrained due to an extremely sparse geological record, but life on Earth may have arisen during this eon (Bell et al. 2015; Nutman et al. 2016). Hadean Earth may have experienced significantly higher rates of meteoric infall compared to today. Similar young exoplanets may likewise be enshrouded by thicker interplanetary dust disks produced by planetesimal destruction (exozodiacal dust).

At the onset of the **Archean** (4–2.5 billion years ago), the Earth likely had a flourishing anaerobic biosphere. The Sun was only about 75–80% as luminous as today, so a robust inventory of greenhouse gases—which likely included carbon dioxide ($CO_2$) and methane ($CH_4$)—would have been required to keep Earth hospitable. There is disagreement on the level of atmospheric Archean $CO_2$: estimates have ranged from approximately modern Earth-like levels to orders of magnitude more than modern Earth (e.g., Kanzaki & Murakami 2015; Rosing et al. 2010). Methane levels may have ranged between 2–3 orders of magnitude higher than today (Pavlov et al. 2000), produced by biological and geological processes. These high, biologically-mediated Archean methane levels may have occasionally triggered formation of an atmospheric organic haze (e.g., Arney et al. 2016; Zerkle et al. 2012), which could itself be considered a biosignature (Arney et al. 2018). Spectral features from $CH_4$, $CO_2$, and haze are strongly apparent in the Archean Earth spectrum (**Figure 3.2**). Note high amounts of $CH_4$ generate spectral features that overlap with the strongest $H_2O$ features in the near-infrared (NIR), so disentangling these gases will likely require measurements of multiple bands and/or high signal-to-noise measurements. A dramatic difference between modern and Archean Earth is the lack of $O_2$ in the Archean atmosphere (and in its spectrum).

In the **Proterozoic** (2.5 billion years ago–541 million years ago), significant atmospheric oxygen accumulation occurred, irreversibly altering the oxidation state of the atmosphere and generating a powerful UV shield in the form of the ozone layer ($O_3$ can be seen in the UV part of the spectrum). It has been suggested that mid-Proterozoic oxygen levels may have been as low as 0.1% of the present atmospheric level, precluding directly detectable $O_2$ spectral features (Planavsky et al. 2014), although $O_2$ might be inferred by detecting $O_3$ in the UV. Thus, the Proterozoic teaches us that $O_2$, while present, may still be spectrally invisible. Biologically modulated season variations in $O_3$ may be detectable at UV wavelengths for the lower oxygen levels of the Proterozoic (Olson et al. 2018), but not for modern Earth whose higher levels of $O_3$ mean its UV spectral feature is saturated. Also, the presence of $O_2$ and oceanic sulfate may have limited mid-Proterozoic $CH_4$ to low (i.e., approximately modern) levels, diminishing methane's role as a greenhouse gas and its spectral features (Olson et al. 2016). However, other biogenic greenhouse gases that could act as biosignatures, such as $N_2O$, may have been enhanced in the mid-Proterozoic due to anoxic and sulfidic waters limiting availability of metals required for the enzymatic conversion of $N_2O$ back to $N_2$ (Buick 2007).

**Phanerozoic or modern Earth** (541 million years ago–present) has been heavily studied in the context of remote sensing and mission development studies (e.g., Sagan et al. 1993). Key gases in modern Earth's atmosphere include $O_2$, $O_3$, $CH_4$, and $CO_2$.

**Why LUVOIR?**

**[Why direct imaging? / Why from space? / Why NUV-Optical-NIR? / Why so big?]**

The most powerful way to characterize potentially habitable worlds is a space-based telescope with a large aperture, which can conduct direct spectroscopy of habitable planets over a broad wavelength range.

Understanding the surface and near-surface environment of a potentially habitable planet is the key to discovering whether that planet can support a liquid water ocean. Direct imaging provides a direct, and short, line of sight down through the planet's atmosphere to probe the planetary surface. Transmission spectroscopy will observe a much longer path through the planetary stratosphere and is therefore sensitive to atmospheric trace gases. However, it cannot see the planetary surface, and due to the effects of aerosols and refraction, it will have difficulty sampling the near-surface atmosphere. Transit observations are also not sensitive to planets around Sun-like stars (long orbital periods, shallow transit depths), so they do not allow us to characterize planets more similar to our own.

LUVOIR's direct imaging spectroscopy will probe deep into planetary atmospheres, potentially all the way to the surface. This will provide crucial information on near-surface physical conditions, water vapor abundance, and biosignature gases, including those that are not abundant at higher altitudes. Because LUVOIR will be observing from space, it will not have to stare through Earth's atmosphere, which contains many of the gases LUVOIR will search for. Those gases include molecular markers of habitable surface conditions (e.g., water vapor, $CO_2$, $CH_4$, liquid surface water) and biomarkers (e.g., $O_2$, $O_3$, $CH_4$, biological surface features). Absorption features of all these molecules lie in the near-UV, optical, and near-IR wavelength ranges.

Of the space-based telescopes being considered for the next decadal survey, LUVOIR has the largest collecting area and the smallest inner working angle. This will allow LUVOIR to detect and spectrally characterize a greater number of planets, and characterize a given planet in a shorter amount of observing time. The combination of tighter IWA and large collecting area will permit spectral characterization from the near-UV into the near-IR for most targets.

Observations over multiple, short time intervals will allow for time-dependent observations that can map planetary surface inhomogeneity and rotational periods. This combination of advantages means LUVOIR will provide mapping and in-depth spectroscopic characterization of a large number of rocky planets around other stars, increasing the probability that we will discover habitability and life, and improving the robustness of the interpretation of any biosignatures that we may find.

Modern Earth can be used to validate and constrain models of the remote observable properties of Earth as an exoplanet (e.g., Robinson et al. 2010, 2011), and its biosignatures and chemical disequilibria inform the search for life (Hitchcock &

Lovelock 1967; Krissansen-Totton et al. 2016a).

*The co-evolution of life and its environment.* Over Earth's history, life has co-evolved with its environment, and it is often impossible to separate the history of the two. As a result, the signs of habitability and life are occasionally the same signatures. Life has impacted its global environment in many ways. For example, oxygenic photosynthesis caused major changes to the chemical state of the planetary atmosphere and oceans, which caused the eventual rise of atmospheric oxygen (Lyons et al. 2014). This rise of oxygen created an ozone UV shield, which may have allowed life to spread from lakes and oceans (where water is an effective UV shield) onto land surfaces where it might produce detectable surface reflectance biosignatures. Earlier in Earth history, the evolution of methane-producing microbes (methanogens) may have generated large atmospheric quantities of $CH_4$, providing an important Archean greenhouse gas, in addition to possibly triggering periods of global organic haze that would have dramatically impacted the climate and spectral properties of Earth (Arney et al. 2016). If LUVOIR finds signs of life, it will advance our understanding of how life interacts with its host planets.

Because these interactions between life and the atmosphere are complex, and can occur on diurnal, seasonal, and much longer timescales, studying them will require sampling a large number of potentially habitable exoplanets, and time-resolved high-fidelity observations over the longest possible baseline for any planets for which LUVOIR has found signs of life. LUVOIR is uniquely suited to pursue these goals. The large aperture of LUVOIR enables a survey of potentially dozens of HZ exoplanets that offers the unique and unprecedented opportunity to witness worlds with a variety of properties. This includes habitable worlds at different ages and stages of evolution, allowing us to compare nascent worlds to ones much older. LUVOIR will also allow us to compare planets around a variety of stellar spectral types, informing our understanding of how star-planet interactions shape biospheres. LUVOIR's large survey has the potential to deliver profound insights into how life emerges and survives over time, and the longer-term impact it can have on its environment. LUVOIR can search for planetary variability on diurnal and seasonal timescales, and serviceability allows for longer baseline observations (e.g., decadal timescales). This may allow LUVOIR to test hypotheses related to the response of the biosphere to time-dependent phenomena such stellar activity or insolation variations due to the planetary orbit. LUVOIR will be capable of unlocking the story of life in the cosmos, and in so doing will improve our understanding of the history of life on Earth.

We emphasize that a wide wavelength range, spanning from the near-UV (NUV) to the NIR, is needed to characterize Earth and detect signs of life during different epochs during Earth history. One of the goals of LUVOIR is to be able to detect life on Earth-analog worlds for as much of Earth history as possible. Furthermore, this is a specific example of a more generic goal, which is to detect biospheres different from Earth. This is especially important when considering the history of oxygen in our planet's atmosphere (**Figure 3.3**). $O_2$ is regarded as the canonical biosignature of Earth, and today it can be detected at visible and NIR wavelengths. However, it was undetectable in the Archean eon, so alternative biosignatures like methane would have to be relied upon for similar anoxic exoplanets. Methane in the Archean likely produced stronger absorption bands than today and may have been detectable even at visible wavelengths (**Figure 3.2**); abundant methane is a much stronger biosignature if it can be detected together with $CO_2$, which absorbs in the NIR and would indicate an atmosphere in chemical disequilibrium most easily explained by biology (Krissansen-Totton et al. 2018). During the mid-Proterozoic, which encompasses roughly 30% of Earth's history, the NUV offers the only wavelengths (from the UV through the thermal IR) where oxygen can be detected via its photochemical byproduct $O_3$ (Schwieterman et al. 2018). This is especially significant given that $CH_4$ levels may have been particularly low during this same time period (Olson et al. 2016; Reinhard et al. 2017), removing $CH_4$ as a significant biosignature at this time and highlighting the need for NUV observations to detect biosignatures over a significant fraction of Earth history.

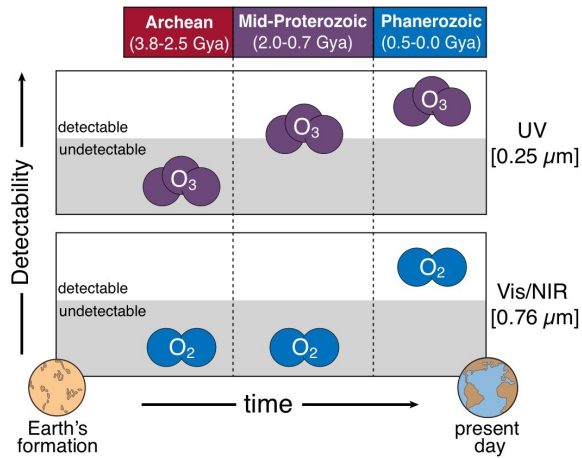

**Figure 3.3.** *O₂, the canonical biosignature of modern Earth, has not always been detectable during Earth's history (bottom row). Therefore, a wide wavelength range from the NUV to the NIR is insurance against false negative detections of life. During the Archean eon, other biosignatures such as methane must be relied on because oxygen is undetectable. During the mid-Proterozoic, oxygen can only be detected via its photochemical byproduct ozone at NUV wavelengths (top row). Credit: E. Schwieterman (UCR) / S. Olson (UCR) / C. Reinhard (GA Tech) / T. Lyons (UCR)*

In the following sections, we describe two signature science cases in detail, covering the science rationale, target selection, measurement requirements, and planet observations.

## 3.2  Signature science case #1: Which worlds are habitable?

Planetary habitability is the outcome of the complex interplay of processes between the planet, planetary system, and host star. Understanding which planets can be, and are, habitable, will require multiple pieces of information including the nature of the host star, a census of greenhouse gases, observed or modeled estimates of planetary surface temperature and pressure, and—most importantly—evidence of surface liquid water. LUVOIR can provide this information for multiple exoplanets. Because it is observing in reflected light, it can probe a planet's deep atmosphere and surface. LUVOIR's direct imaging observations can therefore provide better sensitivity to surface habitable conditions than is possible with upper-troposphere and stratosphere-skimming transmission observations that cannot observe below even thin cloud layers.

We now know that small exoplanets with R < 1.6 $R_{Earth}$ are most likely to indicate a rocky world (e.g., Rogers 2015) that is able to support a surface liquid water ocean. Compelling Kepler statistics indicate these small planets are common (e.g., Dressing & Charbonneau 2013). However, a planet that exists within the HZ may not support surface liquid water due to lack of initial volatile delivery (e.g., Lissauer 2007; Raymond et al. 2006) or subsequent loss of atmosphere and volatiles. This may be either due to its star (Luger & Barnes 2015; Ramirez & Kaltenegger 2014; Tian 2015) or its small size (e.g., Mars). Consequently, although HZ planets are common, the frequency of small, temperate planets with habitable surface conditions remains unconstrained, and can only be determined via observations of exoplanetary environmental characteristics. LUVOIR will search for planets orbiting in or near the habitable zones of their host stars (Kopparapu et al. 2013) and then conduct detailed investigations of their atmospheric compositions, energy budgets, surface properties, and seasonal variability to assess whether any of those potentially habitable worlds can support a surface liquid ocean. This study will assess the frequency of habitable worlds in our local solar neighborhood and measure the diversity of habitable planetary environments. No currently planned mission—including TESS, JWST, and WFIRST—can carry out this census of habitable environments, or characterize planets orbiting Sun-like stars.

In the solar system, the discovery of subsurface liquid water oceans on several icy moons has sparked a hope of finding habitable conditions in the icy moons of our cosmic backyard. LUVOIR can monitor plumes and surface features discovered on Enceladus and Europa over long temporal baselines, providing valuable information on the timing and locations of eruption activities. LUVOIR could also measure reflected sunlight in Titan's near-infrared spectral windows, which provide access to the surface and sub-haze environment. These studies will provide much needed remote sensing data on nearby worlds that can be used in tandem with spacecraft measurements to gain a holistic picture of these environments.

### 3.2.1 Habitable planetary environments

The search for habitable exoplanets (**Figure 3.2**) is a search for worlds with liquid surface water. Water is central to the search for life beyond Earth, and while subsurface liquid water may also provide a habitable environment (e.g., on Europa and Enceladus within the solar system), the focus on *surface* water for exoplanets is driven by detectability concerns. A biosphere that can readily exchange gases with the atmosphere can produce habitability markers and biosignatures that are more easily detected over interstellar distances compared to a subsurface biosphere. In addition to being a proven solvent for life for a vast diversity of metabolisms on Earth, water has many fundamental chemically favorable attributes for life (Pohorille & Pratt 2012), and it is the most common polyatomic molecule and solvent in the universe. Earth, with its surface water ocean, is the only planet known to support life. Elsewhere in the solar system, habitable environments may also exist where liquid water flows over rock in the subsurface of Mars (Ehlmann et al. 2011), and in the subsurface oceans of the moons of Jupiter and Saturn (Carr et al. 1998; Roberts & Nimmo 2008). Clues to habitable sub-surface environments may be revealed by detailed studies of these bodies' surfaces and any outgassing, atmosphere, or plumes of ejecta.

LUVOIR's survey of exoplanets will produce the first empirical constraints on the limits of the HZ for a variety of stellar types and broaden our understanding of the processes that shape habitability. In general, Earth-like atmospheres are considered those composed of a background of $N_2$, and greenhouse gases such as $H_2O$ and $CO_2$. The failure of these principle greenhouse gases (and others, like $CH_4$, $N_2O$, and $O_3$ cf. **Section 3.1**) to provide a clement temperature range as a function of incident stellar radiation defines the limits of the HZ. While LUVOIR may test this classic habitable zone definition, it will also explore and test alternate modes of habitability that may be possible such as: a) worlds where dense $CO_2$ atmospheres may be formed by the early extreme luminosity of the host star for planets orbiting M dwarfs (e.g., Meadows et al. 2018); b) planets with low water abundance (so-called Dune worlds; Abe et al. 2011); c) planets with dense $H_2$-dominated atmospheres (Gaidos & Plerrehumbert, 2013).

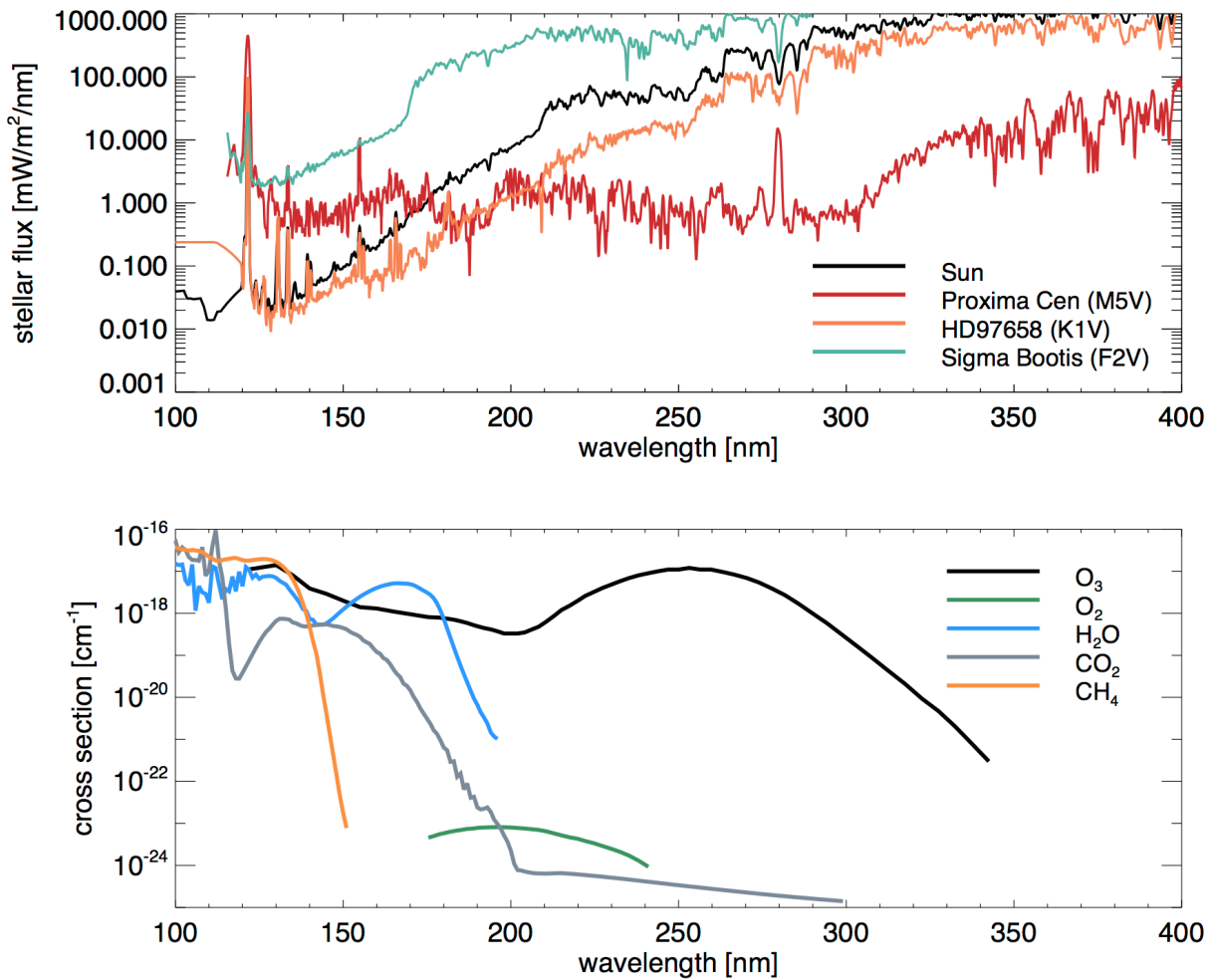

**Figure 3.4.** *LUVOIR can interpret the atmospheric compositions of planets orbiting a variety of stars. Top: UV spectra of the Sun, Proxima Centauri, a K1V dwarf, and a F2V dwarf. Bottom: UV absorption cross-sections for key gases. Atmospheric photochemistry is effectively controlled by the product of the stellar UV and the molecular absorption cross-sections. Thus, different UV spectra from different stars, which could be obtained with LUMOS, will lead to different photochemical outcomes. Credit: G. Arney (NASA GSFC) / Proxima Cen: Meadows et al. (2018); HD97658: France et al. (2016), Youngblood et al. (2016), Loyd et al. (2016); Sigma Bootis: Segura et al. (2003).*

LUVOIR will observe planets in the HZs of F, G, K, and M dwarfs, potentially allowing a comparison of the evolution of habitability on planets orbiting stars with different luminosity evolution (Baraffe et al. 1998) and activity levels. The differing extreme UV (XUV; $\lambda$=10–124 nm) and UV ($\lambda < 400$ nm) spectra of host stars will lead to different atmospheric loss and photochemical processes (**Figure 3.4**), both of which affect planetary atmospheric composition. LUVOIR will characterize how planet atmospheric composition varies with and responds to the host star spectrum.

***FGK dwarfs.*** LUVOIR will primarily target F, G, and K dwarf stars, which offer the best chance of discovering planets with evolutionary histories similar to Earth's (compared to planets orbiting M dwarfs which may undergo substantially different evolutionary paths as described below).

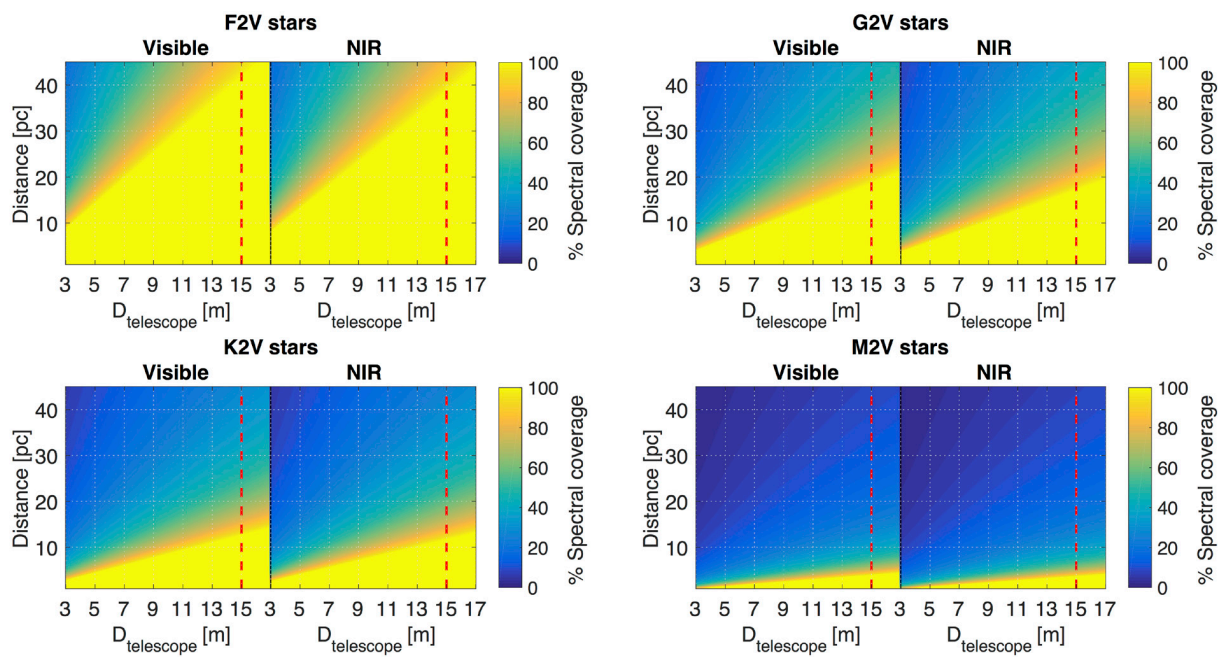

**Figure 3.5.** *Larger telescopes can observe more of an exoEarth's spectrum. The colors in each panel show spectral coverage for stars at different distances from the Sun as a function of telescope diameter for the visible (515–1030 nm) and NIR (1–2 μm) channels. Spectral coverage of 100% means the entire wavelength range in each channel can be observed for an Earth-like exoplanet orbiting at the inner edge of the conservative habitable zone (0.95 AU for a G2V star). We assume $IWA_{VIS} = 3.5\lambda/D$, $IWA_{NIR} = 2\lambda/D$, and $OWA = 64\ \lambda/D$. A larger fraction of the spectrum can be observed for hotter stars, whose HZ planets orbit further away and are less prone to being cut off by the IWA (although the contrast ratio will be less favorable for these stars). Although the cool M dwarfs have low spectral coverage because of their compact HZs, they have a higher occurrence rate than hotter stars. The vertical dashed line indicates LUVOIR-A (15-m). Credit: R. Juanola-Parramon (NASA GSFC)*

About 90% of the LUVOIR stellar targets will be FGK stars, which may represent something more akin to the Earth-Sun system, with moderate stellar evolution. These systems represent our best chance of finding true Earth analogs to set our home world in context. FGK dwarfs, compared to M dwarfs, host HZ planets at large enough orbital separations that they can be seen outside the coronagraph inner working angle (IWA) in direct imaging for a large number of targets. The IWA measures the smallest planet-star angle at which a planet can be resolved, and it scales with λ/D where D is the telescope diameter. **Figure 3.5** shows how the IWA truncates the observable spectrum for planets orbiting a variety of stellar types. HZ planets orbiting FGK stars are less accessible to transit observations due to the long orbital periods of these worlds and small transit depths. If transits can be observed, refraction will likely inhibit observations of their lower atmospheres where water vapor is most abundant on a habitable planet (Misra et al. 2014a). Initial ground-based characterizations of HZ terrestrials with high-resolution spectroscopy will likely focus their observations on M dwarf targets, which offer the best planet-star contrast and best prospects for transit observations. Space-based direct imaging observatories, therefore, provide the first and best opportunity to image terrestrial planets in reflected light and characterize Earthlike worlds orbiting truly Sun-like stars.

*M dwarfs.* M dwarfs comprise 75% of the main sequence stars in our galaxy, so the question of whether their planets can be habitable is critical to understanding the distribution of life in our galaxy. While M dwarf habitable planets are on close-in orbits around their dim stars, LUVOIR's large aperture will provide an IWA that allows direct observations of planets orbiting nearby M dwarfs, and they will comprise about 10% of the stars LUVOIR will survey. LUVOIR can also conduct transit transmission observations of M dwarfs in a wavelength range complementary to JWST, but direct imaging of M dwarf HZ planets will be the most effective way of detecting water vapor in their deep atmospheres. For HZ terrestrial planets, measurement of deep atmosphere water vapor will likely elude transmission spectroscopy measurements of the handful of M dwarf targets that JWST will observe, as transmission probes higher stratospheric altitudes with much lower water abundances (Meadows et al. 2018; Cowan et al. 2015). However, recent work suggests that "habitable moist greenhouse" states with wet stratospheres are possible for slowly rotating planets with thick substellar cloud decks that cool the surface environment. High altitude water can still be lost to space, but if loss occurs slowly, this may result in a novel type of habitable state with observable stratospheric water vapor in transit observations (Fujii et al. 2017; Kopparapu et al. 2017).

Although F, G, and K stars may harbor habitable planets with histories and characteristics more like Earth's, M dwarf planets provide a fascinating complement from a habitability standpoint as they likely undergo a significantly different evolutionary sequence compared to planets orbiting F, G and K stars. This different evolution can include possible extreme water loss during the super-luminous pre-main sequence stellar phase (Luger & Barnes 2015), tidal locking (Heath et al. 1999), and extreme M dwarf activity (Tarter et al. 2007). However, there are mechanisms that could allow for habitable conditions even in the face of these challenges: a) later migration into the habitable zone or late delivery of volatiles can mitigate the damage of the super-luminous pre-main sequence phase (Morbidelli et al. 2000); b) atmospheric circulation on tidally locked worlds can prevent atmospheric collapse on the perpetual night side and generate thick substellar cloud decks that extend the habitable zone closer to the star (Kopparapu et al. 2016; Yang et al. 2013); and c) oceans can shield marine life from the damaging radiation produced by energetic flaring (Segura et al. 2010)—provided, of course, that the planet can retain its atmosphere (e.g., Garcia-Sage et al. 2017).

LUVOIR could directly observe non-transiting exoEarth planets around the closest M dwarfs (e.g., Proxima Centauri b), and it could also observe the same transiting M dwarf planets targeted by JWST or large ground-based telescopes in a shorter complementary wavelength range for transit transmission spectroscopy. Uniquely, LUVOIR's access to the UV could test for atmospheric escape processes by detecting escaping H in transit (Bourrier et al. 2017; Ehrenreich et al. 2015). **Figure 3.6** shows possible transit transmission spectra of TRAPPIST-1 b and e for a variety of scenarios, including habitable and non-habitable states (Lincowski et al, in prep). The "false positive $O_2$" planet (green line) represents a world with abiotic photochemical $O_2$ (**Section 3.3.2**) that can be can be distinguished from truly a habitable planet by its lack of $H_2O$ features and through the appearance of strong $(O_2)_2$ dimer features. These are pressure-induced features that appear in atmospheres with extreme amounts of $O_2$ characteristic of certain abiotic photochemical processes. Also included are a Venus-analog world (red line) that exhibits a relatively flat spectrum broken only by a few $CO_2$ bands in the NIR and $SO_2$ absorption in the UV. The modern Earth-analog planet shows spectral indicators of $H_2O$, $O_2$, and $O_3$, although the $H_2O$ bands are weak since most of the water is trapped in the lowest atmospheric scale height not sensed in transit.

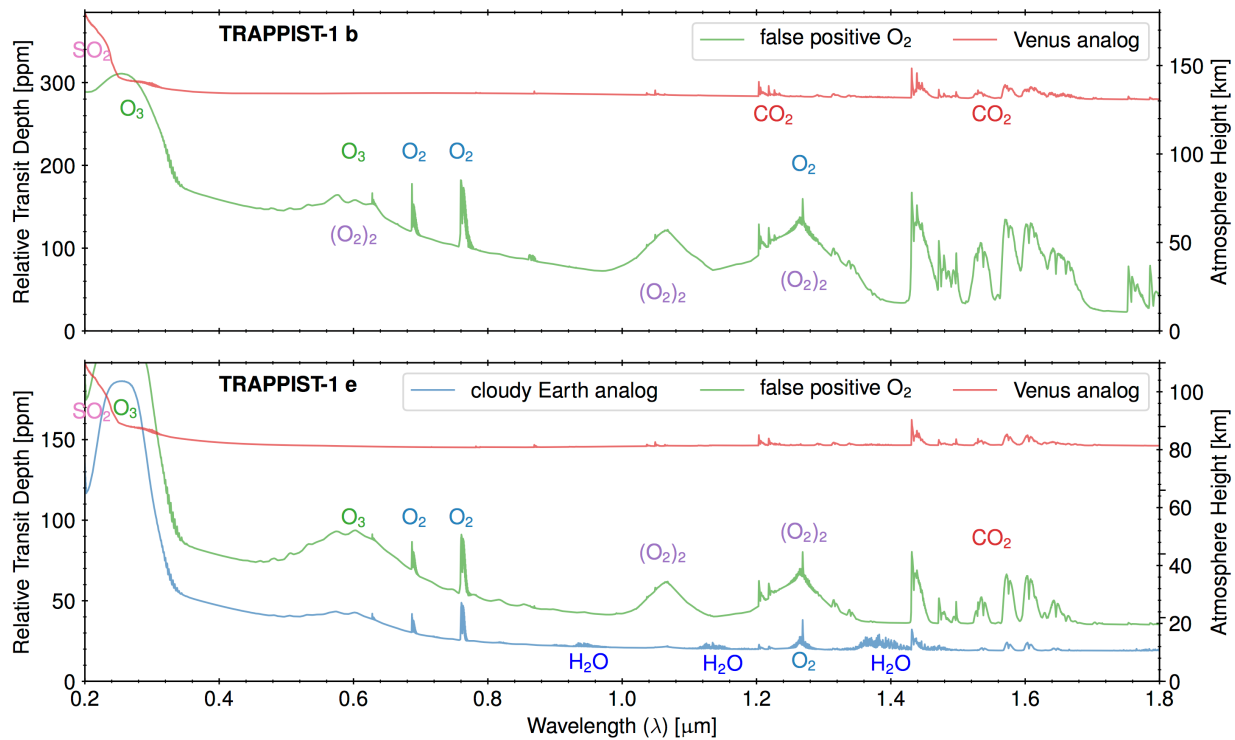

**Figure 3.6.** *LUVOIR could observe planets transiting M dwarfs in the UV-Visible-NIR, complementing longer wavelength observations of these same worlds with JWST and ground-based telescopes. Shown here are possible transmission spectra of TRAPPIST-1 b and TRAPPIST-1 e for several atmospheric scenarios, including habitable and non-habitable states. Non-habitable states include a planet with "false positive" $O_2$ produced by abiotic photochemical processes and a Venus-analog. Also included is an Earth analog for TRAPPIST-1 e. Credit: A. Lincowski (University of Washington).*

***Atmospheric composition of habitable worlds.*** To comprehensively characterize the atmosphere of a detected planet in reflected light, the presence or absence of as many gases of interest as possible must be constrained. It is also important to measure the continuum flux level and slope set by Rayleigh and cloud/haze scattering. **Table 3.1** summarizes several gases of interest, and the wavelengths at which they absorb.

In practice, observations out to only ~1.8 μm will be obtainable for most Earth-like targets, because thermal radiation from the warm telescope mirror (270 K) swamps the planet signal at longer wavelengths; LUVOIR is not a cryogenic telescope to prevent UV-absorbent contamination from condensing on the mirrors (see **Appendix D**). For a modern Earth-like or early-Earth-like HZ terrestrial planet, at wavelengths from 0.2–1.8 μm, trace gas absorption will be primarily from important greenhouse gases $H_2O$ and $CH_4$. Absorption from $CO_2$, $NH_3$, $O_3$, $O_2$, $CO$, and $O_4$ also occurs in this wavelength range (**Table 3.1**) and can be used to characterize diverse terrestrial planet atmospheres, and also provide key information to discriminate terrestrial planets from $H_2$-dominated sub-Neptune planets. The longer the wavelength range and the larger number of absorption bands that can be detected, the greater the fidelity with which atmospheric composition can be discerned. Detecting multiple features improves our confidence in the unique detection of a specific gas by ruling out similar, overlapping absorption features produced by different gases. For example, $CH_4$ and $H_2O$ have several overlapping absorption features at visible and near-infrared wavelengths, including near 1.1 and 1.4 µm, which could be confused in spectra of low to moderate resolution. LUVOIR's NIR channel can obtain spectra at R=200 to help break this degeneracy (it can also obtain spectra at R=70). However, in this example, water can be identified cleanly near 0.94 µm, and $CH_4$ cleanly near 1.7 µm (also near 1 µm at Archean-like abundances), allowing the degeneracy to be broken, and improving model fitting and retrieval for both gases. The smaller 8-m aperture (LUVOIR-B) will reduce the number of planets that this full wavelength range can be obtained for, but the study of the 8-m architecture, and its final design, are not completed. We will ensure that our major science questions can still be addressed with LUVOIR-B and will discuss this in detail in our Final Report.

**Table 3.1.** *Desired spectral features for habitability assessment.*

| Habitability Markers | | |
|---|---|---|
| Molecules/Feature | 0.2–1.0 µm | 1.0–2.0 µm |
| $H_2O$ | 0.65, 0.72, 0.82, 0.94 | 1.12, 1.4, 1.85 |
| $H_2$ | 0.64 – 0.66, 0.8 – 0.85 | |
| $CO_2$ | | 1.05, 1.21, 1.6 |
| $CH_4$ | 0.6, 0.79, 0.89, 1.0 | 1.1, 1.4, 1.7 |
| $S_8$ | 0.2-0.5 | |
| $H_2S$ | < 0.3 | |
| $SO_2$ | < 0.3 | |
| Ocean glint | 0.8 – 0.9 | 1.0 – 1.05, 1.3 |
| Rayleigh scattering | ≲ 0.5 | |

Atmospheric composition and pressure affects habitability, as the stability of surface liquid water depends on both surface temperature and pressure from the overlying atmosphere. Both will be challenging to constrain. A Rayleigh scattering slope, or the detection of dimer (i.e., a molecular complex consisting of two identical molecules linked together) or pressure-induced absorption features (Misra et al. 2014b), could allow measurements of atmospheric pressure, at least to the lowest observable atmospheric layer. The lowest observable layer may not be the surface due to cloud cover, scattering aerosols, or molecular opacity. Detection of Rayleigh scattering requires observations at visible wavelengths and does not place strict requirements on spectral resolution. It will not be possible to measure temperatures directly in the absence of planetary thermal radiation (although thermal detection of exoplanets orbiting Sun-like stars using ground-based observations with ELTs could provide excellent complementary information on planetary temperature; e.g., Quanz et al. 2015). However, it may be possible to constrain temperature, and also habitability, by retrieving a water vapor profile consistent with condensation (i.e., a condensation curve). Constraints on the water vapor profile could be obtained by observing water bands at different wavelengths, which sense different depths in the atmospheric column. Water bands between 0.6 and 1 µm probe the near-surface environment, while NIR water bands at 1.15 and 1.4 µm sense the middle and upper troposphere.

*Other atmospheric features.* A wide spectral range permits discovery of unexpected atmospheric absorbers and provides a longer lever arm to constrain cloud particle sizes and composition through Mie scattering effects. Blue and UV wavelengths are particularly important for observing these effects. Observations of exoplanets may reveal the presence of hydrocarbon, sulfuric acid or water vapor hazes in terrestrial atmospheres, especially at UV-visible wavelengths (e.g., Arney et al. 2016; Hu et al. 2013; **Figure 3.2**). Spectral estimates of the composition, particle size, and optical depth of the planetary aerosols could also constrain the albedo of the planet (whether it is more likely to be reflective, or dark), helping to improve the inherent size-albedo degeneracy for directly imaged, non-transiting planets. Spectral features or scattering slopes of volcanically-generated species such as $CO_2$, $H_2S$, $SO_2$, or $H_2SO_4$ aerosols, especially if these vary on monthly or yearly timescales, could indicate active volcanism (Hu et al. 2013; Kaltenegger & Sasselov 2010; Misra et al. 2015). Geological activity is important for maintaining planetary habitability on Earth over long timescales (Walker et al. 1981) through plate tectonics, the carbonate silicate cycle, and seafloor weathering (Krissansen-Totton & Catling 2017). Elemental sulfur ($S_8$) particles, produced by photochemical reactions involving volcanic gases such as $H_2S$, can produce broad absorption features at $\lambda < 0.5$ µm (Hu et al. 2013). $H_2S$ and $SO_2$ could be detected in the UV at wavelengths $< 0.3$ µm. Some gases, which are highly soluble in water like $SO_2$, could also serve as desiccation markers, hinting at a lack of a significant surface ocean.

*Direct detection of liquid surface water.* While gas phase water may be detected through its atmospheric spectral features, direct detection of surface liquid water may also be possible for a subset of the most observable planets through detection of specular reflectance from liquid water oceans (the "glint" effect). This phase-dependent phenomenon can be sought in either reflection or polarization. These observations are best made in the near-infrared where the relative lack of Rayleigh scattering enhances the detectability of the planetary surface and the glint phenomenon (Robinson et al. 2010; Zugger et al. 2011). Ocean glint could be detected with photometric observations at continuum wavelengths (i.e., between strong atmospheric absorption bands), and especially those near 0.8–0.9, 1.0–1.05, and 1.3 µm; Robinson et al. 2010), at multiple planetary phases between quadrature and crescent. Near-infrared observations also minimize glint false positive signatures from polar ice (Cowan et al. 2012) which is less reflective at longer wavelengths. For an Earth-twin, clouds and hazes may make the polarization signature from surface oceans challenging to detect (Zugger et al. 2011), and the reflective glint signal may prove the best option for the direct detection of surface liquid water. For Earth, glint is most pronounced near a star-planet-observer (phase) angle of 150°, where a glinting planet would be nearly twice as bright as a non-glinting planet. The ability to measure planetary phase functions at such close planet-star separations will depend on the IWA of the high-contrast field-of-view and cannot be achieved for planetary systems with inclinations below about 60 degrees. An IWA of 2 $\lambda$/D in the NIR channel allows glint detection in the 1.33 µm continuum out to

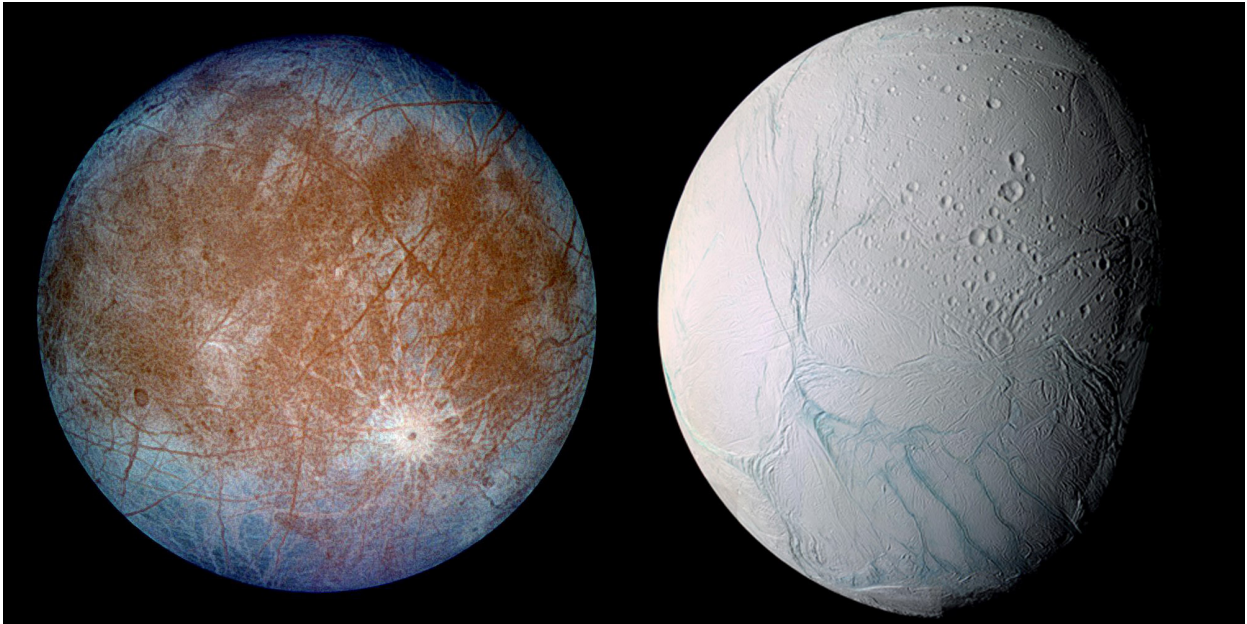

**Figure 3.7.** *Icy moons of our solar system like Europa (left) and Enceladus (right) offer an incredible opportunities to find habitable conditions and life in our cosmic backyard. The relative paucity of craters on Europa testifies to its young surface; the dark linear features point to cracks in the ice shell leading to the deep ocean. On Enceladus, plumes erupt from the "Tiger Stripes" region visible around the south pole. Credit: NASA*

almost 10 pc for habitable planets around G dwarfs, and nearly 5 pc for K dwarfs.

### 3.2.2    Ocean/icy worlds of the solar system

Within the solar system, the icy moons of the giant planets remain some of the most intriguing observational targets, motivated strongly by their potential to harbor life in subsurface oceans. While dozens of moons have or have had oceans, a few special cases are considered the best targets in the search for life and prebiotic conditions: Europa, Enceladus, and Titan. These moons represent end members of a different kind of habitable world from the terrestrial planet: worlds whose icy surfaces, deep oceans, and interactions of endogenic and exogenic energy sources give rise to the physical and geochemical processes that might create and maintain life. LUVOIR will conduct game-changing science for the solar system's ocean worlds, via multi-epoch imaging at high spatial resolution (~25 km resolution at 500 nm for Jupiter at opposition for LUVOIR-A, comparable to imaging with the Juno spacecraft) and sensitive spatially resolved spectroscopy at UV through NIR wavelengths.

For solar system icy ocean worlds, the question is not whether these oceans exist, but rather: "how may we study them?" and "how do these potentially habitable systems operate?" Europa orbits Jupiter between Io and Ganymede in a Laplace resonance, and it has maintained geologic activity over the lifetime of the solar system as a result of early accretional heat, radiogenic activity, and for at least the past 2.5 billion years, immense tidal heating from Jupiter. Europa's surface is churned over by tectonic and resurfacing processes and displays so few craters that it can only be constrained to an age of 40–90 Myr (**Figure 3.7**). The young age of the surface implies ongoing surface-ocean communication, and materials in the cracks

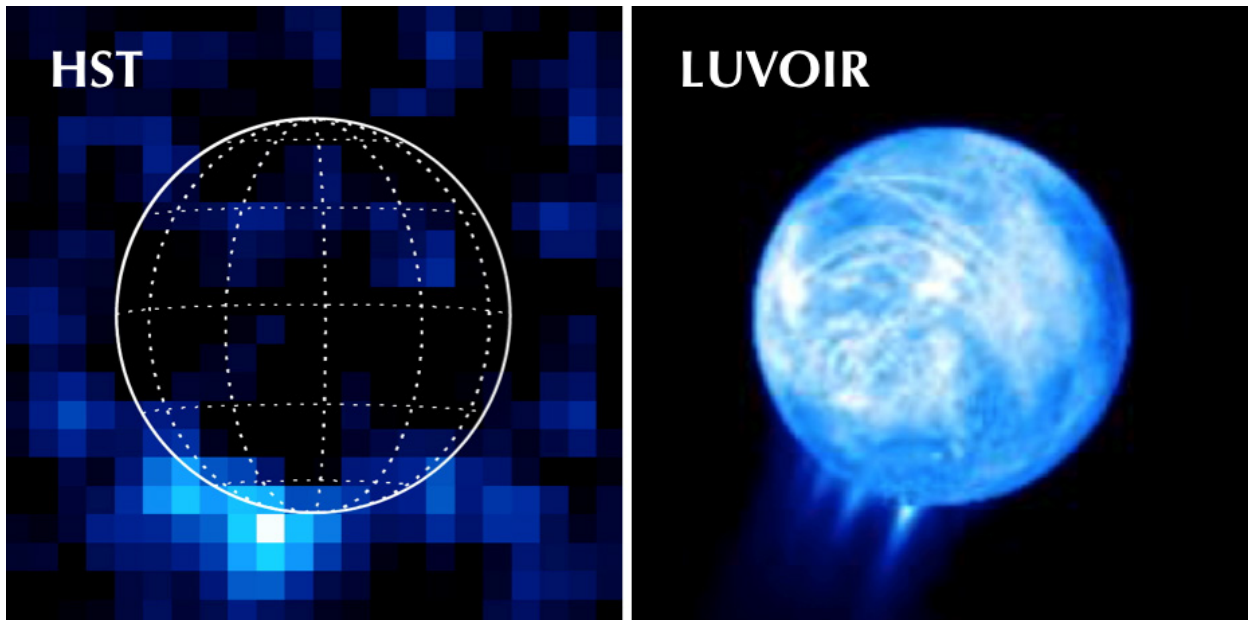

**Figure 3.8.** *Spectroscopic imaging of Europa and its water jets. The left panel shows an aurora on Europa observed with HST (Roth et al. 2014). This UV hydrogen emission (Lyman-alpha) comes from dissociation of water vapor in jets emanating from the surface. The right panel shows a simulation of how this hydrogen emission from Europa would look to a 15-m UV telescope. The moon's surface is bright due to reflected solar Ly-alpha emission. With LUVOIR, one could monitor the ocean worlds of the outer solar system for such activity and image the individual jets. Credit: G. Ballester (LPL)*

and fissures on the surface may represent compounds that have upwelled from the deep sea and/or could be transferred down into the sea.

Spectroscopic measurements of the surface of Europa have revealed the presence of a variety of surface materials. Oxidants such as hydrogen peroxide ($H_2O_2$) and solid state $O_2$ are generated by charged particle bombardment from the environment (Delitsky & Lane 1998). Such oxidants may be important for fueling metabolisms in the Europan ocean if they can be transported through the icy shell (Chyba 2000). Hydrated salts and other materials have been detected and absorb between 1 and 2 µm, accessible to LUVOIR. These include magnesium sulfate hydrates ($MgSO_4 \bullet nH_2O$), bloedite ($Na_2Mg(SO_4)_2 \bullet 4H_2O$) (McCord et al. 1999), and sulfuric acid ($H_2SO_4 \bullet nH_2O$) (Carlson 1999). Typically, these materials occur along ridges, in chaos terrain, and in craters. Their composition does not vary across the moon, suggesting upwelling from a globally mixed ocean. LUVOIR could monitor for variations in the surface distribution of these species that could indicate active upwelling of fresh material from the underlying ocean.

Plumes have been reported for both of the ice-covered ocean moons Europa and Enceladus, which may allow direct access to the chemistry and composition of the deep ocean. Two different groups have employed the HST STIS instrument to find evidence for transient water plumes emanating from its surface via observations of H and O emission (**Figure 3.8**; Roth et al. 2014) and in UV continuum absorption while Europa transited the bright surface of Jupiter (Sparks et al. 2016, 2017). In both cases, the detections were at the limit of the HST instrumentation, yielding detections at 3–6 σ levels of significance (for a water column of ~2 x $10^{20}$ m$^{-2}$ to 2 x $10^{21}$ m$^{-2}$). While some

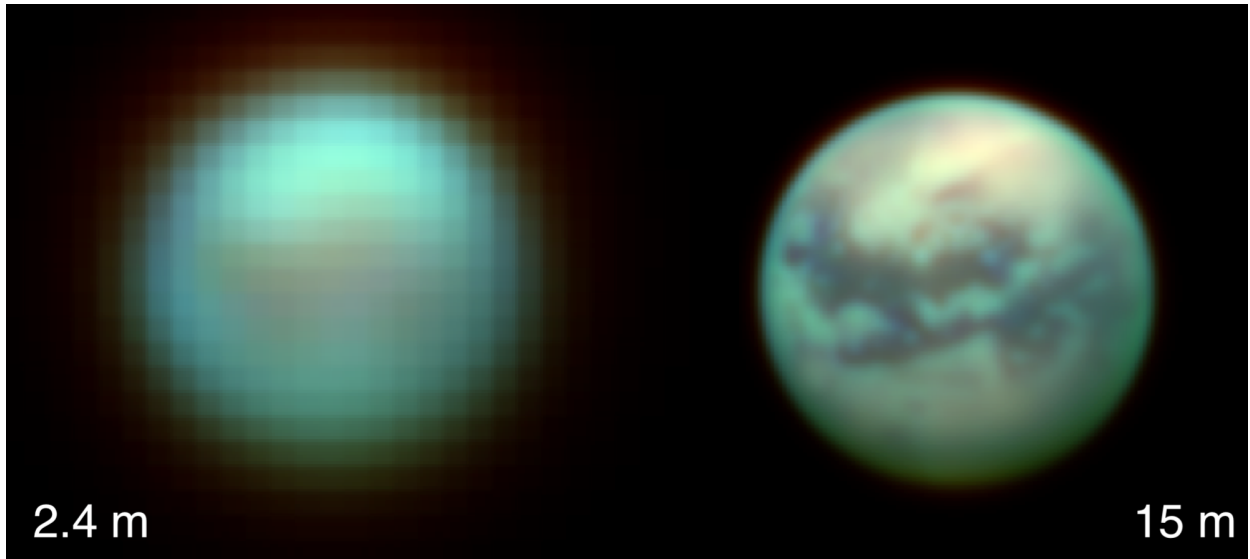

2.4 m                                                          15 m

**Figure 3.9.** *The surface of Titan (false color) observed in NIR spectral windows as seen by a 2.4-m telescope versus a 15-m LUVOIR telescope. Credit: NASA Cassini / R. Juanola-Parramon (NASA GSFC)*

of the observations were repeated, no definitive cyclic timing could be discerned; thus, the frequency and dynamics of these sporadic plumes is unclear. With the planned spatial resolution and collecting power of LUVOIR-A, emission from individual plumes could be resolved in FUV spectroscopic imaging observations with the LUMOS instrument (**Figure 3.8**). Such observations would constrain plume frequency, dynamics, and plume-magnetosphere interactions. ELT ground-based observatories will be able to perform complementary NIR studies, yet strong attenuation of the water signatures in the infrared due to telluric opacities and not having access to the UV will severely hinder the sensitivities obtained using these facilities.

On Enceladus, plumes are known to erupt near the south pole from the "Tiger Stripes" region (**Figure 3.7**). The mass of the ejected material varies following a diurnal cycle (Hedman et al. 2013), responding to openings and closings of the fissures (Nimmo et al. 2014). The plumes themselves operate stochastically, possibly caused by variations in tidal stresses (Teolis et al. 2017). Cassini's

Visual and Infrared Mapping Spectrometer (VIMS) has revealed that the Tiger Stripes region is characterized by water, organics, carbon dioxide, and amorphous and crystalline water ice (absorbing at 1.5, 2.0 μm; Brown et al. 2006). While the plumes of Enceladus have so far only been detected with the Cassini spacecraft, the reported column densities of plume material ($\sim 1 \times 10^{20}$ m$^2$; Hansen et al. 2006) are not much less than those inferred for Europa, making it highly likely that the more frequent Enceladus plumes will be observable with LUVOIR-A in a similar fashion to those of Europa. The possible detection of other materials within the plumes with LUVOIR—for example water vapor, ice particles, or other more intriguing trace species from the deep ocean—will be investigated.

Titan, in addition to its sub-ice internal ocean, harbors a thick methane- and haze-rich atmosphere as well as surface reservoirs of liquid ethane-methane mixtures (**Figure 3.9**). Titan's hazes are transparent in the NIR and allow access to the surface and lower atmosphere for monitoring of surface features. These "spectral windows,"

---

**Program at a Glance – Potentially Habitable Solar System Worlds**

**Goal:** Monitor potential habitable solar system worlds.

**Program details:** Observations of solar system icy moons to monitor surface morphological and compositional changes, study plume activity, and measure atmospheric properties (Titan)

**Instrument(s) + Configuration:** HDI multi-band imaging; LUMOS FUV spectroscopy; ECLIPS open mask IFS spectroscopy

**Key observation requirements:** Long temporal baseline observations of icy moons for monitoring surface changes and plume activity. Spatially resolved FUV (R~30,000) and NIR (R~70, 200) spectroscopy.

---

accessible to LUVOIR, have previously been used to discover methane clouds in the lower atmosphere (Griffith et al. 1998) and to find evidence of hydrocarbon lakes (Stephan et al. 2010). Meanwhile, the UV provides key diagnostics of the hazes and the atmosphere. With LUVOIR's spectroscopy of the complete observable disk accessible over a long baseline, we will better understand the processes acting on the organics / water reservoirs of Titan.

### 3.2.3 Strategy for identifying habitable exoplanets

LUVOIR will revolutionize exoplanet science by searching hundreds of stars for potentially Earth-like exoplanets and discovering hundreds of diverse exoplanets in the process. Precisely estimating the expected exoplanet science yield necessitates modeling the execution of such a mission, which in turn requires constraints on several key astrophysical parameters as well as a high-fidelity simulator of exoplanet direct observation missions. A decade ago, such modeling was not possible. Now, the Kepler Mission has constrained the frequency of Earth-sized potentially habitable planets around Sun-like stars, the Keck Interferometer Nuller and the Large Binocular Telescope Interferometer have constrained the presence of warm dust around nearby stars, and mission simulators have advanced dramatically.

Here we describe an observational procedure to find habitable exoplanet candidates, then establish whether they may in fact be habitable (i.e., have signs of water). For planets that remain promising for habitability and life, the steps for characterization continue in **Section 3.4**. **Figure 3.10** shows a graphical summary of the key questions and the observational means of answering them. Blue steps shown in **Figure 3.10** are discussed here for planet detection, while green steps are discussed in **Section 3.4** for planet characterization.

*1. Establish target list.* LUVOIR's list of target stars is already known. Using a benefit-to-cost metric, we will optimize this list of stars to maximize the probability of detecting an exoEarth based on existing information available in the 2030s—such as the amount of exozodiacal dust around the star, stellar multiplicity, and the presence of other known planets.

*2. Determine which point sources are planets.* Once LUVOIR obtains an image of a potential exoplanet system, metrics are needed to distinguish planets from background stars, galaxies, brown dwarfs, and other confusion sources (**Figure 3.11**). A combination of methods could be used to counter background confusion. Multi-epoch observations will reveal targets with the same proper motion as the parent star, an excellent metric for source confusion discrimination.

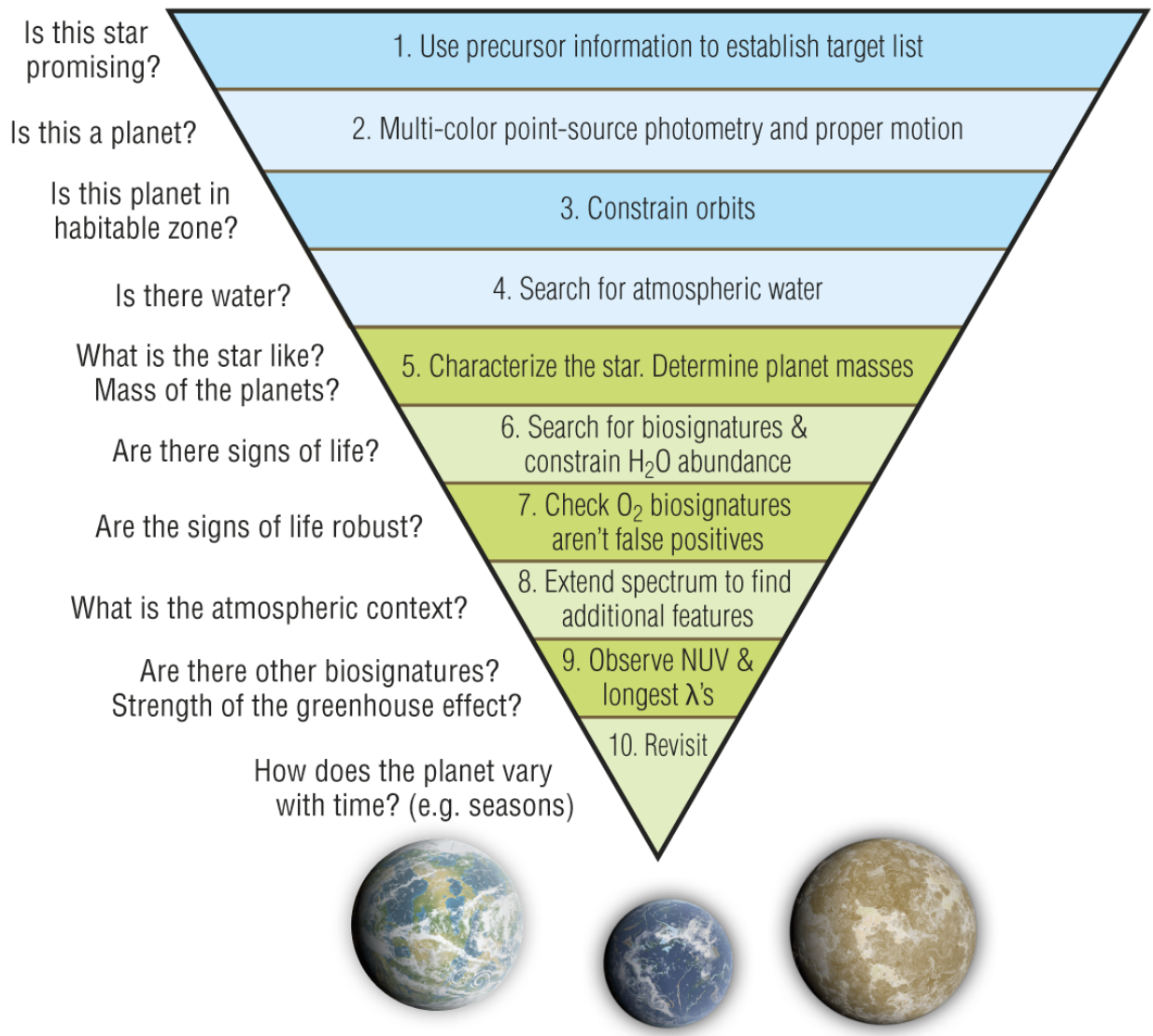

**Figure 3.10.** *Science questions and observational strategy in the search for habitable planets and life. Blue steps at the top of the figure refer to identifying habitable exoplanets; green steps at the bottom of the figure refer to characterizing habitable exoplanets. Credit: T. B. Griswold (NASA GSFC)*

During these same observations, LUVOIR's coronagraph for Architecture A has a camera mode that collects photometric data across a given channel for the whole field-of-view. Thus, multi-color point-source photometry will be obtained during these multi-epoch observations, providing additional information to discriminate planets from background objects.

**3. Constrain orbits.** Multi-epoch observations will reveal targets with the same proper motion as the host star and place constraints on orbits, with observations likely in the V band. About four detections of the planet spaced out over the orbit are needed to constrain the orbit. Some observations will miss the planet, so on average of 6 visits per star may be needed for orbital determination.

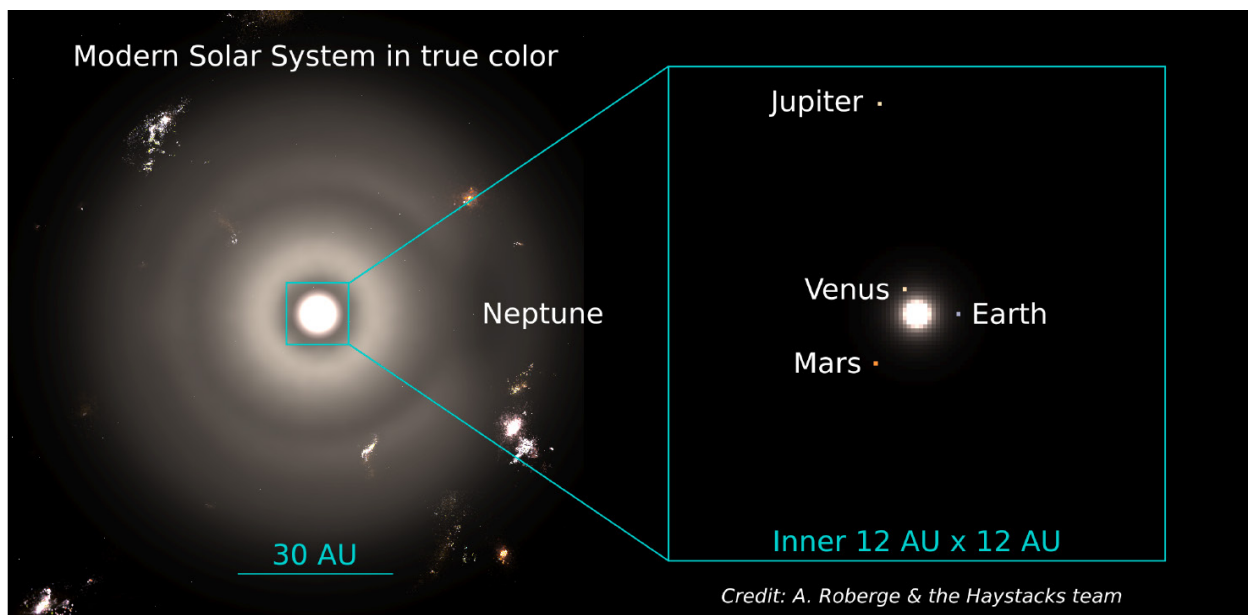

**Figure 3.11.** *Is that a planet? A high resolution simulated view of the solar system at a distance of 10 pc including expected noise and confusion sources: scattered light from interplanetary dust (aka. exozodiacal dust), background galaxies, and background stars. Credit: Roberge et al. (2017)*

*4. Search for water.* A useful wavelength range for initial habitability assessment (i.e., the search for atmospheric water vapor) is 0.87–1.05 μm for two reasons. First, this channel is less likely to be cut off by inner working angle constraints compared to longer wavelength water features. Second, the 0.94 μm $H_2O$ feature is accessible, as is the 1.0 μm $CH_4$ feature, which could provide evidence of an atmosphere rich in $CH_4$ (e.g., an Archean Earth analog or a Jupiter analog). The $H_2O$ and $CH_4$ features do not overlap significantly in this wavelength range, which will remove possible degeneracies for the type of planet under observation. The $H_2O$ band at 1.12 μm, although stronger, overlaps closely with the 1.1 μm $CH_4$ feature, leading to potential misinterpretation of planets with abundant $CH_4$ but no water.

### 3.2.4 Exoplanet yields for the baseline LUVOIR-A concept

Using new exoplanet yield estimation methods (e.g., Stark et al. 2014), we have estimated the quantity and quality of exoplanet science that the LUVOIR mission concept could produce. Here we summarize the exoplanet yield that would result from a habitable planet search with the baseline 15-m LUVOIR-A concept. This initial 2-year campaign includes a survey of hundreds of stars that would take us through Step 4 in **Figure 3.10**, establishing a target list for more detailed characterization to be carried out subsequently (Steps 5–10 in **Figure 3.10**), following the strategy outlined in **Section 3.4**. Note that although this report discusses extending the LUVOIR mission lifetime via servicing, the whole strategy shown in **Figure 3.10** is designed to fit within a 5-year program. In practice, planets to receive more detailed characterization would be selected through competed proposals because LUVOIR is planned as a guest observer facility. A detailed description of the exoplanet science yield calculations, the techniques used, assumptions made, and justification for the adopted observing strategy appears in **Appendix B**.

ExoEarth candidates are defined to be on circular orbits and reside within the conservative HZ, spanning 0.95–1.67 AU for a solar-twin star (Kopparapu et al. 2013). We only consider planets with radii smaller than 1.4 $R_{Earth}$ and radii larger than or equal to $0.8a^{-0.5} R_{Earth}$, where $a$ is semi-major axis for a solar-twin. The lower limit on our definition of the radius of exoEarth candidate comes from an empirical atmospheric loss relationship derived from solar system bodies (Zahnle & Catling 2017). The upper limit on planet radius is a conservative interpretation of an empirically measured transition between rocky and gaseous planets at smaller semi-major axes (Rogers 2015).

We estimate the yield of detected exoEarth candidates for the baseline LUVOIR 15-m mission to be $51^{+75}_{-33}$, where the range is set by uncertainties on $\eta_{Earth}$ and finite sampling uncertainties (see **Appendix B**). If the frequency of habitable conditions is 10%, then 29 exoEarth candidates would be required to guarantee seeing one true exoEarth at 95% confidence. Thus, LUVOIR should be able to detect at least a small number of true "Earths"—or discover that habitable conditions on planets in their stars' habitable zones are rare. Following the detection of habitable planet candidates (i.e., a positive water vapor detection), we will continue to characterize the rest of the spectrum following the strategy in **Section 3.4**.

While searching for and characterizing exoEarth candidates, LUVOIR will detect dozens of other planets, from hot rocky worlds to cold gaseous planets. These other planets are essentially detected "for free" during the exoEarth search. **Figure 3.12** shows the expected exoplanet yields of all planet types when optimizing the observation plan for the detection of exoEarth candidates (green bar). Following the planet classification scheme in Kopparapu et al. (2018), the other planet class types in this figure are: rocky planets (radii of 0.5–1 $R_{Earth}$), super-Earths

(radii of 1–1.75 $R_{Earth}$), sub-Neptunes (radii of 1.75–3.5 $R_{Earth}$), Neptunes (radii of 3.5–6 $R_{Earth}$), and Jupiters (radii of 6–14.3 $R_{Earth}$). ExoEarth candidates are a subset of the rocky and super-Earth planets within their stars' habitable zones. "Hot" planets (red bars) receive 182 to 1x Earth's insolation; "warm" plants (blue bars) receive 1 to 0.28x Earth's insolation; "cold" planets (ice blue bars) receive 0.28 to 0.0035x Earth's insolation.

**Figure 3.13** summarizes the expected target list LUVOIR's search program. LUVOIR will observe ~260 stars covering a wide variety of spectral types, primarily FGK stars with some M dwarfs and a few A dwarfs. Target stars will be within 25 pc; most will be brighter than 7th magnitude. LUVOIR will achieve a high average HZ completeness, and the majority of its time would be spent observing planets around high completeness stars.

In reality, yields may vary from the expected values shown in **Figure 3.12** due to astrophysical uncertainties and the actual distribution of planets around nearby stars. The yield uncertainties shown in **Figure 3.12** are estimated as the root-mean-square combination of the NASA Exoplanet Exploration Program Analysis Group SAG13[1] 1-σ exoplanet occurrence rate uncertainties and the uncertainty due to the random distribution of planets around individual stars. The latter was estimated by assuming that planets are randomly assigned to stars, such that multiplicity is governed by a Poisson distribution, and that each observation represents an independent event with probability of success given by that observation's completeness. Error bars do not include uncertainty in the median exozodi level, any uncertainties in mission performance parameters, or uncertainty in observational efficiency.

1 https://exoplanets.nasa.gov/exep/exopag/sag/

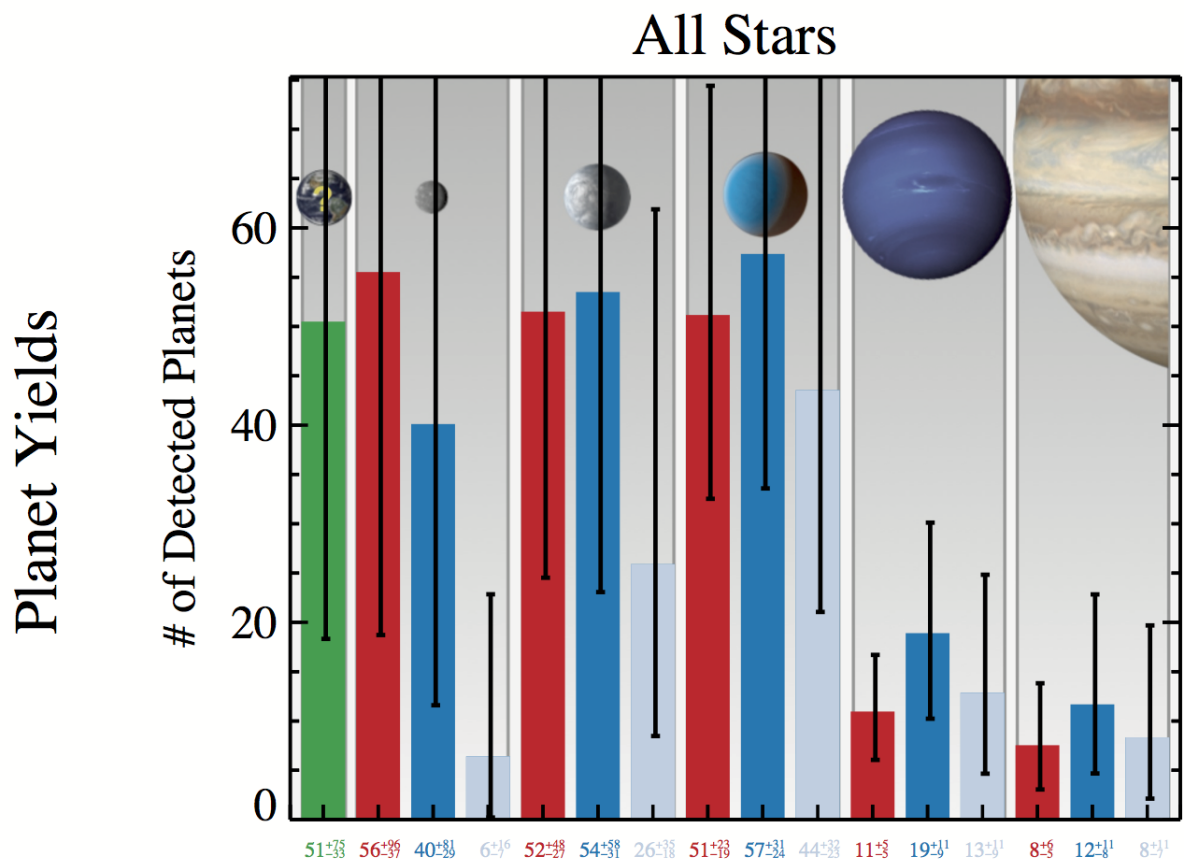

**Figure 3.12.** *Exoplanet detection yields for different classes of planets from an initial 2-year habitable planet survey with the LUVOIR-A concept. Red, blue, and ice blue bars indicate hot, warm, and cold planets, respectively. The green bar shows the expected yield of exoEarth candidates, which are a subset of the warm rocky and super-Earth planets. Planet class types from left to right are: exoEarth candidates, rocky planets, super-Earths, sub-Neptunes, Neptunes, and Jupiters. Planet types other than exoEarth candidates are detected "for free" during the 2-year search campaign. Color photometry and orbits are obtained for all planets. Two partial spectra are obtained for all planets in systems with exoEarth candidates. Credit: C. Stark (STScI)*

The uncertainties for the cold planet yields are likely underestimated, as the SAG13 occurrence rates are purely extrapolations in this temperature regime.

***The role of precursor/follow-up knowledge for exoplanet direct imaging.*** Counter-intuitively, precursor knowledge of the host stars and orbits of exoEarth candidates does not greatly increase the total number of such planets discovered and characterized, in the case where a direct observation mission is target-limited rather than time-limited. This is the situation for both LUVOIR observatory concepts, A and

B. However, such precursor knowledge is of great value for increasing the observing efficiency of exoplanet direct observation missions. A full explanation of this issue will be provided in Stark et al., in preparation. Here we summarize the main conclusions about exoplanet precursor and supplemental studies for future missions like LUVOIR and HabEx.

1) Orbit determination is required to quantify the energy fluxes incident on an exoplanet from its host star. This is particularly necessary to show that an exoplanet is located in its star's HZ. If not done in

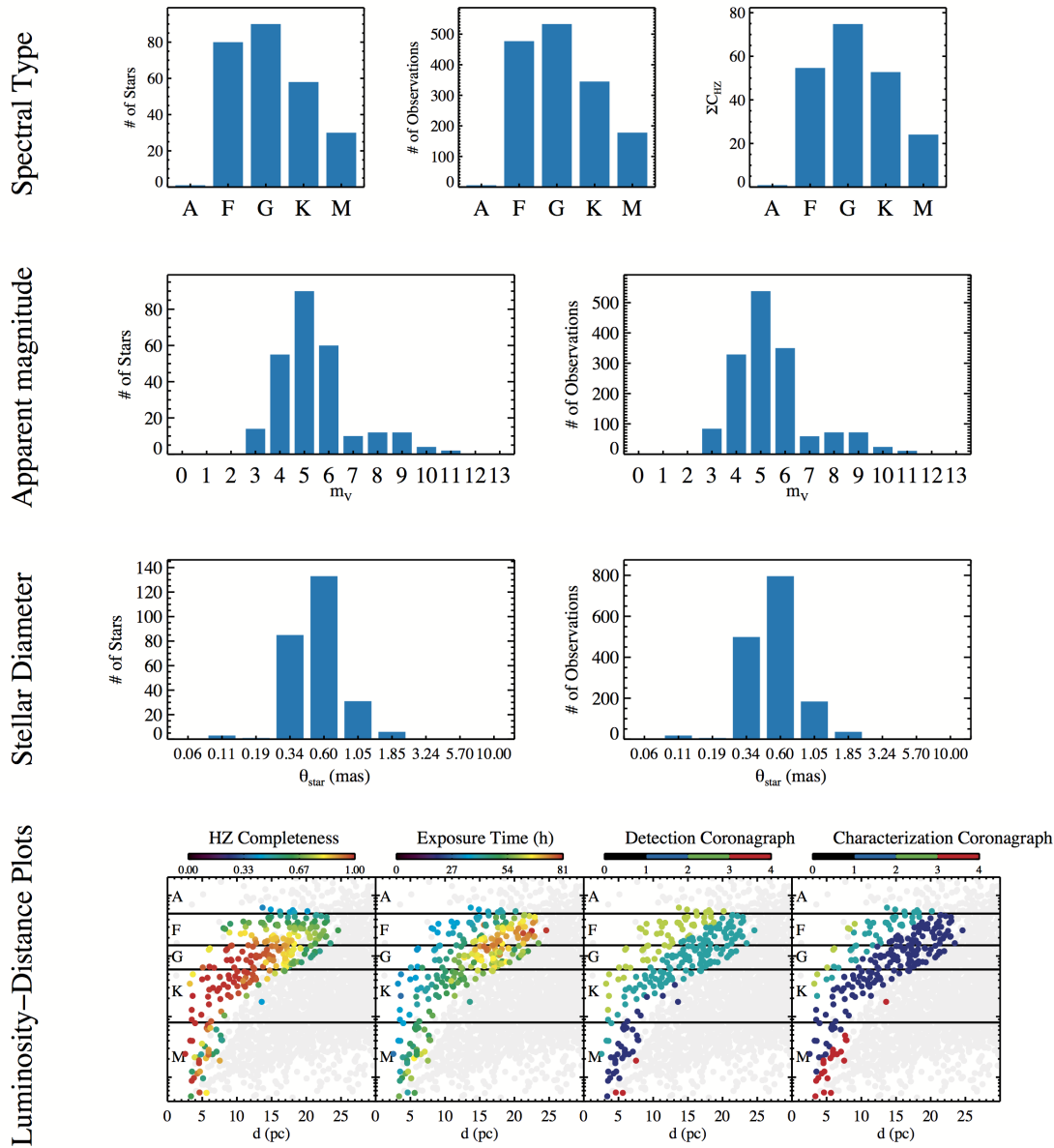

**Figure 3.13.** *Summary of targets selected for observation in the exoEarth search campaign for the LUVOIR 15-m concept. The bottom right panels show which coronagraph was used for each star: 1=Small-IWA APLC, 2=Medium-IWA APLC, 3=Large-IWA APLC, 4=Vector Vortex. Credit: C. Stark (STScI)*

advance of direct observations, then it should be done concurrently.

2) Mass measurements are highly desirable in order to understand an exoplanet's

bulk properties and to constrain the effects of surface gravity on the observed atmospheric spectrum. This information could be obtained either through precursor, concurrent, or follow-up work.

**Program at a Glance – Detecting Habitable Exoplanet Candidates**

**Goal:** Detect habitable exoplanet candidates around a range of FGKM stars.

**Program details:** Observations of nearby stars in target list to search for water-bearing habitable zone rocky exoplanets

**Instrument(s) + Configuration:** ECLIPS coronagraphic imaging and spectroscopy

**Key observation requirements:** Contrast < $10^{-10}$; photometry near 500 nm to detect planets; spectroscopy at R=70, SNR=5 to detect water near 900 nm

3) Precursor observations that identify stars with detectable exoplanets (especially those in the HZ), and which develop orbital ephemerides sufficient to predict when the targets could be best observed, are desirable for mission planning. This information could reduce the mission resources (e.g., integration time) devoted to exoplanet searches and thus save them for other uses.

4) The number of HZs (or ice lines) accessible to a given imaging mission architecture is a 1:1 function of the inner working angle of that architecture. When the mission has sufficient observing resources to study all the exoplanets of interest outside its design IWA, it is target-limited and the effect of prior knowledge on mission yield is small. When there are more accessible targets outside the IWA than mission resources to study them, the effect of prior knowledge on mission yield may be substantial.

5) The primary targets of direct observation missions are stars with HZs outside the design IWA. Precursor knowledge showing the absence of a HZ planet may be used to de-prioritize certain target stars. However, a target's value to comparative planetology and planet formation studies should be taken into account before excluding it from observation.

The scope of the observing program(s) required to address Items 1–3 depends on whether these measurements are done prior to a mission, concurrently, or afterwards. Precursor observations would need to survey all primary targets accessible to the mission design, whereas concurrent or follow-up observations would only need to study the subset of stars with planets of interest. Exoplanet orbits may be determined in advance with ground-based radial velocity instruments or concurrently from a series of exoplanet direct imaging observations with LUVOIR/HabEx. Either ground-based radial velocity instruments or a dedicated space astrometry mission could make exoplanet mass measurements. It may also be possible to make the necessary mass measurements by equipping the direct imaging mission with the appropriate astrometric or radial velocity instrument. A dedicated astrometry mission would likely be the most expensive option to fulfill Items 1–3, while ground-based radial velocity would likely be the least expensive (with other options falling in between). However, a factor of 5–10 improvement in radial velocity measurement accuracy on Sun-like stars will be needed before this method, either from ground or space, could measure the masses of rocky exoplanets.

## 3.3    Signature science case #2: Which worlds are inhabited?

LUVOIR will provide an unprecedented capability to search for life beyond the solar system by characterizing the surfaces and atmospheres of a large sample of terrestrial planets in the HZ and searching for signs of life (**Figure 3.14**). A planet-wide biosphere can modify its environment to leave characteristic indications of its presence such as specific gases (Des Marais et al. 2002; Sagan et al. 1993; Schwieterman et al. 2018), atmospheric chemical disequilibrium (Hitchcock & Lovelock 1967; Krissansen-Totten et al. 2018, 2016a), surface reflectance features (Gates et al. 1965; Hegde et al. 2015; Schwieterman et al. 2015; Seager et al. 2005) and time-dependent modification of environmental characteristics caused by biological processes (Meadows 2008, Olson et al. 2018). Not all features produced by life will be detectable; LUVOIR will only be able to detect the subset of biosignatures that produce absorption features in its wavelength range. In addition to detecting potential biosignatures, it is important to gather additional information so that biosignatures can be interpreted in the context of the whole planetary and stellar environment, a process that can help rule out abiotic processes called "false positives" that can mimic biosignatures. Biosignature assessment is then completed with a search for other corroborating contextual signs of habitability and life. Detection and interpretation of biosignatures and environmental features requires high-contrast, high sensitivity, moderate resolution spectroscopy, and a broad wavelength range to detect multiple spectral features from both potential biosignatures and telltale false positive indicators. A large mirror enables a wider wavelength range to be obtained for any given IWA, and the IWA (not the OWA) is what limits the wavelength range that can be observed for most habitable zone planets at distances accessible to LUVOIR. LUVOIR's survey of possibly dozens of nearby exoEarths offers the best chance in the coming decades to answer at long last

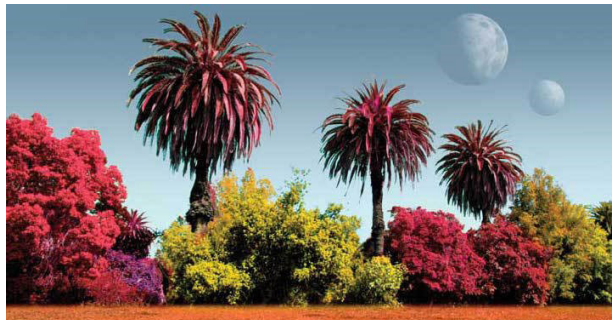

**Figure 3.14.** *LUVOIR will provide a statistical search for life on nearby exoplanets. In the 2040s, LUVOIR will offer the best chance of answering the question "Are we alone in the universe?" Exotic photosynthetic life may generate novel surface reflectance biosignatures, as suggested in this image, and LUVOIR's direct observing capabilities can search for these types of novel photosynthetic pigments. Credit: NASA*

whether we are alone in the universe, and to place Earth's biosphere in context.

### 3.3.1 The search for biosignatures

There are several target biosignatures to be sought, and also signs of disequilibrium chemistry that may allow us to constrain surface fluxes of gases that are unlikely to be produced by abiotic planetary processes. Absorption features from biosignatures and biosignature false positive discriminators are provided in **Table 3.2**. Key environmental characteristics to be sought for interpretation of biosignatures include planetary mass/size/density, atmospheric mass and composition, the nature of the planetary surface, and stellar characteristics including spectral type, UV activity, and age.

While the specific target gases identified for Earth (**Table 3.2**) will be searched for, a more generic search for biosignatures is to determine fluxes (i.e., flow rates) of gases into the environment and constrain sources and sinks for those gases. This can be done by searching for environmental characteristics that are out-of-equilibrium, implying a constant replenishment of a gas against chemical or photochemical destruction. For instance, $O_2$ is out of kinetic equilibrium with $CH_4$ on Earth, implying significant surface fluxes for these gases that cannot readily be explained by abiotic processes (Etiope & Sherwood Lollar 2013; Hitchcock & Lovelock 1967). If no other (abiotic) plausible explanation can be found, the gases in question could represent unexpected biological processes and should be explored further. Below we describe specific target biosignature gases.

**Table 3.2.** *Desired spectral features for biosignature assessment. Note some gases relevant to habitability are also biosignatures and so also appear in* **Table 3.1** *(e.g., methane is a greenhouse gas important to habitability that can also be produced by life).*

| Biosignatures & False Positive Discriminants (indicated with *) | | |
|---|---|---|
| Molecules/Feature | 0.2–1.0 μm | 1.0–2.0 μm |
| $O_2$ | 0.2, 0.63, 0.69, 0.76 (strong) | 1.27 |
| $O_3$ | 0.2–0.3 (strong), 0.4–0.5 | |
| $O_4$ $(O_2-O_2)$* | 0.345, 0.36, 0.38, 0.45, 0.48, 0.53, 0.57, 0.63 | 1.06, 1.27 (strong) |
| CO* | | 1.6 |
| $CH_4$ | 0.6, 0.79, 0.89, 1.0 | 1.1, 1.4, 1.7 |
| $N_2O$ | | 1.5, 1.7, 1.78, 2.0 |
| Organic haze | < 0.5 | |
| Vegetation Red Edge | 0.6 (halophile), 0.7 (photosynthesis) | |

*Oxygen ($O_2$).* The highest priority biosignature gas is molecular oxygen (Meadows 2017). $O_2$ is the byproduct of oxygenic photosynthesis, currently the dominant metabolism on our planet and possibly the most productive metabolism for any planet (Kiang et al. 2007b, 2007a). Oxygenic photosynthesis uses cosmically abundant water, atmospheric $CO_2$, and sunlight to create biomass and power life. Oxygen is unusual for biogenic products in having its strongest features in the visible and near-infrared. Because its high abundance results in even mixing throughout the atmospheric column, it produces strong spectral features even above a planet-wide cloud deck. Oxygen (or its photochemical byproduct ozone) may have been detectable in Earth's spectrum since the Proterozoic period (2.5 billion years ago–541 million years ago; Segura et al. 2003; Kaltenegger et al. 2007), although recent geochemical results may indicate a significant rise in $O_2$ as recently as 0.8 Gyr (Planavsky et al. 2014). Although the photochemical destruction of $H_2O$ and $CO_2$ may generate abiotic $O_2$, along with telltale spectral signatures of these processes (Meadows 2017; **Section 3.3.2**), discrimination between abiotic and biological sources of $O_2$ can also hinge on its production rate and the nature of $O_2$ sinks in the environment. Consequently, constraints on atmospheric and surface composition are required to determine if the biosignature gas has the measured concentration because there is a large surface flux of the gas in an environment that has strong sinks for it, or if there is a weaker, possibly abiotic source of the $O_2$ in an environment with fewer sinks.

*Ozone ($O_3$).* Ozone is a photochemical byproduct of $O_2$. Because it produces UV spectral features that are detectable at low $O_2$ levels that render $O_2$ itself spectrally invisible (e.g., Domagal-Goldman et al. 2014), it can be used to infer atmospheres with low (e.g., low Proterozoic-like) oxygen concentrations (**Figure 3.2**). As a photochemically produced gas, its production rate also depends on the UV spectrum of the host star in addition to the amount of $O_2$ in the atmosphere. The very strong $O_3$ Hartley-Huggins band can be detected between 0.2–0.35 μm. In atmospheres with low oxygen levels, $O_3$ concentrations are highly responsive to changes in $O_2$ levels. Over the course of a planet's orbit, changes in seasonally-dependent biological $O_2$ production would induce changes to strength of the UV ozone spectral band as a type of temporal biosignature (Olson et al. 2018). The weaker Chappuis band extends between 0.4–0.7 μm and is prominent in the spectrum of modern Earth that has abundant $O_3$ (Krissansen-Totton et al. 2016b).

**Methane ($CH_4$).** Methane can be produced by biological and geological processes, with biological processes dominating on Earth. The ratio of biological to abiotic production rates on Earth is possibly close to 65:1 (Etiope & Sherwood-Lollar 2013; IPCC 2007). Methanogenesis is a simple anaerobic metabolism that produces $CH_4$ from $CO_2$ and $H_2$, and is primordial on Earth (Ueno et al. 2006; Woese & Fox 1977). Almost all abiotic mechanisms that produce $CH_4$ on an Earthlike planet involve water-rock reactions (Etiope & Sherwood-Lollar 2013), and so methane in an atmosphere can also indicate habitability by implying the existence of liquid water. Methane's 1.4 and 1.7 μm bands are weakly detectable at modern $CH_4$ concentrations, and it has other shorter wavelength bands that are possibly detectable at higher Archean-like (e.g., ~0.1% of the atmosphere; 4–2.5 billion years ago) concentrations (**Figure 3.2**). $CH_4$ also shows strong bands between 0.8–1.4 μm for a modern Earth-like planet orbiting M dwarf stars like Proxima Centauri b (Meadows et al. 2018), and possibly also for planets orbiting later K dwarfs (Arney et al. in prep.) due to the longer photochemical lifetime of methane around these types of stars. Interpretation of methane as a biosignature, even more so than oxygen, hinges on understanding its sources and sinks, which will require thorough characterization of the environment coupled to modeling. Large methane fluxes may imply biology (Guzmán-Marmolejo et al. 2013). Simultaneous detection of $CH_4$ and $O_2$ will strengthen the biosignature kinetic disequilibrium argument. For anoxic planets, simultaneous detection of abundant $CH_4$ and $CO_2$ (i.e., $CH_4$ mixing ratios > $10^{-3}$) indicates an atmosphere in chemical disequilibrium, and therefore a potential biosignature (Krissansen-Totton et al. 2018).

**Nitrous Oxide ($N_2O$).** Nitrous oxide is produced by life, and it may have been present in large quantities during parts of the Proterozoic eon (Buick 2007; Roberson et al. 2011). Known abiotic sources are minor, and include lightning and reactions involving dissolved nitrates in hypersaline ponds (Samarkin et al. 2010). Nitrous oxide's strongest bands occur longward of the LUVOIR wavelength range; other bands exist between 1.4 and 2 μm, but these bands are weak and tend to overlap with other gases that are likely to be present (such as $H_2O$ and $CO_2$) and would therefore require either very high abundances, or high signal-to-noise data to recover.

**Other potential biosignatures.** To increase the chances of detecting life, especially given the likely diversity of terrestrial planetary environments and considering the varied dominant life forms throughout our own planet's history, other biological impacts on planetary environments should be sought in the UV-visible-NIR wavelength range. These may include atmospheric disequilibria, hazes, surface reflectivity, and temporal variations in atmospheric composition or surface albedo due to life processes.

**Hazes.** Hazes are photochemically produced particles. Organic haze produces a strong, broadband UV-blue absorption feature. Organic haze is generated primarily via the photolysis of $CH_4$, but can only form in the presence of Earth-like $CO_2$ amounts if $CH_4$ source fluxes are at vigorous levels consistent with biological production rates (Arney et al. 2018). Additionally, in the presence of other methyl-bearing biological gases such as dimethyl sulfide ($C_2H_6S$) and methanethiol ($CH_3SH$), organic haze can form at a lower $CH_4/CO_2$ ratio than for atmospheres where the methyl groups are solely sourced from $CH_4$, strengthening the interpretation of biological involvement. Similarly, the formation of sulfur aerosols such as $S_8$ (which produces a blue-UV absorption feature akin to organic haze) is impacted by the $H_2S/SO_2$ ratio (Hu et al. 2013), and $H_2S$ is produced by biological and geological processes. Constraints on the $H_2S$ flux required to produce an $S_8$ haze in a given atmosphere may suggest the involvement of biology if high production levels are inferred.

**Surface reflectivity.** Leaf structure produces a steep increase in reflectivity near 0.8 µm known as the "red edge" (Gates et al. 1965; Seager et al. 2005), which produces a <10% modulation in Earth's disk-integrated brightness at quadrature. Other biological pigments can also produce strong reflectivity signatures (Hegde et al. 2015; Schwieterman et al. 2015). Reflectance edges, which are discontinuities in planetary reflectance as a function of wavelength, could occur through the visible and NIR, depending on the pigmented and/or photosynthetic organism producing the signature. Unexpected reflectance signatures that do not match abiotic compounds could therefore suggest life.

**Temporal changes.** On Earth, life produces seasonal changes in vegetation coverage and albedo, and periodic changes in gas abundances. $CO_2$ concentrations change seasonally (~10 ppm) due to changes in temperature and sunlight intensity, which modulate photosynthetic drawdown of $CO_2$ in the spring and summer, and its release back to the atmosphere with vegetation decomposition in autumn and winter (Keeling et al. 1976). Methane also changes seasonally by ~10–20 ppm. At low $O_2$ levels, seasonal $O_2$ variations might be discerned from variability in $O_3$ (Olson et al. 2018). Seasonal variations like these could also be sought in exoplanet spectra (Meadows 2008; Schweiterman et al. 2017), but would require very high precision, high signal-to-noise spectral data to detect changes on this Earth-like scale, as well as observations at multiple points during the planet's orbit to capture any seasonal variability.

### 3.3.2   False positives for life

Any claims of evidence of life beyond Earth will be met with appropriate scrutiny from the scientific community. Attempts will be made to determine ways in which the claimed evidence for life could be produced without invoking extraterrestrial biology. The exoplanet astrobiology community has spent considerable effort simulating ways in which non-biological processes could mimic proposed exoplanet biosignatures, in particular atmospheric $O_2$ and $O_3$. These efforts have identified multiple such "false positives" for biologically produced $O_2$ and $O_3$, as well as the observations required to discriminate between them and the "true positives" associated with global biological $O_2$ production (see Meadows 2017 for a review).

All of the known abiotic $O_2$ and $O_3$ generation mechanisms rely on photolysis of water vapor or $CO_2$ and a dramatic shift in the redox state of the planet's atmosphere. These changes to the bulk atmospheric composition could be driven by H loss caused by photolysis of high-altitude $H_2O$ on planets without a water "cold trap" (Wordsworth & Pierrehumbert 2014) or by massive H loss from the planet as a result of high stellar luminosity early in the star/planet history during the pre-main sequence phase (Luger & Barnes 2015). This latter process can potentially build up atmospheres with hundreds of bars of abiotic $O_2$, but is probably only relevant to planets orbiting M dwarf stars.

Preferential loss of H from the planet's atmosphere raises the planet's O/H ratio. This sets up favorable conditions for photochemical accumulation of abiotic $O_2$ and $O_3$, as the lack of H atoms will decrease the rate at which $O_2$ and $O_3$ are chemically destroyed. These slow destruction rates allow $O_2$ and $O_3$ to accumulate despite the relatively slow production rates. In an extreme case, where the planet is devoid of $H_2O$ but relatively high in $CO_2$, both $O_2$ and $O_3$ can build up in the atmosphere, as the recombination of photolyzed $CO_2$ is slow (Hu et al. 2012; Tian et al. 2014; Harman et al. 2015; Gao et al. 2015). Planets with $H_2O$ can also accumulate potentially detectable amounts of $O_3$, if they have limited reservoirs of reduced (H-bearing) gases such as methane ($CH_4$) in the atmosphere and are orbiting F- or active M-type stars (Domagal-Goldman et al. 2014; Harman et al. 2015).

The most direct ways to rule out all of the false positives for biological $O_2$ production within LUVOIR's wavelength range are to 1) constrain the total abundance and atmospheric pressure of $O_2$ to rule out the ocean loss scenario (Schwieterman et al. 2016) and 2) constrain the atmospheric redox state by detecting the presence of H-containing species in the atmosphere, specifically $H_2O$ and $CH_4$. The detection of significant quantities of $H_2O$ would indicate an atmosphere with at least some H atoms and would make many of the abiotic $O_2$ accumulation scenarios unlikely. The detection of $CH_4$ would indicate substantial atmospheric H, a significant flux of reduced gases to the atmosphere, rapid chemical destruction of $O_2$ and $O_3$, and therefore kinetic disequilibrium, which is thought to be one of the strongest remotely detectable biosignatures (Hitchcock & Lovelock, 1967).

In some circumstances, detection of $CH_4$ together with $O_2$ may be difficult. The detection of modern-day Earth's $CH_4$ concentrations will be challenging if it is not possible to access wavelengths longer than 1.5 μm due to IWA constraints. Other host stars, such as M dwarfs (Rugheimer & Kaltenegger 2017; Segura et al. 2005) and late K dwarfs (Arney et al. in prep), may increase detectability of planetary $CH_4$ and $O_2$ simultaneously at wavelengths shorter than 1.5 μm, due to the longer photochemical lifetime of $CH_4$ around these stars. For situations where $O_2$ and $CH_4$ cannot be accessed simultaneously, the different mechanisms for generating $O_2$ and $O_3$ abiotically would each have to each be ruled out individually.

To rule out all known abiotic $O_2/O_3$ generation processes, LUVOIR will:

1. Search for the presence of other O-bearing molecules in the atmosphere (e.g., $CO_2$, $SO_2$) to estimate the potential for these gases to be photolyzed and generate $O_3$ abiotically (**Figure 3.15**).

2. Measure the spectral properties of the star in the ultraviolet to understand the energy available to photolyze O-bearing gases and drive H escape.

3. Search for the byproducts of the photolysis of O-bearing gases (e.g., CO).

4. Search for spectral features associated with water and water clouds, to determine if the planet has a cold trap that would prevent H-loss associated with high stratospheric water vapor content.

5. Constrain the potential for a cloud deck by estimating the pressure of the atmosphere. This could be done by accurately constraining the Rayleigh scattering slope and presence of aerosols, modeling the shapes of well-resolved absorption features, or detecting the presence of collisionally induced absorption (CIA) features (e.g., $O_2$-$O_2$; Misra et al. 2014b).

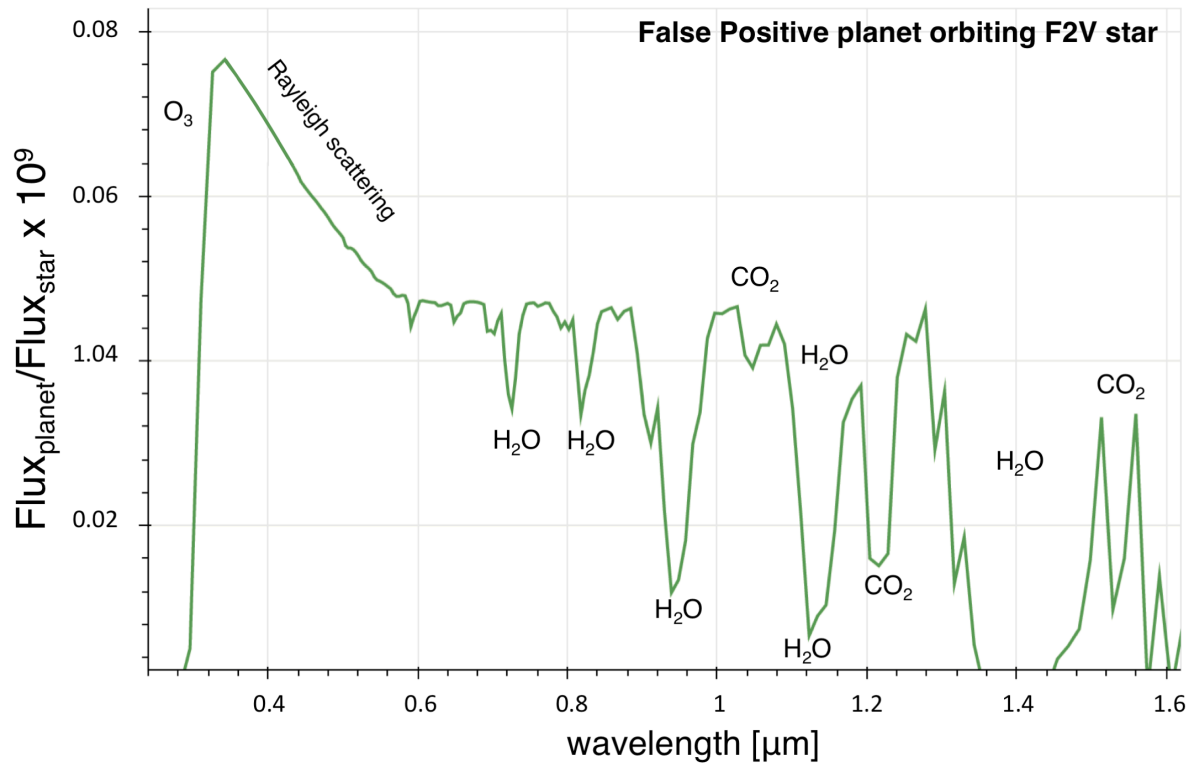

**Figure 3.15.** *A planet with detectable ozone but no life. This spectrum shows a model of an Earth-size exoplanet orbiting a F2V star. The abundant $O_3$ is photochemically produced from non-biological $O_2$ (Domagal-Goldman et al. 2014). Such a world could be mistaken for an inhabited Proterozoic Earth-like planet (**Figure 3.2**). However, the strong $CO_2$ features present in this spectrum suggest the abiotic nature of the oxygen, as a $CO_2$-rich atmosphere has a high likelihood of generating oxygen through photochemistry around this type of star. Credit:* LUVOIR tools */ S. Domagal-Goldman (NASA GSFC)*

6. Look for $O_2$-$O_2$ CIA features near 1.06 and 1.27 μm that would indicate a massive $O_2$ atmosphere more likely to be due to ocean loss than a biosphere.

This combination of measurements will be able to identify or rule out all of the known false positive mechanisms by detecting observable features of these processes and estimating the efficiency with which they could lead to $O_2$ and $O_3$ accumulation.

The other way to strengthen a potential biosignature is to identify independent, secondary features from life. These will be particularly strong if they are associated with the same biological process as the first biosignature. For example, the $O_2$ and $O_3$ in Earth's atmosphere come from oxygenic photosynthesis. LUVOIR could also detect the seasonal drawdown and release of $CO_2$, and/or surface reflectance features like the vegetation red edge and pigments, both of which are associated with organisms that utilize oxygenic photosynthesis. No abiotic spectral mimic is known to exist for Earth's vegetation red edge (Schwieterman et al. 2017). However, this "edge feature" may exist at other wavelengths on exoplanets,

**Earth twin at 5 pc with LUVOIR-A, 15 hours per coronagraphic bandpass**

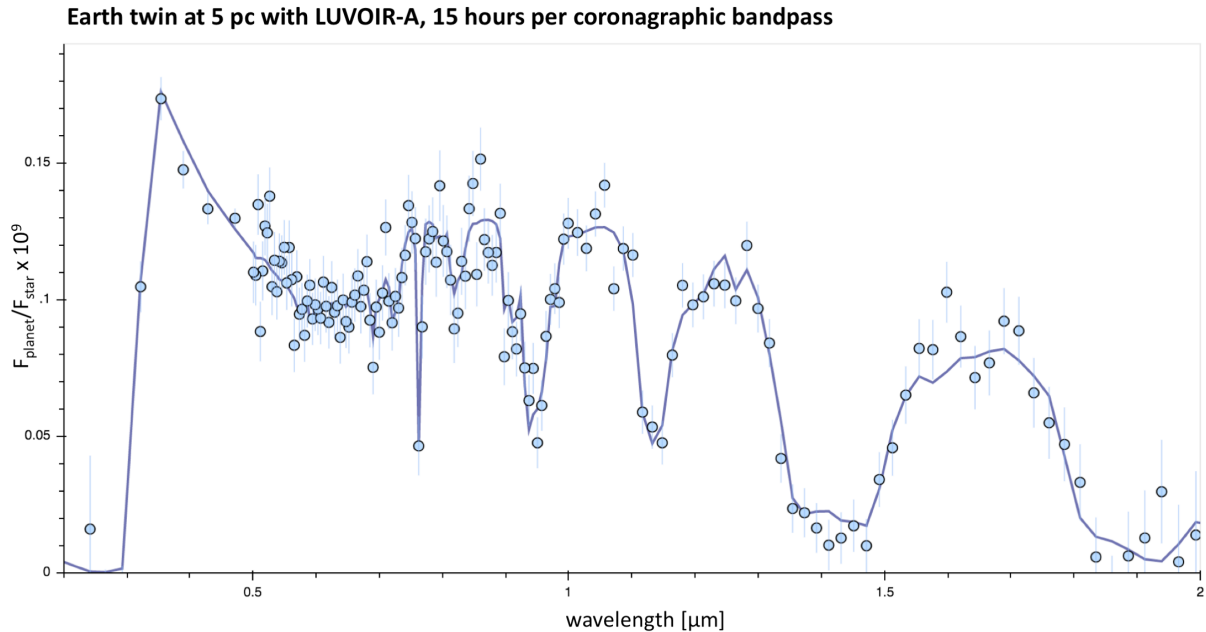

**Figure 3.16.** *The spectrum that could be obtained with LUVOIR-A for an Earth twin orbiting a Sun-like star at 5 pc in 15 hours per coronagraphic bandpass, or roughly four days of total observing time, assuming that two channels can be observed concurrently. This integration time is sufficient to obtain SNR = 10 on the continuum in the V band. Although the planet is dim in the UV, low-resolution photometry in the NUV channel (as shown here) allows the drop off from $O_3$ absorption to be detected. Credit:* LUVOIR tools */ G. Arney (NASA GSFC) / T. Robinson (Northern Arizona University) / J. Lustig-Yaeger (University of Washington).*

based on the spectral properties of the stars (Kiang et al. 2007a), and mineral surfaces could mimic these features if they occur at other wavelengths. Cinnabar and sulfur, for example, produce steep reflectance features at 0.6 and 0.45 μm that may mimic shorter wavelength reflectance "edges" (Schwieterman et al. 2017). Any seasonal variations would also need to be distinguished from the effects of abiotic photochemistry.

LUVOIR's large, in-depth survey of numerous potentially habitable exoplanets provides an additional unique pathway to increase the confidence of a detection of extraterrestrial life: the integrated confidence level that biosignatures have been detected on at least one of LUVOIR's targets will rise if the same signals are detected on multiple worlds. For example, if LUVOIR detects the simultaneous presence of oxygen and methane on one rocky world in the habitable zone, this could be explained by a transient phenomenon or by a binary planet system, with one $CH_4$-rich world and one $O_2$-rich world (e.g., Rein et al. 2014). Because both of these scenarios are highly unlikely, detection of multiple planets showing the simultaneous presence of $O_2$ and $CH_4$ would provide significantly stronger evidence that we are not alone in the universe than the detection of $O_2$ and $CH_4$ in a single planet.

## 3.4    The strategy for habitability and biosignatures confirmation

LUVOIR will enable the comprehensive assessment of planetary habitability, and a search for signs of life, on dozens of *potentially* habitable exoplanets. However, many potentially habitable planets may be uninhabitable, and some habitable planets

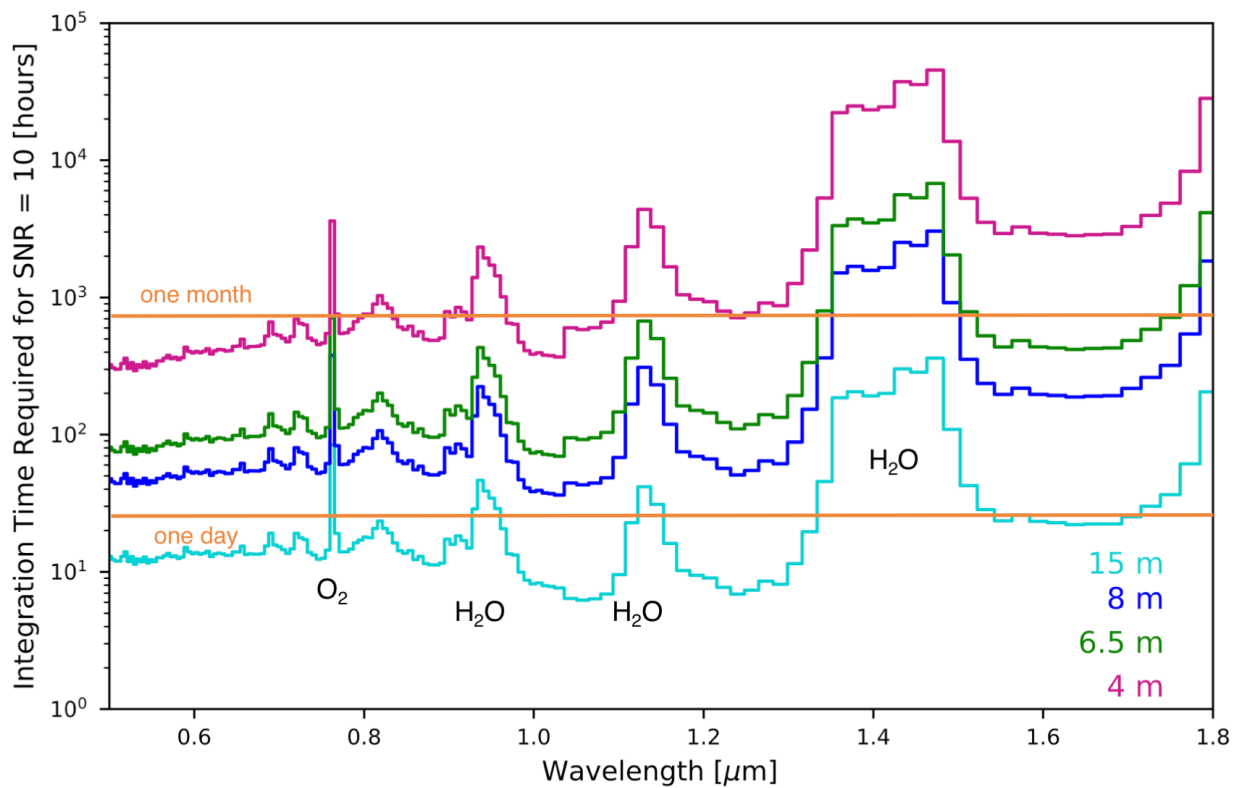

**Figure 3.17.** *Integration times required to obtain SNR=10 on an exoEarth orbiting a Sun-like star at a distance of 5 pc for different-sized coronagraphic telescopes. The calculations assume total system throughput representative of LUVOIR-A with $R_{VIS}$ = 140 and $R_{NIR}$ = 70. Credit: G. Arney (NASA GSFC) / T. Robinson (Northern Arizona University) / J. Lustig-Yaeger (University of Washington)*

may be uninhabited. A high-fidelity spectrum, over a wide wavelength range is insurance against misinterpretation of planetary characteristics.

For planet characterization, we assume that SNR=10 on the planet continuum is required to constrain gas abundances, based on the results of spectral retrieval work using R=140 in the visible channel (Brandt & Spiegel 2014; Feng et al. 2018) and R=70 in the NIR. The R=140 spectral resolution in the visible is driven by the requirement to accurately characterize the $O_2$ A-band, and also to sufficiently retrieve $H_2O$, $O_2$, and $O_3$ abundances (Feng et al. 2018). The NIR resolution is chosen to measure the broad NIR features like $H_2O$ and $CH_4$. Note that LUVOIR-A also includes an option for R=200 in the NIR to resolve the narrow $CO_2$ feature near 1.5 μm. **Figure 3.16** shows what an Earth-twin orbiting a solar analog at 5 pc might look like to LUVOIR-A in 15 hours of integration time per coronagraphic bandpass to obtain SNR > 10 on the continuum for most of the spectrum. **Figure 3.17** shows the exposure times needed to obtain for SNR=10 with a coronagraph as a function of wavelength for an Earth-like exoplanet with different sized coronagraphic telescopes with performance like LUVOIR-A for planets orbiting solar-analog stars 5 pc away.

As a proof-of-concept for the types of observations LUVOIR will be able to perform, here we outline a series of steps (summarized in **Table 3.3**) that could be followed to characterize Earth analogs. The order of these observations may change depending on adjustments to the wavelength ranges of the coronagraph channels and according to prior spectral information that may lead to prioritization of specific gas detections. We continue our discussion from the numbered steps outlined in **Section 3.2.3** for detecting potentially habitable exoplanets (**Figure 3.10**), beginning at Step 5 to pick up where the previous discussion left off. These steps are shown in green text to match the color coding of **Figure 3.10**. Detailed spectral yield calculations are ongoing and will be presented in our Final Report. Good targets may allow for full spectral characterization in approximately several days of observing time (**Figure 3.17**), with additional time for follow-up, higher resolution observations, and/or searches for seasonal variability.

**Table 3.3.** *This table presents a series of spectral observations that could be followed for habitable exoplanet characterization with LUVOIR-A, with increasing wavelength coverage for each step and cumulative additions to the science return. Values in the third column show the number of stars in our target list for which the entire wavelength range needed for a given step is accessible to LUVOIR-A given our IWA and OWA assumptions.*

| Wavelength Range | Purpose | Number of stars in target list with complete wavelength coverage over this range for an orbiting exoEarth |
|---|---|---|
| Characterization Steps | | |
| 0.675–1 µm (Step 6) | Constrain water, search for $O_2$, $CH_4$ | 201 |
| 0.565–1.1 µm (Step 7) | $O_3$, false positive $O_2$ indicators, additional $H_2O$ and $CH_4$ features | 191 |
| 0.4–1.32 µm (Step 8) | Organic haze, $S_8$ particles, additional $O_3$, $O_2$, $H_2O$, $CH_4$, $CO_2$ at high levels | 190 |
| 0.2–1.7 µm (Step 9) | UV absorbers like $O_3$, Earth-like $CO_2$ concentrations, additional $CH_4$, $H_2O$ features | 184 |

*5. Characterize the star and determine planet masses.* A well-characterized host star spectrum is important for understanding the characteristics of orbiting planets. In particular, the UV spectrum of the star and constraints on stellar activity will be important for understanding the potential for planetary atmospheric loss and photochemical processes. The LUMOS instrument on board LUVOIR could be used for these stellar observations.

It will be important to estimate the mass of the planets observed by LUVOIR. If not available via ground-based radial velocity observations, estimates of planet masses could be obtained with astrometry using the HDI instrument on board LUVOIR. We will require extremely precise astrometry (≲ 0.1 µas) to measure the masses of Earth-like exoplanets, and the design of HDI is baselined to meet this goal.

*6. Search for $O_2$ and abundant $CH_4$ and constrain water abundance.* To constrain water abundance and search for $O_2$ and $CH_4$, observations can be obtained over 0.675–1 µm. The presence of $H_2O$ will already be known from initial observations, but higher SNR follow-up will be required to constrain abundance. Key features in this spectral range include: $H_2O$ (0.65, 0.72, 0.82, 0.94 µm), $O_2$ (0.69, 0.76 µm), and $CH_4$ (0.79, 0.89, 1.0 µm, for high Archean-like levels). We consider this a critical wavelength range for spectral characterization due to the availability of habitability markers (water) and key biosignatures (oxygen, methane).

---

**Program at a Glance – Confirming Habitability and Biosignatures**

**Goal:** Detect and characterize inhabited exoplanets.

**Program details:** Obtain NUV/Visible/NIR spectra of potentially habitable planets and search for biosignatures. Measure abundances of biosignature gases and search for surface reflectance biosignatures. Characterize host stars.

**Instrument(s) + Configuration:** ECLIPS coronagraphic spectroscopy to characterize planets; LUMOS UV spectroscopy for host star characterization

**Key observation requirements:** Contrast $< 10^{-10}$; SNR=10 & R=140 to measure $O_2$ at 0.76 μm; R=70 in the NIR channel to measure water and $CH_4$; R=200 in the NIR channel to detect $CO_2$; NUV photometry to detect $O_3$ and other UV absorbers (SNR=5)

---

**7. Search for abundant ozone and false positive $O_2$ indicators.** Next, the spectrum could be extended to bluer and redder wavelengths to include the 0.565–0.685 μm bandpass and the 1.0–1.1 μm bandpass. The extreme case of massive $O_2$ accumulation from high stellar activity in the system's early history (Luger & Barnes 2015) could be ruled out by the absence of $O_2$–$O_2$ CIA features, which become stronger at high $O_2$ levels (e.g., > 10 bars $O_2$). Such features can be present in this spectral range at 0.57, 0.63, and 1.06 μm, in both our visible and NIR bandpasses. Very high $CO_2$ levels that might indicate abiotic photochemical $O_2$ generation may also be detectable through the $CO_2$ feature near 1.05 μm. The long wavelength side of the broad ozone Chappuis band (0.5–0.7μm) can also be seen in the visible bandpasses for planets with modern Earth-like oxygen amounts. Additional important features that could be detected in this wavelength range are $H_2O$ (1.12 μm), and $CH_4$ at high Archean-like abundances (0.6, 1.1 μm).

**8. Search for blue wavelength absorbers, Rayleigh scattering, and other NIR absorbers.** Next, the spectral coverage could be extended to include 0.4–0.565 μm range in the visible, and also to include the 1.1–1.32 μm range in the NIR. The short wavelengths could show signs of blue absorbers including organic haze for methane-rich atmospheres (< 0.5 μm, and haze can be a biosignature), $S_8$ particles for planets exhibiting high levels of volcanic outgassing (0.2–0.5 μm), and the full $O_3$ Chappuis band. In the NIR, an indicator of high $CO_2$ atmospheres is its 1.21 μm band, which is critical for constraining $CO_2$ photolysis as a potential abiotic source of $O_2$ and $O_3$ (Schwieterman et al. 2016). Additional constraints on $O_2$ could come from 1.27 μm band for modern $O_2$ levels.

**9. Observe the UV and search for Earth-like $CO_2$ in the NIR.** By extending the spectrum in the NIR to include the 1.31–1.6 μm bandpass, additional $H_2O$ (1.4 μm) and $CH_4$ (1.4 μm) features can be observed, and importantly, a $CO_2$ band detectable at modern Earth-like concentrations and greater is present near 1.6 μm. This $CO_2$ band can be used to compare to $CH_4$ abundances to determine whether haze or $CH_4$ can be interpreted as biosignatures (Arney et al. 2018; Krissansen-Totton et al. 2018). Methane may be detected near 1.7 μm at modern Earth-like abundances. While NIR observations are obtained, the NUV could also be observed. Planets will generally be dim in the UV, but the spectral falloff due to NUV absorbers could be detected; the low-resolution photometry LUVOIR will

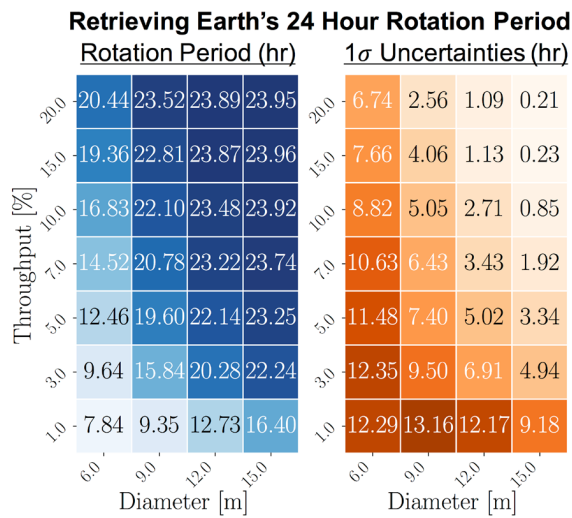

**Retrieving Earth's 24 Hour Rotation Period**

**Figure 3.18.** *Larger telescopes enable more accurate planet rotation rate measurements. This chart shows retrieved rotation periods (left) and estimated errors on the period (right) for an Earth-Sun analog at 5 pc observed across a grid in telescope aperture diameter and total system throughput (Lustig-Yaeger et al. 2018, in prep). Credit: J. Lustig-Yaeger (University of Washington).*

obtain in the NUV improves signal-to-noise. For some atmospheres, like the Proterozoic Earth, weak $O_3$ may be detectable and even quantifiable in the UV because the bottom of the band is not near zero flux and could therefore be measured (**Figure 3.2**, middle panel).

**10. Revisit: Time-dependent and multi-epoch observations.** Photometry of the brightest imaged planets at time-resolved intervals of hours during a single visit can be used to search for time-dependent changes in planet brightness or color which could be indicative of surface or cloud inhomogeneity (e.g., Ford et al. 2001; Cowan et al. 2009). These data could either be taken as multiple photometric images (Step 2 of our characterization strategy; **Figure 3.10**), or extracted from long spectral integrations obtained for other purposes, as long as discrete time intervals in the longer integration could be retrieved at sufficient S/N. Rotational mapping and unmixing models (e.g., Cowan and Strait 2013) can infer the colors and geographic distribution of uniquely identifiable surfaces that contribute to the observed variability with 2–3% hourly multiband photometry spanning a single planetary rotation (see also Cowan et al. 2009). Once a rotational period is determined, variability that deviates from this periodic signal could indicate transient phenomena such as clouds (Pallé et al. 2008). Longer time baseline observations that span a substantial portion of the orbital period will enable searches for seasonal variability (e.g., Olson et al. 2018) and the construction of more precise (two-dimensional) surface maps (Kawahara and Fujii 2010; Fujii and Kawahara 2012).

Large aperture telescopes are essential for time-variability studies. To resolve the rotation period of the planet (**Figure 3.18**), the individual integrations that comprise a time-series observation must sample at significantly shorter time intervals than the rotation period. To that end, the rotation period of Earth can be determined to within 10% for a 15-m telescope with 5% throughput. Smaller aperture telescopes must sacrifice time-resolution for S/N, in effect, eliminating the ability to study the time-variability of rapidly rotating planets.

Obliquity determination can be performed with excellent photometry at a few distinct orbital phases (Schwartz et al. 2016), or with nearly continuous low S/N monitoring (Kawahara 2016). Observations of this quality require a large space telescope like LUVOIR.

As a final note, the combination of the planet's orbit determined in Observational Strategy Step 3 (**Figure 3.10**) and the photometric/spectral observations of the planet throughout as much of its orbit as possible, will provide a comprehensive and powerful dataset to characterize the planetary environment. With these observations, one

### Synergies During the LUVOIR Era

In the next 5–10 years, we may be able to obtain transmission spectra of habitable zone planets orbiting M dwarfs, which are observationally favored for transmission spectroscopy due to their relatively large transit depths and frequent transits. This will constitute our first search for potential biosignatures on potentially habitable exoplanets. This search will likely continue in the LUVOIR era, and this technique may be augmented with a second generation of adaptive optics systems and ELT instrumentation. These advances also may enable ELTs to conduct direct spectroscopy measurements, albeit limited to planets in orbit around M dwarfs.

In addition to a search for signs of life, this research on M dwarf planets will provide intriguing insights into terrestrial planet evolution, as the high stellar activity and early high luminosity of their host stars create a different photochemical environment. However, this same energy source may also cause ocean loss for planets in the habitable zone (see **Section 3.2.1**). This presents a double-edged sword for a biosignature search, as this energy may accentuate the signals from some biosignature gases (such as $CH_4$), but also make atmospheric and ocean retention less likely.

Regardless of the result of these future observations of M dwarf habitable planets, LUVOIR will provide complementary observations on these targets. If biosignatures are detected with ELTs, LUVOIR will confirm the presence of these gases, by detecting them from space and with a complementary shorter wavelength range tied to other molecular lines. LUVOIR can observe transits of the same M dwarf planets targeted by JWST and ground-based telescopes, and it can observe the nearest M dwarf habitable zone worlds in direct imaging. And if planets around M dwarfs are found to suffer catastrophic volatile loss, LUVOIR's UV transit spectroscopy capabilities will probe the atmospheric loss process (**Chapter 4**).

Additionally, LUVOIR will conduct a biosignature search on a complementary set of targets, by observing terrestrial planets orbiting stars like our own Sun. Should prior searches for life prove unsuccessful, LUVOIR will expand our search for life to planets in more familiar stellar environments. And if prior searches for biosignatures prove successful, LUVOIR will expand the diversity of stellar environment for which we can find biosignatures. Either way, LUVOIR will represent a dramatic leap forward in our ability to understand how life is a function of its stellar environment.

can constrain planetary obliquity, determine atmospheric changes in response to long-term variations in incoming and outgoing radiation, and search for longer term variations in surface albedo, cloud patterns and gas abundances that may be indicative of planetary atmospheric dynamics, seasonal changes in climate, planetary volcanism, and other environmental processes.

## 3.5   Summary

LUVOIR is a history-making telescope and offers humanity's best chance of answering the ancient question "Are we alone in the universe?" If there is a global biosphere on a planet orbiting a star in the Sun's neighborhood, LUVOIR will find it by identifying atmospheric and surface features that can only be produced by life. If life exists within our own solar system, LUVOIR will help other missions search for it. In both cases, LUVOIR will place the search for life inside a global context, and within an expanded understanding of planetary processes.

**Table 3.4.** *Summary science traceability matrix for Chapter 3*

| Scientific Measurement Requirements | | | Instrument Requirements | | |
|---|---|---|---|---|---|
| Measurement | Observations | Instrument | Property | Value | Value |
| Find habitable exoplanets | Direct detection of habitable planet candidates and search for water vapor | Direct imaging and spectroscopy of rocky exoplanets in habitable zones to S/N=5 | ECLIPS | High contrast imaging near 500 nm, spectroscopy near 900 nm | Contrast < $10^{-10}$; $R_{vis}$ ~ 70 ($H_2O$) |
| Explore habitability of solar system moons | Long temporal baseline monitoring of icy moons to detect surface changes, study plumes, and measure atmospheric properties | Direct imaging and spatially resolved spectroscopy of moons and plumes | HDI | NUV to NIR imaging | NUV, UBVRI (UVIS channel); JH (NIR channel) |
| | | | LUMOS | FUV multi-object spectroscopy | FUV G120M, G150M, G180M; R ~ 30,000 |
| | | | ECLIPS | Open mask NIR IFS spectroscopy | $R_{NIR}$ ~ 70 *or* 200 |
| Detect and characterize inhabited exoplanets | Abundances of biosignature gases, surface reflectance biosignatures | Direct imaging (S/N=5) and spectroscopy (S/N>10) of exoplanet atmospheres | ECLIPS | High-contrast NUV/VIS/NIR imaging and spectroscopy | Contrast < $10^{-10}$; $R_{vis}$ ~ 140 ($O_2$); $R_{NIR}$ ~ 70 ($H_2O$/ $CH_4$); $R_{NIR}$ ~ 70 ($CO_2$); NUV photometry ($O_3$) |
| Characterize exoplanet host stars | Measure UV spectrum and activity of host stars | Point-source spectroscopy of exoplanet host stars | LUMOS | UV spectroscopy | LUMOS G120M, G150M, G180M, G155M, G300M; R ~ 30,000 |

For worlds in our solar system, LUVOIR's imaging and spectroscopic capabilities will allow global mapping of the atmospheric and mineral composition for worlds large and small. For exoplanets, LUVOIR's large aperture and wide spectral range will provide a rigorous assessment of exoplanet biosignatures as well as their global environmental contexts. This environmental context will include a search for key markers of global habitability, including signs of a hydrological cycle, and the presence of clouds and oceans. LUVOIR will conduct this search for signs of life and habitability on dozens of worlds orbiting a diversity of Sun-like (FGK) stars, representing an expansion on prior and concurrent searches for biosignatures on M-dwarf stars. Finally, LUVOIR will also complement these prior observations of potentially habitable planets around M-dwarfs with space-based observations that will expand the wavelength coverage for transit spectroscopy of these worlds.

Overall, LUVOIR will conduct observations of dozens of potentially habitable worlds around a wide variety of stars. These observations will complement many other facilities, including ground-based telescopes and orbiters/landers to solar system targets. And each of these investigations will provide scientific context for the others; what we learn from LUVOIR's observations of the solar system will influence our interpretation of LUVOIR's observations of exoplanets, and vice versa. As such, LUVOIR will serve as the flagship observatory for a new area of scientific research: *comparative astrobiology*.

# Chapter 4

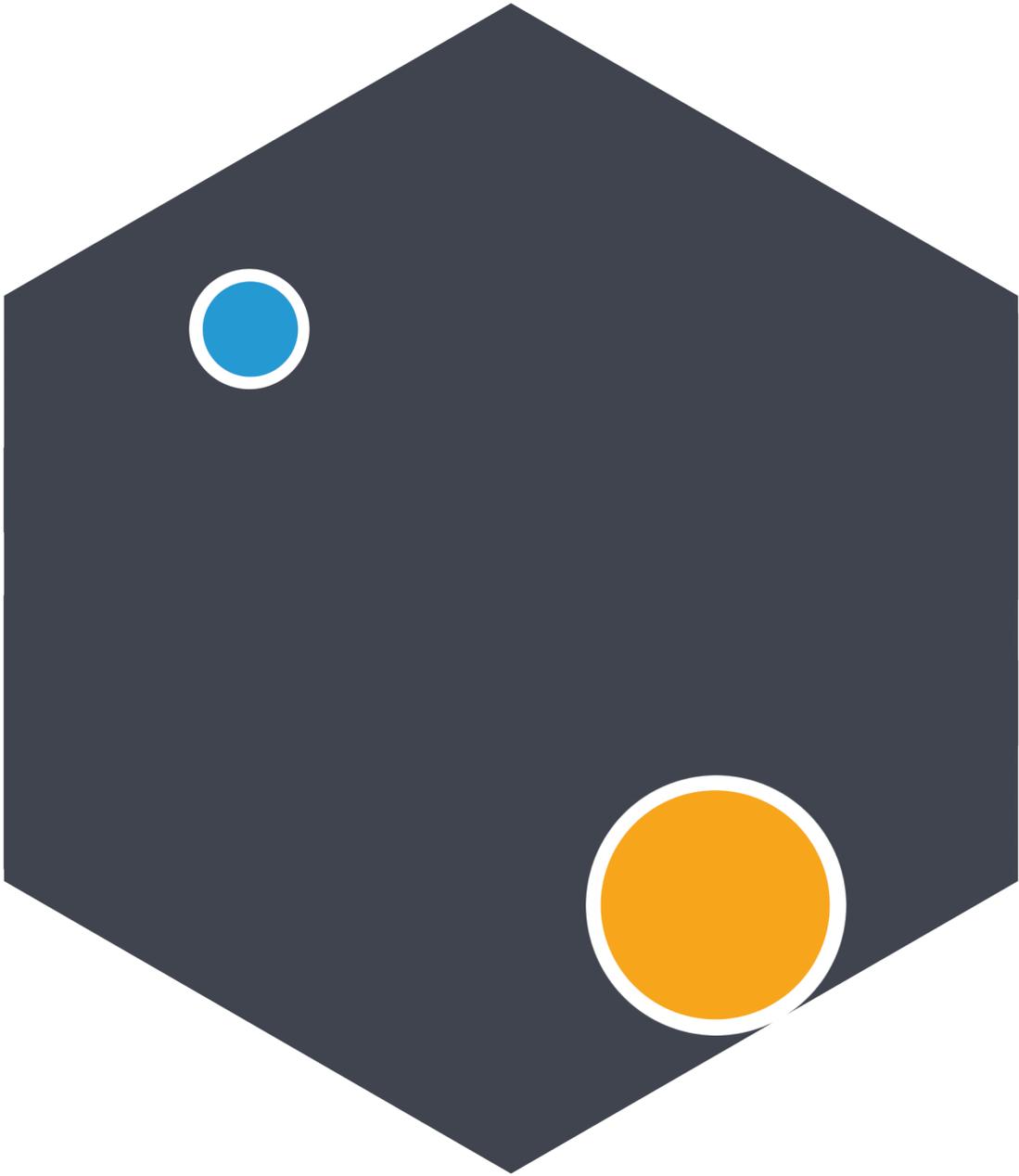

How do we fit in?

## 4  How do we fit in? Planetary systems and comparative planetary science

We learn about planets by comparing them to each other. The atmospheres of Venus, Earth, Mars, and Titan, for example, feature varying mixtures of atmospheric gases, clouds, and hazes. While only one is habitable, our understanding of how planetary surface temperature and climate are influenced by the interaction of atmospheric constituents with incident solar flux and gross planetary properties has been profoundly informed by comparative studies of all these worlds. Atmospheric escape processes, exemplified by the loss of water from Venus, sculpt planetary and exoplanetary atmospheres alike. Recognizing such synergies illuminates our understanding of individual solar system worlds and informs our interpretation of all types of exoplanets, thereby revealing how the processes we understand well from studies of the Earth connect with those of other planets. By studying the processes at work among the stunning diversity of exoplanets (**Figure 4.1**) LUVOIR will exponentially increase the opportunities for comparative exoplanet science.

We also learn by watching as planets are formed and observing the outcomes, including the detritus, of this process. The architecture of planetary systems, how planets are arranged by mass and composition, and the characteristics of debris disks and remnant planetesimals all

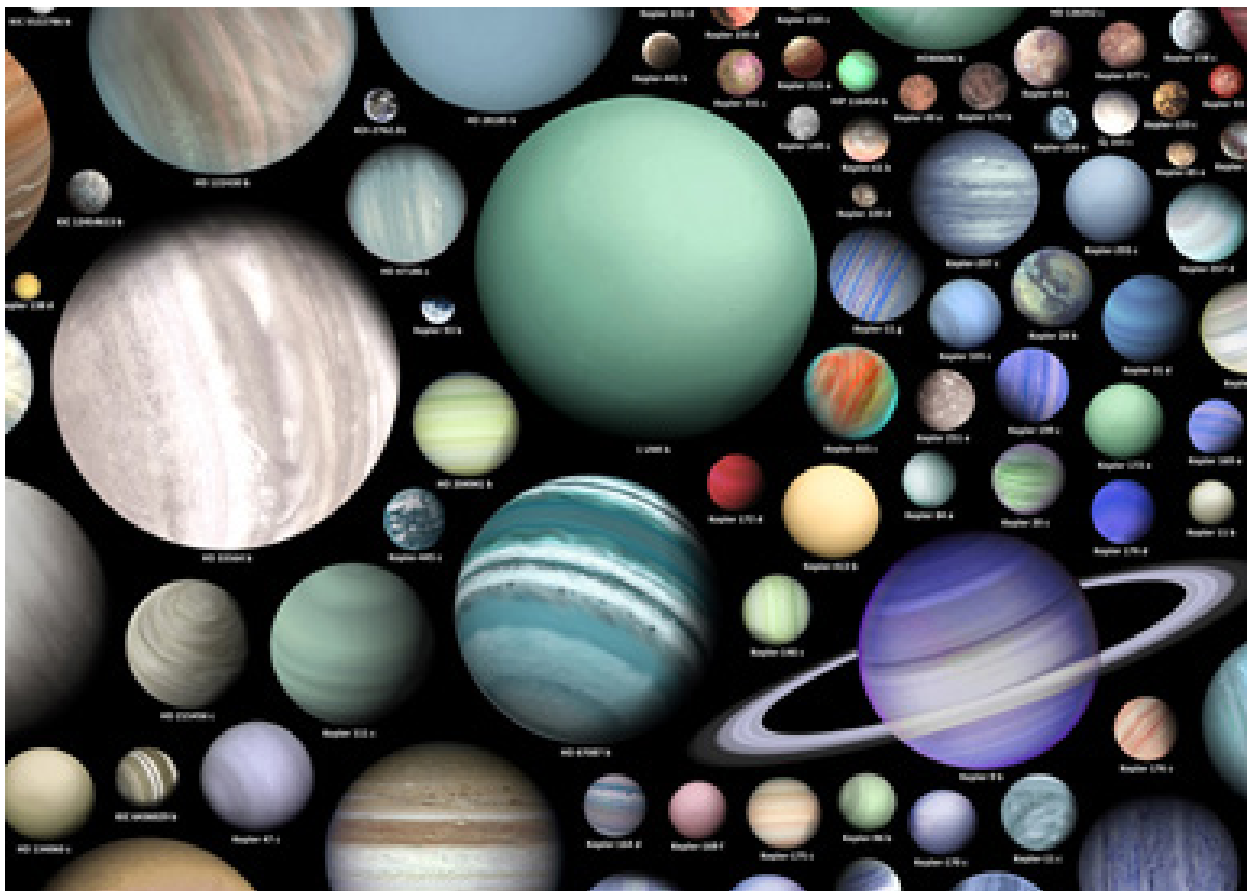

**Figure 4.1.** *Artist's conceptions of some of the 3,706 confirmed exoplanets (as of March 16, 2018). Credit: M. Vargic*

### State of the Field in the 2030s

**Radial velocity:** We expect a factor of ten improvement in ground-based radial velocity measurement precision, to ~10 cm/s. This is size of the signal that the Earth imparts on the Sun when the solar system is observed edge on. Depending on details regarding the stellar jitter and instrumental noise, precise radial velocities provide a possible, but not assured, path to finding and measuring the masses of terrestrial planets in the habitable zones of FGK dwarfs in the LUVOIR era.

**Astrometry:** The ESA Gaia astrometry mission is expected to discover ~21,000 planets between 1–15 Jupiter masses in long period orbits (Perryman et al. 2014). The nearest discovered giant planets, with good mass and orbit constraints, will be excellent targets for LUVOIR comparative planetary studies.

**Millimeter/submillimeter observations:** The Atacama Large Millimeter/Submillimeter Array (ALMA) will study all phases of planet formation, with spectroscopy of molecular and atomic gas and high spatial-resolution (but still 2 to 3 times less than LUVOIR) mapping of cold dust in protoplanetary and debris disks.

**Transit photometry:** Starting in 2018, NASA's all-sky Transiting Exoplanet Survey Satellite (TESS) is expected to detect thousands of relatively nearby transiting planets. The ESA Characterizing ExOPlanets Satellite (CHEOPS) will follow up previously detected exoplanets. The ESA PLAnetary Transits and Oscillations (PLATO) mission, planned for a 2026 launch, has the goal of detecting transiting terrestrial planets.

**Transit spectroscopy:** NASA's James Webb Space Telescope (JWST) will yield detailed characterization of 10–100 close-in exoplanets ranging from super-hot gas giants around all types of stars to temperate terrestrials around low-mass stars. Both NASA and ESA are considering dedicated transit spectroscopy missions for the 2020s (FINESSE and ARIEL, respectively). These missions would build on JWST by enabling a homogeneous and statistical census of the atmospheres of hundreds of close-in giant planets. However, a large sea of Earth- to Neptune-mass planets around solar-type stars will not yet be characterized by the 2030s.

**Microlensing:** The Wide-Field InfraRed Space Telescope (WFIRST) microlensing survey will detect thousands of exoplanets further from their host stars, leading to a statistical census of planets with masses > 0.1 $M_{Earth}$ from the outer habitable zone out to free floating planets.

**High-contrast imaging:** WFIRST will also develop high-performance coronagraphic imaging and spectroscopic technology in space but will only be able to detect a few large exoplanets (gas and ice giants). Second generation instruments on the planned thirty-meter class ground-based telescopes (ELTs) will image thermal emission from young giant planets at much smaller spatial separations and around fainter stars than currently possible, will image and characterize some giants in reflected light in the near-IR, and will detect terrestrial planets in the habitable zones of a handful of nearby low-mass stars.

provide clues to the formation processes that produce the variety of observed planets and their atmospheres. Within the solar system, the bodies of the Kuiper belt contain a frozen record of the formation of the planetary system we understand most thoroughly, our own. By characterizing all types of planets, discovering and observing Kuiper belt objects, and observing circumstellar debris disks, the users of LUVOIR will dramatically enhance our understanding of the process of planet formation.

This chapter explores some of what we can learn from the exoplanetary and planetary science LUVOIR will enable beyond the search for life. We discuss and present a few exemplary LUVOIR science cases relating to the atmospheres of non-habitable planets, planet formation, atmospheric escape, and solar system science, focusing on the value of science enabled by large sample sizes. There are of course countless additional opportunities for exoplanet science with LUVOIR. Many such additional investigations are comprehensively documented in the ExoPAG Science Analysis Group 15 report "Science Questions for Direct Imaging Missions" (Apai et al. 2017), which is an excellent reference for understanding the exciting diversity of exoplanet science beyond the search for habitable worlds. Some additional exoplanet science cases, including transit observations beyond the UV, are described in **Appendix A** of this document.

The "Signature Science" cases discussed in this chapter represent some of the most compelling types of observing programs on a wide range of exoplanets that scientists might do with LUVOIR. As compelling as they are, they should not be taken as a complete specification LUVOIR's future potential in these areas. We have developed concrete examples to ensure that the nominal design can do this compelling science. We fully expect that the creativity of the community, empowered by the revolutionary capabilities of the observatory, will ask questions, acquire data, and solve problems beyond those discussed here—including those that we cannot envision today.

## 4.1   Signature science case #1: Comparative atmospheres

The most compelling exoplanetary characterization discoveries so far have arisen from comparative studies of planets. By revealing the diversity of exoplanet radii, including the stunning ubiquity of sub-Neptune sized planets, as well as unexpected trends in planetary system architectures, the Kepler mission inaugurated era of systematic comparative exoplanetology. LUVOIR will build on these results by thoroughly characterizing the atmospheres of hundreds of directly imaged and transiting exoplanets, over a much wider range of conditions than exist in the solar system, vastly improving our understanding of the fundamental atmospheric processes affecting habitable and inhospitable planets alike.

There will be plenty of opportunities for such studies. The search for exoEarth candidates will uncover hundreds of other types of planets (Stark et al. 2014). **Figure 4.2** illustrates the yield of detected planets of various sizes and temperatures expected to be found in the LUVOIR Architecture A habitable planet search. These additional exoplanets are ice and gas giants at a range of orbital distances from a variety of stellar primaries as well as non-habitable terrestrial and super-Earth sized planets. About 150 sub-Neptune sized planets around a variety of stellar types and at various equilibrium temperatures are expected. For some planets, follow up characterization could be done simultaneously with long exposures to obtain spectra of habitable zone terrestrial planets. In other cases, alternate coronagraph masks and additional exposure time might be needed. For all of these worlds, LUVOIR will enable a host of science programs, ranging from studies of atmospheric dynamics in gas giants to searches for evidence of volcanic gases in the atmospheres of terrestrial

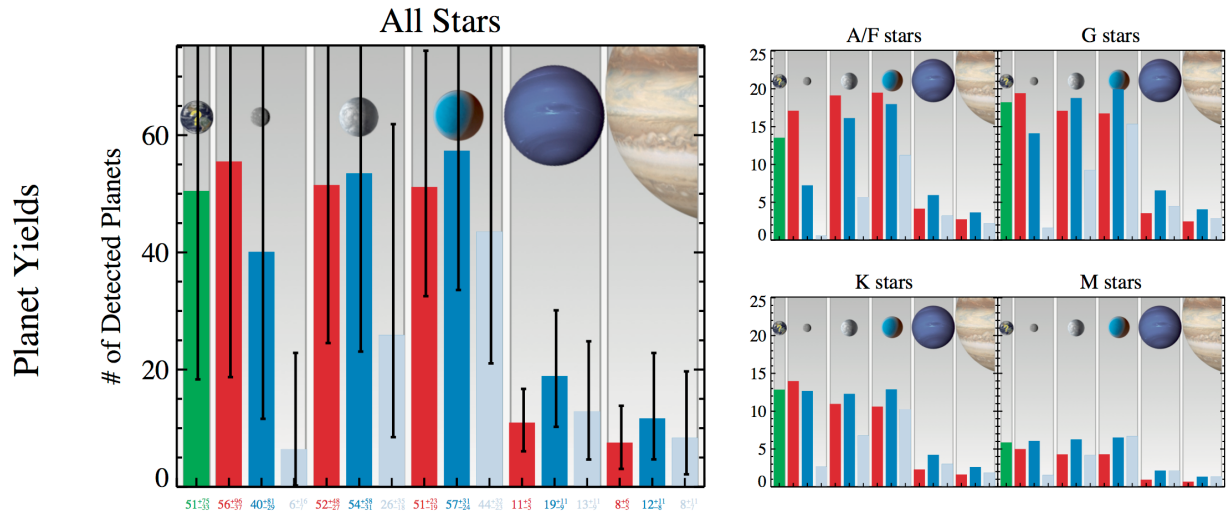

**Figure 4.2.** *Planets of all types detected by LUVOIR Architecture A during a 2-year habitable planet survey. The histograms show the number of planets detected in six size bins: exoEarth candidates (defined in Chapter 3; green bar), rocky planets (0.5–1 $R_{Earth}$), super-Earths (1–1.75 $R_{Earth}$), sub-Neptunes (1.75–3.5 $R_{Earth}$), Neptunes (3.5–6 $R_{Earth}$), and Jupiters (6–14.3 $R_{Earth}$). Red, blue, and ice blue bars indicate hot, warm, and cold planets, respectively. Right panels show the breakdown by stellar type. A detailed description of the exoplanet science yield calculations can be found in* **Appendix B**.

planets. The SAG15 report (Apai et al. 2017) discusses such programs and many others.

The most promising avenue among the wealth of science made possible by this diversity of worlds will be systematic, comparative studies of planetary atmospheres. Many of the more massive planets will have been detected by RV surveys or GAIA, thereby providing masses, while astrometry with HDI will measure or constrain the masses of smaller planets. Planets with known masses, incident fluxes, and measured atmospheric composition will provide a laboratory for testing theories for planet formation, atmospheric evolution, photochemistry, and cloud processes. In this section we briefly summarize some of the highlights of such investigations made possible by LUVOIR.

### 4.1.1 Diversity of composition

LUVOIR will characterize atmospheric diversity among the planets within each system that it observes, systematically measuring atmospheric composition of gas and ice giants, sub-Neptunes, super-Earths, and small rocky planets. Atmospheric composition is of special interest as the particular species present depend on atmospheric temperature and pressure regulated by the incident stellar flux and, for giant planets, emergent planetary flux from the deep interior. The atmospheres of all solar system giants are enhanced over solar abundance, a fingerprint that is generally attributed to the details of planetary formation process **(Figure 4.3**; Mordasini et al. 2016). For terrestrial planets, composition is a signature of initial atmospheric formation as modulated by escape, crustal recycling, and other endogenic and exogenic processes (such as impacts and volcanos).

The visible atmospheres of the relatively cool giant planets probed by direct imaging are directly connected to their deep interiors by a continuous convection zone and the bulk composition of their atmospheres will reflect that of the planet as a whole (Fortney et al.

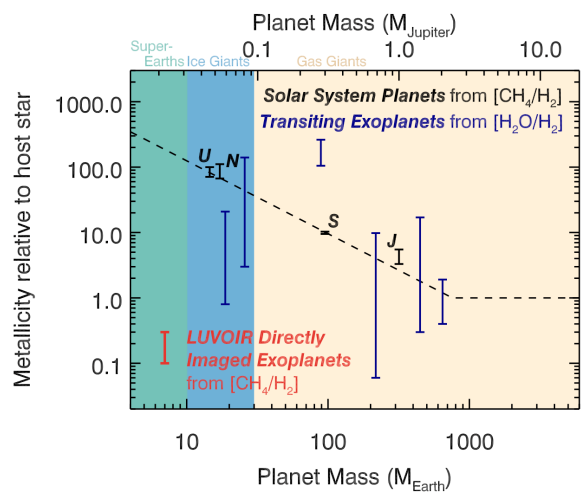

**Figure 4.3.** *Metallicity of Solar System and extrasolar giant planets. Transiting planet metallicity derived from retrievals of $H_2O$ abundance inferred from transit spectra. Solar system abundances are from remote and in situ measurement of $CH_4$. The dashed line shows the notional trend of decreasing [M/H] with increasing planet mass, a prediction of the core accretion gas giant formation mechanism. The red error bar shows expected precision of LUVOIR retrievals of gas giant $CH_4$ and $H_2O$ atmospheric abundances based on retrieval studies for WFIRST (Lupu et al. 2016). Credit: J. Bean (U of Chicago).*

2007). This is different from the atmospheres of the hot transiting giant planets, which are disconnected from their deep interiors by a radiative zone that extends to 100 to 1000 bar or more. Thus, LUVOIR will characterize a distinct population of planets with notably different atmospheric structure and composition than most of the transiting planets expected to be observed with JWST, although note LUVOIR can also observe planetary transits in a wavelength range complementary to JWST.

Understanding how atmospheric composition varies with planet mass and stellar properties is vital, since planet formation models predict a wide range of enrichments in elements compared to stellar abundances (e.g., Oberg et al. 2011; Fortney et al.

2013; Madhusudhan et al. 2016; Mordasini et al. 2016; Thorngren et al. 2016; Espinoza et al. 2017). Observations of transiting giant planets suggest a trend of decreasing atmospheric heavy element enrichment with increasing mass that mirrors the one seen in the solar system (**Figure 4.3**). LUVOIR users will test whether such trends hold for the cooler, more distant population of planets that more closely resemble solar system gas and ice giants. LUVOIR will also constrain the atmospheric composition of planets at the sub-Neptune/super-Earth boundary to understand if these are distinct planet types or simply a continuum of composition.

LUVOIR will obtain reflected light spectra of dozens of gas and ice giants, down to sub-Neptune/super-Earth sizes, at a variety of orbital distances around stars of varying mass and metallicity, thus balancing the population of planets expected to be characterized with JWST transit studies. Optical photometry and spectra of detected planets will readily constrain the presence or absence of atmospheres from the distinctive appearance of Rayleigh and cloud particle scattering. Atmospheric composition can be derived from optical and near-IR spectroscopy, as the wavelength ranges and spectral resolutions required for detecting gases in terrestrial atmospheres (**Chapter 3**) will also permit measurement of $H_2O$, $NH_3$, $CH_4$, Na, and K in gas giants and additional species in the atmospheres of lower mass planets. Atmospheric metallicity can be determined using $CH_4$, $H_2O$, and $NH_3$ bands in optical (**Figure 4.4**) and near-IR spectra. For planets warm enough that atmospheric water is not condensed into clouds, such spectra will also constrain carbon-to-oxygen ratios, a key indicator of formation and migration history. For systems with potentially habitable planets, atmospheric characterization of any giant planets will also provide valuable insights into the formation history of the system.

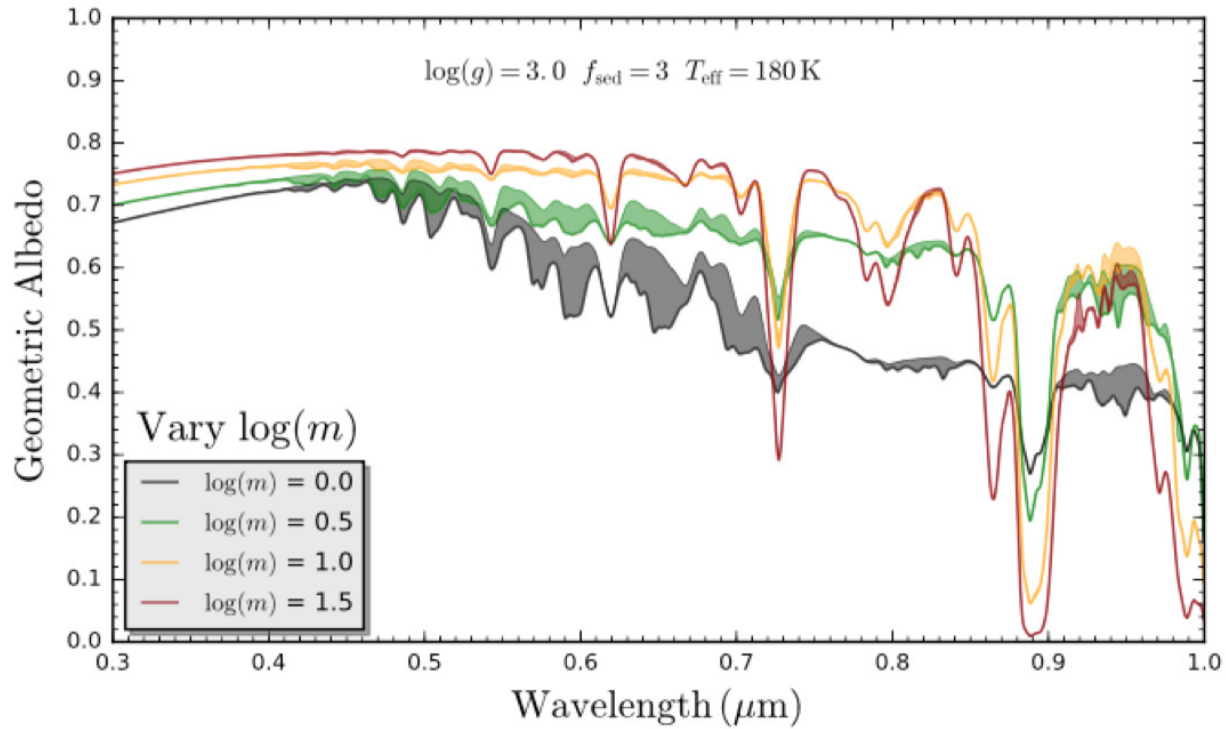

**Figure 4.4.** *Model albedo spectra of gas giant planets with atmospheric metallicity (log(m)) varying from solar to about thirty times solar. The most distinctive absorption features are those of $CH_4$. The shaded regions show the influence of gaseous $H_2O$ opacity, which, counterintuitively, is stronger at lower metallicity because of the feedback in this model between cloud height and water vapor abundance. Observations by LUVOIR users will inform our understanding of such water vapor/cloud feedback effects. Credit: R. Macdonald (Cambridge University).*

While atmospheric chemistry among the hydrogen-helium dominated giant planets can likely be grossly predicted, the atmospheric diversity of terrestrial planets discovered by LUVOIR will doubtless exceed our imagination. By carrying versatile integral field spectrographs covering a wide wavelength range, LUVOIR is well equipped to characterize the atmospheres of all types of terrestrial planets. A few examples of the diversity of worlds LUVOIR will be able to search for and characterize include:

a)  Hot exoEarths / exoVenuses are predicted to be the end-state of planetary evolution for rocky worlds with high $O_2$ or $CO_2$ atmospheres and rocky planets interior to the habitable zone of the star (Berta-Thompson et al., 2016; Luger & Barnes,

2015; Schaefer et al., 2016). Worlds undergoing catastrophic ocean and atmospheric loss may exist in the habitable zone (Meadows et al., 2016) and their characteristics will vary with stellar spectral type (Segura et al., 2003, 2005, 2007; Meadows et al., 2016). Optical and near-IR spectra will constrain $H_2O$, $O_2$, and $CO_2$ abundances, providing insight to this critical class of planets that define the inner edge of the habitable zone.

b)  Water-rich atmospheres may also exist in the usual habitable zone, either on the path to the "exoVenus" end-state, or as a long-term stable configuration of the planet's atmosphere (Goldblatt et al., 2016).

**Program at a Glance – Comparative Atmospheres**

**Goal:** Measure atmospheric composition for a wide range of non-habitable exoplanets.

**Program details:** Optical and near-IR spectra of giant through terrestrial planets at multiple separations around variety of stellar types. Acquired via a mix of dedicated observations and concurrently during long integrations on habitable zone planets.

**Instrument(s) + Configuration:** High contrast spectroscopy with ECLIPS IFS

**Key observation requirements:** Spectral bandpass from 400 nm to 1400 nm; R ~ 100; Continuum SNR > 10

c) Extreme water loss can lead to "Dune-like" worlds with little surface water reservoirs but temperate climates (Abe et al., 2011). Such worlds would be identified by their Rayleigh scattering slope and lack of bright clouds or water vapor.

d) Terrestrial planets of different ages and atmospheric redox states may mimic a similar evolution that occurred throughout Earth's history (Arney et al., 2017; Lyons et al., 2014); this evolution may occur on worlds with or without life (Catling et al., 2018). Such evolution can be driven by atmospheric loss, which when scaled-up can lead to evaporated core rocky worlds that are the remnants of massive atmospheric loss from larger planets (Gillon et al., 2017; Luger et al., 2017).

e) Terrestrial-sized planets with dense envelopes that were not lost can exist beyond the habitable zone, and due to the greenhouse effect from their $H_2$ envelopes, lead to planets with stable oceans well beyond the "traditional" habitable zone (Owen & Mohanty, 2016). Pressure-induced absorption from high-pressure $H_2$ atmospheres will be detectable in the optical and near-IR by ECLIPS.

Comparing the nature of all such terrestrial planets to each other and to any habitable worlds discovered by LUVOIR will place these planets in context. For example, if exoVenus-type planets turn out to be ubiquitous, the relative importance of the runaway greenhouse mechanism will be apparent. However, the diversity of such worlds (do they always sport bright sulfuric acid clouds?) will only emerge through comparative studies of multiple planets. Since we cannot readily predict likely atmospheric compositions for all of the terrestrial planets LUVOIR will discover, the most important capability for characterizing these worlds will be flexibility. As discussed more fully in **Chapter 3**, LUVOIR's spectral range for coronagraphic spectroscopy provides the ability to detect and measure the abundances of a large array of possible gases, including $H_2O$, $CH_4$, $O_2$, $O_3$, and $CO_2$. **Figure 4.5** shows simulated ECLIPS spectra, of both giant and terrestrial planets, that give a flavor of LUVOIR's powerful capabilities for atmospheric characterization.

### 4.1.2   Photochemistry and hazes

Photochemical processes play key roles in shaping the atmospheres of all solar system planets. In the solar system giant planet stratospheres, $CH_4$ photochemistry generates hydrocarbons such as $C_2H_6$ and $C_2H_4$; these molecules can polymerize into more complex hydrocarbon species, some of which condense into aerosols. These hydrocarbons strongly absorb UV and blue light on Jupiter and Saturn. On Venus, complex sulfur chemistry transforms $SO_2$ and $H_2O$ into $H_2SO_4$ that condenses into the thick

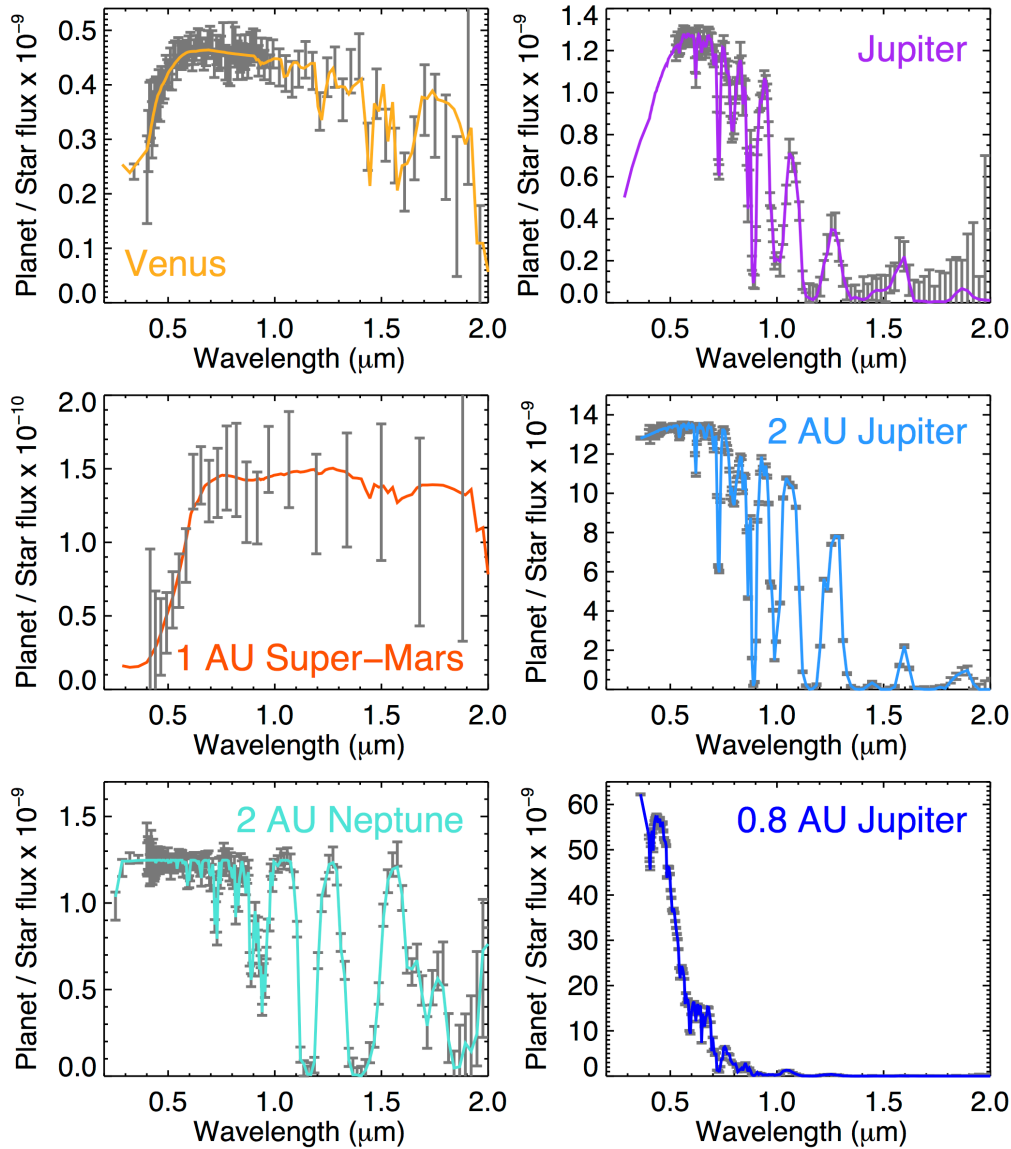

**Figure 4.5.** *Simulated LUVOIR spectra of planets orbiting a Sun-twin star at 10 pc. The simulated data plotted with grey bars were generated using the online LUVOIR coronagraphic spectroscopy tool, assuming the LUVOIR-A architecture (15-m telescope), the current ECLIPS-A instrument parameters, and exposure times of 96 hours per coronagraph band. Input model spectra are over-plotted with colored lines. The Venus model spectrum was provided by the Virtual Planet Laboratory (VPL), generated using the Spectral Mapping Atmospheric Radiative Transfer (SMART) model (Meadows & Crisp 1996; Crisp 1997). The super-Mars model was created by scaling a Mars model spectrum from VPL to a radius of 1 R_{Earth} and an orbital separation of 1 AU. The super-Mars data were rebinned to increase S/N. The warm Jupiter model spectra are from Cahoy et al. (2010) and the warm Neptune model spectrum is from Hu & Seager (2014). Image credit: LUVOIR Tools.*

cloud deck (Yung and Demore 1982). On Earth, the UV-shielding ozone layer is the result of oxygen photochemistry, and sulfur aerosols like $H_2SO_4$ can be produced through volcanic outgassing; such aerosols can produce detectable spectral signatures that may be remotely identified as a sign of vigorous volcanism on terrestrial planets

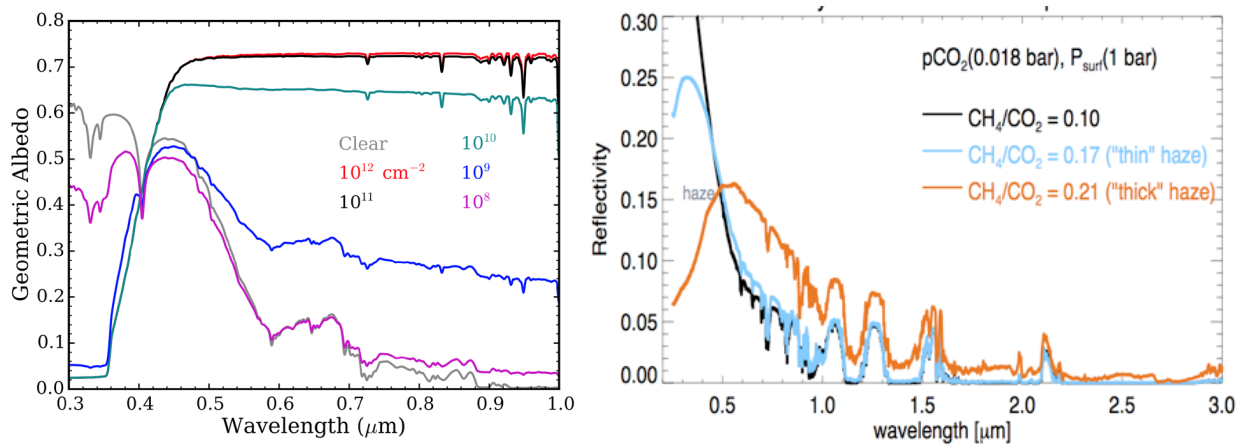

**Figure 4.6.** *Two examples of the impact of photochemical hazes on planet spectra. (Left) Model albedo spectra for a cloud-free 0.8 AU Jupiter-like planet (grey) along with spectra for various column abundances of sulfur ($S_8$) haze particles. In addition to markedly raising the red continuum flux level, note the strong NUV absorption of these hazes. Credit: Gao et al. (2017). Right panel: Model spectra of clear and hazy Archean Earth atmospheres. Credit: G. Arney (NASA GSFC)*

(Misra et al 2015; Hu et al 2013). The hazy disk of Titan, a consequence of $CH_4$ transport and photochemistry is another distinctive example of the importance of hazes. Likewise, no characterization of an exoplanet atmosphere—potentially habitable or not—will be complete without an understanding of the role photochemical hazes play in its atmosphere.

Evidence of hazes has already been inferred from the featureless transit transmission spectra of several exoplanets (Bean et al. 2010, Kreidberg et al. 2014, Knutson et al. 2014a, 2014b, Sing et al. 2014). Exoplanet atmospheres observed by LUVOIR will sample compositional and incident UV flux conditions not found in the current solar system, yielding a diversity of haze types. Examples of the potential impact of two plausible types of hazes, sulfur hazes on a warm giant planet and hydrocarbon hazes on an Archean Earth-like terrestrial planet, are shown in **Figure 4.6**.

Understanding the diversity of photochemical outcomes and how they relate to atmospheric and, for terrestrial planets, surface composition, temperature, and pressure will be a complex task, but one that will greatly inform our understanding of such processes as a whole. For multi-planet systems

---

**Program at a Glance – Atmospheric Hazes**

**Goal:** Constrain presence of atmospheric hazes all planets detectable in each system observed.

**Program details:** UV and optical low-resolution spectroscopy and/or photometry of a wide range of planet classes (giant through terrestrial) at multiple separations around variety of stellar types. Measure UV spectra of host stars.

**Instrument(s) + Configuration:** ECLIPS high contrast low-resolution spectroscopy and/or multi-color photometry. LUMOS UV spectroscopy.

**Key observation requirements:** Spectral bandpass from 250 nm to 1000 nm; R ~ 10; Continuum SNR > 10

---

understanding photochemistry in one planet's atmosphere will inform our understanding of the other planets, including potentially habitable ones. Measurements of the UV spectrum of the host star will also be critical for understanding these processes and interpreting exoplanet atmospheres. LUVOIR observations will illuminate how haze opacity varies with planetary and stellar characteristics, clarify the role hazes play in potentially habitable planets, and improve our ability to model photochemical processes in all types of planetary atmospheres

The UV capability of ECLIPS will be crucial for constraining photochemical hazes as they often absorb strongly in the blue and UV. $S_8$ particles, expected in warm giant planet atmospheres as a consequence of sulfur photochemistry, and organic hazes, which form from methane photochemistry and cause Titan's characteristic orange color, both produce strong UV-blue absorption features. The organic hazes could be detected at low spectral resolution at wavelengths shorter than 0.5 µm for hazy Archean Earth although thinner hazes will require extending to shorter wavelengths (< 0.4 µm; see blue spectrum in right panel of **Figure 4.6**).

Although WFIRST may image some exoplanets, the bluest WFIRST photometric band cuts off at about 0.55 µm, too red a wavelength to usefully characterize hazes. Studies have also showed that hazes and clouds that are opaque in transit transmission observations—thus obscuring other atmospheric features—can still allow observations of deep atmospheric features in direct imaging (Charnay et al 2015; Morley et al 2015). Thus, a UV-coronagraph equipped LUVOIR would provide an outstanding platform for detecting and characterizing photochemical hazes in a great variety of planetary atmospheres. The understanding LUVOIR observers will gain from understanding hazes in a diversity of atmospheres will doubtless aid the interpretation of habitable planet hazes as well.

### 4.1.3   Clouds

All Solar System planets with an atmosphere harbor condensate cloud layers. Clouds influence the atmospheric pressure-temperature structure of a planet, the reflected, emission, and transmission spectra, as well as planetary albedo and energy balance. Because clouds can play such a defining role in controlling the emitted and reflected flux, no characterization of an atmosphere—in particular abundance measurements—can be complete without an understanding of the nature of its cloud decks. To date clouds have also been studied in hot Jupiters via transmission and emission spectroscopy, spectral phase mapping, and in reflected light. Brown dwarfs and directly imaged planets are strongly impacted by their cloud layers and there have been robust studies of clouds in these objects.

Most of our current knowledge on cloud properties and compositions still comes from studies of Solar System planets (most importantly, Earth and Jupiter). Water vapor and water ice clouds in Earth can be studied in-situ and via remote sensing; models developed to explain their behavior and properties are often used as a starting point for models of extrasolar planet clouds (e.g., Ackerman & Marley 2001), although the full diversity of cloud processes in all types of exoplanets has not yet been explored.

Clouds may also be used as temperature indicators and tracers of atmospheric dynamics (circulation, mixing, turbulence). The atmospheres of giant planets sport a layered set of clouds, which respond to atmospheric temperature. Cold giants, like Jupiter, are characterized by the presence of ammonia clouds. In warmer giants the

<div style="border">

## Program at a Glance – Clouds

**Goal:** Constrain presence and composition of the cloud layers all planets detectable in each system observed.

**Program details:** UV through low-resolution spectroscopy and/or near-IR photometry of a wide range of planet classes (giant through terrestrial) at multiple separations around variety of stellar types.

**Instrument(s) + Configuration:** ECLIPS high contrast low-resolution spectroscopy and/or multi-color photometry.

**Key observation requirements:** Spectral bandpass from 250 nm to 1000 nm; R ~ 10; Continuum SNR > 10

</div>

evaporation of first the ammonia, and then the water cloud decks will alter the visible spectra, as a given mass planet is found progressively closer to its star (**Figure 4.5**). Clouds may also mask the presence of specific atmospheric absorbers even if present at large abundances below a thick cloud deck.

A systematic survey of cloud properties across all types of planets will be transformative for building better models of cloud behavior under varying conditions. In particular the characterization of most habitable planets will likely ultimately rest on cloud models, so a broad, systematic survey of cloud properties in all types of planets with LUVOIR will directly impact habitable planet studies. An understanding of how global cloud properties (altitude, thickness, patchiness of various species) vary with observable planetary properties will be one of the most important legacies of LUVOIR exoplanet science.

LUVOIR users will aim to understand which particular cloud species are present in which atmospheres, the processes that control cloud particle sizes and vertical distributions, the dynamical and radiative feedbacks between clouds, atmospheric thermal structure, and global dynamics. For a few favorable planets, LUVOIR observations will shed light on how clouds respond to global planetary diurnal and seasonal cycles. Time-resolved high-precision observations (photometric and spectroscopic light curves) have been obtained with HST and ground-based telescopes for some young giant planets. Extending such studies to cooler atmospheres in reflected light will be possible for bright planets for which photometry can be obtained in an hour or so.

The capabilities of ECLIPS to obtain UV to near-IR reflected light spectra are well matched to illuminating cloud processes. Cloud particle sizes are best constrained through large wavelength coverage, permitting size identification through Mie scattering effects which vary with wavelength. Cloud vertical thickness and number density also strongly influence planetary reflection spectra in the optical. In addition, for a few well-placed planets, the large aperture permits high cadence photometry, allowing searches for rotational modulation from clouds, tracing atmospheric dynamics and measuring rotation periods.

### 4.1.4   Atmospheric evolution and escape

Atmospheric escape is a fundamental physical process that leads atmospheric constituents to become unbound from a planet and alters the composition of the remaining atmosphere. Understanding the relative roles of escape, outgassing, and accretion in a variety of exoplanet atmospheres is critical to understanding the origin and evolution of planetary atmospheres. LUVOIR UV transit observations will both identify the escaping species and constrain the physics of escape.

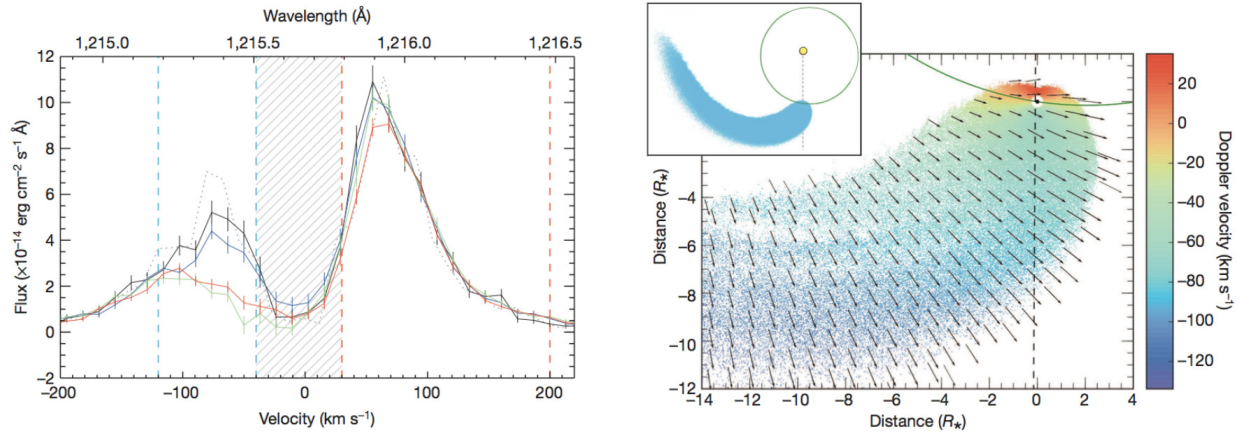

**Figure 4.7.** *Atmospheric escape. Left: Average Lyman alpha line profiles of the M-type star GJ436 that hosts a warm Neptune-mass planet. The solid lines correspond to out-of-transit (black), pre-transit (blue), in-transit (green), and post-transit (red) observations from individual spectra. The line core (hatched region) cannot be observed from Earth because of the interstellar medium absorption along the line of sight. Right: Polar view of a three-dimensional simulation representing a slice of the comet-like cloud coplanar with the line of sight of the GJ436 system. Hydrogen atom velocity and direction in the rest frame of the star are represented by arrows. Particles are color-coded as a function of their projected velocities on the line of sight (the dashed vertical line). Inset: zoom out of this image to the full spatial extent of the exospheric cloud (in blue). The planet orbit is shown to scale with the green ellipse and the star is represented with the yellow circle. Credit: Ehrenreich et al. (2015).*

The first observations of escape were obtained by Vidal-Madjar et al. (2003), who obtained STIS far-ultraviolet transmission spectra of the close-in giant planet HD209458b revealing that the planet possesses a highly extended hydrogen atmosphere due to heating by stellar X-ray and EUV photons. Subsequent HST observations have also detected various metals in the exospheres of giant planets including carbon, oxygen, and magnesium (Vidal-Madjar et al. 2004; Linsky et al. 2010), as well as escaping hydrogen from the warm Neptune-size planet GJ436b, which showed a large tail of escaped planetary material (**Figure 4.7**; Ehrenreich et al. 2015).

Atmospheric escape is a key factor shaping the evolution and distribution of low-mass close-in planets (e.g., Owen & Wu 2013) and their habitability (e.g., Cockell et al. 2016). Indeed, many highly irradiated rocky planets (e.g., CoRoT-7b, Kepler-10b) might be the remnant cores of evaporated Neptune-mass planets (e.g., Lopez et al. 2012). As a consequence, atmospheric escape also has a major impact on our understanding of planet formation (e.g., Van Eylen et al. 2017).

Upcoming missions will provide us with a large number of transiting planets covering the whole parameter space necessary to thoroughly study escape from an observational perspective. For example, the TESS mission is expected to find about 1700 planets orbiting nearby stars, some of which (~15) will be Earth-size planets in the habitable zone of M-dwarfs. The PLATO mission will greatly exceed these numbers, particularly by extending the search to longer-period planets (Sullivan et al. 2015; Rauer et al. 2014). LUVOIR is the only mission capable of characterizing the extended atmospheres of a statistical sample of these worlds, essential for understanding the global properties of atmospheric escape and understanding how the physics of UV-driven atmospheric mass loss determines the long-term stability of all types of planetary atmospheres.

The observational census of atmospheric escape requires the acquisition of UV transmission spectroscopy observations for about 100 transiting planets orbiting stars (later than spectral type A5V). The complete sample of planets is necessary to understand how atmospheric escape correlates with stellar and planetary system parameters. The observations should be carried out at both FUV and NUV wavelengths for the brightest/ nearest systems (for Lyman-alpha in particular it is necessary to observe systems within 60 pc due high ISM absorption). Moreover, because stellar UV emission from active stars can be highly variable from one transit to another, it is important to obtain high S/N detections within an individual transit (Bourrier et al. 2017).

*LUVOIR observing program.* The target planetary systems will be well distributed on the sky and the identification of the best targets will be possible only after the results of the CHEOPS, TESS, and PLATO missions are in hand (still before the launch of LUVOIR). The observations will need to cover UV wavelengths (100–400 nm) where a large number of strong resonance lines of various elements are present. These features, plus a large number of weaker resonance lines of several metals will be the primary target of the observations.

The observations will aim not only at measuring transit depths and shapes, but also at measuring the velocity profile of the absorbing material along the line of sight.

In this way, it will be possible to provide a set of strong observational constraints on models of atmospheric escape. From the few relevant HST observations of transiting planets obtained so far, it appears that the typical velocity of the escaping material is of the order of 10–20 km/s. This requires instruments with a resolving power 40,000–60,000 to resolve the velocities, readily achievable with LUMOS.

Obtaining these observations simulta- neously for a range of species is particularly valuable since different species probe dif- ferent regions in a planet's photo-evapora- tive wind. For example, while Lyman-alpha can trace material beyond the planet's Hill sphere as it interacts with the stellar wind, metal lines like CII trace the inner bound por- tions of a planet's upper atmosphere, where the photo-evaporative wind is launched and Lyman-alpha is heavily extincted by the ISM. Additionally, detection of these metal lines provides a way to probe atmospheric com- positions, which is not subject to the effects of clouds lower in the atmosphere.

The need to reach high-enough signal- to-noise ratios (S/N) for multiple species in a single transit drives the specifications on the effective area, which in turn places constraints on the telescope diameter and component efficiencies. The exposure times are then driven by the transit durations and the need to get enough out-of-transit coverage to make sure transit asymmetries can be unambiguously detected. Additionally, high time resolution (~ 1 second) is required to be able to detect and characterize flaring events of the host stars that need to be taken into account in the data analysis and interpretation. The transit durations observed will cover a wide range of transit durations depending on stellar type and the planet's irradiation, ranging from ~1 hour for planets close-in planets around very late M dwarfs like those in the TRAPPIST-1 system (Gillon

---

### Program at a Glance – Evaporating Exoplanets

**Goal:** Fully constrain the physics of atmospheric escape for planets across parameter space.

**Program details:** High S/N FUV and NUV spectra to detect outflows in ~100 transiting planets and fully map multiple escaping species for ~25 transiting planets.

**Instrument(s) + Configuration:** LUMOS G120M, G150M, G300M

**Key observation requirements:** Large aperture to obtain S/N $\geq$ 20 per R ~ 30,000 resolution element spectra of the cores of stellar chromospheric emission lines in single transits in the NUV and FUV.

---

et al. 2017) up to 15 hours for planets in the habitable zones of Sun-like stars.

As an example of an observing program to characterize the mass-loss characteristics of low-mass planet, we reconsider GJ 436b, a 4.3 Earth radius warm Neptune on a 2.6-day orbit around M2.5 dwarf with a V-band magnitude of 10.7. Although, this planet has been studied carefully in Lyman-alpha (e.g., Bourrier et al. 2016), it is not currently possible to detect other species with current facilities (e.g., Loyd et al. 2017). LUVOIR, using the LUMOS instrument's G150M mode will achieve a 5-sigma detection of the velocity-resolved absorption profile in the 133.5 nm C II line in just 20 minutes. This would provide carbon detections in ~10 points as the planet's Hill sphere transits, probing all the way into the inner bound portions of the upper atmosphere which are inaccessible to current observations. Moreover, while GJ 436 b is an ideal example, upcoming surveys will uncover a large sample of planets around bright nearby stars that can be characterized to a similar level, enabling us to repeat these observations for dozens of planets spanning the full range of planetary parameter space. The Sullivan et al. (2015) simulated TESS catalog predicts that about 25 TESS planets will be amenable to transit maps of the densities, abundances, and velocities for multiple species in their escaping atmospheres.

## 4.2   Signature science case #2: Planetary system architectures

The history of a planetary system is encoded in its architecture, the system-wide configuration of planets and asteroid/comet belts. Architectural information includes the number, masses, spacing, and orbits of the planets; as well as the structure, mass, and composition of any accompanying debris from asteroids and comets (aka. planetesimals). Our own solar system's architecture—including the orbits and compositions of our asteroid and Kuiper belts (e.g., DeMeo and Carry 2014), the spacing and eccentricities of the terrestrial planets (e.g., Chambers & Wetherill 1998), and the orbits of the giant planets (e.g., Raymond et al. 2004)—informs us about the formation, early environment, and habitability of the Earth. Similarly, exoplanetary system architectures provide context for exoEarths; for example, an Earth-size exoplanet in a system with a nearby, highly eccentric Jupiter grew up in a very different environment from one tightly packed together with seven other rocky planets.

The large number of mature and young planetary systems LUVOIR will observe allows fundamental, systematic investigations of the environmental influences on planet formation and evolution. These studies are enabled

---

by LUVOIR's ability to simultaneously map each exoplanet system in detail across the field of view of ECLIPS. Such observations complement techniques that characterize a portion of each planetary system (e.g., transit spectroscopy) and surveys that provide planet occurrence rates (e.g., microlensing). Furthermore, LUVOIR will catalog the chemical outcomes of planet formation around a range of stars by measuring the bulk composition of young extrasolar comets and asteroids, measured through outgassing arising from collisions. LUVOIR also has a unique role to play in filling out our inventory of the smallest and coldest bodies in the solar system, which hold vital clues to its early history.

### 4.2.1    Mature exoplanet systems

The formation and evolution of even mature planetary systems (>1 Gyr) is imprinted in their residual architectures. By characterizing system architectures for multiple planetary systems, LUVOIR will open a window to understanding how planetary formation proceeds under a diversity of environments.

### 4.2.1.1 Orbital dynamics aid determination of planet masses and radii

The attributes of mass and radius are fundamental for characterizing exoplanets. The determination of bulk density allows us to distinguish rocky worlds from planets with substantial gas-envelopes or ice or water worlds. Masses and radii constrain surface gravity and are needed to interpret spectra of planetary atmospheres. Circa 2018, most planet masses are measured via the radial velocity technique, with a small number determined from transit timing variations. Neither of these methods can currently provide mass measurements for true analogs of our Earth around solar-type stars. Therefore, the capability of measuring exoplanet masses with extremely precise astrometry (~0.1 μas) is a key attribute of the High Definition Imager (HDI). Such astrometric measurements will yield reliable orbits and masses for planets with M ≥ 1$M_{Earth}$ orbiting ≥ 1 AU from G2V stars within 10 parsecs (the mass measurement limit is smaller for lower mass stars). Additionally, LUVOIR's immense light-gathering power, short integration times, large instantaneous field of regard, and rapid repointing will enable deep imaging of each exoplanet target star many times, ensuring both an accurate orbital solution as well as a high chance of planet discovery. The combination of direct imaging using ECLIPS and precision astrometry with HDI will robustly disentangle multi-planet systems (Guyon et al. 2012).

Since a planet's brightness in reflected light is a function of the planet radius, albedo, and orbital phase, determining the radius is difficult. Repeat imaging in multiple filters at different orbital phases is key to untangling these degeneracies. As an example, for a planet with a substantial atmosphere observed at crescent phases, the forward lobe of Rayleigh scattering tends to dominate over atmospheric and surface absorption at blue wavelengths, reducing the albedo uncertainty. Exploratory retrieval studies show that at phase angles greater than ~100, degrees measurement of planet brightness in the blue offers improved constraints on planet radii compared to more gibbous phases (Nayak et al. 2016). Further retrieval studies of simulated directly imaged planets observed in reflected light will identify the optimum strategy for constraining radii, and hence bulk density and composition, of both terrestrial and giant planets.

### 4.2.1.2 Orbits, compositions, and formation

Linking information on planet orbits, compositions, and dynamical history of mature exoplanet systems can provide deep insights into their past formation and evolution. For example, giant planets can affect volatile delivery to the inner solar system, leading to potential differences in composition between terrestrial planets in systems with and without giant planets. Atmospheric composition of giant planets also provides clues to their formation and evolution. Gas giant planets formed by the core accretion mechanism are expected to have atmospheric heavy element abundances in excess of those of their parent stars. Available data on hot transiting planets already point to a trend of atmospheric heavy element enrichment with mass (**Figure 4.3**). LUVOIR will extend such comparisons to cooler, more solar system-like giant planets, and uncover systematic deviations from the trend that might indicate other formation mechanisms (like gravitational collapse) for subsets of planets.

Furthermore, differences in atmospheric bulk composition (recorded as C/O ratios) for planets at similar orbital distances may indicate migration has moved some of them from their birthplaces to the locations we see today. Planet migration can also establish planets in orbital resonances and alter their eccentricities. Thus, measurement of orbits, masses, and atmospheric composition for all the planets in any system together permit a much richer understanding than any of these measures alone. Integral field spectroscopy at R=70 will be able to measure precise volatile inventories for a hundreds of gas and ice giant planets, enabling comparisons of C/O ratios over a wide range of orbital separations and identifying relationships between composition, mass, and orbit.

### 4.2.1.3 Mature debris belts

Even in old planetary systems, asteroids, comets, and Kuiper Belt objects (KBOs) left over from the planet formation phase persist. Erosion of these planetesimals produces interplanetary dust, which can actually be the most easily observed feature of a planetary system, due to the large total surface area of the dust grains. Dust in the warm inner region of the solar system is called zodiacal dust, which largely comes from comets (Nesvorný et al. 2010). The spatial distribution of interplanetary dust in the solar system—with the signs of planets imprinted in it—is shown in **Figure 4.8**.

Although sufficiently high levels of exo-zodiacal dust can obscure exoplanets from view, planetesimal belts and the dusty debris they produce serve as records of the system's early history and provide constraints on present day orbital properties. A belt located near a planet yields constraints on its mass and orbit: for example, a sufficiently massive, nearby, and/or elliptical planet would disturb the belt, which provides information on how the planet's orbit has (or has not) changed

---

**Program at a Glance – Planetary System Architectures**

**Goal:** Catalog the architectures of planetary systems around older main sequence stars

**Program details:** High contrast direct observations and astrometric measurements of nearby main sequence stars to measure bulk properties (masses, orbits, atmospheric composition) of all detectable planets and debris belts

**Instrument(s) + Configuration:** ECLIPS broadband coronagraphic imaging and integral field spectroscopy, HDI high precision astrometry

**Key observation requirements:** Contrast < $10^{-8}$; OWA $\geq$ 48 $\lambda$/D; astrometric precision = 0.1 $\mu$as

---

over time. Gravitational interactions between a planet and planetesimals can trap some of the latter into dynamical resonances, which can then be revealed through clumps in the interplanetary dust. Such trapping can serve as a record of the planet's migration history (e.g., Malhotra 1993). Furthermore, if we assume a planetesimal belt is roughly coplanar with the planetary system, we obtain an immediate estimate of the inclination of the system and can constrain a planet's orbit with fewer observations. High resolution, high contrast imaging of exoplanets will simultaneously reveal interplanetary dust, providing dynamical and contextual information without additional observational demands.

### 4.2.2 Young exoplanet systems

Young exoplanet systems offer valuable windows into the later stages of planet formation. Building rocky worlds and the final sculpting of planetary systems involve dynamic and sometimes violent processes. The young comets and asteroids that are the building blocks of larger planets are colliding and fragmenting, exposing dust and gas from their interiors. The orbital distribution of dusty debris from those collisions will tell us about the timescales for rocky planet formation and the properties of young planets. Spectroscopy of the gas provides a unique opportunity to measure the bulk composition of planetary material as a function of stellar type. These studies of terrestrial planet formation via debris in young systems will complement studies of giant planet formation through observations of younger protoplanetary disks (**Chapter 7**). LUVOIR's small inner working angle will be needed to probe within the ice lines of young planetary systems, which in theory divide inner rocky planet

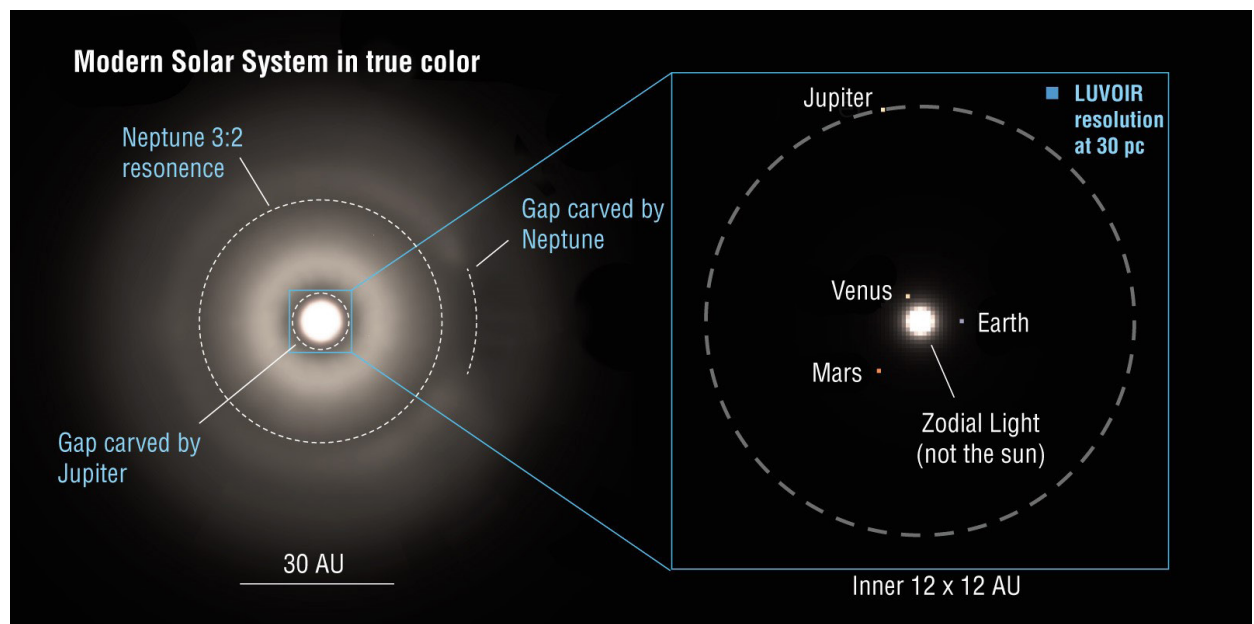

**Figure 4.8.** *The architecture of the modern Solar System, showing the interplay between planets and debris dust. The left panel shows a model of the entire system out to a radius of 50 AU from the Sun, while the inset panel zooms in on the inner system. For ease of viewing, the Sun and possible astrophysical background sources are not included. The bright region at the center of the image is emission from warm debris dust (aka. exozodiacal dust). Two circular gaps in the dust are visible, the inner one caused by Jupiter and the outer one marking the 3:2 mean motion resonance with Neptune. Neptune also creates a partial gap in its immediate vicinity. Credit: Roberge et al. (2017)*

formation regions from the cold icy regions where giant planets form.

### 4.2.2.1 Young debris belts: morphology and architecture

Young debris disks are a powerful probe of the formation and early dynamical evolution of planetary systems. On a massive scale, countless numbers of extrasolar planetesimals—the building blocks of both terrestrial and giant planets—are colliding with each other and producing disks of dust and gas debris that can be observed in detail (**Figure 4.9**). Since the ages of most debris disks span a plausible range of timescales for terrestrial planet formation (~ 10 to hundreds of Myr), the process can be observed in action, thereby constraining when and where it occurs, and shedding light on how processes vary from system to system and as functions of stellar properties.

Space- and ground-based instruments—including HST, ALMA, GPI, and SPHERE—have imaged debris disks containing planet-sculpted dust structures that help constrain the orbital properties and dynamical history of giant planets in young systems (e.g., HR8799; Booth et al. 2016). However, current instruments lack the spatial resolution, sensitivity, and inner working angle to image warm material in the innermost regions of debris disks. The following features enumerate the types of formation signatures that LUVOIR can detect as a function of separation from the host star and stellar age, mass, and metallicity.

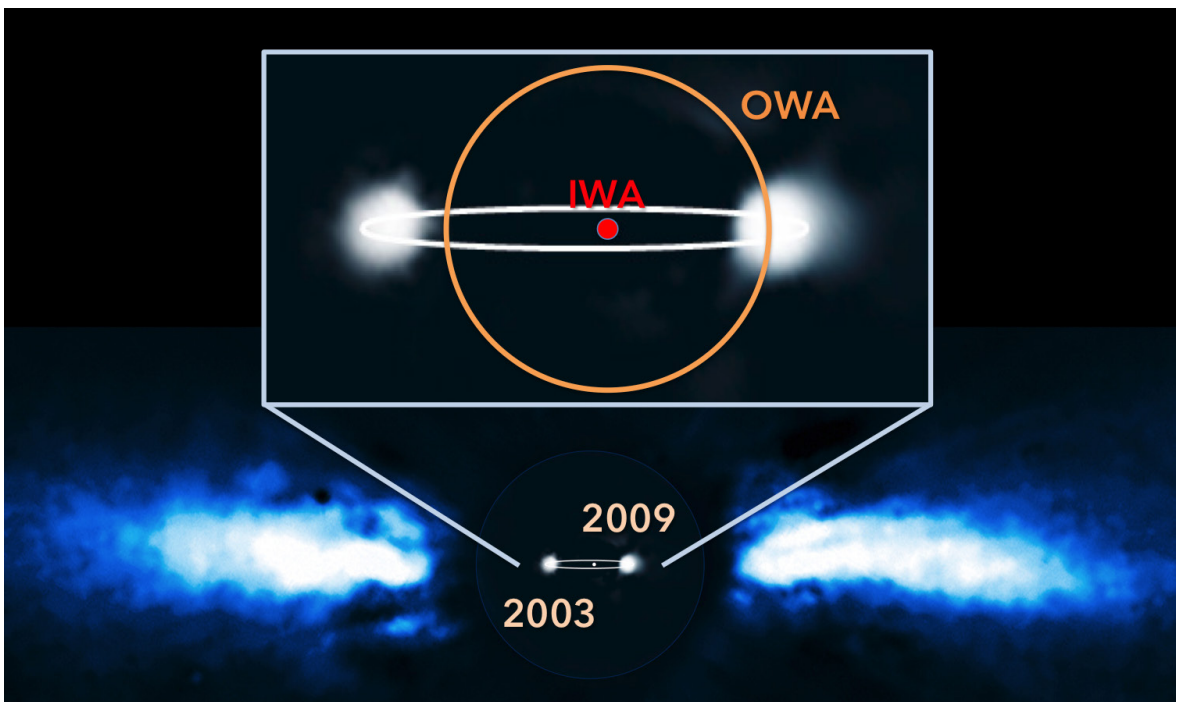

**Figure 4.9.** *LUVOIR can peer into the unseen inner regions of young planetary systems. This image shows the famous Beta Pictoris debris disk, with the gas giant exoplanet Beta Pic b imaged at two epochs in its orbit (2003 and 2009). The semi-major axis of the planet's orbit is ~ 10 AU (Lagrange et al. 2010). This planet is likely responsible for the long known "warp" in the inner portion of the dust disk (e.g., Heap et al. 2000). The inset panel shows a blow-up of the innermost region with the LUVOIR-A coronagraph inner working angle (red circle) and outer working angle (orange circle) overlaid. LUVOIR can search for warm dust and additional planets in the region between the two circles. Credit: ESA / A.-M. Lagrange*

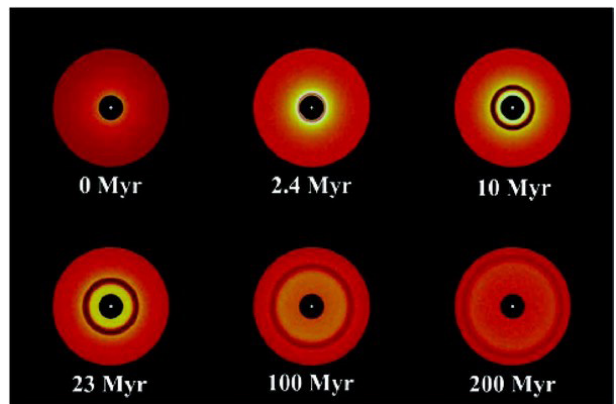

**Figure 4.10.** *Signposts of terrestrial planet formation—dust rings moving outwards with time. Credit: Kenyon & Bromley (2004).*

**_Dust from sites of ongoing terrestrial planet formation._** As terrestrial planets grow in a disk of planetesimals, they dynamically stir nearby bodies, generating collisional cascades that produce copious dust concentrated in rings. The dust is subsequently removed largely by radiation pressure. The dynamical simulation in **Figure 4.10** shows that the region of active formation—marked by bright dust rings—moves outward with time. The growth timescale is expected to depend on radial distance from the star and the local solid disk mass, which may be correlated with the host star mass and metallicity.

These theories can be tested with high resolution, high contrast images of young debris disks with a range of ages. Current high contrast instruments, with their large inner working angles and relatively low contrast (> $10^{-6}$), can only obtain such images for the outer regions of bright disks around early-type stars. Millimeter/submillimeter facilities like ALMA are insensitive to warm dust in inner regions of disks. **Figure 4.11** shows that ECLIPS can obtain high-resolution images reaching within the ice lines (~ 3 AU for the solar system; e.g., Martin & Livio 2012) of nearby young disks within about 135 pc. LUVOIR also has the sensitivity to extend such imaging studies to fainter disks

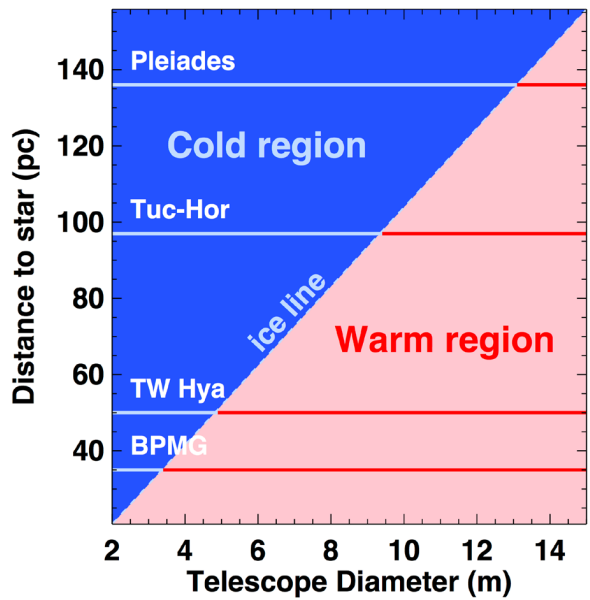

**Figure 4.11.** *Observing the terrestrial planet forming regions of young exoplanet systems with LUVOIR. The plot shows the distance out to which a coronagraph can observe the warm region inside the ice line of a Sun-like star (~ 3 AU) as a function of telescope diameter. The coronagraph IWA = 3.5 $\lambda/D$ and the assumed observation wavelength is 400 nm. In the blue region, the ice line is inside the IWA, so only cold material can also be observed. In the pink region, the ice line is outside the IWA, so warm material can be observed. The mean distances to a few nearby young ($\leq$ 100 Myr) stellar associations are indicated with horizontal lines (BPMG = Beta Pic Moving Group, TW Hya = TW Hydrae Association, Tuc-Hor = Tucana-Horologium Association, Pleiades = Pleiades Open Cluster). Observing inside the ice lines of Sun-like stars in the Pleiades requires a telescope with D $\geq$ 15 m. Credit: A. Roberge (NASA GSFC).*

around lower mass stars than are currently possible.

**_Gravitationally sculpted features associated with terrestrial planets._** Dust structures produced by dynamical interactions between planetesimals and planets can provide constraints on young planets' masses and orbits. The Beta Pic dust

---

**Program at a Glance – Terrestrial Planet Formation**

**Goal:** Map the spatial distribution of dust coming from destruction of planetesimals during the late stages of planet formation

**Program details:** High resolution, high contrast direct imaging of dust in debris disks around solar-type and low-mass stars in nearby young moving groups with a range of ages

**Instrument(s) + Configuration:** ECLIPS broadband coronagraphic imaging

**Key observation requirements:** Contrast $< 10^{-8}$; OWA $> 48\ \lambda/D$

---

disk displays an example of a planet-induced warp (**Figure 4.9**). Other types of structures include gaps cleared of planetesimals by resonances with a planet (**Figure 4.8**) and dust clumps marking planetesimals captured into resonance during a planet's migration (e.g., as invoked for Beta Pic in Dent et al. 2014). By comparing properties of planets within younger protoplanetary disks to those inferred within older debris disks, the evolution of planetary systems, planet migration, growth of planets through giant impacts after the dissipation of the primordial gas disk, and planet-planet gravitational interactions can be traced. However, as for the observations described immediately above, the smaller inner working angle and greater sensitivity of ECLIPS will be needed to see the subtler dust structures created by terrestrial planets.

### 4.2.2.2 Bulk composition of planetary material: extrasolar planetesimals

While the surfaces and atmospheres of exoplanets can be directly studied, their interiors remain hidden. Measuring planetesimal compositions offers the opportunity to infer the bulk composition of planetary material that formed around various types of stars. Compositional differences between relatively unprocessed gas in younger protoplanetary disks vs. gas produced from planetesimals in older debris disks set expectations for giant planet atmospheres accreted purely from relatively unprocessed primordial gas vs. those contaminated by planetesimal impacts. These measurements also inform the possible make-up of sub-Neptune atmospheres accreted during the protoplanetary phase vs. those outgassed from a planet's interior. But even in the solar system, to probe the interior composition of planetesimals, one needs to break them open, as was done by NASA's Deep Impact mission to comet Tempel 1. Fortunately, this breaking open is exactly what is happening to planetesimals in debris disks.

The goal is to determine the bulk composition of planetesimals around a variety of young stars by measuring elemental abundances of material in debris disks. Determining the composition of dust in debris disks is and will remain difficult, because the grains are relatively large and their spectral features are broad, weak, and ambiguous. Spitzer Space Telescope infrared spectra of many debris disks showed silicate emission features in very few cases, and little could be learned about the dust composition even when the emission was present (e.g., Chen et al. 2006). Instead it is the gas component that can provide abundances of all the major elements.

Gas in debris disks has been hard to detect, since the gas masses are low relative to those in younger protoplanetary disks (e.g., Dent et al. 2005). In contrast to those young gas-rich disks, debris gas appears to be primarily atomic rather than

---

**Program at a Glance – Composition of Planetesimals**

**Goal:** Measuring bulk composition of gas coming from young planetesimals in edge-on debris disks.

**Program details:** Spectra of stars with a range of masses in nearby young moving groups.

**Instrument(s) + Configuration:** POLLUX and LUMOS UV / optical point-source spectroscopy

**Key observation requirements:** Spectral bandpass from 100 nm to 400 nm; R ~ 65,000 to 120,000; Continuum SNR > 10

---

molecular, since the dust masses in debris disks are too low to shield molecular gas from rapid photodissociation. The advent of sensitive radio spectroscopy with ALMA has provided the opportunity to survey a limited number of gas species in many debris disks (specifically, CO and neutral carbon). But such observations cannot provide a full compositional inventory of gas in debris disks, as emission lines of other important species (water, atomic oxygen, and metals) are not accessible.

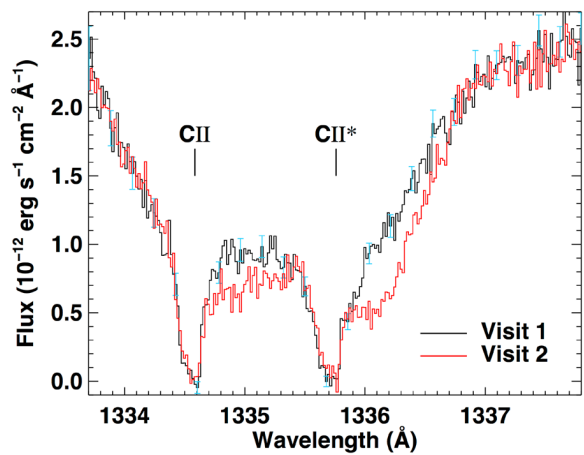

**Figure 4.12.** *Ionized carbon gas coming from planetesimals in the 49 Ceti debris disk, observed via UV absorption spectroscopy using HST. The additional redshifted gas apparent in the second observation (Visit 2) comes from vaporization of star-grazing planetesimals transiting the central star. Credit: Miles et al. (2016).*

To access a wide range of atomic species in debris disks, astronomers have instead employed sensitive UV / optical absorption spectroscopy, using the host stars as the background light sources. An example appears in **Figure 4.12**, which also shows time-variable gas coming from star-grazing extrasolar planetesimals. Far-UV spectroscopy is vital for these studies, as the strong absorption lines of many atomic and molecular gases are only available in that bandpass. This is the technique that was used to inventory the gas in the Beta Pic and 49 Cet disks, both around intermediate mass A-type stars (Roberge et al. 2006, Roberge et al. 2014). In both cases, the gas elemental composition appears extremely and unexpectedly carbon-rich. The reasons for this are not fully understood at this time. For this technique to work, the disks must be edge-on to our line of sight to the central stars (like doing transit spectroscopy of an exoplanet, but all the time). But even more importantly, current UV spectroscopy with the Hubble Space Telescope is not sensitive enough to employ this technique on anything but a handful of UV-bright, nearby A-type stars.

The UV spectrographs on LUVOIR (LUMOS and/or Pollux) will extend this powerful observational technique to gas in disks around later types of stars. For the first time, we will be able to look for differences

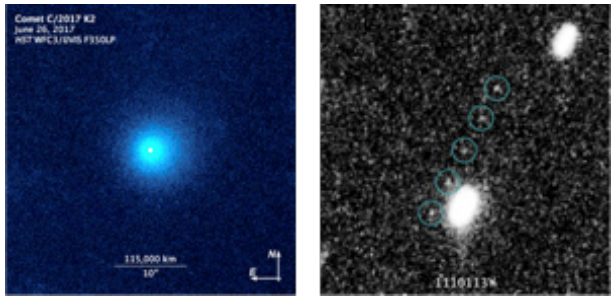

**Figure 4.13.** *At left, an HST image captures the nucleus and dust coma of C/2017 K2 at a heliocentric distance of 15.7 AU (Jewitt et al. 2017). At right is the series of discovery images of MU69, the next New Horizons mission target, obtained with HST WFC3*

in the volatile and refractory composition of planetesimals as functions of stellar mass and metallicity. To do this, we will use high-resolution point-source absorption spectroscopy at UV / optical wavelengths. In some cases, high spectral resolution (R > 100,000) will be needed to separate circumstellar absorption lines from interstellar ones. The instruments' bandpasses cover many strong absorption lines of critical species, including volatiles (e.g., H, C, N, O, CO, and OH), lithophiles (e.g., Na, Mg, Al, Si), and siderophiles (e.g., Mn, Fe, Ni). Furthermore, access to absorption lines from more than one energy level and ionization state is important for determining the gas excitation temperature and the ionization balance of the gas. Both are needed to calculate the total elemental abundances. Similar observational techniques applied to metal polluted white dwarf stars can be used to measure the bulk composition of planet fragments after tidal disruption at the end of their lives (details in **Appendix A**).

## 4.2.3 Solar System comets, asteroids, and KBOs

The vast collection of icy bodies in the outer solar system contains an echo of the primordial processes that led to the formation and modern distribution of the giant planets.

This population is extremely diverse, ranging from geologically active, atmosphere-bearing dwarf planets to small, porous undifferentiated planetesimals that are structurally reflective of primordial condensation and aggregation processes (**Figure 4.13**). Studies of the composition, size distribution, organization, and orbital dynamics of these objects offers our best window into the epoch of planet formation, from which we are able to constrain the structure of the solar system's protoplanetary disk, the mechanisms of small body accretion, and the process of planetary migration. Furthermore, as the only remnant sub-planetary population available for direct study, the observed properties of these objects are broadly applicable to our understanding of planetary system development in other star systems.

It is only over the last 25 years (Jewitt and Luu, 1993) that we have begun to map the distribution of Trans-Neptunian Objects (TNOs) in the Kuiper-Edgeworth Belt (KEB), the inner of two icy body reservoirs. Over 1800 TNOs have been directly observed, ranging in diameter from ~25 km (1999 DA8, Gladman et al, 2001) to >2300 km (e.g., Pluto, Eris). Prior to this, the existence of orbital reservoirs was inferred entirely from comets, icy bodies that are gravitationally perturbed into the inner solar system where they become visible by virtue volatile sublimation and proximity to Earth. Comet orbital properties reflect their original reservoir, with short period (P < 20 years) comets (SPCs) in the Jupiter and Encke families following prograde orbits centered on the ecliptic in a manner consistent with an origin in the KEB, while long (P > 200 years) period comets (LPCs) display an isotropic distribution of inclination and direction consistent with a more spherically distributed Oort cloud. Orbital simulations suggest that the evolution from the stable reservoirs to an active comet proceeds by slow accumulation

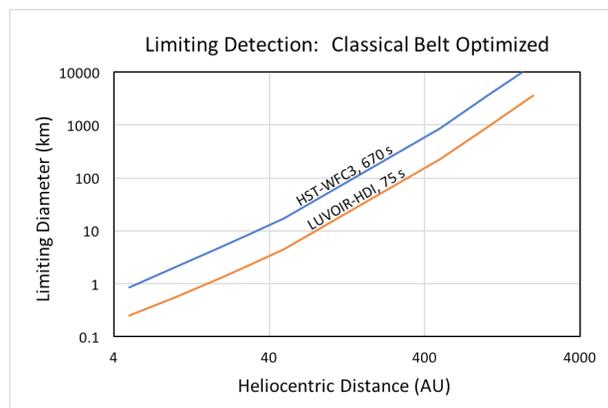

**Figure 4.14.** *The limiting diameter detectable at SNR ~ 5 is shown as a function of distance for a single image that is depth-optimized to the Classical Kuiper-Edgeworth Belt. The results from HST (in 670 sec) and LUVOIR (in 75 sec) are compared. Credit: W. Harris (Arizona LPL)*

of gravitational perturbation by galactic tides (for LPCs) and successive orbit-crossing events (for SPCs) over time scales of up to 0.1 Gyr, with the majority being lost to ejection before entering the 'activity zone' inside 2.5 AU (e.g., Duncan, Levison, and Dones, 2005). With projected lifetimes of order $10^5$–$10^6$ years, the typical active comet is therefore a short-lived remnant of a much larger population of evolving objects in the region beyond Jupiter.

The major factor limiting the study of icy planetesimals is detection efficiency for small bodies. These objects are only visible due to reflected sunlight that experiences inverse square dilution in both the outbound and reflected directions, resulting in a $R^{-4}$ brightness function with heliocentric distance. This effect is magnified in the gas coma, because the rate of sublimation from a nucleus decreases rapidly with temperature, resulting in a lower scattering column density with increasing heliocentric distance. This combination of factors results in a severe observational bias against smaller and less active bodies that is reflected in the clear difference in the observed size distributions of TNOs and SPCs, where even the largest

SPC nuclei (e.g., 109P/Swift-Tuttle) are smaller than every known TNO, and the paucity of gas production measurements from comets beyond 2.5 AU.

HST is currently the deepest telescope for detection of small bodies by virtue of the much lower sky-background and its use of diffraction limited imaging. WFC3 broadband (F350LP filter) imaging can currently identify (SNR~5) a 15 km diameter object in a circular orbit at 40 AU in ~600 s (the crossing-time for a 0.08" resel). The LUVOIR HDI will extend the HST advantage as a detection tool, even when compared with the next generation of 30-m class ground based telescopes. Over the same bandpass, LUVOIR-HDI would achieve SNR~5 detection of 3.5 km diameter objects over a ~70 s resel-crossing time, a 4-fold reduction in limiting diameter in 1/8th the time (**Figure 4.14**). In addition to point source detection, the HDI and LUMOS instruments would provide a 100x increase in sensitivity to the surface ice composition and low surface brightness UV emission from the gas comae of TNOs (**Figure 4.15**). LUVOIR is essential for the following types of studies.

*Population statistics.* The deepest wide-field surveys available are currently unable to detect objects <10 km in diameter beyond the orbit of Saturn, yet dynamical models identifying the KEB as the source of SPCs and Centaurs require that such bodies must be extremely common. LUVOIR-HDI is capable of extending the detection threshold down to 1–10 km sizes typical of SPCs. A campaign targeting size distributions in the classical KEB between 40–50 AU would identify any objects in this region ≥ 3 km in diameter at a rate, including time-sequenced returns to establish motion, of 1 sq-deg per day of integration. In addition to small KEBs, these observations would obtain trailed images of Centaurs between 6–30 AU to depths of 0.5–2 km and detect Sedna-sized

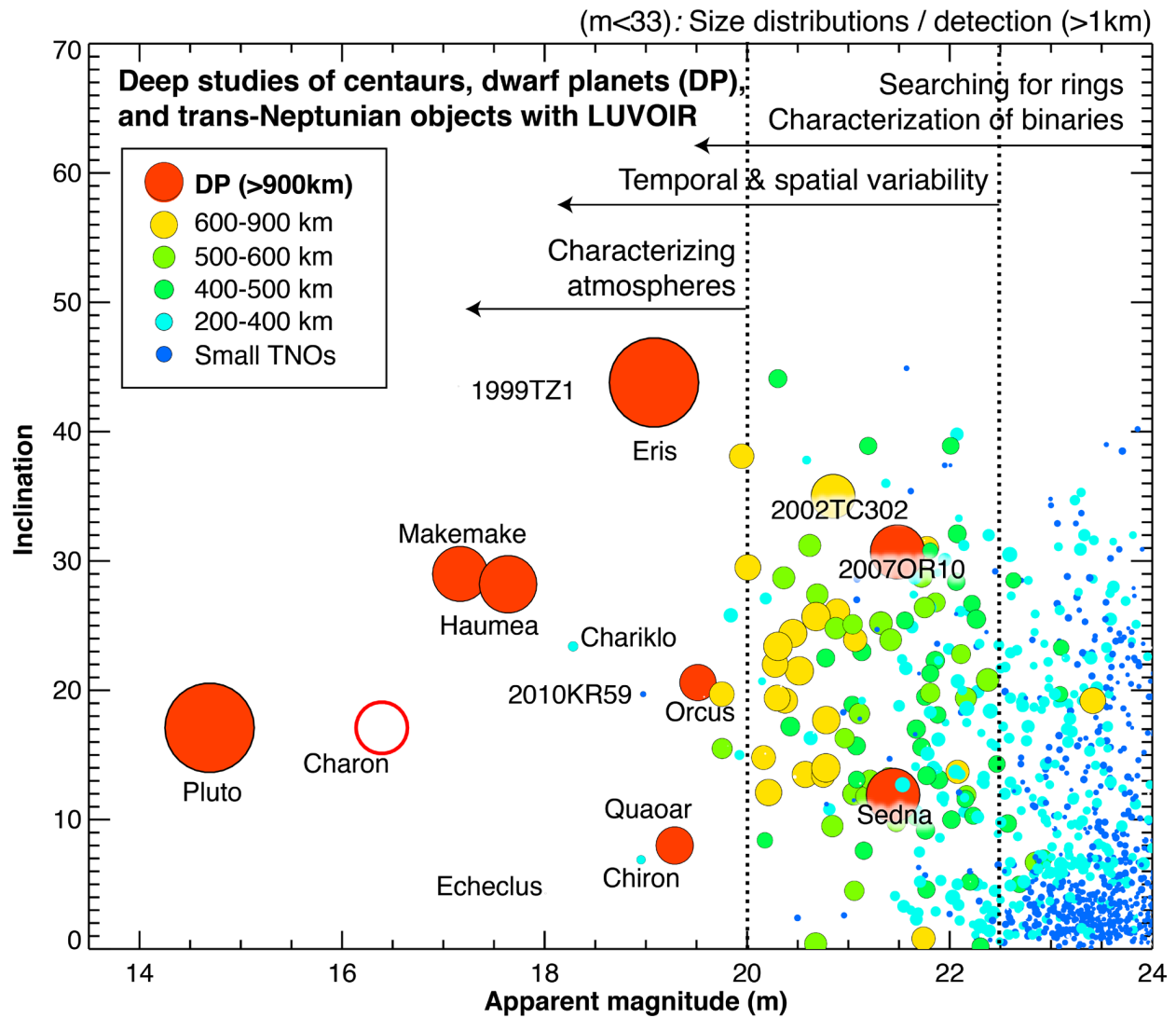

**Figure 4.15.** *Brightness of centaurs and trans-Neptunian objects compared to the capabilities of a LUVOIR-class observatory with sensitive spectrometers in the UV/Optical/NIR spectral range. LUVOIR will permit a unique characterization of the deepest regions (40–50 AU) of our solar system (objects with m<33), also permitting unprecedented studies of their atmospheres (m<20), surface composition / variability (m<22) and their origin and evolution via studies of their rings and binary configuration (m<24). Credit G. Villanueva (NASA-GSFC).*

Dwarf planets at distances up to 1000 AU (Sedna's aphelion).

*Moons and rings.* The cold classical Kuiper Belt's binary population is thought to be a primordial remnant produced during the epoch of planetesimal formation. With negligible binding, binary pairs of 100 km objects have shown separations >10⁵ that sharply constrain subsequent perturbations. Binary occurrence increases rapidly with decreasing separation to the limit of HST's angular resolution. LUVOIR will be able to characterize the occurrence rate of Kuiper Belt binary systems down to the Roche limit for objects with diameters as small as 120 km (**Figure 4.15**), approximately the size at which the Kuiper Belt size distribution breaks from a steep to shallow slope. Two of the largest Centaurs, Charliko and Chiron, are thought to support ring systems. Their origin is a mystery, and determining the rate at which they occur, along with understanding the processes that create, maintain, and destroy them, has implications for the occurrence rate of small satellites, sub-surface volatile reservoirs, and collision rates in the outer solar system.

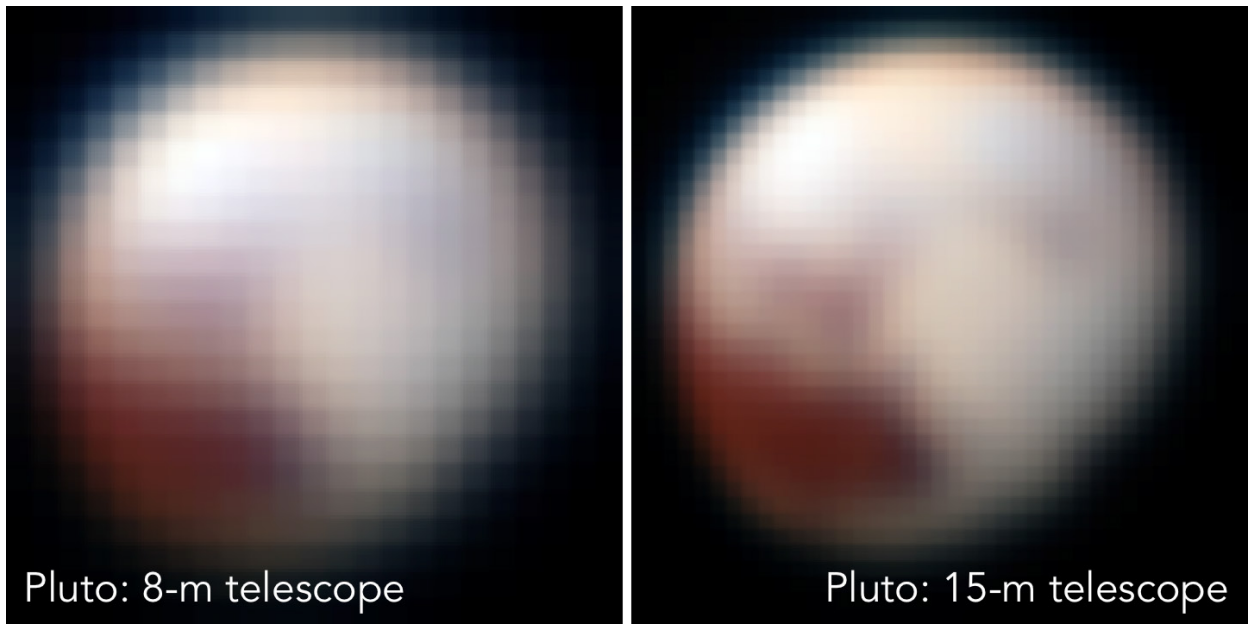

**Figure 4.16.** *LUVOIR high spatial resolution and spectroscopy in the 0.5–2.5 μm range can provide spectral characterization ranging from high fidelity spatial, diurnal, and seasonal coverage of larger Dwarf Planets to more complete identification of surface composition on smaller bodies. Credit: NASA/ New Horizons/R. Juanola-Parramon (NASA GSFC).*

*Surface composition.* NASA's New Horizons mission has revealed the complex cryogenic world Pluto (**Figure 4.16**), which, like Neptune's moon Triton, has a spatially heterogeneous surface composition shaped by a variety of poorly understood processes including both geologic activity and atmospheric transport of CO, $N_2$, and $CH_4$ ices. More remote (e.g., Eris) and smaller (e.g., Sedna, Makemake) TNOs have spectra dominated by $CH_4$ absorption. Surface and seasonal variations, including sublimation/ condensation of atmospheres, on all of these worlds are poorly constrained and can be directly addressed by LUVOIR spectroscopic study in the 0.5–2.5 μm range. In contrast to the hydrostatically organized dwarf planets, smaller TNOs are spectrally featureless or present surfaces with small amounts of non-volatile water ice. This likely represents their primitive, more homogeneous structure. LUVOIR's high sensitivity, spectroscopic capabilities and spatial resolution will permit a deep characterization of the surface composition on smaller bodies.

*Volatility.* Comets are the primary means through which we obtain volatile composition information from icy planetesimals. However, the sample is skewed by the strong temperature dependence of CO, $CO_2$, and $H_2O$ sublimation, the $R^{-4}$ reduction in scattered surface brightness, and a historical emphasis on wide-field observations of photochemical fragments (e.g., CN instead of HCN or HNC). These factors combine bias study toward the global properties of the most active nuclei at small heliocentric distances. This emphasis lacks perspective on factors including compositional heterogeneity, seasonal patterns in surface temperature, the characteristics of low-activity objects

---

**Program at a Glance**

**Goals:** (1) Measure the size & spatial distributions and the rate & scale of binary/ring systems among the Centaurs and in the Kuiper belt. (2) Determine the orbital distribution of Trans Neptunian Objects. (3) Obtain the spectral properties of small (diameter < 100 km) bodies beyond Neptune. (4) Measure cometary volatiles.

**Program details:** Deep broadband optical/NIR imaging surveys with HDI; optical/NIR point-source spectroscopy; UV spatially resolved spectroscopy.

**Instrument(s) + Configuration:** HDI multi-band imaging; HDI grism spectroscopy or ECLIPS spectroscopy (without coronagraph masks); LUMOS multi-object spectroscopy.

**Key observation requirements:** Tiled imaging with repeats to obtain orbits; non-sidereal drift matched to orbital motion; spectral bandpasses 0.5 – 2.5 μm and 100 nm – 200 nm.

---

(e.g., Main Belt Comets, the Jupiter Trojans), the characteristics of sublimation beyond the distance where $H_2O$ ice contributes, and the role of exotic processes such as thermal electron impact excitation (Feldman et al., 2015). For objects beyond the orbit of Mars, the dominant volatile ice is either CO (e.g., Jewitt et al. 2017) or an exothermic reaction such as the amorphous to crystalline $H_2O$ ice conversion (Prialnik and Bar Nun, 1990). Particularly for distances beyond the orbit of Jupiter, the composition of a comet coma has been partially sampled for only a few objects, and, in several cases, is detectable only from an associated dust coma. The broad spectral coverage of LUVOIR will combine observations of atomic species (O, C, S, H, and D) in the UV with the classical visible band fragment molecules (OH, CN, $C_2$, etc.) and their near-IR parents between 1–2.5 microns. Moreover, it will have the sensitivity to detect, or place far more stringent limits on gas production from low activity, volatile rich bodies including some Trojan asteroids and the Main Belt Comets and to monitor the characteristics of ice sublimation for nuclei in the CO dominated region of comet orbits.

## 4.3　Summary

Before the Kepler mission flew we knew of only a few hundred exoplanets, most discovered by the radial velocity method and thus lacking information on planetary radii. Kepler added thousands of confirmed and candidate planets to the list of known exoplanets, but perhaps the greatest contribution to exoplanet science was the discovery of exceptional diversity among planets. Super-earth and sub-Neptune sized planets, not represented by any solar system objects, turned out to be exceptionally abundant. This diversity of planet types broadened our vision and led to a revolution in efforts to understand planet formation and evolution.

LUVOIR will enable an equivalent new era of comparative exoplanet science. Instead of comparisons of planet sizes and masses, LUVOIR will empower its users to conduct comparative studies of planetary atmospheres, including their composition, chemistry, clouds, dynamics, photochemistry, and escape processes for hundreds of planets. By improving our understanding of fundamental planetary processes on all types of planets our ability to interpret observations of and deeply understand habitable planets will only be magnified. Likewise, by providing a new view into the process and outcomes of planet formation LUVOIR has the potential to revolutionize our understanding of how planetary systems form and evolve. Equipped with a flexible, powerful observatory designed to characterize exoplanets of all stripes, the next generation of exoplanet astronomers will achieve a comprehensive view into the nature of the diversity of planets—particularly including their atmospheres—and planetary systems.

**Table 4.1.** *Summary science traceability matrix for Chapter 4*

| Scientific Measurement Requirements | | | Instrument Requirements | | |
|---|---|---|---|---|---|
| Objectives | Measurement | Observations | Instrument | Property | Value |
| Explore the diversity of planetary atmospheres | Direct measurement of exoplanet atmosphere compositions | Direct spectroscopy of exoplanets at multiple separations from a variety of stars; S/N>10 | ECLIPS | High-contrast VIS, NIR spectroscopy | Contrast < $10^{-10}$; $R \sim 100$ |
| Constrain the physics of atmospheric escape | Detection, mapping, and characterization of exoplanet outflows | Stellar point-source spectroscopy during transit for ~100 planets | LUMOS | FUV, NUV spectroscopy | G120M, G150M, G300M; $R \sim 30,000$ |
| Examine exoplanet atmospheric hazes and clouds | Haze/cloud detection and characterization | Direct spectroscopy and/or photometry of exoplanets at multiple separations from a variety of stars; UV stellar spectroscopy | ECLIPS | High-contrast UV, VIS low-resolution spectroscopy and/or imaging | Contrast <$10^{-10}$; $R \sim 10$ |
| | | | LUMOS | UV spectroscopy | G120M, G150M, G180M G300M; $R \sim 30,000$ |
| Determine planetary system architectures | Masses, orbits, atmospheric composition of planets and debris belts | Direct imaging and spectroscopy; precision stellar astrometry | ECLIPS | High-contrast VIS, NIR spectroscopy | Contrast < $10^{-8}$; OWA > 48 λ/D; $R \sim 100$ |
| | | | HDI | VIS astrometry | < 0.1 μas precision |
| Determine morphology and architecture of debris disks | Spatial distribution of dust in debris disks | Direct imaging of debris disks around solar-type and low-mass stars over a range of ages | ECLIPS | High-contrast VIS imaging | Contrast < $10^{-8}$; OWA > 48 λ/D |
| Measure bulk composition of extrasolar planetesimals | Chemical composition of gas from planetesimals in debris disks | Point-source spectroscopy of stars with edge-on debris disks; S/N > 10 | POLLUX / LUMOS | UV spectroscopy from 100–400 nm | $R$=65,000–120,000 |
| Characterize the full population of solar system minor bodies | Population statistics, morphology, and composition of minor bodies | Deep imaging for faint object detection at S/N > 5, time series imaging for orbits, spectroscopy of minor bodies | HDI | Multi-band imaging, grism spectroscopy | Tiled imaging with repeat visits; moving object tracking |
| | | | ECLIPS | Open mask VIS/NIR spectroscopy | $R \sim 70$ to 200 |
| | | | LUMOS | FUV multi-object spectroscopy | |

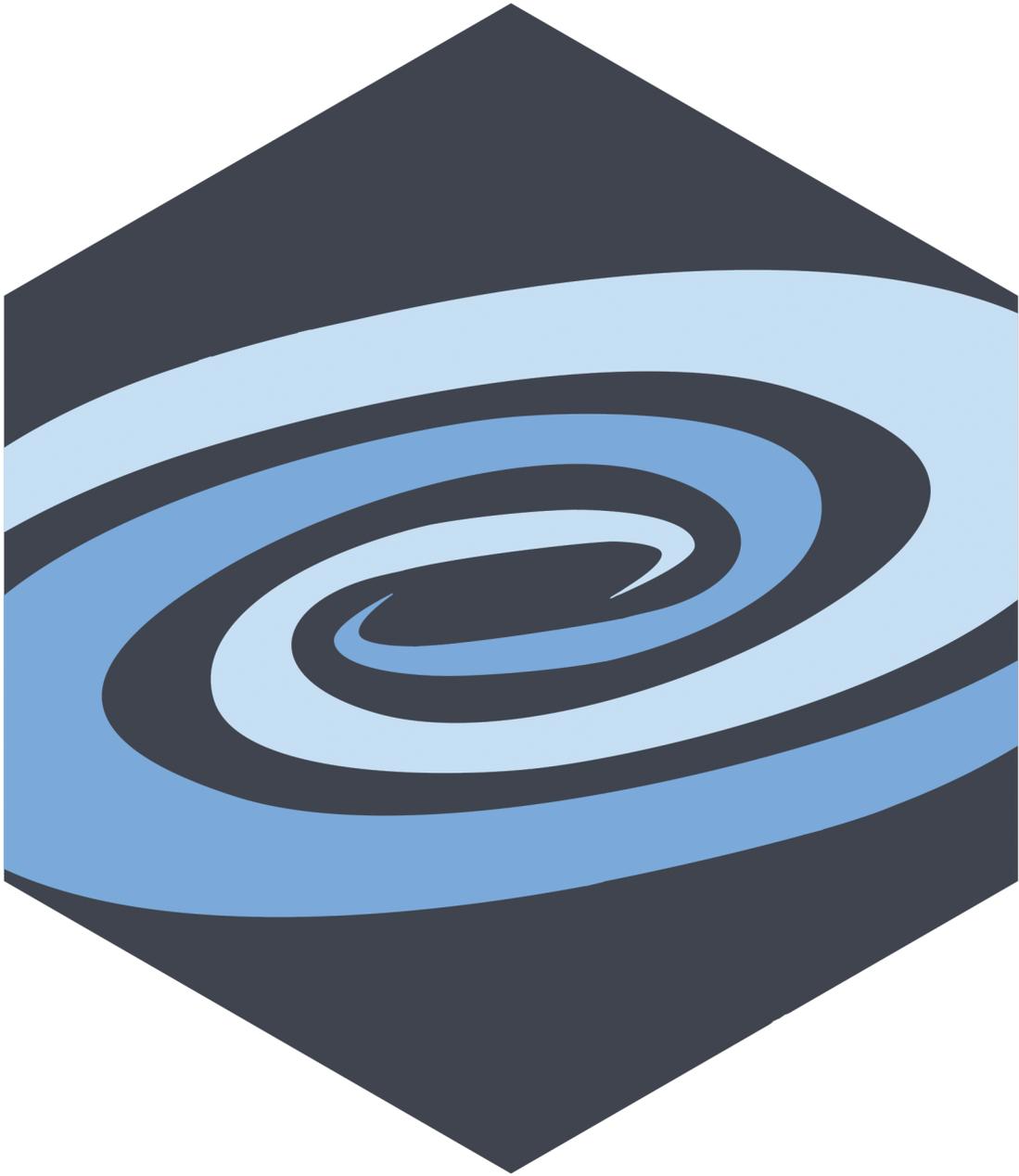

How do galaxies evolve?

# 5   How do galaxies evolve?

This question is one of the oldest in astrophysics, dating from the very first realization a century ago that galaxies are "island Universes" unto themselves. The quest to answer it has motivated some of astronomy's brightest minds to build its most ambitious telescopes. Over decades of discovery spanning the electromagnetic spectrum, astronomers have painted a rich picture of how galaxies grow. They begin as small seed fluctuations in the large-scale structure to assemble "hierarchically" into self-bound dark matter halos. These potential wells draw baryons together, and this gas becomes galaxies and the stars, and in some places, life emerges. We know that supermassive black holes lurk in the hearts of nearly every galaxy, and for brief periods might govern their evolution.

Yet our understanding of the physics behind these emergent patterns lags behind our ability to characterize them. We can model the tree-like hierarchical assembly of dark matter, but how baryons populate the branches of this tree is mapped out only for the largest pieces at late times, and we do not know what sets the minimum scale for galaxy formation. We do not know why some galaxies are dominated by their bulges while others have none: is it an effect of the hierarchy, gas accretion, or activity from their supermassive black holes? We do not know how galaxies sustain their star formation for much longer than their present gas supply allows. We do not know why the most massive galaxies cease forming stars at any significant level (called "quenching"), and stay that way. These questions, and many more, comprise the leading edge of galaxy formation studies today (Somerville & Davé 2015; Madau & Dickinson 2014).

Building on today's state-of-the-art, and anticipating developments in the next decade, this chapter will pose questions that we expect will remain unsolved even in the 2030s, because they require observations at wavelengths, sensitivity, and spatial resolution that are well beyond those of other foreseeable facilities. We will focus on two broad domains of galaxy evolution as the "Signature Science" that motivates LUVOIR's UV coverage, high-resolution imaging, and multiplexing capabilities, and in turn its aperture, wavelength range, and instrumentation. To understand "The Cycles of Matter," astronomers using LUVOIR will be able to deploy multi-object spectroscopy in the ultraviolet, with up to 50-fold sensitivity gains compared to Hubble and the multiplexing ability to observe more than a hundred objects at once. To map "The Multiscale Assembly of Galaxies," astronomers using LUVOIR will employ extremely stable 10 mas imaging with an unprecedented combination of depth and wavelength coverage.

**The Cycles of Matter:** How do galaxies acquire the gas they use to form stars? How do they sustain star formation over billions of years when they appear to contain much less gas than this requires? How does feedback from star formation and active galactic nuclei (AGN) expel gas and metals, and to what extent is this feedback recycled into later star formation? What happens to a galaxy's gas when it quenches? Is it used up, ejected, or hidden?

Much of galaxy formation is still hidden from view because it involves gas at temperatures that are too high and densities that are too diffuse to observe directly, or physical scales that are too small to resolve. All these scales, from < 100 pc to more than 1 Mpc, are relevant when we consider how gas flows drive galaxy evolution. With its uniquely powerful UV capability at high spatial resolution, LUVOIR will directly observe these gas flows to trace the raw materials of galaxies, the ejected products of star formation, and the recycling between them over the latter half of cosmic time.

**The Multiscale Assembly of Galaxies:** We know empirically that galaxy formation is a multiscale process spanning at least seven orders of magnitude in mass and three in size that unfolds over the 13 billion year sweep of cosmic time. The brightest galaxies we know about are 10 billion times brighter than the faintest, and the densest are 1 million times more crowded than the most diffuse. In between these extremes, galaxies follow orderly scaling relations between mass, size, star formation, and metal content. This remarkable regularity spanning a vast range of diversity challenges our current theories and even our imaginations to understand how Nature does it. Part of the story is that even the largest galaxies of today began as smaller seeds at the dawn of time, and gradually built up into massive giants by acquiring gas and merging with other galaxies. Along this "merger tree" path, galaxies grow by accreting gas and merging with their neighbors large and small. By reaching depths of 33–34th magnitude, LUVOIR will detect the early building blocks of galaxies like the Milky Way, and fill out the merger tree that leads to galaxies at the present time. It will also "see inside galaxies" to unprecedented limits, resolving their internal building blocks at < 100 pc scales to unravel the processes inside galaxies that drive their evolution. All these themes motivate and exploit LUVOIR's unique spatial resolution and broad wavelength coverage.

The "Signature Science" cases here represent some of the most compelling types of observations astronomers might do with LUVOIR at the limits of its performance. But, compelling as they are, they should not be taken as a complete specification of the LUVOIR program. We have developed concrete examples to ensure that the nominal design can do this compelling science, and so that the astronomical community has detailed examples they can adapt to their own interests. We expect that the creativity of the community, harnessing the revolutionary capabilities of the observatory, will ask questions, acquire data, and solve problems that we cannot envision today. That is as it should be, as a flagship should always reach far beyond current capabilities and break the current limits of our imagination. With this in mind, let us ask vital science questions that we believe will remain unsolved until LUVOIR, and consider how they drive the requirements for its aperture, resolution, and wavelength coverage. We hope these cases will stimulate the reader to imagine their own science cases and to begin applying these tools to design their own future.

## 5.1 Signature science case #1: The cycles of matter

Much of the still-unknown story of how galaxies evolve comes down to how they acquire, use, and recycle their gas. Inflows and outflows of gas likely shape the evolution of star formation within a galaxy. Inflows ultimately arise from the intergalactic medium (IGM) to provide fuel for continued star formation over the lifetime of a galaxy. Outflows from the interstellar medium, driven by stellar radiation and explosions or by AGN, or some combination of them, keep galaxies from forming too many stars in their central regions. The balance of these flows set the rate at which galaxies accumulate heavy elements.

All these flows meet in the circumgalactic medium (CGM), an extremely diffuse gaseous medium spanning roughly 30 times the radius and 10,000 times the volume of the visible stellar disk (**Figure 5.1**; Tumlinson, Peeples, & Werk 2017), made up of clouds, streams, and bubbles embodying galactic accretion, feedback, and recycling. We

---

### State of the Field in the 2030s

LUVOIR's science aims must be considered in light of progress that is expected between now and its launch. What will the field of galaxy formation look like after the advances of the 2020s, when we will have operational JWST, ALMA, and one or more extremely large telescopes on the ground? Of course, we cannot know what surprises Nature will supply, but experience teaches that new facilities break new ground and are forced by competitive peer review to address the most demanding problems of their time.

First, we can expect that essentially the entire sky will have been surveyed at seeing-limited resolution in the optical bands, e.g., Sloan Digital Sky Survey (SDSS), PanSTARRS, and the Large Synoptic Survey Telescope (LSST). LSST's accumulated co-add will reach ~28th magnitude in the optical over 10 years, with 0.8–1" resolution. These maps will be joined by the massive fiber-based spectroscopic surveys that started with SDSS and the 2dF survey and will expand deeper and wider with the Subaru PFS survey and its contemporaries. All-sky imaging provides accurate photometry, and spectroscopy provides additional physical diagnostics of stellar population age and metallicity, and galactic dust content. These vast datasets support rich multivariate analyses with statistical precision and excel at determining galaxy population statistics and galaxy/galaxy correlations in large-scale structure and halo substructure. After these surveys, there might be very little about the large-scale distribution of galaxies that remains to be learned. These massive spectroscopic surveys also detect millions of strong gas absorbers, such Mg II and Ca II lines, which probe dense interstellar medium (ISM) and CGM gas. However, as we will show, they cannot access the key UV physical diagnostics over most of cosmic time.

The other major theme of the 2020s will be high-resolution imaging in the IR, sub-millimeter, and radio. JWST and WFIRST will be the prime space observatories, bringing 50–100 mas resolution to imaging and multi-object spectroscopy. WFIRST will provide Hubble-quality imaging to a large sky area, and JWST will detect galaxies to AB ~ 32. JWST is optimized to reach "the first galaxies," seeing the first seeds of modern galaxies at z > 15. It will also be revolutionary in its mid-IR capability, able to observe the rise of dust, ice, and molecules in galaxies over most of cosmic time. Long-wavelength facilities such as ALMA and SKA will bring their powers to bear on the neutral and molecular gas content of galaxies at high resolution. Finally, 20–30 m telescopes on the ground will, if successful with adaptive optics, bring < 50 mas imaging and spectroscopy at IR wavelengths, in a comparable resolution and depth space as JWST.

What observational parameter space remains after all this? Space observatories are the platform of choice for high-resolution optical imaging, UV imaging and spectroscopy, highly repeatable precision photometry, ultrastable astrometry, and high-performance optical coronagraphy. LUVOIR's "Signature Science" motivates and uses these capabilities, while acknowledging the areas where ground-based telescopes or space facilities at other wavelengths do their best. And in the end, it is the complementarity and collaboration between these domains that drives the science forward.

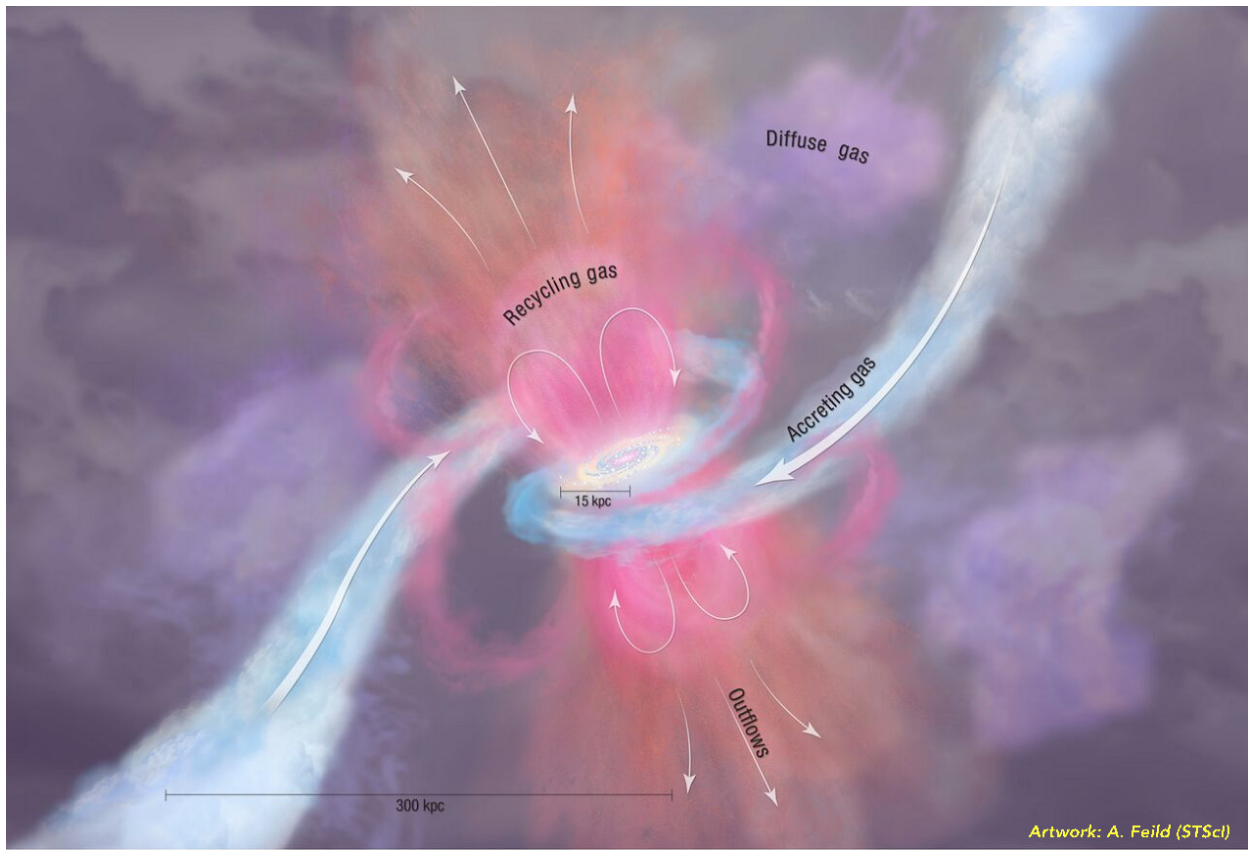

**Figure 5.1.** *Galaxies are much more than they appear to be in starlight. Normal galaxies are surrounded by a massive reservoir of diffuse gas that acts as their fuel tank, waste dump, and recycling center: the circumgalactic medium. It is fed by accretion out of the cosmic web and by outflows from the galaxy. Recent evidence indicates that the CGM is a major factor in galaxy evolution. Unraveling the CGM and all its associated flows is major driving force for LUVOIR and its instruments. Adapted from Tumlinson, Peeples, & Werk (2017).*

now understand that the CGM has a total mass comparable to the stellar masses of galaxies and is richly structured in density, temperature, and chemical enrichment. No picture of galaxy formation and evolution is complete without understanding this major galactic component.

To organize our discussion, we will proceed into galaxies and back out again, starting from the IGM and how it moves into the CGM as galactic fuel. From there we will consider the flows within the CGM and how gas from inside it and outside it are processed and recycled. Finally we will consider the disk/halo interface, where accretion becomes the ISM and the ISM becomes feedback. At each stage we will see how astronomers using LUVOIR could make definitive measurements of processes that have been hidden from view. First, we must stop to investigate why observing these processes makes LUVOIR's UV sensitivity so essential to its design.

### 5.1.1 The essential ultraviolet

Life as we know it is acutely sensitive to energetic radiation from our Sun: UV photons cause rampant cellular damage and genetic mutations. It is possible that life would not exist were it not for the protective ozone layer that keeps these harmful photons out. We can have life on Earth, or UV astronomy from the ground, but it seems we cannot have both. In this state of affairs, we must place our telescopes in space if we are to make a full characterization of the gas in halos and the IGM over the last 10 Gyr of cosmic time.

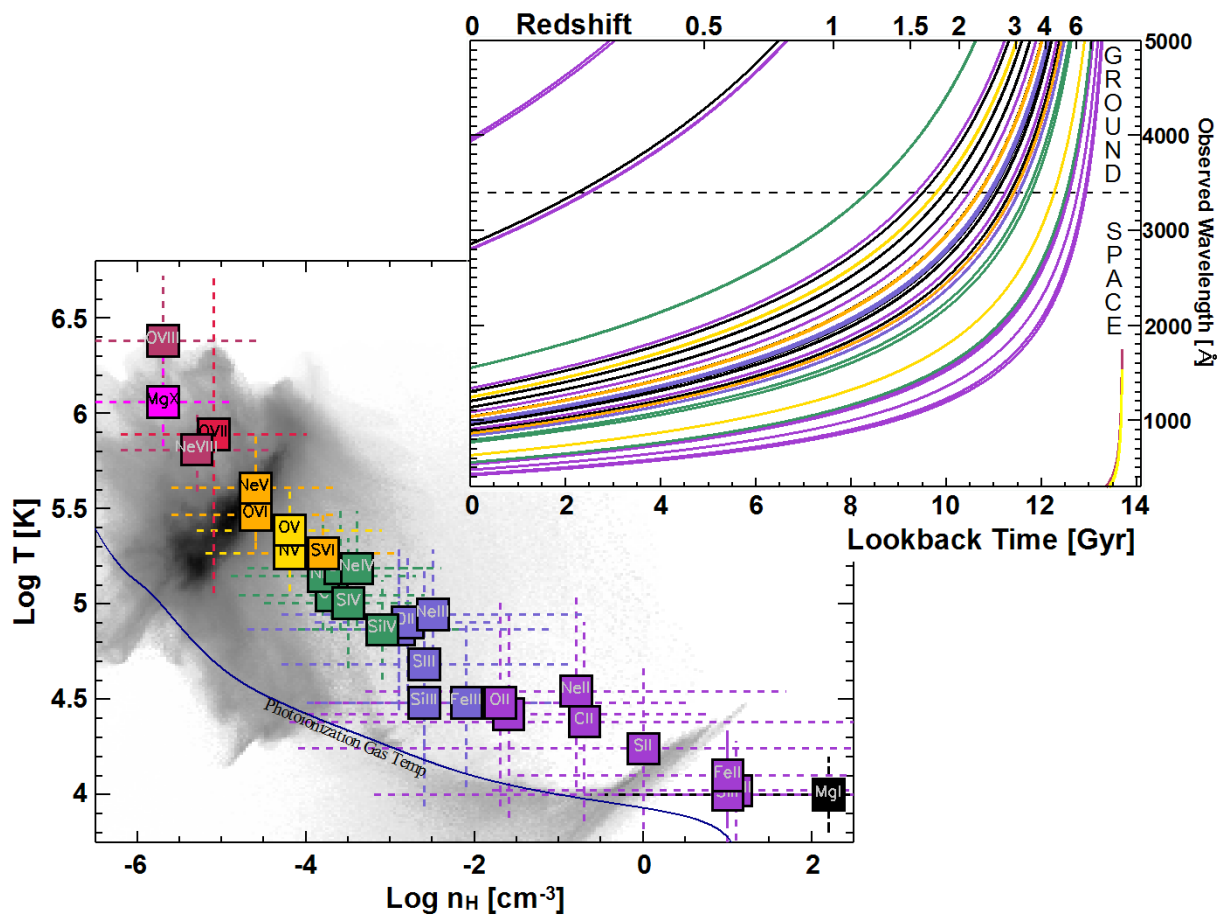

**Figure 5.2.** *Diffuse gas surrounding galaxies requires UV capability for most of cosmic time. The greyscale image below shows that most CGM gas lies at T > 10⁵ K, where atomic physics places its absorption and emission in the rest-UV. The colored squares mark diagnostic ions used to study this gas, placed at the temperature/density combination where they are strongest.*

**Figure 5.2** shows that the ability to cover UV wavelengths is essential to explore the full range of temperature and density where we can find the CGM and IGM baryons—most of the normal matter in the universe. In the panel at lower left, the greyscale intensity map shows the distribution of gas in the halo of a simulated L* galaxy from the EAGLE project (Oppenheimer et al. 2016). This gas is all in the CGM of the simulated galaxy, yet it exists in multiple "phases" from 1 million K and densities near the cosmic mean (~$10^{-5}$-$10^{-6}$ cm$^{-3}$) to denser gas close to the galaxy's disk at $10^4$ K and $\log n_H > -1$. In these conditions, the hydrogen and metals in the gas will exist in high states of ionization: 1% or less of the H may be neutral, while common metals like C, N, O, S, Mg, and Fe may be ionized multiple time to the stages indicated on the figure (e.g., O VI or O+5). In these conditions, the quantum mechanical rules of energy transfer dictate that the gas will emit and absorb energy primarily at UV wavelengths. Detailed calculations indicate that 60–80% of photons emitted and absorbed by diffuse CGM gas are UV photons (Bertone et al. 2013). These lines appear in the upper right panel where their observed wavelengths are plotted versus cosmic lookback time. We conclude that the UV species that best trace the temperature and density regime of diffuse galactic gas flows lie in the UV over the last 10 Gyr of cosmic time. This inescapable piece of physics means that, if we are interested in resolving questions about how galaxies acquire, process, eject, and recycle their gas, over the last 10 Gyr of cosmic time, we need access to UV wavelengths that are observable only from space.

### 5.1.2   Gas flows in absorption: More sources, deeper limits, earlier times

The IGM and the CGM—the lowest density matter in the cosmos—are intrinsically difficult to study. Astronomers have spent decades trying to grasp the nature and impact of this material, with modest success. Absorption-line spectroscopy has been the primary tool of choice, because it excels at detecting diffuse gas down to the very lowest densities (even lower than the cosmic mean density), especially if those regions occupy large regions of space. Absorption lines can be detected in gas spanning a huge range of temperature and densities, providing a lever arm on physical diagnostics for almost all relevant phases of the gas.

The availability of background sources sets a critical limit on the use of absorption lines as probes of diffuse gas. The most throughput-optimized UV spectrograph at the moment is Hubble's Cosmic Origins Spectrograph (COS), which can perform high S/N spectroscopy of sources only down to about 18th magnitude (AB mag in the FUV). There are only about 1000 such objects on the entire sky (**Figure 5.3**), making UV spectroscopy of intervening gas a game of choosing sources carefully to yield the desired population of foreground galaxies. Using this technique, Hubble users have explored modest samples of mainstream L* galaxies (N = 42; Tumlinson et al. 2013), dwarf galaxies (N = 40; Bordoloi et al. 2014), radio-selected star forming disks (N = 45; Borthakur et al. 2015), galaxies in dense environments (Burchett et al. 2017; Johnson et al. 2016) and galaxies hosting AGN (N = 13; Berg et al. 2018). As a result of the paucity of sources, the quasar absorption lines (QSOALS) method can make only crude statistical maps by sampling many halos with one absorption-line path through each and then aggregating the results. The upshot is a set of essentially one-dimensional maps, with only a few forays beyond, and, with Hubble, limited mainly to galaxies at z < 0.5.

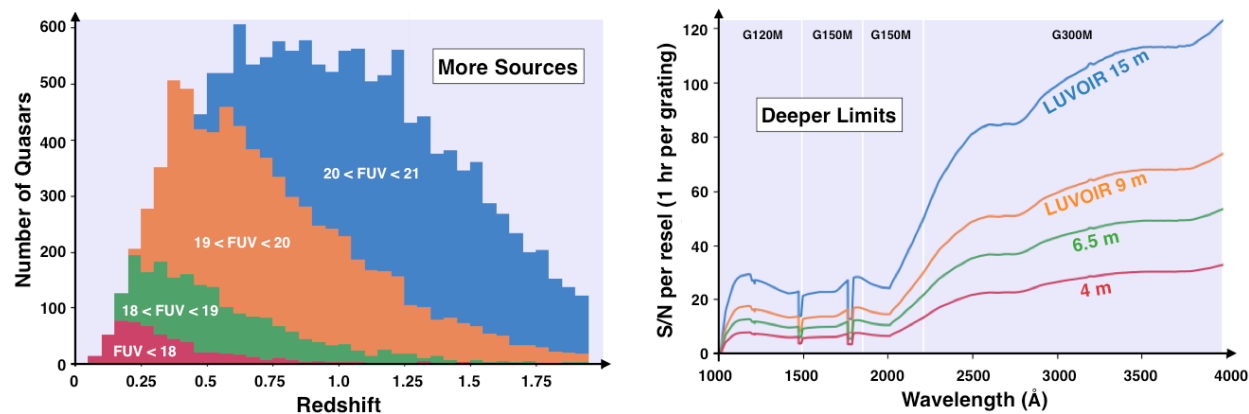

**Figure 5.3.** *Left: Distribution in redshift for SDSS quasars of various GALEX FUV magnitudes. For S/N = 20 spectroscopy, Hubble/COS is limited to the objects in the red wedge, which barely reaches past z ~ 0.7. With its 3 mag deeper grasp, LUVOIR/LUMOS users will have many more QSOs to choose from, with enough in the z = 1–2 interval to finally build precise maps of how gas feeds galaxies at that epoch. Right: LUVOIR+LUMOS S/N for a flat-spectrum FUV = 19 source in one hour of integration for each of the G120, G150, G180 and G300M gratings (R = 30–40,000). The LUVOIR limits were derived from the LUMOS ETC available at luvoir.stsci.edu .*

The LUVOIR Ultraviolet MultiObject Spectrograph (LUMOS) is designed for point-source spectroscopy 30–50 times more sensitive than Hubble/COS, at about double its highest resolution (R ~ 40,000). LUMOS users will be able to choose from any of the quasars counted in **Figure 5.3**, including thousands of choice objects at z > 1. This is a critical time in the history of the universe: star formation rates begin declining from their z ~ 2 peak, AGN have passed their epoch of maximum activity and are turning off, the "red sequence" of passive galaxies is beginning to emerge, and the first large concentrations that we call galaxy clusters are about to mature. To understand the almost totally unknown role of IGM and CGM gas at this epoch, we must be able to observe tens or even hundreds of QSO/ galaxy pairs. The areal density for such pairs is non-linear with magnitude (and so with aperture), given the QSO and galaxy UV luminosity functions. With a 15-m aperture, roughly 1% of all L* galaxies at z ~ 0.2 will have close-separation QSOs for which high-S/N, high-resolution spectroscopy is feasible in an hour. For higher-redshift galaxies at z ~

0.7, the number is lower by a factor of 10, but that still gives access to a respectable 1/1000 of all L*-type galaxies. With such large samples, LUMOS users will be able to address a wide range of critical open questions about the flows of matter.

**Where are the missing baryons?** Normal galaxies possess only a few percent of their expected budget of baryonic matter (Fukugita, Hogan, & Peebles 1998; McGaugh 2005). A large mass fraction exists in the CGM, but this measurement has so far been performed only near L* galaxies and only at z < 0.2 (Werk et al. 2014). The wavelengths accessible to LUMOS provide coverage to an extensive range of rest-frame extreme-UV ions that redshifts into the FUV for z > 0.5 (Tripp 2013). This includes nearly every ionization state of the most abundant heavy element, oxygen (O I though O VI). Performing a baryon census in which most of the gas is traced by a single element will eliminate some of the most serious systematic errors in ionization models that plague these measurements. The remaining oxygen ions—O VII and O VIII—reside in very low density, high temperature gas, generally accessed today through a small number of X-ray sightline in which megaseconds of Chandra or XMM time have been invested (the lines lie at ~20 Å in the rest-frame). Fortunately, other ions visible in the UV at z ~ 1 probe that same regime. One example

---

**Program at a Glance – LUVOIR QSOALS Key Program (Figures 5.3 and 5.4)**

**Program details:** 100 QSOs at z = 1–1.5, approximately 190 hours of exposure time.

**Instrument(s) + Configuration:** LUMOS FUV and NUV spectroscopy, R ~ 30–40,000, G120M, G150M, G180M, G300M. One hour per grating.

**Key observation requirements:** S/N = 20 or higher from 1100 to 4000 Å (blue curve in **Figure 5.3**)

**Synergies with other ground/space facilities:** 30-m+ ground-based optical/IR telescopes obtain data on associated galaxies. ALMA and radio telescopes can constrain their ISM gas and metal content.

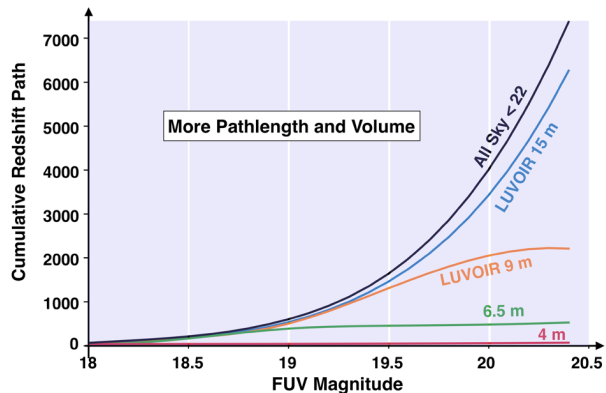

**Figure 5.4.** *Redshift search path to Ne VIII absorption for various apertures assuming a LUMOS throughput. Observations of 1 hour each are assumed, with only those pixels at S/N > 10 contributing to the path. Sources are drawn from the GALEX catalog having FUV magnitudes AB < 22 mag, cross-matched with SDSS DR7 for redshifts. The total path available from this sample is shown in black.*

is Ne VIII ($\lambda\lambda$ 775,780 Å), which to date has only been observed a handful of times with Hubble owing to the lack of bright sources at z > 0.5. As shown in **Figure 5.3** and **Figure 5.4**, LUVOIR can not only provide spectra with S/N = 10 or greater in an hour, it can do so for *thousands* of sources, with a total available Ne VIII redshift path, for single observations of one hour per sightline, more than a factor of 100 larger than that available to a 4-meter aperture with a LUMOS-like throughput. A program targeting Ne VIII would also search for T > $10^6$ K gas via Mg IX, Si XII, S XIV, and Fe XVI, in effect making LUVOIR a powerful X-ray telescope, while retaining access to all the lower-ionization diagnostics in **Figure 5.2**. Every sightline at z ~ 1 thus offers an opportunity to observe baryons across the full spectrum of their phase space, from neutral, colder gas (e.g., O I, H I), through to warm, ionized gas (e.g., O VI), as shown in **Figure 5.2**. This Signature Science forms a synergy with ground-based telescopes, which are needed to measure the redshifts, masses, star formation rates, and metallicities of the associated galaxies. As Hubble and Keck have cooperated to advance this field in the 2020s, so can LUVOIR and the next generation of extremely large telescopes on the ground.

**Follow the metals to map feedback.** Heavy elements are "Nature's tracer particles"—the equivalent of a message in a bottle that tells us where the products of star formation have been carried over time. These messages spread far and wide, clearly showing that some galaxies have transported their metals over vast distances—even megaparsecs—away from their birthplace. If we can piece together all the messages in all the bottles from all the island universes, we should come to understand how galactic feedback actually works.

Hubble's UV capability has extensively probed the extent of metals around galaxies, culminating in the finding that only 20% of all metals produced are still retained in the galaxies that made them (Peeples et al. 2014). The rest are elsewhere, many in the CGM. It is striking how wide a range of metallicity is seen in normal galaxy halo gas at z < 1, with ~50% of the gas having metallicities around 5% Solar, while the other half is roughly Solar (Lehner et al. 2013; Wotta et al. 2016). The lowest-metallicity CGM gas may be associated with infalling material, either from the IGM or accreting from a hot corona, while the higher metallicity gas could be ongoing enriched feedback. Extensive work with numerical simulations indicates that the physical extent of metals around galaxies, and their relative mixtures in the phases of the gas, can provide critical tests of poorly understood feedback models. Understanding the origins of this gas, whether it will collapse onto the galaxy or will be subsumed back into the corona, is critical to understanding how and from where galaxies replenish their fuel supply for forming stars. Or, indeed, whether they are able to do so at all.

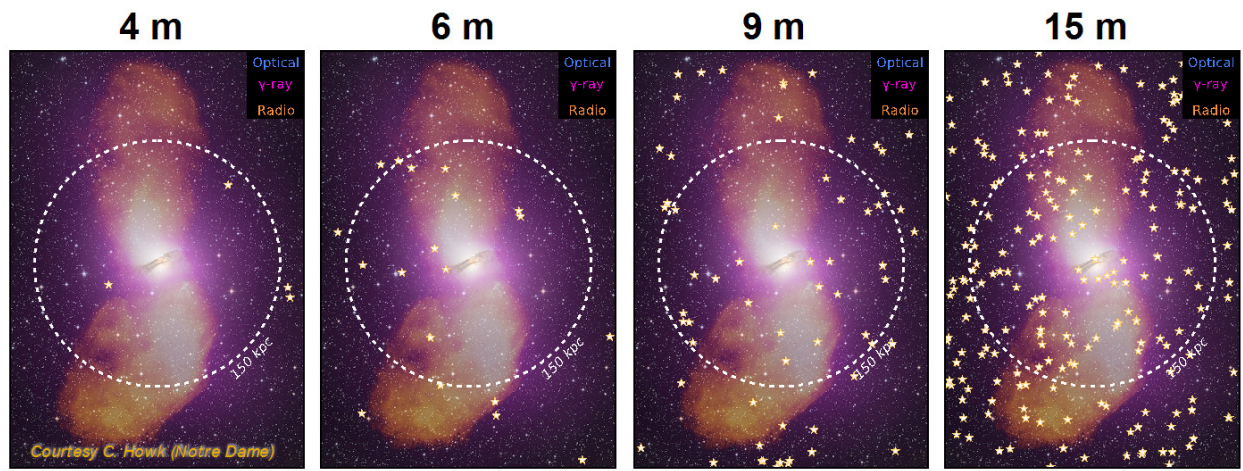

**Figure 5.5.** *A realization of the sky about the radio galaxy Cen A demonstrating the density of sources available for high-resolution, high-S/N spectroscopy against background QSOs behind low-redshift galaxies assuming different telescope apertures. Such observations are critical for understanding the driving of jets and their interaction with the CGM, which may dictate the rate at which CGM gas could cool to fuel further star formation. Each star represents a QSO down to FUV = 22 mag that could be observed to S/N > 10 in ~0.5–1 hour assuming a LUMOS spectrograph. A 9-m telescope has a source density smaller by ~5x than that of the 15-m. The nearest QSO for which Hubble spectroscopy is achievable in <20 orbits is at impact parameter R = 300 kpc, two times further from the core of this galaxy than the circle representing R = 150 kpc. Background galaxies, which can also be used to map the gas, are ten times more numerous.*

LUVOIR will allow us to measure the prevalence of low-metallicity gas about galaxies of different masses, star-formation rates, and environments using UV absorption spectroscopy. This is currently not feasible now, because of the relative low density on the sky of UV-bright sources accessible to Hubble's spectrographs. Moreover, LUVOIR will be able to do this over the critical z = 0.5–1.5 epoch of galaxy formation. Performing a census of metals over this redshift range can exploit the simultaneous coverage of almost every UV ion of oxygen, sidestepping any systematic problems with using C, Si, O, and N variously, as is done at z < 0.2. This is only possible with UV coverage and redshift that places UV lines of O II, O III, and O IV into the space UV (Tripp 2013).

Using this capability, LUVOIR's users can complete the low-z metals census,

constrain feedback from galaxies, assess the importance of galactic recycling, and ultimately tell an important early part of the story in the rise of the elements and life. Another critical example is shown in **Figure 5.5**. AGN feedback is thought to play an important role in keeping hot halos from cooling in quiescent galaxies, though it is difficult to probe the jet and radiation interactions with the surrounding CGM gas. Increasing the density of sources to which we have access with UV spectroscopy will allow us to probe that interaction directly. The key needs for such science are high spectral resolution and sensitivity (aperture) to provide high enough source density that individual AGN hosts can be studied.

**Precision cosmology with the IGM:** Though it lies outside galaxy halos, the IGM as traced by the Lyman alpha forest provides unique constraints on cosmological structure formation. As shown in **Figure 5.3**, the high throughput of the G300M grating on LUMOS provides S/N > 30 coverage of the forest in a single hour. To date, less than five quasars have been observed covering the forest at S/N ~ 10 or greater at z ~ 1, at the cost of nearly 100 Hubble orbits. The baryon census and feedback studies described above require access to redshifted Lyman alpha (HI 1215), so a program targeting metal absorption obtains unprecedented Lyman alpha forest "for free." A LUVOIR program observing 100 quasar sight lines at S/N > 30 would perform three major measurements: (1) constrain the matter power spectrum P(k) to better than 5% accuracy on scales of 0.1 to 100 Mpc at z ~ 1, bringing precision cosmology to a new epoch in time; (2) measure the H I column density distribution function $f(N_{HI})$, which at the low H I end measures the IGM, but at the intermediate and high HI end directly maps onto the CGM and ISM of galaxies; (3) measure the equation of state of the IGM, $T = T_0(\rho/\rho_{avg})^{\gamma-1}$, to accuracies approaching 10%. The thermal history of the universe bears the fingerprints of major heating events such as reionization of Hydrogen and Helium and the influence of the Hubble expansion, but also can be used to trace the influence of other hydrodynamic processes more closely linked to galaxy formation and evolution.

   **A comprehensive LUMOS QSOALS campaign:** All of these studies of CGM missing baryons, metals, and the IGM can be carried out with a single, comprehensive QSO survey. For definiteness, and for scaling to other architectures, we scope this to use the 100 brightest QSOs identified in the SDSS DR7 QSO catalog, using matched GALEX FUV fluxes. We set the S/N goals such that 1 hour in each of the LUMOS High resolution gratings will yield the blue curve in the upper right of **Figure 5.3** for an FUV = 19 mag QSO. For brighter objects, we scale the exposures by FUV magnitude to maintain the same S/N. This survey of 100 QSOs will require just under 200 hours of integration time (and, perhaps ~300 hours with overheads). Spectroscopy for 100 background quasars at z~1 represents a fundamentally game changing prospect for the study of the formation and evolution of the baryons in, around, and in-between galaxies and the larger cosmological context galaxies live in. No mission current or planned other than LUVOIR could complete such a program in a treasury scale program allocation of time. Any claim of understanding the history of baryons in the universe demands a study of this epoch of transformation in cosmic time, and LUVOIR rises to the challenge.

### 5.1.3   Mapping the cycles of matter with the faintest light in the universe

Quasar absorption lines are optimal for detecting and characterizing the IGM and CGM without regard to gas density, but they are limited in what they can teach us because they reveal little about the 3D distribution of the absorbing gas. To improve this situation, we must attempt to "take a picture" of a galaxy's gas flows in two dimensions, using emission from the gas itself that reveals its density, temperature, metallicity, and kinematics. With its millions of ~0.1" apertures, customizable to nearly any source, LUMOS will be able to probe physical processes—shocks, accretion, ejecta, and recycling—at scales that even simulations today can barely reach. This capability will enable LUVOIR's users to investigate small-scale structures within the CGM, especially in those vital regions where outflows meet inflows and gas is being recycled. We will consider two specific applications of this capability: large maps of small-scale gas dynamics in nearby galaxies, and detection of the extremely diffuse CGM gas that fills halos.

   Ground-based IFU spectrographs such as MUSE at VLT and KCWI at Keck are pioneering the search for CGM gas emission at z > 2, where the relevant diagnostic lines pass into the visible bands (**Figure 5.2**). Gas reservoirs extending over hundreds of kpc appear to be illuminated by radiation from the stars and AGN in the galaxies (Cantalupo et al. 2014) and some structure can be resolved (Martin et al. 2016). These instruments can access some of the important UV diagnostic lines, most readily Lyα, but only at high redshift. They are also limited still to relatively low spatial resolution, which translates to physical scales of > 10 kpc. They are limited in sensitivity by bright and time-variable sky backgrounds that will make detection of the much weaker lines of key metals exceedingly difficult in individual halos.

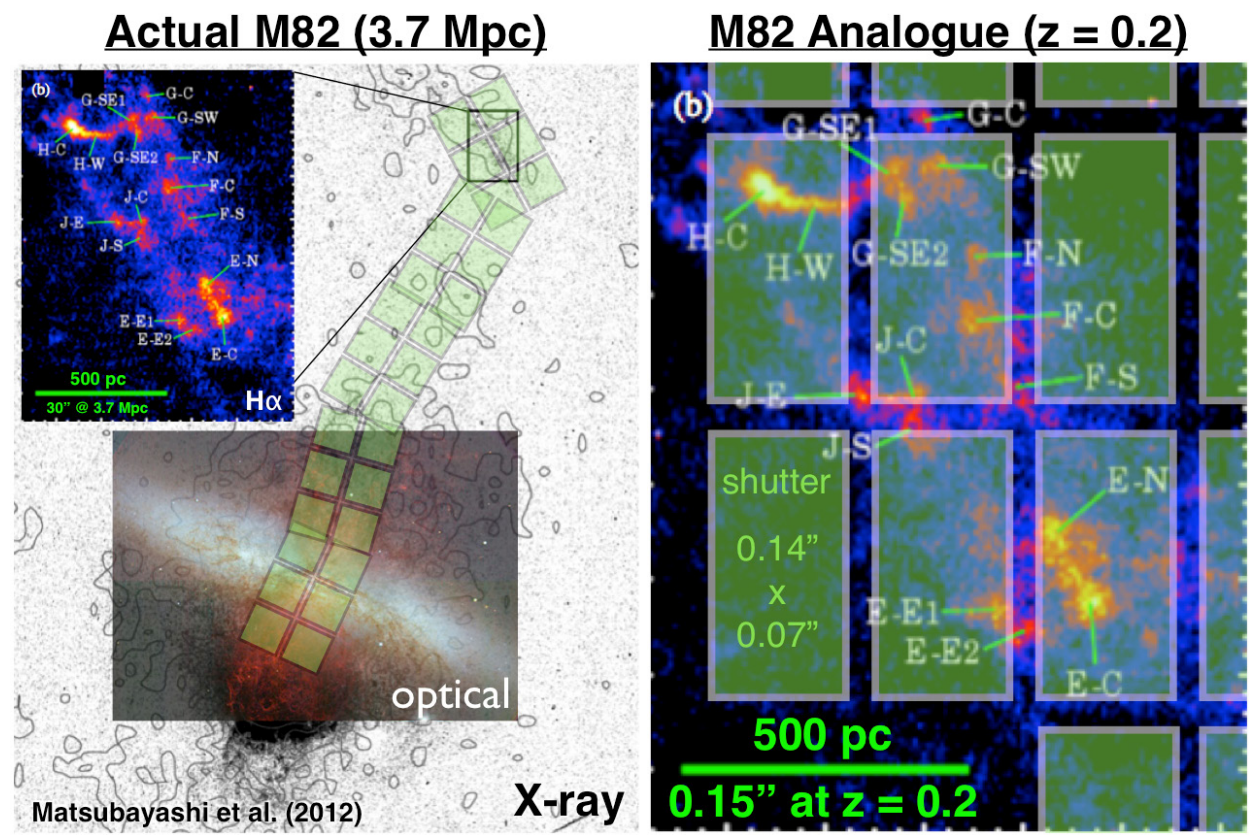

**Figure 5.6.** *Two examples of employing LUVOIR / LUMOS to examine the small-scale physics of galactic outflows. At left, X-ray and optical images of the M82 starburst galaxy are then zoomed in to the false color inset, where one microshutter of 0.14x0.086" matches well to the bright orange clumps. The full LUMOS field of 2x3 microshutter arrays could be used to tile the full scale of the outflow. At right, we use the zoomed in region from at left to show how these clumps of interacting gas can be mapped at the level of individual shutters, offering access to the appropriate physical scales with unique UV diagnostics and low sky backgrounds.*

LUMOS will bring at least three unique abilities—UV coverage, high spatial resolution, and extreme sensitivity—to bear on this important problem, in ways that ground-based telescopes cannot. With this instrument, astronomers will be able to map the density, temperature, and mass flow rates of the CGM, directly, using the UV radiation emitted by CGM gas as it cycles in and out of galaxies. **Figure 5.6** shows how the LUMOS microshutters map to small-scale (< 1 kpc) clouds that carry gas and metals away from the prototypical starburst galaxy,

---

**Program at a Glance – High-Definition View of a Local Galaxy CGM (Figure 5.6)**

**Program details:** 30 QSOs within 2.2 deg of Cen A (R < 150 kpc at 3.86 Mpc).

**Instrument + Configuration:** LUMOS FUV and NUV spectroscopy, G120M, G150M, G300M.

**Key observation requirements:** S/N = 10 at O VI (G120M), 20 at Ly$\alpha$, S/N = 10 at C IV (G150M), S/N = 10 at Mg II (G300M), for a total of approximately 100 total hours of integration.

**Synergies with other ground/space:** Radio and X-ray telescopes trace the cold and hot components of the outflow, while the UV traces warm gas and determines accurate kinematics.

---

M82. Four separate footprints of the 2 x 3–array microshutter field of view are overlaid in light green on the galaxy and its outflow. The false color zoom shows small clumps that trace the interaction of the galaxy's outflow with the CGM. Did these clumps cool and form where they are? Or is this material directly ejected? If so, what does it imply about the mass ejection rate, and the mass rate of recycling? This is galactic feedback in action, but only at UV wavelengths can we prove the relevant energy scales to answer these questions. Multiply this small region many-fold, and it becomes clear that we must be able to observe hundreds of such places in this complex flow to understand its true dynamics. At the distance of M82 and other nearby galaxies, the 0.14 x 0.07" shutters subtend parsec-scale sizes and so may be ganged together into larger "virtual apertures" to map the faint light emitted by gas entering and leaving galaxies, resolving these complex flows at sub-kiloparsec scales. At z < 2, where the relevant diagnostics are still in the UV, each shutter still reaches sub-kpc scales, as shown by a rescaling of the M82 outflow clumps in the right panel of **Figure 5.6**. The ability to resolve gas flows at parsec to kiloparsec scales in the key UV diagnostics lines is a unique ability of a large UV-sensitive space telescope.

Using the same array of shutters binned into larger "virtual apertures," LUVOIR can also seek the extremely faint emission from the widely distributed diffuse CGM and structure within it. **Figure 5.7** shows such a

---

**Program at a Glance – Emission Line Maps
of Nearby Galaxy Gas Flows (Figure 5.6)**

**Program details:** Observe M82 outflows. 10 multi-shutter array tiles, 100 objects / tile, 10 hours per tile, 100 hours. Exact shutter placement determined from LUMOS UV pre-imaging. Total ~100 hr.

**Instrument(s) + Configuration:** LUMOS FUV and NUV spectroscopy, R ~ 30-40,000, G120M, G150M, G180M, G300M

**Key observation requirements:** S/N ~ 5 or higher in key lines, with exposure times tuned to achieve this in multi-shutter apertures of varying size.

**Synergies with other ground/space facilities:** Ground-based (Ha) or LUMOS UV maps can provide additional diagnostics; radio detections might be achieved for dense clumps in nearby galaxies.

---

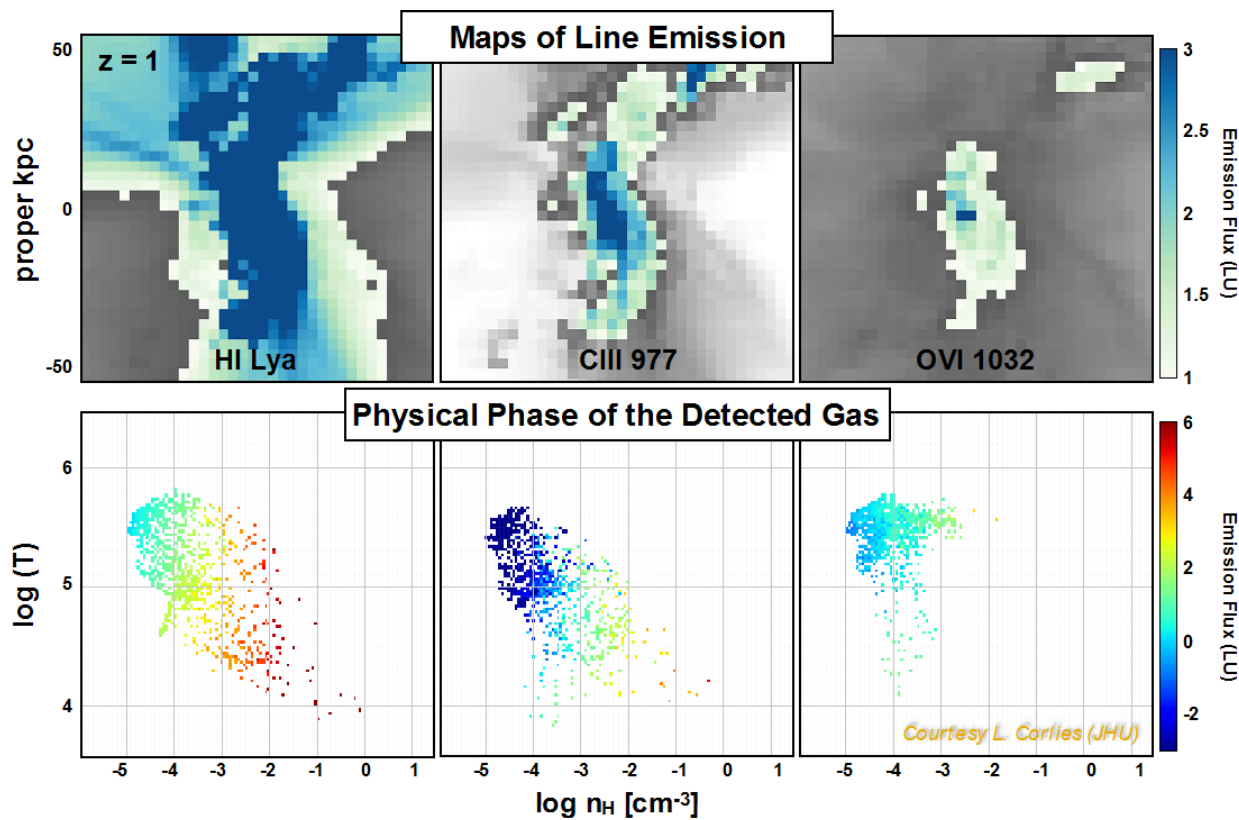

**Figure 5.7.** *A rendering of simulated CGM emission in three key UV lines for a Milky-Way progenitor galaxy at z = 1. These lines range from 1000-1200 Å in the rest frame, so they are uniquely visible in the UV at z < 2, or half of cosmic time. The 15-m LUVOIR+LUMOS will directly detect bright emission (blue/dark green regions) and resolve bright knots in the light green regions. Significantly smaller telescope apertures will lack the sensitivity for all but the brightest knots, and will not resolve them even if they are detected. Credit: L. Corlies (JHU).*

hypothetical map from a new hydrodynamical simulation of a Milky Way progenitor galaxy at z = 1 (FOGGIE; Corlies et al. 2018). This diffuse gas will be challenging to detect even for LUVOIR, but appropriate integration times could be fitted in as parallels to the week-long exoplanet visits, for example. In deep LUMOS exposures, the structure of the CGM can be detected in multiple lines, allowing us to count up the heavy element content of this gas, to watch the flows as they are ejected and recycled, and to witness their fate when galaxies quench their star formation, all as a function of galaxy type and evolutionary state. LUVOIR could map galaxies in fields where deep imaging identifies filaments in the large-scale structure, and where ground-based ELTs have made deep redshift surveys to pinpoint the galactic structures and sources of metals to be seen in the CGM. Because this radiation is far weaker than local foreground radiation (SB ~ 100–1000 photons cm$^{-2}$ s$^{-1}$ sr$^{-1}$) ground-based telescopes seeking it at redshifts where it appears in the visible (z > 2) must perform extremely demanding sky foreground subtraction to reveal the faint underlying signal. These foregrounds are considerably lower from space (by factors of 10–100), shortening required exposure times by an equivalent factor. Using this technique and binning up 0.5–1" regions of the microshutter array (a few kpc at z < 2),

LUMOS users can examine the large scale distribution of the filaments and extended disks they feed, from the peak of cosmic star formation down to the present.

When put together, the three essential ingredients of UV coverage, spatial resolution, and extreme sensitivity will enable mapping of gas flows through the CGM in its rich detail. LUVOIR accesses the key lines at z < 2, a 10 Gyr era encompassing the peak of cosmic star formation, the rise of the brightest quasars, the widespread quenching of massive galaxies, and the emergence of galaxies as we know them today. Only an optimized UV mission such as LUVOIR has the sensitivity and resolution to apply these powerful diagnostics to this cosmic era when so much happens.

### 5.1.4    Gas flows and quenching

Galaxy quenching is a prime target for LUVOIR's unique power. How galaxies quench, and remain so, is a major open question. The number density of passive galaxies has increased 10-fold over the 10 billion year interval since z ~ 2 (Brammer et al. 2011). Galaxies undergoing quenching are the ideal laboratories to study the feedback that all galaxies experience: the galactic superwinds driven by supernovae and stellar radiation, the hot plasma ejected by black holes lurking in galactic centers, and the violent mergers that transform galaxy shapes while triggering the consumption or ejection of pre-existing gas. LUVOIR will have the collecting area to support deep, wide-field UV multi-object spectrograph searches for CGM gas at the line emission fluxes that are expected, and with the spatial resolution to observe the transformation of star forming disks to passive spheroids at 50–100 pc spatial resolution and closely examine the influence of AGN on this process. For galaxies identified as quenching, emission maps of the surrounding CGM will determine the fate of the gas that galaxies must consume or eject and powerfully elucidate the physical mechanisms that trigger and then maintain quenching. Only a diffraction-limited space telescope with an aperture of at least 10–12 meters can achieve such spatial resolution in the optical and observe the rest-frame UV light necessary to witness the co-evolution of stars and gas in galaxies undergoing this transition. As most of the development of the present-day red sequence occurred since z ~ 2, and the key diagnostics are rest-frame UV lines, this critical problem is a unique and compelling driver for LUVOIR's aperture, UV coverage, and capabilities for multi-object spectroscopy.

### 5.1.5    Mapping inflow and outflow with down-the-barrel spectroscopy

We can now consider the final steps in our view of galactic gas flows with LUVOIR: how CGM gas turns into ISM, and how ISM gas returns to the CGM. What drives the flows that transport mass from the CGM into galaxies, and then back? The inflows are driven primarily by gravity and cooling acting on and within gas that enters the halo in filaments, strips off satellites, or becomes thermally unstable while orbiting in the halo. Gas cooling and gravity are straightforward to implement in models and simulations, even if the emergent behavior they create is hard to simulate. "Feedback"—the general term for mass, momentum, and energy from stars and AGN that influence the course of galactic evolution—is far more complex, because the underlying physical mechanisms span a huge range of density, temperature, energy, and physical scales. Feedback flows such as supernovae and AGN winds originate on parsec scales, but propagate to hundreds of kiloparsecs while interacting energetically, hydrodynamically, and radiatively with everything they encounter, including current inflows and past generations of outflows.

Understanding how feedback operates physically to influence galaxies as they grow is an active and abiding challenge in the astrophysics of galaxies. Theoretical models of how flows develop and propagate for various sources and physical mechanisms, including winds and radiation pressure from main-sequence OB stars, the collective effects of correlated supernovae, and jets and winds from AGN. At their best these models make specific predictions for the mass and energy transport, velocity and acceleration profiles, and time evolution of these flows as a function of the source properties (e.g., Murray et al. 2011, Thompson et al. 2011). These trends are often then implemented as "sub-grid" prescriptions in large-scale numerical simulations that attempt to recover realistic galaxy populations and internal properties. These prescriptions are labeled "sub-grid" because they occur at sub-kiloparsec scales that cosmological simulations still cannot resolve.

Observers can test these physical models of feedback, but with a major limitation. Some bulk properties of galactic outflows can be revealed by spectroscopy that uses the driving sources themselves—whether AGN or star-forming regions—as background sources, in a so called "down-the-barrel" spectrum. Inflow is detected as redshifted absorption and/or emission, while outflow is blueshifted. Often the observed profiles include absorption and emission from separate portions of the gas, which must be teased apart to reveal the details of the flow. The major limitation of current "down-the-barrel" measurements is that they typically cover most or all of a galaxy's disk, so the observed profiles average over a large number of individual sources, erasing the source-by-source variations in energy and mass that trace the key physical variables. Nevertheless, this technique has successfully demonstrated a correlation between gas velocity, star formation rates, and star formation surface density (Martin 2005) and star formation surface density (Kornei et al. 2012), with galaxies with higher star formation rates exhibiting higher outflow velocities. These signatures get stronger as the sightline approaches the galaxy's semi-minor axis, suggesting that the flows are biconical in shape and emerge up out of the disk (Bordoloi et al. 2014), as also seen in hydrodynamical simulations. Down-the-barrel measurements of small "green-pea galaxies" have examined the escape of ionizing radiation from their star-forming regions using UV/optical line ratios (Henry et al. 2015).

To resolve the physics at scales closer to the actual sources, astronomers using Hubble's COS instrument are making pioneering measurements of 16 individual clusters in the face-on nearby galaxy M83 (D = 4.6 Mpc). This program (Program 14681, PI Aloisi) is observing 16 UV-bright clusters across the face of M83, to map out the gas flows emerging from them individually. This program requires 40 orbits to execute, so doing 10 times as many clusters or a few galaxies would be a large or very large allocation of Hubble time.

**Figure 5.8** shows the potential of the LUMOS multi-object mode to transcend these limitations and fully resolve these outflows. In particular, LUMOS could probe the launch-points for winds across the face of a spiral galaxy by observing its individual OB associations, providing a measure of the energy injection needs for launching winds. This would resolve the flows at the physical scales of individual star-forming regions, allowing us to correlate outflow properties with their driving sources in detail, rather than averaged over the whole disk. Outflows can be examined as a function of the star-formation region that produced them, as a test of specific predictions for how flow velocities and mass transfer rates depend on time and energy input (Murray et al. 2011, Thompson et al. 2005). Not only will LUMOS enable the dissection of flows at small scales, but the multiplexing of the microshutter arrays provides a huge efficiency gain that will enable maps of resolved flows for a wide range of galaxies in the nearby Universe. It is critical to understand the wind-expulsion history of galaxies as a function of mass, as the winds are likely critical for shaping the stellar mass-halo and mass-metallicity relationships of galaxies, which depend strongly on galaxy mass. This is best accomplished by high-S/N, high-resolution spectroscopy with broad UV wavelength coverage. The wavelength coverage, notably to wavelengths as short as λ~1000 Å, is critical. If we are to understand galaxy transformation and the role that winds may play in it, the ability to observe these flows at the relevant small scales is needed.

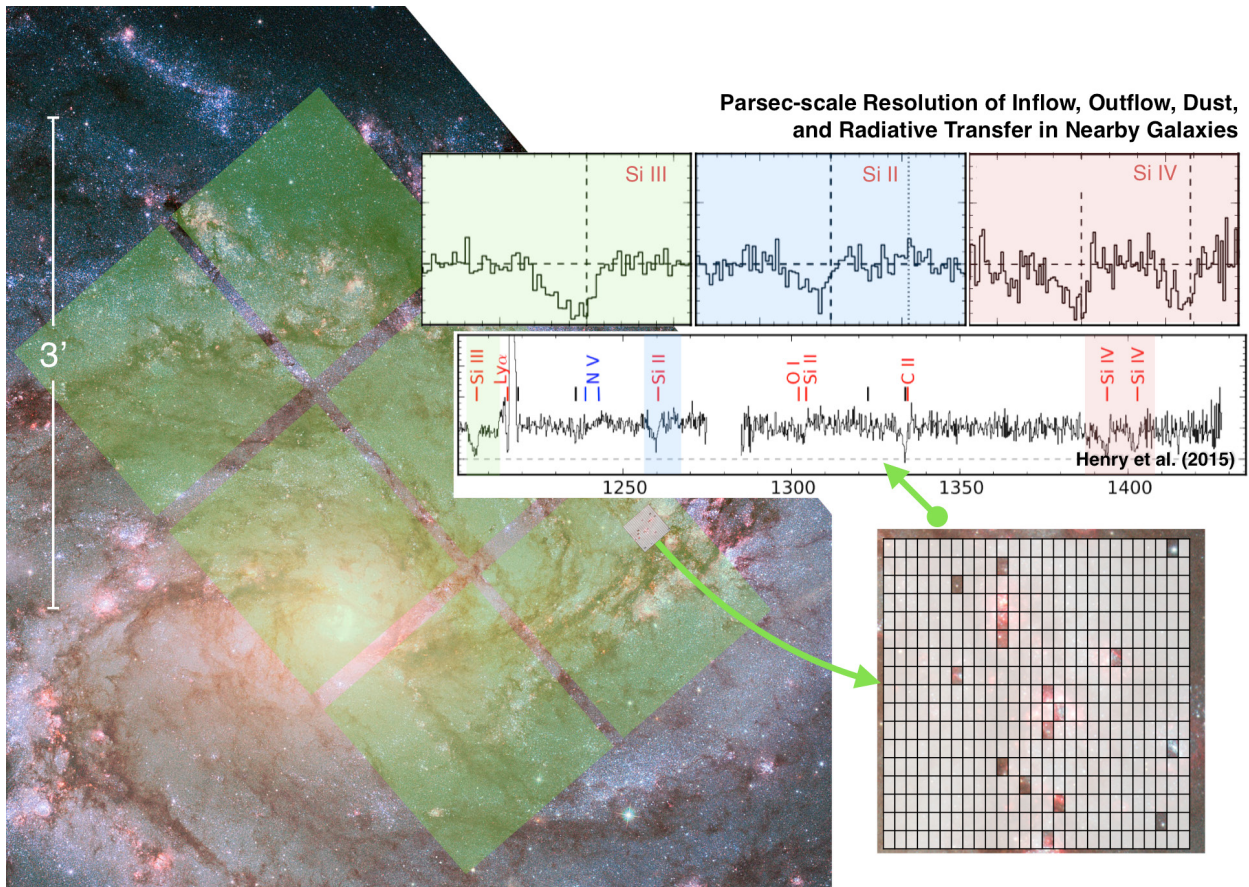

**Figure 5.8.** *LUMOS will perform intensive multi-object spectroscopy of star-forming regions, ISM gas, and galactic outflows in nearby galaxies. Here we show the footprint of the LUMOS multi-object mode overlaid on the nearby galaxy M83 at 4.6 Mpc. At this distance, the LUMOS microshutters subtend 2-3 parsecs. At the top, we show three silicon lines that trace multiphase gas in outflows, as proxied by Hubble/COS spectra of low-z "green pea" galaxies by Henry et al. (2015). LUMOS users will be able to examine a wide range of ionization, metallicity, kinematics, and dust diagnostics down to 1-3 parsec scales at the positions of hundreds of individual stellar clusters and ISM simultaneously, and for many nearby galaxies in a single program.*

## The LUVOIR Deep and Wide Galaxy Fields

Since the original Hubble Deep Field (Williams et al. 1996), large-area surveys at the deepest limits have been a mainstay of galaxy evolution studies; the Ultra Deep Field, CANDELS, and the Frontier Fields have followed. Much of the LUVOIR's "Signature Science" will follow the same model, in which multiple scientific objectives are enabled by a single set of deep exposures over a large area.

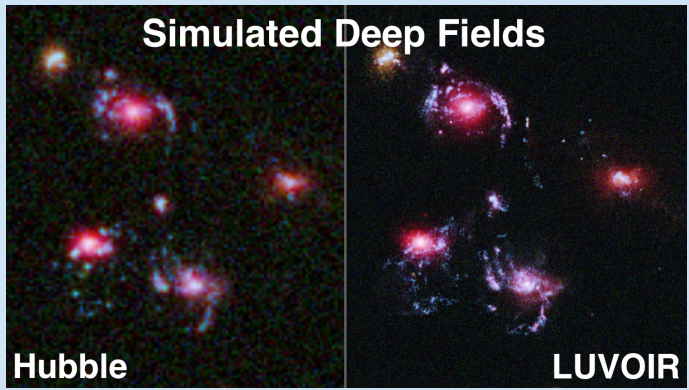

**Simulated Deep Fields**

**Hubble**                    **LUVOIR**

The nominal High Definition Imager (HDI) has a field of view of 2 x 3 arcmin, or 6 sq. arcmin, sampled by two focal plane arrays with Nyquist sampling that view the sky at the same time. Detailed calculations show that HDI mounted behind LUVOIR's 15-meter A architecture will reach 5σ photometric limits of AB = 33–33.5 for point sources in integrations of 10 hours per band. The most basic "LUVOIR Deep Field" is a single field of view deep integration, much like Hubble's Ultra Deep Field (AB ~ 29; Beckwith et al. 2006) or Extreme Deep Field (Illingworth et al. 2013), with these 10 bands taking 100 hours of integration to reach the tabulated limits.

A LUVOIR "Wide" Field, inspired by Hubble's CANDELS program (Grogin et al. 2011), exploits LUVOIR's high mapping speed to cover 720 arcmin$^2$ in 120 tiled fields of view. If we limit this program to 1200 hours integration time, or 10 hours per tile, 1 hour per band, the limits are AB ~ 31–32 mag, which surpasses Hubble's deepest limits by > 2 mag and matches JWST's deep limits.

These limits are unique to LUVOIR: not even 30-meter class telescopes on the ground will reach such depths, owing to time-variable sky backgrounds. This capability enables LUVOIR to detect (1) a single Sun-like stars (AB = 4.72) out to 5.5 Mpc, (2) a main sequence O star to 500 Mpc or redshift z ~ 0.1, nearly the entire volume covered by the SDSS spectroscopic survey, and (3) a 0.001 L* galaxy at z= 6, deep enough to detect the early seeds of galaxies like our own Milky Way. This is a broadly applicable capability that will advance the many areas of science we contemplate as "Signature" for LUVOIR.

| Per band | The LUVOIR Deep Field: 6 arcmin$^2$ in 100 or 1000 hours | | | | | | | | | |
|---|---|---|---|---|---|---|---|---|---|---|
| | F225W | F275W | F336W | F475W | F606W | F775W | F850W | F125W | F160W | F220W |
| 10 hr | 33.0 | 33.2 | 33.5 | 33.6 | 33.4 | 33.0 | 32.6 | 33.2 | 33.0 | 29.7 |
| 100 hr | 34.4 | 34.6 | 34.9 | 34.9 | 34.7 | 34.3 | 33.9 | 34.5 | 34.2 | 31.0 |
| | The LUVOIR Wide Fields: 720 arcmin$^2$ in 1200 hours | | | | | | | | | |
| 10 hr | 31.2 | 31.4 | 31.8 | 31.9 | 31.8 | 31.4 | 31.0 | 32.0 | 31.7 | 28.5 |

<div style="border">

**Program at a Glance – Gas, Radiation, and Dust Feedback in Nearby Galaxies (Figure 5.8)**

**Program details:** 1000 sources / galaxy (100 sources per config, 10 configs / galaxy, 4 hr / config) on 5 nearby galaxies over a range of mass / metallicity. Total 200 hours of integration.

**Instrument(s) + Configuration:** LUMOS multi-object FUV and NUV spectroscopy, R ~ 30–40,000, G120M, G150M, G180M, G300M, 1000–4000 Å.

**Key observation requirements:** S/N > 30 in UV continuum over 1000–4000 Å (FUV mag < 19).

**Synergies with other ground/space facilities:** ALMA and radio telescopes can constrain ISM gas, dust, metal content, and cluster age/metallicity.

</div>

## 5.2    Signature science case #2: The multiscale assembly of galaxies

The galaxies we see in our neighborhood today have spent billions of years growing and evolving from their first faint seeds at Cosmic Dawn. Over this time, some grow by forming stars and merging with their neighbors to become big galaxies with prominent bulges and massive disks hosting violently active black holes. Those that grow too big too fast can "quench," ceasing to form stars and then evolving passively as their stars age. Other galaxies remain small, perhaps eventually merging into larger ones. Nature makes galaxies of these kinds and every type in between, and no theory of origins will be complete without a full understanding of how this happened.

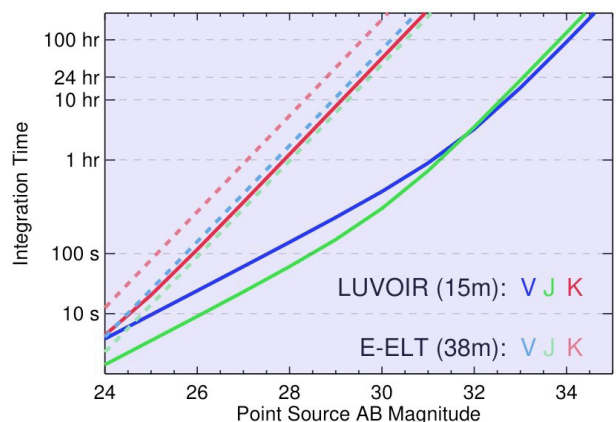

**Figure 5.9.** *Integration time required to achieve S/N = 10 for point-source photometry in three broad bands for the 15-m LUVOIR A (solid lines) and the 38-m E-ELT (dashed lines). If we take the 10-hour exposure as a nominal "large program" to which many nights or orbits would be devoted, LUVOIR reaches AB = 32.5 mag in V and J, 3.5 magnitudes deeper than E-ELT. The LUVOIR limits were derived from the HDI ETC available at* luvoir.stsci.edu/hdi_etc

Why will galaxy formation remain Signature Science in the era of LUVOIR, even after JWST and the ELTs? In short, none of these highly capable facilities will be able to achieve LUVOIR's unique combination of depth, resolution, and mapping efficiency. Depth allows us to see the smallest building blocks of galaxies. With enough resolution we can look inside galaxies at small physical scales, breaking down their formation at a level of detail that isolates the key physical processes. Finally, any observational insights must be backed by statistically significant samples collected with a high mapping efficiency or speed. Optimizing for these three figures of merit will enable LUVOIR's users to make revolutionary advances in the Signature Science of galaxy formation and evolution.

### 5.2.1    Galaxy assembly at the faint frontier

Pushing back the faint frontier has been a constant theme of galaxy formation from the earliest days, through the Hubble Deep Field, and into the present. LUVOIR will expand the faint frontier to the smallest relevant scales over nearly the whole of cosmic time. In the modern universe, galaxies occupy an enormous range of mass from giant ellipticals at $M^* > 10^{12} M_\odot$ to ultra faints at $M^* < 10^4 M_\odot$. In the prevailing hierarchical paradigm, all galaxies of any substantial size grow from the steady accumulation of gas and by the successive mergers of many smaller galactic components in what is known as a merger tree. Every giant galaxy has many dwarfs in its past; our own Milky Way has evidently accumulated many such dwarfs over its history. Within the next few hundred million years, it will acquire the two Magellanic Clouds in a minor merger with the even more dramatic major merger with Andromeda to follow. Telling the full story of galaxy formation starts with being able to work backwards through this tree of galactic origins to see the roots, and this requires pushing back the faint frontier.

The brightest galaxies at any redshift will tend to be the seeds of the brightest galaxies at any later time. The galaxies we see at the dawn of time with Hubble, and soon with JWST, are not the earliest seeds of galaxies like our Milky Way. While we can probably reach Milky Way-like progenitors as far back as z = 6 with JWST (Okrochkov & Tumlinson 2010), what we see will be the main trunk of the tree, not the hundreds of other branches that existed at that time. But what is the minimum threshold for galaxy formation that LUVOIR should try to reach?

The faint frontier expanded in an unexpected direction with the discovery of ultra-faint dwarf (UFD) galaxies in the halo of the Milky Way. These galaxies are gravitationally bound but possess only $\sim 10^5 M_\odot$ of stars, or even less, with characteristic radii of only 100–300 parsecs. They were first detected as slight over-densities in the all-sky star map produced the Sloan Digital Sky Survey (Willman et al. 2005). The UFDs are now believed to be the ancient "fossils" of tiny galaxies that formed 80–90% of their stars before or during the epoch of reionization (Ricotti & Gnedin 2005; Brown et al. 2012; Weisz et al. 2014), like those first galaxies that preceded our own Milky Way. If so, then the ultimate faint frontier lies where we can detect the UFD scale over the full sweep of cosmic time.

We demonstrate the power LUVOIR will bring to such a study by deriving the plausible mass limits reached as a function of redshift in a single very deep exposure. **Figure 5.10** shows the stellar mass range that defines the dwarf populations of galaxies, with the UDFs at the minimum threshold (as we currently know it). The figure also shows the limits achievable by telescope of varying size, including Hubble, JWST, and the two LUVOIR architectures. The High Definition Imager will enable LUVOIR users to reach the extreme low mass end ($M^* < 10^{5.5} M_\odot$) of the stellar mass limit at many redshifts in a deep survey (AB ~ 33–34) and wider surveys of $10^6$–$10^7 M_\odot$ systems in much shorter exposures.

Only LUVOIR will be able to detect galaxies at the scale of the Ultra-Faints when they are still forming stars. This capability will reveal the entire grand sweep of galaxy formation to the earliest times, permitting full reconstruction of mass functions and merger trees. LUVOIR will also be able to trace the relationships between massive galaxies and their faintest satellites, which respond to details of the gas and star formation physics in complex ways that are still poorly understood. LUVOIR can, for instance, probe the relationship between L* galaxies and their faintest satellites, looking for evidence of when and now dwarf galaxies can be quenched by their larger, quenched neighbors ("conformal quenching") for indications that galaxies at the UFD scale continue to form stars after reionization.

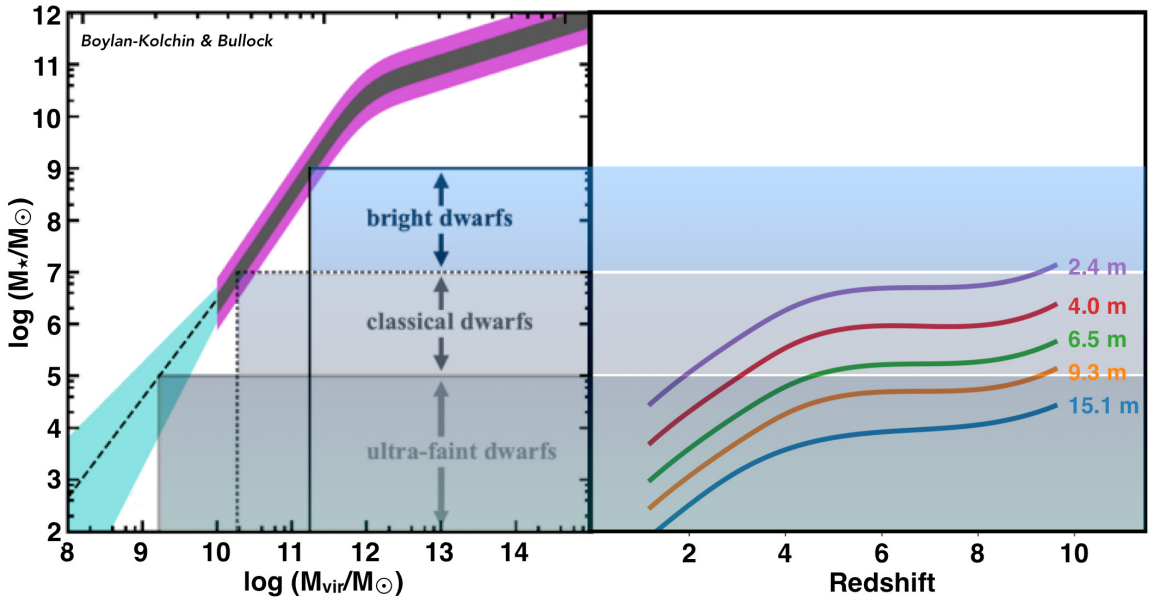

**Figure 5.10.** *At left, the stellar mass from Boylin-Kolchin & Bullock. Three stellar mass ranges for dwarf galaxies are shown. At right, the limits for detection by four observatories in a 500 ksec observations, assuming a 200 pc extended source and S/N = 5. LUVOIR's Architecture A can reach the UFD scale out to redshift z ~ 10.*

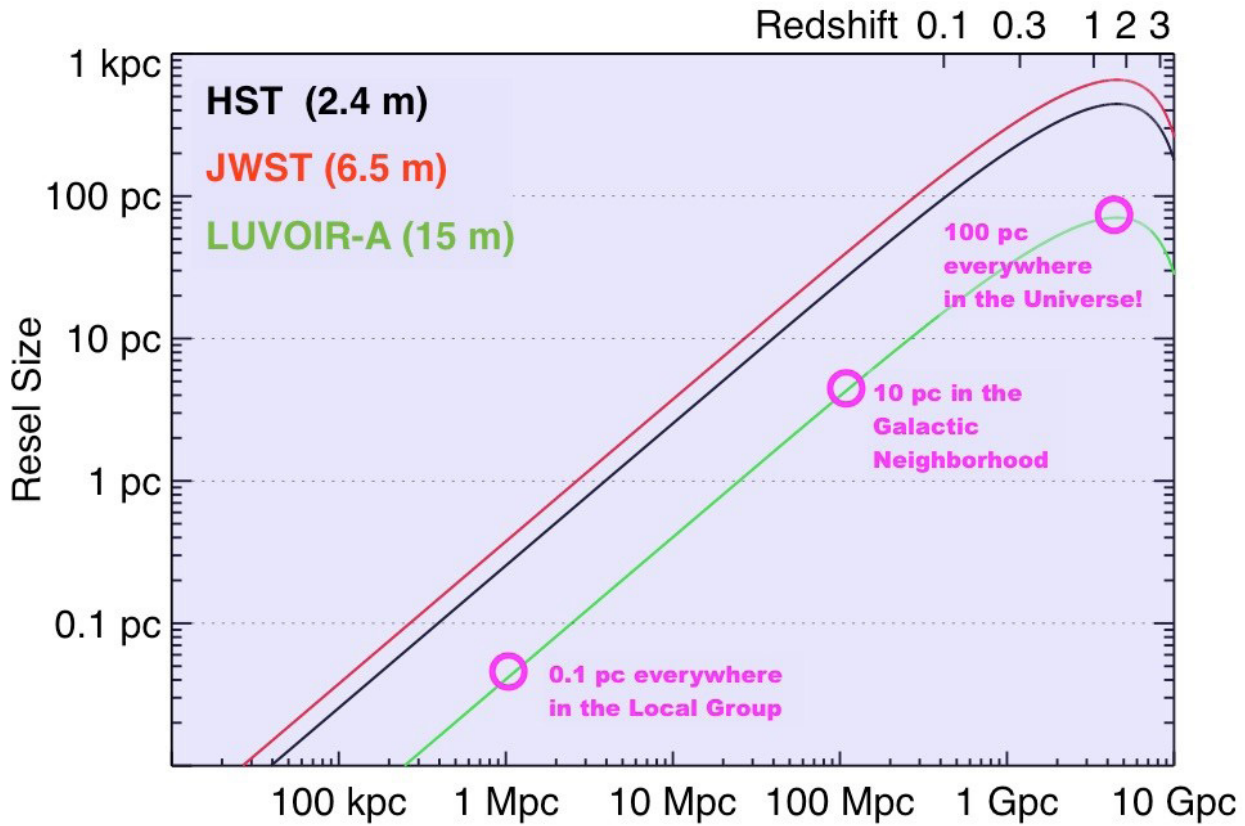

**Figure 5.11.** *Physical size corresponding to the spatial resolution element for Hubble (black, 0.5 micron), JWST (red, 2 micron), and LUVOIR-A (green 0.5 micron), which has a resolution of 80 pc or smaller at all redshifts.*

### 5.2.2 Seeing inside galaxies as they form and transform

Galaxy assembly is a messy, complex process. Using Hubble, astronomers have surveyed large samples of galaxies during the rise and fall of cosmic star formation (Madau & Dickinson 2014). These studies have mapped out the relationship between stellar mass and star formation rate (Whitaker et al. 2012), and traced the rise of quenched galaxies on the red sequence almost back to its beginnings (Kriek et al. 2009). Perhaps the greatest surprise from this work is how small galaxies appear to be at early times. **Figure 5.12** shows that the red sequence is already identifiable at z ~ 2–3, but with most of its quenched galaxies occupying only a few kpc in size. Indeed, star forming galaxies at these redshifts average several times larger than passive galaxies at the same mass. How do these galaxies quench, and do they actually get smaller as they do?

Hubble itself has struggled to address what happens inside galaxies at these early times because its diffraction limit corresponds to physical scales of 300–400 pc at z = 2–3. Hubble images place only a few pixels across these small yet massive galaxies at high redshift. This is sufficient to marginally resolve galaxy disks, but not the smaller-scale processes at work within them. It is also sufficient to detect ~1 kpc star-forming clumps (Elmegreen et al. 2007, 2009; Forster Schreiber et al. 2011), but large clumps may be blended collections of smaller star-forming regions (Tamburello et al. 2017; Rigby et al. 2017; Bordoloi et al. 2016), even if the clumps are like the largest and brightest in our immediate neighborhood, 30 Doradus in the LMC (diameter d ~200 pc) and the Carina Nebula (d ~ 140 pc).

We know that galaxies have structure on 100 pc scales because of evidence from the few cases where natural gravitational telescopes are provided by foreground galaxy clusters (**Figure 5.13**). In these rare cases, Hubble probes individual star-forming regions in ways that illustrate the power of LUVOIR. **Figure 5.13** shows an example, where Hubble reveals two dozen star-forming clumps in a lensed galaxy, with radii of 30–50 pc. These clumps have the sizes and luminosities of the brightest star-forming regions in the nearby universe, and none would be resolvable to Hubble without lensing. The lensing reconstruction of this lensed galaxy provides a rare view of what distant star-forming galaxies actually look like. Convolving this truth image with the scaled empirical Hubble PSF shows which features LUVOIR can recover, as a function of aperture. LUVOIR's users will be able to recover almost all the star-forming clumps in the F390W filter. This simulation shows that for LUVOIR, any distant galaxy can be imaged with the sharpness that Hubble can now achieve for the most favorable gravitationally lensed systems, which cover only a tiny fraction of the sky. LUVOIR would therefore be able to survey star formation down to 100 pc scales for thousands of galaxies, which is *terra incognita*.

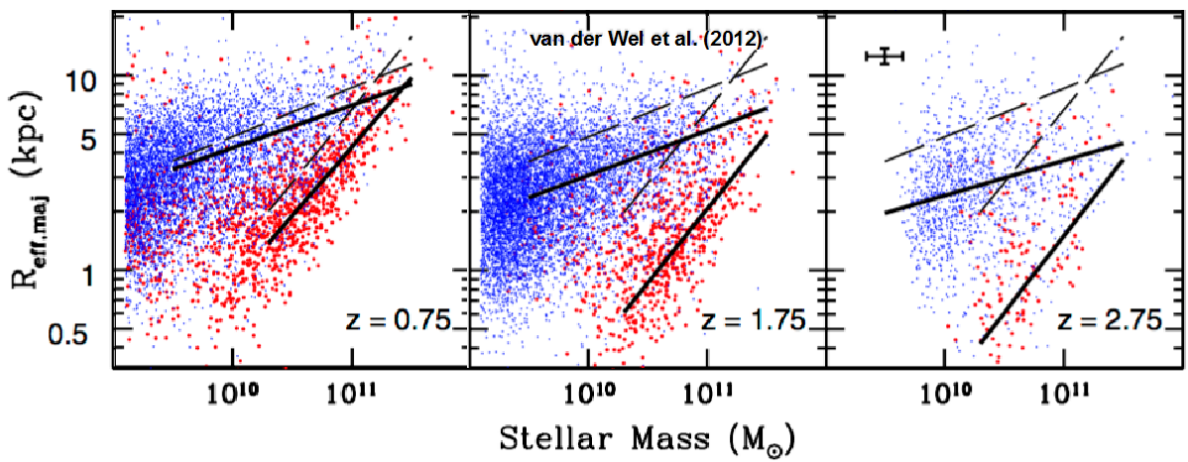

**Figure 5.12.** *Galaxy sizes and masses from the Hubble CANDELS survey. Even massive galaxies at z > 1 have sizes of only a few kpc. Adapted from van der Wel et al. (2014).*

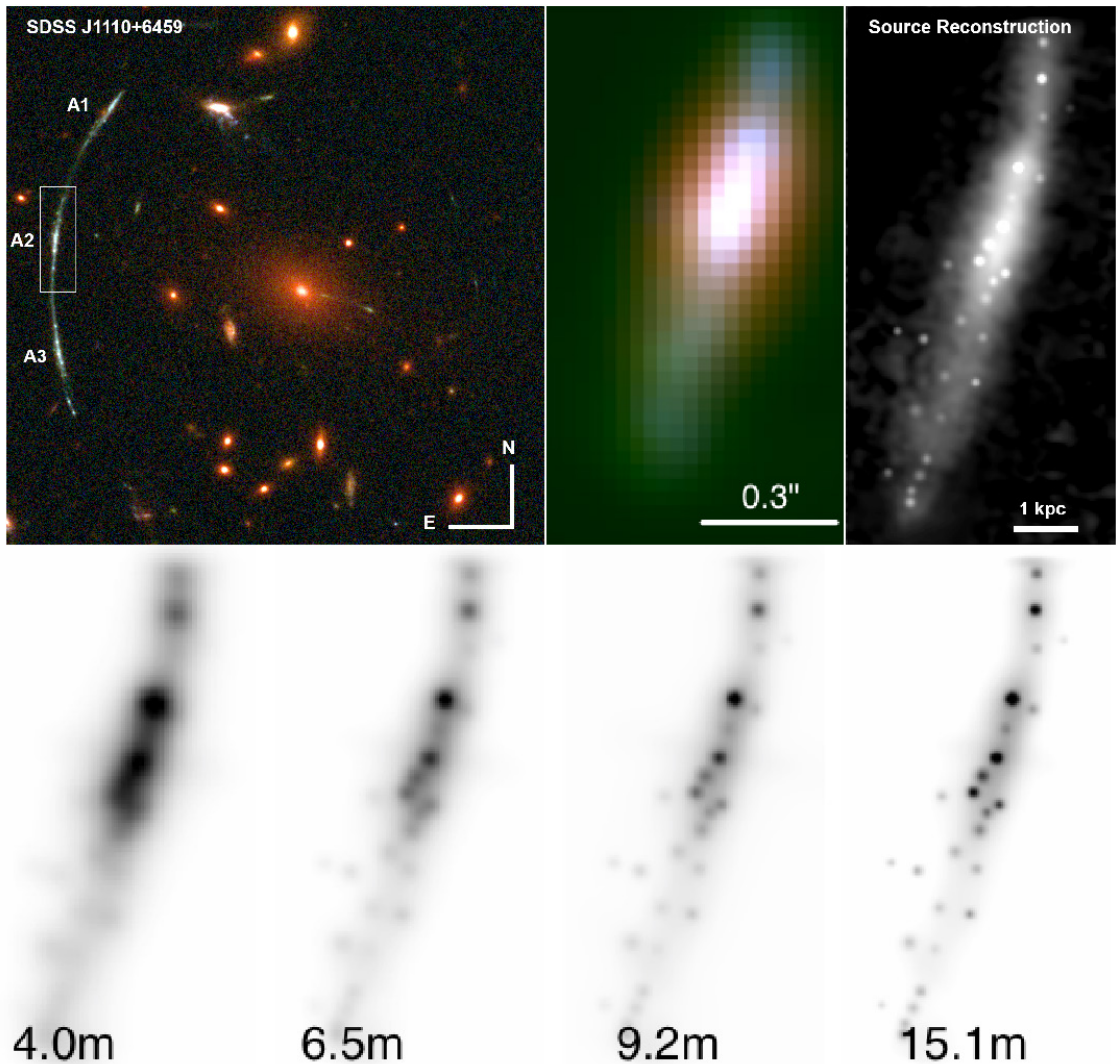

**Figure 5.13.** *A gravitationally lensed galaxy at z = 2.481, as imaged by Hubble/WFC3 in F105W, F606W, and F390W (top panels), as reconstructed in the source plane (middle panel), and as it would look to Hubble were it not lensed (right panel). The left panel shows the source reconstruction, blurred to the normal spatial resolution of Hubble. The far right panel shows the source reconstruction with lensing. The lower panels simulate images from a large space telescope, without lensing, using a PSF scaled from an empirical PSF for Hubble WFC3-UVIS F390W. Adapted from Johnson et al. (2017b) and Rigby et al. (2017).*

**Galaxy death, or quenching and compactification:** The mystery of how galaxies "die"—or cease to form stars at any significant level—is one of the most abiding

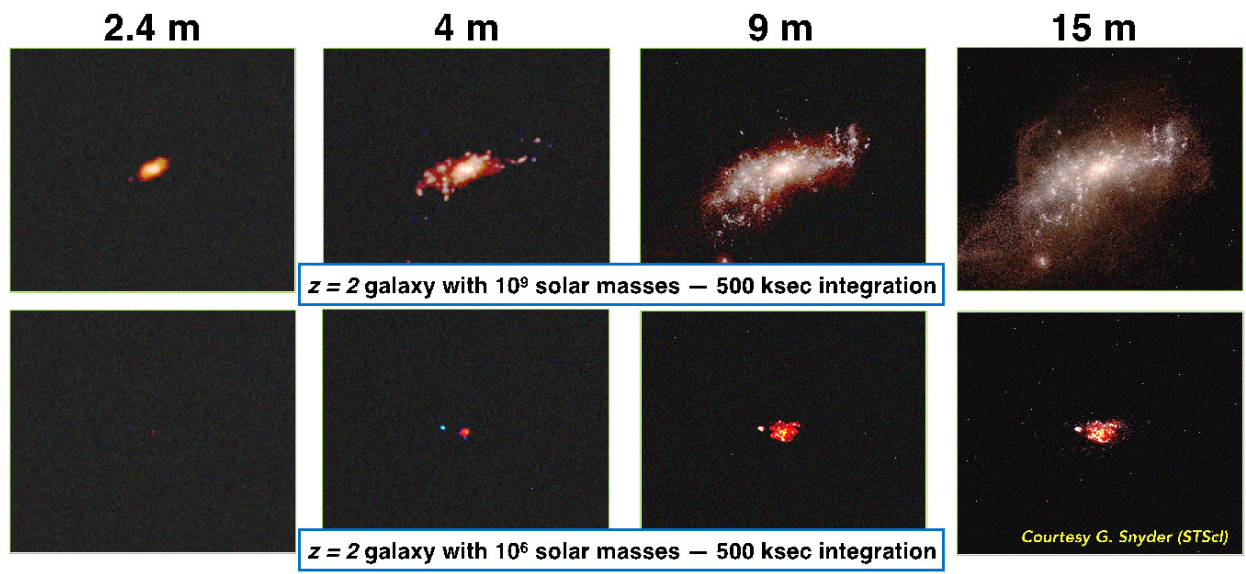

**Figure 5.14.** *Simulated galaxies showing the rich interior detail that is visible in deep exposures with telescopes of varying size. In all cases we assume extremely deep 500 ksec exposures. The $10^9$ $M_\odot$ galaxy is at the top and $10^6$ $M_\odot$ galaxy is at the bottom. Credit: G. Snyder (STScI)*

in astrophysics. Edwin Hubble's original tuning fork (1926) captured the basic truth that some galaxies are disky, blue, and star-forming, while others are spherical, redder, and quiescent. Numerous complex and overlapping mechanisms from stellar feedback to AGN and mergers have been proposed to turn star-forming galaxies into quenched ones, but there is still no definitive physical understanding of this basic phenomenon.

The presence of fully quenched galaxies at high redshifts z > 2 is surprising enough, but even more so are the extremely compact sizes of many of these galaxies. Ultra-compact galaxies pack a Milky Way's worth of stars, $10^{11}$ $M_\odot$ or more, into a kiloparsec or less. These galaxies are only marginally resolved by Hubble, with > 50% of their total light falling onto just one or a few pixels. Yet these galaxies are possibly the earliest progenitors of today's massive ellipticals, and if so they are a key part of the galaxy formation puzzle.

To unravel the origins of these mysterious galaxies, we must be able to resolve their internal structures, to measure stellar content and ages of their stellar populations, to look for internal gradients in age, to follow internal dynamics, and to trace all these quantities over time as this population arises. This is Signature Science for LUVOIR owing to its exquisite spatial resolution, 100 pc or better at all redshifts, which will allow us to map the internal structures and permit age dating of stellar populations at small scales. LUVOIR's broad UV/optical coverage and collecting area are also essential to this problem: rest-frame UV and optical colors are much more sensitive indicators of population age than are the optical/IR colors available with JWST. LUVOIR will surpass JWST on both counts, both in terms of physical resolution and the diagnostics that can be measured at that resolution. The same is true of ground-based telescopes, over small fields, which will exceed LUVOIR's raw spatial resolution with extreme NIR extreme AO, but will struggle to reach the same depths and cannot apply rest frame UV star formation diagnostics.

## 5.3 Dissecting galaxies one star at a time

Our ability to determine when galaxies assemble their stellar populations, and how that process depends on environment is a fundamental component to any robust theory of galaxy formation. By definition, the dwarf galaxies we see today are not the same as the dwarf galaxies and proto-galaxies that were disrupted during assembly. Our only insight into those disrupted building blocks comes from sifting through the resolved field populations of the surviving giant galaxies to reconstruct the star-formation history, chemical evolution, and kinematics of their various structures (Brown et al. 2010). Resolved stellar populations are cosmic clocks and assay meters that can assess the age and metallicity of their galaxies using well-defined relationships obeyed by stellar luminosity and color. Their most direct and accurate age diagnostic comes from resolving both the dwarf and giant stars, including the main sequence turnoff. As demonstrated by the Panchromatic Hubble Andromeda Treasury program (PHAT, Dalcanton et al. 2012), fitting stellar evolution models to CMDs can reveal not only when star formation occurred in Andromeda, but how episodic it is, and how it varies across the disk (e.g., Williams et al. 2014). This work highlights both the need for the survey depth required to obtain a CMD accurate enough to differentiate between models and the ability to map across as much of a galaxy as possible.

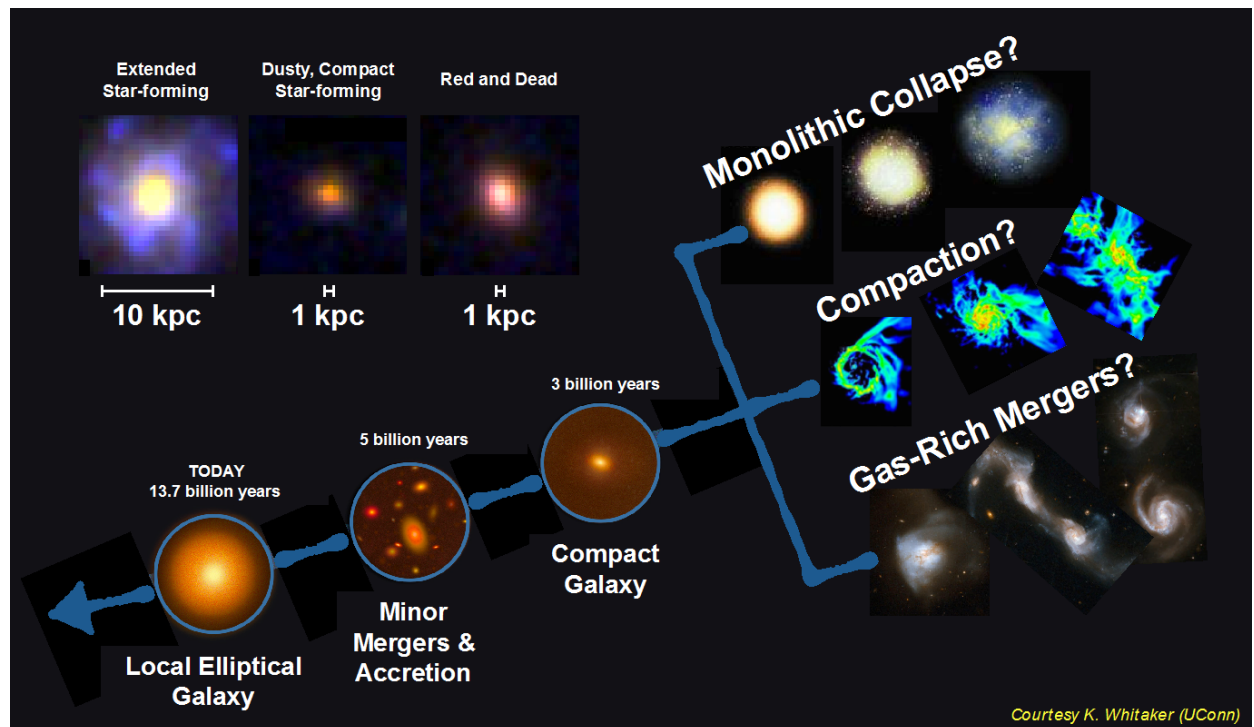

**Figure 5.15.** *At z ~ 2, massive galaxies of surprisingly small size are already "red and dead." These "ultracompact" red galaxies are likely the progenitors of today's massive ellipticals. They may have formed in a monolithic collapse, by compaction of star-forming progenitors, or gas rich major mergers are possibilities. Why did they stop forming stars so early, and why have they not resumed in the 10+ billion years since? At z ~ 2, these galaxies are very poorly resolved by Hubble, and JWST will do little better. LUVOIR will resolve these galaxies' internal structures to measure age gradients and dynamics that reflect their origins.*

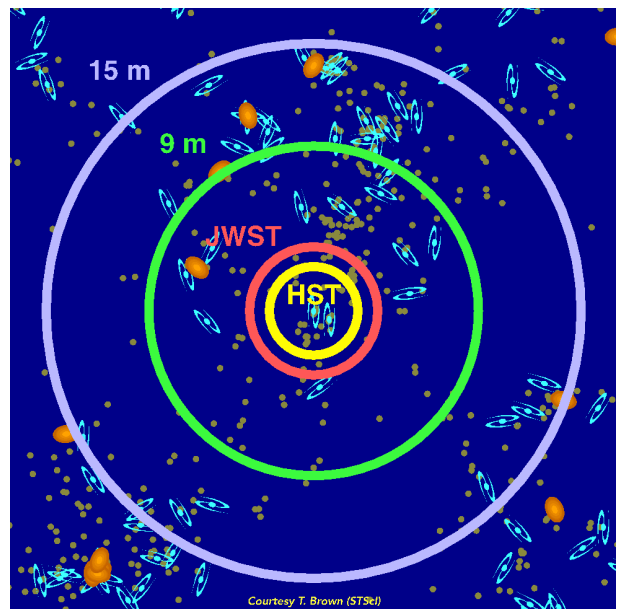

**Figure 5.16.** *Map of local universe (24 Mpc across) shown with the distances out to which HST (yellow), JWST (orange), and LUVOIR (9-m (green), 15-m(purple)), can detect the main sequence turnoff from a CMD in V and I passbands at SNR=5 in 100 hours. Giant spirals, like M31, are indicated by the blue galaxy symbols, giant ellipticals as orange blobs, and dwarf galaxies as small green dots. Derived using the HDI exposure time calculator at luvoir.stsci.edu/hdi_etc.*

Unfortunately, the main sequence turnoff is too faint to detect with any existing telescope for galaxies beyond the Local Group, and beyond this the increasingly severe effects of crowding become an even more fundamental limit. Both effects greatly hinder our ability to infer much about the details of galactic assembly because the galaxies in the Local Group are not representative of the galaxy population at large. LUVOIR will transform our ability to determine stellar histories by leveraging both light gathering power and spatial resolution, extending our reach beyond the Local Group, and thus significantly increasing the diversity of galaxies we can study.

We consider two types of surveys by way of example. First, we consider the distances that can be reached for characterization of diffuse (i.e., non-crowded) stellar populations in nearby galaxies. **Figure 5.16** shows a cartoon of the local Universe with galaxies in their actual positions and marked by type. In this view, 24 Mpc across, a 15-m LUVOIR can reach the main-sequence turnoff at SNR = 5 in 100 hours (V and I). For diffuse populations, this capability can be used to probe faint dwarfs or age-date the outer regions of massive galaxies. At closer distances, these observations can be collected over time to build up the timeline of proper motion and work out galaxy motions in 3D (see **Chapter 6**).

For a second type of survey, we consider regions where crowding comes into play. LUVOIR's spatial resolution will enable usable photometry at much higher stellar densities than smaller telescopes. The PHAT program has empirically determined that usably accurate photometry can be obtained in UVOIR images up to a surface density of ~15 stars per square arcsecond with Hubble. Scaling this limit up by $1/D^2$, we find that a 15-meter LUVOIR could detect the main sequence turnoff out to 5 Mpc, working with up to 400 stars per square arcsecond before crowding becomes prohibitive (**Figure 5.17**).

LUVOIR will work in concert with 30m-class ground-based telescopes expanding our reach to other well-populated galaxy groups, with LUVOIR obtaining photometry of G dwarfs stars down to V~34 mag, and the ground obtaining kinematics of much brighter giants out to the Coma Sculptor Cloud. The dwarf stars in the Coma Sculptor Cloud are effectively inaccessible to the ground, requiring gigaseconds of integration even for an isolated star.

This capability for studies of resolved and semi-resolved stellar populations has a wide range of applications from mapping the history of star formation in galaxies to assessing the impact or reionization at the smallest scales.

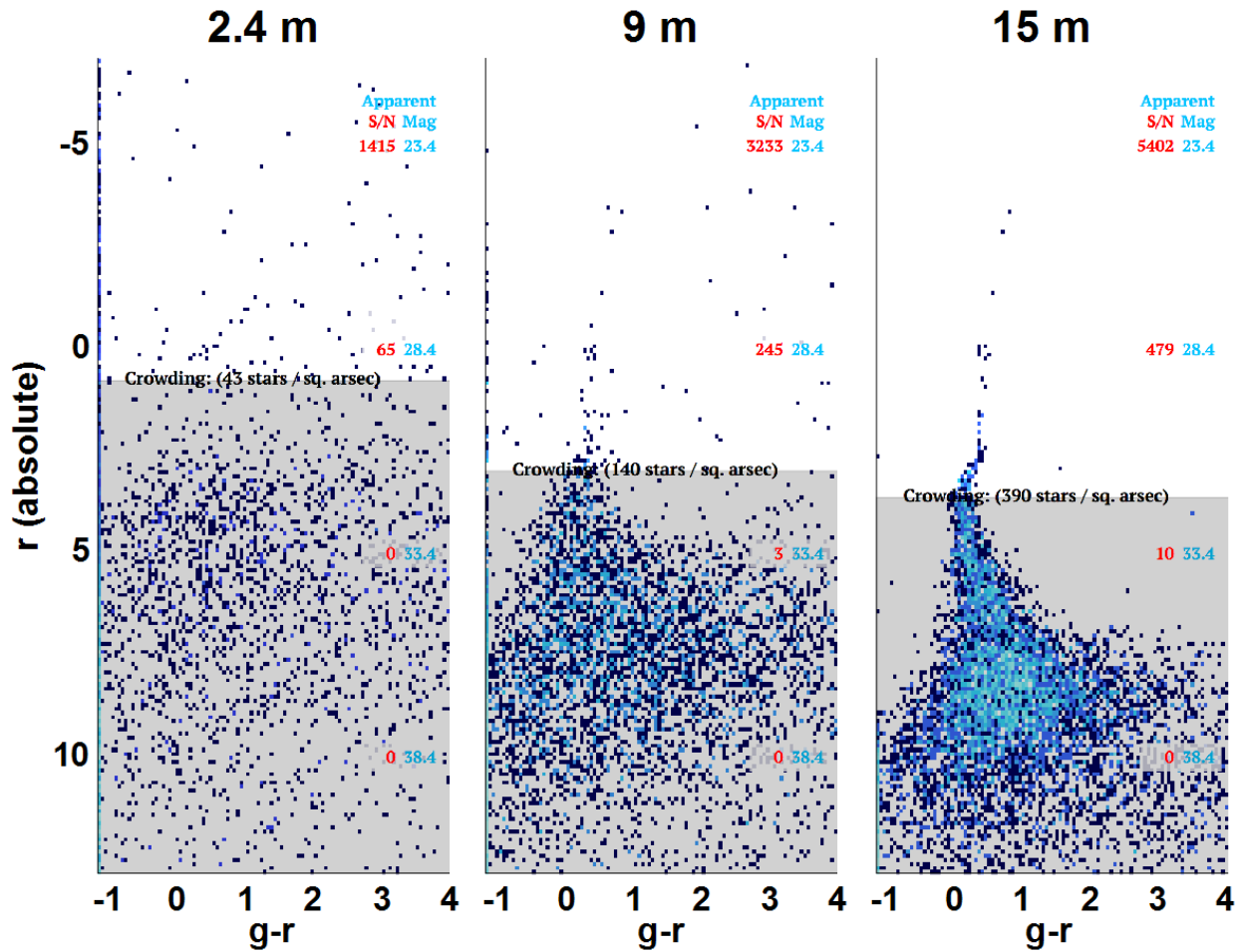

**Figure 5.17.** *Simulated CMDs for a single-age solar-metallicity stellar population. The simulation uses a 10 Gyr-old isochrone from FSPS (Conroy et al.), placed at 5 Mpc and "observed" in 50 hours each in the g and r bands. To compute the crowding limit, we assume AB = 23 per square arcsecond. The Hubble / WFIRST analogue (2.4 meters) cannot overcome the crowding and poor photometry. The 9-m LUVOIR can resolve the giant branch before crowding becomes too severe at the sub giant branch. The 15-m LUVOIR can reach to the top of the main sequence to apply population age and metallicity diagnostics. These figures were made with the LUVOIR CMD simulation tool at luvoir.stsci.edu/cmd.*

The LUVOIR HDI ETC and the CMD tool, linked here, provide prospective users with the opportunity to derive the requirements for their own applications.

**Table 5.1.** *Summary science traceability matrix for Chapter 5*

| Scientific Measurement Requirements | | | Instrument Requirements | | |
|---|---|---|---|---|---|
| Objectives | Measurement | Observations | Instrument | Property | Value |
| Measure the baryons over cosmic time from 10–10$^7$ K | Column density, velocity width, and redshift for multiple ions | S/N >=20 spectra of 100 QSOs from 1000 to 4000Å | LUMOS | R~30,000 Gratings | G120M, G150M G180M, G300M |
| Provide a high definition exploration of the CGM of local galaxies | Column density, velocity width, and redshift for multiple ions | S/N >=10 spectra of 30 QSOs within R=150 kpc of the galaxy 1000 to 3000Å | LUMOS | R~30,000 Gratings, MOS | G120M, G150M G180M, G300M |
| Trace the CGM in emission | Emission line maps of key ions (e.g., HI, CIII, OVI) | S/N >5 emission spectra in multi-shutter apertures 1000 to 3000Å | LUMOS | R~30,000 Gratings, MOS | G120M, G150M G180M, G300M |
| Characterize nearby galaxy inflows and outflows in detail | Spatially map inflowing/ outflowing gas and dust in absorption | S/N >=30 spectra of 1000 sources/galaxy for 5 nearby galaxies from 1000 to 4000Å | LUMOS | R~30,000 Gratings, MOS | G120M, G150M G180M, G300M |
| Explore star formation histories across the Hubble sequence | Color-Magnitude diagrams for stars in galaxies out tens of Mpc across the full range of galaxy type | S/N > 5 imaging of resolved stellar populations | HDI | Multi-band UVIS imaging | UBVRI filters |

# Chapter 6

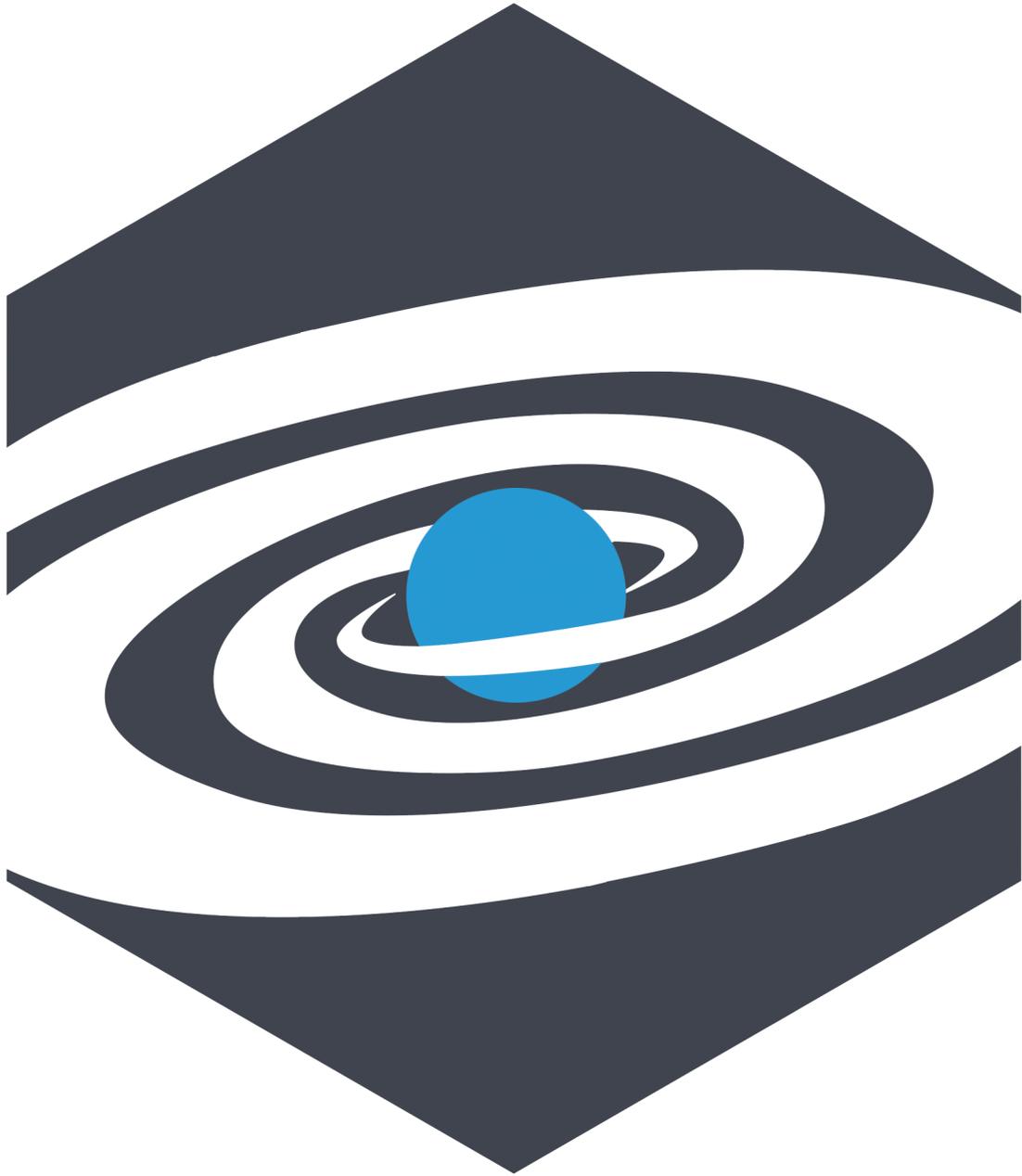

What are the building blocks of cosmic structure?

# 6    What are the building blocks of cosmic structure?

Our understanding of how the universe works is sophisticated but incomplete. On cosmological scales, the total mass-energy density is almost precisely equal to the critical value. Most of the mass-energy density is in the dark energy component and one quarter of the total is a cold dark matter (CDM) component. There is a complex interplay between the cosmic expansion history, dark matter, radiation, and baryons, which yield many of the observable elements in the universe. Yet we are still working towards ascertaining the true nature of dark energy and dark matter. Furthermore, our models for galaxy formation and evolution from the birth of the first stars through the era of reionization, and on to the present, remain rudimentary.

Using ground- and space-based observatories through the early 2030s, we will make significant inroads in our understanding of the cosmic "dark sector" and the properties of galaxy building blocks in the early universe. This drives many of the science objectives of JWST, Euclid, LSST, WFIRST, and the planned 20-m to 40-m ground-based telescopes. Yet these telescopes will not reach beyond a specific frontier that we already recognize as critical for a comprehensive understanding of structure formation in the universe. This frontier exists at the scales of the lowest mass galaxies (with total masses $\leq 10^6$ M$_\odot$ that we know are there, from the first sparks of galaxy formation at z>10 to the least luminous dwarf galaxies in the present day. These galaxies map to physical scales as small as 10 kpc and with stellar concentrations as small as 100 parsecs. At these size and mass scales, we enter regimes where competing scenarios for the evolution of the dark matter density field, and its associated baryonic structures, make distinguishable predictions that can be robustly tested with observations that reach fainter than AB = 33 mag.

To explore this ultra-faint regime, we require LUVOIR, which (for the 15-m architecture) will be capable of resolving < 100 parsec scales at all redshifts while reaching a 5-σ point source limiting AB magnitude of 33 (0.23 nJy) in just 10 hours and ~35 mag (0.04 nJy) in ~10 days. Three novel observational regimes that require LUVOIR are described in this chapter:

1) **The dark matter power spectrum on scales below 100 kpc:** Between the universe's horizon scale and galactic scales, the structure we measure is consistent with dark matter being entirely non-relativistic and non-interacting. This does not uniquely identify a specific fundamental particle for dark matter, however. Indeed, several candidate particle classes are allowed by existing observations. This has motivated work on dark matter particle candidates that, on astronomical scales, have ensemble behaviors consistent with being "warm" (mildly relativistic) and/or "self-interacting." The microphysics of the interactive or radiative properties of dark matter influences the nature of structure on various scales and epochs in time. Specifically, the imprint of dark matter physics manifests at scales below 100 kpc (corresponding to the scale of self-gravitating halos of a few million solar masses) in the statistics and shapes of these structures as functions of size and mass over cosmic time.

2) **The limits of galaxy assembly in the pico-Jansky regime:** Associating galaxies with the underlying dark matter structure requires modeling the interplay of cooling mechanisms, the epoch of reionization, and feedback effects from star formation and AGN. The most massive dark matter halos have long baryonic cooling times, with stars

## The State of the Field in the 2030s

By the year 2035, HST, Euclid, WFIRST, and JWST will have completed their missions and LSST will have achieved its 10-year survey depth. At least two 20–40 meter class ground-based optical-NIR observatories will be operational. SKA should be online and creating deep maps of the neutral hydrogen distribution out to high redshifts. LUVOIR will be following on the success of these facilities by bringing unparalleled sensitivity and angular resolution in the ultraviolet and optical regime.

**James Webb Space Telescope (JWST):** Deep NIRCam surveys with JWST will exist down to AB = 31.5 mag (for 0.1" diameter source with 500 ksec exposure time). This will allow JWST users to discover many new ultra-faint sources but will still not be deep enough for many of the luminosity regimes we need to reach to fully map the assembly of structure from small-scale components at $6 < z < 10$.

**LSST, Euclid, and WFIRST:** These sky survey telescopes will detect many new faint dwarf galaxies in the local universe but will not have the resolution to characterize their morphological structure in detail nor measure their proper motions. However, these targets can be studied with LUVOIR.

**Extremely Large Telescopes on the Ground:** These facilities will have similar angular resolution as LUVOIR in the NIR but will have difficulty detecting sources fainter than AB ~ 31 mag due to the brightness of the night sky. The spectroscopic capabilities of such 20–40 meter telescopes will be a superb complement to the imaging power of LUVOIR.

**Square Kilometer Array:** The SKA will perform wide-area deep imaging of the neutral hydrogen gas at cosmological distances ($z < 10$) and will provide the knowledge of where to target deep imaging with LUVOIR to study the role of reionization suppression of star formation.

**LUVOIR:** In the context of studying the building blocks of galaxies, LUVOIR will reach stellar mass limits typical of ultra-faint dwarf galaxies *at all redshifts*. LUVOIR observers will detect sources with fluxes fainter than AB = 33 mag, where reionization suppression of star formation is predicted to be important. LUVOIR will also extend dwarf galaxy detection out to 25 Mpc in the local universe, and allow observers to detect directly the sources of ionizing radiation in the cosmos and characterize the energy distribution of that ionizing radiation. Its high precision astrometric stability and calibration, will allow the measurement of transverse velocities of stars within Local Group galaxies to directly measure mass density profiles and to map the gravitational fields around nearby galaxies by measuring satellite orbits within their central halos. This information will provide additional constraints on the kinematics and the nature of dark matter.

subsequently forming late. Very low mass halos are not able to retain a steady pool of star-forming baryons due to ionizing radiation that is keeping the universe transparent to this light. The source of the ionizing photons likely comes from a great abundance of star-forming, low mass galaxies. The bulk of the physical processes that dominate the early formation and evolution of cosmic structure all take place on scales below 10 kpc. This corresponds to minuscule volumes within massive galaxies, or in dwarf galaxies, the smallest galaxies that are able to form and retain stars. Observing these scales across a full range of galaxy properties from the current epoch to the end of the era of reionization will require sensitivities down to a few tens of picoJanskys (AB ≈ 35 mag). Such observations will be transformational as they will reveal star and galaxy populations not yet seen with current or planned telescopes.

3) **The history of ionizing light from observations shortwards of rest-frame 900 Å:** To complete our understanding of the emergence of structure, we must characterize the influence that ionizing radiation has on early galaxies. Directly observing ionizing photons at the epoch of reionization is effectively impossible due to the opacity of the intergalactic medium to the ionizing photons of that epoch. LUVOIR will leverage its FUV and UV sensitivity and resolution to directly detect weak Lyman continuum radiation escaping from z<1 galaxies in a *spatially resolved* manner for multiple objects per pointing, revealing the environmental factors that favor the escape of radiation, providing crucial data on how light escaped galaxies during the epoch of reionization at z ≥ 7.

With LUVOIR, the community acquires a unique and unmatched combination of *sensitivity* and *volume* to observe, across most of cosmic time and in many different intergalactic environments, what the universe actually does on spatial scales where the confluence of the matter power spectrum, dark matter physics, and baryonic processes *all interplay*. LUVOIR will accomplish this by extending our census of dwarf galaxies to much fainter limits in the current epoch and extend that census, at brighter limits, across the last 13 billion years. This will allow LUVOIR users to establish the direct connection between the current-epoch matter power spectrum with its early progenitors. With such information, we can directly distinguish between competing dark matter models that show significant differentiation on scales below 100 kpc. LUVOIR investigators will also directly observe the impact of reionization on early star formation by measuring the galaxy luminosity function well below limits achievable with current or planned telescopes. Finally, LUVOIR will enable a direct characterization of that ionizing radiation by acquiring the spectra of UV light escaping from galaxies at wavelengths below rest-frame 900 Å, something only a large-aperture UV-sensitive telescope can accomplish.

The Signature Science Cases discussed in this chapter represent some of the most compelling types of observing programs on the building blocks of structure that scientists might do with LUVOIR at the limits of its performance. As compelling as they are, they should not be taken as a complete specification LUVOIR's future potential in these areas. We have developed concrete examples to ensure that the nominal design can do this compelling science. We fully expect that the creativity of the community, empowered by the revolutionary capabilities of the observatory, will ask questions, acquire data, and solve problems beyond those discussed here—including those that we cannot envision today.

## 6.1 Signature Science Case #1: Connecting the smallest scales across cosmic time

Dwarf galaxies are the smallest luminous, dark-matter dominated objects known. The least luminous dwarf galaxies contain just dozens of stars, although their host dark matter halos have total masses around $10^7$ $M_\odot$. A small number of these ultra-faint dwarf galaxies have been detected in orbit around our Milky Way Galaxy to date. The SDSS Segue-1 system, at a distance of 23 kpc, has an absolute magnitude of $M_V = -1.5$

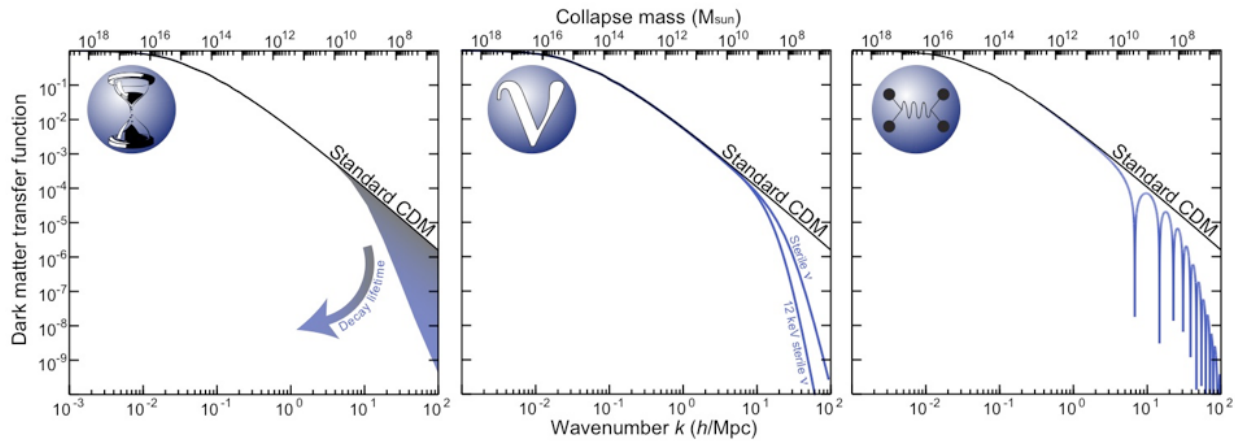

**Figure 6.1.** *The matter power spectrum as a function of spatial scale, as depicted by the "transfer function" for different dark matter particle scenarios at the current epoch (z=0). In each panel, smaller mass and physical scales are to the right. The standard cold dark matter (CDM) model prediction is shown for comparison. Left panel: Decaying dark matter model. Middle panel: A warm dark matter (WDM) model. Right panel: A self-interacting dark matter model (e.g., Buckley & Peter 2018). Credit: R. Massey*

mag (Simon et al. 2011). The Virgo-I dwarf galaxy has $M_V$ = -0.8 mag and is 87 kpc away (Homma et al. 2016). Even at this distance, we are only reaching one third of the distance to our Galaxy's virial radius (Walsh et al. 2009), and are far from having a complete survey of the accessible area above and below the Galactic plane.

There is evidence that the ensemble of dwarf galaxies is primarily responsible for the reionization of the universe and for maintaining its transparency to ultraviolet radiation. The space density and internal structure of these galaxies are closely connected to the fundamental properties of the dark matter particle, the level and duration of reionization, and the granular limits of the galaxy formation process. To date, limits on the thermal signature of dark matter have been placed by studies of the intergalactic medium (IGM) through Lyman-alpha forest statistics (e.g., Viel et al. 2013), strong gravitational lensing (e.g., Li et al. 2016), and dwarf-galaxy-scale statistics (e.g., Kim et al. 2017). One of the most constraining measurements, by Jethwa et al. (2018), places a 95% confidence limit on dark matter particles having a mass greater than 2.9 keV. Dark matter particles may be significantly colder (more massive) yet be interacting (either self-interacting or with a dark matter-baryon interaction mechanism), or have more nuanced properties that fall outside of the currently popular prescriptions for dark matter candidates (e.g., Buckley & Peter 2018). Each type of dark matter property manifests itself on astronomical scales through a gravitational signature, in the statistical description of dark matter structure. Our measure of this signature is through the matter power spectrum today, as well as at earlier cosmic times.

**Figure 6.1** illustrates the matter power spectrum behavior for several dark matter cases compared to CDM. At wave numbers above k ≈ 10h/Mpc, corresponding to scales well below that of the Milky Way Galaxy, the distinctions between these power spectra are significant. The k ≈ 10 h/Mpc scale corresponds to total mass scales in the $10^9$ to $10^{10}$ $M_\odot$ range. These correspond to classical dwarf galaxies, which have stellar masses of less than $10^5$ to $10^6$ $M_\odot$ and are fainter than $M_V$ = -10 mag. As we approach smaller scales, near k ≈ 100 h/Mpc, we reach the regime of $10^7$ $M_\odot$ (total mass) halos. This approaches the lower mass limit where gravitationally bound halos can host star formation, before the necessary radiation cooling for forming stars becomes too inefficient. Halos of total mass of $10^7$ $M_\odot$ have stellar masses as low as just tens of solar masses. In the current epoch, such systems have visual absolute magnitudes between -2 and 0. This defines the capability level LUVOIR targets.

By the early 2020s, the Dark Energy Survey and LSST will have completed the census of dwarf galaxies in the vicinity of the Milky Way Galaxy. In the spirit that each massive galaxy's satellite galaxies provide just one "draw" of the underlying physics that gives rise to this population, we must expand on that sample by surveying the satellite dwarf population associated with massive galaxies similar to our own Milky Way. By the late 2020s, JWST and ELTs will have the sensitivity to extend the dwarf satellite galaxy search out to several Mpc. Dwarf galaxies are identified through detection of their constituent stars, as long as the membership of those stars can be confidently determined, through colors or spectroscopy over an extended field of view. LUVOIR will extend this sensitivity, using horizontal branch stars, to far beyond the Virgo Cluster's 16.5 Mpc distance. This has multiple advantages. First, a sample of massive and Milky Way-analog galaxies provides multiple "draws" similar to our well-studied Galaxy. Second, this much larger volume gives us access to an enormous dynamic range of environments, so that dwarf galaxies can be studied in both volatile and quiescent states.

LUVOIR's sensitivity, angular resolution, and Nyquist sampling will deliver transformative measurements on the smallest galactic scales. We highlight the potential breakthroughs in this field by presenting two dramatically different (but connected) periods in the history of the universe: the extended local volume in the current epoch (out to a luminosity distance of ~30 Mpc, or z = 0.008), and the early universe, 13 Gyr in the past (at z~7), about half-way through the reionization transition era from a visually opaque to a transparent universe.

### 6.1.1   Dwarf galaxies in the local universe: testing dark matter models

Dwarf galaxies are predicted to exist around all massive galaxies and in all environments. Galaxies out to 40 Mpc from us, with masses comparable to the Milky Way, are being systematically surveyed for dwarf galaxies as faint as $M_V$ = -12.3 mag (comparable to the Leo-I Galactic dwarf; e.g., Geha et al 2017). These surveys are identifying dwarf galaxies that are current-epoch analogs to the early galaxies responsible for re-ionizing the universe (e.g., Boylan-Kolchin et al 2015). The potential connections are complex, however. Many of the faintest local dwarf galaxies detected are chemically enriched relative to their early-universe counterparts. The stars in these low-mass neighbors of our Galaxy are created, in part, through inefficient molecular cooling while, at the same time, experience additional quenching effects from an intense ionizing radiation field. Exploring the connections between local dwarf galaxies and corresponding properties in the early universe suggests that even JWST will not be able to detect the full complement of dwarf galaxies at early times by several magnitudes of sensitivity (see **Figure 6.6** and **Figure 6.8**).

To make a major breakthrough in understanding the connection between dwarf galaxies and the mass-scales of their dark matter halos will require the detection and characterization of significant samples of faint and ultra-faint dwarf galaxies in a range of environments over a wide span of cosmic time. Then, with subhalo-abundance match-ing techniques, the luminous components can be connected with their dark matter halo mass properties and, in turn, connected to the underlying matter power spectrum that they are ultimately drawn from. LUVOIR will allow us to do this both locally and at high redshift.

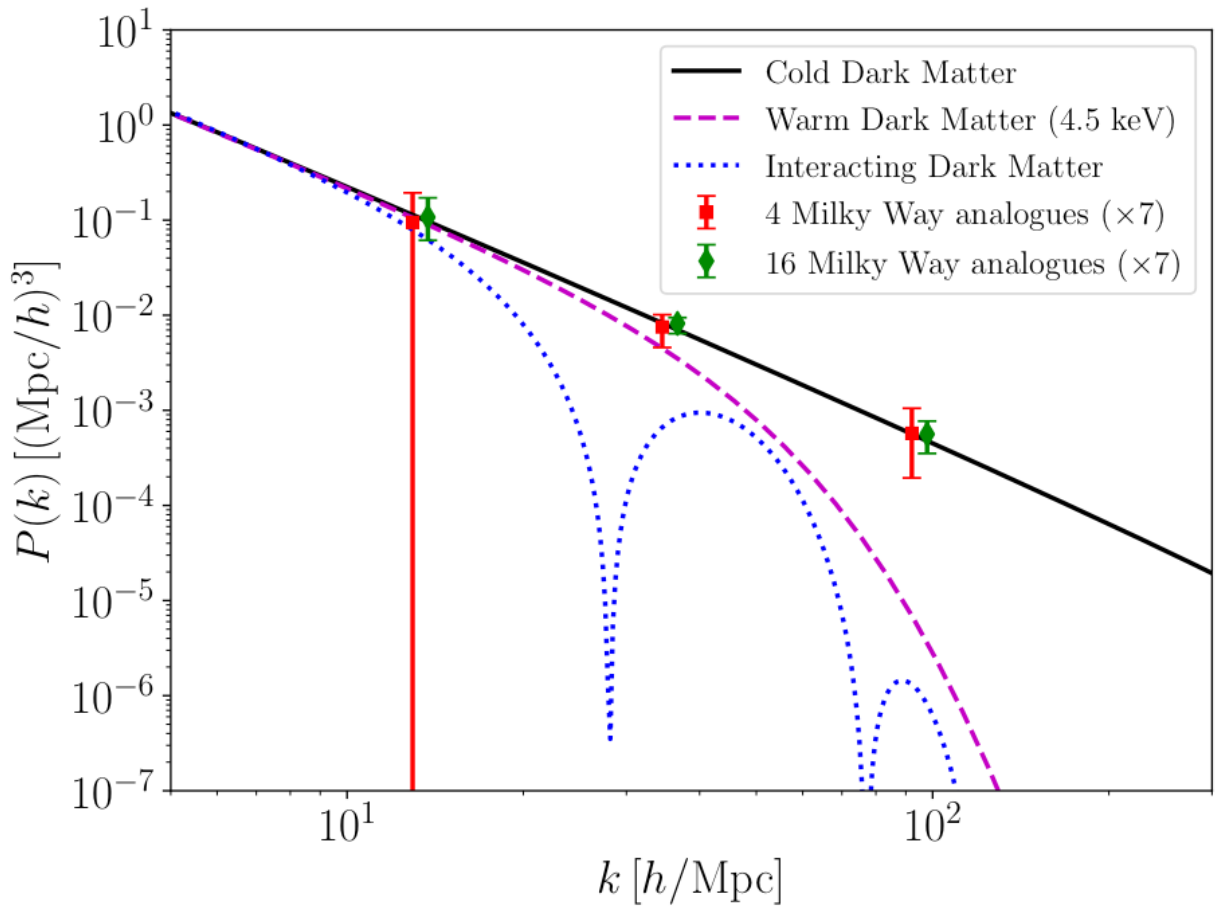

**Figure 6.2.** *The matter power spectrum for samples of four and sixteen Milky Way analog systems. The error bars shown are magnified by a factor of 7 to make them visible in this figure. The abscissa positions of the points have been shifted slightly for clarity. For illustration, the underlying power spectra for a 4.5 keV warm dark matter model and a self-interacting dark matter model (Vogelsberger et al. 2016) are shown to illustrate the scales where these different scenarios exhibit detectable effects.*

We can calculate the fidelity of the power spectrum measurements that will be possible from a completeness-corrected sample of dwarf galaxies around several Milky Way Galaxy-scale systems. The results are shown in **Figure 6.2** for sample sizes of 4 and 16 Milky Way-like satellite systems. The error bars shown in **Figure 6.2** have been magnified by a factor of 7 to make them visible in this plot. These calculations begin by adopting an empirical relation between dark matter halo masses and their stellar mass (Bullock & Boylan-Kolchin 2017) (see **Figure 6.3**), although how reliable this is in the regime of the true ultra-faint galaxies is the focus of ongoing work (e.g., Kim et al. 2017). At the low-mass end of the galaxy distribution, the effects of star formation suppression due to an intergalactic ionizing background flux can be significant. Barber et al (2014) develop a semi-analytic description of this effect (see also Dooley et al. 2017). In a rigorous treatment of the physics of galaxy formation, many other evolutionary effects can affect the survival and apparent properties of individual dwarf galaxies. However, as we are considering the possibility of acquiring broadly complete samples of dwarf galaxies around massive systems, including those that extend towards the parent halos' virial radii, we begin by including only this dominant suppression signature of reionization. Finally, a relation between the adjusted differential mass function and the corresponding power spectrum within specified scale-ranges is needed. While aspects of this connection have long been explored (e.g., Eisenstein & Hu 1999), the method derived by Schneider (2015) is particularly practical and that is what we have used for this calculation.

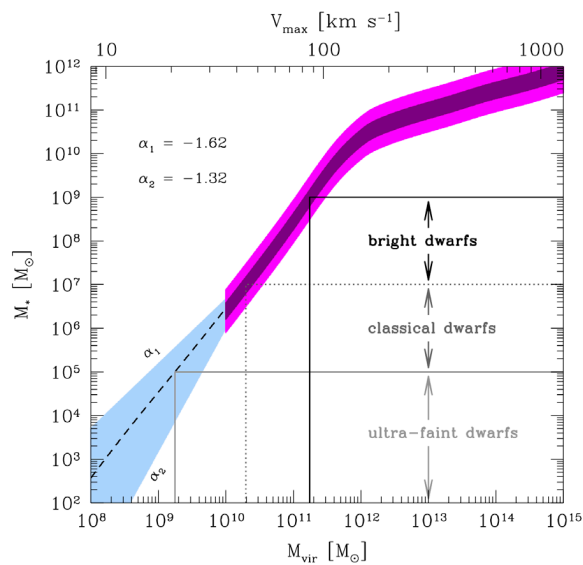

**Figure 6.3.** *The correspondence between stellar mass and halo (virial) mass can be made through abundance matching in N-body simulations. The stellar mass ranges for typical dwarf galaxy populations are shown in this reproduction from Bullock & Boylan-Kolchin (2017).*

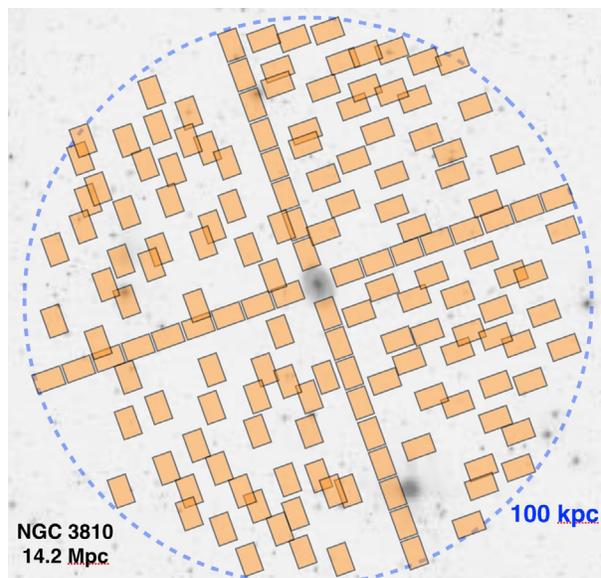

**Figure 6.4.** *The LUVOIR image tiling strategy (orange rectangles) to sample the dwarf galaxy population around the Milky Way analog NGC 3810 at a distance of ~14 Mpc. The 156 HDI pointings cover 50% of a region that extends out to half the virial radius (~100 kpc). Full sampling is performed along the major and minor axes of NGC 3810.*

A LUVOIR program to obtain the needed constraints on the small-scale matter power spectrum would begin with two contiguous chevrons of HDI observations for each target. Each strip extends from the central galaxy out to about half the virial radius (~100 kpc); one strip is along the analog's major axis, and the other is along its minor axis. In addition, we would observe a number of random positions within the central 100 kpc to ensure we achieve a 50% filling fraction of the survey region (see **Figure 6.4** for a survey layout centered on NGC 3810). This survey approach samples one quarter of the total virial-radius volume.

The dwarf galaxies discovered from these observations are then used to estimate the census of the complete population within the entire virial volume of each central galaxy targeted in the survey. The advantage of LUVOIR for this study is that the combination of its sensitivity and survey efficiency is unachievable by any current or planned platform, on the ground or in space, by orders of magnitude. With these representative surveys of luminous dwarfs in each system, we then use the abundance-matching framework to transform this to a mass function. The mass function is then compared with that predicted for the CDM-derived matter power spectrum (e.g., Schneider 2015). By targeting multiple systems that are as similar to the Milky Way as possible, and assuming that each one is an effective "draw" on the properties of the underlying physics that results in the populations we observe, we can estimate both the power spectrum's amplitude and its measurement uncertainty (see **Figure 6.2**) on several spatial scales corresponding to those in the derived mass functions.

A LUVOIR survey to characterize the dwarf galaxy mass function around Milky Way analogs will provide the degree of accuracy we need to distinguish between different dark matter scenarios at >4σ significance as **Figure 6.2** demonstrates. The targeted galaxies will be in hand as several Milky-Way scale galaxies have already been identified out to 40 Mpc (e.g., Danieli et al. 2017; Geha et al 2017). Indeed, dwarfs within the nearer systems have been identified to $M_V$=-9 mag, and to $M_V$=-12 mag in the more distant systems. A LUVOIR deep imaging survey around Milky Way analog systems within 15 Mpc will reach down to be to the horizontal branch ($M_V$=0 mag) in dwarf galaxies detected in these systems, reaching the limits of dwarf galaxy formation. The LUMOS instrument could be operated in parallel during this type of survey to simultaneously map the absorption features from the circumgalactic medium of the central galaxy.

**Figure 6.5** demonstrates the potential of a Nyquist-sampled imager to detect and characterize different stellar populations on space telescopes of various aperture sizes. The individual horizontal branch or red giant branch stars (depending on distance) will be detectable within each candidate dwarf galaxy. Many of the dwarf galaxy systems identified with LUVOIR will be superb targets for ELTs, given the precise positions and properties to aim for. The least luminous candidates in many target Milky Way analogs will be fainter than AB = 31 mag and would likely require additional LUVOIR observations.

The internal structural profiles of dwarf galaxies in the local universe will also be feasible as these systems will be extremely well resolved with LUVOIR. For example, the FWHM of the PSF of the 15-m LUVOIR HDI instrument in the r-band (650 nm) is 9 milli-arcseconds. An individual dwarf galaxy analogous to the Virgo-I dwarf with a half-light radius size of ~20 parsecs would, at the Virgo Cluster's distance, subtend ~0.50

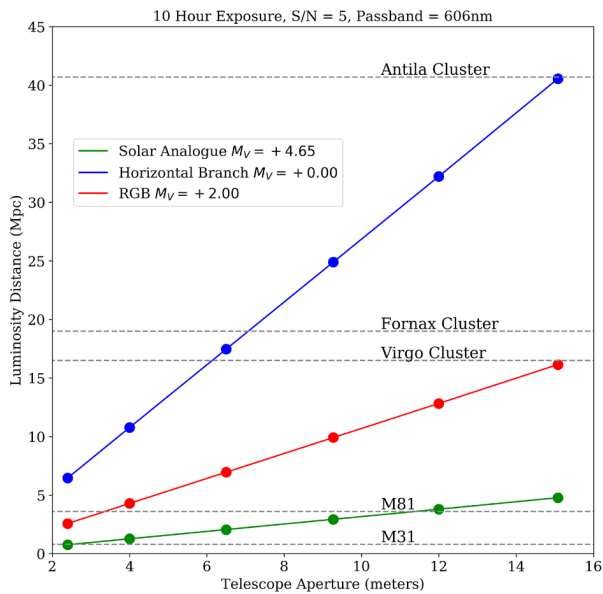

**Figure 6.5.** *The distance to which three representative stellar types can be detected, versus aperture size of LUVOIR+HDI-like platforms. The distances shown here are for a S/N = 5 optical detection with a fiducial 10-hour integration in the V-band. In such an observation, observers using the 15.1m LUVOIR could detect RGB stars out to 16 Mpc and horizontal branch stars out to 40 Mpc. With a 100-hour integration, the 15.1-m LUVOIR will have the ability to detect individual RGB stars out to 29 Mpc.*

**Program at a Glance - Dwarf Galaxies and the Matter Power Spectrum**

**Goal:** Measure matter power spectrum on small scales (< 100 kpc) to discriminate between standard cold dark matter, interacting dark matter, and several warm dark matter models at a statistical significance of 4-$\sigma$ or better (at k = 40 h/Mpc).

**Program details:** Observations of 4 Milky-Way analogs in the 5–15 Mpc distance range to gain a representative sample of the entire luminosity range of dwarf galaxies by detecting their constituent horizontal branch stars at high significance. Total time required is ~390 hours.

**Instrument(s) + Configuration:** HDI instrument on LUVOIR-A (15-m) observatory. Optional: LUMOS MOS operated in parallel to study CGM.

**Key observation requirements:** Imaging in V, R passbands in multiple pointings to cover 50% of the central 100 kpc region. Depth required corresponds to reaching S/N=5 per horizontal branch star with $M_V$ = 0.0. Survey fill factor is set to 50% to ensure reliable estimates of completeness.

arcsecond and would therefore be sampled with ~3000 distinct resolution elements. This means we will be able to fully characterize the dwarf galaxy's stellar light profile, its stellar populations, and its internal structure—feats that would not be possible with any other current or upcoming telescopes.

### 6.1.2  Dwarfs in the distant universe: probing the impact of reionization

The reionization era is the point in cosmic history where the universe transitions from being opaque to UV radiation to its present largely transparent state. The faint-end slope of the galaxy luminosity function becomes quite steep by z~7 (e.g., Bouwens et al. 2015; Finkelstein et al. 2015), which implies that if galaxies exist at luminosities below the *current* detection limits, then these ultra-faint dwarf galaxies should be the dominate sources for reionizing the universe. Gravitationally magnified Hubble Frontier Field data do show high-redshift galaxies, which are at least 100 times fainter in luminosity than previously observed (Livermore, Finkelstein and Lotz 2017). However, the steep faint-end slope cannot continue to indefinitely faint galaxies—there must be a turnover or cutoff. As such, the behavior of the luminosity function at the faint end should reveal the degree to which faint galaxies powered cosmic reionization. As in the local universe, the dark matter power spectrum defines the scaffolding upon which high-redshift dwarf galaxies are built. Star formation within dwarf galaxies then plays multiple roles, not the least of which is as a source of ionizing photons. While there are not yet enough identified z $\geq$ 7 sources of ionizing photons to account for the universe's far UV transparency at z < 6 (e.g., Schenker 2015), the balance would tip in favor of high-z galaxies if there are large numbers of them that are fainter than what we presently can detect.

After reionization, the UV background should suppress star-formation and gas accretion onto halos at log (M/M$_\odot$) < 9 (see e.g., Iliev et al. 2007; Mesinger & Dijkstra 2008; Alvarez et al. 2012, Jaacks et al. 2013). Additionally, halos with virial temperatures less than $10^4$ K (log [M/M$_\odot$] < 8; Okamoto et al. 2008; Finlator et al. 2012) will not be able to cool gas via atomic line emission, and so are not expected to host efficient star formation. These processes should result in a turnover in the number density of very faint galaxies at z~7 (e.g., Jaacks et al. 2013). Existing Hubble observations of the Frontier Fields clusters have found no evidence for such a turnover down to $M_{UV} \approx -14$ mag, which corresponds to log $(M/M_\odot) \approx 9.5$. A direct test requires pushing fainter, down to at least $M = -13$ mag. The apparent luminosity of such faint sources at z ~ 7 corresponds to tens of picoJanskys. LUVOIR will be capable of detecting extended sources at such low flux levels with a survey of several hundred hours per pointing, which would directly test the hypothesis that the UV background suppresses star formation and produces a significant turnover in the low-mass galaxy luminosity function.

Finkelstein (priv. comm.) and Jaacks et al. (2013) have placed important constraints on the shape of the faint-end of the z=7 luminosity function. Using the modified form of the Schechter luminosity function with a turnover as described by Jaacks et al. (2013), the rest-frame UV luminosity function may exhibit a turnover that is detectable at, and fainter than, an absolute UV magnitude around $M = -13.5$ mag. Abundance-matching with current best estimates of the UV luminosity function implies that this corresponds to halo masses log $(M_h/M_\odot) \approx 9$ at z = 7, consistent with theoretical expectations for the suppression mass. Observationally, this deviation would start to be seen in NIR passbands at AB $\approx 33$ mag. LUVOIR-enabled deep imaging surveys that reach the turnover in galaxy LF that is predicted to occur due to UV background feedback suppression, will directly test this hypothesis. However, there are two additional observational constraints that must be accommodated. The first is that while the luminosity function must turn over, it cannot go to zero, at least not until well below the regime of dwarf galaxy progenitors. The second is the emissivity of ionizing photons in the intergalactic medium (IGM), which has been estimated using absorption features in z > 6 quasar spectra (e.g., Bolton et al. 2007; Becker & Bolton 2013). These two constraints set limits on the number of ionizing photons emanating from galaxies at a particular epoch.

**Figure 6.6** shows the predicted number of galaxies per square arcminute per magnitude detected in scenarios with and without a turnover in the differential UV luminosity function. Such a survey would deliver 40% more galaxies (down to -13.5 mag) if no turnover is present. Users of the 15-meter LUVOIR could detect this difference in the number of observed galaxies in a single HDI pointing at a 3-σ confidence level in about 340 hours of total exposure time, which includes enough time to observe each field in 3 passbands and is within the time allocation range of a large or multi-cycle treasury program. We have assumed that these ultra-faint z = 7 galaxies are resolved (~0.1 arcsec) with radii in the range of 200–300 parsecs (e.g., see Ono et al. 2013). The 9-m LUVOIR would require about 950 hours per pointing to achieve a similar detection.

Ideally, LUVOIR users would want to probe to these depths across both ionized and neutral regions (which in the 2030s, we should know of from SKA 21cm mapping, and WFIRST Lyman-α grism mapping; and/or potential NASA Probe mission Lyman-α intensity mapping). Current theories predict that we should see this turnover in ionized regions, but not in neutral regions. Models of reionization are *extremely* sensitive to this, as the galaxies near this limit dominate the ionizing emissivity. They dominate the UV luminosity density due to their numbers, but they are also more likely to have larger ionizing escape fractions than their more massive counterparts. A survey consisting of such multiple pointings may only be feasible with the LUVOIR-A design.

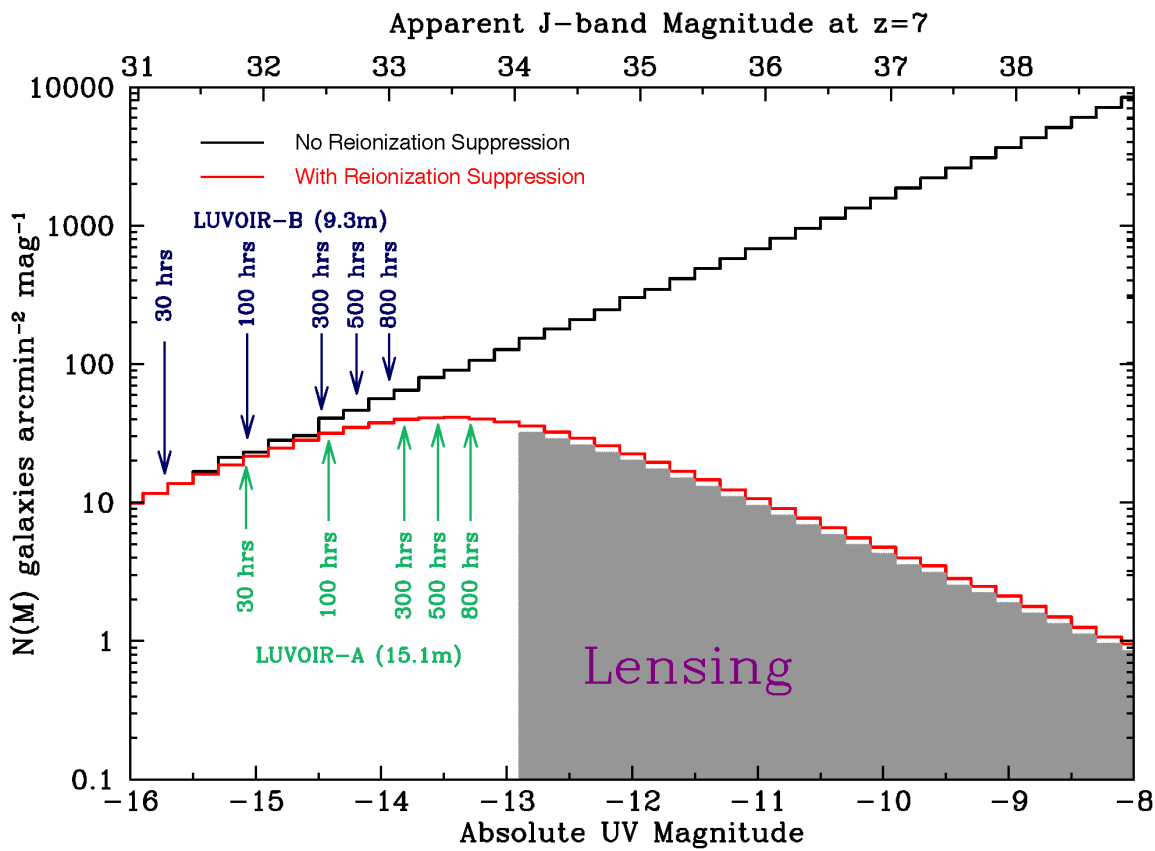

**Figure 6.6.** *Predictions for the number of z=7 galaxies per square arcminute per magnitude with reionization suppression (red histogram) and without reionization suppression (black histogram). The effects of reionization suppression manifest as a turnover in the luminosity function at an absolute UV magnitude around M ≈ -13.5 mag. The depths reached with the two LUVOIR architectures are shown as a function of survey time. Such a survey could deliver up to 40% more galaxies if no turnover is present.*

To observe directly the build-up of the extreme low mass end of the galaxy mass function from the epoch of reionization onwards LUVOIR users will require a telescope that can conduct a census of low-luminosity galaxies over a broad range of redshifts. Dwarf galaxies are faint even in the nearby universe and become much more so at higher redshifts. LUVOIR, however, will probe remarkably small stellar mass systems across a wide redshift range. To demonstrate this, we employ the Gonzalez et al. (2012) stellar mass–luminosity (M-L) relation for high-redshift galaxies and derive the expected stellar mass scales that can be probed for space telescopes of different apertures.

**Figure 6.7** shows the results for a 500 ksec exposure with a Nyquist-sampled imager. LUVOIR architectures enable the study of the extreme low mass end ($M^* < 10^{5.5}$ $M_\odot$) of the halo mass function at many redshifts in a single deep survey and enables wider surveys of $10^6$–$10^7$ $M_\odot$ systems in much shorter exposures.

Studying the faint end of the luminosity function at high-redshift also allows users of LUVOIR to seek the progenitors of the local dwarf galaxy populations, and identify their role in the process of reionization, in the context of the underlying dark matter power spectrum properties that allows them to form in the first place. There must be self-

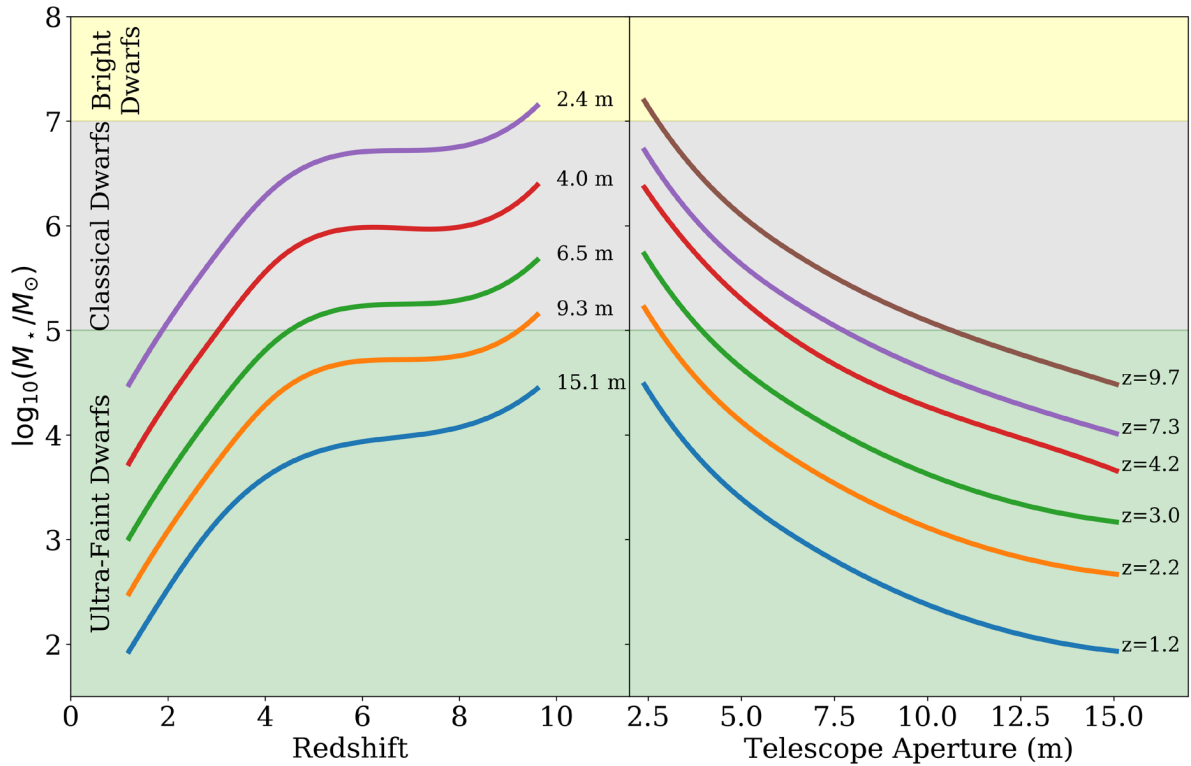

**Figure 6.7.** *The sensitivity to detecting galaxies of a particular stellar mass (y-axes) as a function of redshift and telescope aperture for a fiducial 500 ksec observation that returns a S/N = 5 detection of a 200-parsec diameter source. The notional stellar mass ranges for bright, classical, and ultra-faint dwarf galaxies are indicated.*

consistency between the galaxy populations expected at early times with the total galaxy populations that we detect subsequently. Part of the consistency check is against the dwarf galaxy census LUVOIR will perform in the local universe. As noted above, the faint end of the UV luminosity function can have a steep slope; a low-luminosity cutoff; or a combination of these. At z~7, the ultraviolet luminosity of local dwarf galaxy progenitors can be calculated (Boylan-Kolchin et al. 2015). This is shown in **Figure 6.8**. The most luminous local dwarf is the Large Magellanic Cloud, and its progenitor at z~7 has a rest-frame absolute ultraviolet magnitude of -15.8. A LUVOIR ultra deep field (500-hour exposure) will reach AB = 33.5, corresponding to $M_{UV}$ = -13.5 mag at z~7 for resolved dwarf galaxies enabling not only the detection of

LMC progenitors but progenitors of the Small Magellanic Cloud (SMC) as well.

## 6.2  Signature Science Case #2: Constraining dark matter via high precision astrometry

Dwarf spheroidal galaxies (dSph) in the local universe are extraordinary sites to explore the properties of non-baryonic dark matter. First, their mass is dominated by dark matter—they have mass-to-light ratios 10 to 100 times higher than the typical L* galaxy like the Milky Way (Martin et al. 2007; Simon & Geha 2007; Strigari et al. 2008). Second, they are relatively abundant—over 40 dSph galaxies have been found in the Local Group (Simon et al. 2011; Torrealba et al. 2016) and more will be discovered in the LSST era. Third, and perhaps most striking, is that all well-studied dSph galaxies, covering more than four orders of magnitude in luminosity, inhabit dark matter halos with the same mass (~$10^7 M_\odot$) within their central 300 pc (Strigari et al. 2008; see **Figure 6.9** here).

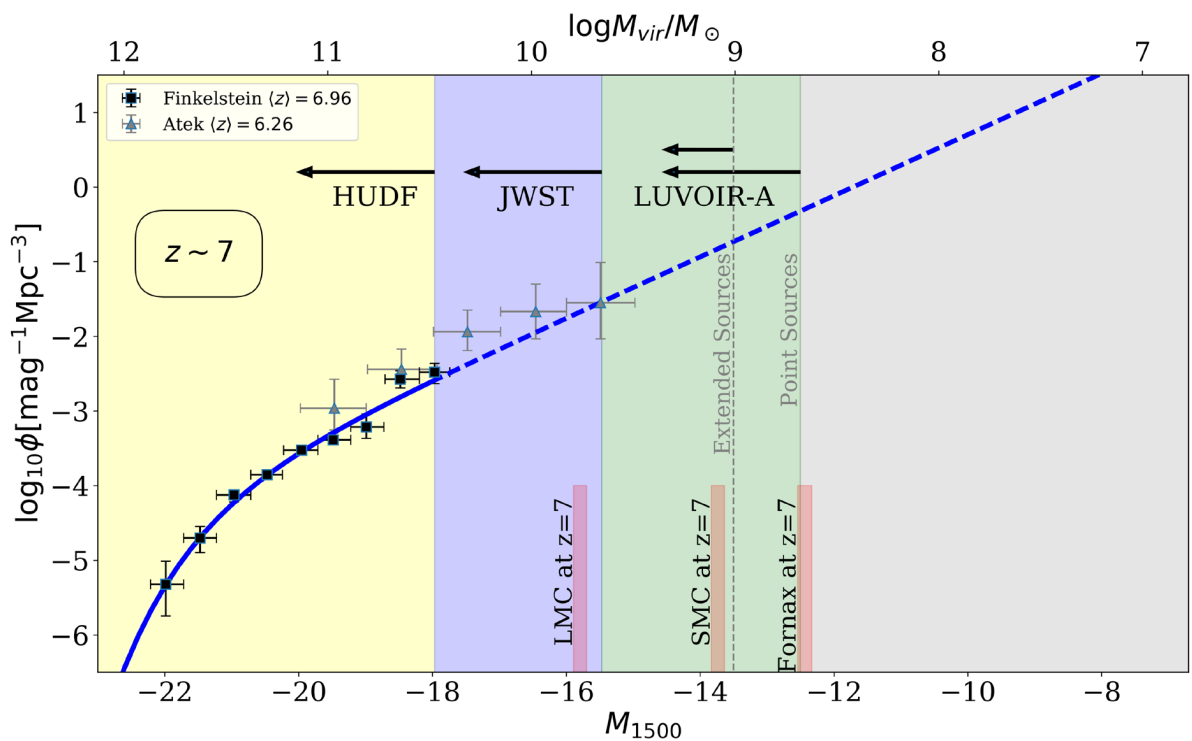

**Figure 6.8.** *The luminosities of Local Group dwarf galaxies, extrapolated to z~7, based on Figure 4 of Boylon-Kolchin et al. (2015). The observed z~7 ultraviolet luminosity function and the best fit to the data (solid line) and an extrapolation of that fit (dashed line) are also shown. The arrows show limits that can be reached with unlensed deep HST imaging (HUDF), anticipated JWST deep fields with a limiting magnitude of AB = 31.5 mag, and with LUVOIR-A 15-m which, in 100 hours, reaches a 5-σ limiting magnitude of AB = 33.5 mag (corresponding to $M_{UV}(z~7) = -13.45$ mag) for extended 0.10" sources and AB = 34.4 mag for point sources.*

The ability of dark matter to cluster in phase space is limited by intrinsic properties such as mass and kinetic temperature. Cold dark matter particles have negligible velocity dispersion and very large central phase-space

---

### Program at a Glance - The High Redshift Luminosity Function

**Goal:** Detect the turnover in z = 7 galaxy luminosity function predicted by models for the impact of reionization on early star formation.

**Program details:** Observations of 4 fields, 2 in regions with HI emission detected via SKA or other deep radio survey. LUVOIR imaging must be deep enough to allow detection of turnover at 3-sigma confidence level. Total time required for 4 fields is 1400 hours.

**Instrument(s) + Configuration:** HDI instrument on LUVOIR-A (15-m) observatory.

**Key observation requirements:** Broadband imaging in I, J, H passbands for a total of ~340 hours per field (S/N = 5 per object at a mag limit of 33.5). The time per field is 70% in I-band and 15% each in the J and H-bands. Galaxies at z~7 selected using Lyman break method.

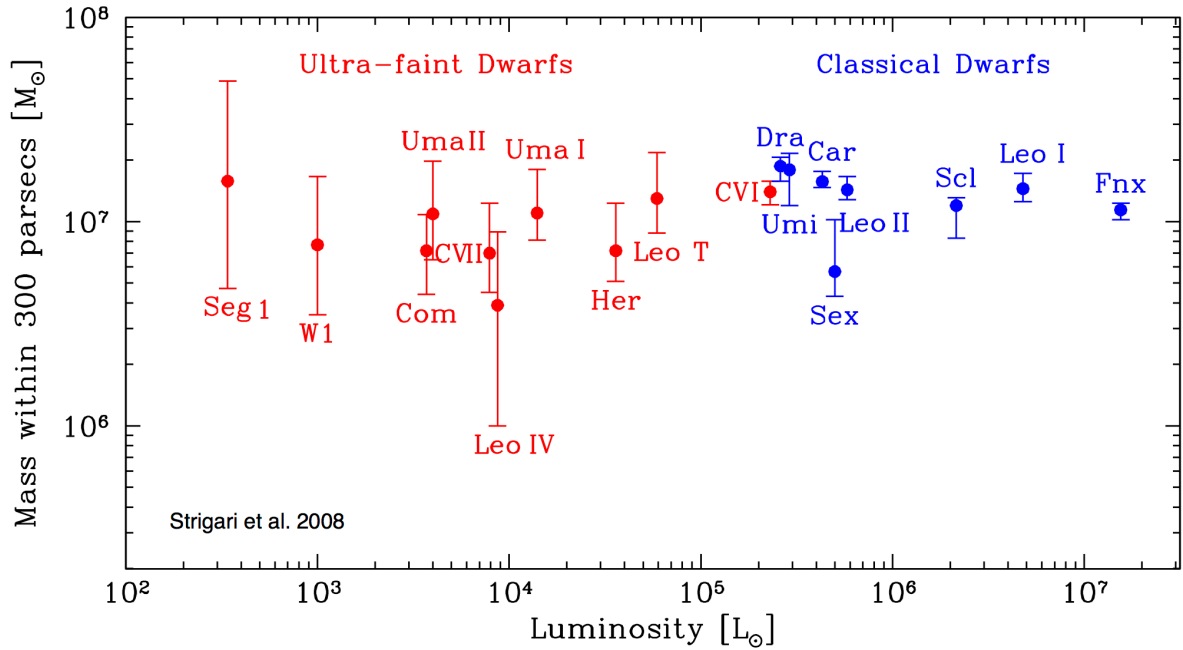

**Figure 6.9.** *The integrated mass within the inner 300 pc of local dwarf spheroidal galaxies as a function of their luminosity. The central density of these galaxies remains relatively constant over many orders of magnitude in luminosity. Figure credit: Strigari et al. (2008).*

density, resulting in steep central density profiles. In contrast, dark matter halos with highly relativistic particles (i.e., warm dark matter) have smaller central phase-space densities, so that density profiles saturate to form systems with flat central density profiles (constant central cores). Owing to their small masses, dSphs have the highest average phase space densities of any galaxy type, which implies that for a given dark matter model, phase-space limited cores will occupy a larger fraction of the virial radii. Hence, *the mean density profile of dSph galaxies is a fundamental constraint on the nature of dark matter*.

Current observations are unable to measure the density profile slopes within dSph galaxies because of a strong degeneracy between the inner slope of the dark matter density profile and the velocity anisotropy of the stellar orbits. Radial velocities alone cannot break this degeneracy even if the present samples of radial velocities are increased to several thousand stars (Strigari et al. 2007). The only robust way to break the anisotropy—inner slope degeneracy is to combine proper motions with the radial velocities. The required measurements include proper motions for ~100 stars per galaxy with accuracies better than ~10 km/sec (i.e., < 40 μas/yr at 60 kpc) and ~1000 line-of-sight velocities.

Furthermore, the orbital motions of dwarf satellite galaxies within the halo of their more massive central galaxy can reveal important properties of the dark matter distribution of galaxies like our own Milky Way and M31 (e.g., see Massari et al. 2018). While such studies have been done with existing telescopes for our Galaxy and M31, LUVOIR users will be able to extend such investigations out to several Mpc and thus expand the range of galaxies whose gravitational potential is mapped in this way. Measuring orbital parameters for dwarf satellite galaxies requires proper motion measurements with accuracy similar to that noted above in order to see typical transverse orbital velocities of about 10 km/sec.

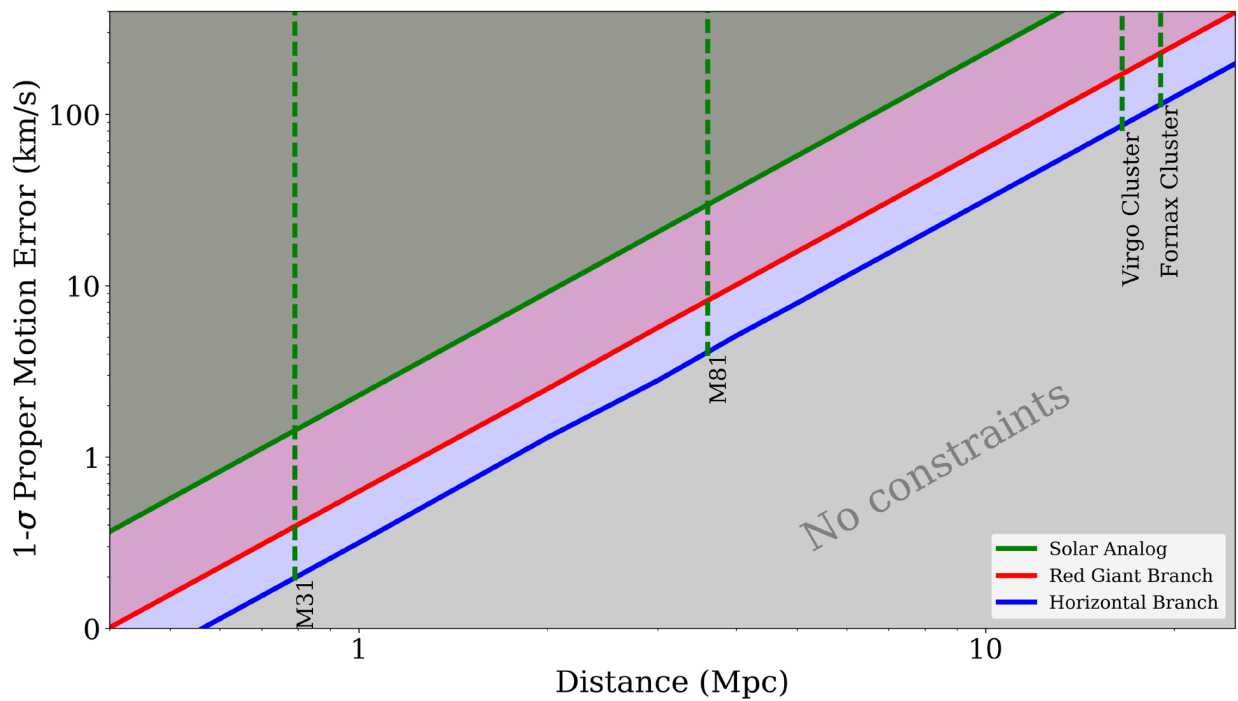

**Figure 6.10.** *The transverse velocity error from proper motion measurements achievable with the 15-m LUVOIR as a function of galaxy distance. Green, red, and blue lines show the accuracies expected for solar-type main sequence stars, RGB stars, and HB stars, respectively, assuming a 5-year baseline and 100 stars per epoch.*

In the case of the brightest of these dSph galaxies such as Fornax and Sculptor, sufficient velocities and proper motions can be obtained using stellar giants. Ground-based large-aperture (~10–m) telescopes could measure the spectra, and Gaia or JWST can measure the proper motions. For the less massive dwarfs, where the dark matter dominance is the greatest, main sequence stars will have to be used to obtain the numbers of velocity and proper motion measurements needed. This will require larger (30-m class) ground-based telescopes for the velocities. The 30-m class telescopes may also be able to obtain the necessary proper motions but it will be extremely challenging: it will require precisely stitching many fields together, most of which are unlikely to contain enough background quasars of sufficient brightness to be useful as astrometric references.

However, the necessary astrometric precision can be readily achieved by LUVOIR users, given its multi-arcminute field of view, high stability, and the pixel geometry calibration system in the HDI instrument. **Figure 6.10** shows the predicted transverse velocity error as a function of distance for LUVOIR-A. With a baseline of 5 years and 100 stars per galaxy, a transverse velocity error of 10 km/s can be achieved out to 4 Mpc (corresponding to a proper motion error of 0.5 μas/yr) using RGB stars as tracers and out to 5.5 Mpc (~0.4 μas/yr) using horizontal branch stars as tracers. Thus, LUVOIR and ELTs working together will provide some of the best constraints on the density profiles of satellite dwarf galaxies and the mass distribution around central galaxies

---

**Program at a Glance - Connecting Proper Motions to Dark Matter**

**Goal:** Measure proper motions of stars in 20 Local Group Dwarf galaxies.

**Program details:** Imaging of 20 dwarf galaxies done in conjunction with ELT ground-based spectroscopy. Total baseline will be 5 years with 3 epochs of observations for each target.

**Instrument(s) + Configuration:** HDI instrument on LUVOIR, with pixel geometry calibration using the onboard solid-state laser interferometer.

**Key observation requirements:** Broadband imaging in V and I passbands. The total exposure time per dwarf galaxy will depend on its distance but expected to be about 20 hours per epoch. With up to 20 such galaxies and a total of 3 epochs per galaxy, we estimate this program would require about 1,200 hours of LUVOIR time spread over 5 years (i.e., ~400 hours per epoch). A companion spectroscopic survey using a 20 to 40-m ground-based telescope will be needed to acquire ~1000 line of sight stellar velocities within each dwarf galaxy

---

in the nearby universe, revealing important knowledge about the kinematic nature of dark matter.

## 6.3 Signature Science Case #3: Tracing ionizing light over cosmic time

During the epoch of reionization, somewhere between a redshift of z=12 and z=6, corresponding to 360 Myr to 920 Myr after the Big Bang, the universe underwent a phase change from being radiation-bounded to density-bounded. The cause of this phase change was ionizing radiation[1] emitted by the first collapsed objects. In the density-bounded phase the neutral material became enveloped in an expanding meta-galactic ionizing background (MIB). The emergence and sustenance of the MIB along with the accompanying evolution of large-scale structure are dependent on a key parameter, for which we have very little theoretical or observational guidance, namely the fraction of ionizing radiation, $f^e_{LyC}$, escaping from these first and subsequent collapsed objects. Uncertainty in the evolution with redshift of $f^e_{LyC}$ is a major systemic unknown (Finkelstein et al. 2015; Ellis 2014; Haardt & Madau 2012), impeding our understanding of the strength of ionizing radiation feedback on the formation of structure on hierarchical and secular timescales from reionization through to the present day.

At low redshifts, the uniformity of the MIB appears to be tied to the low end of the neutral hydrogen column density distribution of Ly$\alpha$ forest absorbers, $(12.5 < \log[N_{H\,I}(cm^2)] < 14.5)$, effecting the "ionizing photon underproduction crisis," where a paucity in the number of low column absorbers found by Danforth et al. (2016) requires a MIB $\approx 2$ to 5 times higher than theoretical estimates (cf. Haardt and Madau 2012; Kollmeier et al. 2014; Shull et al. 2015; Puchwein et al. 2015; Khaire et al. 2015; Gaikwad et al. 2017; Tonnesen et al. 2017). At higher redshifts, it is an open question whether the primary source of the MIB was the first black holes or the first stars (Madau & Haardt 2015). However, it is estimated that small galaxies could have dominated the ionizing radiation budget provided: the faint end slope of the galaxy luminosity function was steep, that it extended to absolute UV magnitudes $M_{1500}$ < -13 mag, and that the escape fraction was 5% < $f^e_{LyC}$ < 40% (Bouwens et al. 2015; Finkelstein et al. 2015; Khaire et al. 2016, and references therein). As shown above, LUVOIR provides the ability to determine the faint-end shape of the luminosity function at z~7. Here, we discuss a different, complementary role for LUVOIR: characterization of the ionizing radiation itself.

---

1 By ionizing radiation, we mean photons in the Lyman continuum (LyC) with wavelengths below the hydrogen ionization edge at 912 Å

Although JWST is designed in part to discover the source(s) behind the reionization epoch, it cannot directly observe the escape of ionizing radiation because of attenuation along the line-of-sight by overlapping HI clouds in the IGM with large neutral hydrogen column densities ($\log[N_{H I} \, (cm^2)] > 17.2$, a.k.a. Lyman limit systems). The cosmic evolution of the IGM opacity is stark: the mean transmission as a function of redshift at z = [0.5, 1, 2, 3, 4, 5, 6] is $T_{IGM}$ = [0.97, 0.96, 0.8, 0.5, 0.3, 0.08, 0.01] (Inoue et al. 2014; McCandliss & O'Meara 2017), leaving less than 1% transmission by the IGM at the epoch of reionization. Therefore, the low-z universe, accessed through the UV, has a clear advantage for direct detection of escaping LyC, requiring only minuscule corrections for IGM attenuation. The ultimate goal is a statistically significant determination of LyC luminosity function evolution across cosmic time as envisioned by Deharveng et al. (1997) and Shull et al. (2015), providing a full accounting of the LyC escape budget from star-forming galaxies, AGN, and quasars of all types. These observations will allow determination of the sources and sinks of the MIB.

LUMOS will provide LUVOIR users with the ability to perform spatially resolved sampling of a large number of sources in a single observation over an extended field, allowing an in-depth characterization of the environmental factors that favor LyC escape. Spatial resolution is an important diagnostic, as models indicate that LyC escape exhibits gross variations, depending on the line-of-sight to an emitting region with respect to intervening neutral and ionized material in surrounding disks, superbubbles and circumgalactic streams (Dove & Shull 1994, Bland-Hawthorn & Maloney 1999, Dove et al. 2000, Shull et al. 2015).

McCandliss & O'Meara (2017) have calculated the intensities of LyC leakage expected from star-forming galaxies with characteristic UV $L^*_{1500(1+z)}$, as provided by Arnouts et al. (2005) (z < 3), attenuated by various H I column densities. Examples are shown in **Figure 6.11** for three galaxies spanning the range of redshifts accessible by the low resolution grating in LUMOS. The spectra show the characteristic sharp drop in flux at the Lyman edge followed by a recovery in transmission that is proportional to $(\lambda/911.8)^3$, which we refer to as a Lyman "drop-in." McCandliss & O'Meara (2017) find that the absolute escape fraction, integrated over the entire EUV bandpass, is significantly greater than the relative escape fraction, $f^e_{900}$, measured in a narrow window close to the Lyman edge, for $\log[N_{H I} \, (cm^2)] > 17.25$. An effort to spectroscopically resolve the Lyman drop-in signature with deep observations in the redshifted EUV will provide enhanced fidelity to determinations of $f^e_{LyC}$. Complementary high spectral resolution observations at longer wavelengths will be especially important for objects at progressively higher redshift to account for IGM attenuation due to Lyα forest absorbers along the line-of-sight.

GALEX has shown that UV emitting sources are intrinsically faint and are widely separated in angle. LUMOS, with its 5 arcmin$^2$ ($\Omega_{msa}$ = 3.0 x 1.6 arcmin$^2$) FoV, high effective area and low background equivalent flux is the ideal instrument for carrying out a survey of LyC emitting objects. France et al. (2017) have outlined the expected science return from a one hundred-hour pilot observing program devoted to probing LyC leakage in star-forming galaxies at redshifts $z \leq 1$ in the far-UV. These expectations are informed by the estimated effective area for the G145LL grating of LUMOS (R ≈ 500) in the 15-m architecture (shown in **Section 9.3.2**), where a peak $A_{eff}(1150Å) \approx 10^5$ cm$^2$ and low dispersion produce a background limit of AB ≈ 34 mag.

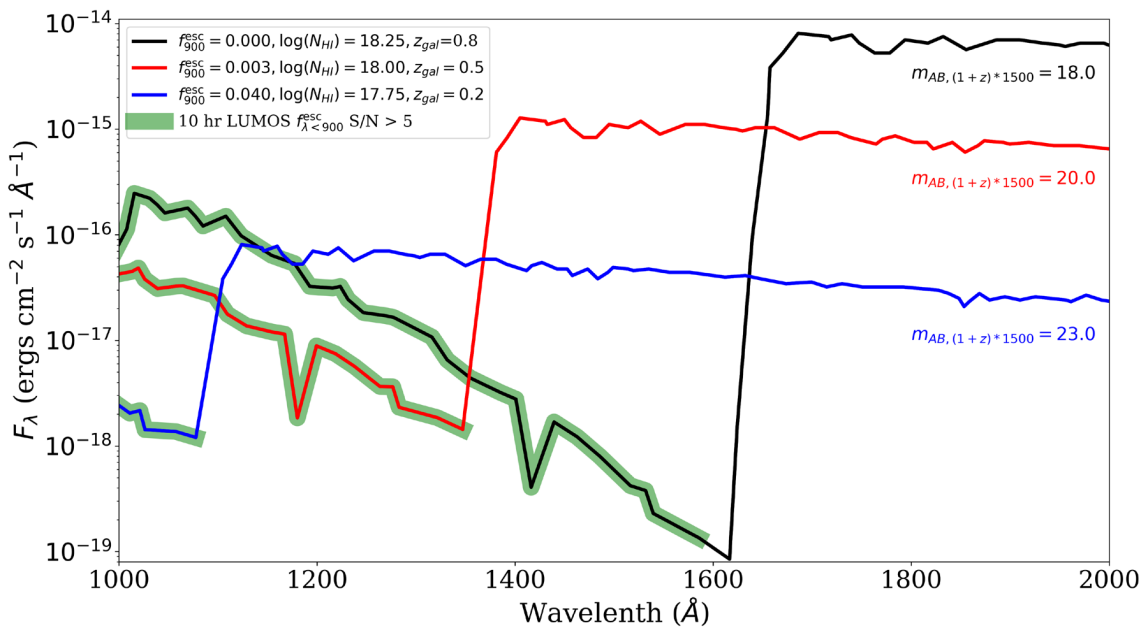

**Figure 6.11.** *Attenuated SEDs at redshifts z = [0.2,0.5,0.8] for star-forming galaxies with characteristic apparent UV magnitudes of $M_{1500}$ x (1+z) AB = [23, 20,18] mag, attenuated by various HI column densities. The green overlays on the curve indicate regions where a 10 hour LUMOS integration returns a S/N of 5 or greater.*

The top panel of **Figure 6.12** shows the cumulative number of star-forming galaxies observed to the indicated escape fraction, $f^e_{900}$, at the Lyman edge ($5\sigma$ detection limit) as a function of the apparent UV magnitude, $m_{(1+z)900Å}$. The result is also summarized in **Table 6.1**. The bottom panel shows the cumulative number of galaxies sampled for LyC leak as a function of the rest frame absolute UV magnitude, $M_{1500Å}$.

**Table 6.1.** *Number of observable galaxies in FOV 3.0 x 1.6 arcmin$^2$ as a function of redshift interval $\Delta z$ to the indicated, $f^e_{900}$ limit at $5\sigma$, $\Delta t = 10$ hours, $\Delta\lambda = 30$ Å. A one hundred-hour program will observe 10 such fields.*

| | N($\Delta z$) | | | | | Totals |
|---|---|---|---|---|---|---|
| $f^e_{900}$ | $\Delta z=(0-0.2)$ | $\Delta z=(0.2-0.4)$ | $\Delta z=(0.4-0.6)$ | $\Delta z=(0.6-0.8)$ | $\Delta z=(0.8-1.2)$ | |
| 1.00 | 9 | 72 | 292 | 358 | 295 | 1028 |
| 0.64 | 8 | 64 | 225 | 270 | 217 | 786 |
| 0.32 | 6 | 54 | 152 | 175 | 135 | 525 |
| 0.16 | 5 | 46 | 102 | 113 | 82 | 350 |
| 0.08 | 4 | 38 | 68 | 72 | 48 | 232 |
| 0.04 | 3 | 31 | 45 | 45 | 27 | 153 |
| 0.02 | 3 | 25 | 29 | 27 | 14 | 100 |
| 0.01 | 2 | 20 | 18 | 15 | 6 | 64 |

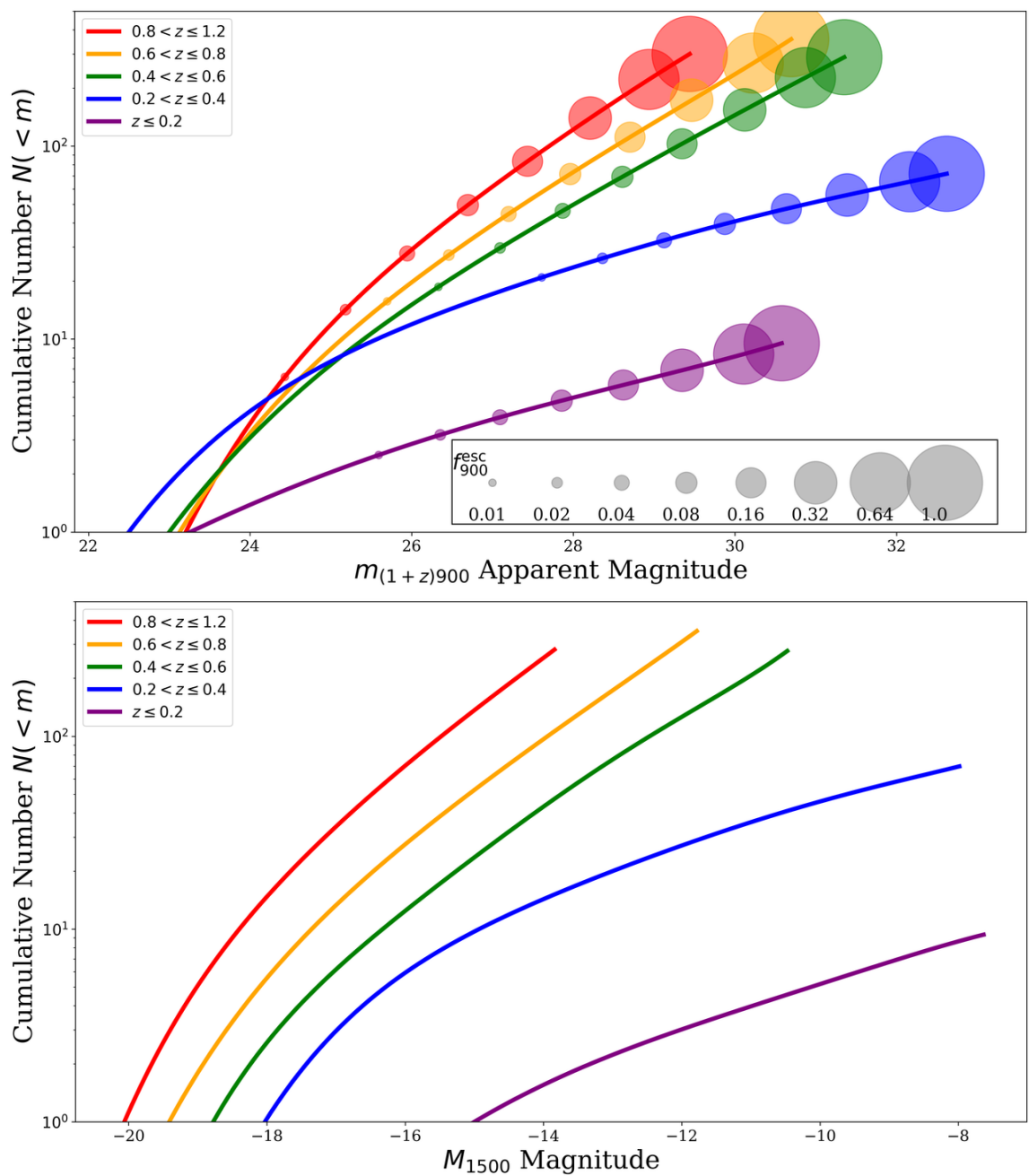

**Figure 6.12.** *Top: The cumulative number of galaxies detected in a single LUMOS field of view as a function of redshift (colored lines) and escape fraction (different size circles) as a function of the 5-σ limiting flux in a 10-hour observation over a 30 Å interval shortward of (1+z) \* 911.8 Å. Bottom: The cumulative number of galaxies detected in a single LUMOS field of view as a function of redshift (colored lines) and the absolute rest-frame 1500 Å magnitude.*

The construction of high-fidelity luminosity functions places a requirement on sample size, where 25 objects per luminosity bin per redshift interval will yield an approximate rms deviation of ~20% for each point. For the z < 1 redshift range proposed here, the luminosity range should be 2 to 3 orders of magnitude with ~20 to 40 bins covering 5 redshift intervals. Redshifts are required for each object. These considerations suggest an observing program with 500 to 1,000 objects per redshift interval, yielding 5,000 to 10,000 objects total. **Table 6.1** and **Figure 6.13** demonstrate that LUVOIR users can meet this goal with 10 LUMOS fields observed for 10 hours each. Such a program would be a definitive exploration of LyC leakage at the faint end of the luminosity function for the 0 < z < 1 regime.

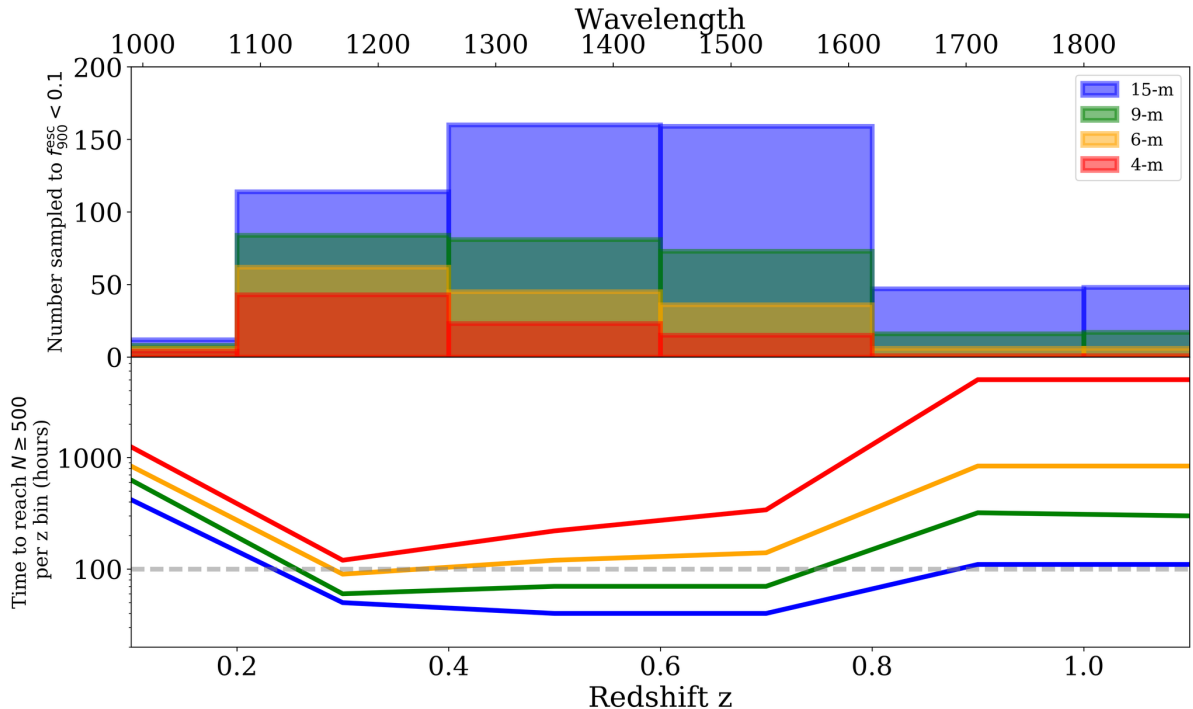

**Figure 6.13.** *The upper panel shows how the number of detectable objects with low (< 10%) escape fraction rises with increasing telescope aperture. The lower panel shows the amount of time required to obtain at least 500 galaxies per redshift bin over the range 0.1 < z < 1.1. A 100-hour LUMOS program on LUVOIR meets this threshold for nearly all redshifts.*

Quantification of how $f^e_{900}$ changes with $M_{1500Å}$ is of the highest priority, especially for those objects with intrinsically low $f^e_{900}$. In **Figure 6.13**, we show an estimate of how the number of detectable objects with $f^e_{900}$ limits would be affected by a change in aperture, under the assumption of a 10-hour observation in a $\Omega_{msa}$ = 5 arcmin² FOV. In the top panel, we plot the number of objects sampled to $f^e_{900}$ < 0.1 as a function of redshift for the indicated telescope diameters. In the bottom panel, we display the total time required to populate the redshift bins of

---

**Program at a Glance - Precision Measurements of Lyman Continuum Radiation**

**Goal:** Detect and quantify the evolution of ionizing radiation of galaxies at 0 < z < 1

**Program details:** Low resolution spectroscopy of galaxies below the Lyman break over extended fields with LUMOS.

**Instrument(s) + Configuration:** LUMOS G145LL

**Observations requirements:** 500-1000 galaxies per z=0.2 bin detected at S/N = 5 or greater over a 30 Å window below the Lyman break. These requirements translate to acquiring 10 separate LUMOS fields, each with an exposure time of 10 hours.

**Table 6.2. Summary science traceability matrix for Chapter 6**

| Scientific Measurement Requirements | | | Instrument Requirements | | |
|---|---|---|---|---|---|
| **Objectives** | **Measurement** | **Observations** | **Instrument** | **Property** | **Value** |
| Determine the turnover in the high redsfhift galaxy luminosity function | Incidence frequency of z~7 galaxies at faint magnitudes | 3-band imaging at 5σ to J > 33.5 for at least 4 fields | HDI | UVIS & NIR filters | I (UVIS) J, H (NIR) |
| Constrain the nature of dark matter via dwarf galaxies | Matter power spectrum on small scales | Multi-band imaging down to $M_V$ = 0.0 at 5σ to cover 50% of central 100 kpc region around N ≥ 4 Milky Way analogues | HDI | UVIS filters | V, R |
| Constrain the nature of dark matter via astrometric methods | Proper motion determination for stars in Local Group galaxies | Visible band imaging of 20 Local Group galaxies at high astrometric precision over 5 years i+ ground for line of sight velocity spectra. | HDI | UVIS filters, ultra-precise astrometry | V, I band. HDI precision astrometric mode. |
| Determine the escape fraction of ionizing radiation out to z~1 | Amount of flux below the Lyman break for multiple galaxies | FUV spectroscopy for N>5000 0.2 < z < 1.2 galaxies from 1000 to 2000 Å | LUMOS | R~500 Gratings | G145LL |

**Table 6.1** with at least 500 galaxies, also as a function of telescope aperture. Easily seen is the precipitous drop in the number of objects sampled for the smaller telescope diameters at the highest redshifts and at the lowest levels of $f^e_{900}$. These figures emphasize that a thorough quantification of the LyC escape phenomenon will require the largest telescope possible.

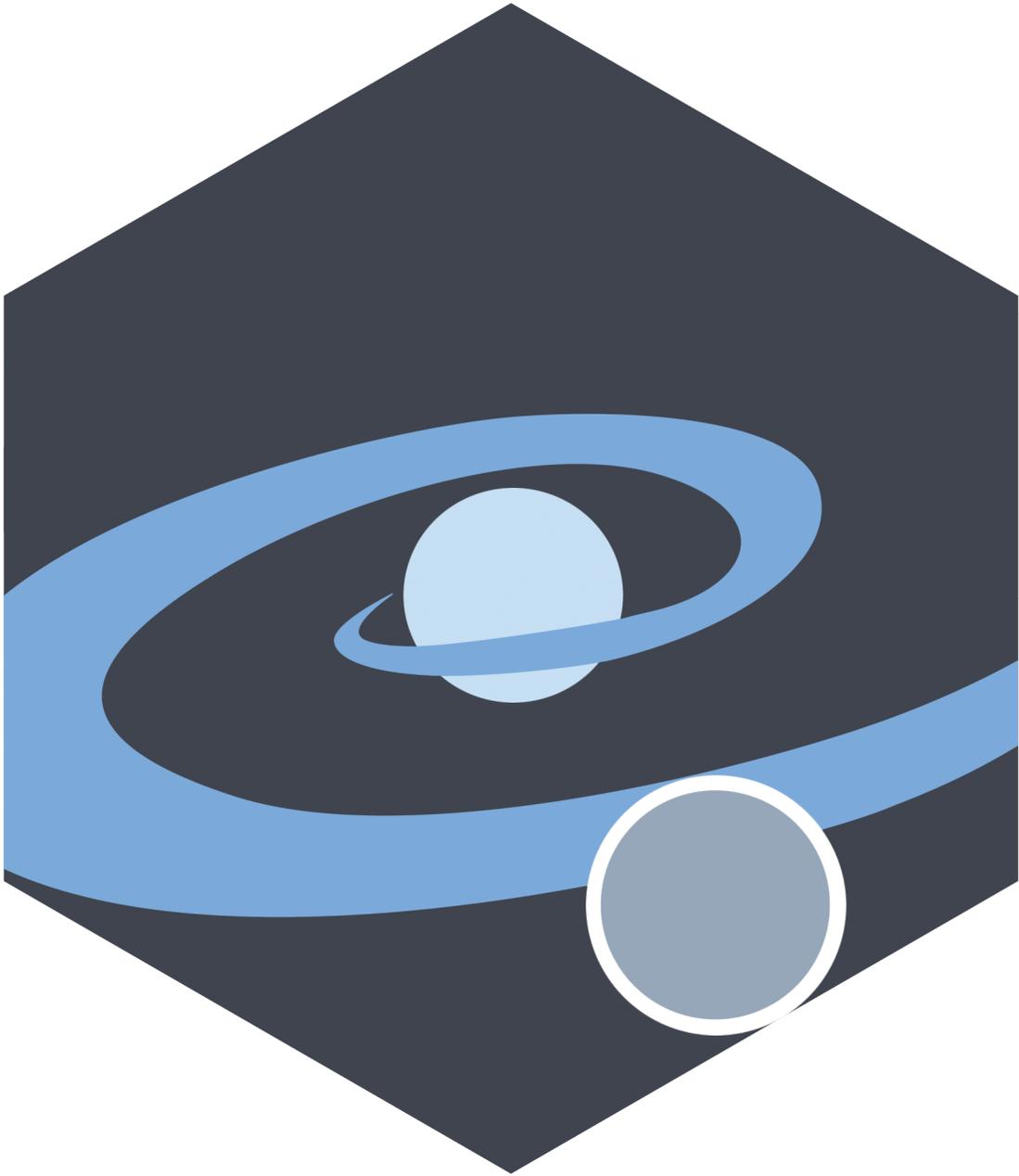

How do stars and planets form?

## 7   How do stars and planets form?

The Signature Science Cases discussed in this chapter represent some of the most compelling types of observing programs on stars, stellar evolution, and the local universe that scientists might do with LUVOIR at the limits of its performance. As compelling as they are, they should not be taken as a complete specification LUVOIR's future potential in these areas. We have developed concrete examples to ensure that the nominal design can do this compelling science. We fully expect that the creativity of the community, empowered by the revolutionary capabilities of the observatory, will ask questions, acquire data, and solve problems beyond those discussed here—including those that we cannot envision today.

### 7.1   Signature science case #1: How do stars form?

*The shape, nature, origin, and variability of the stellar initial mass function*

Stars form from the fragmentation of the dense parts of molecular clouds, in regions referred to as "cores." For all the simplicity of this description, significant challenges remain for a quantitative formulation of star formation. The relation between cores and stars is still matter of debate, as is the origin of the distribution of stellar birth masses (the stellar initial mass function, or IMF), the influence of the surrounding environment, and the effects of early dynamical evolution. The reasons for these debates are the significant observational challenges faced by measurements of the relevant quantities (cores and stars at different stages of formation and early evolution), and the fact that *we do not have yet a predictive theory covering all stages of star formation*. Extant models are able to describe individual observational results, but not link them. For instance, mas-sive stars are notoriously difficult to form in simulations, and observations have not been able to determine how (fragmentation, coalescence) and where they form. Mass segregation has not been tested as a function of environment, nor is known whether it is a dynamical or primordial effect (Kryukova et al. 2012; Kirk et al. 2012; Pang et al. 2013; Elmegreen et al. 2014). Star clusters eject stars via N-body relaxation; the frequency of these runaways, which is around 4% in the Milky Way, has neither been established in other galaxies, nor studied as a function of a galaxy's properties and environment. We are also still struggling to establish whether age spreads exist in young clusters, although some evidence for this exists for the R136 cluster in the Large Magellanic Cloud (LMC), with a spread of a few Myr.

The two problems feed into each other: without a theory, we cannot unequivocally interpret the observations of resolved or unresolved stars; and without unambiguous measurements, we cannot formulate a theory. In order to break this circular ambiguity, we need to perform unambiguous measurements of the early stages of star formation and the stellar IMF in a vast range of environments and parent galaxy properties, to guide models and finally achieve one of the key goals of modern astrophysics: a theory of "how stars form."

Accomplishing the characterization and quantification of star formation and the stellar IMF will require the synergistic combination of observations from a broad wavelength range, UV to millimeter. This combination is partially offered by current (Hubble, ALMA, VLT, etc.), upcoming (JWST), and planned (30m-class ground-based telescopes) facilities. Among the capabilities that are neither available nor planned, and which will be re-

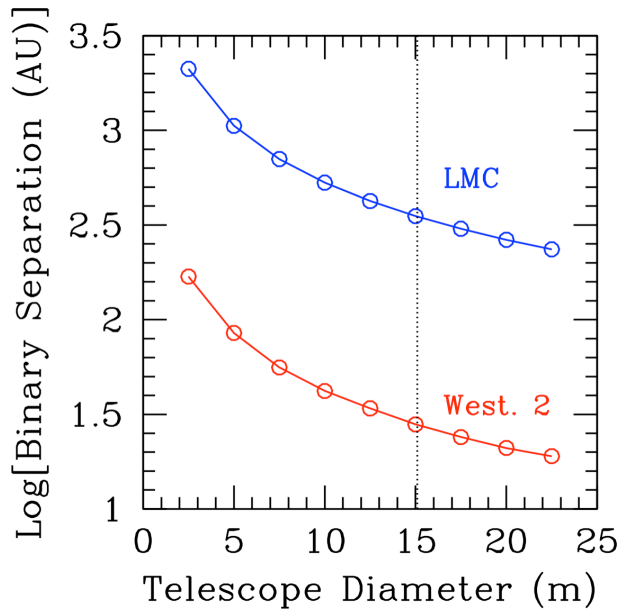

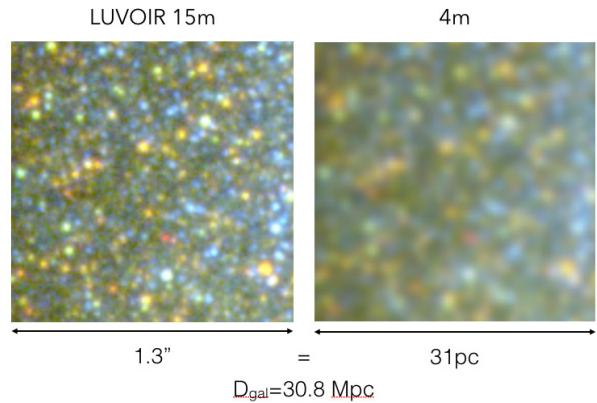

**Figure 7.2.** *An inner 8.4" by 8.4" region of the galaxy IC 4247 from Calzetti et al. (2015), as viewed by a 15-m LUVOIR and a 4-m telescope if IC 4247 was a distance of 30.8 Mpc. At this distance, each box is 1.3" on a side, corresponding to 31 pc. The 4-m image would need ~16 times the exposure time as the LUVOIR image.*

**Figure 7.1.** *The resolvable separation, in AU, between two stars, as a function of telescope diameter, for binaries in the Westerlund 2 star-forming region (at a distance of 4 kpc) in the Milky Way and in the Large Magellanic Cloud (50 kpc). A 15-meter telescope fully separates binary systems down to ~30 AU in the star-forming region and down to ~350 AU in the Large Magellanic Cloud. Massive binary stars at different stages of evolution are separated by distances between a few tens and a few thousand AU (Portegies Zwart et al. 2010), and a 15-m telescope will resolve binaries at critical stages of evolution. With this telescope, spectral signatures of very massive stars within dusty star clusters will be detectable in ~200 luminous infrared galaxies and ~20 ultra-luminous infrared galaxies out to about 150 Mpc.*

quired to accomplish the overarching goal of formulating a theory of star formation, are:

1. Angular resolution for imaging: 0.007" at 500 nm (**Figure 7.1** and **Figure 7.2**), with stable point spread function (PSF) and stable astrometry across the entire field of view (FoV).

2. UV and optical imaging spectroscopy over 1'–3' FoV, with 0.01"–0.02" angular resolution.

3. High-sensitivity UV spectroscopy with resolving power R > 3000, to resolve atomic and molecular lines from winds and photospheres in stellar spectra.

Determining the nature, average properties and relevant physical parameters, and potential environmental dependencies of star formation (and its corollary: the stellar IMF) across the full stellar mass range, from the high to low mass ends, will require synergy across multiple wavelength regions, from the UV to the mm. Important characteristics are poorly constrained, including merger rates, binary fractions below O stars, multiplicity, fraction of runaway stars, and whether very massive stars (VMSs, i.e., stars with masses > 150 $M_\odot$) exist and are common. In addition to its primary and unique role in achieving a full understanding of star formation, LUVOIR will play a complementary role in characterizing the shocks, jets, and outflows from the individual protostars and protoclusters, and their natal cores, identified by ALMA.

---

**State of the Science in 2030**

By 2030, JWST and WFIRST will have observed the stellar populations down to 0.3 solar masses for all galaxies within the local 0.5 Mpc, with lower limits for nearer galaxies. This will test theories of low-mass star formation, and constrain variations of the low-end of the stellar initial mass function in low-density environments. ALMA will have characterized the populations of gas cores in several nearby galaxies, and will have shed light on the relation between the core mass function and the stellar initial mass function around the low mass turnover. Improved models will help interpret the complex chemistry of the pre-stellar cores. In parallel, adaptive optics-assisted 30-m class telescopes will have assembled large samples of binary stars at a range of masses out to the Magellanic Clouds, using both radial velocity (RV) and proper motion techniques. Combinations of these facilities will have increased the number of candidate galaxy hosts of very massive stars, i.e., stars with masses > 150 $M_\odot$, by at least tenfold, up from the current census of a handful.

By 2030 NASA's TESS mission will have identified a thousand planets smaller than Neptune with orbital periods less than 40 days around nearby stars (Ricker et al. 2014); we can expect that many will have been further characterized using ground-based (e.g., ELTs) and space-based facilities (e.g., JWST). NASA's WFIRST and ESA's PLATO are expected to launch around 2024 and will extend exoplanet detections out to and beyond the habitable zone, with the latter mission focusing on bright solar-type stars. Ground-based RV monitoring will have extended the statistics of planets larger than Super-Earths out to a few AU (e.g., Pepe et al. 2010). Thus, by 2030 the exoplanet population of nearby stars within a few AU will be fairly well characterized. In relation to planet-forming disks, ALMA will have delivered hundreds of continuum disk images at a resolution of ~5–10 AU in nearby star-forming regions, thus fully characterizing the mid-plane dust emission in the giant planet-forming region. Many of these disks will also have AO-assisted ground-based and/or WFIRST scattered light imagery at similar spatial resolution, probing the population of small grains and the overall disk structure. JWST will have surveyed hundreds of these young stars at a resolution of > 20 AU, and probed material ejected in jets/outflows as well as warm (~500 K) disk gas inside the snowline via medium-resolution ($\Delta v$~100 km/s) spectroscopy.

---

### 7.1.1    *Very massive stars*

VMSs are stars that exceed the standard limit of 150 $M_\odot$, and one needs to observe very young (<~1.5 Myr), massive star clusters (>$10^5$ $M_\odot$) in order to detect their presence, due to small number statistics at the high end of the IMF, and rapid evolutionary timescales. Yet these stars can heavily influence their surrounding environment; for instance, they can provide between 25% and 50% of the ionizing photon flux from the host cluster.

A solid case for the presence of VMSs in R136, the central cluster in 30 Dor in the Large Magellanic Cloud, has been made by Crowther et al. (2010, 2016). UV spectroscopy has provided evidence for the potential presence of VMSs in an additional two unresolved star clusters in NGC 3125 (Wofford et al. 2014) and NGC 5253

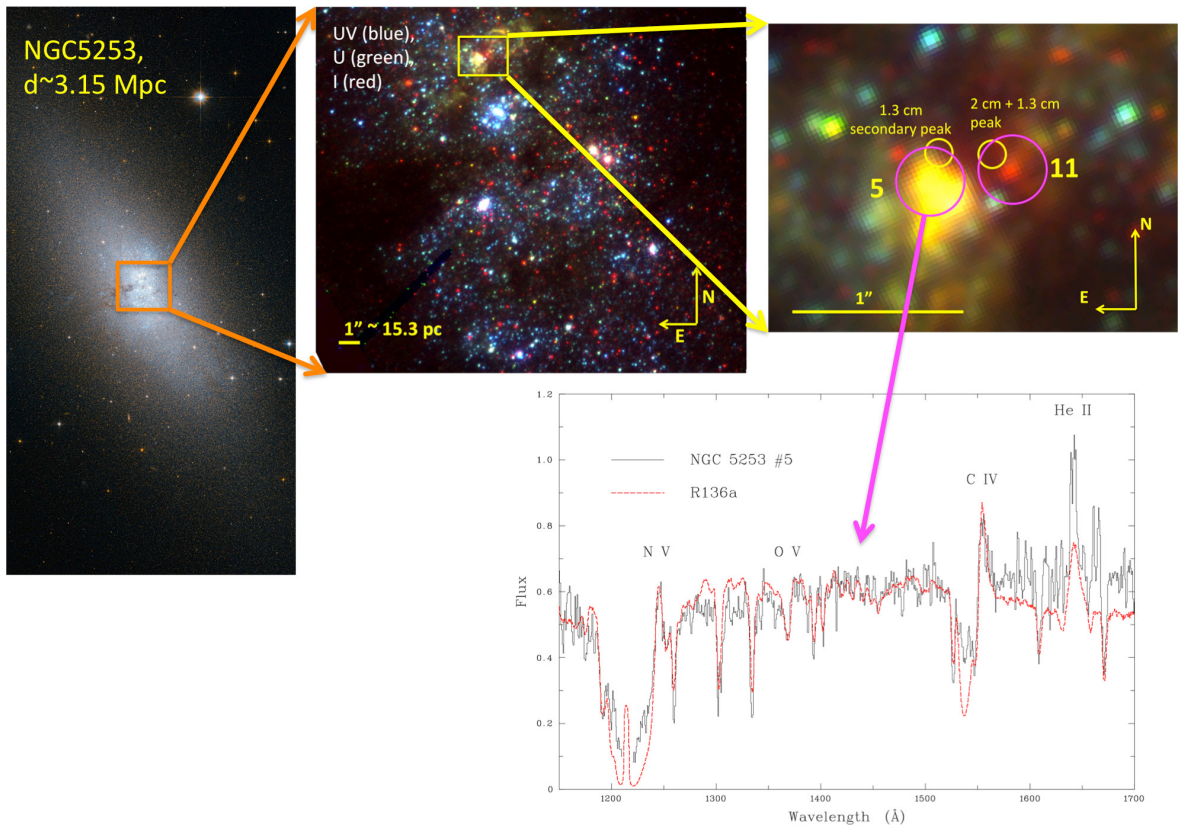

**Figure 7.3.** *Identifying very massive stars (VMSs) in star clusters. NGC 5253 (left panel) is an amorphous dwarf located at 3.15 Mpc distance, and hosting a starburst within its central ~250 pc region (orange rectangle in the left panel, and central-top panel, which shows a three-color combination from HST imaging). Most of the dust and molecular gas in this galaxy are concentrated in a ~20 pc region located at the north end of the starburst (yellow rectangle in the center panel, and right-top panel). The two main radio peaks are located (yellow circles in the top-right panel; Turner et al. 2000). The two radio peaks coincide with two clusters identified in the HST images (magenta circles in the top-right panel; called 5 and 11 by Calzetti et al. 2015). UV-to-K HST photometry shows that the two clusters are ~1 Myr old, with masses 1–3 x $10^5$ $M_\odot$, and with $A_V$ ~2–50 mag of dust. UV spectroscopy of cluster 5 further reveals presence of VMS signatures: P Cygni NV (1240 Å) and CIV (1550 Å) profiles, broad He II (1640 Å) emission, blue-shifted OV (1371 Å) wind absorption, and absence of SiIV (1400 Å) P Cygni emission/absorption (bottom panel; Smith et al. 2016), as confirmed from the comparison with the UV spectrum of the LMC cluster R136a, also containing VMSs (Crowther et al. 2016). The HST UV spectrum of cluster 5 required 3 hours of exposure. The same amount of time with LUMOS on a 15-m will detect with S/N~10 a $10^6$ $M_\odot$ cluster with similar dust properties as cluster 5 out to a distance of 80 Mpc (25 times further away than NGC 5253). LUVOIR will increase by at least tenfold the parameter space in distance and galactic properties (morphology, dust content, star formation rate, etc.) of the systems whose VMS content can be identified and characterized.*

(**Figure 7.3**; Smith et al. 2016). Detection of supernovae that have characteristics typical of pair-instability SNe provide indirect evidence for the presence of VMSs. Whether the VMSs are the result of birth conditions or of mergers is still matter of debate, but models can successfully reproduce many observational properties using a rotating single VMS (Yusof et al. 2013, Kohler et al. 2015). Data on the VMS numbers and frequency in young star-forming regions are lacking, due to observational limitations: they can only be recognized in the UV, and the cluster needs to be resolved into individual stars in order to obtain a census of VMSs. Individual cluster stars can only be resolved to the distance of the LMC with Hubble (Crowther et al. 2016). Reaching beyond the Magellanic Clouds, to resolve into individual stars the young populations in the star-forming galaxies of the Local Group out to ~1 Mpc (e.g., NGC 6822, IC 1613, NGC 3109, Gr 8, the unusual nearby bright starburst IC 10) will offer a significant increase in the potential number of individually detected VMSs.

In addition to a census of VMSs within individual clusters, a census of the frequency of VMSs in different environments is required to gauge their overall impact on the evolution of galaxies, and to constrain models for their formation and evolution. Ideally, one would sample a wide range of very young star clusters in as wide a range of galaxies as possible, including luminous infrared galaxies (LIRGs) and ultra-luminous infrared galaxies (ULIRGs). Presence of VMSs can be inferred using unique UV spectral signatures: P Cygni NV (1240 Å) and CIV (1550 Å) profiles, broad He II (1640 Å) emission, blue-shifted OV (1371 Å) wind absorption, and an absence of SiIV (1400 Å) P Cygni profiles. In the optical, a VMS would be easily confused with WR emission from a lower mass O-star. Thus, UV spectroscopy is a key requirement for obtaining a census of VMSs and other massive stars in the nearby Universe, together with quantitative measurements of their properties.

UV multiplexing (multi-object spectrograph or integral field spectrograph) to obtain spectra of resolved massive stars in clusters out to 0.7–1 Mpc, and spectra of unresolved young, massive star clusters out to 150 Mpc (Arp 220, the prototype ULIRG, is at 77 Mpc distance) is a minimum requirement, and will secure about 200 LIRGs and ~20 ULIRGs. Large detector formats increase efficiency by covering entire star clusters/star-forming regions and/or entire star cluster populations in a single pointing. A large telescope aperture, ~15-m, will ensure both enough sensitivity (~2 hours exposure to reach S/N = 10 at 160 nm on a compact cluster with $m_{AB}$(FUV) = 23 mag; this is a 1 Myr old, $10^6$ M$_\odot$ cluster at a distance of 80 Mpc and embedded in 20 mags of extinction, as expected in Arp 220) and angular resolution for resolving individual stars within clusters out to 0.7–1 Mpc (0.01" resolution in the UV corresponds to a separation of 0.03 pc at 0.7 Mpc distance). As such, LUVOIR+LUMOS represents an ideal combination for VMS studies.

### 7.1.2    Stellar multiplicity

Stellar multiplicity and binary frequency constrain models of star formation and the IMF (Offner et al. 2014). Hydro-dynamical simulations show that massive stars require dense and massive accretion disks to form, as these are needed to overcome the radiation pressure barrier. The disks tend to be unstable and break into complex systems. The resulting properties of the multiple systems (number of companions, distribution of separations [i.e., short vs. long orbital periods], distributions of mass ratios, and their dependence on the stellar mass) are model-dependent. Furthermore, the dynamics of binaries drive the evolution of star clusters while, simultaneously, the combination of cluster dynamics and internal stellar processes determine the internal evolution of each binary (Portegies Zwart et al. 2010), but, again, there is enormous dependency on uncertain parameters.

Short-period (spectroscopic) binaries will remain the domain of ground-based telescopes, especially the upcoming integral field spectrographs on adaptive optics-assisted (AO-assisted) 30-m+ class telescopes. Long-period binaries require high angular resolution, a very stable PSF, and high precision photometry across the entire FoV, which needs to be of a few arcminutes, in order to increase efficiency by targeting each cluster with as few pointings as possible. Upcoming AO-assisted red/near-IR cameras (e.g., ELT/MICADO) may achieve the required level of stability over large FoVs, and deliver proper motions as accurate as 10 µas/yr over 4 years. However, the physical properties of the stars need to be characterized in order to constrain models. This demands measuring the **resolved** massive stars' winds and photospheric parameters (including bolometric luminosities and masses), which can be accomplished only with spectroscopy of individual stars in the 100–200 nm range (Wofford et al. 2012). Binaries at different stages of evolution within a star cluster have mean separations that change between a few tens and a few thousand AUs (Portegies Zwart et al. 2010). Resolving binaries with UV spectroscopy *at all stages of evolution* requires resolving mean separations across a wide range from tens to hundreds of AU; probing a range of environments requires reaching at least the distance of the Magellanic Clouds. LUMOS on LUVOIR-A will resolve binary stars with separations down to ~40 AU in the iconic high-mass star cluster Westerlund 2 (~4 kpc from the Sun) and ~500 AU in the LMC.

The diffraction-limited angular resolution HDI on LUVOIR-A (0.007″) will resolve and image all binary stars down to ~30 AU separation in Westerlund 2 (**Figure 7.1**), and down to smaller separations in the many dozens of intervening star-forming regions spanning a wide range of densities and other properties. A 30 AU separation is found for a system of two equal-mass late-type B stars (~3 $M_\odot$ each) with a 25-year orbital period. The tens-AU regime is interesting for the investigation of the progenitors of peculiar objects and GRBs, as models suggest massive stars may be interacting in the post-MS phase. This will also address numerous additional related questions, including the IMF of binary stars, and the role of binary stars in determining the upper end of the IMF (**Figure 7.4**).

### 7.1.3   The low-end of the IMF in unresolved populations

Recent observational results suggest variations at the low mass end of the stellar IMF (< 1 $M_\odot$), although small number statistics and degeneracies plague all current studies. Analysis of spectroscopic features of the integrated light from elliptical galaxies indicates that the slope of the low-end of the IMF becomes steeper, i.e., more bottom heavy, for increasing velocity dispersion (e.g., van Dokkum & Conroy 2010; Conroy & van Dokkum 2012; Spiniello et al. 2014, 2015; van Dokkum et al. 2017), in agreement with results from analyses of the mass-to-light ratio in the same systems (Cappellari et al. 2012). However, measurements of lensed early type galaxies and of IR spectral features have challenged the robustness of IMF-sensitive spectral features and of mass-to-light ratio determinations (Smith et al. 2014, 2015).

Direct IMF determinations, i.e., counting stars in resolved stellar populations, yield a cleaner result, although current facilities cannot reach far enough to probe a rich range of galaxy properties. Geha et al. (2013) have investigated the stellar content of two ultra-faint dwarf galaxies (UDGs) using HST multi-color imaging, concluding that their IMFs are significantly flatter than either those of Kroupa or Salpeter in the 0.5–0.75 $M_\odot$ range. When evaluated together with direct low-mass IMF determinations in the Magellanic Clouds and the Milky Way, Geha et al. conclude that there is a systematic trend for lower velocity dispersion and lower metallicity systems to have flatter IMFs, in agreement with the trends observed for the early type galaxies. Conversely, Wyse et al. (2002) determine that the low-mass IMF down to 0.3 $M_\odot$ in Ursa Major, another of the Milky Way satellites, is similar to the IMF of dense Milky Way globular star clusters. These contradictory results may stem from the fact that the mass range directly probed in the Milky Way satellites is too close to the turnover of a Kroupa/Chabrier IMF for an unambiguous determination of the faint-end slope, and the measurements need to be pushed below 0.2–0.3 $M_\odot$ to yield reliable results (Offner et al. 2014).

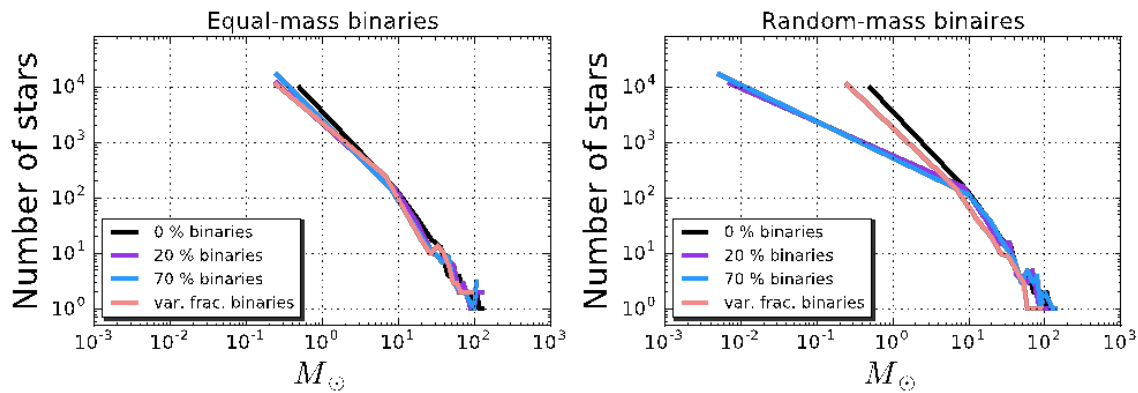

**Figure 7.4.** *Simulations of binary systems under different assumptions of binary fractions and mass ratios. In all cases the goal is to quantify the input IMF of the resolved stars, in order to reproduce, for the unresolved binary systems, the observed Salpeter stellar IMF (slope = -2.35) in the mass range 0.5–150 $M_\odot$. The left panel shows the case of equal mass binaries, while the right panel shows the case in which the binaries have randomly selected mass ratios (down to stellar masses of 0.01 $M_\odot$). In both panels, the black line shows the case of 0% binary fraction, i.e., the output ('observed' for unresolved binaries) IMF, expressed as number of stars per mass bin. The purple and blue lines show the cases of fixed binary fractions, 20% and 70%, respectively. The orange line assumes mass-dependent binary fractions: 70% for M > 10 $M_\odot$, 20% for 2 $M_\odot$ < M < 10 $M_\odot$ and 10% for M < 2 $M_\odot$. Equal mass binaries (left panel) are drawn from an intrinsic IMF that is not dissimilar from the output one, with the exception of a mass break at a few $M_\odot$. Conversely, binaries with random mass ratios (right panel) need to be drawn from intrinsic IMFs that are potentially very different from the observed one: the cases with fixed binary ratios display a dramatic flattening of the intrinsic slope for masses below 10 $M_\odot$, while the case with variable binary fractions could be drawn from an intrinsic IMF with a steeper high-end slope than the observed one.*

Near-future facilities (JWST and WFIRST) will make in-roads for direct star counting in nearby galaxies, mostly in the satellites of the Milky Way. JWST will resolve individual stars down to 0.3 $M_\odot$ for all Milky Way dwarf satellite galaxies out to ~0.5 Mpc, and down to the hydrogen burning limit for the nearest dwarfs, thus enabling testing of models of low mass star formation and probing variations of the low-end of the IMF. The IR capabilities on JWST will provide a factor of ~2 improvement in efficiency over HST for detection of low-mass stars (Geha 2014), and repeat observations to separate populations (foreground from external ones) via proper motions will further increase the efficacy of the approach.

Conversely, the central regions in elliptical galaxies will remain unresolved even with the most powerful telescope we can conceive; so indirect (integrated population) methods will remain the only option. The central regions of ellipticals appear to be special places in many respects: they host supermassive BHs; show high velocity dispersion/deep potential wells and high [$\alpha$/Fe] abundance patterns; and are the centers of the first collapsing structures, with very high inferred star formation rates when the bulk of the stars formed, etc. Several physical models for the IMF seem to suggest that these would be prime sites for IMF variation. Among these galaxies are archetypes like M87, M60, and M49 (all Virgo ellipticals), but also the nearby field ellipticals Centaurus A and NGC 3377.

One of the most promising avenues for constraining the low-end IMF is the exploitation of the surface brightness fluctuation technique at the spectroscopic level, called "fluctuation spectroscopy" (e.g., van Dokkum & Conroy 2014). The basic idea is the following: for nearby elliptical galaxies the number of giants per pixel is small and so their numbers will fluctuate (via Poisson statistics) from pixel to pixel. One can therefore identify pixels containing relatively few giants. By focusing on "low fluctuation" pixels one can obtain a relatively unobstructed view of main sequence stars, which are easier to model in integrated light than the entire population. The breakpoint is when the mean number of stars per resolution element is < $10^5$, which is where the pixels are dominated by MS stars, and there are very few, if any at all, giants. These resolution elements need to be observed with multi-object spectroscopy between 400 nm and 1000 nm (an extension to the LUMOS bandpass currently under consideration), in order to separate M dwarfs from F and G dwarfs, and infer variations at the low-mass end of the IMF.

Currently, HST delivers a mean number of stars per resolution element (~0.2") of ~$10^7$ in the central regions of Virgo ellipticals, implying that an improvement of about a factor 100 in diffraction-limited region area is required. This demands ~0.01" angular resolution spectroscopy over arcminute-sized fields. Lack of blue coverage (JWST) or large FoV (e.g., HARMONI on ELT) makes these future facilities inadequate for achieving this goal. For future AO-assisted integral field spectrographs, the main challenge will be to control the stability of the PSF and plate scale over a large FoV at blue wavelengths. To accomplish a major breakthrough, 400–1000 nm multi-object spectroscopy over several arcminutes FoV, with a 10x improvement

---

### Program at a Glance - The Nature of the IMF

**Goal:** Constrain massive star formation and the stellar initial mass function.

**Program details:** UV and optical spectral signatures of individual stars and star clusters / groups in nearby galaxies (d < ~150 Mpc). Imaging of stellar systems to identify multiplicity, and follow-up multi-object spectroscopy in the 100–200 nm range to characterize stars in multiple systems. NUV-optical multi-object spectroscopy of the centers of ellipticals to constrain variations at the low-end of the IMF.

**Instrument(s) + Configuration(s):** HDI NUV and optical imaging; LUMOS multi-object UV spectroscopy (with extension into optical)

**Key observation requirements:** About 250 targets; Imaging resolution of 0.007"; Imaging SNR = 10–300; Multi-object spectroscopy angular resolution of 0.01"–0.02"; Spectroscopy SNR ≥ 10 over 100–1000 nm; Spectroscopy resolution R > 3000 to resolve stellar wind and photospheric lines.

in spatial resolution over HST at 500 nm is required. Independent resolution elements are 2 PSF widths in size, translating to a resolution requirement of ~0.01" at 500 nm.

## 7.2  Signature science case #2: How and in what environments do planets assemble?

### Detecting protoplanets and tracing the composition and evolution of planet-forming material

One of the main science objectives of LUVOIR will be to detect and characterize Earth-size planets in the habitable zone of nearby stars (**Chapter 3**). Understanding how and in what environment planets assemble contributes to this science objective in two critical ways: i) by determining which nearby planetary systems are most likely to host life-bearing planets and ii) by constraining planet properties that are not directly observable (e.g., bulk composition), which are important in interpreting the spectra of potential Earth-like planets.

Recent high-resolution images of circumstellar disks around young (1–10 Myr)

stars revealed complex structures (e.g., ALMA Partnership 2015; Wagner et al. 2015; Andrews et al. 2016; Perez et al. 2016; Pohl et al. 2017; Hendler et al. 2018), some of which point to advanced planet formation. A few candidate giant planets have been recently detected around Myr-old stars (e.g., Sallum et al. 2015). Short timescales to assemble asteroid-size objects and giant planets are well in line with the evolution and dispersal of gas and dust in the solar nebula (Pascucci & Tachibana 2010). Thus, 1–10 Myr-old circumstellar disks provide an opportunity to study planet formation in action.

Constraints on planet formation also come from the properties of Gyr-old exoplanets and planetary systems (**Chapter 4**). Indeed, scaling relations in disks and exoplanets with the mass of the central star strongly suggest that sharp edges and gaps in young disks remain imprinted in the exoplanet population (e.g., Alexander & Pascucci 2012; Mulders et al. 2015). In addition, the composition of early planetary atmospheres should inform the formation history of planetary systems (e.g., Cridland et al. 2017).

At the distances of typical star-forming regions (e.g., Taurus-Auriga or Chamaeleon I), 1 AU corresponds to an angular scale of ~10 mas and gas probing ~1 AU radii will have FWHMs of ~30 km/s. Therefore, the high angular and spectral resolution of LUVOIR is a unique tool for mapping the inner disks where terrestrial planet formation occurs (**Figure 7.5**). Such high-resolution observations will directly link birth environments to the final architecture of the exoplanetary systems. LUVOIR observations of young stars will detect and characterize protoplanets (**Section 7.2.1**), trace the composition, evolution, and dispersal of planet-forming material (**Section 7.2.2**), and map protoplanetary material dispersed via disk winds (**Section 7.2.3**).

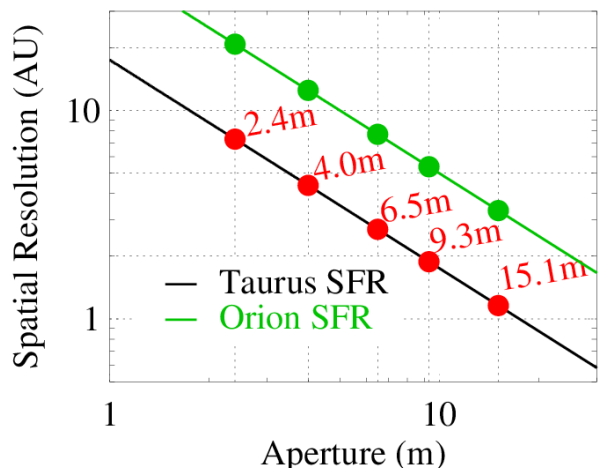

**Figure 7.5.** *Terrestrial planet-forming radii resolved at 500 nm as a function of telescope aperture for nearby star-forming regions with a range of massive star content.*

### 7.2.1    Detect and characterize protoplanets

As protoplanets are likely accreting gas through a circumplanetary disk, multi-wavelength observations are necessary to separate disk emission from that produced in an accretion shock and thereby constrain the properties of forming planets. Accreting giant planets will glow in U- and B-band as well as UV and Balmer emission lines (e.g., Hα at 656.3 nm; **Figure 7.6**), while the disk will start to dominate at ~1000 nm (e.g., Zhu 2015). 30m-class ground-based telescopes with extreme AO systems will be diffraction limited at near-infrared and visible wavelengths over small field of view (~few square arcseconds) and will mostly probe re-processed emission from one circumplanetary disk at a time.

Jovian mass planets at 5–50 AU separations will carve deep gaps in their natal disks that can be readily resolved in LUVOIR coronagraphic scattered light images in the visible. UV/visible imaging in HDI filters sensitive to these accretion diagnostics will enable the direct detection of protoplanets within these gaps. **Figure 7.6** shows an example of a simulated face-on disk + actively accreting protoplanet, at the distance of the Taurus star-forming region, in units of the signal-to-noise per resolution element of HDI's narrow-band Hα filter. Conservative estimates of the accretion luminosity indicate that such intermediate distance Jovian/Saturnian mass planets should be detectable in all nearby star-forming regions. Since these accreting planets can be detected without coronagraphy in narrow-band Hα imagery, nearby star-forming regions can be surveyed with HDI's 3' × 2' field-of-view, enabling a statistical study of protoplanets in both mass and semi-major axis. Furthermore, the accreting protoplanet survey will identify the brightest targets for follow-up spectroscopy in the FUV and NUV with LUMOS—enabling us to characterize the circumplanetary accretion environment in detail, a capability unique to LUVOIR.

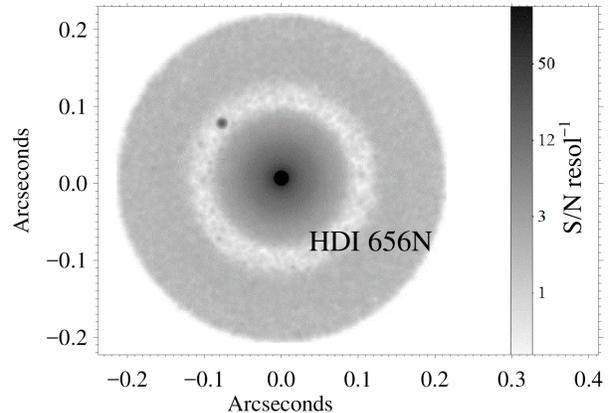

**Figure 7.6.** *Simulated protoplanetary disk (α = 10⁻³; h/r = 0.04) scattered light image and accreting protoplanet (located in the dust gap at the upper left) at the distance of the Taurus star-forming region. The 1 Jupiter mass protoplanet is located 30 AU from the host proto-K star, and is accreting with 1/100th the Hα luminosity of more massive, large-separation protoplanets (Zhou et al. 2014). In a 9000 s exposure with the HDI narrow-band Hα filter, the protoplanet is detected at S/N~20, the outer disk is detected at S/N between 1–2 per resolution element (easily binned to increase contrast), and the inner disk is detected at S/N between 2–12. Disk simulation courtesy of R. Dong (Steward Observatory/University of Arizona), based on work published in Dong & Fung (2017).*

Neptune-mass planets may be too faint to be directly detected but can dynamically carve detectable narrow (~1–10 AU) gaps in circumstellar disks. High-resolution scattered light images probe the population of small (sub-micron) grains that are coupled to the gas and, as such, can trace dynamical perturbations induced by planets. The narrow gap at ~80 AU in the nearby disk of TW Hya imaged with HST is a candidate for a sub-Jovian carved gap (Debes et al. 2013). JWST will detect similar gaps at large radial distances (> 30 AU), but LUVOIR's observational capability is unique for the study of planet forming regions inside ~10 AU while still capturing structures at large distances thanks to its field of view.

While star-light suppression techniques (e.g., coronagraph, angular differential imaging) will be necessary for all instruments to detect exoplanets at ~1 AU, only LUVOIR can extend these studies into the UV, fully probe the accretion component, and help determine the properties of forming planets at Solar System scales. By comparing where exoplanets form in disks and their properties at birth with the location and properties of exoplanets around Gyr-old stars, we will be able to assess which processes shape planetary architectures and constrain the luminosity evolution of giant planets. In addition, we can test if the pile-up of single RV giants at ~1 AU results from initial conditions in the disk (e.g., the location of the snowline; Ida & Lin 2008) or rather disk dispersal processes (e.g., migration at the gap created by photoevaporation; Alexander & Pascucci 2012). These studies will clarify where and how planets form, constrain the mechanisms for gap formation and disk dispersal, and provide the initial conditions for the direct exoplanet characterization performed by LUVOIR.

### 7.2.2   Trace the composition, evolution, and dispersal of planet-forming material

UV spectroscopy is a unique tool for observing the molecular gas in the inner disk; the strongest electronic band systems of $H_2$ and CO reside in the 100–170 nm wavelength range (e.g., Herczeg et al. 2002; France et al. 2011). UV fluorescent $H_2$ spectra are sensitive to gas surface densities lower than $10^{-6}$ g cm$^{-2}$, making them an extremely useful probe of remnant gas at r < 10 AU. In cases where mid-IR CO spectra or traditional accretion diagnostics (e.g., H$\alpha$ equivalent widths) suggest that the inner gas disk has dissipated, far-UV $H_2$ observations can offer unambiguous evidence for the presence of a remnant molecular disk (Ingleby et al. 2011; France et al. 2012; Arulanantham et al. 2018).

LUVOIR's multi-object, high-resolution capability (R > 30,000, 4.8 square arcminutes per LUMOS field) enables emission-line surveys and absorption-line studies of high-inclination (i > 60°) disks. Absorption line spectroscopy through high-inclination disks, currently limited to a small number of bright stars (e.g., Roberge et al. 2000, 2001; France et al. 2014), are especially important because the large wavelength coverage of LUVOIR provides access to important molecular species in the UV and NIR, such as $H_2$, CO, OH, $H_2O$, $CO_2$, and $CH_4$. UV absorption line spectroscopy is the only direct observational technique to characterize co-spatial populations of these molecules with $H_2$, offering unique access to absolute abundance and temperature measurements without having to rely on molecular conversion factors or geometry-dependent model results as with emission-line spectroscopy.

Uniform spectral surveys of local star-forming regions are required for a systematic determination of disk abundances (a direct measurement of the initial conditions for planet-formation) and gas disk lifetimes (the timescales for gas envelope accretion and migration of planetary cores through their natal disks). For star-forming regions beyond Taurus-Auriga, the efficiency of these surveys goes up dramatically with the introduction of multi-object spectroscopy, greatly reducing the total observing time and increasing the survey efficiency in a census of star-forming regions at distances less than 1 kpc. Combining these surveys with the detailed characterization of debris disks discussed in **Chapter 4** enables us to trace the physical and chemical evolution of the bulk of the disk mass, constraining the composition of the gas that can be accreted onto the core of giant planets as well as that of the planetesimals that could constitute the bulk of the icy planets.

### 7.2.3   Map protoplanetary material dispersed via disk winds

Recent theoretical work suggests that disk evolution and dispersal are driven by a combination of thermal and MHD disk winds (e.g., Alexander et al. 2014; Gorti et al. 2016; Ercolano & Pascucci 2017). Disk winds can affect all stages of planet formation: from planetesimal formation, by reducing the disk gas-to-dust mass ratio (e.g., Carrera et al. 2017); to the mass of giant planets, by starving gas accretion onto late-forming planet cores (e.g., Shu et al. 1993); to planetary orbits, by dispersing gas from preferential radial distances and halting planet migration (e.g., Alexander & Pascucci 2012; Ercolano & Rosotti 2015). However, computational challenges and poorly constrained input parameters make it difficult to predict how basic disk properties evolve and to ascertain the role of different winds in dispersing protoplanetary material.

On the observational side, high-resolution (R > 30,000) optical and infrared spectroscopy yields growing evidence of slow (<40 km/s) disk winds around Myr-old stars (e.g., Pascucci & Sterzik 2009; Sacco et al. 2012; Natta et al. 2014; Simon et al. 2016). Modeling of the line profiles and line flux ratios suggest that emission arises at disk radii between 0.1 and 10 AU. On similar spatial scales, jets (ejected material moving at ~100 km/s) are also expected to be accelerated and collimated (e.g., Ray et al. 2007). Many of the brightest jet diagnostics are accessible to high-resolution imaging spectroscopy at UV and optical wavelengths (see Frank et al. 2014 for a recent review).

The high spatial resolution and wavelength coverage of LUVOIR are therefore needed to understand the origin of disk winds, the relative roles and interplay of MHD and photoevaporative winds, as well as to clarify the physical mechanism by which jets are launched and collimated. In addition to high-resolution spectroscopy, LUVOIR slitless spectroscopy (i.e., open microshutter array) and narrow-band images in UV and optical forbidden lines will map for the first time the launching region of jets and disk winds, and reveal their interaction. POLLUX spectropolarimetry will complement the science case discussed above by uniquely constraining the strength and topology of stellar magnetic fields during pre-main sequence evolution, with important implications for stellar and disk formation and evolution (e.g., Gómez de Castro et al. 2016).

### 7.2.4   Star and planet formation in Orion

The Orion complex, at a distance of ~400 pc (e.g., Kounkel et al. 2017), includes rich star-forming clusters that are more typical birth environments for stars than nearby low-density star-forming regions such as Taurus. Most stars in the Galaxy, including our own, formed in rich clusters (e.g., Lada & Lada 2003; Adams 2010) that contain massive stars whose high-energy UV and X-ray photons contribute to disperse planet-forming material around young stars (e.g., Johnstone et al. 1998; Clarke & Owen 2015). The Orion complex covers different external UV fields and the critical 1–10 Myr age range over which planet-forming disks disperse. It represents one of the best sites to study planet formation and several of its regions have already a rather complete stellar and disk census (e.g., Fang et al., in prep.).

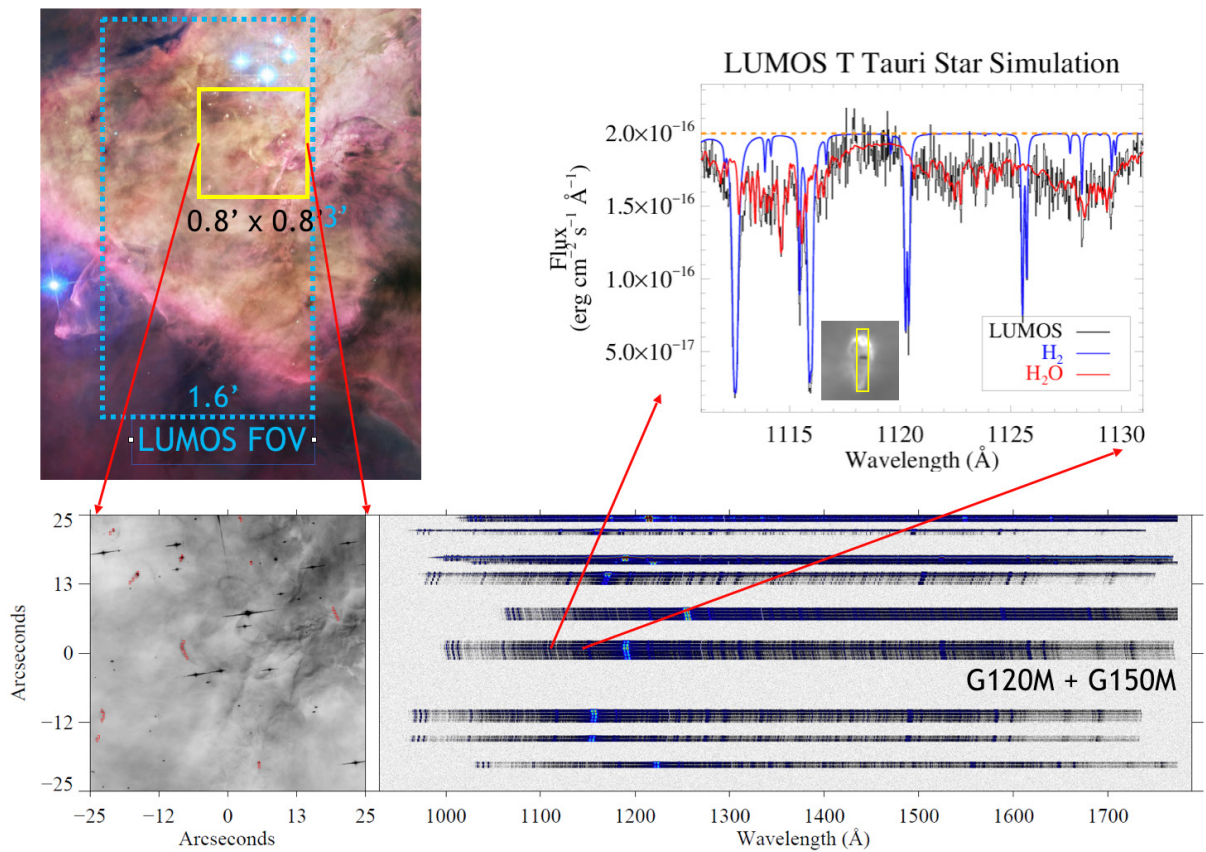

**Figure 7.7.** *Multi-object FUV spectroscopy of protoplanetary disks in the Orion Nebula. Upper left: HST-ACS image of the Orion Nebula (color credit, Robert Gendler), showing the full FoV of the LUMOS spectrograph (France et al. 2017). Lower panels zoom in on a ~0.8 x 0.8 arcmin region showing ~30 protostellar/protoplanetary systems (Bally et al. 1998) and the apertures of the LUMOS microshutter array overplotted (slits are oversized for display). The two-dimensional spectra of the protoplanetary disk and accreting protostar are shown at right. In the upper right, we show a zoom in on the 1111–1132 Å spectral region containing strong lines of H$_2$ and H$_2$O. The combination of spectral coverage, large collecting area, and multiplexing capability make LUVOIR ideal for surveying the composition of the planet-forming environments around young stars.*

LUMOS is ideal for ultraviolet spectral surveys of the Orion complex (**Figure 7.7**). First, LUMOS has 40 times higher sensitivity and 2–3 times higher spectral resolution than HST/COS. Second, it carries a multi-object imaging spectroscopy mode that enables the simultaneous observation of ~tens-to-hundreds of targets (depending on source field density). *Each LUMOS FUV pointing in dense stellar environments like Orion would collect a dataset comparable to the medium-resolution UV spectroscopic archive of*

*protoplanetary disks observations from the entire 28 years of HST observations*. In about an hour, LUMOS will reach S/N = 10 per resolution element (R = 40,000) over a 3' x 1.6' field to F$\lambda$ = F(1100 Å) = 2 x 10$^{-16}$ erg/cm$^2$/s/Å, typical of the continuum flux of young stars in Taurus scaled to the distance of the Orion complex. This means that in about 2–3 hours per field LUMOS will cover the entire FUV and NUV spectral range of Orion stars enabling the detection of the most abundant molecular species in disks,

---

**Program at a Glance - The Environments of Planet Assembly**

**Science goal:** Trace the main molecular carriers of C, H, and O during planet assembly

**Program details:** Multiple pointings per star-forming region to study different ages and UV environments in the Orion complex

**Instrument(s) + configuration(s):** LUMOS high-resolution (R=40,000) multi-object spectroscopy

**Key observation requirements:** S/N of 10 per spectral resolution element. FUV imaging parallels offer complementary information.

---

e.g., $H_2$, CO, $H_2O$, and OH. With pre-defined maps of targets, LUMOS can cover most stars in the Orion Nebula Cluster (~1 Myr, ~24' size), 1980 (~1–2 Myr, ~16' size), σ Ori (~3–5 Myr, ~33' size), λ Ori (4–8 Myr, 49' size), 25 Ori (~7–10 Myr, 33' size) in < 150 hours at FUV and NUV wavelengths. This program would allow us to trace the evolution and dispersal of the main molecular carriers of C, H, and O during planet assembly in the terrestrial and giant-planet forming regions, trace molecular and low-ionization metals from disk winds, and determine the absolute abundance patterns in the disk as a function of age. This survey will reveal how the changing disk environment impacts the size, location, and composition of planets that form around other stars.

## 7.3  Signature science case #3: The rise of the periodic table

### 7.3.1  The first stars and the first metals

The nucleosynthetic signatures of the first stars and supernovae are imprinted in the compositions of the most metal-poor stars found today. No first-generation (Pop III) stars are known at present, but dozens of candidate second-generation stars are known. These stars have iron abundances less than $10^{-4.5}$ times the solar abundance (i.e., [Fe/H] < -4.5), or are iron-poor but highly enhanced in carbon. Hundreds more are expected to be found among ongoing and future surveys (e.g., LAMOST, LSST). When compared with predicted model yields, these abundance patterns reveal the nature of the elusive Pop III stars, providing the only direct tests of the evolution and end states of individual Pop III stars (e.g., Frebel & Norris 2015). Their locations and kinematics reveal the nature of the environments and the epochs when they formed and released the first metals into the Universe.

Only a few tens of absorption lines are commonly found in the optical spectra of these second-generation stars, so only ~5–10 elements are regularly detected. Many others (Be, B, Si, P, S, Sc, V, Cr, Mn, Co, Ni, and Zn) are expected to be present but are rarely detected, and the upper limits derived from their optical non-detections are often uninformative. The UV part of the spectrum is an unexplored window that would allow all of these elements to be detected if present in the most metal-poor stars known (**Figure 7.8**). Key lines include those of boron (B; atomic number Z=5; BI lines at 2088, 2089 Å), phosphorus (P; Z=15; PI lines at 2135, 2136 Å), sulphur (S; Z=16; SI lines at 1807, 1820, 1826 Å), chromium (Cr; Z=24; CrII lines at 2055, 2061, 2065 Å), and zinc (Zn; Z=30; ZnII lines at 2025, 2062 Å) (e.g., Roederer et al. 2016).

A large-aperture telescope with a high-resolution UV spectrometer enables observations of these elements in all known

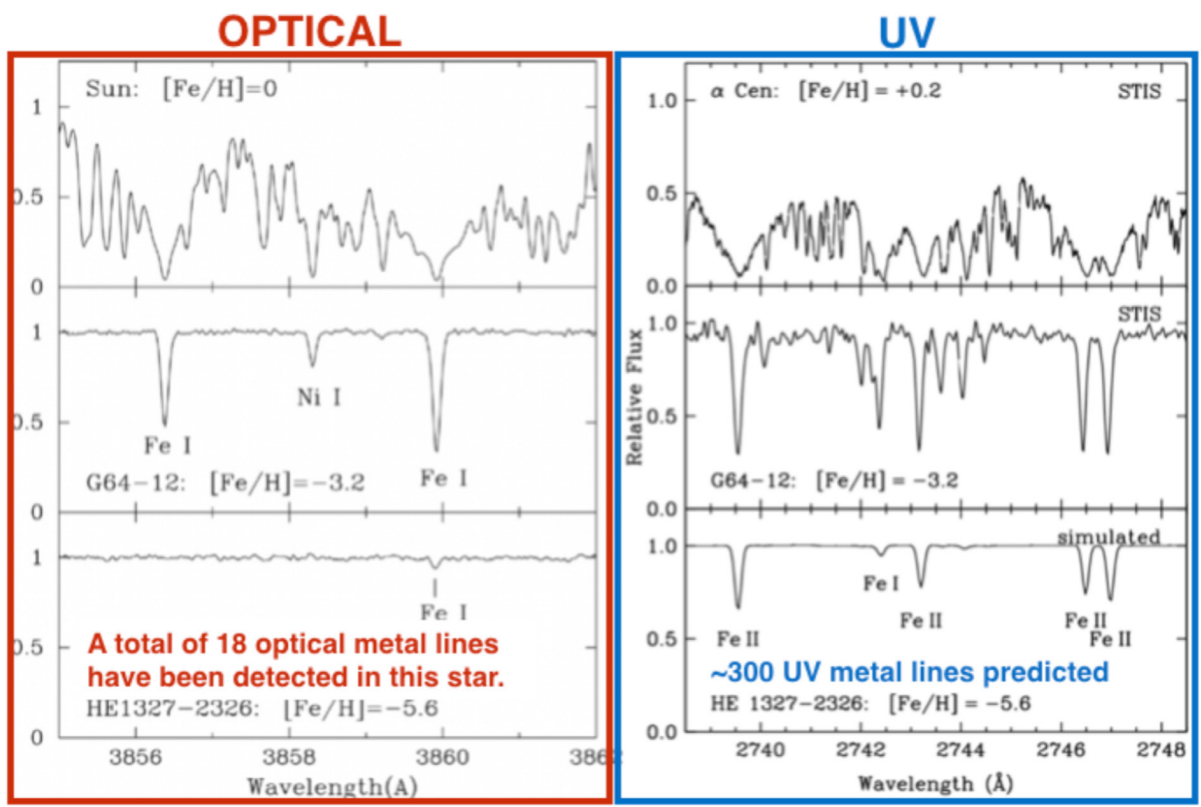

**Figure 7.8.** *Comparison of the optical and UV spectral domains of a normal G dwarf star at solar metallicity (top row), a typical metal poor G subgiant (middle row), and one of the candidate second generation stars (bottom row). The red panel is from Aoki et al. (2006). LUVOIR could observe the UV spectrum of every candidate second-generation star in the field, and many more in nearby dwarf galaxies, obtaining data similar to the lower right panel across a much wider spectral range.*

second-generation stars in the halo field, and many more found in nearby dwarf galaxies. Presently, with COS on HST, only one or two of the brightest candidate second-generation stars can be observed, but even in these cases the data quality is insufficient to detect all elements that could, in principle, be detected. With LUVOIR, however, high-quality UV spectra (R~30,000; S/N~80; 1800 < λ < 3100 Å) could be obtained for F or G-type dwarfs or subgiant stars at V~15.5 mag (9-m) or V~16.6 mag (15-m) in ~1 hour each. This advance would enable us to increase the stellar sample sizes by about two orders of magnitude, revolutionizing our understanding of the first stars, the first supernovae, and the first metals in the

Universe. In other words, LUVOIR+LUMOS would enable the acquisition of high-quality UV spectra of the vast majority of stars whose optical spectra can be observed today from the ground.

### 7.3.2 The origin of elements heavier than iron

The elements heavier than iron, which have been detected in the ancient stars of the Galactic halo, in the ISM, dust grains, meteorites, and on Earth, are formed by neutron-capture reactions (e.g., Sneden, Cowan, & Gallino 2008). Relatively high neutron densities (~$10^{22}$–$10^{28}$ cm$^{-3}$) lead to heavy-element nucleosynthesis via the rapid neutron-capture process (r-process) in supernovae or neutron star mergers. Lower neutron densities ($\sim 10^7$–$10^{10}$ cm$^{-3}$) lead to nucleosynthesis via the slow neutron-capture process (s-process) in AGB stars or the late evolutionary stages of massive stars. Neutron densities intermediate between these two extremes ($\sim 10^{15}$ cm$^{-3}$) lead to nucleosynthesis via the intermediate neutron-capture process (i-process), which may occur in a variety of sites including super-AGB stars, post-AGB stars, He-core and He-shell flashes in low- metallicity low-mass stars, and massive stars.

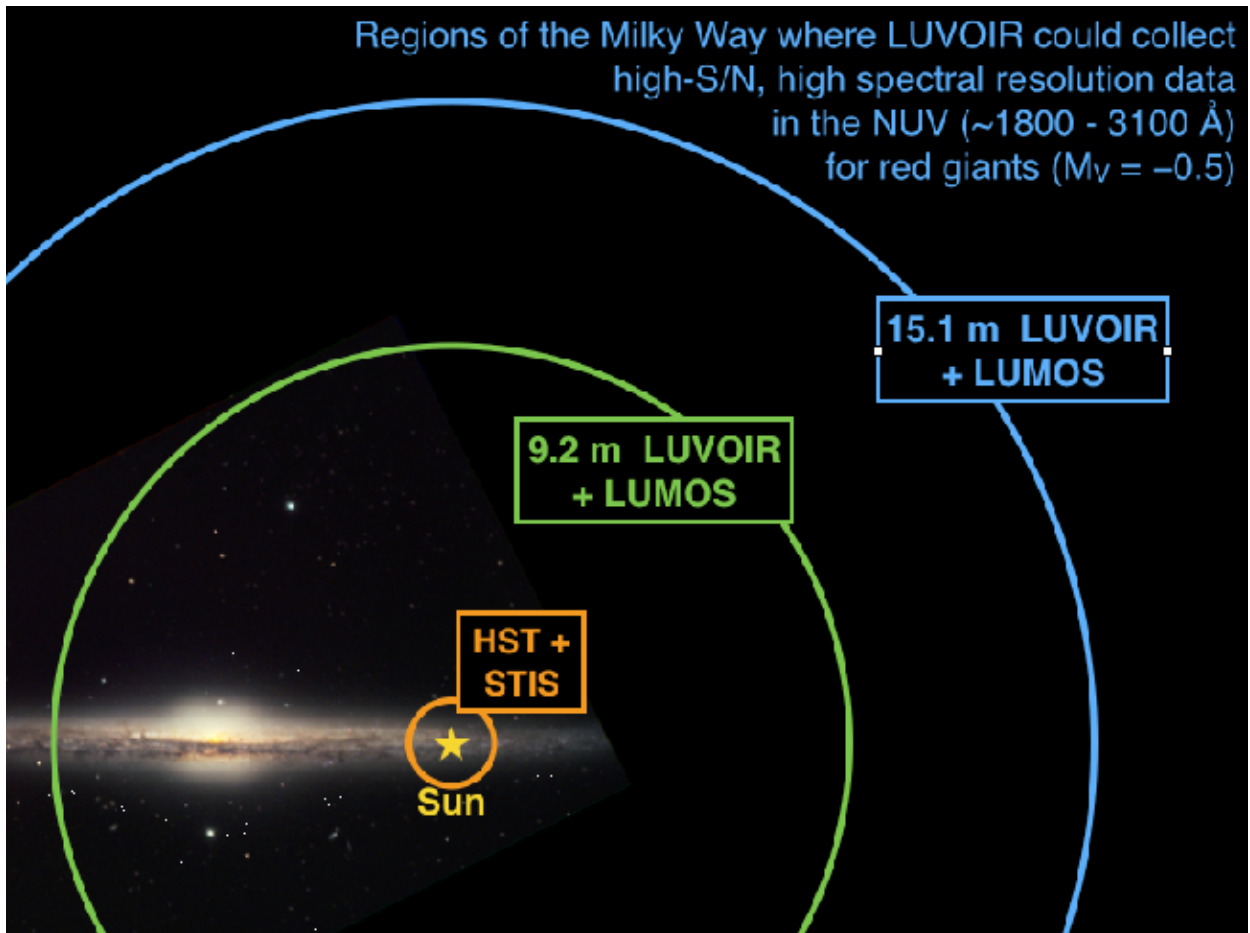

**Figure 7.9.** *Comparison of distances at which LUVOIR can obtain high-precision chemical abundances in metal-poor giants to examine the origins of the first heavy elements. These elements are best studied in the UV; many of these stars are so metal poor that the key lines in the optical are intrinsically too weak, but stronger UV lines of Fe, Mg, etc. can still be detected. These are challenging, expensive observations for Hubble since the stars are ancient and not bright in the UV. LUMOS will push these limits to ~20 kpc, a significant fraction of the total mass of the galactic halo.*

Optical spectra obtained from the ground can reveal the nature of the enrichment (e.g., the r-process), but key elements useful to discriminate between models have no absorption lines in the optical domain. High-resolution UV spectroscopy enables a 40% improvement, compared to optical/near-IR spectra, in the number of elements ($\sim 15$ to 20) that can be detected in the atmospheres of late-type (FGK) stars that retain the chemical signatures of nucleosynthesis in earlier generations of stars. These include elements like germanium (Ge; atomic number Z=32;

---

**Program at a glance - New Insights on Nucleosynthesis**

**Science goal:** Determine elemental abundances in very metal poor stars to constrain Pop III nucleosynthesis. Determine elemental abundances in late type stars to constrain R-, S-, and I-process nucleosynthesis.

**Program details:** High-resolution FUV and NUV spectroscopy of stars in the Milky Way, predominantly in the outer Milky Way halo.

**Instrument(s) + configuration(s):** LUMOS high-resolution (R=40,000) multi-object spectroscopy. G180M and G300M gratings.

**Key observation requirements:** S/N of >80 per spectral resolution element.

---

GeI line at 3039 Å), selenium (Se; Z =34; SeI line at 2074 Å), cadmium (Cd; Z=48; CdI line at 2288 Å), tellurium (Te; Z=52; TeI line at 2385 Å), platinum (Pt; Z=78; PtI line at 2659 Å), and mercury (Hg; Z=80; HgII line at 1942 Å), among others (e.g., Roederer & Lawler 2012). These abundances provide new, critical constraints on stellar nucleosynthesis mechanisms like the r-, s-, and i-processes.

A large-aperture telescope with a high-resolution UV spectrometer would enable observations of these elements in samples of stars best suited to revealing the complex physics that drives the nucleosynthesis of the heaviest elements. Presently, with STIS or COS on Hubble, only the handful of brightest stars in the solar neighborhood can be observed, and these are sub-optimal representatives of stars reflecting dominant contributions from the r-, s-, or i-processes. With LUVOIR, high-quality UV spectra (R~30,000; S/N~80; 1800 < λ < 3100 Å) could be obtained for metal-poor red giants, which typically present the richest line pattern for analysis, at V~15.0 (9-m) or V~16.1 (15-m) in 10 hours, mimicking an observing mode and strategy routinely used with HST/STIS (e.g., Roederer et al. 2012). Examples of all kinds of enrichment patterns are found within this magnitude limit, enabling for the first time observations of the most scientifically advantageous stars for the task. In other words, LUVOIR+LUMOS would enable the acquisition of high-quality UV spectra of the vast majority of stars whose optical spectra can be observed today from the ground.

**Table 7.1.** *Summary science traceability matrix for Chapter 7*

| Scientific Measurement Requirements | | | Instrument Requirements | | |
|---|---|---|---|---|---|
| Objectives | Measurement | Observations | Instrument | Property | Value |
| Constrain massive star formation and determine the stellar IMF | Imaging and spectra of individual stars and star clusters | Imaging S/N=10-300; Spectroscopy S/N > 10, R>3000 | HDI LUMOS | UVIS imaging, 0.007" resolution FUV, NUV gratings, MOS | UVBRI filters G120M, G150M, G180M, G300M |
| Detect and characterize the composition of planet-forming material | Spectroscopy of molecules in protoplanetary disks | Spectroscopy S/N > 10, R=40,000 | LUMOS | FUV, NUV gratings, MOS | G120M, G150M, G300M |
| Follow the first metals created by stars | Spectroscopy of extremely metal poor stars in the Milky Way halo | Spectroscopy S/N~80, R=30,000 | LUMOS | FUV, NUV gratings, MOS | G180M, G300M |
| Trace the history of R-,S-, I- process material | Spectroscopy of late-type stars in the Milky Way halo | Spectroscopy S/N ~80, R=30,000 | LUMOS | FUV, NUV gratings, MOS | G180M, G300M |

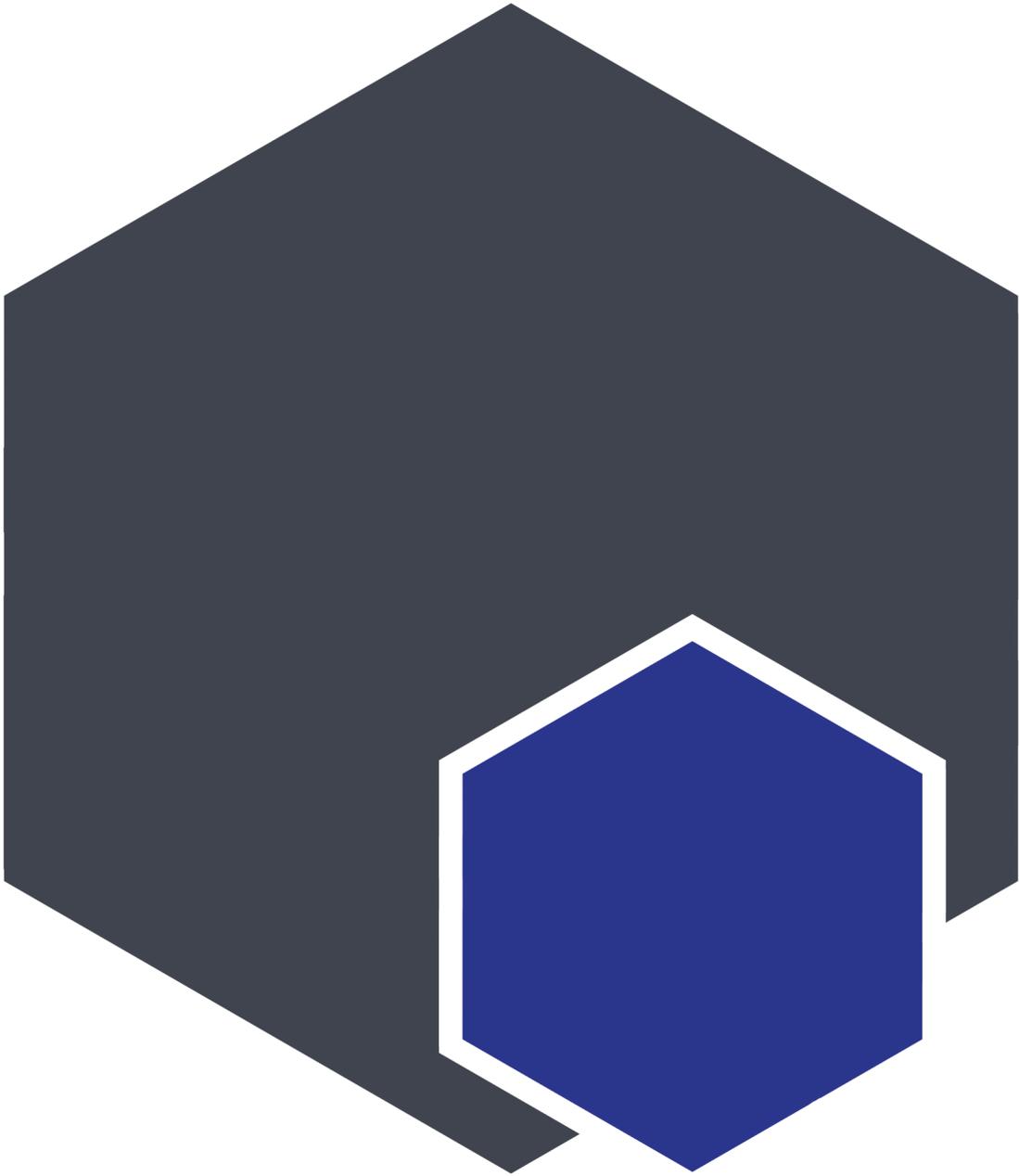

The LUVOIR architectures

# 8   The LUVOIR architectures

LUVOIR's compelling science objectives define a set of high-level mission capabilities: sensitivity, resolution, flexibility, high-contrast imaging, and mission duration. The LUVOIR Study team has used these capabilities to define a set of fundamental performance goals that include:

- a large (8–15 meter) aperture;
- broad wavelength sensitivity from the far-ultraviolet to the near-infrared;
- a suite of instruments with imaging, spectroscopic, and high-contrast capabilities necessary to achieve the science observations discussed in this report; and
- a potentially multi-decade mission lifetime enabled through on-orbit serviceability and upgrading of instruments.

The LUVOIR Study Team is working to develop two classes of feasible and executable mission architectures that meet these goals at either end of the aperture range: LUVOIR-A is a 15-meter segmented-aperture telescope, while LUVOIR-B is an 8-meter segmented-aperture telescope. Both architectures will have a suite of modular instruments and a total facility wavelength range from 100 nm to 2.5 μm. Four instrument concepts are being developed as part of these architecture studies:

1. The Extreme Coronagraph for Living Planetary Systems (ECLIPS), a near-UV / optical / near-IR coronagraph capable of $10^{-10}$ contrast at inner working angles as small as 2 λ/D.

2. The LUVOIR UV Multi-Object Spectrograph (LUMOS), a low- and medium-resolution multi-object imaging spectrograph spanning the far-UV (~100–200 nm) to the near-UV (~200–400 nm), with a separate far-UV imaging channel.

3. The High Definition Imager (HDI), a near-UV / optical / near-IR imager with a wide field-of-view and high spatial resolution.

4. POLLUX, a high-resolution UV spectropolarimeter, contributed by a European consortium under the leadership of the Centre National d'Etudes Spatiales (CNES).

In this chapter we will describe the overall design philosophy, as well as mission-level aspects of the architectures. The following chapter goes into greater detail on the science payload, including the optical telescope element and science instruments. The POLLUX instrument, and its science case, are described in **Chapter 10**.

## 8.1   Study philosophy

The goal of the LUVOIR design study is to develop a mission concept that achieves the science objectives described in this report, subject to traditional engineering and programmatic constraints of mass, volume, cost, schedule, and technical risk. Ultimately, the 2020 Decadal Survey Committee will evaluate LUVOIR and the other mission concepts on the following merits:

- Is the science compelling?
- Is the mission executable and feasible?
- Does the science yield justify the cost and risk?

The study team faces several challenges when trying to plan and scope a mission that may not launch for two decades. At the time the Decadal Committee performs their evaluation, many of the constraints listed above will be unknown.

> LUVOIR's two design concepts demonstrate a scalable mission architecture with a range of science yield, cost, and technical risks that can be adapted to an uncertain future.

For example, mass and volume are limited by the capability of the launch vehicle. However, the fleet of available launch vehicles is currently in flux. Existing heavy-lift vehicles like the Delta IV Heavy are being phased out in favor of new vehicles like Space X's Falcon Heavy, Blue Origin's New Glenn, and NASA's own Space Launch System. While it is certain that these vehicles will be in operation within the next decade or so, what is less certain are the limits of their capability, the cadence of their launch schedule, and the cost of their service.[1]

Another unknown constraint is the mission cost. An independent cost estimate of all large mission concepts will be completed as part of the Decadal review process; each is likely to be a multi-billion-dollar large-scale mission, similar in scope to HST, the James Webb Space Telescope (JWST), and the Wide-Field Infrared Survey Telescope (WFIRST). What will not be well understood is how those costs will fit into the future budgetary environment at NASA. While the science cases may fully justify the cost, the available budget may require long and protracted development schedules, potentially reducing the value and increasing the risk to the missions.

Finally, the Decadal Committee will be performing their evaluation just as JWST is nearing launch, WFIRST is being developed, and a new generation of ground-based observatories preparing to come online. It is impossible to predict what discoveries these facilities will make that may change the scientific landscape in the next two decades.

In response to these uncertainties, the LUVOIR STDT has decided to put forth for evaluation two mission architectures that bracket a range of capability, cost, and risk. LUVOIR-A is the larger of the two concepts with a 15-m primary mirror diameter that maximizes science yield while accepting moderate technical and programmatic risk. LUVOIR-A is the largest, most capable observatory that can be deployed from an anticipated 8.4-m diameter fairing, launched with NASA's Space Launch System Block 2 heavy lift vehicle. Conversely, LUVOIR-B represents a more conservative approach where science yield is balanced with technical and programmatic risks. LUVOIR-B will be designed to fit within an industry standard 5-m-class fairing.

At the time of writing this report, the study team has completed an initial design iteration for the complete LUVOIR-A concept and has begun exploring a number of design alternatives in order to maximize system performance. The team has also begun to evaluate several designs for LUVOIR-B. To be clear, both mission concepts will continue to evolve as each architecture gains definition. Over the course of the next year, detailed designs for both concepts will be finalized for evaluation by the Decadal Committee. Both of these final point designs will be presented in the Final Report; for this Interim Report we focus primarily on LUVOIR-A and its design evolution.

---

1 See **Appendix D** for more details on current and anticipated launch vehicles.

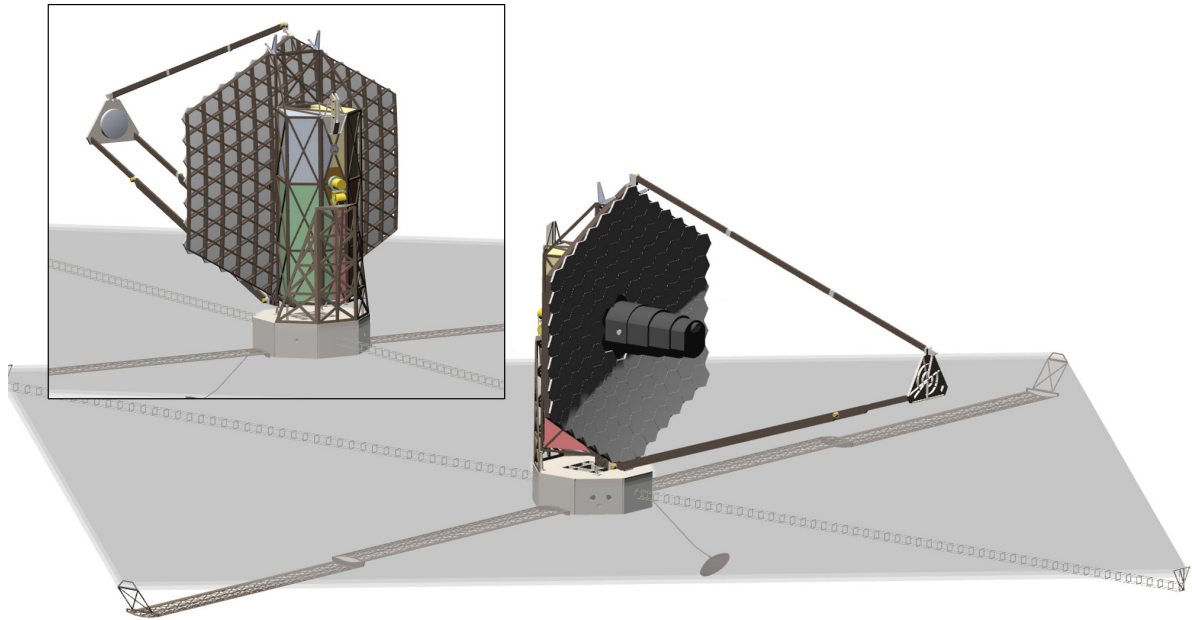

**Figure 8.1.** *The LUVOIR Observatory. The front view shows the primary mirror, secondary mirror support structure, and the deployable aft-optics support structure. The sunshield is rendered transparent here in order to show the spacecraft below. The inset shows a rear view, highlighting the backplane support frame holding the four instruments, depicted here as simple colored volumes: LUMOS (green, lower left), HDI (red, lower right), ECLIPS (blue, upper left), and POLLUX (yellow, upper right). The two-axis gimbal is also visible near the middle of the backplane support frame.*

## 8.2    Architecture A

### 8.2.1    Design overview

An initial concept for LUVOIR-A, shown in **Figure 8.1**, builds upon the legacies of HST, JWST, and WFIRST. Similar to HST, LUVOIR covers a bandpass spanning the UV to the near-IR, has a suite of imagers and spectrographs, and is designed to take advantage of potential future servicing capabilities that are currently being developed by NASA and other agencies. Like JWST, LUVOIR has a segmented primary mirror that deploys from a stowed configuration, along with the secondary mirror support structure, the sunshield, and a number of other subsystems. Finally, like WFIRST, LUVOIR has a high-contrast coronagraph instrument, capable of directly imaging and spectroscopically characterizing exoplanets. In this sense, LUVOIR builds upon the technical legacy of the great space observatories that will have preceded it.

LUVOIR-A is divided into three primary segments: The Flight Segment consisting of the observatory itself, the Ground Segment consisting of ground stations and mission and science operations centers, and the Launch Vehicle.

#### 8.2.1.1 Flight Segment

The Flight Segment, or observatory, is shown in **Figure 8.2** and is divided into the payload and the spacecraft. The payload consists of the optical telescope element, the backplane support frame, the Vibration Isolation and Precision Pointing System (VIPPS), the gimbal system, and four serviceable instrument modules. The spacecraft consists of the payload interface tower, the sunshield, and the spacecraft bus. The payload is articulated with respect to the spacecraft using a two-axis gimbal system. Repointing the observatory involves rolling the entire observatory about the sun-earth axis through 360°, while simultaneously pitching the payload with the gimbal from 0° to 90° (0° corresponds to the telescope boresight normal to the sunshield, while 90° corresponds to the telescope boresight parallel to the sunshield). With these two degrees of freedom, the payload can point anywhere in the anti-sun hemisphere. An additional second axis on the gimbal allows just the payload to roll about the telescope boresight from 0° to 90°. The boresight roll allows targets to be aligned to instrument-specific apertures, and also enables point-spread function roll subtraction. Details on the specific elements of the payload are provided in **Chapter 9**.

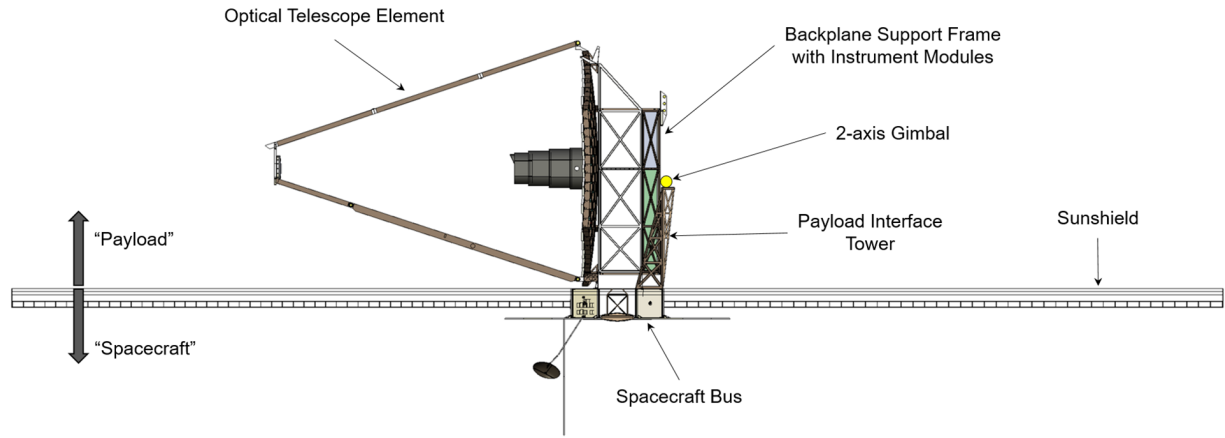

**Figure 8.2.** *The LUVOIR Flight Segment, with identified subsystems. Everything above the sunshield (with the exception of the payload interface tower) is considered part of the payload. The payload interface tower, the sunshield, and everything below the sunshield are collectively considered the spacecraft. The observatory is shown in the Gimbal-90° orientation.*

The payload and spacecraft are physically separated from each other via the Vibration Isolation and Precision Pointing System (VIPPS), a non-contact interface that is controlled in six degrees-of-freedom by voicecoil actuators. The VIPPS is located between the gimbal and the backplane support frame and is discussed in more detail in **Section 8.2.2.1**.

The spacecraft, shown in **Figure 8.3**, must provide basic services to the payload: power, communications, attitude control, command and data handling, propulsion, and structural support during launch. The spacecraft also thermally isolates the payload from solar heat loads via the sunshield. Each of these subsystems are designed to not only meet the minimum 5-year mission needs, but also anticipate future evolution of the LUVOIR observatory through serviceability.

### 8.2.1.1.1   Spacecraft structure

The spacecraft bus is octagonal in shape. A central bay holds the propellant tanks, which are refuelable. The central bay is surrounded by the primary spacecraft structure that supports the mass of the payload and transfers loads to the launch vehicle during launch and ascent. Additional spacecraft subsystems are supported by the secondary spacecraft structure around the perimeter of the primary structure and are accessible for serviceability.

### 8.2.1.1.2   Sunshield

A central component of the LUVOIR thermal management architecture is the deployable sunshield. The sunshield isolates the payload from solar thermal loads and provides a cold environment in which the payload can be thermally stabilized with active heaters, and in which system radiators have sufficient cold-sink temperatures.

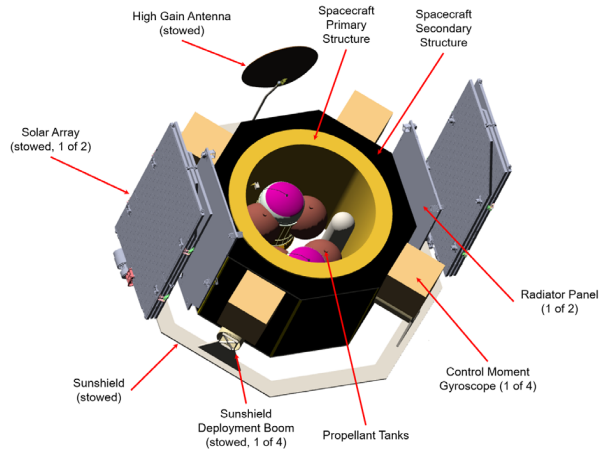

**Figure 8.3.** *The LUVOIR-A spacecraft in its stowed configuration. To deploy, the two solar arrays would first fold down to a position perpendicular to their current orientation, and the individual panels would further unfold. The high gain antenna assembly would also fold down to a nominal position below the spacecraft. Four deployable booms would then pull the sunshield out from its folded configuration in the four cardinal directions. The propellant tanks are refuelable from below the spacecraft. The four control moment gyroscopes are mounted to the exterior of the spacecraft for accessibility during potential servicing.*

Nominally, the sunshield remains in a fixed attitude normal to the Sun-observatory axis, maintaining a constant thermal load on the sunward side of the sunshield as the payload slews to different targets on the leeward side of the sunshield. The payload always sees a constant thermal environment, regardless of its pointing attitude. Two science observations necessitate exceptions to this nominal attitude and require the entire observatory to pitch toward the sun (while still keeping the payload in shadow). Exoplanet revisit observations must be scheduled at critical time intervals to maximize observational completeness and may require access to sunward portions of the sky. Also, observations of specific solar system bodies, such as Venus and comets, require pointing the observatory up to 45° from the Sun-Earth axis. **Figure 8.4** shows several different pointing scenarios for the observatory.

Although the sunshield appears similar in concept and design to that of JWST, it differs in two important aspects. First, it is larger. While JWST's sunshield is tennis-court sized, the LUVOIR-A sunshield is closer in size to a football field, spanning nearly 80-m tip-to-tip. And whereas JWST's sunshield is an extremely complex system of 5 thin layers that need to be precision deployed to tight tolerances on the angle and separation between the layers, LUVOIR's sunshield is simpler. A minimum of two sheets of single-layer insulation are needed to meet the thermal performance requirements, although three layers will likely be needed for redundancy against micrometeroid strikes. The spacing and angle between these layers is not critical, so long as they do not touch.

The manner in which the sunshield is deployed also differs between LUVOIR and JWST. JWST uses two rigid pallets to hold the sunshield system. These pallets are stowed up around the telescope and instruments, cocooning the payload during launch and ascent. After launch, the pallets fold down, and port and starboard deployable booms pull the sunshield layers out from the pallet, before spreaders and tensioners fine-position the five sunshield layers. In contrast, LUVOIR will package the entire sunshield at the base of the spacecraft. Four deployable booms will pull the folded sunshield out in each of four directions: fore, aft, port, and starboard. Details of the deployment are still being worked by the Study Team, but we anticipate several advantages over JWST, as well as several new challenges. Advantages include a smaller relative mass and volume for the stowed sunshield system, the use of high TRL deployment mechanisms, such as AstroMasts®, and overall lower risk to the deployment due to fewer mechanisms. New challenges include venting of the stowed sunshield during ascent, and the fact that the payload will be exposed once the fairing is jettisoned. We expect to have more detail regarding the sunshield deployment advantages and risks available in the final study report.

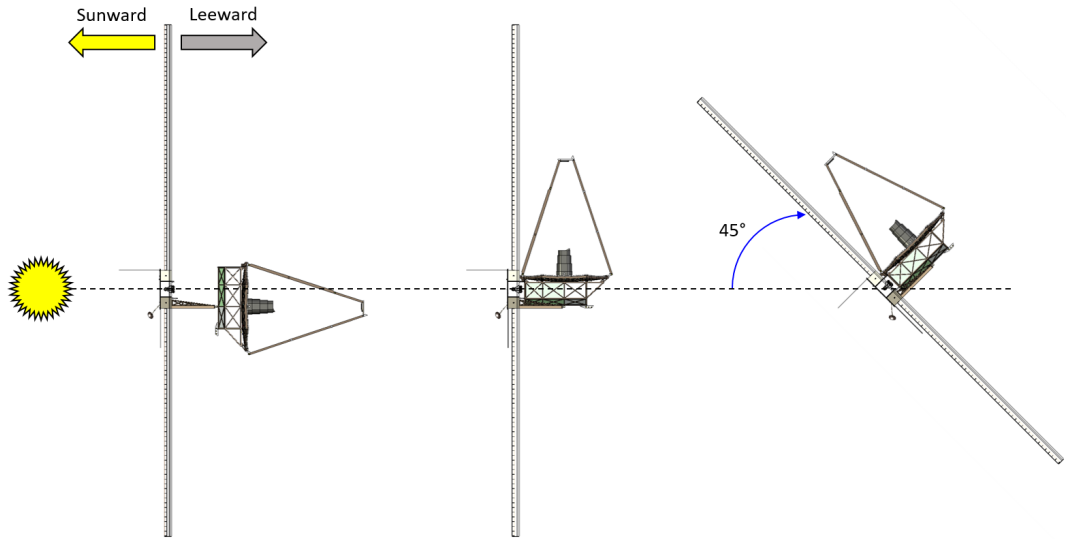

**Figure 8.4.** *Example payload pointing attitudes. Nominal science observations maintain the sunshield normal to the Sun-observatory axis, as shown in the left two figures. The thermal load on the sunward side of the sunshield remains constant generating a thermally-stable environment on the leeward side, regardless of payload pointing. Exoplanet revisit observations and solar system targets of opportunity require the entire observatory to pitch towards the sun to a 45° angle with the Sun-observatory axis.*

### 8.2.1.1.3   Attitude control system (ACS)

The spacecraft is three-axis stabilized and inertially fixed, with the spacecraft axial vector parallel to the sun-spacecraft axis. The ACS is responsible for countering external disturbances (primarily solar pressure torques), as well as changing the yaw axis of the ob-servatory and reacting against gimbal pitch changes during payload retargeting maneuvers. The ACS is also responsible for pitching the entire observatory sunward for specific observations (see **Section 8.2.1.1.2**).

The ACS sensor suite serves two functions: star trackers, gyroscopes, accelerometers, and coarse sun sensors determine the attitude and inertial position of the spacecraft relative to the sun, while proximity sensors on the VIPPS determine the relative attitude between the spacecraft and the payload. ACS actuators, consisting of four control moment gyroscopes (CMGs) and sixteen 5-lb. thrusters respond to the sensor suite to maintain the absolute attitude and inertial position of the spacecraft. The four CMGs are arranged to provide full three-axis stabilization with redundancy. The thrusters are arranged in two sets: the first is used during momentum dump maneuvers, the second is used for stationkeeping and orbit maintenance.

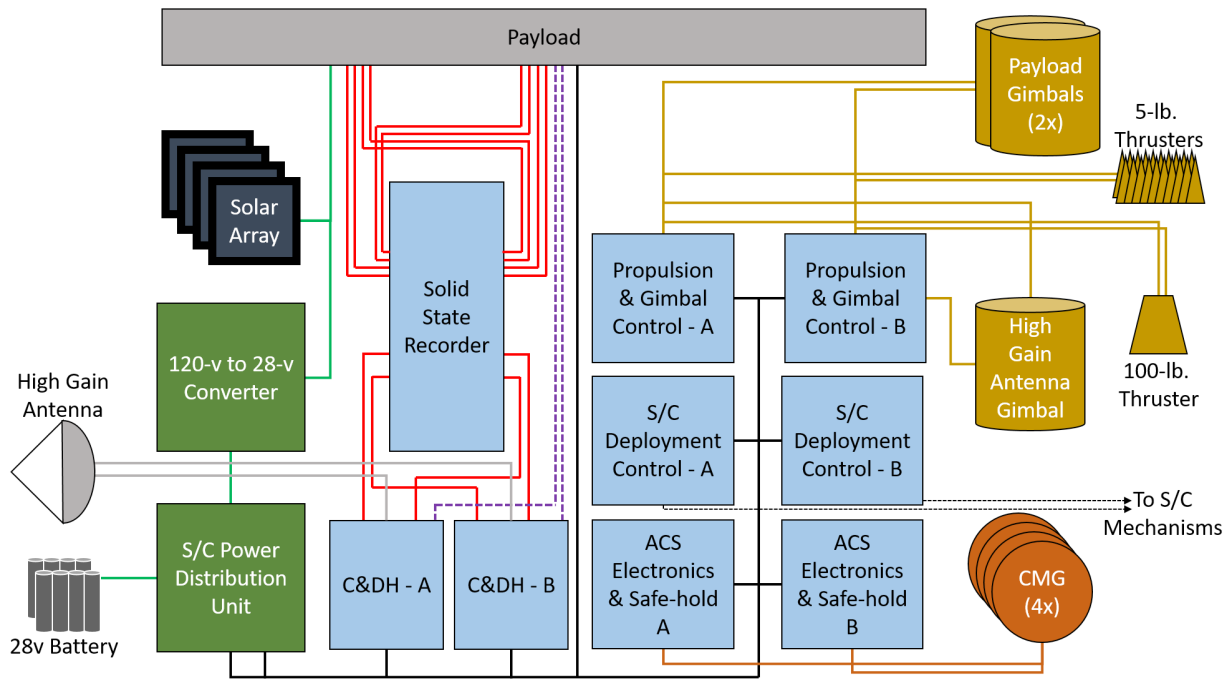

**Figure 8.5.** *Block diagram of the spacecraft (S/C) avionics systems and interface to the payload. Solid black lines represent a MIL-STD-1553 bus, solid red lines represent individual SpaceWire links, dashed purple lines represent timing (1 pps) links, and solid green lines represent power cables. ACS, propulsion, communication, deployment mechanisms, and gimbal control electronics are connected to their respective subsystems via appropriate links (RS-422, 28-v, 1553, etc.). The solid-state recorder provides a buffer for a single downlink's-worth of data and is connected to the payload via redundant and cross-strapped SpaceWire. The solar array and electrical power subsystem supplies 120-v power directly to the payload, and 28-v power (via a down converter) to the S/C. A 28-v battery provides power during launch and ascent.*

### 8.2.1.1.4    Command & data handling (C&DH)

The C&DH system consists of a solid-state recorder, a board-level computer for command and control, and dedicated controller boards for spacecraft deployment and gimbal mechanisms. To ensure the payload never sees direct sunlight, a separate ACS safe-hold processor takes over in the event of a C&DH system failure to maintain a safe attitude. Each instrument contains enough internal storage for two day's-worth of data. Therefore, the solid-state recorder on the spacecraft only needs to act as a buffer for each downlink and is sized to hold only a single downlink's worth of data (~10 Tb, including margin). **Figure 8.5** shows a block diagram of the spacecraft avionics and their interface to spacecraft subsystems and payload.

### 8.2.1.1.5    Communications

The communications system uses Ka-band for data downlink and S-band for telemetry and housekeeping. A 1.8-m high gain antenna is used to connect with ground stations. Commands and telemetry during launch and ascent will use S-band omni antennas. A link budget is provided in **Table 8.1**.

### 8.2.1.1.6    Electrical power systems

Approximately 105 m$^2$ of solar panels provide more than 31 kW of power needed by the LUVOIR observatory, allocated by system in

**Table 8.1.** *Link budget for LUVOIR science data downlink.*

| LUVOIR Science Data Downlink @ 26,500MHz to White Sands, New Mexico, 20° elevation | | | |
|---|---|---|---|
| Link Type: | Space-Ground | 1/2 Angle Half-Power Beam Width (°): | 0.22 |
| Science Type: | Space Science | Polarization: | Circular |
| Modulation Scheme: | Quadrature Phase-Shifting Key (QPSK) | Ground Station Latitude: | 32.54 |
| Coding: | Low-Density Parity Check Rate 1/2 | Ground Station Longitude: | -106.61 |
| Bit Error Rate: | $10^{-6}$ | Ground Station Altitude: | 1.45 |

| | | |
|---|---|---|
| Transmit Frequency | 26,500 | MHz |
| Transmit Power | 20 | dBW |
| Antenna Diameter | 1.8 | m |
| Antenna Efficiency | 55 | % |
| Antenna Gain | 51.38 | dBi |
| Spacecraft Passive Loss | 3 | dB |
| Spacecraft Pointing Loss | 0.5 | dB |
| Transmitter Equivalent Isotropically Radiated Power (EIRP) | 67.88 | dBWi |
| Slant Range | 1,800,000 | km |
| Min. Elevation Angle | 20 | deg |
| Total Propagation Loss Effects | 3.025 | dB |
| Power Received | -181.256 | dB |
| Ground Station G/T | 46.9 | dB/K (18.3m antenna) |
| Received Carrier to Noise Density (C/No) | 94.24 | dB |
| Modulation Loss | 0 | dB |
| Information (Data) Rate | 430 | Mbps |
| Total Information (Data) Rate | 86.33 | dB |
| Differential Encoding/Decoding Loss | 0 | dB |
| User Constraint Loss | 0 | dB |
| Received Eb/No | 7.91 | dB |
| Implementation Loss | 3 | dB |
| Required Eb/No At Decoder | 1.89 | dB (LDPC Rate 1/2 @ 10^-6) |
| Margin | 3.02 | dB |

**Table 8.2**. The solar array is divided into two deployable wings with three panels each, and is located at the base of the spacecraft, below the sunshield. A 24 amp-hour battery is included for providing power during launch and ascent. The power system provides 28-v to spacecraft subsystems, and 120-v directly to the payload. A 120-v payload bus was chosen primarily to minimize the power harness across the VIPPS interface. The higher voltage allows for fewer, thinner conductors, and therefore more flexible cables, to bridge the non-contact interface.

LUVOIR adopts a Class-A mission risk posture throughout the observatory (both the spacecraft and the payload). All electrical system boards are side-A / side-B or box-level redundant, all mechanisms have dual-windings, and all electrical power and data interfaces are either cross-strapped, redundant, or both.

**Table 8.2.** *Power Budget for the LUVOIR-A Observatory. Current Best Estimate (CBE) values are grass-roots estimates based on instrument designs. Maximum Expected Values (MEV) include a 40% growth allowance that accounts for the early-stage nature of the designs. Maximum Permissible Values (MPV) include an additional 25% margin as required by NASA Goddard Space Flight Center GOLD Rules. We note that the ECLIPS instrument includes the Control System Processor, which is responsible for performing all closed-loop sensing and control for the payload, and therefore requires more power than the other instruments.*

| System / Subsystem | CBE Power (W) | Growth Allowance | MEV Power (W) | GOLD Rules Margin | MPV Power (W) |
|---|---|---|---|---|---|
| Optical Telescope Element (OTE) | 3,000 | 40% | 5,000 | 25% | 6,667 |
| Backplane Support Frame (BSF) | 2,650 | 40% | 4,417 | 25% | 5,889 |
| ECLIPS | 1,640 | 40% | 2,733 | 25% | 3,644 |
| HDI | 480 | 40% | 800 | 25% | 1,067 |
| LUMOS | 650 | 40% | 1,083 | 25% | 1,444 |
| POLLUX | 550 | 40% | 917 | 25% | 1,222 |
| Spacecraft | 5,260 | 40% | 8,767 | 25% | 11,689 |
| Total: | 14,230 | - | 23,717 | - | 31,622 |

### 8.2.1.1.7   Propulsion

A single 100-lb. thruster is used for the first mid-course correction maneuver, which occurs ~12 hours after launch and corrects the trajectory for variability in the launch vehicle performance. A set of four 5-lb. long-duration thrusters are used for subsequent mid-course correction maneuvers, as well as for the orbit insertion burn to enter the SEL2 halo orbit.

### 8.2.1.1.8   Flight dynamics

LUVOIR will orbit the second Sun-Earth Lagrange (SEL2) point in a quasi-halo orbit that does not exceed 25° from the Sun-Earth axis at its maximum, or 5° from the Sun-Earth axis at its minimum. A notional delta-v budget is provided in **Table 8.3**. Station-keeping and momentum dumps will occur nominally once every three weeks and will be scheduled to coincide to minimize science observation down-time.

**Table 8.3.** *Delta-v budget for the LUVOIR-A mission concept. MCC: mid-course correction; LOI: L2 orbit insertion.*

| Maneuver | ΔV (m/s) | Mission Time | Burn Time (min) | Notes |
|---|---|---|---|---|
| MCC-1 | 45 | L +12h | 48 | Dependent on launch vehicle performance reliability |
| MCC-1a | 10 | MCC-1 + 5h | 11 | Clean up for MCC-1 efficiency |
| MCC-2 | 20 | L +22h | 106 (1.6 hr) | Clean up for MCC1 and solar radiation pressure effects after Sunshield deployment Using 4 5lb thrusters (Sunshield is deployed) |
| MCC-3 | 5 | L +132h | 27 | Using 4 5lb thrusters |
| LOI | 30 | L + 98day | 159 (2.65 hr) | Using 4 5lb thrusters |
| Orbit Maintenance | 60 | ~every 3 weeks after LOI | >1 | ΔV for 10 years ~ 4 m/s for stationkeeping and 2m/s for Momentum Unloading per year |

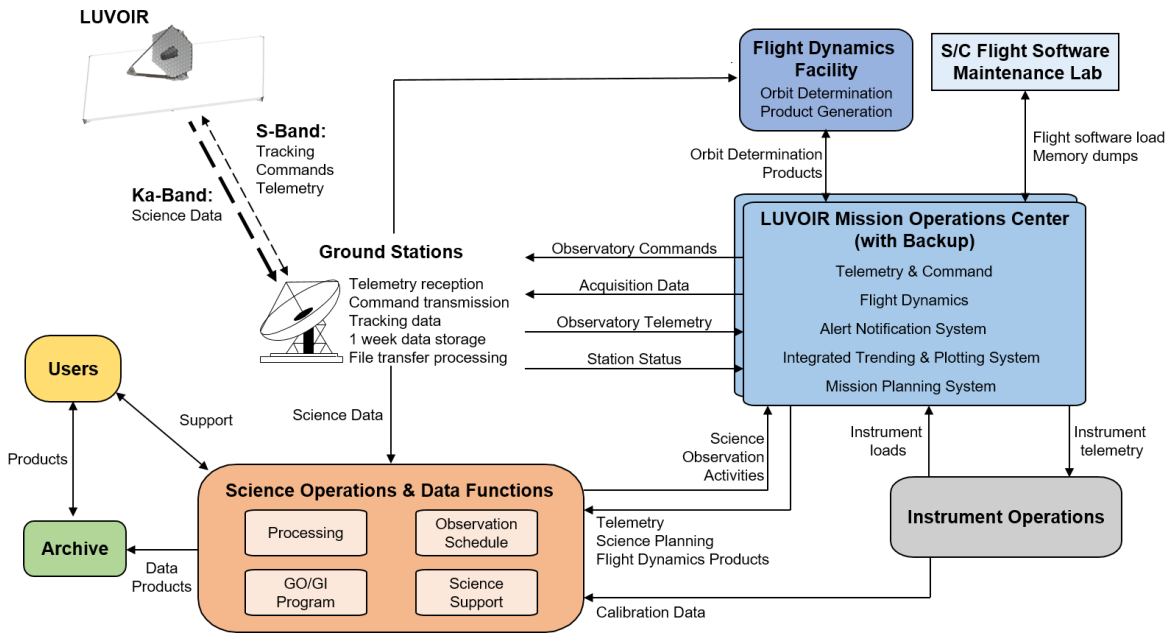

**Figure 8.6.** *A notional LUVOIR ground system. Two ground stations in White Sands, NM and South Africa allow for twice-daily downlink passes. The Mission Operations Center (MOC), backup MOC, and the Science Operations Center would be located at suitable institutions with the experience and infrastructure necessary to support a large-scale mission with a broad user base.*

## 8.2.1.2 Ground System

The LUVOIR Ground Segment will be based on the models of other large observatories such as HST and JWST. Primary and backup mission operations centers will be responsible for overall control of the observatory, orbit determination, telemetry, and mission planning. Similarly, a science operations center will be responsible for planning the observation schedule, supporting a guest-observer program, processing science data, and distributing and archiving the data for the user base. Two Ka/S-band ground stations are also baselined to provide two downlink passes per day; a third station would provide additional margin against missed passes due to weather or other conditions but is not fundamentally necessary to meet mission requirements. The ground stations consist of 18-m dishes in White Sands, NM, and in South Africa. **Figure 8.6** shows a schematic of a notional ground system.

## 8.2.1.3 Launch Vehicle

As discussed in **Section 8.1**, LUVOIR-A has been designed to take full advantage of NASA's Space Launch System (SLS) Block 2 vehicle with an 8.4-m × 27.4-m "long" fairing. No other anticipated launch vehicle is capable of accommodating both the mass and volume of the LUVOIR-A design. However, the scalability of the architecture, as will be demonstrated by LUVOIR-B, allows for the mission design and science objectives to be tailored to whichever heavy-lift launch vehicles may be available or preferred in the 2030s. Additional details on current and anticipated launch vehicles are available in **Appendix D**.

### 8.2.1.4 Concept of operations

LUVOIR's concept of operations will be similar to other large observatories of its class. There are three general observation scenarios, related to the three key elements of LUVOIR's science campaign: general astrophysics observations, local solar system observations, and high-contrast exoplanet observations. Each scenario may impose different operational constraints on the observatory. Regardless of the scenario, each observation will begin with a slew to the new target field-of-view.

#### 8.2.1.4.1 Slew and target acquisition

Following a science observation, wavefront maintenance routine, or stationkeeping/momentum dump maneuver, LUVOIR will be ready to begin a new science observation. The observatory will slew to a new target based on previously uploaded coordinates. Slews will incorporate pitch-angle changes using the gimbal system, and yaw adjustments using the spacecraft ACS. For special sunward observations, the spacecraft ACS may also need to pitch the entire observatory. During a slew, the VIPPS will be actively isolating the payload from the spacecraft, reducing the dynamic inputs to the payload and thus reducing the amount of settling time that will be required once the final attitude is achieved.

Star trackers will be used to coarsely align the observatory to the target field, after which fine guiding will be done by the HDI instrument. Acquisition of suitable guide stars known to be in the field-of-view will be performed autonomously using the control system processor. Once fine guidance has begun, the observation can proceed, depending on the type of observation.

#### 8.2.1.4.2 General astrophysics observations

General astrophysics observations using HDI, LUMOS, or POLLUX may begin shortly after a slew is complete, once system dynamics and thermal drift have settled to a level commensurate with their required wavefront stability (on the order of nanometers). While waiting for this settling to occur, internal instrument calibrations may be performed, if necessary, as well as instrument "set-up": selecting appropriate filters and channels, or in the case of LUMOS, aligning the microshutter array to the target field.

During general astrophysics observations, parallel observation with other instruments is supported. It is possible to have HDI, LUMOS, and POLLUX collecting data simultaneously under the condition that secondary instrument operations do not impact those of the primary instrument. For example, if a specific LUMOS observation is being executed, HDI may collect science data on whatever field it happens to be pointed at, but HDI would not be able to request pointing dithers to fill in focal plane gaps during this time. Similarly, if HDI were operating as prime, LUMOS would not be able to request roll maneuvers to align targets to the micro-shutter array.

#### 8.2.1.4.3 Solar system observations

Most solar system observations will likely require engaging LUVOIR's target tracking capability (up to 60 mas/s). Following a slew to target, the target would first be acquired by HDI using in-field guide stars. A trajectory would be computed, and tracking would begin. For cases in which it is desired to observe the object with LUMOS, ECLIPS, or POLLUX, an additional offset would be computed to place the object in their respective fields-of-view. Depending on the speed and duration of the observation, tracking may be accomplished by the fast steering mirror, VIPPS, gimbal system, spacecraft ACS, or any combination thereof.

#### 8.2.1.4.4   Exoplanet observations

High-contrast exoplanet observations will be the most demanding of LUVOIR's capabilities. Following a slew, it is likely that a longer settling period will be required for the active control systems to help stabilize the wavefront error before acquisition of high-contrast images can begin. Once engaged, the "dark-hole" acquisition algorithm will use images from the coronagraph focal planes to generate updates to the deformable mirrors, gradually increasing the contrast in the focal plane to the desired level (~$10^{-10}$ for most observations). Once the contrast is achieved, science observations will begin by taking long integrations to reveal targets of interest. During this time, onboard metrology systems and low-order and out-of-band wavefront sensors will work to maintain the wavefront error and high-contrast image. This entire process is executed autonomously by the onboard control system processor.

It is during the dark-hole acquisition process and science integrations that picometer-level wavefront error stability is required. If the wavefront error is drifting faster than the dark-hole wavefront control algorithm can keep up, the system will never achieve $10^{-10}$ contrast. Similarly, once the contrast is achieved, if the system drifts faster than the low-order or out-of-band wavefront sensors can keep up, then the contrast will degrade and long science integrations will not be possible. LUVOIR incorporates several techniques to achieve the necessary stability, discussed in more detail in the next section.

Parallel observations are also possible during exoplanet science observations, again assuming the secondary instrument operations do not impact the high-contrast performance. Given the picometer-level stability requirement, this may preclude even the actuation of filter wheels within the other instruments.

### 8.2.2   Design drivers

#### 8.2.2.1 High-contrast imaging

While LUVOIR is intended to be a multi-purpose, multi-user observatory, the desire to perform high contrast ($10^{-10}$) direct imaging of habitable exoEarths around solar-type stars with a coronagraph drives almost every aspect of the architecture. In addition to the need for coronagraph instruments capable of $10^{-10}$ contrast, it has been shown that achieving and maintaining such contrast requires wavefront stability of the end-to-end optical system on the order of tens of picometers RMS over the spatial frequencies corresponding to the dark-hole region in the focal plane (Lyon & Clampin 2012; Shaklan et al. 2011). This stability is needed over time periods corresponding to the wavefront control bandwidth—on the order of tens of minutes for systems using light from the target star to generate the wavefront estimates. The approach to achieving high-contrast imaging on LUVOIR is three-fold:

*High-contrast through ultra-stable systems:* This approach attempts to design a system that is as stable as possible. Thermal stability is achieved using materials with near-zero coefficient of thermal expansion at the nominal operating temperature of 270 K: Corning ULE® glass for the mirror segment substrates and composite material for most of the structure (Eisenhower et al. 2015; Park et al. 2017). These materials are coupled with milli-Kelvin-level thermal sensing and control of the mirrors, structures, and interfaces. Material property characterization is critical to understanding the behavior of structures and interfaces and modeling their performance at the picometer level. The choice of materials and the material properties work together to create a system that is both thermally stable and controllable at the observatory operating temperature. Ground-based metrology systems that can verify model predictions are currently being developed and have demonstrated the measurement of displacements of ~20 pm (Saif et al. 2017).

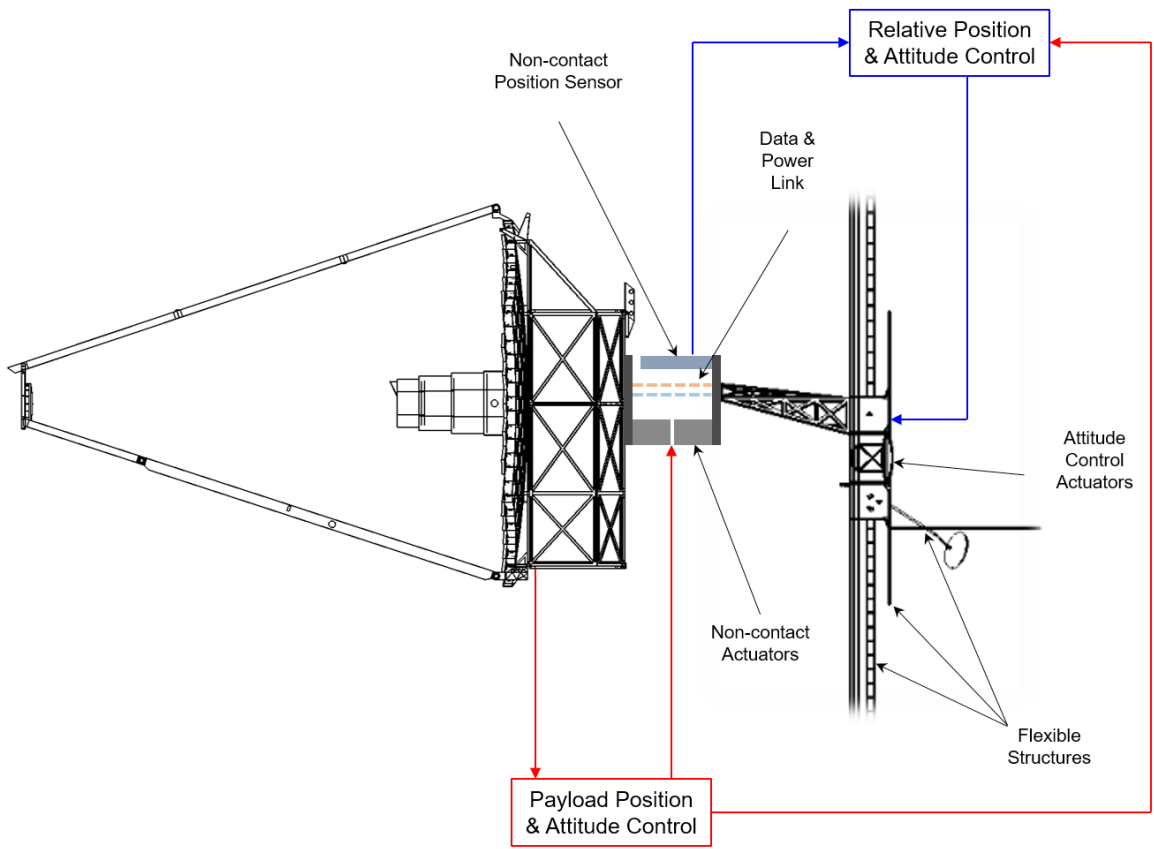

**Figure 8.7.** *Block diagram of the VIPPS control system showing two separate position and attitude control loops. The payload provides a line-of-sight fine-guidance signal from the instrument suite, and commands the non-contact actuators to maintain a fixed pointing vector for the payload. The spacecraft attitude control actuators respond to commands from the non-contact position sensor to maintain an attitude relative to the payload. Dynamic disturbances from the attitude control actuators and flexible structures on the spacecraft are isolated from the payload via the non-contact interface, with the exception of power and data cables that must bridge the gap. The size of the VIPPS interface is exaggerated for clarity.*

Dynamic stability is achieved with stiff mirrors and structures, passive isolation at the disturbance sources, and the non-contact VIPPS that actively isolates the payload from the spacecraft (Dewell et al. 2017). The VIPPS "floats" the telescope and controls the payload attitude relative to the interface plane via six non-contact voicecoil actuators. The VIPPS effectively isolates any dynamic disturbances from the spacecraft attitude control system from transmitting to the payload and exciting resonances that contribute to wavefront instability. The VIPPS also provides fine pointing control of the payload during science observations (Dewell et al. 2017). **Figure 8.7** shows a notional block diagram of the VIPPS control loops.

While the VIPPS provides ideal mechanical isolation, power and data signals must still be transmitted between the payload and spacecraft. Using a 120-v bus for the electrical power system allows for fewer, smaller-gauge conductors to be used, minimizing the number of cables and cable stiffness that must bridge the non-contact gap. Similarly, by performing as much of the data-processing on the payload side as possible, the cabling requirements for data transfer between the payload and spacecraft are minimized. Work is being done to both determine the stiffness of cables that will be bridging the non-contact interface, as well as modeling those stiffness effects as part of the system-level integrated modeling effort.

**Figure 8.8.** *Pointing and wavefront control loops baselined for the LUVOIR-A architecture. Payload line-of-site pointing is sensed at the HDI focal plane via fast centroiding on selected guide stars. Pointing is then controlled by three actuators with different bandwidths: the fast steering mirror controls the highest bandwidth errors, the VIPPS controls lower bandwidth errors, and the spacecraft ACS controls the lowest. In addition to low-order wavefront error terms, the low-order wavefront sensor inside of the coronagraph also senses line-of-site pointing errors specific to the coronagraph channels and uses the deformable mirrors to correct slow drifts in both pointing and wavefront during exoplanet observations. A segment edge sensor and laser metrology system maintains phasing and alignment of the primary and secondary mirrors.*

*High-contrast through wavefront control:* This approach reduces the time period over which the system must be stable by increasing the temporal bandwidth of the wavefront control system. **Figure 8.8** shows a block diagram of the pointing and wavefront control loops that are anticipated for LUVOIR. Pointing errors are sensed at the focal plane of the High Definition Imager instrument and drive high-bandwidth corrections at the fast steering mirror, and lower bandwidth corrections at the VIPPS. Slow, low-order pointing and wavefront drifts are sensed internal to the coronagraph with a low-order wavefront sensor (LOWFS), similar to the one to be used on WFIRST (Shi et al. 2017) and corrected at the coronagraph deformable mirrors. The LOWFS system, however, is fundamentally limited in which wavefront terms are sensed, and how fast they can be sensed. At the same time, fast dynamic motions of the primary mirror segments will be measured using edge sensors and corrected with piezo-electric (PZT) actuators (Feinberg et al. 2016). A laser metrology system is also being evaluated for sensing and controlling secondary mirror position errors (Nissen et al. 2017). Closed-loop control will be achieved with both feedback and feedforward systems (Shi et al. 2017).

*High-contrast through wavefront tolerance:* This approach attempts to relax the fundamental stability requirement from tens of picometers to hundreds of picometers, or even nanometers, through new coronagraph architectures that are designed to be less sensitive to the wavefront error modes that are most prominently excited in the LUVOIR architecture (Jewell et al. 2017; Ruane et al. 2016; Zimmerman et al. 2016). Additionally, spatial filtering techniques are being explored that may enable additional orders-of-magnitude in contrast, further relaxing the stability requirement. While many of these approaches show promise, they are still very early in development and it is unclear how much the stability requirement may be relaxed.

The technologies and design principles described above represent an array of tools that will be applied to the LUVOIR concept to achieve the required stability for high-contrast imaging of exoEarths. The current state-of-the-art of these technologies, and their paths to development are described in more detail in **Chapter 11**.

It is also important to note that picometer-level wavefront stability is only necessary during high-contrast exoplanet observations. It is not intended that LUVOIR will maintain picometer stability 100% of the time while on-orbit. During slews, stationkeeping maneuvers, and momentum dumps, the wavefront of the system will surely drift due to mechanical disturbances and changes in thermal loads, even with active control systems like the VIPPS, edge sensors, and thermal control always operating. However, most general astrophysics observations do not require the system to achieve its targeted wavefront stability before observations can begin. A more detailed concept of operations is provided in **Section 8.2.1.4**.

### 8.2.2.2 Thermal management

The thermal architecture of the observatory plays a critical role in enabling the broad wavelength range of LUVOIR's science goals. The nominal 270 K operating temperature represents a balance between operating at a cold enough temperature to achieve low thermal backgrounds in the NIR science bands, while operating warm enough to help mitigate contamination concerns in the UV science bands. Evidence from JWST thermal testing shows that below 260 K, molecular species rapidly begin to deposit on surfaces, which would degrade UV throughput (see **Appendix D**). Thus, 270 K optimally maximizes NIR science while not significantly impacting UV science.

Separate from science considerations, 270 K was also chosen for a number of engineering reasons. At warmer temperatures, damping of dynamic disturbances is increased. Also, the selection of materials that have both near-zero coefficient of thermal expansion and that are also well characterized is larger near room temperature. Finally, a warmer temperature was chosen to help reduce the complexity of fabrication, integration, and test campaigns. By fabricating components at a temperature near where they are to be operated, modeling errors can be reduced, as well as the complexity, number, and duration of thermal cycle tests.

While most of the payload maintains an operating temperature of 270 K, several optical benches and detectors operate at colder temperatures of 170 K and 100 K, achieved through passive cooling. The thermal management system must work to simultaneously hold the optical telescope at 270 K with 1-mK stability while enabling the passive cooling. A large sunshield, similar to JWST's, provides a stable, cold-biased environment for the payload. Heaters are then used to precisely control the temperature throughout the payload.

### 8.2.2.3 Serviceability

Drawing on the legacy of HST, LUVOIR is being designed from the outset to be serviceable and upgradeable to enable a multi-decade science mission that is responsive to a changing scientific landscape. We note however, that while the Study Team is designing LUVOIR to be serviceable, we are not designing the infrastructure to actually execute the servicing. Nor is it required to service LUVOIR for its users to achieve all of the science presented in this report within the first five years of the mission.

Other programs within NASA are currently developing autonomous and tele-robotic servicing capabilities.[2] NASA's Human Exploration and Operations Mission Directorate is also investigating the viability of using a Deep Space Gateway in cis-lunar space for the servicing of scientific assets, such as LUVOIR.[3] If this infrastructure con-

2 For example, the NASA GSFC Satellite Servicing Projects Division (**https://sspd.gsfc.nasa.gov**).

3 See **https://exoplanets.jpl.nasa.gov/exep/tech-nology/in-space-assembly/**

tinues to be developed through investments elsewhere in NASA and by other government agencies, then LUVOIR could leverage these resources to extend its mission beyond the initial 5-year requirement. Thus, the reliability, lifetime, and consumables have been designed to meet a 5-year minimum lifetime for all mission elements, a 10-year goal lifetime for all serviceable or replaceable mission elements, and a 25-year goal lifetime for all non-serviceable/replaceable elements. An additional advantage to designing for serviceability can be realized much earlier in the mission lifetime: the modular nature of LUVOIR's systems can facilitate testing and re-testing at multiple stages during integration.

The serviceable elements on the observatory include the four instrument modules themselves. **Figure 8.9** shows a concept for how the instrument modules could be removed from the BSF and replaced. Key spacecraft subsystems are also designed to be replaceable, including the control moment gyroscopes, the avionics, and the solar array. The main elements of the observatory that have not been designed to be serviced include the optical telescope element and backplane support frame, the VIPPS and gimbal system, and the sunshield.

### 8.2.2.4 Launch Vehicle

As the capabilities of the SLS Block 2 and its user base are still evolving, so is the certainty of which of several fairing options will be developed: an 8.4-m × 19.1-m "short" fairing, an 8.4-m × 27.4-m "long" fairing, and a 10.0-m × 27.4-m fairing. The LUVOIR team opted to be conservative in the fairing diameter but chose the 8.4-m "long" fairing as the "short" version does not provide enough height under the ogive to fit a > 7-m aperture. The lift capacity for the Block 2 configuration with the 8.4-m "long" fairing is 44.3 metric tons (NASA 2017). The deployable LUVOIR architecture is designed to maximize the science capability that can be fit within these volume and mass constraints. **Figure 8.10** shows the LUVOIR-A concept stowed in the 8.4-m "long" fairing and **Table 8.4** shows an allocated mass budget.

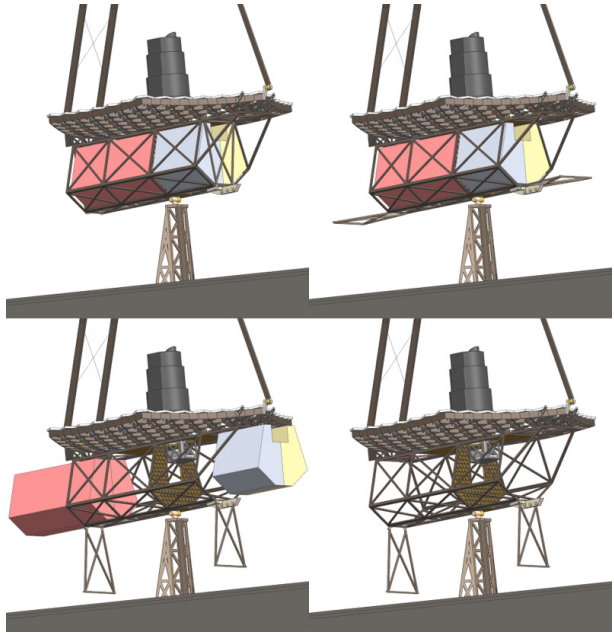

**Figure 8.9.** *Notional servicing sequence. Top Left: The payload is gimbaled to the 0° angle. Top Right: Servicing doors at either end of the BSF are opened. Bottom Left: Instrument modules are extracted. Bottom Right: The payload is ready to receive new instrument modules. In practice, only one instrument would be serviced at a time. While this model doesn't have enough detail to show all of the mechanisms required, the study team has included rails, latches, grapple fixtures, and mate/demate fixtures in more detailed models, as well as the Master Equipment List and mass allocations.*

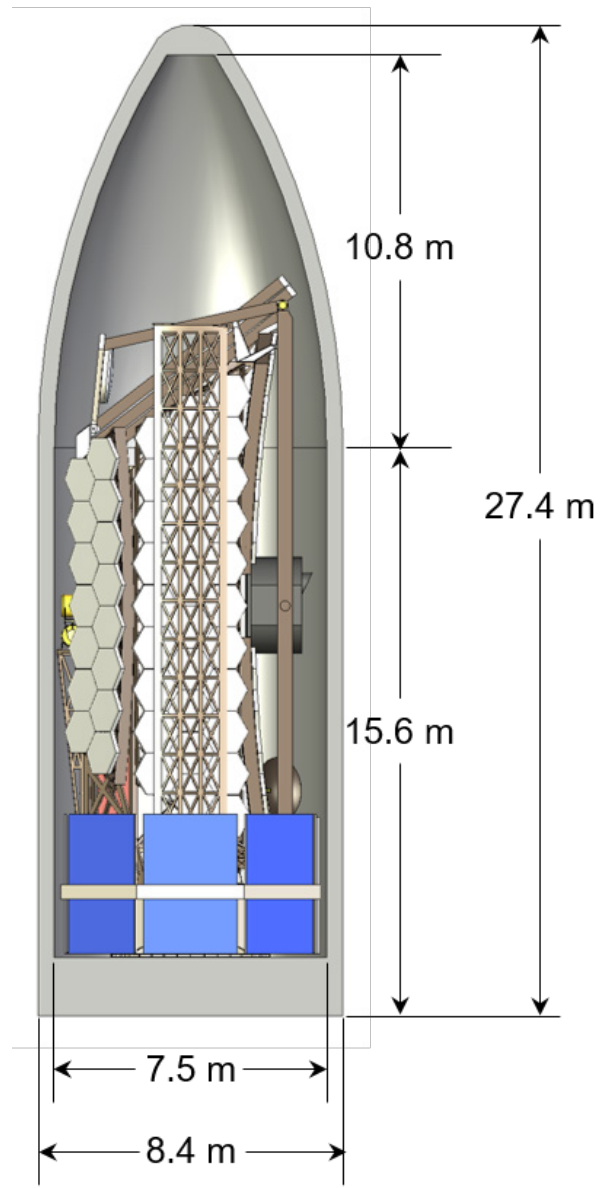

**Figure 8.10.** *LUVOIR-A stowed in the SLS Block 2 8.4-m "Long" fairing.*

### 8.2.2.5 Integration and test

The integration and testing (I&T) of LUVOIR requires careful thought and planning to minimize both the programmatic and technical challenges a mission of this magnitude will present. We understand that a long and drawn out integration and test campaign will directly impact the total cost and schedule of the mission development. A lesson learned from the JWST program is that costs are generally driven by the expense of having a "marching army" to staff the project, including managers, scientists, engineers and technicians among others. Once the final I&T of the observatory is underway, more work is serialized rather than in parallel. When there is a problem in this phase of integration, a large contingent of the marching army may have to stand down while the problem is resolved.

**Table 8.4.** *Mass Budget for the LUVOIR-A Observatory. Current Best Estimate (CBE) values are a combination of grass-roots estimates based on instrument designs and top-down allocations based on mass fractions from JWST. Maximum Expected Values (MEV) include a 30% growth allowance that accounts for the early-stage nature of the designs. Maximum Permissible Values (MPV) include an additional 25% margin as required by NASA Goddard Space Flight Center GOLD Rules. We note that the OTE and BSF growth allowance factors are lower than would typically be applied as their designs are based heavily on flight-qualified prototypes. For the OTE, the primary mirror segment assembly is based on the Advanced Mirror System Demonstrator (Matthews et al. 2003), and the BSF is based on the JWST BSF design.*

| System / Subsystem | Mass (kg) | Growth Allowance | MEV Mass (kg) | GOLD Rules Margin | MPV Mass (kg) |
|---|---|---|---|---|---|
| Optical Telescope Element (OTE) | 7699 | 19% | 9505 | 25% | 12673 |
| Backplane Support Frame (BSF) | 4436 | 26% | 5995 | 25% | 7993 |
| ECLIPS | 1200 | 30% | 1714 | 25% | 2286 |
| HDI | 1060 | 30% | 1514 | 25% | 2019 |
| LUMOS | 830 | 30% | 1186 | 25% | 1581 |
| POLLUX | 780 | 30% | 1114 | 25% | 1486 |
| Spacecraft | 7,483 | 30% | 10690 | 25% | 14253 |
| Propellant | 1,502 | 0% | 1502 | 25% | 2003 |
| Total: | 25,028 | - | 33,276 | - | 44294 |

This is distinctly different from earlier phases of a project where lower levels of assembly can be integrated and tested in parallel. A problem in one parallel path is rarely a reason to stand-down the majority of project staff in all paths. Thus, problems are generally less costly in terms of the marching army the earlier they occur. This is justification alone to do as much early prototyping and testing as possible. The less "new" information that is learned during Observatory I&T, the better.

At the time of preparing this interim report, the Study Team has only begun to think about the I&T flow for LUVOIR at the most basic levels: trying to determine which levels of assembly may be tested in parallel versus which must be done serially; which requirements must be verified by test versus which must be verified by analysis; and the logistics of transporting, assembling, and testing systems as large as LUVOIR. We will report here our initial thoughts on what will drive the LUVOIR I&T campaign, however a more detailed discussion will be presented in the final report, once a more complete picture of the LUVOIR architecture is available.

There are three primary challenges to the I&T of LUVOIR:

- the sheer size of the observatory,
- the verification of picometer-level stability, and
- the control of contamination impacting the far-UV science.

Overcoming these challenges will require a combination of leveraging processes and techniques that have been developed for missions such as HST and JWST and developing new methods and facilities specific to the LUVOIR architecture. We address each challenge separately below.

### 8.2.2.5.1 Observatory size

Like the International Space Station (ISS), LUVOIR-A will be a space system of extraordinary size and complexity. Unlike the ISS, however, LUVOIR will be assembled on the ground and launched as a single system. This presents a logistical challenge: how does one verify that the fully integrated system meets all performance requirements and is capable of surviving the extreme environments encountered during launch? Traditionally, NASA has adhered to a golden rule of "test as you fly, fly as you test," subjecting fully integrated flight systems to a series of environmental, performance, and functional tests to ensure their viability on-orbit. However, missions such as LUVOIR are so large and complex that it is impossible to exactly replicate the flight conditions in a test environment for the entire observatory. This necessitates the "test as you fly" paradigm to evolve into a process combining lower-level testing with using high-fidelity, validated models to verify the expected performance of the observatory. Projects such as JWST have had to obtain waivers for the "test as you fly" requirement: e.g., while the optical telescope and instruments (a.k.a. "OTIS") underwent cryogenic optical performance testing, the test itself was not a full-aperture end-to-end optical test that verified image quality. Instead, the test confirmed the ability of the system actuators to respond to commands based on analysis of image data collected at the instrument focal planes. Most importantly, this test was the culmination of many subsystem tests and model validation efforts building on top of each other – all used to create a system-level model that is capable of accurately predicting on-orbit performance.

Similarly, once OTIS is integrated to the spacecraft and sunshield, the full observatory will not be tested at cryogenic temperatures in a thermal vacuum chamber. JWST will only see a portion of the tests that most space systems would normally see at this level, including mechanical testing followed by functional tests to verify assembly workmanship. Instead, the individual subsystems underwent rigorous performance testing at lower levels of assembly. Validated models, tracking of requirements, and verification of workmanship will be relied upon to ensure the fully integrated observatory will perform as expected on-orbit.

LUVOIR will require a similar I&T methodology for performance verification, though it is likely that the division between verification by test and verification by analysis will need to happen at even lower-levels of assembly. For example, even though LUVOIR-A's 15-m diameter primary mirror would physically fit inside of the Johnson Space Center Chamber-A (where the JWST OTIS was tested, see **Figure 8.11**), it would not fit through the chamber door fully deployed. Once deployed, there would be very little room left for optical, mechanical, and thermal ground support equipment, all critical to executing a performance test. The Plum Brook test chamber at NASA's Glenn Research Center (see **Figure 8.11**) is larger and would accommodate LUVOIR with some room for ground support equipment, but its remote location raises another challenge: transporting the observatory prior to launch. The stowed JWST OTIS barely fits in a C-5 Galaxy aircraft—the largest cargo transport operated by the United States—and the fully integrated observatory must be barged to its final launch site in South America. Given that JWST has already stretched the available transport, integration, and testing infrastructure to the limit, LUVOIR will need to develop new options, including but not limited to:

- New integration and test facilities co-located with the final launch site. These facilities would allow components and subsystems that are assembled at remote locations to be transported by traditional means before being integrated and tested at the final launch site.

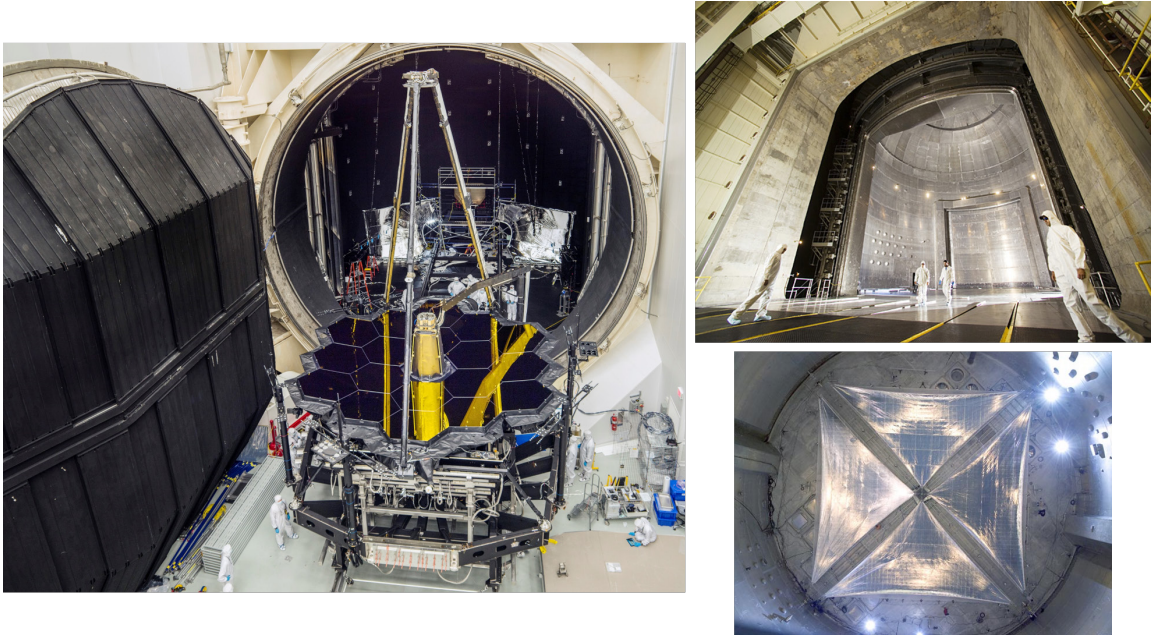

**Figure 8.11.** *Left: JWST emerges from the Johnson Chamber A test facility following cryogenic optical testing. It is clear from this photo that LUVOIR-A would not fit through the door fully deployed. If deployment inside the chamber were possible, the height of the chamber would limit the ability to perform optical testing. Right Top: The Plum Brook Station Space Power Facility at NASA's Glenn Research Center could accommodate a fully deployed LUVOIR-A. Right Bottom: A 20-m solar sail being deployed inside of the Plum Brook Station chamber. Credit: NASA*

- New transport vehicles, such as the Stratolaunch aircraft,[4] or a modified 747 that carries cargo outside its hull similar to the Shuttle Carrier Aircraft, or easy access to waterways may be necessary to move the largest components, including the fully integrated observatory.

- A modular design of LUVOIR would enable the disassembly and reassembly of large portions of the system. Individual modules can undergo environmental and performance testing before being reassembled. Designing for serviceability is already a step in this direction, as at least the instruments and spacecraft subsystems are developed to be easily and repeatably inserted and removed.

This leads to the issue of qualifying the hardware for the launch environment. While it seems plausible, perhaps even advanta-geous, to argue that performance verification can be broken into smaller pieces, it is hard to make that same argument when it comes to verifying assembly workmanship in preparation for launch. Without subjecting the observatory to mechanical launch environments, it is hard to believe that a launch vehicle vendor will accept the risk of a catastrophic failure. At this point in time, the design of LUVOIR is not mature enough to know whether existing vibration and acoustic testing facilities will be able to accommodate it. Two of the largest vibration tables in the world are located at NASA Glenn Research Center's Plum Brook Station[5] and the European Space Agency's European Space Research and Technology Centre (ESTEC)[6]. At this time, it is unclear if either has the capacity to shake something

4 See http://www.stratolaunch.com/

5 See https://www1.grc.nasa.gov/facilities/spf/

6 See http://www.esa.int/Our_Activities/Space_Engineering_Technology/Test_centre/Hydrau-lic_shaker

with as much mass and inertia as LUVOIR. It appears that both large acoustics chambers at Plum Brook Station and ESTEC would fit an observatory of LUVOIR's size with some ground support equipment to protect the stowed observatory.[7]

Regardless of the approach taken, the LUVOIR I&T program will require careful attention to logistics to ensure a smooth flow. And while much of the hardware and ground support equipment that was developed for JWST won't be directly useable by LUVOIR, the concepts and methodologies developed for JWST I&T will be. Key among them is the concept of using validated models at every level of assembly to verify requirements, especially those pertaining to ultra-stability.

### 8.2.2.5.2    Verifying picometer-level stability

Many of the challenging requirements for LUVOIR create similarly challenging requirements for the test facilities and ground support equipment, which may very well need to be more stable or sensitive than the flight hardware under test in order to verify system performance. LUVOIR's ultra-stability requirements will demand micro-gravity-level vibration isolation and sub-milliKelvin-level thermal stability for at least some testing. While these levels of sensitivity are being demonstrated in lab facilities today over small scales (Saif et al. 2017), these levels are beyond the state-of-the-art for larger class facilities and may well be beyond what is physically possible for ground-based facilities. Verification by analysis may be the only way to demonstrate that LUVOIR is capable of achieving picometer level wavefront stability on-orbit.

### 8.2.2.5.3    Controlling contamination for far-UV performance

The LUVOIR Study team recognizes that conducting science in the far-UV means that the hardware will be very sensitive to molecular contamination. At this time, our general plan is to use standard processes used on previous UV and X-ray instruments and systems. These include, but are not limited to, material controls, component bake-outs both on ground and on-orbit, and certifications. The 270 K operating temperature was selected in part for contamination control reasons. It is slightly above the temperatures at which quartz crystal microbalance monitors have measured molecular buildup during JWST testing, and it is above the temperature at which known molecules of concern would stick to critical surfaces. Mirror heaters are also being sized to allow for occasional baking out of contaminants on-orbit, similar to the annealing done on flight detectors on HST. These bake-outs would be done infrequently, perhaps once a year, to minimize the impact on science operations. Finally, LUVOIR will use techniques developed for JWST to deal with particulates that are on the mirror substrate. For example, JWST has cleaned the entire primary mirror using a brush technique with excellent results on large particulates.

### 8.2.3    Design alternatives

While the architecture presented in **Section 8.2.1** represents one possible approach to a feasible 15-m-class LUVOIR mission, there exists a number of design alternatives that provide different approaches to achieving the ambitious science goals identified by the STDT. At the time of this writing, the Study Office has completed some of the trade studies identified below to develop the current LUVOIR-A architecture and is in the process of considering the others in an attempt to improve overall system performance.

---

7 See https://www1.grc.nasa.gov/facilities/aapl/ and http://www.esa.int/Our_Activities/Space_Engineering_Technology/Test_centre/Large_European_Acoustic_Facility_LEAF

### 8.2.3.1 Coronagraph vs. starshade

There are two fundamental approaches to achieving the necessary starlight suppression for exoplanet science: a coronagraph, and a starshade (sometimes called an external occulter). Each has its own advantages and disadvantages.

A coronagraph works by rejecting on-axis starlight that has entered the telescope while allowing off-axis planet light to be imaged onto a focal plane. The starlight rejection is usually achieved through some combination of occultation, diffraction control, and destructive interference. Since a coronagraph is located inside the observatory payload, it allows for very nimble observations; as quickly as the telescope is able to slew from one target to the next, a new exoplanet system can be observed. The two primary drawbacks to a coronagraph instrument are the extraordinary wavefront stability required by the optical system to achieve $10^{-10}$ contrast, and limited throughput and bandpass that result from the complex optical systems that implement the starlight suppression. Coronagraphs are efficient at finding target exoplanets, but less efficient at characterizing them.

Conversely, starshades work by rejecting the on-axis starlight before it ever enters the telescope, removing the need for extreme diffraction and wavefront control from the observatory. Instead, the challenges of a starshade include the need to precisely deploy an extremely large (>70-m diameter) structure and fly that structure in formation with the observatory at separation distances of tens to hundreds of thousands of kilometers. Starshades enable high-throughput, broadband observations of target stars making them effective at spectroscopic characterization of exoplanets. But the need to reposition the starshade over large distances for every telescope pointing vector means that starshades are inefficient at observing many systems and finding new targets. **Figure 8.12** shows some of the technology challenges and demonstration efforts that are ongoing for starshades.

LUVOIR's science case requires the ability to survey hundreds of potential planetary systems in order to guarantee enough exoEarths to statistically constrain the frequency of habitability. A coronagraph provides the necessary agility to perform such a survey. Ideally, once a target of interest is discovered, a starshade could be moved into place to enable efficient spectral characterization of the target. However, LUVOIR's large aperture would require starshades in excess of 70 meters at very large separations, well beyond the starshade technologies that are currently being developed. For now, LUVOIR has decided to rely on a coronagraph instrument, equipped with a spectrograph, to perform the exoplanet discovery and characterization. However, the observatory is being designed to be starshade-compatible and a follow-on starshade rendezvous with LUVOIR may be a viable option; see **Appendix E** for more details.

### 8.2.3.2 Off-axis, unobscured vs. on-axis, obscured apertures

An off-axis, unobscured telescope design has the advantage of increasing throughput by removing the central obscuration, while also improving compatibility with several mask-based coronagraph architectures, providing substantial gains in coronagraph core throughput. However, an unobscured aperture would not leverage the approach used on the JWST for deploying the secondary and would require developing and demonstrating new packaging concepts for the observatory. Such repackaging is feasible given the large fairing volume provided by the SLS Block 2 vehicle, but likely constrains the total aperture diameter to be smaller than would be possible with an on-axis design. It is possible the gains in coronagraph throughput exceed the reduction in aperture size in terms of exoplanet science yield, however the smaller aperture also negatively impacts collecting area and resolution for the general astrophysics program. An off-axis optical design also requires the secondary mirror to be nearly twice as far from the primary mirror than a comparable on-axis design in order to maintain the same angles-of-incidence across the pupil, a requirement driven by the need to mitigate polarization aberrations that can significantly reduce coronagraph performance.

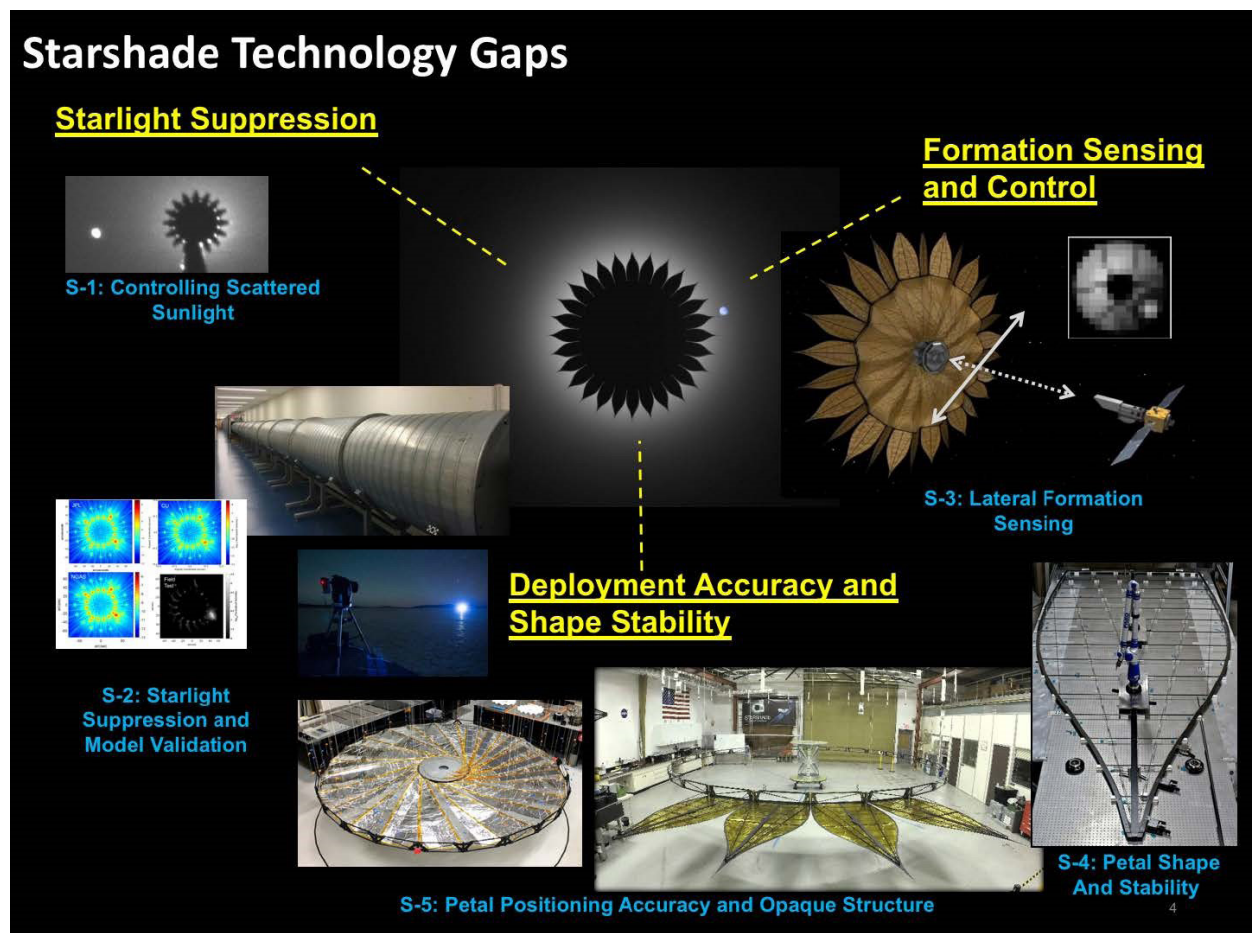

**Figure 8.12.** *Starshade technology development challenges (Figure courtesy NASA/ExEP[8]). Three main challenges remain for starshade technology development: validating the starlight suppression of the starshade design, precision deployment of a 70-m-class structure, and formation flying between the starshade and telescope.*

The specific geometry of the on-axis segmented pupil is related to this trade as well. The current LUVOIR-A baseline represents an extension of the JWST segment geometry with four additional rings of slightly smaller segments (1.15-m flat-to-flat vs. 1.32-m flat-to-flat). The first optical design for LUVOIR required the central ring of segments to be removed to accommodate both a larger secondary mirror obscuration as well as the optical path between the secondary, tertiary, and fine-steering mirrors. The resulting large

8 See https://exoplanets.nasa.gov/exep/technology/gap-lists/

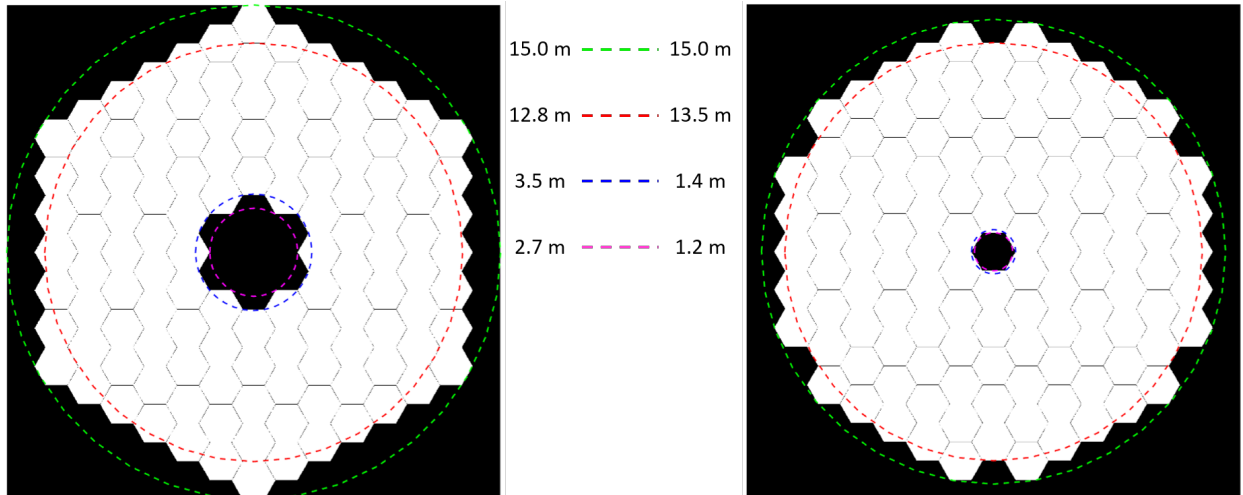

**Figure 8.13.** *A comparison between the initial LUVOIR-A aperture (left) and the new aperture (right). The new aperture achieves the same 15.0-m outer diameter while maximizing the inscribing diameter and minimizing the central obscuration. Both apertures have 120 segments, although the segment size increased from 1.15-m flat-to-flat (left) to 1.223-m flat-to-flat (right).*

obscuration negatively impacts coronagraph performance. At the time of writing this report, the team has studied a number of new aperture geometries and optical designs to mitigate this effect. Although many of the figures in this interim report show the original aperture design with a large central obscuration, the team has since chosen a new aperture geometry and optical design that is more conducive to coronagraph performance. **Figure 8.13** shows the old LUVOIR-A aperture compared to the new aperture.

### 8.2.3.3 Optical vs. radio communications

The conventional Ka-band system that is currently baselined on LUVOIR-A provides adequate margin for LUVOIR's anticipated data volumes. However, an optical communication system would provide shorter downlink times, as well as additional room for data volume growth as new instruments may be deployed. Optical communication technology has a high Technology Readiness Level (TRL) after having been demonstrated on the Lunar Laser Communications Demonstration (Boroson et al. 2014). The existing limitation is with the number of available ground stations to ensure adequate coverage. Should there be investment in new optical communications ground systems, LUVOIR could easily adopt this solution. Thermal impacts of increased power consumption for the optical communication system also needs to be considered.

### 8.2.3.4 Disturbance isolation vs. disturbance reduction

The current LUVOIR baseline design uses several layers of both passive and active isolation to prevent disturbances from reaching the opto-mechanical payload. Passive isolators are used at the disturbance source (control moment gyroscopes, mechanisms, etc.) and a non-contact active isolation system separates the payload from the spacecraft. Initial analysis by the LUVOIR engineering team shows that this architecture contributes to a dynamically ultra-stable system capable of achieving high-contrast observations.

An alternative to isolating disturbance sources from the optical system is to remove those disturbances altogether. A primary source of dynamic disturbance comes from spinning mass used as part of the attitude control system, such as reaction wheel assemblies or control moment gyroscopes. Microthrusters are a relatively newer technology that may be used in place of these mechanisms to control and maintain the attitude of the system while introducing very little or no disturbance to the system (Chapman et al. 2011).

For the current LUVOIR design, the articulated nature of the payload necessitates the use of control moment gyroscopes; the force required to react against the gimbal mechanism as it slews the payload to a new pointing far exceeds the capabilities of microthrusters. However, a hybrid approach may be possible in which control moment gyroscopes or reaction wheels are used during slews of the payload and then stopped. Microthrusters would then take over and maintain the observatory attitude during an observation. It is unclear whether this approach is more efficient than the current architecture using a non-contact isolation system, but it does provide an alternative technology approach should the non-contact system's development be delayed.

### 8.2.3.5 Approaches for on-orbit metrology and wavefront control

As this and subsequent chapters describe, many tools are being applied to the challenge of achieving picometer-level stability on a system as large and complex as LUVOIR. Stable structures and materials, active and passive dynamic isolation, thermal sensing and control, wavefront sensing and control, and on-board metrology all contribute to the goal of ultra-stability. Within each technology there are several options for specific implementations. For example, metrology systems may consist of segment edge sensors, a laser truss, or a combination of both. Edge sensors themselves could be capacitive, inductive, or optical. Systems-level modeling is being performed to determine the highest-performance, highest-TRL configuration that is capable of achieving the metrology requirements.

The specific implementation of wavefront sensing and control is another area of study. Image-based techniques such as phase retrieval are high TRL after development for JWST but may not be sufficient for active control during coronagraph operations. Low-order and out-of-band wavefront sensing techniques are under development by WFIRST for maintaining wavefront stability as part of coronagraph instrument designs. All of these techniques suffer from the need to collect enough stellar photons to estimate the wavefront error, a process that can take many minutes to tens of minutes for the types of stars that LUVOIR will observe.

**Chapter 11** goes into more detail on the specific technologies that enable the LUVOIR-A architecture as described here and in Chapter 9. There are many tools in the toolbox, of varying capability and technology readiness. At the time of writing this interim report, the Study Team is still evaluating the optimal combination of wavefront sensing, metrology, isolation, and control that will provide picometer-level stability to enable high-contrast imaging.

## 8.3   Looking ahead

The LUVOIR study team has begun a second design iteration on the LUVOIR-A architecture, with the primary goal of further improving the exoplanet science yield. The effort focuses mostly on optimizing the segmented aperture geometry to be more compatible with coronagraph mask designs, specifically reducing the central obscuration size and maximizing the inscribing diameter of the aperture. In parallel with this effort, coronagraph design teams are continuing to develop new masks and instrument architectures that continue to push the state-of-the-art on segmented-aperture coronagraphy to further improve the performance of LUVOIR.

The study team is also developing a structural-thermal-optical integrated model of the design that can be used to evaluate the thermal and dynamic stability of the system. This is critically important to demonstrating that the design and the incorporated technologies are capable of achieving the picometer-level wavefront stability required for the high-contrast imaging. The integrated models will also be used to inform technology development efforts and focus resources on the most promising technologies.

While completing an update to the LUVOIR-A architecture, the study team plans to develop a concept for LUVOIR-B: an 8-m observatory that is capable of being launched in a conventional 5-m diameter fairing. We expect that not all elements of the observatory will scale accordingly and that some level of redesign will be needed on LUVOIR-B. For example, while the size and mass will scale proportionally, the power consumed by the instruments most likely will not, requiring updates to the thermal management system. The LUVOIR-B design also gives the team an opportunity to explore other options for some of the trades outlined in **Section 8.2.3**, including an off-axis unobscured primary mirror option.

In **Chapter 9**, we describe in detail the payload of the LUVOIR-A design, including the optical telescope element, the backplane support frame, and the three US-led instrument designs: ECLIPS, LUMOS, and HDI. Element-specific design drivers and design alternatives are also provided. **Chapter 10** provides additional detail on the CNES-contributed POLLUX instrument design, and **Chapter 11** provides greater detail on the technologies that enable or enhance the LUVOIR science mission, and what additional development is still needed.

# Chapter 9

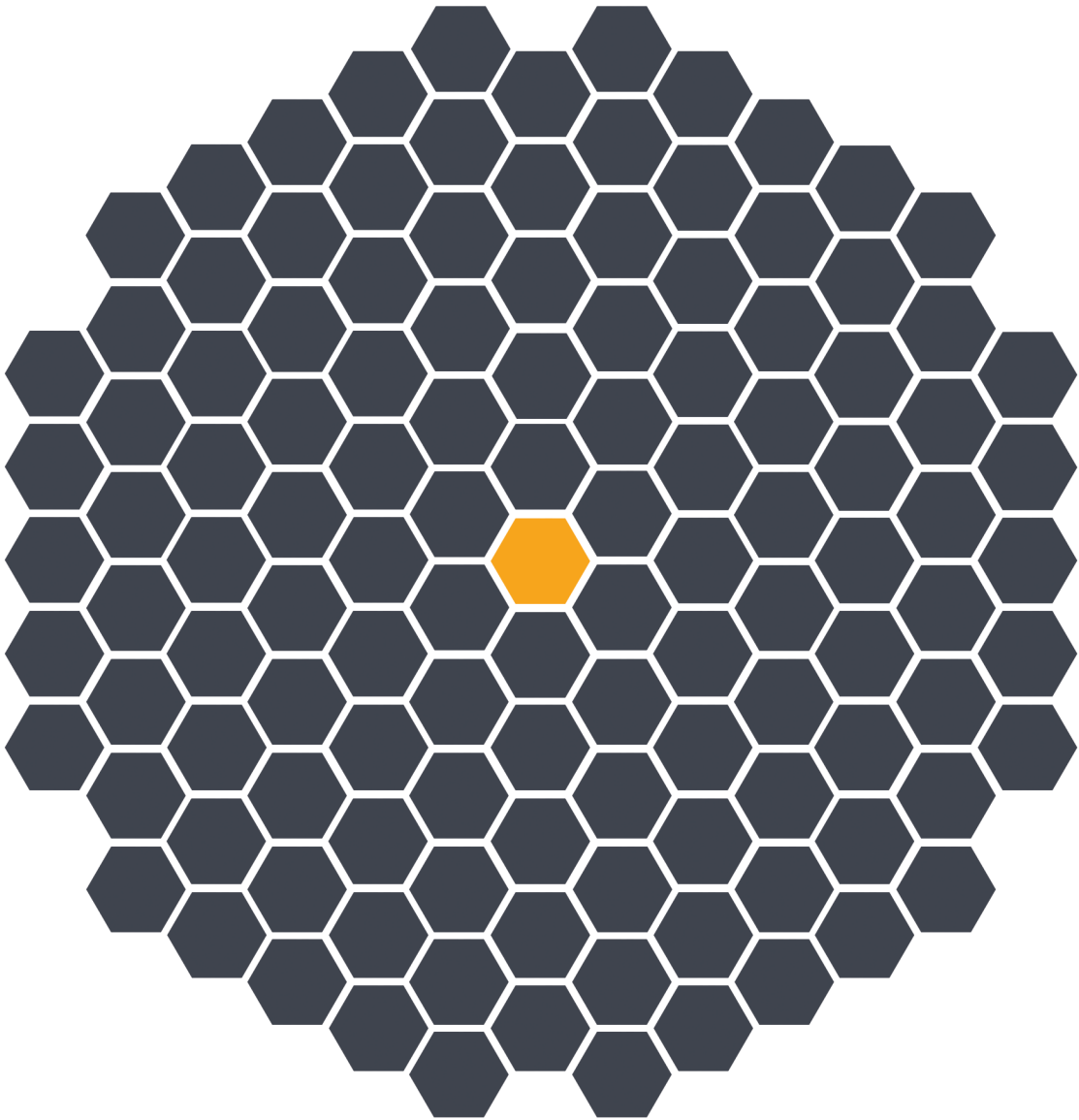

The LUVOIR telescope and instruments

# 9   The LUVOIR telescope and instruments

LUVOIR is composed of two distinct elements: The Payload and the Spacecraft. The Payload is composed of several components, highlighted in **Figure 9.1**, which will be discussed further in this chapter:

- The optical telescope element (OTE)
- The backplane support frame (BSF)
- The science instruments
  - Extreme Coronagraph for Living Planetary Systems (ECLIPS)
  - LUVOIR Ultraviolet Multi-Object Spectrograph (LUMOS)
  - High Definition Imager (HDI)
  - POLLUX (discussed in detail in **Chapter 10**).

## 9.1   Optical telescope element (OTE)

### 9.1.1   OTE overview

The optical telescope element (OTE), highlighted in **Figure 9.2**, is the optical path every photon follows at the end of its long journey from the astronomical objects being observed to the instrument suite and detectors. The OTE must be designed to achieve all science objectives identified by the LUVOIR Study Team, and yet be capable enough to respond to an ever-changing scientific landscape.

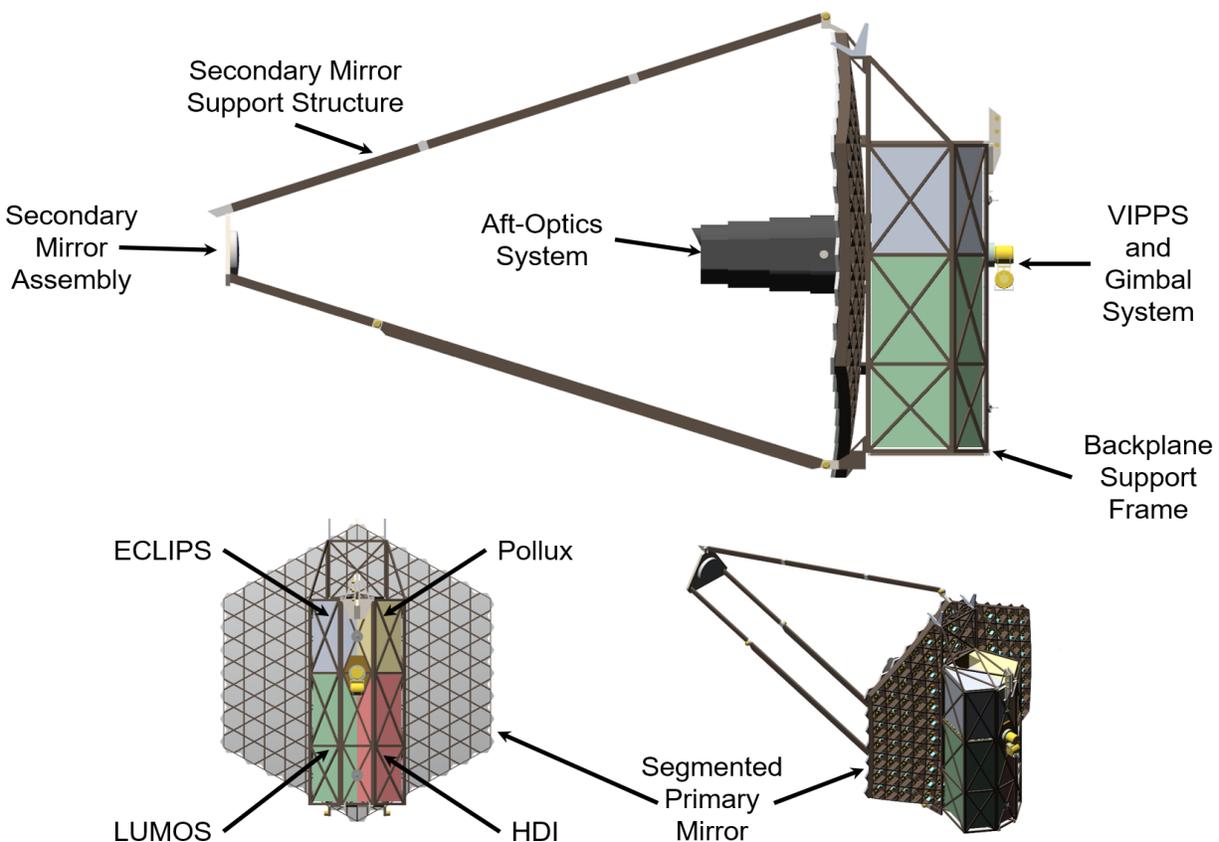

**Figure 9.1.** *Major elements of the LUVOIR payload.*

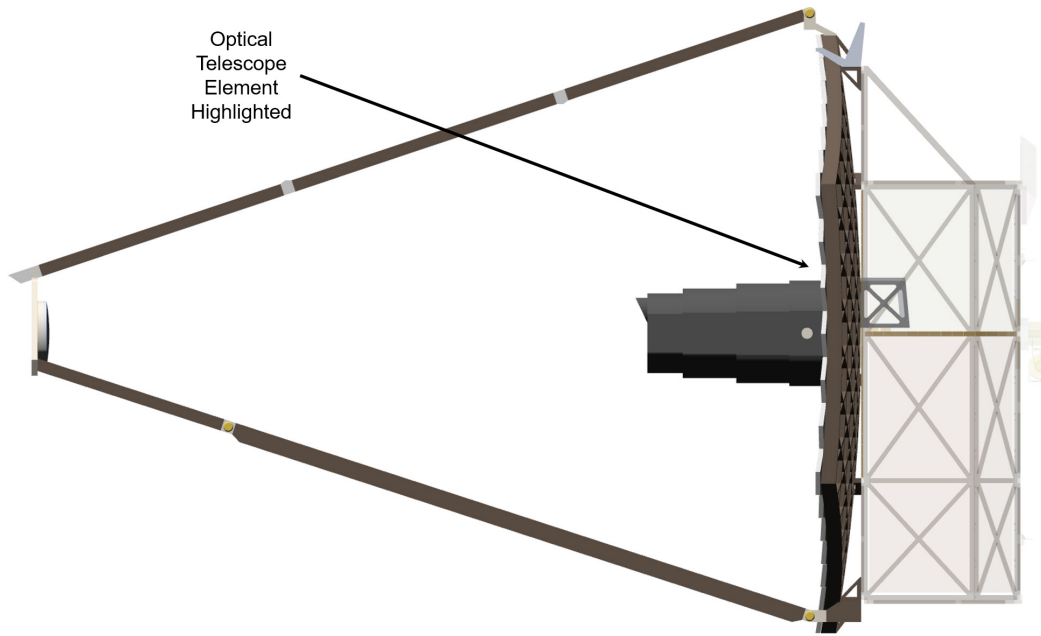

Optical
Telescope
Element
Highlighted

**Figure 9.2.** *The LUVOIR Payload with the optical telescope element highlighted. The small cage shown inside of the backplane support frame volume is the tertiary mirror assembly.*

### 9.1.2 OTE design drivers

Some of the key drivers that shape the implementation of the OTE include:

- High-contrast imaging
- Science Instrument fields-of-view
- Wavelength range
- Wavefront error
- Thermal management
- Launch vehicle mass-to-orbit capability
- Launch vehicle volume
- Contamination control

As discussed in **Chapter 8**, the wavefront error stability needed to achieve the high-contrast exoplanet science is a critical requirement that drives the OTE design. The material choices, thermal control system, segment-level architecture, segment phasing concept, and secondary mirror support structure design have been driven by the need to maintain wavefront stability at the picometer level (Feinberg et al. 2017).

Unlike the science instruments and significant portions of the spacecraft, the OTE is not being designed for modularity nor potential serviceability and as such, it must be reliable and robust.

### 9.1.3 OTE design implementation

The breadth of the science case described in earlier sections implies a challenging set of high-level capabilities for the OTE: sensitivity, resolution, stability, and wavelength range. It must be agile enough to repoint quickly and maximize observation time, and the whole system must be flexible enough to adapt to ever-evolving science objectives. Combining these capabilities with constraints such as mass, volume, and technical maturity, the Study Team derived the OTE performance specifications, shown in **Table 9.1**.

### 9.1.3.1 Optical design

The LUVOIR-A OTE is designed as a three-mirror anastigmat system, with a fourth fast steering mirror located at the real exit pupil of the OTE. The advantages of this system include a wide field-of-view that can be accessed by a number of instruments,

**Table 9.1.** *Performance specifications for the LUVOIR-A OTE, derived from science capabilities.*

| Specification | Value | Mapped Capabilities |
|---|---|---|
| Aperture Diameter | 15 meters | Sensitivity, Resolution, Flexibility |
| Field-of-View | 15 arcmin × 8 arcmin | Flexibility, Agility |
| Wavelength Range* | 100 nm – 2.5 μm | Sensitivity, Flexibility |
| Reflectivity | > 60% at 105 nm<br>> 90% at 115 nm<br>Average ~ 90% between 400 and 850 nm<br>Average > 95% between 850 nm and 2.5 μm | Sensitivity, Flexibility, Agility |
| Static Wavefront Error† | < 38 nm RMS | Resolution, Stability |
| Pointing Stability‡ | ± 0.3 mas 1-σ per axis over an observation | Resolution, Stability |
| Object Tracking | 60 mas / sec | Flexibility, Agility |
| Slew Rate | Required: Repoint anywhere in anti-sun hemisphere in 45 minutes<br>Goal: Repoint anywhere in anti-sun hemisphere in 30 minutes | Agility |

*The blue end of the wavelength range is limited by the coating performance, while the red end is limited by the telescope thermal background; longer wavelengths can be observed for sufficiently bright objects. †The static wavefront error is the end-to-end optical performance, including that of the instrument. To achieve an end-to-end 38 nm RMS wavefront error, the contribution from the OTE itself would necessarily need to be lower. ‡The pointing stability here is the requirement at the OTE focal plane and is achieved by a tiered approach incorporating the fast steering mirror, VIPPS, and spacecraft attitude control system. °Object tracking and slew rates levy requirements on the VIPPS and attitude control system, however we include them here for a complete picture of the OTE's capabilities.

with spherical, coma, and astigmatism aberrations corrected. The inclusion of the fast steering mirror also allows for ultra-fine pointing stability to be achieved by all of the instruments and relaxes some of the requirements on the Vibration Isolation and Precision Pointing System (VIPPS) to point the massive science payload.

The biggest disadvantage of a three-mirror anastigmat system is the two additional reflections that are needed (tertiary mirror, fast steering mirror) before light enters any instrument. This throughput reduction is compensated by both the collecting area of the telescope and the high reflectivity of the coating at UV wavelengths. The two extra reflections also present two additional surfaces that can be contaminated and further degrade the throughput in the UV science bands. That being said, eliminating the aberration correction offered by the tertiary mirror, and the pointing control of the fast steering mirror before light is directed to each of the instruments may result in significantly more complex instruments, each potentially requiring their own corrective optics and fast steering mirror system. The Study Team has weighed the throughput gains achieved by changing the OTE design against the resulting complexity of the individual science instrument designs and has decided to maintain the three-mirror anastigmat design of the OTE.

**Figure 9.3** shows a ray trace of the current OTE system, and **Figure 9.4** shows a wavefront error map over the telescope's field-of-view, with each instrument's field-of-view superimposed. **Figure 9.5** shows an example wavefront error allocation for the OTE+ECLIPS instrument system. Note that even with the ~25 nm RMS design residual for the OTE, the total system wavefront error still achieves diffraction-limited performance with margin at the system-level. All four mirrors are coated with a protected, "enhanced" Al+LiF coating to enable observations at wavelengths as short as 100 nm. A thin overcoat of $MgF_2$ or $AlF_3$ reduces the hy-groscopic sensitivity of the LiF layer (Quijada et al. 2014).

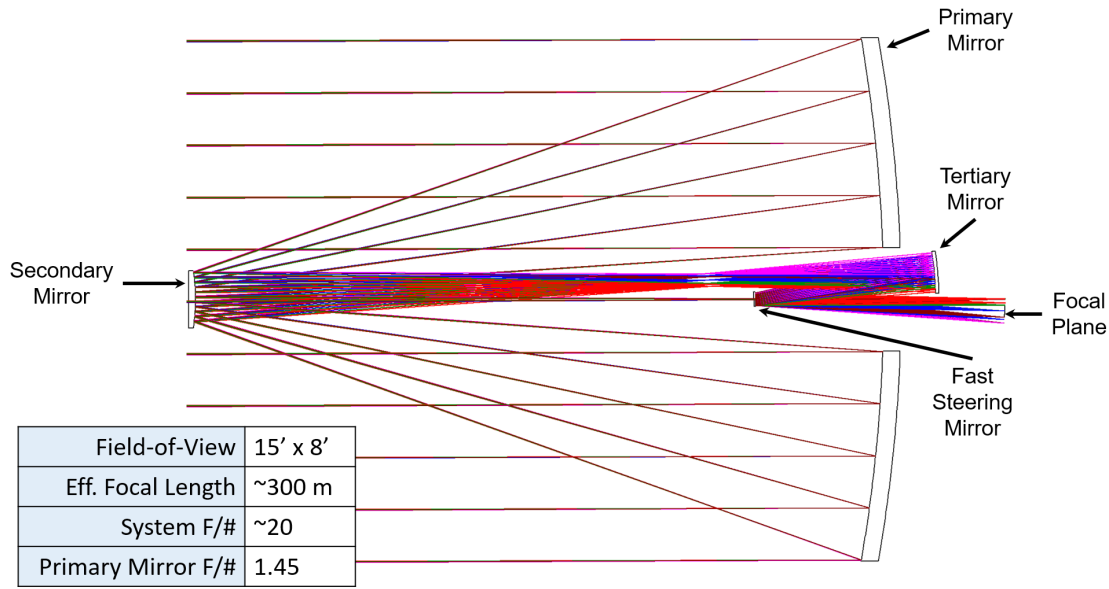

| Field-of-View | 15' x 8' |
| --- | --- |
| Eff. Focal Length | ~300 m |
| System F/# | ~20 |
| Primary Mirror F/# | 1.45 |

**Figure 9.3.** *Ray-trace of the LUVOIR OTE optical design. First-order optical parameters are shown in the inset.*

For this particular optical design, a sizeable central hole in the primary mirror was needed to pass the ray bundle through to the aft-optics system. The minimum achievable clearance hole was 2.7 m in diameter; the secondary mirror obscuration is entirely contained within this diameter. To optimally pack the mirror segments around this hole and minimize unused collecting area, while also maintaining the outer diameter of the primary mirror in the 15-meter range, a segment size of 1.15 m flat-to-flat was selected and arranged in five concentric rings, for a total of 120 segments. **Figure 9.6** shows an early design for the LUVOIR-A aperture, with annotated diameters.

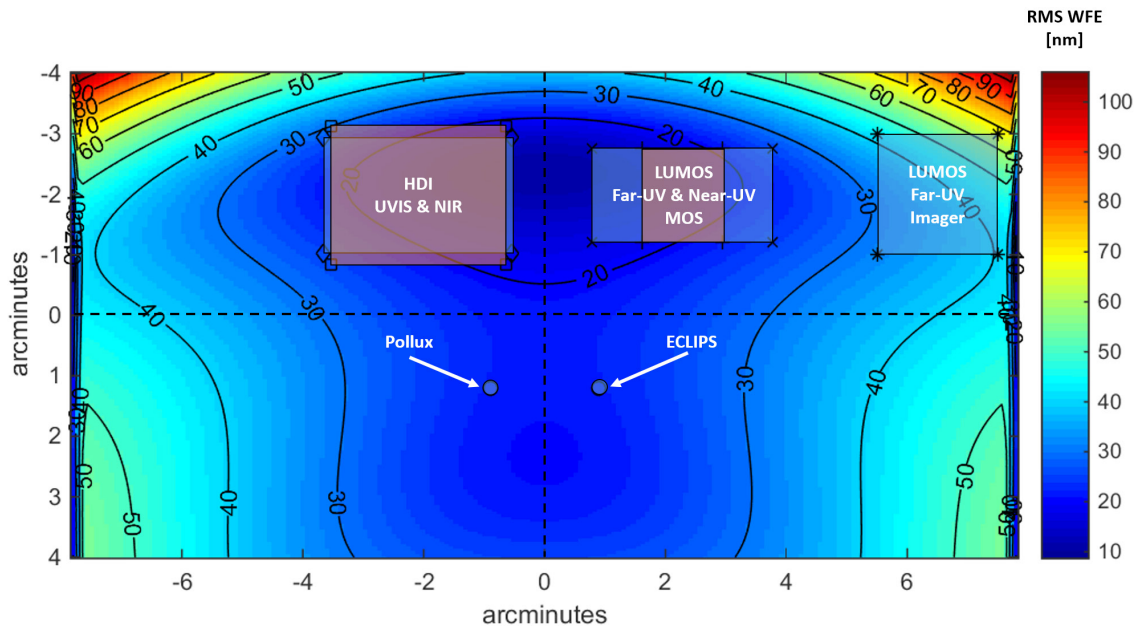

**Figure 9.4.** *The RMS wavefront error over the 15 x 8 arcmin OTE field-of-view. With the exception of the LUMOS far-UV imager, each instrument's field-of-view has a better-than-diffraction-limited wavefront from the telescope. The LUMOS far-UV imager uses its internal optics to correct the wavefront error from the telescope.*

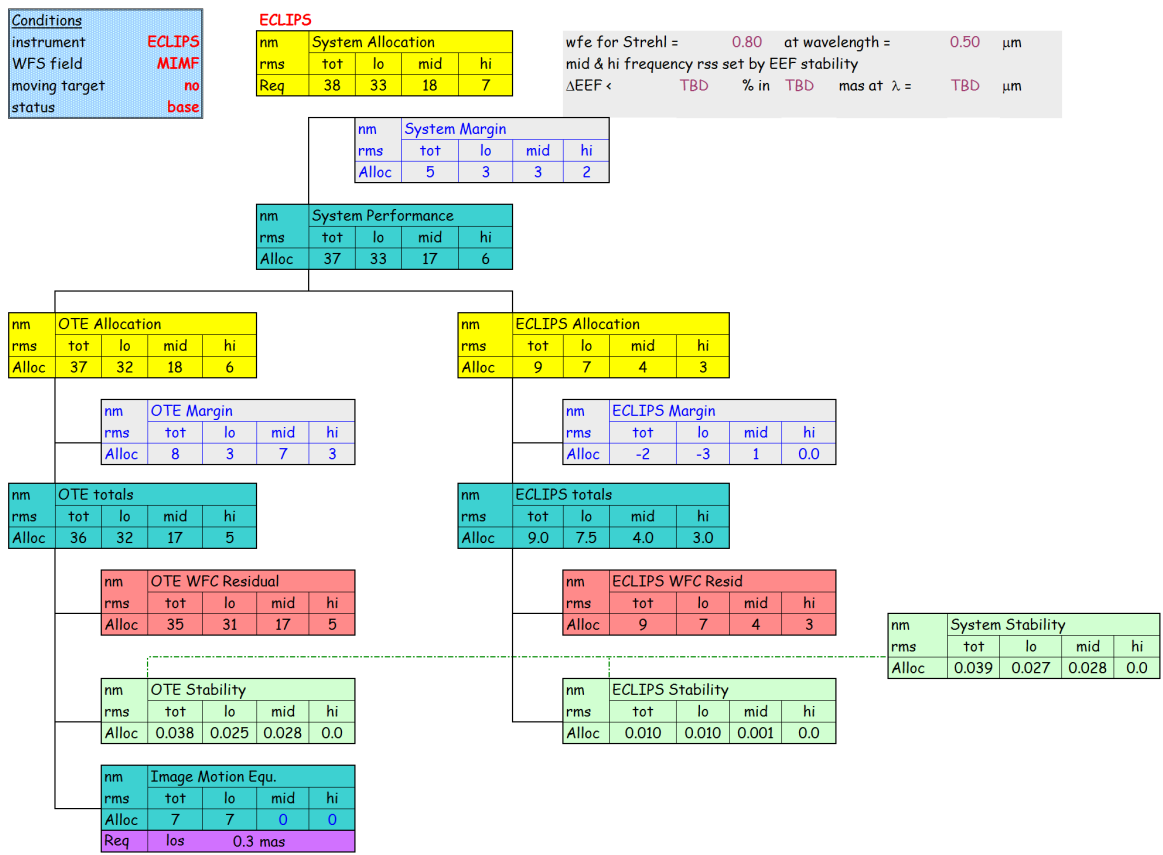

**Figure 9.5.** *Top-level wavefront error budget showing static and dynamic wavefront error allocations. In this example, the ECLIPS instrument is show for a reference. Yellow boxes indicate top-down allocations; teal boxes indicate bottom-up roll-ups. The static wavefront error roll-up is based on optical design residuals and engineering judgment on alignment and fabrication errors. In this example, system level stability is shown to be 40 pm RMS, however additional analysis is necessary to confirm the underlying assumptions.*

As with the number of reflections in the OTE, the central obscuration does little to degrade the general astrophysics science because the obscuration is balanced by the sheer size of the primary mirror. However, we have found that the obscuration substantially degrades the performance of the coronagraph instrument, specifically with respect to the "core throughput," or the percentage of planet photons that are focused into the central core of the planet point-spread function. Evidence indicates that the ratio of the circumscribing diameter of the obscuration to the inscribing diameter of the primary mirror should be on the order of 10–15% to maximize the performance of the on-axis telescope / coronagraph system. Additionally, most coronagraph masks, when optimized, will circularize the outer diameter of the aperture, discarding any portion of the aperture outside of the red-dashed circle in **Figure 9.6**.

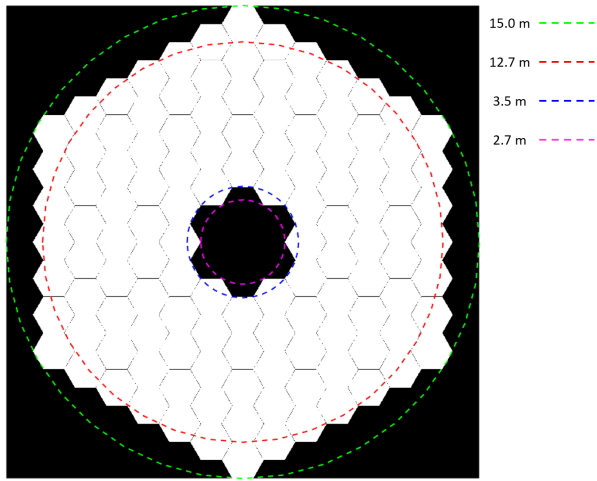

**Figure 9.6.** *The initial LUVOIR entrance pupil aperture. There are 120 segments, each 1.15 m flat-to-flat, with 6 mm gaps. Inscribed and circumscribed diameters are shown for reference.*

Variations on the aperture shown in **Figure 9.6** that address these issues have been explored and the telescope design has been adjusted. **Figure 9.7** shows the new LUVOIR-A aperture geometry, compared to the old one. The optical design of the OTE was also updated to accommodate the reduction in the central obscuration, although it still maintains a three-mirror anastigmat form. At the time of preparing this report, however, detailed opto-mechanical models of the new aperture have not yet been completed, so the figures throughout this report still reflect the old aperture design shown in **Figure 9.6**.

These types of design evolutions are currently morphing LUVOIR-A into a more efficient system that better balances the general astrophysics science and the search for habitable planets.

### 9.1.3.2 Mechanical design

While the design and implementation of the OTE draws heavily from the James Webb Space Telescope (JWST) in many ways, the size of LUVOIR compared to other space telescopes such as the Hubble Space Telescope (HST) and the JWST is immense, as shown in **Figure 9.8**.

The mechanical layout initially focused on maximizing the primary mirror area while minimizing deployment complexity while being constrained by the Space Launch

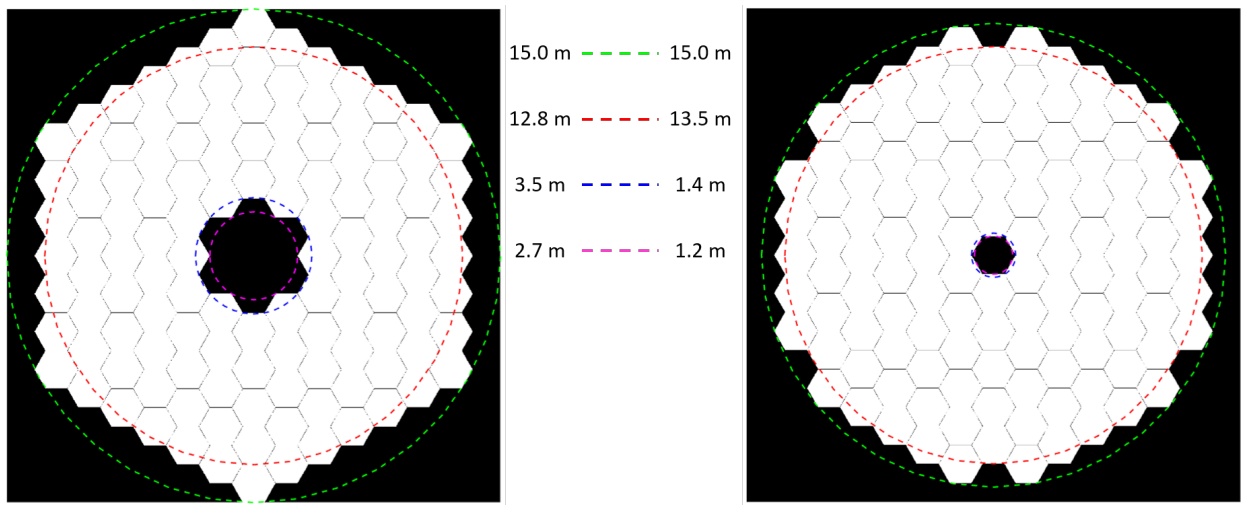

**Figure 9.7.** *A comparison between the initial LUVOIR-A aperture (left) and the new aperture (right). The new aperture achieves the same 15.0-m outer diameter while maximizing the inscribing diameter and minimizing the central obscuration. Both apertures have 120 segments, although the segment size increased from 1.15-m flat-to-flat (left) to 1.223-m flat-to-flat (right).*

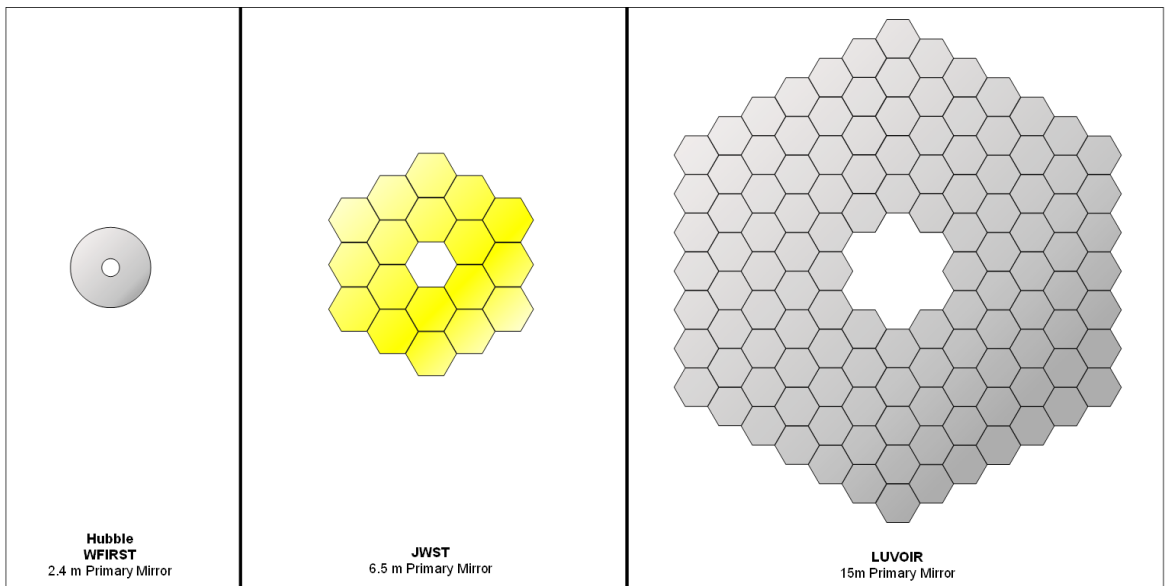

**Figure 9.8.** *The LUVOIR-A primary mirror compared to the HST/WFIRST and JWST primary mirrors.*

System fairing volume. In order to fit in the fairing, a JWST-like wing-fold deployment has been adapted to a dual-wing-fold on each side of the primary mirror as shown in **Figure 9.9**. This allows JWST-like deployment mechanisms, hinges, and latches to be used on LUVOIR. The dual-wing-fold geometry also allows the primary mirror to better conform to the circular diameter of the fairing, allowing more volume in the center of the fairing for the instrument stack. Similarly, the JWST single-hinge secondary mirror support structure (SMSS) deployment was adapted for LUVOIR, with two additional hinge points on the top beam of the SMSS, as shown in **Figure 9.10**. Finally, the aft-optics system was made to be deployable using a nesting-telescope structure to accommodate launch-vehicle envelope constraints.

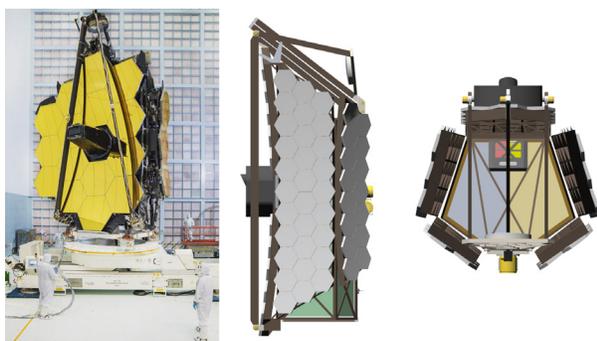

**Figure 9.9.** *Left: JWST "folded" each wing of the primary mirror along one axis in order to fit the observatory into the launch vehicle fairing (only a single wing is shown folded here). Right: LUVOIR builds upon this concept by using two "folds" per wing to fit a much larger primary mirror into the (larger) launch vehicle fairing. JWST photos courtesy NASA/GSFC.*

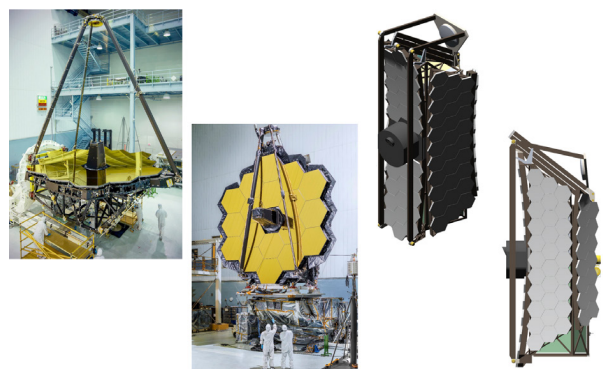

**Figure 9.10.** *Left: JWST's secondary mirror fully deployed. Center-Left: JWST's secondary mirror stowed so that it can fit into the launch vehicle. Right: LUVOIR uses a similar technique to stow the much longer secondary mirror support structure for launch. JWST photos courtesy NASA/GSFC.*

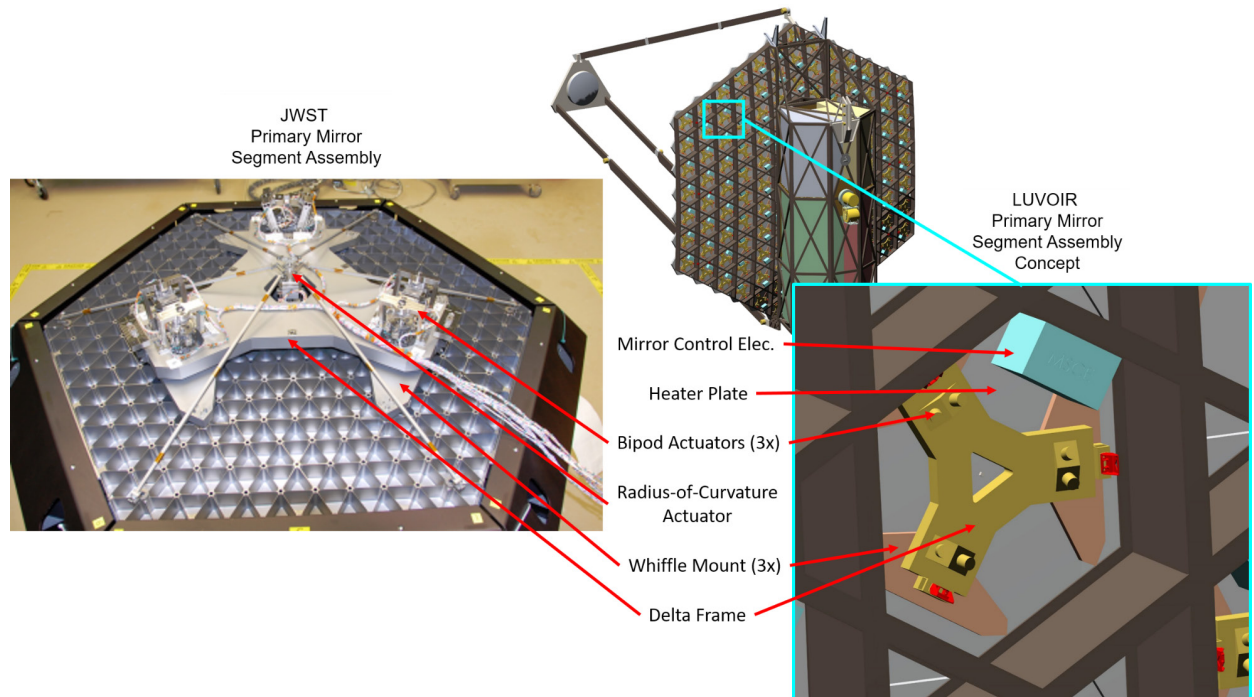

**Figure 9.11.** *Left: JWST primary mirror segment assembly, viewed from the back (Credit NASA/ GSFC). Right: LUVOIR's primary mirror segment assembly concept, showing many shared components. While JWST uses a radius-of-curvature actuator, LUVOIR does not. However, LUVOIR uses a heater plate to control the segment temperature and co-locates the mirror segment control electronics with each segment.*

The design of the primary mirror segment assembly architecture also leverages the JWST design, shown in **Figure 9.11**. A single, stiff mirror segment is mounted on a hexapod for 6 degree-of-freedom rigid body positioning of the segment. A whiffle structure and delta frame interfaces the mirror segment substrate to the hexapod actuators and to the primary mirror backplane support structure.

The LUVOIR segment assembly architecture departs from that of JWST in the materials, optimized for the stability requirements and the operating temperature. Instead of beryllium segments, LUVOIR has baselined Corning's Ultra-Low Expansion (ULE®) glass for the mirror substrate, although other low coefficient of thermal expansion (CTE) materials such as Zerodur may also be used. Additionally, LUVOIR's delta frame and whiffle will be zero-CTE composite material.

JWST required radius-of-curvature actuators on each segment because the mirrors could not hold their shape in a 1-g environment during integration and test, and therefore needed built-in correctability. For LUVOIR, further detailed analysis is required, but it is believed that the stiffness of the ULE

> LUVOIR's optical telescope element uses many design features that were developed for the James Webb Space Telescope such as segmented mirrors, deployable sections of the primary mirror, mirror actuation, and a deployable secondary mirror.

LUVOIR's nominal operating temperature of 270 K provides for several advantages over JWST including the manufacturing of the primary mirror segments as well as the qualification testing of the OTE.

mirrors, with their closed back design, will be substantially stiffer than JWST's open-backed beryllium mirrors, and therefore able to hold their prescribed shape in 1-g. Studies currently being conducted with mirror vendors will help to demonstrate if these assumptions are accurate. For now, LUVOIR is not baselining the use of a radius-of-curvature actuator.

Another critical new component to the segment assembly architecture is the edge sensor and piezoelectric transducer (PZT) actuator control system. Each mirror segment is fitted with one edge sensor per edge (i.e., two edge sensors total per shared edge) for a total of 622 edge sensors across the LUVOIR aperture. While the current design baselines capacitive-based edge sensors, optical or inductive sensors could also be used if sufficiently mature and capable. Each sensor measures local gap, shear, and dihedral angle between the segment edges. Incorporating these three measurements from all 622 sensors allows for a global solution to be found for the six degree-of-freedom position of each segment. The positions of each segment are then fed back to the hexapod fine-stage PZT actuators to control the segment position with picometer resolution (Saif et al. 2017). A discussion regarding both edge sensor and PZT technology development can be found in **Chapter 11**.

### 9.1.3.3 Thermal design

High-contrast imaging necessitates that the mirrors of the OTE be extremely stable. In order to eliminate the effects of thermal expansion and contraction, an ultra-stable thermal environment must be created. As mentioned in **Chapter 8**, the nominal operating temperature of LUVOIR has been set to 270 K. In another departure from JWST, the individual segment, secondary, and tertiary mirror assemblies incorporate an active thermal control system. A heater plate immediately behind the glass substrate radiatively heats the mirrors to 270 K±1 mK. The primary mirror backplane and secondary mirror support structures are also actively heated at key control points in order to maintain the global thermal stability of the OTE (Eisenhower et al. 2015; Park et al. 2017).

The nominal 270 K temperature has several advantages over JWST. First, by fabricating components at a temperature near where they are to be operated, modeling error can be reduced, as well as the complexity, number, and duration of thermal cycle tests. Second, performance testing of LUVOIR will be completed at a much warmer temperature than JWST which should ease the qualification campaign since it won't involve cryogenic tests.

### 9.1.3.4 Electrical design

The payload electrical architecture is shown in **Figure 9.12**. Each of the 120 mirror segment control electronics boxes are located within the OTE at each segment location, as illustrated in **Figure 9.11**. A block diagram of the mirror segment control electronics is shown in **Figure 9.13**. Due to the large number of these boxes (120×, plus one for the secondary mirror assembly), it is desirable to use low power designs and techniques for the individual boards to minimize the multiplying effect on the overall power demand. Circuit board size and mass will be minimized for the same reason. The high data rate of the edge sensor readout exceeds the capability of the 1553 bus, so a Low-Voltage Differential Signaling (LVDS) interface and data router is employed to transport the data to the Control System Processor, located inside the coronagraph instrument.

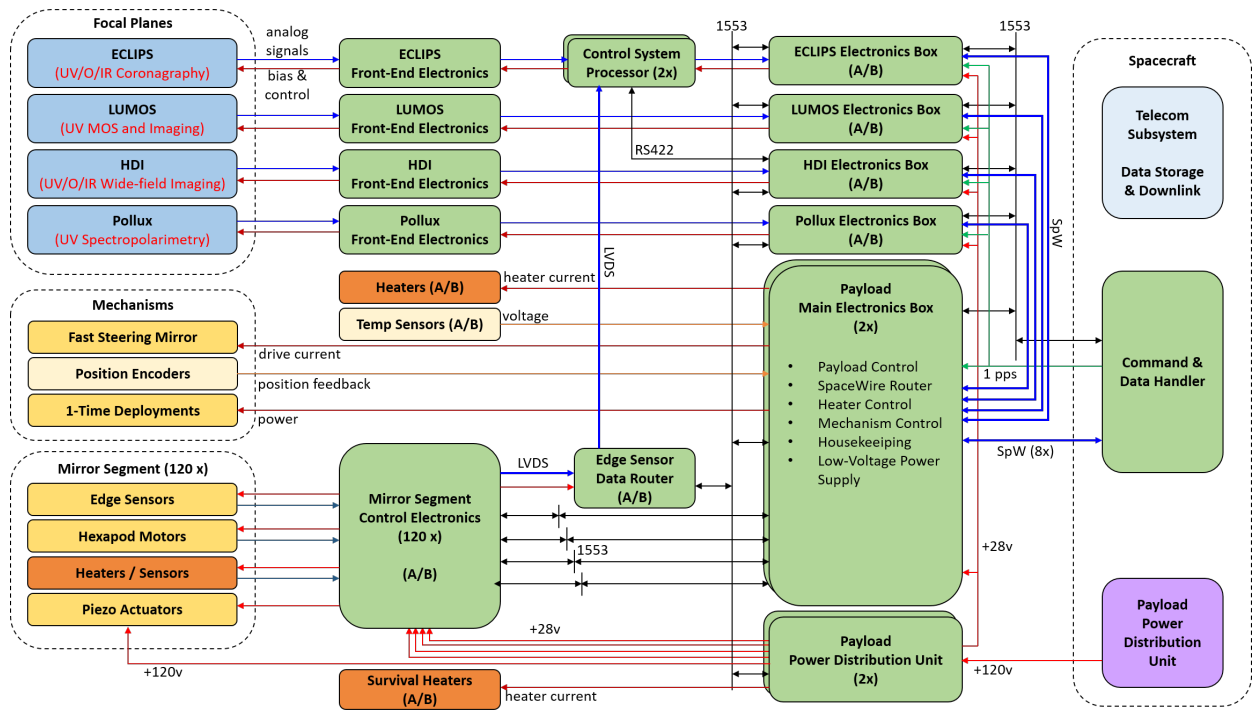

**Figure 9.12.** *The LUVOIR electrical system architecture. Primary data interfaces include 1553, low-voltage differential signaling (LVDS), SpaceWire (SpW), and RS422. Most systems receive 28-v power from the power distribution unit, with the exception of the mirror segment piezo actuators, which require 120-v power. "(A/B)" denotes electrical boxes that are board-level redundant; "(2x)" denotes boxes that are box-level redundant. Cross-strapping is not shown for clarity.*

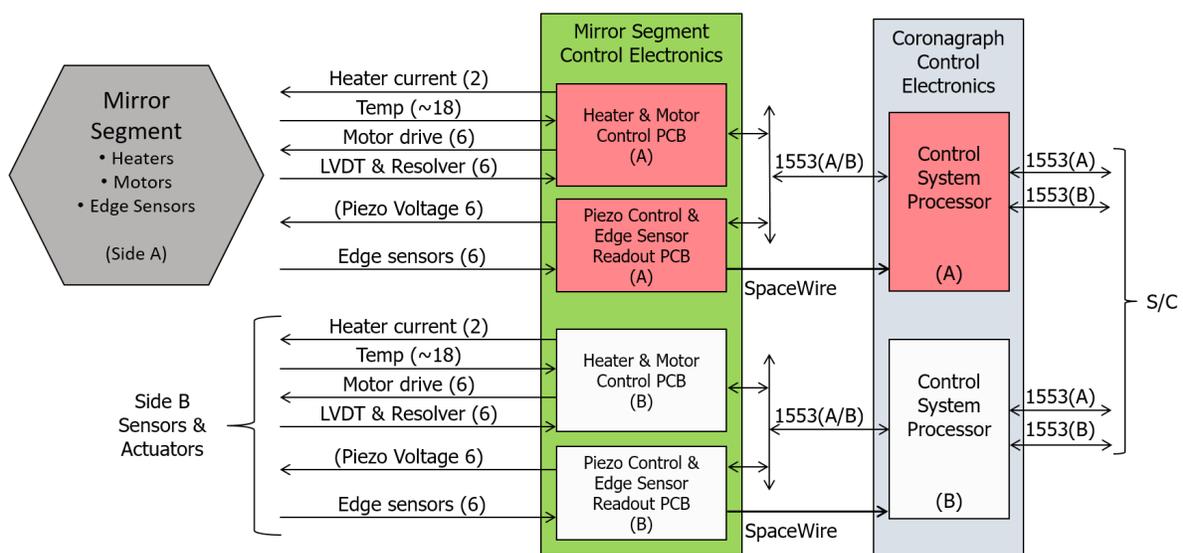

**Figure 9.13.** *Block diagram of the mirror segment electrical system. PCB: printed circuit board, LVDT: linear variable differential transformer (for actuator position measurement).*

## 9.2    Backplane support frame (BSF)

### 9.2.1    BSF overview

The backplane support frame (BSF), highlighted in **Figure 9.14**, is the primary mechanical interface between the instrument suite and the OTE. It must be robust enough to support both the large primary mirror as well as the entire suite of science instruments and their related support hardware. Additionally, it must provide modular interfaces for the science instruments to allow for serviceability.

### 9.2.2    BSF design drivers

Some of the key drivers that shape the implementation of the BSF include:

- Provide needed stiffness to the OTE during ground integration and test and on-orbit

- Provide modular interfaces for all science instruments and related hardware
- Provide housing for the payload main electronics box and the power distribution unit
- Provides launch supports for the stowed primary and secondary mirrors
- Provide an interface between the Payload and the Spacecraft
- Launch vehicle mass-to-orbit capability
- Launch vehicle volume

### 9.2.3    BSF design implementation

#### 9.2.3.1 Optical design

The BSF is primarily a mechanical structure, so there is no optical design related to it. Even so, the BSF serves as the primary metering structure between the OTE and the science instruments. Thus, optical alignment tolerances set requirements on the mechanical and thermal design of the BSF and instrument interfaces.

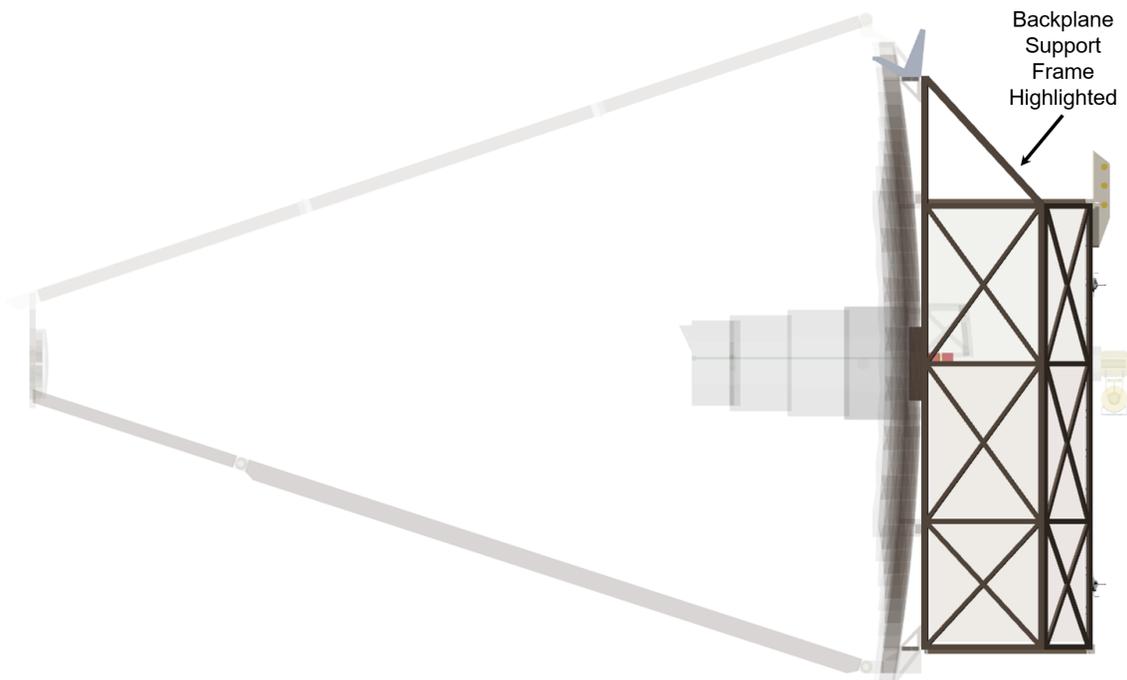

Backplane Support Frame Highlighted

**Figure 9.14.** *The LUVOIR Payload with the backplane support frame is highlighted.*

LUVOIR's BSF is an evolution and combination of design and features from both the Hubble Space Telescope and the James Webb Space Telescope.

### 9.2.3.2 Mechanical design

Like the combination of BSF and the integrated science instrument module (ISIM) structure on the JWST, LUVOIR's BSF would be a bonded composite, precision optical metering structure. The composite material provides a very stiff yet relatively lightweight and thermally stable structure.

LUVOIR is being designed with modularity in mind, which would enable the telescope to be serviceable should the technology exist to do so and should NASA decide to extend the life of the telescope beyond its initial core mission. While serviceability is not necessary to meet any of the science cases described in this report, it may be an attractive way to enhance or extend the life of a flagship observatory. However, because it is not required to meet the science goals, the servicing mission itself is outside the scope of this study and this report.

At this time, the design has two access doors, shown in **Figure 9.15**, which can open away from the science instrument bays

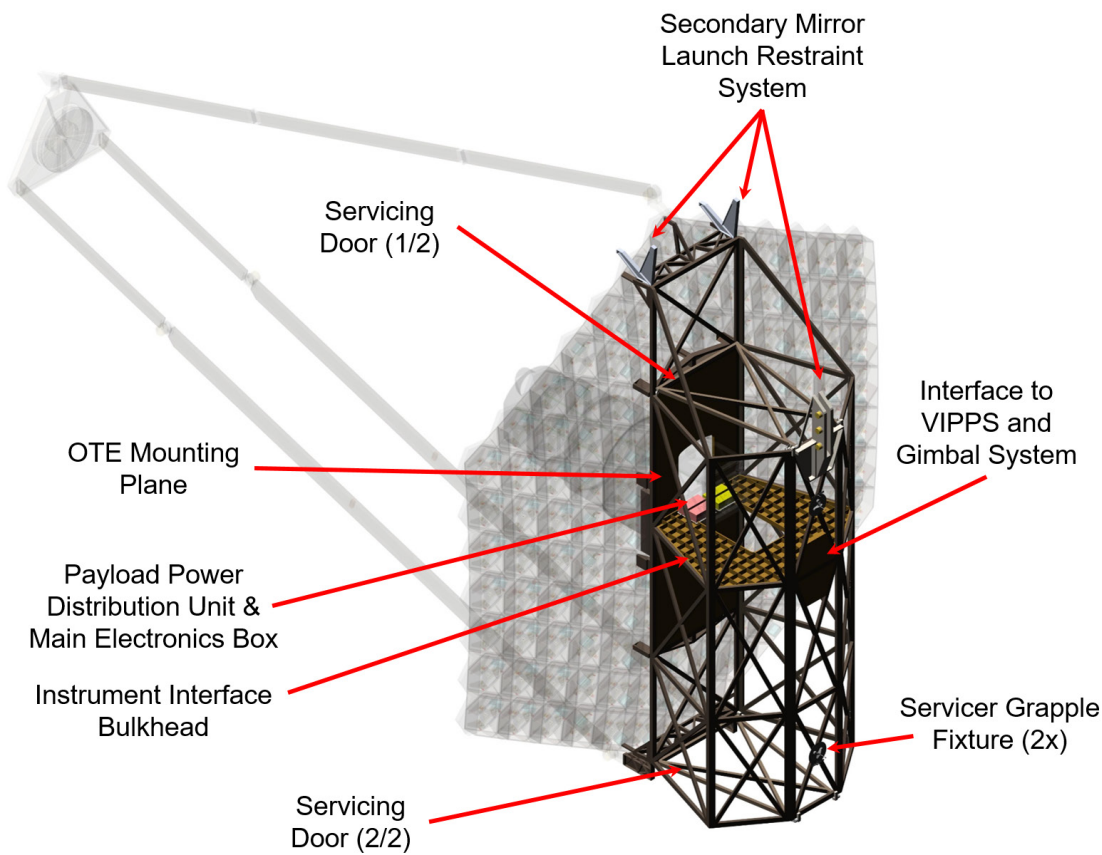

**Figure 9.15.** *Key features of the BSF. Not shown are instrument radiator panels that would be mounted to either side of the BSF, for passively cooling specific components to 270 K or 170 K. An additional radiator, used to cool 100 K components, would be mounted near the top of the BSF.*

Design modularity is a key feature of LUVOIR and is expected to make science instrument integration less complex. Modularity also enables the possibility of servicing LUVOIR, a feature that isn't necessary to achieve the science cases described in this paper. Rather, servicing allows for future upgrades which could enable science cases that haven't been thought of today.

providing access to the science instruments themselves. The instruments would be attached to the BSF at the "instrument interface bulkhead," also shown in **Figure 9.15**, via a guiderail and latching system similar to those used on HST. **Figure 9.16** shows a possible servicing concept for LUVOIR.

Despite LUVOIR's many similarities with the JWST, this modularity is a deviation from the JWST architecture. During the design phase of JWST, modular interfaces were considered for the science instruments and

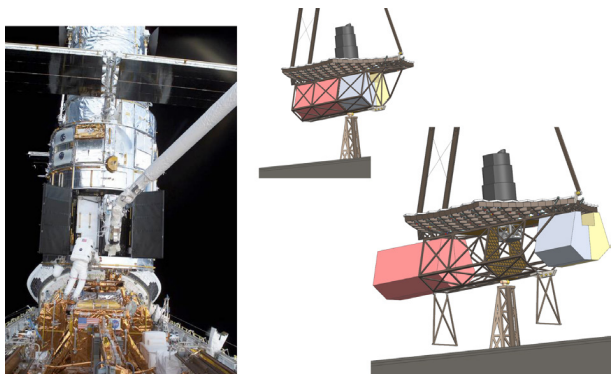

**Figure 9.16.** *LUVOIR's science instrument interfaces will be modular by design. Left: The Hubble Space Telescope was designed to be serviceable. Here, two of the doors to the science instruments are open and astronauts are working to replace an instrument (Photo credit: NASA). Right: LUVOIR's BSF is designed to be modular in nature, using similar interfaces as those on the Hubble Space Telescope. While not necessary to achieve the baseline science described in this report, this allows for the possibility of servicing in the future should the necessary infrastructure be available.*

ISIM structure, with designs that were an evolution of the Hubble latches. However, the modular interface concept on JWST was discarded for three primary reasons:

1. JWST was not designed to be serviceable,

2. The ISIM element integration and test plan—at the time—had only one installation of the flight instruments planned which didn't justify the level of effort for this highly complex interface, and

3. The initial architecture for the ISIM element was very asymmetric which didn't lend itself to easy access for modularity and adding it was non-trivial.

That being said, there is nothing inherent in a composite structure design that would prevent a modular interface from being implemented as long as it was designed to be modular from the start. Trying to retrofit a modular interface to an existing design would be very problematic.

### 9.2.3.3 Thermal design

The BSF must maintain alignment of the instrument modules with the OTE, and therefore must be both structurally and thermally stable. The individual composite beams of the structure will be instrumented with thermal sensors and heaters and wrapped in thermal blanketing, with a goal of maintaining a temperature of 270 K ± 1 K.

The external surface of the BSF structure will also carry radiators for the payload electronics. Portions of either side of the BSF will hold radiators to help maintain the nominal 270 K temperature of most of the Science Instrument systems, as well as radiators that will passively cool certain detectors and optical sub-benches to ~170 K. Finally, a small radiator mounted to the top of the BSF will be used to passively cool specific detectors to ~100 K.

### 9.2.3.4 Electrical design

The BSF holds both the payload power distribution unit and the payload main electronics box. The power distribution unit receives 120-v power provided by the spacecraft and distributes it to the various payload systems. The 120-v high-voltage is passed directly to the PZT actuators on the mirror segments. The power distribution unit also steps the 120-v power down to 28-v for the rest of the instruments and payload systems. The payload main electronics box serves several functions:

- Routes data from the individual storage within each Science Instrument to the spacecraft solid state recorder for buffering prior to downlink,

- Controls heaters and temperature sensors throughout the BSF and OTE,
- Controls all deployment mechanisms on the payload, and
- Provides telemetry and housekeeping data to the spacecraft command & data handler.

**Figure 9.12** shows the payload electrical system architecture and the interfaces between the main electronics box, power distribution unit, and the rest of the payload systems.

## 9.3    Science instruments

### 9.3.1    Overview

Several science instruments have been baselined for LUVOIR by the Study Team. They will all be housed in the BSF, shown in the area highlighted in **Figure 9.17**. Three of the instruments are being studied by the LUVOIR Study Team, while the fourth, POLLUX, is being studied by a European consortium led by the Centre National d'Etudes Spatiales (CNES). The three US-studied instruments are described in this section, while more details on POLLUX and its science objectives can be found in **Chapter 10**.

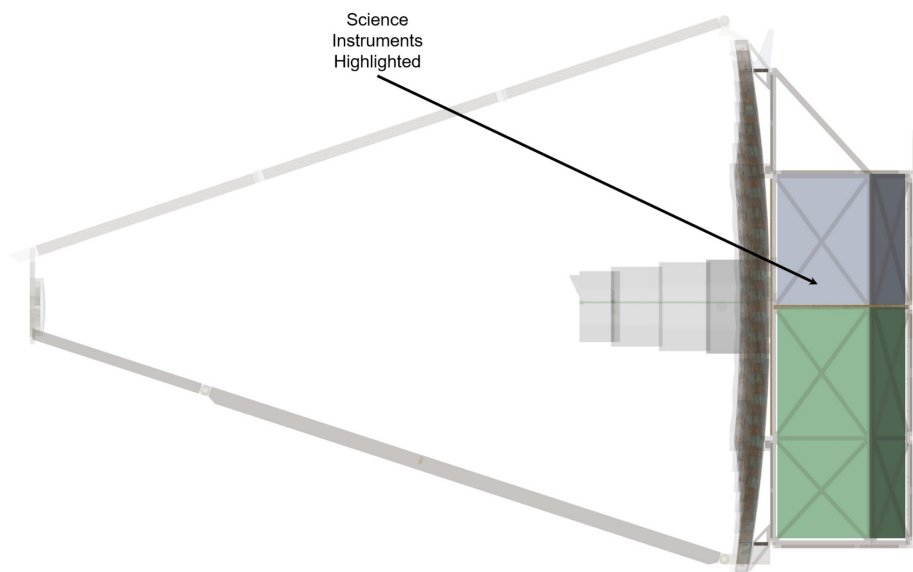

**Figure 9.17.** *The LUVOIR Payload with the science instruments highlighted.*

**Table 9.2.** *Science instrument capability summary.*

| Instrument | Channel | Wavelength Range | | Field-of-View | Angular Resolution | Spectral Resolving Power | Inner Working Angle (max.) | Outer Working Angle (min.) |
|---|---|---|---|---|---|---|---|---|
| | | nm | | arcmin | mas/pix | | λ/D | λ/D |
| ECLIPS | Total | 200 | 2,000 | | | | | |
| | UV | 200 | 525 | 0.006 Ø | 1.4 | 7 | 3.5 | 64 |
| | Vis | 515 | 1,030 | 0.016 Ø | 3.5 | 140 | 3.5 | 64 |
| | NIR | 1,000 | 2,000 | 0.030 Ø | 6.9 | 70, 200 | 2 | 64 |
| LUMOS | Total | 100 | 400 | | | | | |
| | Far-UV MOS | 100 | 200 | 3 × 1.6 | ~25 | 500 – 60,000 | | |
| | Near-UV MOS | 200 | 400 | 1.3 × 1.6 | ~25 | 40,000 | | |
| | Far-UV Imager | 100 | 200 | 2 × 2 | ~12.5 | | | |
| HDI | Total | 200 | 2,500 | | | | | |
| | UVIS | 200 | ~1,000 | 2 × 3 | 2.75 | ~600–700 | | |
| | NIR | ~1,000 | 2,500 | 2 × 3 | 8.25 | ~600–700 | | |

The Extreme Coronagraph for Living Planetary Systems (ECLIPS) is intended to survey sun-like stars in the local neighborhood and search for exoplanets within an annular region around the star defined by the inner and outer working angles (IWA and OWA respectively). It will directly image exoplanets via high-contrast imaging and characterize the atmospheres of those planets via spectroscopy of reflected light. Its emphasis is on the search for biosignatures on earth-like planets within the habitable zone, though all planets will receive some degree of characterization (Pueyo et al. 2017).

The LUVOIR Ultraviolet Multi-Object Spectrograph (LUMOS) is designed to support the UV science requirements of LUVOIR, from exoplanet host star characterization to tomography of circumgalactic halos to water plumes on outer solar system satellites. LUMOS offers multi-object spectroscopy across the UV bandpass, with multiple resolution modes to support different science goals (France et al. 2017).

The High Definition Imager (HDI) instrument is the primary astronomical imaging instrument for observations in the near UV through the near IR. It will be used to characterize stellar populations to rigorously test star formation theories; explore outer planet atmospheres; discover and characterize distant objects in the solar system; reveal the impact of the epoch of reionization on galaxy formation; visualize the evolution of galaxies; map dark matter; measure cosmological parameters with well-calibrated distance indicators out to the distance of the Coma Cluster; and detect 100s of exoEarths via astrometry.

A more detailed discussion of the current iterations of the science instruments follows. **Table 9.2** summarizes the general capabilities of the LUVOIR science instrument suite. It should be noted that the designs continue to evolve as the architecture for LUVOIR-A evolves.

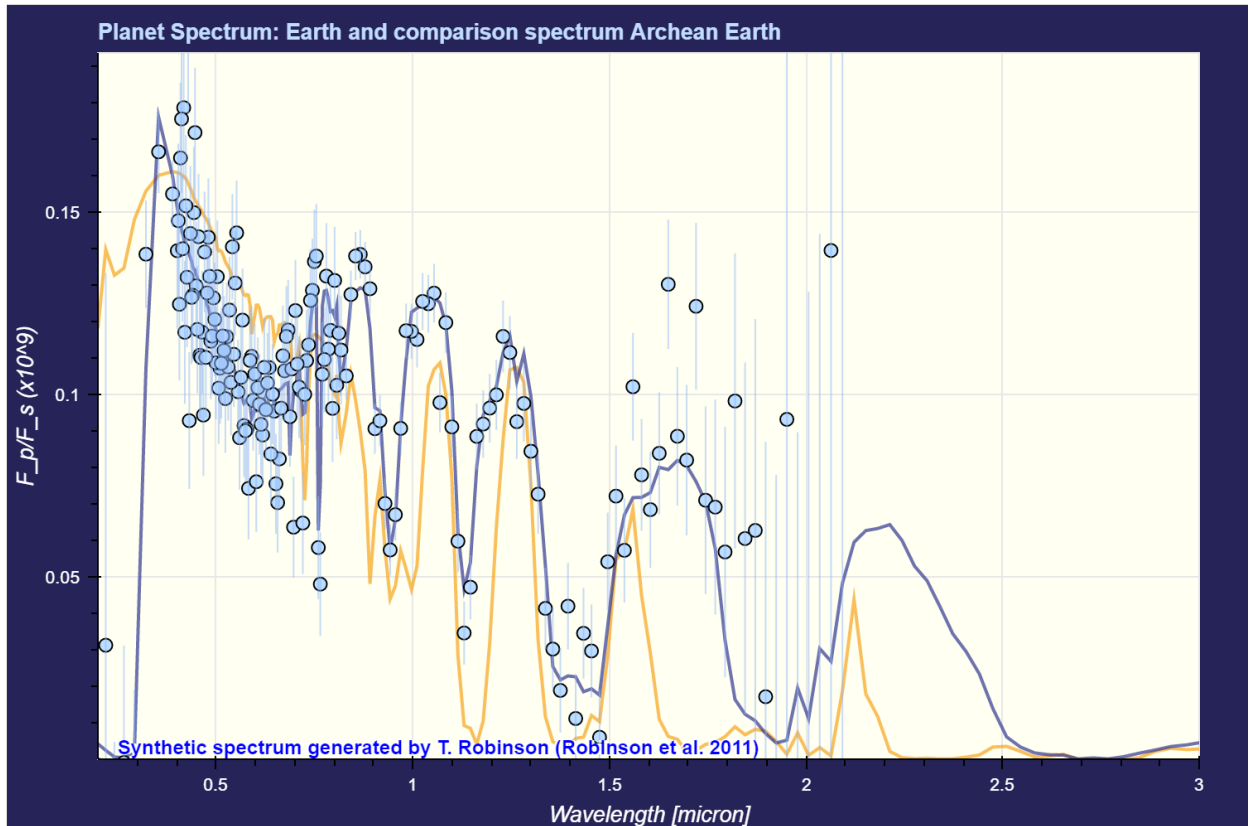

**Figure 9.18.** *Simulations of the spectra from an Earth-like planet around a sun-like star at 10 pc. The orange line depicts a simulated Archean Earth spectrum, the blue line depicts a simulated modern-Earth spectrum, and the points simulate data collected by LUVOIR-A with ECLIPS. Credit: LUVOIR Tools.*

### 9.3.2 Extreme Coronagraph for Living Planetary Systems (ECLIPS)

#### 9.3.2.1 ECLIPS overview and science goals

The scientific goals of the coronagraph instrument are commensurate with the ambitious philosophy underlying LUVOIR-A, and are organized around two key science themes:

(1) measuring the occurrence rate of biomarkers in the atmospheres of rocky planets orbiting in the habitable zone of their host stars, and

(2) studying the diversity of exoplanet systems.

The former science theme is significantly more stressing on the instrument and drives the trades we studied to design ECLIPS. Any mission aimed at measuring the occurrence rate of biomarkers in the atmosphere of nearby habitable-zone rocky planets, ought to first be capable of detecting a statistically significant ensemble of exoEarth candidates (Stark et al. 2015). The detectability of exoplanets in long coronagraph exposures depends on both the level of contrast achieved in the high-contrast region or "dark hole" of the focal plane, and on the throughput of the coronagraph at the apparent separation of the planets, expressed using the Inner and Outer Working Angle (IWA and OWA, respectively) scalar metrics.

The characterization of identified exoEarth candidates is equally as important

> ECLIPS will build on advances from the development of WFIRST's Coronagraph Instrument (CGI).
>
> WFIRST's CGI will mature high-contrast technologies and provide an end-to-end system-level demonstration of advanced starlight suppression in the presence of actual observatory disturbances as well as background sources such as exozodiacal light.

as their detection. **Figure 9.18** shows a simulated spectrum of a mature earth along with a 2-Gyrs-old Archean earth generated with the LUVOIR STDT online exoplanet spectrum simulation tool.[1] This example illustrates the most salient characteristics of the atmosphere of earth analogs which we seek to characterize with great precision using ECLIPS, and translate into the following three requirements on the instrument:

(1) Continuous spectral coverage from 200 nm to 2.0 μm in order to capture the spectral features associated with carbon and oxygen-based molecules, which help discriminate the various atmospheric compositions,

(2) Spectral resolution of R = 140 in the visible, and

(3) Spectral resolutions of R = 70 & 200 in the near-IR.

Spectroscopy of faint exoEarths beyond 1.6 μm will be limited to the closest and brightest targets due to the thermal background from the 270 K telescope. Nevertheless, redder spectral coverage will be invaluable to studying the details of the atmospheres of our nearest neighbors, as well as characterizing larger planets.

### 9.3.2.2 ECLIPS design drivers

ECLIPS is a complex instrument, as captured in **Figure 9.19** and **Figure 9.20**. At the top

level this complexity stems from several design drivers:

- The need to have continuous spectral coverage from 200 nm to 2.0 μm,
- The need to minimize the angle of incidence at all reflective surfaces to mitigate polarization aberrations,
- Packaging constraints due to instrument volume constraints, and
- Ultra-stable wavefront errors to enable long-duration high-contrast observations.

### 9.3.2.3 ECLIPS design implementation

There is currently no flight heritage for a coronagraph such as the one proposed for LUVOIR. There are several ground-based systems that use coronagraphs, however their overall requirements are far less stringent than those imposed on ECLIPS, and their contrast ratios are several orders of magnitude less. The WFIRST Coronagraph Instrument (CGI) is the most similar coronagraph to ECLIPS but it is still under development.

We expect that CGI development will help to advance the state-of-the-art in terms of how to successfully implement a coronagraph instrument on a space telescope, despite WFIRST and LUVOIR being very different observatories. The experience gained from integrating, testing, and operating a high-contrast coronagraph on-orbit will be invaluable. Some of the most critical pieces of information that will be learned by the operation of WFIRST CGI will be how to post-process high-contrast data with positively verifiable targets. Ground data processing has advanced significantly in the past 10 years but operates on large-signal data sources. The WFIRST CGI speckle field will be low signal and relatively uncorrelated. Performing transit spectroscopy with HST is an analogous example: this was a new field of data analysis, and the experience from working with Hubble data is now feeding directly into JWST. Similarly, working with real high-contrast data from WFIRST will inform the design of LUVOIR and ECLIPS.

---

1 See https://asd.gsfc.nasa.gov/luvoir/tools/

In order to accommodate the variety of high reflectivity coatings and detector technologies that span ECLIPS's wavelength range, the instrument is split into three channels that cover the following bandpasses: UV (200 to 525 nm), optical (515 nm to 1030 nm), and NIR (1.0 μm to 2.0 μm). These bands are achieved using a series of dichroics at the entrance of the instrument. The small overlap in wavelength coverage between each channel is used to calibrate detector gains between bands. Each channel is equipped with:

- Two deformable mirrors (DMs) for wavefront control,
- A suite of coronagraph apodizing, occulting, and Lyot stop masks
- A Zernike wavefront sensor to be used as a low-order or out-of-band wavefront sensor (LOWFS, or OBWFS),
- A spectral filter wheel to select the instantaneous science bandpass for the channel, and
- Separate science imagers and spectrographs.

While the three channels can operate in parallel, each channel can only observe in one bandpass at a time. **Figure 9.19** shows a side view of the three spectral channels in their layered configuration.

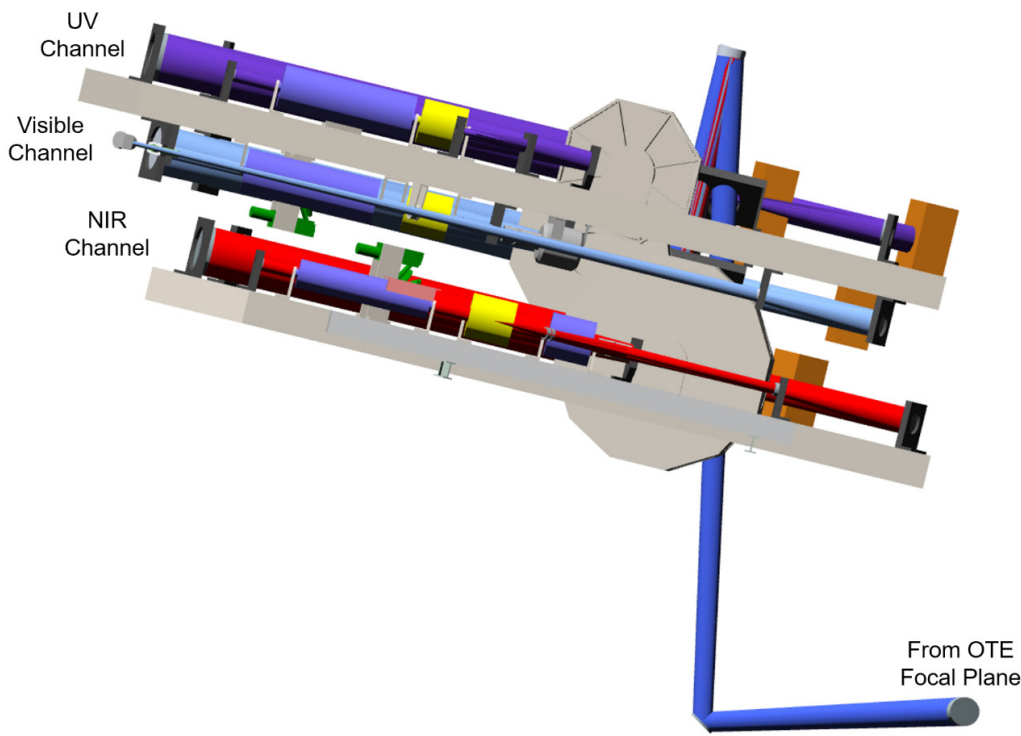

**Figure 9.19.** *Side view of the ECLIPS instrument showing the three spectral channels in a layered configuration. Light enters from the telescope in the lower right corner and a pair of crossed fold mirrors directs the light up to the instrument.*

Technology development for segmented-aperture coronagraphs is ongoing, and there are many promising candidates including shaped/apodized pupil Lyot coronagraphs, vector vortex coronagraphs, phase-induced amplitude apodization coronagraphs, and nulling coronagraphs. The landscape of coronagraph design may look very different at the time that an observatory such as LUVOIR is actually built, and the Study Team recommends that these candidates be re-evaluated then to maximize the possible science yield. At the time of this writing, however, the Study Team has chosen to design the ECLIPS instrument around the apodized pupil Lyot coronagraph (APLC) and vector vortex coronagraph architectures. Models suggest the APLC architecture currently delivers very good performance with LUVOIR-A's segmented, obscured aperture, and the masks and implementation directly leverage the significant investments in the WFIRST coronagraph instrument technology development effort (Eldorado Riggs et al. 2017; Mazoyer et al. 2016; N'Diaye et al. 2014; Zimmerman et al. 2016). Vector vortex coronagraphs provide advantages for distant unresolved stars. More details about each coronagraphs' strengths and weakness are available in **Chapter 11**.

As the name suggests, APLCs work by placing an apodizing mask in the pupil plane to correct for the segment and obscuration geometry of the telescope. The design of these masks is a complex optimization problem. Once the geometry of the focal plane mask and Lyot stop are set, a linear optimization is used to generate binary apodizing masks. Because it is difficult to design apodizing masks that cover the entire possible habitable zone with sufficient throughput, ECLIPS is designed with a suite of focal plane masks with varying inner and outer working angles. The masks will be chosen as a function of the distance to each observed host star.

To make the design robust to stellar angular size and pointing errors, the focal plane masks are designed that the inner edge of the dark hole is set to be smaller than the outer edge of the focal plane mask (N'Diaye et al. 2015). And to make the design robust to pupil misalignments, an input pupil with larger segment gaps and secondary struts than the actual telescope design is assumed. This baseline approach is conservative as it allows for large pupil shear between the plane of the apodizer and the primary mirror. Based on calculations using preliminary geometries similar to the LUVOIR-A primary mirror geometry, we predict that relaxing this constraint once the system stability is better understood will increase the throughput by about a factor of two.

### 9.3.2.3.1   Optical design

In the current design of the ECLIPS instruments, three mirrors are used to pick off the light from the OTE focal plane, collimate it, and re-image the telescope pupil onto the first DM in each channel. Once the light is collimated by these pre-optics, it is separated into each channel via a series of dichroic beamsplitters. The mirrors in each channel are coated according to their bandpass to provide the best reflectivity: the pre-optics and UV channel mirrors are coated in aluminum, the visible channel mirrors are coated in silver, and the NIR channel mirrors are coated in gold. To minimize polarization aberration effects within the instrument, all 90° fold mirrors are used in crossed-pairs, such that one mirror compensates the polarization effects of the other.

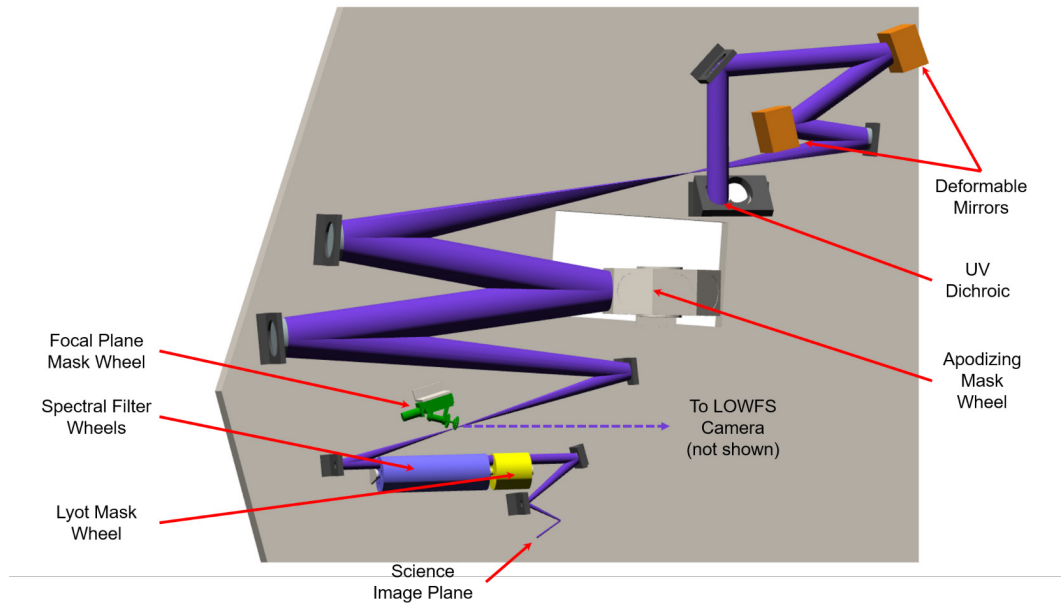

**Figure 9.20.** *This top-down view of the ECLIPS instrument shows the UV channel beam path. Light enters from the telescope in the upper right, and the UV dichroic directs the 200-525 nm light into the channel; longer wavelengths pass to the other channels below. After reflecting from the two deformable mirrors, two image relay mirrors re-image the pupil from the first DM onto the apodizing mask. A second pair of relay mirrors re-image that pupil onto the Lyot mask. A focal plane mask is located at the intermediate focus. A final set of imaging mirrors form an image at the science camera. The visible channel is similar, except a flip-in mirror directs light after the Lyot stop to either an imaging camera or an integral field spectrograph. The NIR channel is also similar, except after the Lyot stop a flip-in mirror directs light either to an integral field spectrograph, or a high-resolution slit spectrograph.*

**Figure 9.20** shows a top-down view of the UV channel opto-mechanical layout; the visible and NIR channels are very similar in design. Light enters from the telescope in the upper right corner and is relayed to the pair of DMs, with a pupil image being formed at the first DM in each channel. A pair of relay mirrors re-image the pupil from the first DM onto the apodizing mask. A second pair of relay mirrors re-images the pupil again onto the Lyot stop, and the coronagraph focal plane mask is located at the intermediate focus. After passing through the Lyot stop, the optical bandpass is further reduced to 10–20% by spectral filters for science observations, or 2% for wavefront sensing.

Depending on the channel, after passing through the spectral filters, the light can be sent to one of several back-end instruments. In the UV channel, the only back-end instrument is an imaging camera. In the visible channel, a flip-in mirror can be used to direct the light to either an imaging camera or an R=140 integral field spectrograph (IFS). In the NIR channel, a flip-in mirror can be used to direct the light to either an R=70 IFS, or an R=200 slit spectrograph. These back-end instruments were chosen to maximize the science capabilities of ECLIPS.

The visible and NIR IFSs are similar to those installed on ground-based instruments (Hinkley et al. 2011; Macintosh et al. 2008; Wildi et al. 2009) and the one envisioned for WFIRST (Demers et al. 2015). A high-resolution (R~1500) fiber-fed spectrograph (Mawet et al. 2017) was also considered.

While this solution is less mature, it was thought that it would better use detector real estate and could potentially yield higher resolution spectra for the brighter planets/ most nearby systems. Unfortunately, at this time the fiber-fed spectrograph has been found to significantly reduce throughput and increase serial characterization times of targets thus reducing efficiency and exoEarth yields.

In addition to the primary science beam path, each channel also supports a low-order /out-of-band wavefront sensor (LOWFS/ OBWFS). In LOWFS mode, the focal plane mask doubles as a Zernike wavefront sensing mask. The core of the point-spread function that is rejected by the focal plane mask is reflected into the LOWFS camera. A divot in the center of the focal plane mask provides the phase reference for the Zernike wavefront sensor (Shi et al. 2015). In OBWFS mode, the system operates much the same way, except instead of reflecting only the central core of the point-spread function, the mask reflects the entire point-spread function. Note, that using the OBWFS in a channel precludes science observations in that same channel. Thus, it is intended that at any given time, one of the three channels operates as an OBWFS for the entire instrument while science observations are being performed in the other two channels.

### 9.3.2.3.2  Mechanical design

ECLIPS is a very dense instrument, requiring a large number of optical elements, mechanisms, and detectors to be packaged in a relatively small volume. A central reference bench serves as the primary metering structure of the instrument. This reference bench holds the pre-optics that pick-off the light from the OTE focal plane and direct it to the individual channels. Two optical benches, aligned perpendicularly to the reference bench, hold the three instrument channels. One optical bench holds the UV and visible channels on either face; the second optical bench holds the NIR channel. **Figure 9.19** shows the two optical benches holding the three channels.

### 9.3.2.3.3  Thermal design

The entire instrument is held at 270 K; the exact required thermal stability of the optical benches and components is still being analyzed but is expected to be on the order of ±10 mK or better. Within the NIR channel, a sub-bench holding all elements after the Lyot stop is passively cooled to 170 K to help reduce the thermal background. The UV and visible channel EMCCD detectors are also passively cooled to 170 K; the NIR channel H4RG detectors are passively cooled to 100 K. As a contamination control approach, all heaters are also sized to allow for infrequent bake-outs of the optical elements on-orbit, similar to the annealing approach used on HST detectors.

### 9.3.2.3.4  Electrical design

The ECLIPS main electronics box also includes the control system processor (CSP). The CSP is the central brain for all control systems on LUVOIR, including the primary mirror segment edge-sensor / PZT control system discussed in **Section 9.1.3.2**, and any other metrology that may be included (see **Figure 8.8** for a control system block diagram that incorporates the CSP). The CSP is responsible for autonomously executing the wavefront control algorithms required to dig the high-contrast dark hole. The CSP uses a hybrid architecture consisting of a central processing unit to handle overall algorithm execution and data I/O, and a battery of field programmable gate arrays to perform dedicated, computationally intensive tasks such as matrix inversions or fast Fourier transforms.

### 9.3.2.3.5   Detectors

As with the coronagraph architectures, the technology development of detectors for exoplanet science is proceeding at a rapid pace. By the time LUVOIR is to be developed, it is likely that detectors with better sensitivity, less noise, and higher radiation tolerance will exist. However, to benchmark system performance and science yields now, the Study Team has selected a suite of detectors that exist today and only require a moderate amount of engineering development to be capable of achieving the science objectives described in earlier chapters. **Chapter 11** provides more detail on other detector technologies that may prove superior with sufficient technology development. The detector technologies are specific to each channel and our baseline choices rely heavily on technologies that have been used on previous missions and/or are planned in the WFIRST baseline instruments.

In the UV and visible channels, electron multiplying CCD (EMCCD) devices will be used, with the UV channel devices being δ-doped for better shortwave performance. Current devices are being developed and tested for the WFIRST coronagraph mission with 1k × 1k, 13 μm pixels. These 1k × 1k devices will be sufficient for the UV and visible channel imagers, however to cover the field-of-view and spectral resolution of the visible channel IFS, a 4k × 4k device will be required. In photon-counting mode, these devices produce <1 $e^-$ read noise, and 1×10$^{-4}$ $e^-$/pixel/second dark current (Nemati et al. 2016).

The baseline technology for the NIR channel is the Teledyne HAWAII 4RG (H4RG) HgCdTe detector. Current devices have a format of 4k × 4k with 10 μm pixels and are three-side buttable for building larger arrays. H4RGs are the baseline detector for the WFIRST wide-field instrument and are an evolution of H2RG detectors used on the JWST NIRCam instrument. Current devices have a median read noise of ~5 $e^-$/pix, and a median dark current: 2 × 10$^{-3}$ $e^-$/pix/sec. It is believed that the read noise can be further reduced to 1–2 $e^-$/pix with additional optimization of the readout circuit.

### 9.3.2.3.6   Deformable mirrors

Our choice of DM technology is based on the need for high-density devices to achieve large outer-working angles. For instance, achieving high contrast for the outer region of the habitable zone around the most nearby stars requires an outer-working angle of ~48 λ/D. Providing wavefront control over a dark hole of that size requires ~96 actuators across the pupil. Imaging of outer giant planets drives the actuator count higher, to 128 actuators across the pupil to achieve a ~64 λ/D outer working angle. In order to keep the optical design compact, we baselined a 128 ×128 actuator micro-electro-mechanical systems (MEMS) DM device. A secondary technical driver is the fact that such compact DMs can achieve small Fresnel numbers, and thus are more amenable to DM-based correction of amplitude errors. While this technology will not be matured by WFIRST at the component level, there are avenues to do so using smaller satellites (Cahoy et al. 2014).

### 9.3.2.3.7   Element select mechanisms

Each of the three channels is equipped with a series of element select mechanisms that accommodate the various apodizing, focal plane, and Lyot masks, as well as spectral filters. The number of elements in these mechanisms is driven by two considerations. First, each channel can only operate over an instantaneous 10–20% spectral bandpass at a time, due to limitations on wavefront control techniques and mask designs, requiring between six and eight spectral filters per channel. Each of those 10–20% bands must further be reduced to three 2% bands for collecting the wavefront control images. With additional neutral density filters, each spectral channel has between 29 and 37 filters alone. The second consideration is that each combination of spectral bandpass, IWA, and OWA, requires a specific set of apodizing, focal plane, and Lyot stop masks. Each channel has between 8 and 11 mask combinations to enable different optimized high-contrast regions on the focal plane.

### 9.3.3 LUVOIR Ultraviolet Multi-Object Spectrograph (LUMOS)

#### 9.3.3.1 LUMOS overview and science goals

A ubiquitous theme that has emerged in the science definition phase has been the study of gas in the cosmos, its relationship to (and evolution with) star and galaxy formation, and how this gas is transferred from one site to another. Understanding the flow of matter and energy from the intergalactic medium to the circumgalactic media, and ultimately into galaxies where it can serve as a reservoir for future generations of star and planet formation, is essentially a challenge in characterizing the ionic, atomic, and molecular gas at each phase in this cycle.

LUVOIR will be capable of characterizing the composition and temperature of this material in unprecedented scope and detail; on scales as large as the cosmic web and as small as the atmospheres of planets around other stars. The common denominator for this science case is that the strongest emission and absorption lines in the gases to be studied—therefore the highest information content for understanding the physical conditions in these objects—reside at ultraviolet wavelengths, roughly 100–400 nm. The LUVOIR Ultra-Violet Multi-Object Spectrograph (LUMOS) instrument is designed to make revolutionary observational contributions to all of the disciplines that call for high-resolution spectroscopy, multi-object spectroscopy, and imaging in the ultraviolet bandpass.

#### 9.3.3.2 LUMOS design drivers

Like the other systems on LUVOIR, the LUMOS instrument design is constrained by launch vehicle mass-to-orbit capability and fairing volume. But the most significant design driver for LUMOS is throughput at far-UV wavelengths. Although recent developments in mirror coating prescription and processes have improved performance at short wavelengths, the reflectivity in the far-UV is still only a fraction of what it is for most metallic coatings in the visible and NIR.

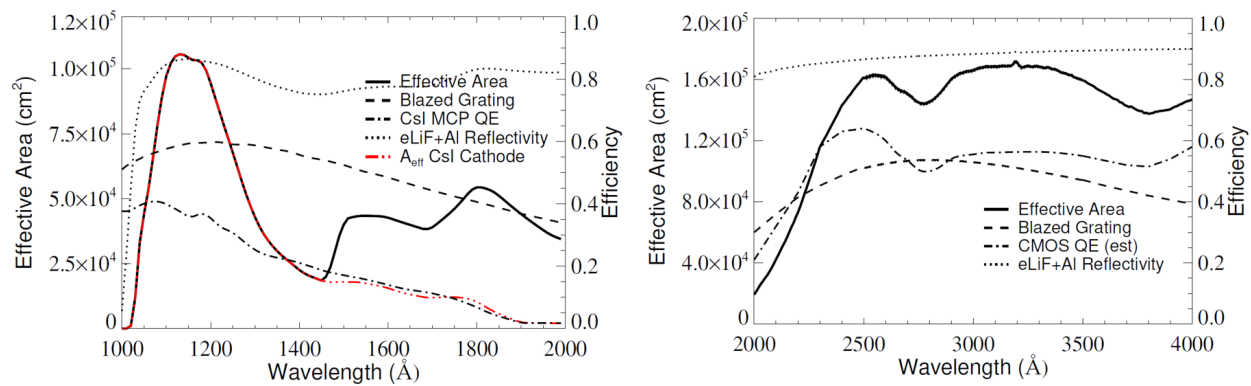

**Figure 9.21.** *Effective area of the LUMOS far-UV (left) and near-UV (right) multi-object spectrograph modes.*

Thus, the optical design must be carefully optimized to reduce the number of surfaces in the system, and contamination must be minimized to ensure as many photons as possible make it through the system to the detector.

### 9.3.3.3 LUMOS design implementation

LUMOS is a highly multiplexed ultraviolet spectrograph, with medium and low-resolution multi-object imaging spectroscopy and far-UV imaging modes. LUMOS can be thought of as an analog to the successful HST Space Telescope Imaging Spectrograph (STIS) instrument, with two orders-of-magnitude higher efficiency, multi-object capability, and a wide-field multi-band imaging channel (France et al. 2016). Coupling the relatively high instrumental throughput with a factor of ~40× gain in collecting area over HST ([15 m / 2.4m]$^2$ = 40), LUMOS can reach to limiting fluxes of order 100–1000 times fainter than currently possible with the HST Cosmic Origins Spectrograph (COS) and STIS instruments. **Figure 9.21** shows predicted effective area curves for the multi-object spectrograph

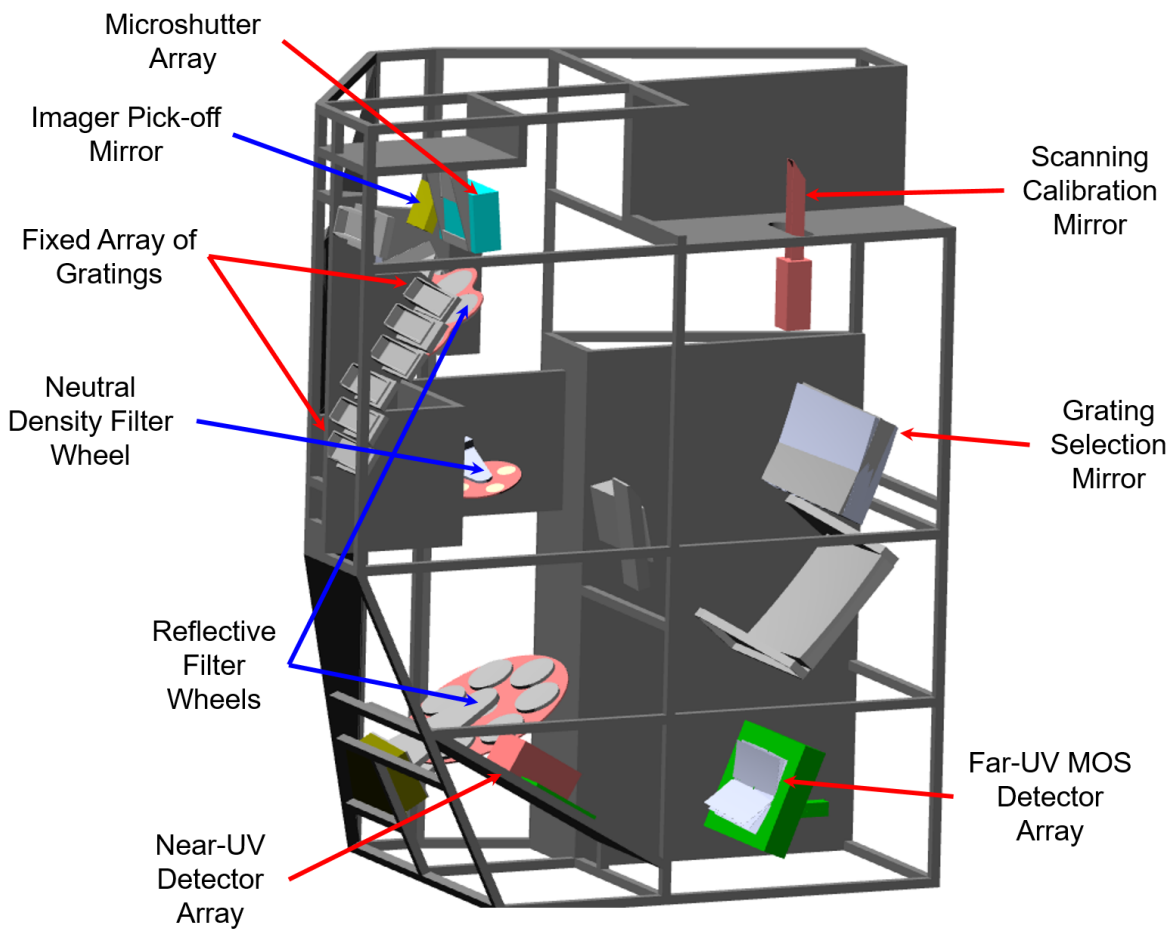

**Figure 9.22.** *Side view of LUMOS showing key components. Items with a red arrow correspond to elements in the multi-object spectrograph channel; items with a blue arrow correspond to elements in the imaging channel. Note that the far-UV imaging channel detector is hidden behind the bulkhead of fixed gratings. The scanning calibration mirror is retractable to move it out of the beam path.*

modes. **Figure 9.22** shows the layout of the LUMOS instrument.

#### 9.3.3.3.1 Optical design

##### 9.3.3.3.1.1 Multi-object imaging spectrograph

**Figure 9.23** shows the light path through the multi-object spectrograph (MOS) channel. Light from the OTE enters from the upper right. The entrance aperture for the spectrograph is a 3 × 2 grid of microshutter arrays (Li et al. 2014). The microshutter arrays build on the design of the Near Infrared Spectrograph instrument on JWST (Kutyrev et al. 2008), with 6 individual tiles of 420 × 840 shutters with individual shutter dimensions of 100 μm × 200 μm. The microshutter array grid defines the field-of-view for multi-object spectroscopy: 3' × 1.6' for the far-UV modes and 1.3' × 1.6' for the near-UV mode. In both modes, light passing through the microshutter array is directed onto a series of six fixed gratings via a fixed convex biconic optic and a second aberration-correcting toroidal steering mirror. Piston, tip, and tilt control of second mirror selects which of the six gratings is illuminated at any given time.

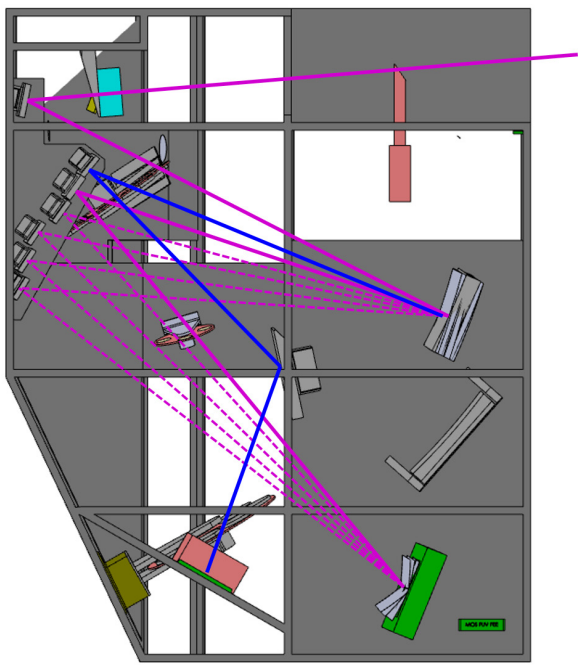

**Figure 9.23.** *Ray path through the multi-object imaging spectrograph channel of LUMOS. Light enters from the telescope at the upper right corner and passes through the microshutter array (teal box), which is located at the OTE focal plane. Light is then directed to the grating selection mirror, which is actuated in piston, tip, and tilt to direct the light at one of six, fixed gratings. The bottom five gratings are used for far-UV observations and direct light toward the far-UV detector array, which is also actuated in tip and tilt depending on the selected grating. The top grating is for near-UV observations and directs light to the near-UV detector via an additional relay mirror.*

Of the six fixed gratings, five are used for far-UV (100–200 nm) spectroscopy, and include medium, low, and very low spectral resolution options. The sixth grating provides a medium-resolution near-UV (200-400 nm) option. **Table 9.3** summarizes the spectral resolution, bandpass, angular resolution, and field-of-view for each mode of the multi-object spectrograph. All of the far-UV MOS modes are focused onto a 2 × 2 array of large-format microchannel plate detectors. The near-UV MOS mode is focused onto a 3 × 7 array of δ-doped CMOS devices, similar to those used in the HDI UVIS channel (see **Section 9.3.4.3.5**).

##### 9.3.3.3.1.2 Far-UV imaging

The majority of the LUVOIR imaging science is addressed through the HDI instrument (200 nm–2.5 μm, see **Section 9.3.4**), and LUMOS will provide a complimentary far-UV imaging capability from 100–200 nm. The LUMOS far-UV imaging aperture is physically offset from the MOS microshutter array aperture, and light from the OTE enters this channel through an unobstructed open aperture. **Figure 9.24** shows the light path through the imaging channel. Two aberration-correcting optics direct the light through two identical reflective filter wheel assemblies that serve to define the imaging bandpass in this mode. A neutral-density filter wheel is also inserted between the two filter wheels to accommodate far-UV bright object protection, target acquisition for the MOS channel, and safe imaging of the target field through the microshutter array for shutter selection. The images are recorded on a single 200 mm × 200 mm microchannel plate detector. **Table 9.3** summarizes the bandpass and angular resolution achieved by the FUV imaging channel.

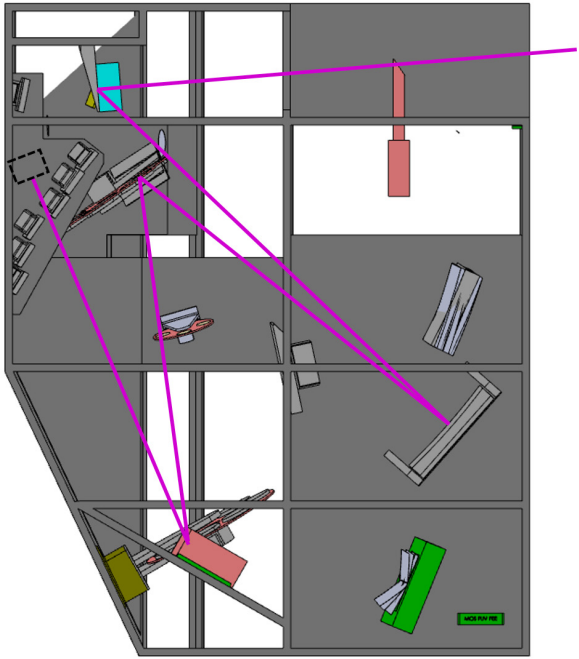

**Figure 9.24.** *Ray path through the LUMOS imaging channel. In this view, the imaging channel is one layer "below" the MOS channel – thus the imaging channel is offset from the MOS channel in the telescope field-of-view. Light enters from the telescope at the upper right and is reflected from a fixed mirror (hidden behind the microshutter array). The light then reflects from another fixed mirror to two reflective filters that define the optical bandpass. Two filters are used in series to improve the out-of-band rejection of the filters. A neutral density filter can be placed via a filter wheel between these two reflective filters for bright object protection. After the second reflective filter, light is directed to the far-UV imaging detector. The location of the detector, hidden behind a bulkhead in this view, is indicated by the dashed box.*

### 9.3.3.3.1.3 Cross-over mode

LUMOS performs bright object protection by pre-imaging the target field through the microshutter array. LUMOS contains a 'cross-over mode' that uses the grating selection mirror to direct light to the first filter wheel assembly in the imaging channel, through a neutral density filter wheel, and to the far-UV imaging detector. In this way, target brightness is quantified prior to spectral imaging acquisition with the MOS (imaging target acquisitions can also be made through the neutral density filter for bright-object-protection in the far-UV imaging mode). Imaging the spectroscopic target field through the microshutter array has a second benefit: it enables autonomous microshutter selection and fine-guidance adjustments to acquire the target through the selected shutter/slit. Operationally, target acquisition is achieved as following:

1. The payload slews to the target field using the VIPPS, gimbal system, and/ or spacecraft attitude control system, as needed.

2. The target field is imaged through the microshutter array (with all shutters open) in cross-over mode with the darkest neutral density filter in place for bright object protection.

3. As needed, the neutral density filter is reduced until adequate signal is achieved.

4. User-supplied celestial coordinates for the target objects are referenced to local detector coordinates, and the instrument autonomously identifies those targets in the imaged field.

**Table 9.3.** *The instrument parameter goals for each LUMOS mode. The target value is shown on top, the average value at the center of the bandpass or field-of-view (FoV) is shown second in parentheses, and the average value over 80% of the bandpass or FoV is shown third in parentheses and italics. This last value demonstrates that LUMOS achieves the spectral and spatial resolution goals across the majority of its spectral and spatial detector area.*

| Instrument Parameter | G120M | G150M | G180M | G155L | G145LL | G300M | FUV Imaging |
|---|---|---|---|---|---|---|---|
| Spectral Resolving Power | 30,000 (42,000) (*30,300*) | 30,000 (54,500) (*37,750*) | 30,000 (63,200) (*40,750*) | 8,000 (16,000) (*11,550*) | 500 (500) | 30,000 (40,600) (*28,000*) | N/A |
| Optimized Spectral Bandpass | 100-140 nm (92.5-147.4 nm) | 130-170 nm (123.4 -176.6 nm) | 160-200 nm (153.4- 206.6 nm) | 100-200 nm (92.0- 208.2 nm) | 100-200 nm | 200-400 nm | 100-200 nm |
| Angular Resolution | 50 mas (11 mas) (*17 mas*) | 50 mas (15 mas) (*19.5 mas*) | 50 mas (17 mas) (*24 mas*) | 50 mas (15 mas) (*27.5 mas*) | 100 mas (32 mas) | 50 mas (8 mas) (*26 mas*) | 25 mas (12.6 mas) (*12.6 mas*) |
| Temporal Resolution | 1 msec | 1 msec | 1 msec | 1 msec | 1 msec | 1 sec | 1 msec |
| Field-of-View | 2' × 2' (3' × 1.6') | 2' × 2' (3' × 1.6') | 2' × 2' (3' × 1.6') | 2' × 2' (3' × 1.6') | 2' × 2' (3' × 1.6') | 2' × 2' (1.3' × 1.6') | 2' × 2' (2' × 2') |

5. Fine pointing adjustments are autonomously computed to align the target objects with microshutters. VIPPS and fast-steering mirror commands are generated to achieve the fine pointing adjustments.

6. Unused shutters are closed, and the instrument switches to the appropriate MOS-mode configuration to begin science observations.

### 9.3.3.3.2 Mechanical design

LUMOS is essentially two instruments in the same housing, and in very close proximity. The mechanical design uses a composite truss structure to hold the optical elements in place without obstructing the beam paths of neighboring channels.

### 9.3.3.3 Thermal design

Contamination control is a primary concern for the LUMOS instrument, thus it is designed to operate at a slightly warmer temperature (280 K) than the surrounding structure and instruments (270 K). Thus, contaminants will be more likely to stick to the colder surfaces, and not the warmer LUMOS mirrors. Additionally, heaters on all mirrors, gratings, and detectors are sized to allow for infrequent on-orbit bake-outs, to help evaporate any accumulated water vapor from the optics.

### 9.3.3.3.4 Electrical design

In addition to the standard mechanism and detector control electronics, a dedicated electronics box is included to control the microshutter array. A high-voltage power supply is also included to drive both the calibration lamp sources, as well as the microchannel plate detectors.

### 9.3.3.3.5 Detectors

The far-UV MOS and imager focal planes consist of large-format microchannel plate detectors, which leverage a long heritage of reliable performance on multiple NASA space and sub-orbital missions. The far-UV imaging channel detector is a single 200 mm × 200 mm microchannel plate with 20 μm

The LUMOS instrument is designed to make revolutionary observational contributions to all of the disciplines that call for high-resolution spectroscopy, multi-object spectroscopy, and imaging in the ultraviolet bandpass. LUMOS will reach limiting fluxes on the order of 100 – 1000 times fainter than currently possible with the HST Cosmic Origins Spectrograph and Space Telescope Imaging Spectrograph instruments.

spatial resolution and employing a cross-strip anode readout system (Ertley et al. 2016; Vallerga et al. 2014), which enhance detector lifetime. The imaging microchannel plate will have an open-face design with an opaque CsI photocathode that has high flight heritage on the HST Cosmic Origins Spectrograph instrument and numerous rocket missions (Hoadley et al. 2016).

The far-UV MOS focal plane consists of a 2 × 2 array of 200 mm × 200 mm microchannel plates, with a 12 mm gap between the tiles. These microchannel plates again use a cross-strip anode readout and have 20 μm spatial resolution. As the MOS will disperse spectra across this tiled array, each side of the array can be tailored to the specific wavelengths that will fall there. The blue side of the array will use the same open-faced CsI photocathodes as the imaging channel, while the red side of the array will use sealed-tube bialkali photocathodes (Ertley et al. 2016). The near-UV MOS focal plane consists of a 3 × 7 array of δ-doped CMOS detectors with 8,192 × 8,192, 6.5-μm pixels per tile.

#### 9.3.3.3.6   Coatings

LUMOS optics will use the same protected "enhanced" LiF coatings as the LUVOIR OTE using a high-temperature Al+LiF deposition technique (Quijada et al. 2014). These coatings have demonstrated > 85% normal-incidence reflectivity between 103 and 130 nm. An additional thin protective layer of $MgF_2$ or $AlF_3$ is added above the LiF layer to reduce the hygroscopic sensitivity of the LiF without degrading the far-UV performance of the coating.

### 9.3.4   High-Definition Imager (HDI)

#### 9.3.4.1 HDI overview and science goals

The LUVOIR observatory will revolutionize the study of the formation and evolution of planets, stars, and galaxies through a combination of very high sensitivity, high angular resolution, and a highly stable and well-calibrated point spread function. A key instrumental capability for LUVOIR is the High Definition Imager (HDI) instrument – the primary astronomical imaging instrument for observations in the near-UV through the NIR. Among the key science cases for HDI are:

- Understand the detailed physics of cosmic reionization by measuring the ionizing radiation escape fraction in galaxies as a function of time and environment. This is accomplished via the detection and measurement of Lyman continuum flux for $z \geq 2$ galaxies.
- Study the mechanisms of cosmic reionization by measuring variations, as a function of position on the sky, of the ultra-faint end of the $z \sim 7$ galaxy luminosity function (down to absolute AB magnitude of -12). A drop in the ultra-faint galaxy number density is predicted.
- Reconstruct detailed and accurate star formation histories in many galactic environments not reachable with other

facilities by directly detecting stars below the main sequence turn off in all major types of galaxies. This requires reaching out to distances of ~10 Mpc.

- Map the growth of substructure and the evolution galaxy morphology in the era where cosmic star formation peaks (2 < z < 4) by observing small-scale structure within z > 2 galaxies, down to spatial scales of 100 pc, in the rest-frame UV and visible. In particular, probe the distribution and properties of sub-galactic stellar systems (<100,000 solar masses) over the redshift range $2 < z \leq 10$.

- Constrain the distribution and properties of dark matter by measuring proper motions of stars in Local Group galaxies and by measuring proper motions of galaxies out to nearest groups and clusters within 15 Mpc of Milky Way.

- Search nearby stars for exoplanets via their induced astrometric wobble signature on their host stars to identify systems with Earth-mass planets in habitable zone regions. This is a potential companion program to the main LUVOIR coronagraphic exoplanet survey.

- Measure the long-term global atmospheric and interior dynamics of the gas and ice giant planets in the outer Solar System. On smaller scales, monitor the changes in surface features and exospheres of small, airless bodies in the Solar System due to volcanic or geyser activity.

These scientific investigations all call for an instrument that can instantaneously observe a field-of-view that spans 6 square arcminutes and provides pixel sampling that takes full advantage of the angular resolution provided by the telescope. The diverse nature of the above science cases (and these are just a few of many), also demand an instrument with an ample range of spectral elements including standard broad, medium, and narrow band filters as well as several dispersing elements to enable low-resolution slitless spectroscopy.

### 9.3.4.2 HDI design drivers

The HDI design is driven by several considerations:

- The need to enable a wide-field-of-view with Nyquist sampling at the focal plane implies a very large number of detector elements and pixels,

- The need for HDI to serve as the primary fine-guidance sensor for the observatory places special requirements on the detector performance and instrument reliability,

- Supporting image-based wavefront sensing for observatory commissioning and periodic wavefront maintenance routines, and

- The astrometric science objectives require on-board calibration systems to achieve the required pixel calibration accuracies.

### 9.3.4.3 HDI design implementation

#### 9.3.4.3.1 Optical design

The optical design provides a 2 x 3 arcminute field-of-view with two channels—an ultraviolet-visible (UVIS) channel covering the range 200 nm–~1000 nm and a near-infrared (NIR) channel covering the range ~1000 nm–2500 nm. The respective focal plane detector arrays provide Nyquist sampled images at 400 nm (2.75 mas/pixel) for UVIS imaging and at 1200 nm (8.25 mas/pixel) for NIR imaging. The choice to achieve Nyquist sampling is motivated by a desire to maximize information extraction from the images and to enable the best possible image quality when HDI is operating in parallel with other instruments (some of which cannot tolerate dithering maneuvers by the observatory pointing system). In addition, the optical design of the HDI instrument gives the UVIS and NIR channels nearly identical fields-of-view and allows for the option to perform simultaneous observations in both channels. A model of the HDI instrument layout is shown in **Figure 9.25**.

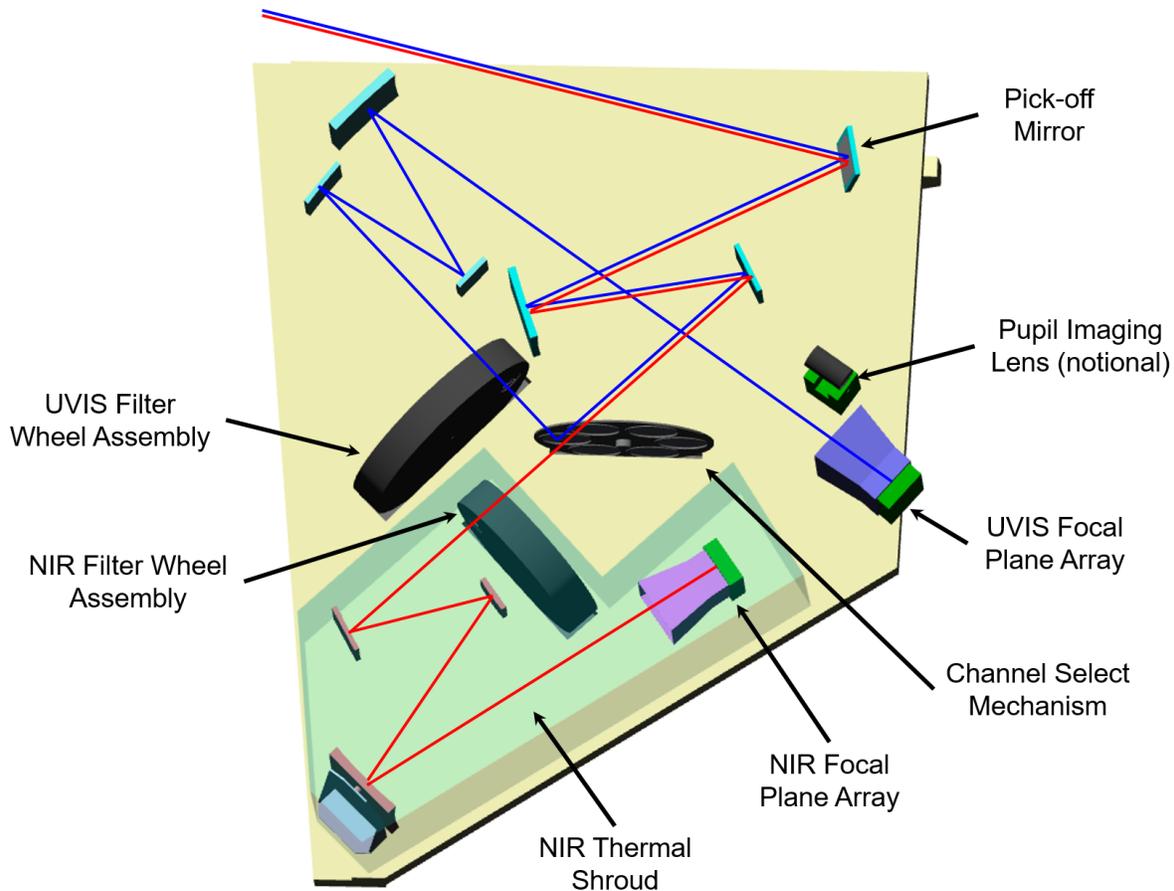

**Figure 9.25.** *Rendering of the HDI instrument. Blue lines trace the UVIS channel ray-path; red lines trace the NIR channel ray-path. Light enters from the telescope at the upper left. For a sense of scale, the UVIS filter wheel assembly is ~1 m in diameter.*

The UVIS channel contains a filter select mechanism consisting of 4 wheels, with each wheel capable of holding 13 elements. The design allows for 41 science spectral elements, 4 clear slots (needed to allow access to each set of filters on each wheel), 1 dark slot, 4 defocus lenses, and 2 Dispersed Hartman sensor elements. The NIR channel filter select mechanism consists of 3 wheels with 10 elements per wheel. The current design allows for 26 science spectral elements, 3 clear slots, and 1 dark slot. For both the UVIS and NIR channel, the filters will be selected based on current user preferences, experience with the imagers currently on board HST, and imagers that will be on JWST.

The channel select mechanism is a single wheel with 6 positions and controls the path of the light from the relay optics to the HDI detectors. The six positions are:

1. Clear – sends all light from the telescope to the NIR channel. This position also allows light from the internal calibration system to be sent into the UVIS channel.

2. Full Reflective – send all light from the telescope into the UVIS channel. The

backside of this mirror is also fully reflective, allowing light from the internal calibration system to be sent into the NIR channel.

3. 50/50 Beam Splitter – broadband beam splitter for simultaneous UVIS and NIR imaging, albeit with each channel receiving half the light.

4. Dichroic Beam Splitter – sends light from the telescope to both the UVIS and NIR channels simultaneously but over a restricted range of wavelengths in each channel, allowing nearly 100% transmission in each of these windows.

5. Optimized UV mirror – send all light from the telescope in a defined UV passband into the UVIS channel. Can supersede the need to use a transmissive UV filters for some wavelength ranges of interest in the 200–390 nm range.

6. A second optimized UV mirror for UVIS only imaging.

### 9.3.4.3.2   Mechanical design

The mechanical design of HDI is straightforward. A single optical bench holds all of instrument opto-mechanical components and serves as the primary interface between the instrument and the BSF.

### 9.3.4.3.3   Thermal design

The bulk of the instrument is held at 270 K, however a passively cooled shroud around the NIR channel maintains the NIR optics at 170 K to reduce thermal background. The visible focal plane array is also passively cooled to 170 K, while the NIR focal plane array is passively cooled to 100 K.

### 9.3.4.3.4   Electrical design

The HDI electrical system is designed to accommodate the very large focal plane read outs. The form factor and placement of the detector front-end electronics in each channel are chosen so that the ASICs (application-specific integrated circuits) are as close as possible to the detector arrays to minimize harness length. The front-end electronics also accommodate the necessary hardware to accommodate the pixel processing algorithms like frame co-adding and sampling-up-the-ramp readouts.

The large focal planes imply large data volumes. For normal science operations, assuming a 200 s average integration time consisting of 40 co-added frames at 5 second intervals, the data volume from both channels combined is ~144 Mbps (assuming 1.6:1 lossless compression). HDI includes enough memory to hold data from two days' worth of continuous observation, or about 25 Tbits. With an expected two downlinks per day, this provides ample margin for data storage within the instrument.

### 9.3.4.3.5   Detectors

The UVIS channel includes a 2.7 Gigapixel imaging array comprised of forty (40) 8k × 8k detectors arranged in a 5 × 8 pattern. The UVIS detectors are currently envisioned to be CMOS-based devices with 5 μm pixels. The 8k × 8k format for each sensor included in the current design has not yet been produced in flight-qualified scientific systems but is within a realistic technology trajectory from current devices. The assumed read noise is 2.5 e$^-$/pixel and the assumed dark current is 0.002 e$^-$/sec/pixel, consistent with current state of the art devices. A final detector decision, of course, would not need to be made for several years allowing ample time for detector development.

The NIR detectors are envisioned to be a 4 × 5 array of HgCdTe-based devises with 10 μm pixels. The 4k × 4k format for each sensor included in the current design is TRL 5, due to the development work on H4RG detectors for the WFIRST mission (Content

The High Definition Imager (HDI) instrument will use Gigapixel arrays and serve as the primary astronomical imaging instrument for observations in the near UV through the near IR.

et al. 2013). We are assuming 2.5 e$^-$ per pixel readout noise and dark current levels of ~0.002 e$^-$/sec/pixel. Current state-of-the-art read noise values are closer to 5 e$^-$, however it is believed a 2× improvement can be achieved with modest improvements to the read-out electronics. More discussion regarding the technology development of these detectors can be found in **Chapter 11.**

### 9.3.4.3.6 Special instrument modes

In addition to simultaneous UVIS and NIR imaging, HDI provides three special modes of operation. HDI's precision astrometry mode enables a range of exciting science not feasible on existing telescopes including astrometric detection of exoplanets and measuring proper motions of extragalactic sources in the Local Group and beyond. To achieve the astrometric accuracy required for these applications (better than ±1 micro-arcsecond), we need to calibrate the position of every pixel in the UVIS detector array. Such a metrology calibration system is needed if one wishes to measure galaxy proper motions out at the 10–15 Mpc distance scale or to detect the stellar wobble induced by Earth-mass exoplanets orbiting their host main sequence stars. An all-fiber metrology system (Crouzier et al. 2016) is included in HDI to allow for the calibration of the UVIS focal plane pixel geometry to a precision of 10$^{-4}$ pixels. We will investigate the astrometry error budget more thoroughly in future.

A second special mode of HDI is to function as LUVOIR's primary fine-guidance sensor. Both the UVIS and NIR focal plane have the capability of defining small regions-of-interest around bright foreground stars. These regions-of-interest can be read-out at high speeds (up to 500 Hz) to provide a pointing signal to the fast steering mirror and VIPPS. This can be done without interrupting regular science operations.

Finally, the HDI focal plane will serve a function similar to that of the Near Infrared Camera (NIRCam) on JWST and provide defocused image data for phase retrieval and wavefront sensing (Dean et al. 2006). This dataset will be used during commissioning of the observatory to align and phase the primary mirror segments after deployment, as well as during routine maintenance of the wavefront as needed. Six elements in the UVIS filter wheel are dedicated to supporting this mode of operation: 4 weak lenses for generating defocused point-spread function (PSF) data, and 2 dispersed Hartmann sensors (DHS) for performing coarse phasing of the primary mirror segments.

## 9.4 Alternate instruments and design implementations

The instrument suite discussed in this chapter is representative of what a set of first-generation instruments might look like but is by no means the only set of instruments LUVOIR could support. The instrument capabilities presented here are necessary to achieve the science objectives discussed earlier in this report, but even that discussion is not exhaustive of all of the science one could imagine doing with a facility as capable as LUVOIR. In the Appendices of this report, you'll find over 20 additional science cases (**Appendix A**), as well as additional instrument ideas (**Appendix E**) for the LUVOIR mission concept.

It should also be noted that the specific details of the instruments presented in this Chapter are not final. For each instrument, there exist different design implementations that may provide similar, or even expanded, science capability. For example, extending the wavelength range of LUMOS would afford high-resolution multi-object imaging spectroscopy in the visible regime. Different concepts for the ECLIPS back-end spectrometers could also be considered to support different types of exoplanet observations. The Study Team continues to explore several options to expand the science capability of the proposed instrument suite.

The payload described here in **Chapter 9** only represents a point design for the LUVOIR-A concept. It is an architecture that closes on all of the driving requirements, achieves the compelling science objectives laid out in this report, and is technologically and programmatically feasible and executable. Design optimizations and alternative design concepts will certainly be explored by future teams, should LUVOIR be prioritized for implementation.

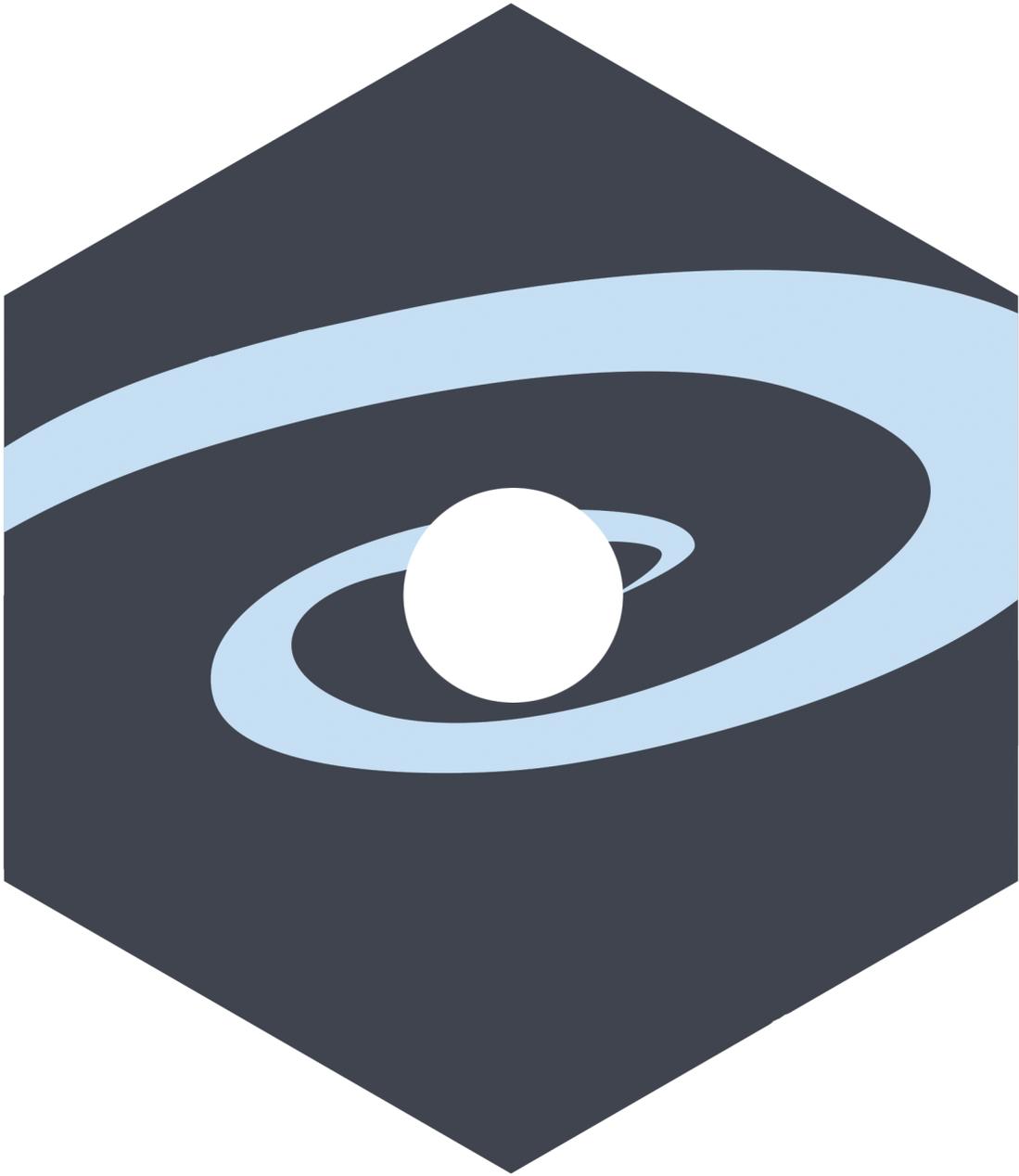

POLLUX: European study of a UV spectropolarimeter

# 10 POLLUX: European Study of a UV Spectropolarimeter

## Study leads and contact information

**Coralie Neiner**
LESIA, Paris Observatory
5 place Jules Janssen
92190 Meudon, France
coralie.neiner@obspm.fr
Phone: +33145077785

**Jean-Claude Bouret**
Laboratoire d'Astrophysique de Marseille
38, rue Frédéric Joliot-Curie
13388 Marseille cedex 13, France
jean-claude.bouret@lam.fr
Phone: +33491056902

## Participating institutes

The Phase 0 study of POLLUX is led by LAM and LESIA with the support of CNES (France), and developed by a consortium of European scientists. The consortium consists of 154 scientists and engineers from 67 institutes in 13 European countries. The main institutes participating in the present document are: Armagh Observatory (Northern Ireland), Bulgarian Academy of Sciences (Bulgaria), CEA (France), Center for Astrobiology (Spain), Complutense University of Madrid (Spain), CNES (France), CSL (Belgium), DESY (Germany), Exeter University (UK), Geneva University (Switzerland), GEPI (France), IAP (France), IMPS (Germany), IPAG (France), IRAP (France), LAM (France), LATMOS (France), LESIA (France), MPS (Germany), Royal Observatory Edinburgh (UK), Space Research Institute (Austria), Stockholm University (Sweden), Strasbourg Observatory (France), Trinity College (Ireland), TU Delft (The Netherlands), UK Astronomy Technology Center (UK), University of Edinburgh (UK), University of Hamburg (Germany), University of Leicester (UK), Catholic University of Leuven (Belgium), University of Liège (Belgium), University of Pisa (Italy), University of Warwick (UK).

## 10.1 Executive summary

The POLLUX instrument concept study started in January 2017. POLLUX is a high-resolution spectropolarimeter operating at UV wavelengths, designed for the 15-meter LUVOIR-A architecture. The instrument will operate over a broad spectral range (**97 to 390 nm**), at high spectral resolution (**R ≥ 120,000**). These capabilities will permit resolution of narrow UV emission and absorption lines, allowing scientists to follow the baryon cycle over cosmic time, from galaxies forming stars out of interstellar gas and grains, and stars forming planets, to the various forms of feedback into the interstellar and intergalactic medium (ISM and IGM), and active galactic nuclei (AGN).

The most innovative characteristic of POLLUX is its **unique spectropolarimetric capability** that will enable detection of the polarized light reflected from exoplanets or from their circumplanetary material and moons, and characterization of the magnetospheres of stars and planets and their interactions. The magnetospheric properties of planets in the solar system will be accessible at exquisite levels of detail, while the influence of magnetic fields on the Galactic scale and in the IGM will be measured. UV circular and linear polarization will provide a full picture of magnetic field properties and impact for a variety of media and objects, from AGN jets to all types of stars. POLLUX will probe the physics of accretion disks around young stars, white dwarfs, and supermassive black holes in AGNs, and constrain the properties,

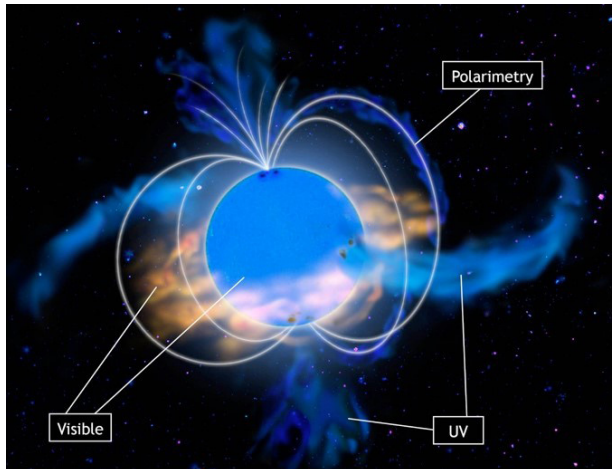

**Figure 10.1.** *Schematic view of the different features and regions traced by observations at different wavelengths around a hot star. Credit: S. Cnudde.*

especially sphericity, of stellar ejecta and explosions.

Since the parameter space opened by POLLUX is essentially uncharted territory, its potential for groundbreaking discoveries is tremendous. It will also neatly complement and enrich some of the cases advanced for LUMOS, the multi-object UV spectrograph for LUVOIR. In this chapter we outline a selection of key science topics driving the POLLUX design. We also introduce the current instrument concept and identify technological challenges that we will address in the coming years toward the advanced design and construction of POLLUX.

## 10.2  Science overview

### 10.2.1  Stellar magnetic fields across the HR diagram

From the ground in the visible domain, we can obtain sophisticated tomographic mapping of structures on the surfaces of stars and their magnetic fields. However, to understand the formation and evolution of stars and their accompanying planets it is necessary to also explore their circumstellar environments. POLLUX will provide a powerful, high-resolution, full-Stokes (IQUV) spectropolarimetric capability across the ultraviolet domain (97–390 nm) to uniquely trace these structures. When combined with the immense light-gathering power of LUVOIR, it will deliver unprecedented views of the disks, winds, chromospheres and magnetospheres around a broad range of stars, as highlighted below (also see **Figure 10.1**).

Magnetic fields play a significant role in stellar evolution, while also being a key factor in both planet and star formation. We need POLLUX spectropolarimetry to identify and characterize the B-fields across all stellar masses, to address central questions such as:

- How do B-fields develop in the pre-main-sequence (PMS) phase? POLLUX will characterize the strong B-fields arising at the sheared interface between star and disk in the PMS phase (via MgII, FeII, and CII UV lines from the interface region), and will trace the flows via polarization measurements of the nearby continuum.
- How does the stellar dynamo build-up and evolve? As stellar rotation decouples from the young planetary disk, the B-field is predicted to get stronger and more complex (Emeriau-Viard & Brun 2017). POLLUX will investigate propagation of magnetic energy through the stellar atmosphere into the uppermost coronal layers and stellar wind.
- How do stellar dynamos form and evolve in cool stars? What is their impact on planet formation and evolution? POLLUX will reveal the magnetically mediated interaction between young cool stars and their planetary disks during dynamo formation and stabilization when planets and planetary atmospheres form.
- A small but significant fraction of massive stars have strong B-fields (~10%; e.g., Fossati et al. 2015). However, very weak

fields could be ubiquitous in massive stars. POLLUX will uniquely detect sub-Gauss B-fields in massive stars, providing a much-needed understanding of weak fields, field-decay mechanisms, and their impact on stellar evolution.

- POLLUX will enable studies of stellar B-fields beyond the Milky Way for the first time. These will open-up observations in the metal-poor Magellanic Clouds (MCs), probing the evolutionary processes linked to B-fields at metallicities equivalent to those at the peak of star-formation in the Universe.

POLLUX will also enable breakthroughs in other areas of contemporary stellar astrophysics, including:

- Jet formation: The mechanism driving the powerful jets from the interaction between disks around PMS stars and their magnetospheres is poorly constrained; POLLUX will transform our understanding of this dramatic part of star formation.

- Winds & outflows: A large uncertainty in evolutionary models of massive stars is the link between rotation and mass lost via their winds, which limits our understanding of the progenitors of γ-ray bursts, pair-instability SNe, and gravitational-wave sources. POLLUX observations in the Galaxy and MCs will map the density contrasts with stellar latitude required to calibrate these models.

- Novae & SNe: Both classical novae and SNe remnants provide excellent opportunities to learn about the production and lifecycle of dust. POLLUX spectropolarimetry is required to identify sites in the ejecta where dust condenses—allowing us to also infer the effects of kicks imparted by core-collapse

SNe—and the shapes/sizes of the grains involved.

- White dwarfs (WDs): A minority of WDs are strongly (>1MG) magnetic (Ferrario et al. 2015), but little is known of the incidence of weaker fields. POLLUX UV spectropolarimetry will explore this regime for the first time to investigate the ubiquity of B-fields in WDs, as well as the processes generating their fields (e.g., amplification of fossil fields and/or binary mergers?)

A 100-hour POLLUX observing program would be transformational in our understanding of B-fields in stellar evolution. For instance, we could include UV monitoring of two low-mass sources (PMS T Tauri, evolved T Tauri) to directly characterize their B-fields and magnetic-energy transport, combined with observations of a first sample of tens of Galactic massive stars to assess the presence/strength of weak B-fields

### 10.2.2  What are the characteristics of exoplanet atmospheres and how do planets interact with the host stars?

The characterization of exoplanets is key to our understanding of planets, including those in the Solar System. POLLUX's unique, simultaneous, high-resolution and polarimetric capabilities in the UV are essential to unveiling the origins of the huge range of chemical and physical properties found in exoplanetary atmospheres (e.g., Sing et al. 2016), and to understand the interaction between planets and their host stars (e.g., Cuntz et al. 2000).

The line and continuum polarization state of starlight that is reflected by a planet is sensitive to the optical properties of the planetary atmosphere and surface and depends on the star-planet-observer phase angle (**Figure 10.2**). Atmospheric gases are efficient scatterers at UV wavelengths and deviations from the expected polarization across the UV would reveal the presence and microphysical properties of aerosol and/or cloud particles. This information is crucial for getting insight into a planet's climate. It would also strongly complement transit observations. POLLUX can significantly detect polarization signatures for close-in gas giants (and brown dwarfs) orbiting stars out to distances of 70 pc, which currently comprises over a dozen targets, two of them transiting.

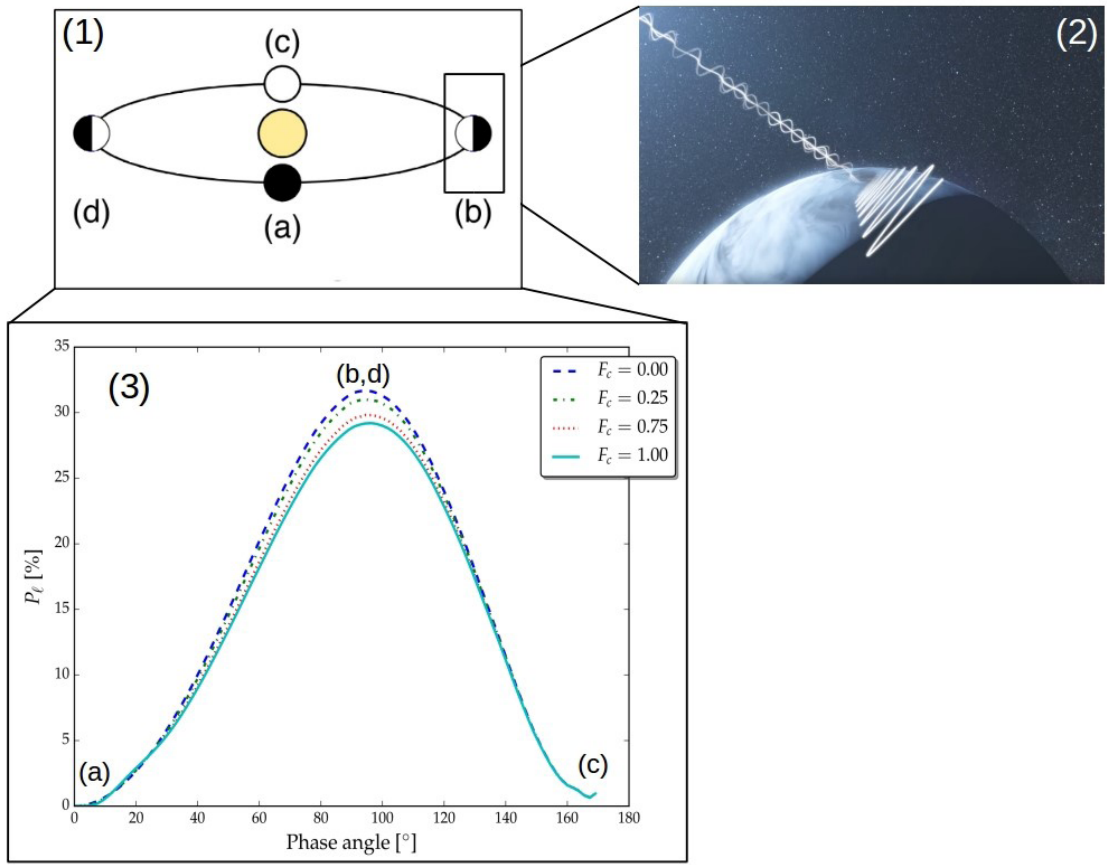

**Figure 10.2.** *Top panels 1 & 2: Schematic of a planetary system in which the unpolarized stellar light becomes polarized through reflection by a planetary atmosphere. Bottom panel 3: Degree of polarization (in %) as a function of planetary orbital phase, labeled as in panel 1, at 300 nm. The different lines indicate different atmospheric cloud coverages, where zero corresponds to the cloudless condition. POLLUX will measure the degree of polarization as a function of wavelength and planetary orbital phase. Credit: L. Rossi*

Star-planet interactions (SPI) could generate detectable signatures in exoplanetary systems. Searching for SPI has developed into a very active field (e.g., Shkolnik et al. 2003; Kashyap et al. 2008; Miller et al. 2015). Evidence of SPI has been observed in UV stellar emission lines (e.g., NV; France et al. 2016). POLLUX will be capable of studying the intensity of UV stellar emission lines forming in a wide range of temperatures as a function of planetary orbital phase, uniquely combining this information with measurements of the magnetic field derived from the Stokes V profiles of these lines. This will allow for the first time identification of the stellar regions mostly affected by SPI and hence to understand their origin.

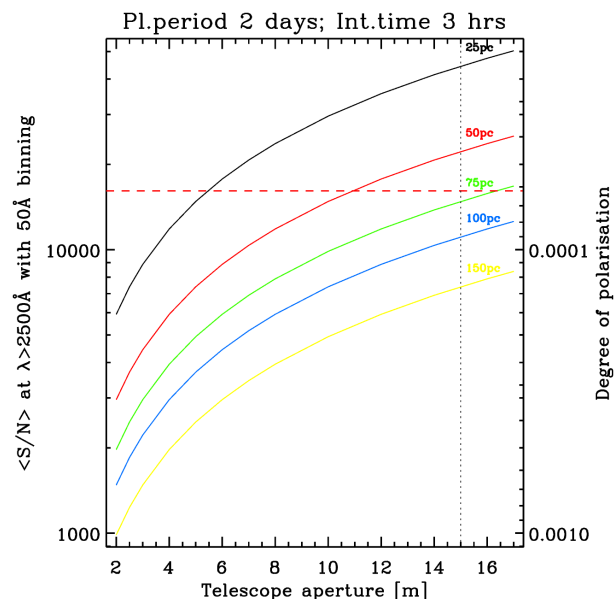

**Figure 10.3.** *Average S/N in the near-UV obtained with 3 hours of POLLUX exposure time as a function of the distance to a target Sun-like star (line colour) and size of the telescope primary mirror. The red dashed horizontal line indicates the maximum near-UV polarization signal of a hot-Jupiter with a 2-day orbital period. The black dotted vertical line indicates the size of the LUVOIR-A primary mirror.*

At UV wavelengths, a hot Jupiter orbiting a Sun-like star with a two-day orbital period presents a maximum polarization of ≈30% (**Figure 10.3**), leading to a maximum polarization signal of the order of 7 x 10⁻⁵, when observed unresolved from its star. With POLLUX mounted on a 15-m LUVOIR telescope and a binning size of 50 Å, the signal-to-noise ratio (S/N) necessary to detect such a signal can be obtained in less than 3 hours of exposure time for systems as distant as 70 pc (**Figure 10.3**). Measuring and modeling the variation of the polarization signal as a function of the planet's orbital phase requires at least eight measurements spread across the orbit. A POLLUX 100-hour observing program would allow the detection and characterization of the UV polarimetric signature for over a dozen targets.

### 10.2.3  The various phases of the ISM and extragalactic IGM

Matter in the interstellar space is distributed in diverse, but well-defined phases that consist of (1) the hot, ionized (T~$10^{6-7}$K) ISM that emits soft X-rays, (2) the warm neutral or ionized medium ($6000 < T < 10^4$ K), and (3) the cold (T~10–200 K) neutral medium and molecular star-forming clouds. Boundaries between the different phases play a fundamental role in the cooling of the gas and by consequence in Galactic evolution. However, it is not yet clear how these different phases trade matter and entropy. We need to address the multi-phase aspect of the ISM and the influence of different factors such as magnetic field, thermal/pressure (non-) equilibrium, metallicity, shocks, etc.

Gaining fundamental insights into the processes that influence the different phases requires the wealth of interstellar absorption features appearing in the UV spectra of hot stars, and requires observing them at R ≥ 120,000 to measure individual clouds and disentangle the contributions from the different phases. POLLUX will, for the first time, give access to all phases at the required high spectral resolution. It will enable simultaneous study of tracers of hot-gas (OVI, CIV), warm gas (OI, NI, and many singly-ionized species), as well as cold-HI gas like CI, and of the molecular phases through $H_2$ and CO lines. POLLUX will resolve the velocity profiles of the different $H_2$ rotational levels, yielding temperature and turbulence, and information on the formation of $H_2$ and its role as a coolant, e.g., in turbulence dissipation.

ISM studies also need to be generalized to external galaxies. POLLUX will enable tomography of chemical abundances in several hundred nearby galaxies using neutral gas absorption lines toward individual stars across the entire (stellar) body of galaxies or toward background quasars. Abundance variations across and between galaxies of different types and metallicities will document the dispersal/mixing spatial/time scales of newly produced elements and their nucleosynthetic origin, the role of infalling and outflowing gas in the metallicity build-up, and the dust production/composition through depletion patterns. Furthermore, cooling rates, physical conditions, and the molecular gas content can be determined spatially in the HI reservoir (with implications for the regulation of star formation and for galaxy evolution at large) and in neutral shells of HII regions (with implications for the star-formation process itself).

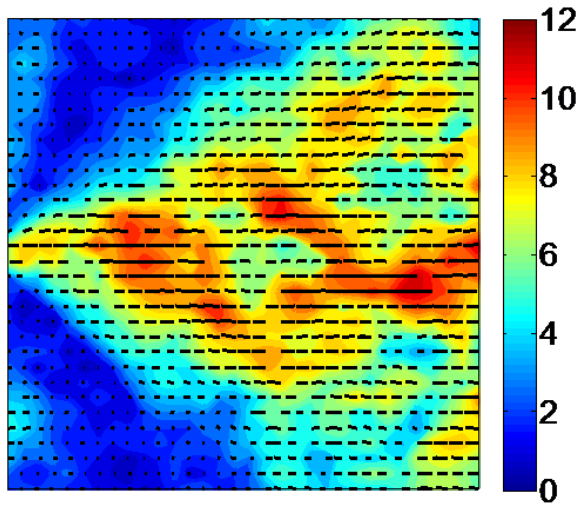

**Figure 10.4.** *Synthetic polarization (%) map of the simulated diffuse ISM (1 pc²). Contour color: percentage of polarization in the SII 1250 A absorption line. Orientation of the bars: polarization direction. The mean magnetic field (3 µG) is oriented along the x-axis. This figure shows that the direction of polarization correctly indicates the magnetic field direction and that the expected polarization is mostly above 5%, within POLLUX capabilities.*

A huge step forward in the knowledge of the Galactic magnetic field in terms of sensitivity, sky coverage and statistics has been made from recent dust polarization measurements. However, they tell us nothing about the distance of the magnetic field and its distribution in the different components or phases. The POLLUX spectropolarimeter will provide this information through its ability to detect linear polarization in UV absorption lines through Ground State Alignment (GSA, Yan & Lazarian 2012). It will open up access to the magnetic field 3-D, by measuring its orientation in different clouds, on limited distances along sight lines and on small scales (**Figure 10.4**). In a 100-hour program, observing several hundreds of hot stars with a S/N of 500 to measure linear polarization in optically thin absorption lines at a level of a few % will help understand the interplay of magnetic field and the different ISM phases. By discerning magnetic fields in gas with different dynamical properties, POLLUX will allow the first study of interstellar magnetic turbulence (Zhang & Yan 2017).

## 10.2.4 Beyond the unresolvable regions of active galactic nuclei: revealing accretion disk physics, dust composition and magnetic field strength with UV polarimetry

POLLUX will offer unique insight into the still largely unknown physics of Active Galactic Nuclei (AGNs), which are believed to arise from accretion of matter by supermassive black holes in the central regions of galaxies. Some key signatures of accretion disks can be revealed only in polarized light, and with higher contrast at ultraviolet wavelengths. Specifically, high-resolution UV polarimetry will provide geometric, chemical and thermodynamic measurements of accretion disks in unprecedented detail. By probing the ubiquitous magnetic fields expected to align small, non-spherical dust grains on scales from the accretion disk out to the extended dust torus, POLLUX will be able to reveal the mechanisms structuring the multi-scale AGN medium. The key information encoded into the polarized light will allow determinations of the mineralogy, structure and alignment of the smallest dust grains, together with line-of-sight magnetic-field strengths. Measurements of magnetic field strengths will also provide constraints on the structure of magneto-hydrodynamic winds in nearby, broad absorption-lines systems accessible to the instrument.

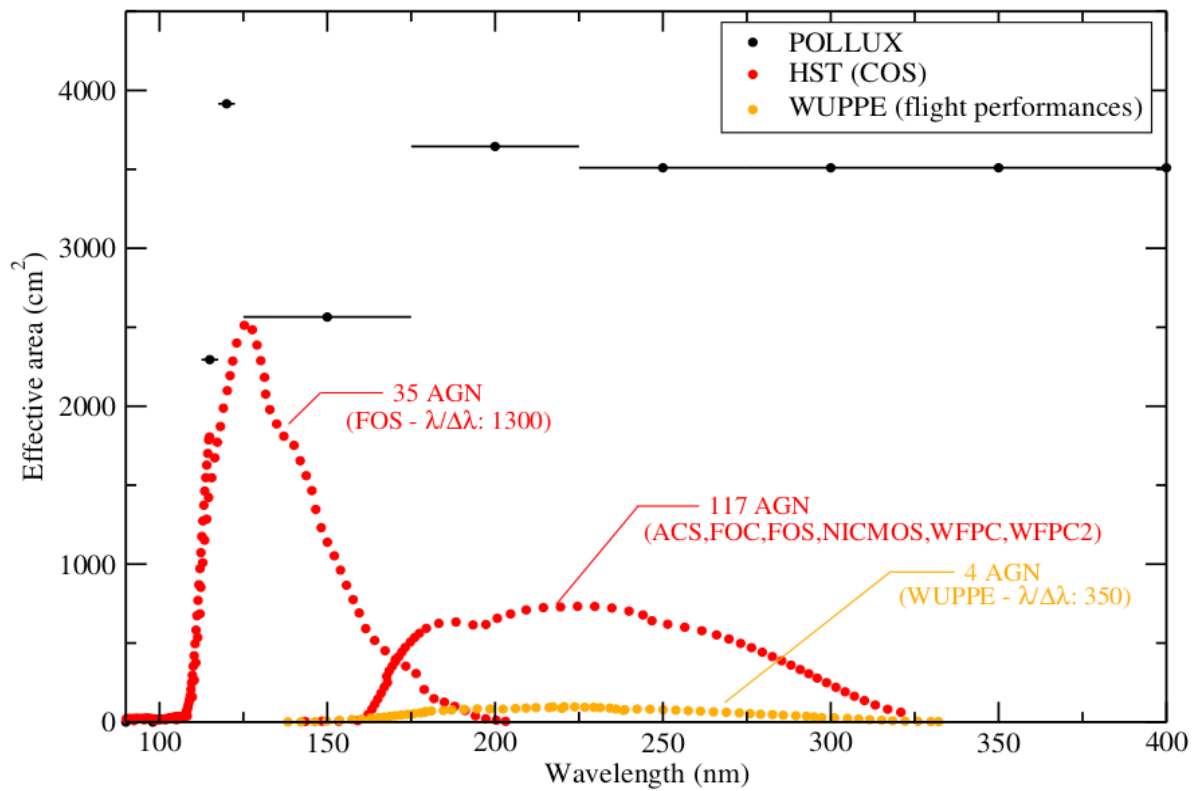

**Figure 10.5.** *Effective area of POLLUX (in black; assuming 135 m² for LUVOIR with a RC telescope), compared to those of previous space-based UV spectrographs, as indicated. Only 4 AGN were observed with WUPPE. A total of 117 were observed with different polarimetric instruments on board HST, out of which only 35 were in the UV with HST/FOS (for which an effective area similar to that of HST/COS was adopted). POLLUX will enable observations of hundreds of AGN over a wide UV spectral domain, and with much greater spectral resolution than ever achieved.*

On larger scales, UV polarimetric studies of young star-forming regions will provide unprecedented insight into the enigmatic relation between onset of star formation and triggering of nuclear activity. Once activated, an AGN is expected to feed vast amounts of energy back into the interstellar medium of its host galaxy through radiation and shocks. This feedback can decrease or increase the star formation activity of the host; our understanding of the physical conditions in this interaction will be greatly improved with UV polarization observations from POLLUX.

In addition to the extended environment of AGNs, POLLUX will provide insight into the nature and lifetime of particles in relativistic jets, which are other key factors to fully understand AGN feedback on star formation in galaxies. By focusing on bright, low-redshift galaxies, it will be possible to obtain, for the first time, high-resolution spectra providing striking details on the structure and physics of AGNs. Complemented by the optical and infrared instruments on board LUVOIR, POLLUX will constitute a groundbreaking means of assessing the important role played by AGNs on galaxy evolution (**Figure 10.5**).

### 10.2.5 Testing fundamental physics and cosmology using absorption lines towards quasars

Physics and cosmology as we know them today have been remarkably successful in reproducing most of the available observations with only a small number of parameters. However, it also requires that 96% of mass-energy content of the Universe is in mysterious forms (dark energy and dark matter) that have never been seen in the laboratory. This shows that our canonical theories of gravitation and particle physics may be incomplete, if not incorrect. Improving the sensitivity of current observational constraints is therefore of utmost importance, irrespective of whether it is consistent with the current standard physics—in which case it will reject other scenarios—or whether it will instead favor new physics.

Absorption-line systems produced by intervening gas in the spectra of background sources provide original sensitive probes of fundamental physics and cosmology. The high UV spectral resolution of POLLUX on LUVOIR will open a unique window on such probes, in particular (1) the measurement of the primordial abundance of deuterium, (2) the stability of fundamental constants over time and space and (3) the redshift evolution of the cosmic microwave background (CMB) temperature. The D/H ratio can be estimated from the DI and HI Lyman series lines. Any change in the proton-to-electron mass ratio ($\mu = m_p/m_e$) translates into a relative wavelength change of the $H_2$ Lyman and Werner lines. Finally, the CMB radiation excites the CO molecules so that the relative population in different rotational levels (measured through the electronic bands in the UV) is an excellent thermometer for the CMB temperature.

We remark that, while these are independent probes, they are intimately related by the underlying physics. For example, models involving varying scalar-photon couplings (e.g., Avgoustidis et al. 2014) also affect the temperature-redshift relation so that constraining this relation is complementary to a search for varying fundamental constants. The Big Bang nucleosynthesis calculations of the D/H ratio are also dependent on the fundamental constants and can be altered if new physics is at play (e.g., Olive et al. 2012).

The baseline specifications of LUVOIR/POLLUX are S/N~100 per resolution element in 1h for $F\lambda = 10^{-14}$ erg s$^{-1}$ cm$^{-2}$ Å$^{-1}$. This means that for known quasars with low-z $H_2$ absorbers as observed by HST/COS (Muzahid et al. 2015), we can reach a precision of a $\Delta\mu/\mu \sim$ a few $10^{-7}$ in 20–30h (**Figure 10.6**). This is about an order of magnitude better than the current best limits (around $5 \times 10^{-6}$ with UVES on the Very Large Telescope), in less observing time. Similarly, the achieved precision on $T_{CMB}$ scales directly with the S/N ratio (0.1 K at S/N~100 and R~100,000 when current limits are $\Delta T$~1K). The same is again true with the D/H ratio, although a very high resolution is less critical

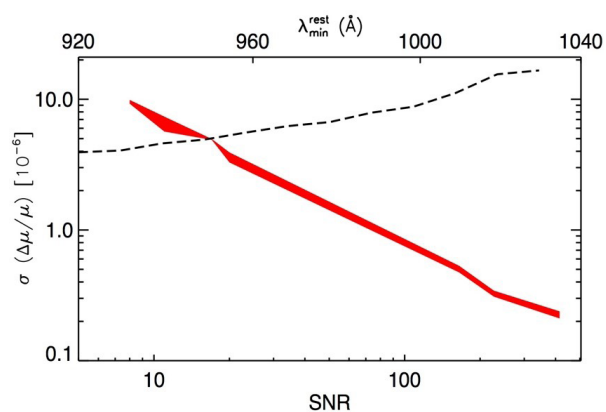

**Figure 10.6.** *Expected error on $\Delta\mu/\mu$ as a function of signal-to noise ratio (red band, lower x-axis) and minimum wavelength (dashed line, absorber's rest-frame, upper x-axis) covered by a R=120,000 spectrum.*

since the corresponding lines are typically thermally broadened by at least 10 km s$^{-1}$.

### 10.2.6 Solar system: Surfaces, dust scattering, and auroral emissions

UV observations uniquely probe the surfaces of telluric bodies of the solar system. They diagnose their volcanic and plume activity, their interaction with the solar wind, and their composition relevant to the space weather and exobiology/habitability fields. Very few UV polarimetric observations were obtained with WUPPE (e.g., Fox et al. 1997), revealing in particular the Io surface as spatially covered by 25% $SO_2$ frost with polarization variations associated to different volcanic regions. POLLUX will primarily characterize volcanism and/or plume activity of icy moons from polarized solar continuum reflected light and spectral UV albedo. Its high sensitivity is necessary to track any organic and ice composition of the crust of comets and Kuiper Belt objects from their UV spectra.

The giant planets' UV aurorae are mainly radiated from H and $H_2$ atmospheric species, collisionally excited by accelerated charged particles precipitating along the auroral magnetic field lines. Aurorae thus directly probe complex interactions between the ionosphere, the magnetosphere, the moons, and the solar wind. Precipitation of auroral particles is additionally a major source of atmospheric heating, whose knowledge is needed to assess the energy budget, the dynamics, and the chemical balance of the atmosphere. The narrow FOV of POLLUX will measure the bright complex aurorae of Jupiter and Saturn, the fainter ones of Uranus, and catch those of Neptune, only seen by Voyager 2 (Lamy et al., 2017). The high spectral resolution will be used to finely map the energy of precipitating electrons from partial spectral absorption of $H_2$ by hydrocarbons (Ménager et al. 2010, Gustin et al., 2017) and the thermospheric wind shear from the H Ly–$\alpha$ line (**Figure 10.7**, left panels; Chaufray et al. 2010).

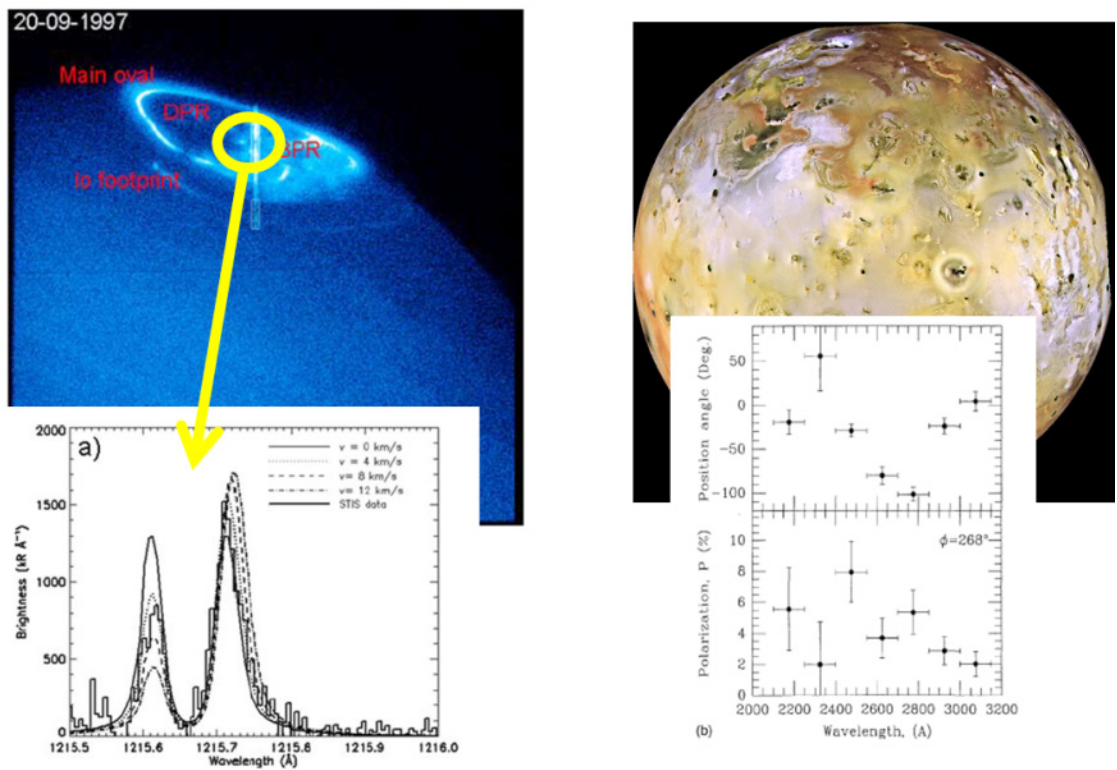

**Figure 10.7.** Left: auroral emissions observed on Jupiter by HST. The Ly-alpha emission spectrum in the polar auroral region was obtained with STIS. Its strong asymmetry can be reproduced with a wind shear reaching 4-8 km/s. Credit: Chaufray et al. (2010). Right: Io observed by Galileo. The linear polarisation of the surface measured by WUPPE between 220–320 nm presents complex variations with wavelength and provide information on the volcanic activity on the surface of Io. Credit: Galileo Project / JPL / NASA / Fox et al. (1997)

The asymmetric profile of the H Ly-$\alpha$ line in the Jovian auroral region (**Figure 10.7**, left panels) should be resolved by POLLUX. For an exposure time of 10 minutes, and considering the brightness of the wings of the Ly-$\alpha$ line measured by HST/STIS, the two wings should produce 68±8 cts/px and 34±6 cts/px respectively, implying a velocity larger than 4 km/s, which can be measured precisely by POLLUX.

The albedo of a region of 100 km$^2$ can be derived with an accuracy of 0.1% with a spectral resolution of 1 nm and a 5-min exposure near 320 nm. The linear polarization of Io's surface varies between 1 and -10% between 220 and 320 nm; near 320 nm, the linear polarized signal observed by POLLUX for a binning size of 1 nm in 5 minutes should have a S/N~10, which is better than WUPPE (**Figure 10.7**, right panels).

## 10.3  Design drivers

The science goals described in the previous section lead to the technical requirements for POLLUX presented in **Table 10.1**, with comments in the last column.

## 10.4  Design overview & implementation

The baseline configuration of POLLUX that we will now present allows fulfillment of all the requirements for the instrument performance. Most of the technologies required for a complete implementation are at TRLs compatible with a Phase 0 study. We did not find fundamental restrictions or physical limitations preventing its implementation. We will discuss what research and development is needed to allow us to realize the baseline configuration by the time of LUVOIR implementation.

**Table 10.1.** *High-level requirements*

| Parameter | Requirement | Goal | Reasons for requirement |
|---|---|---|---|
| Wavelength range | 97–390 nm | 90–650 nm | 97 nm to reach Ly$\gamma$ line. 390 nm to reach CN line at 388 nm in comets |
| Spectral resolving power | 120,000 | 200,000 | Resolve line profiles for ISM, solar system, and cosmology science cases |
| Spectral length of the order | 6 nm | ≥6 nm | To avoid having broad spectral lines spread over multiple orders |
| Polarization mode | Circular + linear (= IQUV) | | |
| Polarization precision | 10$^{-6}$ | | Detect polarization of hot Jupiters |
| Aperture size | 0.03" | 0.01" | Avoid contamination by background stars in Local Group galaxies |
| Observing modes | spectropolarimetry and pure spectroscopy | | |
| Radial velocity stability | Absolute = 1 km/s and relative = 1/10 pixel | | Avoid spurious polarization signature |
| Flux stability | 0.1% | | Probe flux and polarization correlation in WDs |
| Limiting Magnitude | V=17 | | To reach individual stars in MCs |
| Calibration | Dark, bias, flat-field, polarization, and wavelength calibration | + Flux calibration | |

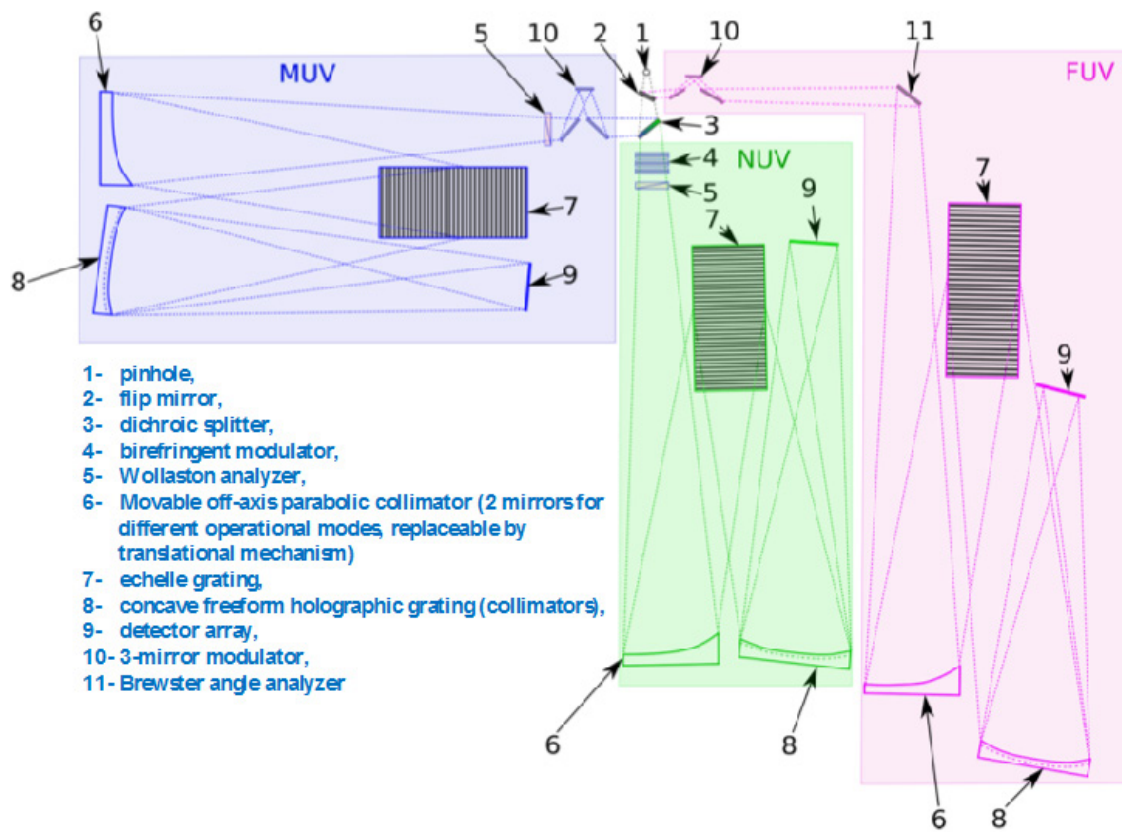

**Figure 10.8.** *POLLUX instrument baseline architecture schematic diagram*

### 10.4.1 Baseline optical architecture and specifications

In the baseline configuration of POLLUX, we adopted the telescope parameters provided by the LUVOIR study. POLLUX is a spectropolarimeter working in three channels. For practical reasons we refer to these as NUV (19–390 nm), MUV (118.5–195 nm), FUV (97–124.5 nm). Each is equipped with its own dedicated polarimeter followed by a high-resolution spectrograph. The spectra are recorded on δ-doped EMCCD detectors. MUV+NUV channels are recorded simultaneously, while the FUV is recorded separately. POLLUX can be operated in pure spectroscopy mode or in spectropolarimetric mode. POLLUX can be fed by the light coming from the telescope or from sources in the calibration unit. We anticipate that POLLUX can operate with a 270 K housing, in line with the requirement of LUVOIR.

We now describe the major assumptions that we adopted to design the optical architecture of this configuration. They are illustrated on **Figure 10.8**:

- The instrument entrance is a pinhole, rather than a slit, for simpler aberration correction.
- The working spectral range is 300 nm. It is split into three channels: far ultraviolet (FUV), medium ultraviolet (MUV) and near ultraviolet (NUV). This allows POLLUX to achieve high spectral resolving power with feasible values of the detector length, the camera optics field of view, and the overall size of the instrument. It also allows to use dedicated optical elements, coatings, detectors, and polarimeter for each band, to obtain gains in efficiency.

- The FUV and MUV boundaries are set relative to the Lyman-$\alpha$ line. The lower limit for the MUV band is set at Lyman-$\alpha$ minus roughly 3 nm, that is 118.5 nm, while the upper one for the FUV is Lyman-$\alpha$ + roughly 3 nm, i.e., 124.5 nm.

- The shortest wavelength for the FUV strongly depends on the main telescope throughput and may be reconsidered in the future.

- The MUV and NUV channel are separated by means of a dichroic splitter. Such splitters can have a high efficiency (e.g., a mean reflectance of 61% over 140–170 nm, most of the MUV band, and a mean transmittance of 83% over 180–275 nm, most of the NUV band; see http://www.galex.caltech.edu/researcher/techdoc-ch1.html). A dichroic splitter allows the instrument to work in two bands simultaneously and use the full aperture, thus achieving high resolving power with relatively small collimator focal length. In the present design, we set the MUV/NUV boundary at 195 nm, to have a maximum of one full octave in each channel (here the NUV).

- Currently there are no dichroic splitters operating in the FUV below the Ly-$\alpha$ line and there is no evidence that such an element will become possible in the future. We have decided to use a flip mirror to feed the FUV channel. The flip mirror is located immediately before the dichroic splitter.

- The splitters are placed as close to the focal point as possible in order to decrease their size and the size of the

polarimeters. Currently the flip mirror is located 20 mm away from the focus and the distance from focus to dichroic is 35 mm.

- In each channel, the beam is collimated by an ordinary off-axis parabolic (OAP) mirror. The off-axis shift and the corresponding ray deviation angle are chosen in such a way that the distance between the entrance pinhole and the echelle grating is large enough to place the polarimeter and corresponding mechanical parts. The MUV and NUV mirrors have identical geometry, though they may have different coatings and have slightly different operation mode due to the difference in each polarimeter's design.

- Echelle grating works in a quasi-Littrow mounting. The exact values of the groove frequency and the blaze angle are computed to obtain the target dispersion and subsequently the spectral resolving power.

- The cross-disperser in each channel operates also as a camera mirror, so it is a concave reflection grating. This approach allows minimization of the number of optical components and increases the throughput. In order to correct the aberrations, the cross-disperser's surface is a freeform and has a complex pattern of grooves formed by holographic recording.

- Adopted coatings on the optical elements of POLLUX are those used for the telescope, except for the polarimeters. In the future, they will be optimized for each element of each channel (see **Section 10.5**).

- Polarimeters are located immediately after the splitters in each channel to avoid instrumental polarization by the spectrograph elements. The polarimeters are retractable in the MUV and NUV to allow the pure spectroscopic mode. In the FUV only the modulator is retractable. The analyzer is kept in the optical path to direct the beam towards the collimator.

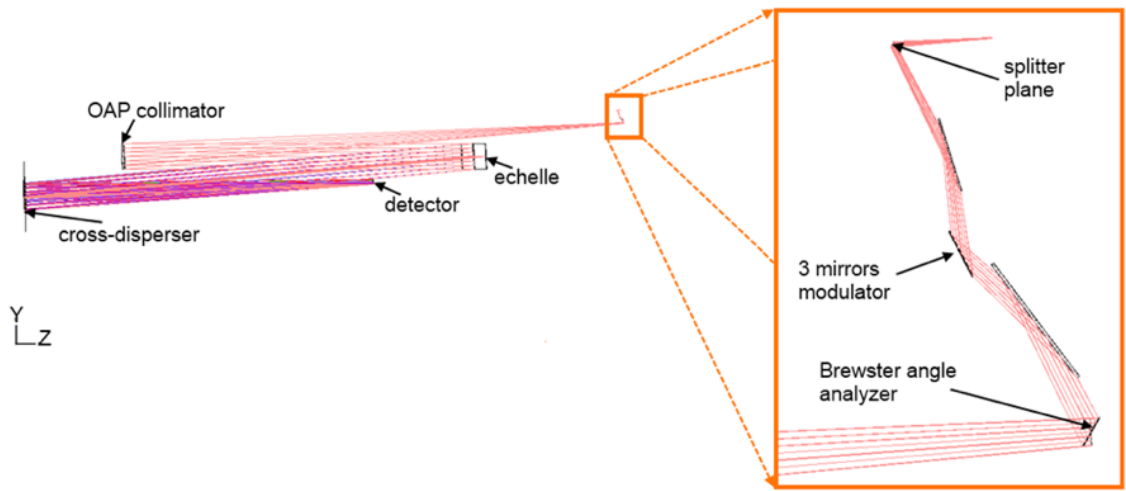

**Figure 10.9.** *An example of a schematic optical scheme, here for the FUV channel, including a zoom on the FUV polarimeter unit.*

- Change of the optical path caused by removing the polarimeter from the beam is compensated by translating the OAP mirror for the three channels.
- The polarimeter design was optimized for each channel accounting for the technological feasibility (see **Section 10.4.2**). The polarimeters should have minimal size in order to decrease their influence on the image quality. Firstly, transparent plates introduce some aberrations. Secondly, due to polarization ray splitting, the collimator may operate in an unusual mode and have considerable aberrations. Thirdly, the shorter the optical path inside the polarimeter, the smaller the difference between the spectropolarimetric and the pure spectral observation modes.

It is necessary to switch beams in POLLUX in order to feed the detectors with light coming from the telescope, or from sources in the calibration unit. Furthermore, in order to compensate the optical path difference and maintain the same beam position and the angle of incidence at the echelle when switching from the spectropolarimetric mode to the spectroscopic mode (done by removing the polarimeters from the optical train), it is necessary to change the collimator mirror (see #6 in **Figure 10.8**). Due to the focal length change, the collimated beam and therefore the theoretical resolution limit also change. On the other hand, the pinhole projection size also changes, so the resolution values found while accounting for the aberrations should be re-scaled.

The optical design is optimized for the conditions described above. In order to account for possible misalignments due to switching from the pure spectroscopic mode to the spectropolarimetric one, the target spectral resolving power was set to 130,000 (i.e., 110% of the requirement).

### 10.4.2  Polarimeters

Below 120 nm, $MgF_2$ is opaque. Above this wavelength, both the birefringence and transparency of $MgF_2$ recover quickly. In order to optimize the throughput in particular below 150 nm, we explored modulation based on reflection rather than transmission for the FUV and MUV modulators.

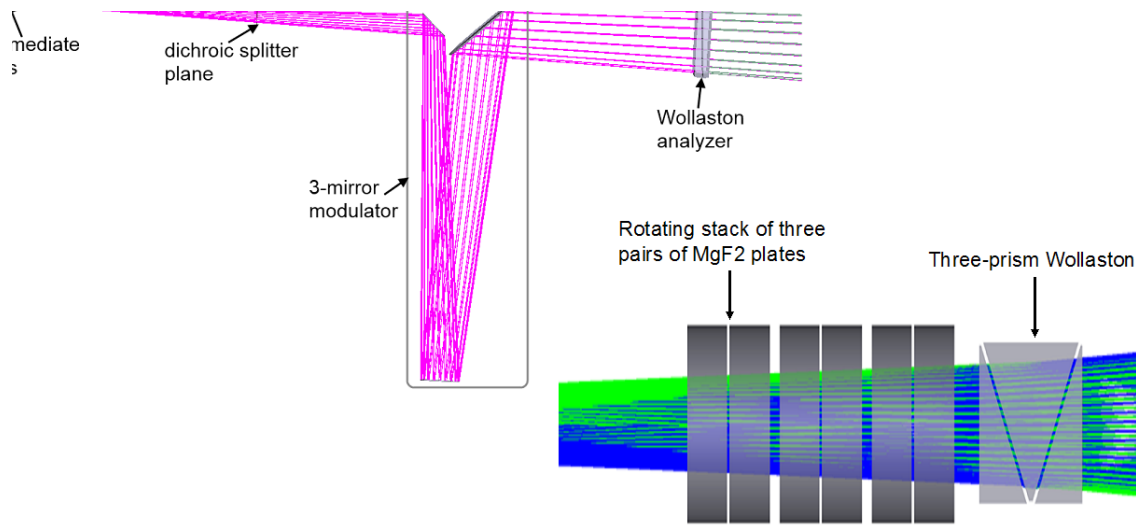

**Figure 10.10.** *Optical schemes for the polarimeter units of the MUV (left) and NUV (right) channels.*

- The FUV polarimeter has a three-mirror modulator with SiC mirrors and high incidence Brewster angle (close to 80 degrees) to record polarization. Only the reflected P beam can be recovered with this technique, hence we cannot use two-beam polarimetry to reduce systematics.

- The MUV polarimeter (see **Figure 10.10**) has a three-mirror modulator coated with Al+LiF. The three mirrors rotate as a whole around the optical axis of the instrument. The first and third mirrors work at an incidence angle close to 47 degrees and the second mirror at the complementary of twice this angle. The choice of three mirrors ensures that the output beam is in the same axis as the entrance beam. A Wollaston prism made of $MgF_2$ is the current baseline solution for the analyzer, but other options will be studied.

- The NUV polarimeter is completely adapted from the ARAGO design (Pertenais et al. 2016), that is a birefringent modulator made of three pairs of $MgF_2$ plates, and a Wollaston prism of $MgF_2$ for the analyzer. However, thanks to its reduced operational range with respect to ARAGO, it will be simplified for POLLUX.

### 10.4.3  Mechanical architecture

The mechanical layout is based on the optical design discussed above (cf. **Figure 10.11**). It is intended to illustrate a feasible structural design concept including the necessary mechanisms, which fit within the available instrument volume.

The major points of this layout can be summarized as:

- The central support structure of POLLUX consists of a central chassis, which provides mechanical interfaces for the MUV, FUV, NUV arms, calibration unit, polarimetry and beam switching mechanisms. This will also provide the mechanical interface to the telescope but has not been represented at this time.

- MUV, FUV, NUV arms are generically similar optical benches constructed of rectangular or triangular box sections. They provide flat mounting interfaces for the collimator, disperser, camera and detector subassemblies. The benches are connected to the central support via box section brackets with angled bolted flanges (**Figure 10.12**).

- The collimator exchange mechanism is implemented by linear slides using

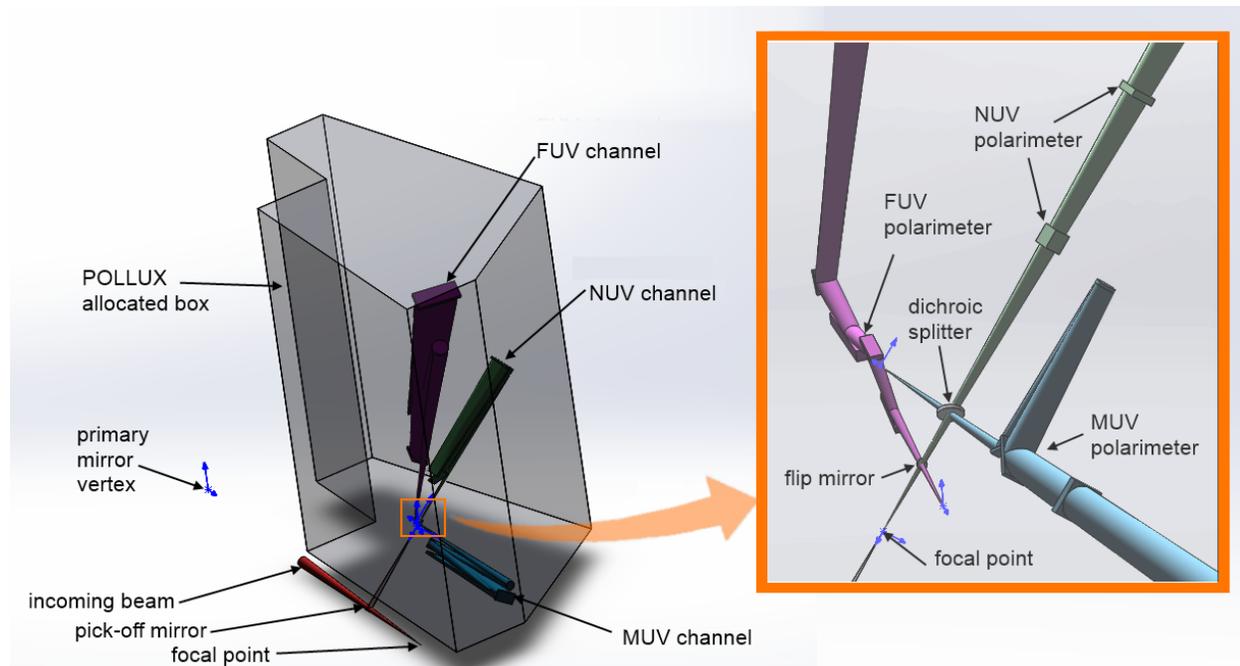

**Figure 10.11.** *Optical system of POLLUX arranged inside the dedicated volume.*

recirculating ball bearing carriages. Two slides are used to better control tilt error and for robustness. The slides are positioned via a leadscrew driven by a stepper motor.

- The calibration insert mechanism rotates a fold mirror into the beam. The mirror is mounted on a radius arm attached directly to the motor shaft. To achieve the required repeatability, the mirror position should be compliantly mounted and the position defined kinematically against a stop.

- The polarimetry insert mechanisms for the three channels are implemented by linear slides using recirculating ball bearing carriages. Two slides are used to better control tilt error and for robustness. The slides are positioned via a leadscrew driven by a stepper motor.

- The polarimetry rotation mechanisms are implemented by worm and gear

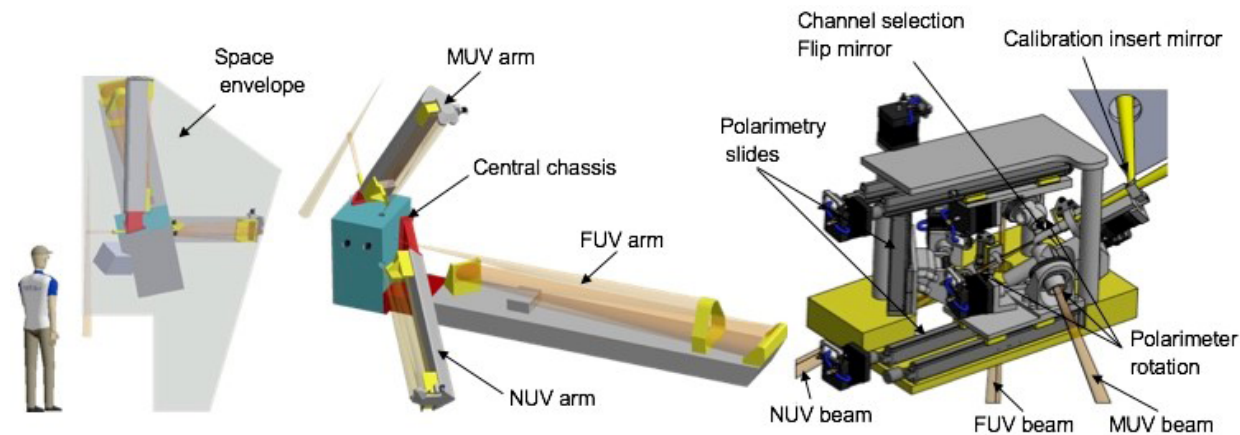

**Figure 10.12.** *Rendering of POLLUX within the allocated volume (left); Opto-mechanical layout (center); Beam switching and polarimetry mechanisms (right).*

mechanism driven by a stepper motor. The axis is defined by a pre-loaded angular contact bearing pair. For the FUV and MUV, the polarimeters are rotated to four positions, while for the NUV the modulator is rotated to six positions.

### 10.4.4 Detectors

The detectors of POLLUX are based on the technology of surface processing of thinned, back-side illuminated EMCCDs (e.g., "δ-doping"). These have now become competitive with micro-channel plates (MCPs) in the FUV to NUV range. They combine the linearity of CCDs with photon-counting ability, which is a key capability enabling detection of faint UV signals. Furthermore, these detectors now deliver high quantum efficiency (e.g., in-band QE > 60%, see http://www.mdpi.com/1424-8220/16/6/927), thus offering the possibility to reach very high signal-to-noise ratios.

Recent developments show that visible-blindness can be achieved with proper treatment (e.g., anti-reflection coatings). Detectors with 13 μm pixels will be used for POLLUX. They may be passively cooled to ~ 120K (to reduce dark current level etc.). In the FUV channel, the detector active area is 203 mm x 19 mm, while for MUV and NUV, the active areas are 131 mm x 19 mm and 131 mm x 24 mm, respectively.

### 10.4.5 Operation modes

POLLUX is designed to work in two science modes, a pure spectroscopic mode and spectropolarimetric mode.

- In pure spectroscopy mode, one measurement produces either one spectrum in the FUV, or two spectra, one in the MUV and one in the NUV. It is expected that in a majority of cases, two measurements (one in the FUV and one in the MUV+NUV) will be obtained for a target.

- In spectropolarimetric mode, one full-Stokes polarimetric measurement requires four spectra in the FUV, or ten spectra in the MUV+NUV channels (six spectra in the NUV and four spectra in the MUV. Here again, we anticipate that in most cases two measurements (one in the FUV and one in the MUV+NUV) will be obtained for a target.

When switching to the spectroscopic mode, the full polarimetric unit for MUV and NUV, or the polarimetric modulator for the FUV, is removed from the optical train. The optical path length and the beam position are then maintained by translating the collimator mirrors (see **Section 10.4.1**).

In addition to the science modes, POLLUX can work in calibration mode. This consists of inserting one of the calibration lamps in the light path, blocking the stellar light, to acquire calibration images.

The calibration unit (CU, **Figure 10.13**) groups the necessary light sources, including a cold redundancy. In order to minimize the impact of the instrumental polarization on the final calibration, the light from the calibration sources will be injected before the polarimeters of each channel as early as possible in the optical chain, i.e., at the level of the flip-mirror mechanism directing the light towards the FUV or MUV+NUV channels.

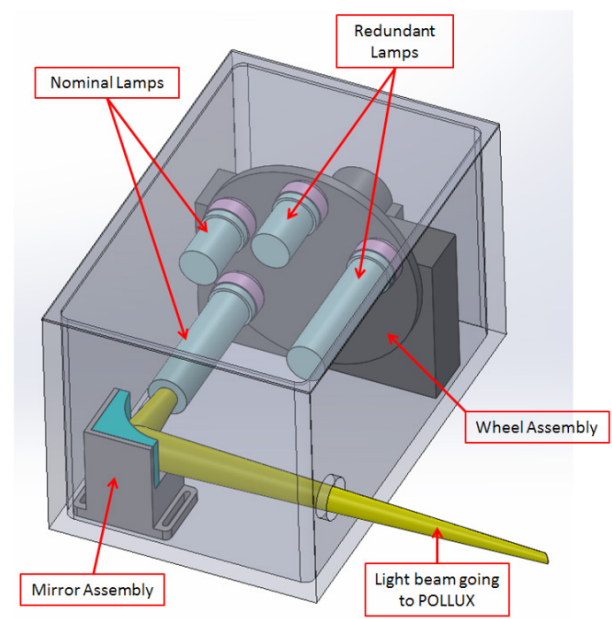

**Figure 10.13.** *Schematic rendering of a possible calibration unit for POLLUX.*

The injection of the calibration beams requires an additional mechanism inside the CU, for the selection of the calibration sources. A wheel, bearing the calibration lamps, is placed facing a mirror assembly that is used to redirect the light from the lamps to POLLUX while reproducing the f-number of the LUVOIR telescope. This way, we avoid any systematic effect resulting from different illumination patterns between the scientific and calibration beams within the instrument. Such wheel mechanisms are usually used as filter wheels inside optical instruments and therefore they have a long flight heritage.

The main calibration purposes for POLLUX internal calibration sources are wavelength calibration and flat fielding. The optical stimulus for the wavelength calibration is a Pt/Ne Hollow Cathode Lamp (HCL), covering the entire wavelength range of the MUV+NUV channel. The flat-field (FF) source is used to calibrate the pixel-to-pixel response variations and to monitor the blaze function and/or the evolution of the relative spectral response function (which will be tied to celestial calibrators at sparser time intervals). The optical stimulus for the FF is a deuterium arc lamp. Given that the FF will not be acquired with every observation, the impact on the overall mission efficiency is marginal. A power supply box will be placed next to the CU and will include a high voltage supply, which is required for the HCL.

The error matrix of polarization will be calibrated on the ground. Any possible evolution or aging will be monitored in flight by regular observations of a set of celestial calibrators, used as reproducibility sources.

We have currently not identified calibration sources covering the full FUV channel. Consequently, in this channel, the ground-based and onboard calibrations will be completed with observations of celestial standards, essentially white dwarfs.

Standard calibration images will be collected once per day for each detector through a fixed sequence: 10 flat-field images, 5 bias images, and 1 dark image. In addition, wavelength calibration should be obtained. In pure spectroscopy mode: 2 wavelength calibration images will be downloaded at each new pointing of the telescope, one once the telescope is pointed and before the science acquisition start, and the other after the science acquisition is finished and the telescope moves away. In spectropolarimetric mode: 1 wavelength calibration image would be obtained not only before and after the acquisition but also between each spectropolarimetric measurement.

### 10.4.6  Performance evaluation

The overall efficiency of POLLUX was computed under the following set of assumptions:

- The pick-off mirror is assumed to be covered with the same broadband coating as the telescope mirrors (Al+MgF$_2$+SiC).
- The dichroic is taken to be identical to that used in GALEX, but its efficiency curves are shifted by 17.1 nm to the red.
- The coating of the flip mirror is single-layer SiC. The coatings of the collimators and the cross-dispersers are single layer SiC in the FUV, and Al+LiF+AlF$_3$ in the MUV and NUV.
- The 3-mirror modulator and analyzer of the polarimeter in the FUV are SiC. For the MUV, the 3-mirror modulator is coated with Al+LiF, while the analyzer is MgF$_2$. For the NUV, both the plates and analyzers are MgF$_2$.

- For each channel, the echelle gratings work under pure Littrow mounting. They are etched into Si substrate, and their profiles are not fully optimized at this stage of the study. The echelle coating is taken to be Al+LiF for all the channels.
- Efficiency of the cross-disperser grooves is that of an ideal blazed profile in Al multiplied by the coating reflectivity.
- The detector quantum efficiency assumes an uncoated δ-doped EMCCD. In the future it will be optimized for each of the channels separately.

The total efficiency (**Figure 10.14**) is the product of efficiencies of all the elements. The result was also converted into the effective area (adopting a geometrical coefficient $A_{geom}$=135 x 10$^4$ cm$^2$ for LUVOIR).

In the final design of POLLUX, we expect that using specifically tailored coatings for each channel will improve the throughput of POLLUX. We also intend to investigate alternative designs reducing the number of reflections. However, the current baseline already shows that POLLUX is feasible and its science goals reachable.

### 10.4.7 Future developments for POLLUX

In 2018, we will continue to improve the optical and mechanical design. We will also study the thermal architecture, the thermo-mechanical stability, the main electronic hardware and software, the data telemetry, the power and mass budget, the AIT/AIV model philosophy, the contamination and cleanliness issues, and the radiation impact.

## 10.5 Engineering and technology challenges

**Table 10.2** presents the current TRLs and heritage of the main components of POLLUX. Here we present the development plan we have prepared for key elements in

**Table 10.2**. Two of the elements have low TRLs, namely the cross-dispersers and polarimeters:

*Cross-dispersers.* Each of the cross-dispersers of the present design is a concave freeform holographic grating. This is a novel type of optical element having high aberration correction capabilities. For the cross-dispersers, computations show that a sufficient aberration correction is possible only if the spectral components at the cross-disperser's surface are separated, although it leads to the grating's aperture increase.

A R&D program has been set up in collaboration with Jobin-Yvon HORIBA (France) to study the feasibility of the required holographic recording geometry parameters, and develop and test prototypes. According to current technology, no showstoppers have been identified.

Examples of clear apertures and asphericities of the cross-dispersers used in the present design are shown in **Figure 10.15**.

*Polarimeters.* An existing R&D program funded by CNES started four years ago on the development of new concepts for UV polarimeters. It was extended for three more years in the fall of 2017. In this frame, a first prototype of the concept proposed for the NUV channel has been built and tested in the visible, and is now being tested in the UV. The prototype was optimized for the ARAGO project and will be specifically tailored for POLLUX in 2018.

Furthermore, the concept with three mirrors proposed for FUV and MUV will be further studied and prototypes will be developed and tested in the framework of the CNES R&D. A PhD student has already been hired to work on this specific point.

The performance of the polarimeters strongly depends on the properties of the material and coatings to be used. Measurements to characterize them (e.g., refractive index) will be performed in collaboration with the Max-Planck institute in Göttingen and the Metrology Light Source of PTB in Berlin (Germany).

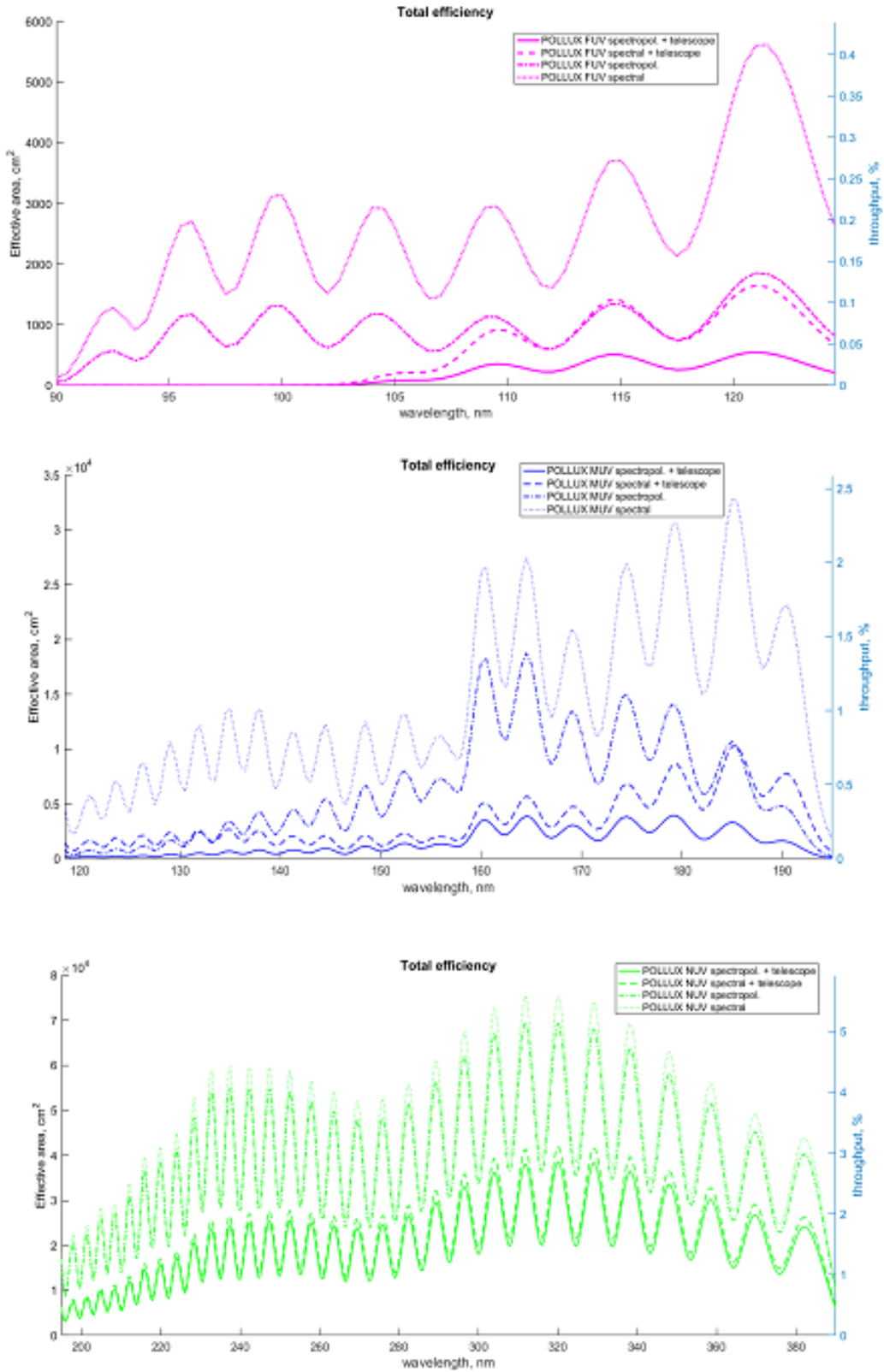

**Figure 10.14.** *Throughput for POLLUX and effective area. For the MUV and NUV, they have been computed with and without the polarimeters. See text for underlying assumptions.*

**Table 10.2.** *TRL and heritage of main POLLUX components*

| | Element | Est. TRL | Status | Comment |
|---|---|---|---|---|
| 🟩 | Flip-Mirror mechanism | 9 | Exist | Similar mechanism has already flown (e.g., HST) |
| 🟨 | Dichroic | 5 | To be tailored | Heritage from GALEX and ARAGO study |
| 🟥 | FUV polarimeter | 2 | Concept defined | New concept |
| 🟥 | MUV polarimeter | 2 | Concept defined | New concept - Analyzer exists and has been tested |
| 🟧 | NUV polarimeter | 4 | Proven concept | Based on HINODE and CLASP heritage, and ARAGO study |
| 🟩 | FUV, MUV and NUV collimator translation | 9 | Exist | Similar mechanism has already flown (e.g., CHEMCAM) |
| 🟩 | FUV, MUV, NUV collimator | 8 | To be tailored | OAP mirror |
| 🟧 | FUV echelle grating | 4 | To be studied | Manufacturer identified. Size exceeds what is presently doable. Prospects are encouraging |
| 🟩 | MUV and NUV echelle grating | 7 | To be tailored | Manufacturer identified. Similar gratings flew on sound rockets |
| 🟥 | FUV Cross-disperser | 2 | Concept defined | To be studied. On-going contact with Jobin-Yvon Horiba. |
| 🟥 | MUV, NUV Cross-disperser | 2 | Concept defined | On-going contact with Jobin-Yvon Horiba. Preliminary studies are encouraging |
| 🟩 | FUV, MUV, NUV detectors | 6 | To be studied and tailored | $\delta$-doped EMCCDs; will rely on heritage from recent or soon to be flown missions (e.g., CHESS, LIDOS, FIREball-2) |
| 🟨 | FUV coating | 5 | To be studied | Choice will depend on the cut-off wavelength of the telescope. |
| 🟨 | MUV, NUV coatings | 5 | To be studied | Several options are identified and will be compared |
| 🟩 | Calibration lamps | 9 | Exist | Heritage from SCIAMACHY, IUE, HST... |

Other elements of POLLUX have higher TRLs but require further development and tailoring:

***Detectors.*** The technology of $\delta$-doped EMCCDs is not fully mature yet. More R&D is required to further demonstrate the dynamic range and how low spurious noise is, in a realistic, end-to-end environment for spectroscopy. We need to assess how the detectors behave around and below Lyman-$\alpha$ for instance. Implementation of antireflection coatings, or of metal-dielectric stacks as visible rejection filters must be improved. A last point will be to demonstrate feasibility of detector wafers large enough to accommodate our needs (typical detector size is 15k x 2k for POLLUX). This size appears achievable in the coming years (communication from Dr. Shouleh Nikzad, JPL). In case tiling of detectors will be needed, tests of the impact on the science cases will be done. These are specific studies that will be led by Leicester University (UK).

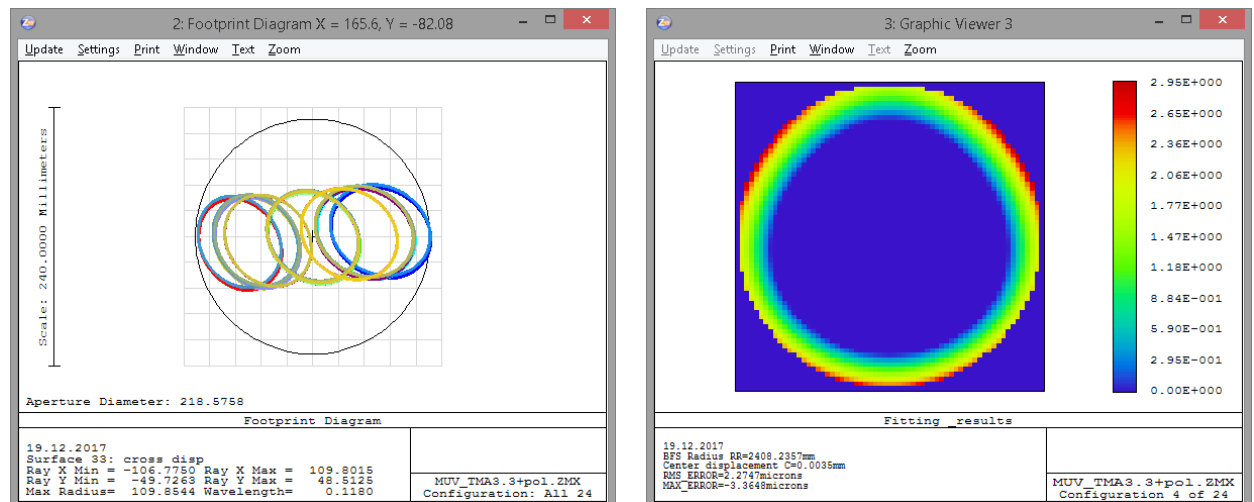

**Figure 10.15.** *MUV cross-disperser. (Left) Footprint diagram. (Right) Asphericity map in microns.*

*Echelle gratings.* In the current design, the MUV and NUV echelle gratings have almost identical sizes and blazing angles. The groove frequencies differ approximately by a factor of two. Both the groove frequencies and the angles are non-standard. For the FUV echelle grating, the groove frequency and blazing angle are even more unusual. We have no evidence that such a grating has been ever produced before for the UV domain. For the three channels however, feasibility is in the range of today's technological limits (communication from Dr. Randall McEntaffer, Penn State University).

Open issues include the accuracy of the groove profiles and the coatings optimization for each of the channels. Another high technological risk is the clear aperture of the gratings. Today's process tools are limited to 200 mm diameter wafers (~140 x 140 mm clear aperture), so at least one of the dimensions on each of these gratings is a challenge (e.g., McEntaffer et al. 2013). These points will be addressed in the framework of a R&D plan between Penn State and University of Colorado, and the POLLUX consortium.

*Coatings.* A CNES R&D program to obtain and select coatings with excellent UV and FUV efficiency, particularly in the 90 nm to 390 nm range, started in October 2017. The goal is to reach above 90% for as much as possible of this spectral range, high-uniformity coatings (< 1%), over a large spectral range with low polarization (< 1%) for high-contrast imaging, and pre-launch stability of ultra-thin coatings.

Materials with the best properties are $MgF_2$, LiF, and $AlF_3$. Single and double layers of these materials have been investigated, e.g., Al/LiF, Al/$AlF_3$, Al/LiF/$MgF_2$, Al/LiF/$AlF_3$, and combinations with different process conditions. At the short wavelength edge, presently a top coating with $AlF_3$ gives the best reflectance results, as has been published most recently by Del Hoya and Quijada (NASA GSFC). A reflectance of more than 30% above 100 nm and more than 90% from 120 nm to 130 nm has been reported (see **Figure 10.16**).

Members of our consortium have investigated in the past protective coatings of $MgF_2$, LiF, and $AlF_3$ for mirrors at 120 nm and studied also their environmental stability and durability. To optimize these coatings at the short wavelength cutoff it is necessary to investigate further the dependence on coating thickness and substrate temperature during the deposition process. We propose to investigate protected aluminum coatings in collaboration between MPS (Göttingen, Germany) and OptiXfab (Jena, Germany). Experimental studies with mirror samples will start in 2018.

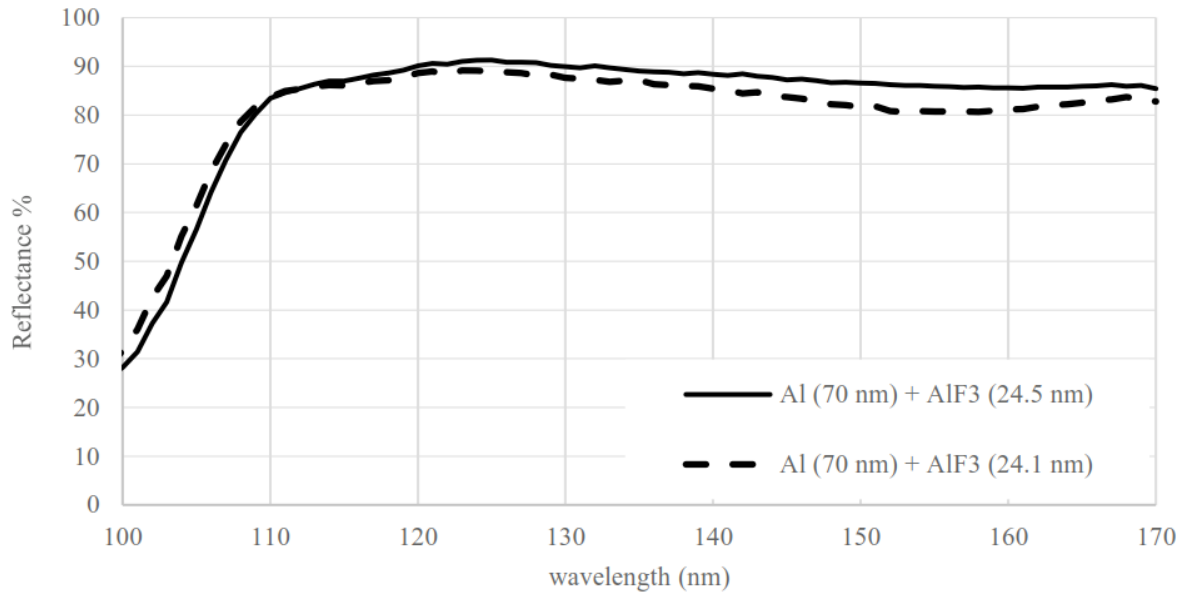

**Figure 10.16.** *Reflectance of Al protected with AlF$_3$ coatings. Credit: Quijada et al. (2017).*

*Dichroic.* The dichroic filter used for beam separation of the MUV and the NUV is based on the GALEX heritage. It provides a mean reflectance of 61% over the 140–170 nm band and a mean transmittance of 83% over the 180–275 nm band. The reflectance was extrapolated down to 118.5 nm, the transmittance up to 390 nm, for the computation of the efficiency calculations presented in **Section 10.4.6**. A study by REOSC performed for ARAGO has demonstrated feasibility down to 119nm. Our development plan is to tailor a dichroic beam splitter to the MUV and NUV, and further increase the efficiency in both channels of POLLUX. This will be done in partnership with the REOSC Company.

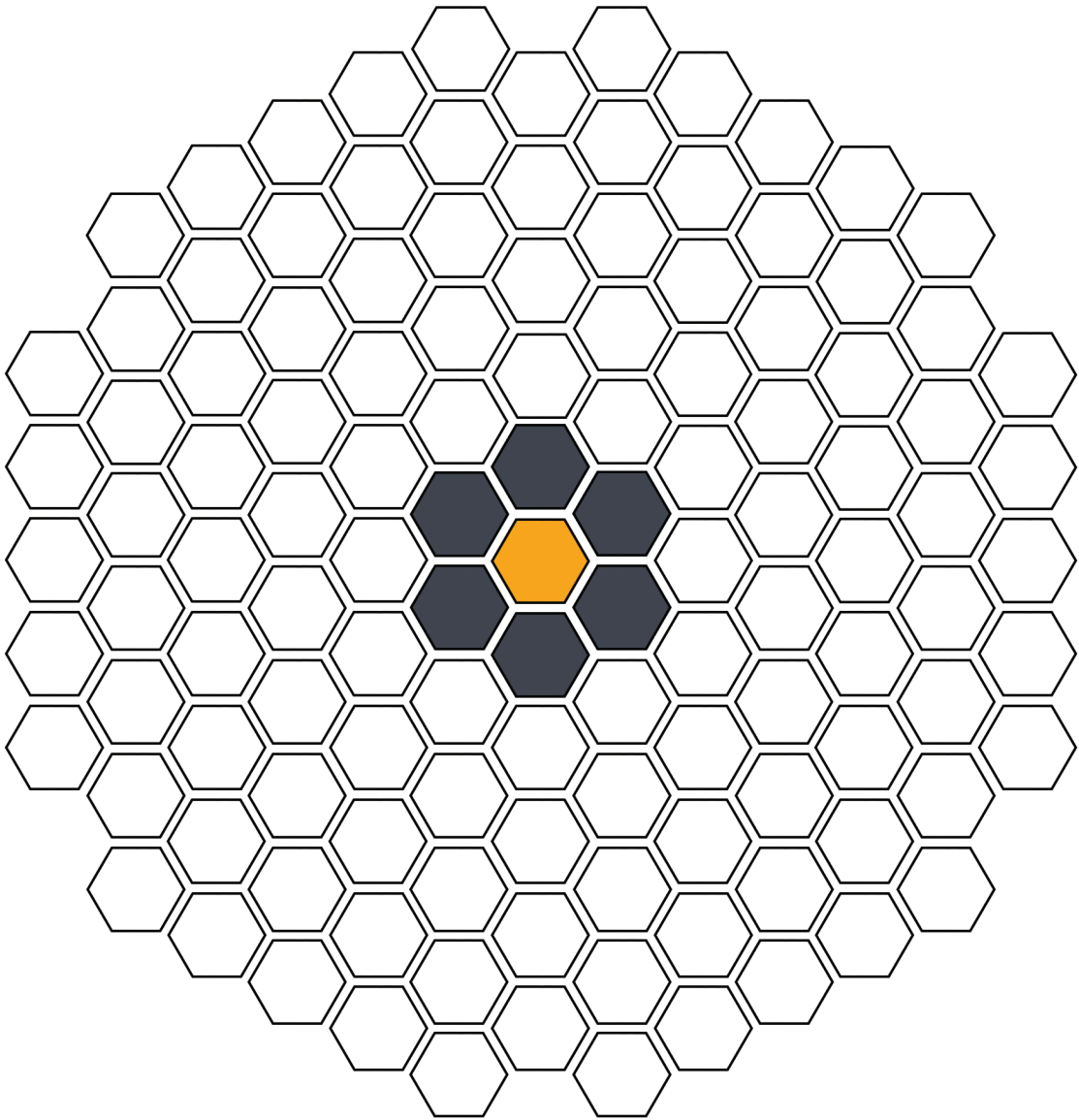

LUVOIR technology development

# 11 LUVOIR technology development

## 11.1 Introduction

The engineering concept behind LUVOIR combines JWST-leveraged telescope design, high-contrast imaging technologies developed and matured by the WFIRST mission, and innovations from the aerospace and commercial sectors. This chapter describes how the critical technological elements of LUVOIR are driven by its major scientific goals and outlines a near-term technology development roadmap. It is distilled from the efforts of the LUVOIR Technology Working Group, a group of subject matter experts in telescopes, structures, mirrors, actuators, wavefront sensing and control, coronagraphy, astronomical instruments, optical devices, mechanisms, detectors, spacecraft attitude control, vibration isolation, and other relevant disciplines. This group is working in concert with the LUVOIR Study Team to identify technologies needed to implement a successful LUVOIR mission. Preliminary assessments were reported in (Bolcar 2017) to provide early identification and prioritization of key LUVOIR technologies. This chapter provides an updated evaluation of these technologies, capturing progress made since that initial report.

LUVOIR's extraordinary capabilities will be driven, first and foremost, by its large aperture. As described earlier, LUVOIR will be launched with several elements stowed or in a folded configuration. These will deploy on orbit, and be locked into a solid, stable structure. Image-based wavefront sensing and control techniques will phase its optical elements to achieve diffraction-limited optical performance. The technologies for this process are derived directly from JWST and will be proven in space by that mission (Dean et al. 2006; Redding et al. 2003; Feinberg et al. 2008; Acton et al. 2012).

Extension of JWST wavefront sensing and control technologies to the shorter LUVOIR wavelengths is a matter of engineering, with the exception of coronagraph control and ultra-stability aspects discussed later in this chapter. LUVOIR will be larger than JWST, but its individual structures and mechanisms scale directly from space qualified JWST structures and mechanisms. Indeed, they may be easier to fabricate and test, since LUVOIR will not be cryogenic, and represent engineering evolution of existing designs that do not require new invention.

The technologies discussed in this chapter are critical to the science objectives of the LUVOIR mission. Some are currently immature, without demonstration by previous flight missions. In NASA parlance, they have a low Technology Readiness Level (TRL), from TRL 2 to TRL 4 (see **Table 11.1** for definition of TRLs). All will need to be matured, through analysis and simulation, then by hardware breadboards and testbeds, then by building engineering models and prototypes for system or subsystem level testing in representative environments, to establish TRL 6 by the mission Preliminary Design Review (PDR). Recommendations for how and when they can be matured are noted throughout the chapter.

There is overlap within the major technology areas: between different coronagraph design families, with differing performance and sensitivity to disturbances; and between the various active optics and instrument technologies. There is a wide range of maturity as well. This provides LUVOIR scientists and engineers with options—and an architectural challenge. What combination of approaches at the system level best balances cost, risk, and complexity, while providing the required performance to achieve the stated science cases? The answer to this question will be a moving target, changing as progress is made in the various technologies. The initial approach is to prefer less complex solutions, while anticipating needs that may arise as understanding of the most challenging requirements improves.

**Table 11.1.** *Technology Readiness Levels, and when they need to be achieved for critical LUVOIR technologies (Hirshorn et al. 2007).*

| TRL # | Maturity Criteria | Needed Prior To… |
|---|---|---|
| 9 | Actual system "flight proven" through successful operations on-orbit | |
| 8 | Actual system completed and "flight qualified" through test and demonstrations (ground or flight) | |
| 7 | System prototype demonstration in a target/space environment | |
| 6 | System/subsystem model or prototype demonstration in a relevant environment (ground or space) | Project Preliminary Design Review (PDR) |
| 5 | Component and/or breadboard validation in relevant environment | |
| 4 | Component and/or breadboard validation in laboratory environment | Project Phase A |
| 3 | Analytical and experimental critical function and/or characteristic proof-of-concept | |
| 2 | Technology concept and/or application formulated | |
| 1 | Basic principles observed and reported | |

The path to TRL 6 is based on performance testing at the component, subsystem, and system levels that will guide design choices, and ground the overall LUVOIR architecture in demonstrated performance. Computer modeling and analysis is an essential part of this process, and model verification at each step of technology demonstration is required to validate the tools that predict system performance. These tools will quantify the key architectural trades and provide an essential guide for LUVOIR development at every project phase, as they have for other observatories in the past.

We begin the discussion by reviewing coronagraph technologies, as they are the source of requirements for other technologies—such as the need for ultra-stable optics and high-performance detectors. Ultra-stability is the second topic of discussion—this includes mirror fabrication and coatings, as well as active optics technologies. Then, instrument technologies, especially detectors for all of the LUVOIR instruments, as well as devices such as microshutter arrays for spectrographs, are discussed.

## 11.2 Coronagraph technology: High contrast imaging with an obscured aperture

LUVOIR's groundbreaking exoplanet observations will be accomplished by high contrast direct imaging, using a coronagraph to block the light from a star, in order to detect and characterize the much fainter light reflected from planets orbiting the star. The optical layout of a canonical coronagraph is sketched in **Figure 11.1**. It is an optical instrument that focuses light captured by the telescope onto an occulting mask. The mask blocks the star light at the very center of the field but passes light from objects even a very small angle off-center—like planets orbiting that star. After the mask, the beam is re-collimated and passed through a pupil stop, to remove starlight diffracted by the occulting mask. Finally, the beam is re-focused to form a starlight-suppressed image of the exoplanet system. This image can be produced either directly on a detector for broadband imaging, or on a lenslet array at the entrance of an integral field spectrograph for wavelength-resolved imaging.

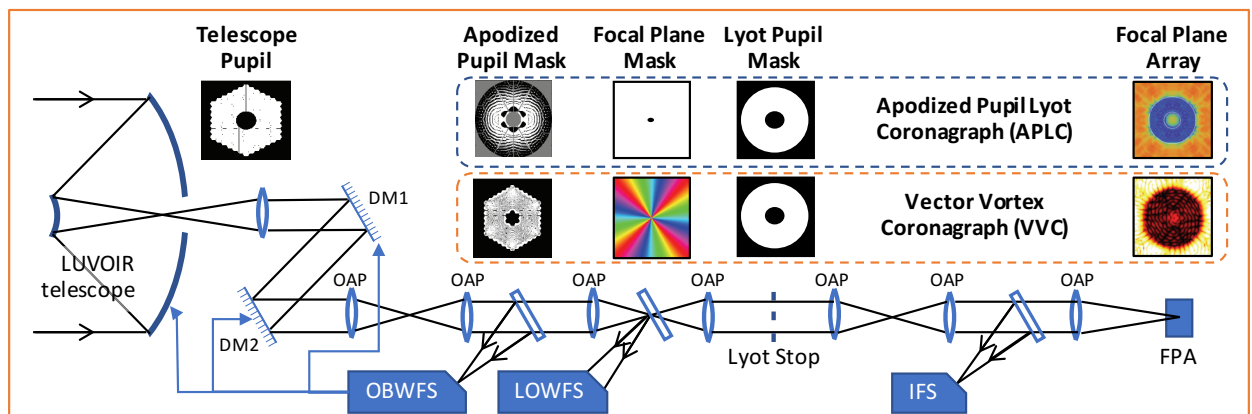

**Figure 11.1.** *Coronagraph elements, showing apodized pupil Lyot coronagraph and vector vortex coronagraph features. The telescope is pointed at a target star, whose light is passed to the coronagraph, through an apodized pupil mask, and then focused on a focal plane mask. The mask suppresses the star image. The rest of the light, including the off-axis planet light, passes by the mask. Downstream, a Lyot stop suppresses residual diffracted starlight, and then the planet light is refocused onto a focal plane array detector or fed into an integral field spectrograph. Two deformable mirrors (DMs) are used to establish high contrast. Out-of-band and low-order wavefront sensors (OBWFS, LOWFS) stabilize wavefront and intensity variations to preserve contrast. OAP: off-axis parabola—a typical reflective mirror used to relay the light from pupil plane to image plane throughout the system.*

### 11.2.1 Coronagraph architectures

Coronagraph designs from the late 2000's were optimized for circular (or elliptical) apertures that do not have any obscurations in the pupil. This will not be the case for LUVOIR's on-axis telescope, with its secondary mirror and support struts, plus gaps between mirror segments introducing discontinuities in the optical beam. The diffraction caused by these features results in a broadened, azimuthally structured point-spread function. To achieve high suppression of the central starlight, these effects must be countered by other elements in the coronagraph, such as:

- an apodizing mask to filter the obscuring pupil features
- a focal plane occulting mask with a complex transmission characteristic

- deformable mirrors (DMs): small mirrors that can be precisely shaped to remap the telescope pupil geometry.

The coronagraph research community is actively exploring how to optimally combine these design elements (Mazoyer et al. 2017).

The key design objectives for a coronagraph are the deep suppression of the on-axis starlight, and the preservation of the planet light immediately adjacent to the center of the field, in the "dark hole" detection region. Starlight suppression is quantified in terms of contrast: the ratio of the intensity of the faintest detectable planet to that of the central star. The inner dimension of the dark hole is the inner working angle (IWA): the angle which quantifies how close a planet can be to the star and still be detected. IWA is commonly expressed in aperture-normalized terms, as a multiple of $\lambda/D$: the ratio of the wavelength of light divided by the aperture diameter, which is about the diffraction-limited half-width of an ideal point-spread function. Other key design metrics are bandpass—typically 10 to 20% of the center wavelength—and throughput, the percentage of the planet light captured by the telescope that makes it through to the core of the point-spread function imaged on the detector. Together, these metrics determine how quickly an observation can reach a high level of integrated signal, and how efficient spectroscopic characterization observations will be.

Early work on advanced space corona-graph concepts emphasized achieving high contrast on telescopes with idealized unob-scured apertures. Using a Lyot design with a band-limited occulter mask together with two DMs, the NASA Terrestrial Planet Finder project demonstrated contrast of $5 \times 10^{-10}$ with 10% bandpass at an IWA of 4 $\lambda$/D. This established that the performance needed to image an Earth analog is possible. It also kicked off a flurry of new coronagraph designs and tests. The best demonstrated contrasts at an IWA of 3.1 $\lambda$/D[1] on an unob-scured aperture currently stand at:

- $1.2 \times 10^{-10}$ at a bandpass of 2%
- $6 \times 10^{-10}$ at a bandpass of 10%
- $1.3 \times 10^{-9}$ at a bandpass of 20%

The addition of the Coronagraph Instrument (CGI) to the WFIRST mission offered a major opportunity to the exoplanet imaging community—the application of state-of-the-art technology on a 2.4-meter diameter space telescope—and also new challenges. The WFIRST telescope has a large central obscuration (30% of diameter) and six secondary support struts. Together, these pose an exceptionally difficult coronagraph design problem. Furthermore, WFIRST will operate in a high-vibration environment that will dynamically tilt and deform the

CGI wavefront during observations. In response to these challenges, coronagraph development has made rapid progress over the past 4 years, leading to the conception and laboratory demonstration of new design solutions for obscured apertures, and the application of Low-Order Wavefront Sensing and Control (LOWFS). LOWFS uses light rejected by the coronagraph mask to actively stabilize the wavefront during coronagraph observations. LOWFS and related active stabilization methods are discussed in **Section 11.3**.

WFIRST CGI recently demonstrated a contrast $<1 \times 10^{-8}$ at an IWA of 3 $\lambda$/D with a bandpass of 10% using 2 different coronagraph designs and LOWFS control. The demonstration was performed on a testbed that includes a WFIRST-like shadowed pupil and deliberately-induced dynamic disturbances that trace to those expected on orbit. As documented in Shi et al. (2017) and Seo et al. (2017), the hybrid-Lyot coronagraph design achieved $5 \times 10^{-9}$ contrast averaged over 3–9 $\lambda$/D with a 10% bandwidth in the dynamic environment. Another major testbed milestone was reached in 2017 when the CGI team integrated a shaped-pupil Lyot coronagraph with an integral field spectrograph and demonstrated the first $1 \times 10^{-8}$ contrast over a broad (18%) spectroscopically-dispersed bandpass (Cady et al. 2015; Groff et al. 2017).

The progress made by the WFIRST project in developing obscured aperture coronagraphs has been driven by rigorous lab testing. A similar ground testing campaign, using a high-quality LUVOIR-traceable testbed, will be needed to demonstrate and debug LUVOIR coronagraphs. One such facility, the JPL Decadal Survey Testbed, will be used to demonstrate coronagraph performance at levels that trace to LUVOIR and HabEx science goals (Crill & Siegler 2017). This testbed will also provide the data for model validation activities, to ensure that the models used to engineer the coronagraphs and testbeds capture the key effects governing coronagraph performance. In parallel, the Space Telescope Science Institute is now commissioning the High Contrast Imager for Complex Aperture Telescopes (HiCAT) testbed, aimed at system level demonstration at modest performance levels (non-vacuum), but realizing the nested loop wavefront control architecture that is necessary for LUVOIR. Caltech is also developing the High Contrast Spectroscopy Testbed, shown in **Figure 11.2**, that will demonstrate high-contrast, high-spectral resolution instrument concepts.

---

1 See the NASA Exoplanet Program Technology Appendix: https://exoplanets.jpl.nasa.gov/exep/technology/technology-overview/

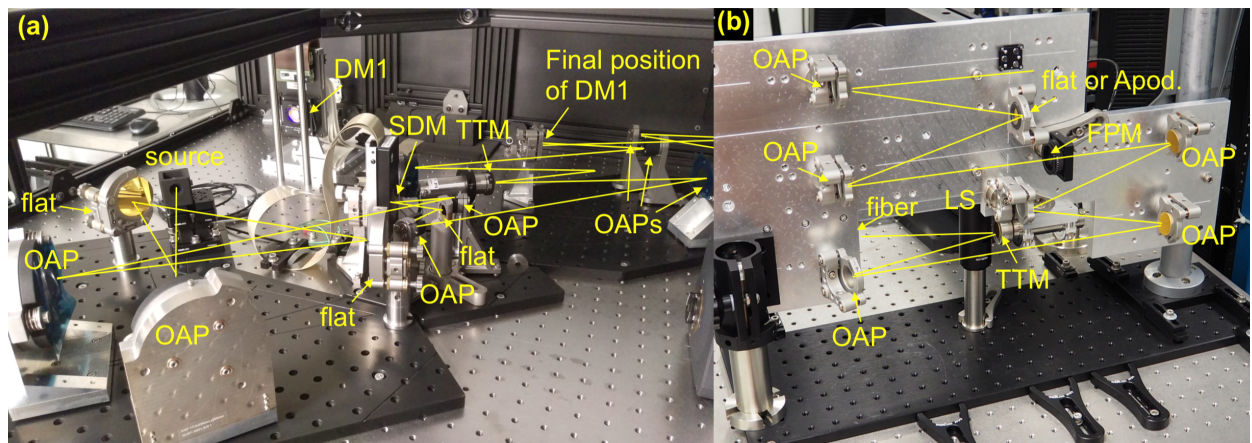

**Figure 11.2.** *(a) The Caltech High Contrast Spectroscopy Testbed telescope simulator and wavefront control modules. A segmented deformable mirror (SDM) simulates the primary mirror of a segmented aperture telescope. A Boston Micromachines kilo-DM (DM1) will be used for wavefront control (not yet installed in photo). (b) A compact coronagraph and fiber injection unit consisting of an optional apodizer (Apod), focal plane mask (FPM), and Lyot stop (LS) followed by a tip-tilt mirror (TTM) used to couple planet light into a single mode optical fiber, which feeds into a spectrograph. The hardware shown will be deployed at W. M. Keck Observatory as part of the KPIC instrument in 2018. Credit: D. Echeverri and J.-R. Delorme, (Exoplanet Technology Lab, Caltech).*

LUVOIR coronagraph instruments will need 10 to 100 times the starlight suppression performance of the WFIRST CGI, requiring continuing development of key coronagraph elements beyond the WFIRST technology freeze. Some of these further developments are already underway, including the

Segmented Coronagraph Design Analysis (SCDA) study, which sponsors independent coronagraph design work at Universities and NASA Centers, exploring alternate design pathways, and comparing performance in the context of telescopes like LUVOIR (Shaklan 2016). SCDA has already yielded several coronagraph designs that theoretically achieve high contrast ($10^{-10}$), with bandpass up to 15%, IWA of 3–4 $\lambda$/D, and with core throughput as high as 20%[2]—all with segmented apertures traceable to LUVOIR. The performance of these coronagraphs can be degraded by numerous effects such as leaked light from spatially-resolved stars, mask fabrication and alignment errors, DM calibration errors, and dynamic wavefront error disturbances. The SCDA design teams

---

2 The "core throughput" of the coronagraph is defined here as the ratio of the energy in the planet point-spread function core to the total incident energy on the obscured primary mirror, ignoring losses due to mirror reflectivity, filter transmission, spectrograph transmission, and detector quantum efficiency.

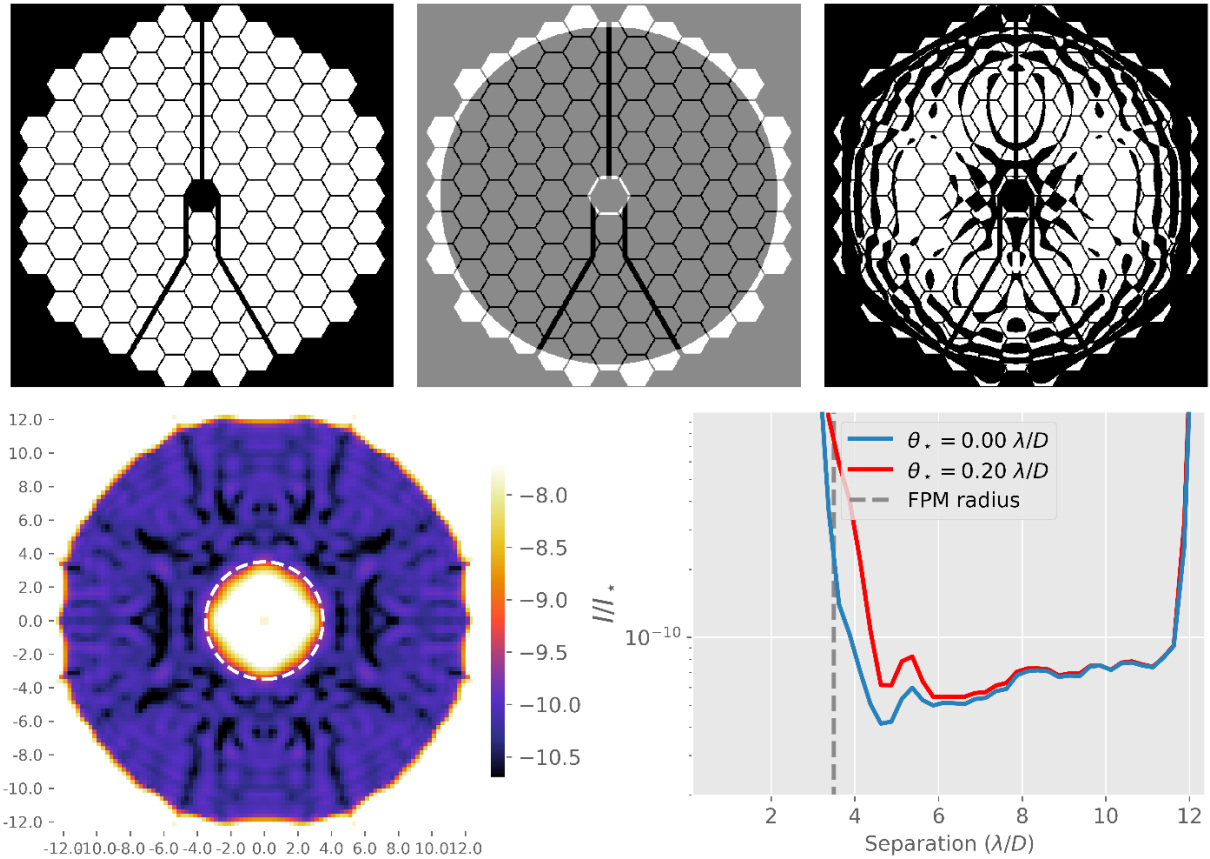

**Figure 11.3.** *Example apodized pupil Lyot coronagraph design. In the upper left is the LUVOIR-A telescope pupil. The top center and top right show the Lyot mask and apodizing mask, respectively. The bottom left shows the annular high-contrast region in the coronagraph focal plane that would result after wavefront control. The bottom right is an azimuthal average of the radial cut through the dark-hole region and shows the effects of finite stellar diameter on the contrast. The vertical dashed line shows the diameter of the focal plane mask occulting spot. Credit: N. Zimmerman (NASA GSFC).*

are pursuing solutions that are robust to these effects.

For LUVOIR-A, the preferred coronagraph designs include the apodized pupil Lyot coronagraph (APLC) (N'Diaye et al. 2015; Zimmerman et al. 2016). The APLC uses a mask technology that is similar to the shaped pupil coronagraph on WFIRST CGI. It has the virtue of being relatively tolerant to resolved stellar diameter (maintaining most of the $10^{-10}$ dark zone at star diameters up to 0.1– 0.2 $\lambda$/D) but has a relatively large IWA (> 3 $\lambda$/D) and a throughput penalty originating in the transmission loss of the apodizer mask.

**Figure 11.3** shows an example of an APLC mask design and performance.

The second preferred design is the vector vortex coronagraph (VVC), which can reach the same contrast as the APLC but with smaller inner working angle (~2 $\lambda$/D) and over a wide bandpass, limited only by the wavefront control system performance (Ruane et al. 2016). For centrally obscured telescope apertures, however, the VVC has significantly more sensitivity to stellar diameter than the APLC.

The best overall solution for LUVOIR might be to include multiple coronagraph design types on board. Like the WFIRST CGI, the mask would be selected according to the demands of each observation. Thus, LUVOIR could use its VVC for smaller, more distant stars, or longer wavelength near-infrared observations in general, whereas the APLC would be used for nearby stars large enough to be spatially resolved by the telescope.

**Table 11.2.** *LUVOIR coronagraph technology status and path ahead. Green shaded options are in the LUVOIR-A baseline design. Yellow options may provide alternative or potentially enhancing solutions given enough development. Red options would only be considered with substantial development and gains in system performance.*

| Technology | Driving Requirements | Technical Challenges | Solution Paths | Current TRL | Path to TRL 4 | Path to TRL 5 | Path to TRL 6 | Note |
|---|---|---|---|---|---|---|---|---|
| Segmented aperture coronagraph architecture | Contrast: < 1e-10  Inner Working Angle: < 4 λ/D  Throughput: > 20%  Bandpass: > 10% | Centrally-obscured aperture  Wavefront drift sensitivity  Resolved stellar diameter  Model maturity  Algorithm convergence time | Hybrid Lyot | 4 | ✓ | WFIRST CGI Development | | WFIRST will demonstrate 1e-8 contrast with an obscured aperture to TRL 9. Shaped Pupil is directly traceable to LUVOIR Apodized Pupil architecture. Hybrid Lyot studies for LUVOIR aperture are underway. |
| | | | Shaped Pupil | 4 | | | | |
| | | | Apodized Pupil | 3 | Testbed Demos | Enhanced Testbeds, including LOWFS | Testbed with subscale segmented telescope and relevant dynamic and thermal disturbance | Under study via SCDA. Apodized pupil is traceable to WFIRST shaped pupil concept. |
| | | | Vector Vortex | 3 | | | | |
| | | | Phase-Induced Amplitude Apod. | 3 | | | | Active funding through TDEM program. |
| | | | Visible Nulling | 3 | | | | |
| | | | Photonic Solutions | 2 | | | | Requires study to determine potential benefits. |
| Deformable Mirrors (DMs) | Accuracy: < 5 pm  Stroke: > 1 µm  Format: > 64 × 64 | Actuator count  Actuator yield  Actuator stability  Electronics packaging  Open-loop commandability  Electrical connections | Electrostrictive DMs | 5 | ✓ | ✓ | | WFIRST will demonstrate 48×48 electrostrictive DM to TRL 9. |
| | | | MEMS DMs | 4 | ✓ | Environmental Testing | | Active funding through TDEM and SBIR programs. |
| High-contrast image post-processing | Improve total image contrast by >10× | Low initial contrast  Speckle coherence considerations | Hubble-derived | 3 | Analysis | Test w/ CGI data | | Current methods provide 10-500× improvement, but at 1e-5 initial contrast. |

The phase-induced amplitude apodization—complex mask coronagraph (PIAA-CMC) is a third Lyot coronagraph design variant that has been developed for high performance on segmented apertures (Guyon et al. 2010). The prospects for this new design family are promising, particularly in terms of inner working angle and throughput. The LUVOIR Study Team anticipates new PIAA-CMC designs tailored to LUVOIR to evaluate within the next year alongside the APLC and apodized VVC.

Yet another coronagraph design family that is a candidate for LUVOIR are nulling coronagraphs, such as the visible nulling coronagraph (VNC) (Lyon et al. 2008). Instead of an apodizing and focal-plane mask architecture, the VNC concept relies on pupil shearing interferometry to destructively interfere on-axis starlight. The VNC offers excellent inner working angle (2 λ/D) and relatively high throughput (no apodizing mask), but at the cost of greater optical complexity. The VNC remains under active development both in terms of design concept and laboratory demonstration (Hicks 2016).

These advances have moved the coronagraph state of the art forward, as summarized in **Table 11.2**. Testbed demonstrations have established TRL 4 for the WFIRST CGI coronagraph designs, against the performance objectives of WFIRST. While successful launch and operation of the WFIRST CGI will be extremely valuable to LUVOIR technology development, it is not a prerequisite for, nor will it fully qualify all aspects of, LUVOIR-specific coronagraphs. Rather, WFIRST has shown the great value of high-quality testbeds for development of mission-specific coronagraphs. This is the path forward for LUVOIR, using rigorous ground testing to complete development of LUVOIR coronagraphs. APLC and VVC designs tailored to LUVOIR have been successfully demonstrated at the proof-of-concept level using computer models, establishing TRL 3—laboratory demonstrations are needed to advance to TRL 4, 5, and 6.

Finally, we note rapid progress in photonic devices, driven by the communications industry. Design and analysis of idealized coronagraphs using devices such as photonic lanterns and universal linear optical processors indicate real potential for new methods to help solve some of the hardest problems in obscured-aperture coronagraphy. Further work is needed in this area, to explore performance with realistic devices and begin experimentation.

The next steps for the LUVOIR-specific coronagraph designs include: more detailed analysis and modeling, emphasizing robustness to noise effects; and then demonstration on testbeds to establish TRL 4. Further testing, culminating in closed-loop operation with a scaled testbed segmented telescope, is needed to establish TRL 6.

**Path to TRL 4:** Continue the SCDA study to model coronagraph performance with realistic noise inputs, including optical errors, dynamic disturbances and stellar diameter. Perform testbed demonstrations of the coronagraph designs and achieve moderate contrasts ($10^{-8}$ or better) over 15% bandpass at relevant inner working angles. Correlate the testbed demonstration with models that predict $10^{-10}$ contrasts.

**Path to TRL 5:** Perform testbed demonstrations of coronagraph designs that achieve flight requirements on contrast, bandpass, IWA, and throughput, in the presence of expected dynamic and thermal instabilities.

**Path to TRL 6:** Develop a prototype coronagraph instrument that incorporates flight-like masks, deformable mirrors, and optics, including a segmented aperture front-end telescope system. Achieve flight requirements on contrast, bandpass, IWA, and throughput in the presence of expected dynamic and thermal instabilities.

## 11.2.2 Deformable mirrors

Deformable mirrors (DMs) are small mirrors with reflective face sheets whose optical figure can be changed over a small range using actuators. When used in coronagraphs, one DM is typically placed at a pupil image, where it strongly affects the phase of the electric field, providing direct correction of the largest source of wavefront error, namely the primary mirror. In this configuration a high-contrast dark hole can only be obtained in one half of the field of view. Adding a second DM at a plane away from the pupil image allows correction not only of the phase, but of the beam amplitude, which varies across the beam. The combination of two DMs also enables grating-like diffraction to null "speckles"—concentrations of light that occur in the dark-hole detection region. This results in a dark hole that is now symmetric with respect to the optical axis and in theory has wider bandpass. In general, the optical shape of the DMs is determined by

- obtaining an estimate of the complex electrical field at the science camera (in order to avoid non-common-path errors),
- using a precise diffractive model of the DM response through the coronagraph in order to find the DM settings that will best cancel the aberrated complex field, and
- iterating this process in order to obtain a solution that is robust to noise in the

estimate or model uncertainty. The estimation is in general done using spatial or temporal modulation of the images at the science camera

The DMs in a coronagraph also set the outer size limit of the dark hole. The outer working angle (OWA) is theoretically defined by the spatial Nyquist frequency of the coronagraph DMs, as the reciprocal of twice the actuator spacing. LUVOIR will require high actuator density, from $64 \times 64$ to $128 \times 128$ actuators, to provide dark-hole diameters between 32 and 64 $\lambda/D$ (440–880 mas at a wavelength of 1.0 μm for LUVOIR-A). Note that the IWA and OWA both scale with wavelength. Coupled with the narrow observing bands required for high contrast, this means that multiple observations at different wavelengths are needed to map out the full detection region.

Once a dark hole is achieved, the DMs must maintain contrast by correcting slow drifts in the rest of the optical system, using picometer-level actuator commands. The measurements that allow this to happen are provided by concurrent wavefront sensing, and/or metrology, discussed in **Section 11.3**. The CGI testbed has shown that a DM feed-forward control can also be useful, to provide fast response based on mechanical measurements (Shi et al. 2017). The CGI testbed has also shown that DM actuation error and actuator creep, though small, can limit coronagraph performance. Ultimately, high spatial resolution and/or repeated rounds of the contrast acquisition algorithm may be required to counter actuator errors not measurable by other means.

**State of the Art:**

(1) Electrostrictive PMN-based (lead-magnesium-niobate) DMs have been used to demonstrate $6 \times 10^{-10}$ contrast at 10% bandwidth with unobscured pupils, and $1 \times 10^{-8}$ contrast with the obscured

WFIRST CGI pupil. These are being baselined for the WFIRST CGI instrument (TRL 5).

(2) Micro-electro-mechanical systems (MEMS) DMs have been used in a laboratory setting to achieve $5 \times 10^{-9}$ contrast in the visible nulling coronagraph (TRL 4). Continuous facesheet MEMS DMs are also routinely used on ground-based observatory systems to achieve moderate contrasts ($10^{-5}$–$10^{-6}$) (TRL 4).

**Path to TRL 5:** Continue to mature DM technology by flight qualifying the devices and electronics. Improve actuator counts to $64 \times 64$ or larger. Increasing the actuator count of electrostrictive DMs is accomplished by mosaicking actuator blocks behind a common facesheet. MEMS DMs actuator counts can be increased with photolithography, but scaling electrical interconnects is a challenge. **Figure 11.4** shows a wiring diagram for an 8,000-actuator device currently under development. Other goals include miniaturization of drive electronics and demonstrating feed-forward control precision and repeatability consistent with LUVOIR wavefront sensing and control requirements. Environmental tests of MEMS DMs will demonstrate robustness to launch vibrations.

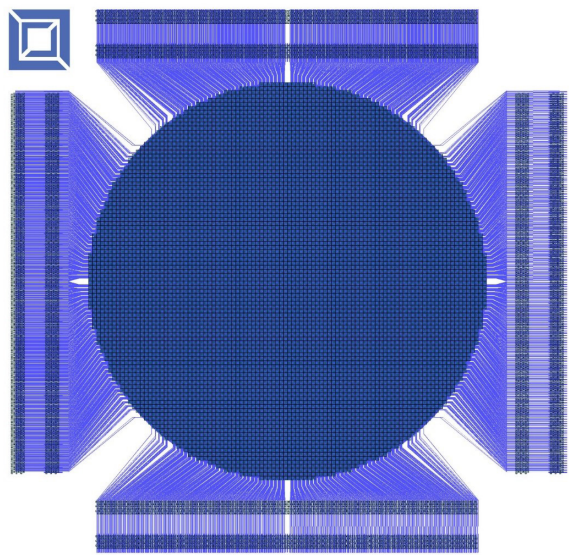

**Figure 11.4.** *Wiring diagram for a ~100 × 100 actuator MEMS deformable mirror device. The final device would have ~8k actuators in the clear aperture. Credit: P. Ryan (Boston Micromachines Corporation).*

**Path to TRL 6:** Use either candidate flight-traceable DMs in a coronagraph $10^{-10}$ contrast demonstration with a segmented aperture telescope front end and closed-loop control.

## 11.3 Active optics: An architecture for ultra-stability

The hardest technical challenge facing LUVOIR is the need to preserve $10^{-10}$ contrast during coronagraph observations. High contrast is established at the beginning of each observation using "speckle nulling" controls, such as the electric field conjugation algorithm, which drives the coronagraph's two DMs to suppress residual starlight in the dark hole, compensating phase and amplitude irregularities in the telescope and coronagraph optics. The DMs must make and hold wavefront adjustments as small as a few picometers RMS. Once contrast is established it must be stably maintained for long periods of time, potentially many days, while the coronagraph accumulates photons from the planets, often at a rate of a few photons a minute.

Analysis shows that the required stability is a function of the spatial frequency of the disturbances, so that a low spatial frequency wavefront error drift can be as large as 100 pm RMS without excessive contrast degradation, while segment-to-segment phasing drift must be held below about 40 pm RMS (Nemati et al. 2017). All of this is required on a spacecraft with vibration disturbances from spinning attitude control reaction wheels, with thermal deformations from an environment with slow temperature changes and using devices such as DMs and materials such as invar, carbon composites and adhesives that, at the picometer level, are prone to creep (Sokolowski et al. 1993).

Against this challenge LUVOIR has several powerful tools. The first is a design that creates a stable thermal environment, even as the telescope is repointed from star to star. This is accomplished with a large flat sunshield, and a gimbal between the sunshield and the telescope, as described in **Chapter 8**. Second is the use of thermally stable materials, coupled with milli-Kelvin-level thermal controls, especially for the primary mirror segments. These minimize thermally driven deformations in LUVOIR mirrors and structures. Third is the use of passive isolation between the spacecraft attitude control system to attenuate high-frequency vibrations before they disturb the telescope.

However, these tools alone will not achieve ultra-stability. For that, LUVOIR will also require active measurement and control of the optics and wavefront during and between observations, to preserve the wavefront established during speckle nulling control in the face of thermal and other drift effects. To keep vibrations from limiting contrast, LUVOIR will also need active disturbance isolation between the payload and the spacecraft as a whole. These capabilities will be provided by active optics technologies described in this section:

- wavefront sensing and control running in parallel with coronagraph observations;
- active, high bandwidth metrology of the optical configuration; and
- active, high-attenuation vibration isolation at the spacecraft/telescope interface.

### 11.3.1 Finding the minimum-cost, maximum-performance solution

LUVOIR has many tools that can be combined in multiple ways to provide the ultra-stability needed for coronagraphy. Which combination will provide the needed performance at the lowest risk, cost, and complexity requires resolving trade-offs that will define the architecture for LUVOIR's active optics system. Final resolution of these design trades requires advancing the TRL of the individual technologies, at the component and subsystem levels, ultimately to achieve TRL 6 for the combined active optics observatory architecture at the mission PDR. This will be done by analyzing, building and testing components and subsystems: first testing components; then testing integrated subscale systems; finally testing at the engineering or prototype level in system-level, subsystem-level assemblies, or testbeds. For the LUVOIR-A point design study, the design team has made choices that emphasize simplicity. More complexity will be added only as needed to meet performance or reduce cost.

This testing is done to verify component and subsystem performance—and in the process, to validate and verify computer models of each component and subsystem. These models can then be combined, intermixing structures, thermal, optics, dynamics, and controls discipline modeling tools in an integrated computing environment: so-called STOP modeling (Structural-Thermal-Optical Performance). Ultimately, with component models grounded in testing, system-level STOP models can accurately predict system-level performance, quantifying the performance aspects of trades between different design options. Even now, at the very earliest stages, integrated modeling is proving essential in identifying architectural approaches, and setting performance allocations for the various components and subsystems, including contrast, image quality, and wavefront error budgets (Redding 2016).

STOP modeling is not a new technology or approach, but LUVOIR will present new challenges, especially the need to accurately predict performance in the picometer wavefront regime, where no material is entirely stable. This is a step beyond current practice. Sources of creep will include DM actuator nonlinearities, desorption of moisture from composite materials, natural creep of materials such as Invar, and epoxy strain relief. These creep effects have the potential to cause changes that degrade contrast, if not measured and controlled at the appropriate time intervals—and the LUVOIR STOP models will need to capture these effects.

### 11.3.2 Wavefront sensing and control concurrent with coronagraph observations: LOWFS, OBWFS, and artificial guide stars

One way to stabilize the coronagraph contrast is by continuously measuring and correcting changes in the optical wavefront, using concurrent wavefront sensing and control (WFSC). This approach has the advantage that it can directly measure all of the changes that occur in the beam train, whether in the telescope or in the coronagraph. It has several challenges, however, such as the long integration times needed to make individual wavefront measurements when observing the dimmest target stars.

The WFIRST CGI testbed is already demonstrating the effectiveness of low order wavefront sensing and control (LOWFS) in measuring and controlling low-order wavefront disturbances (Shi et al. 2016). LOWFS uses starlight rejected by the coronagraph's occulting spot. As shown in

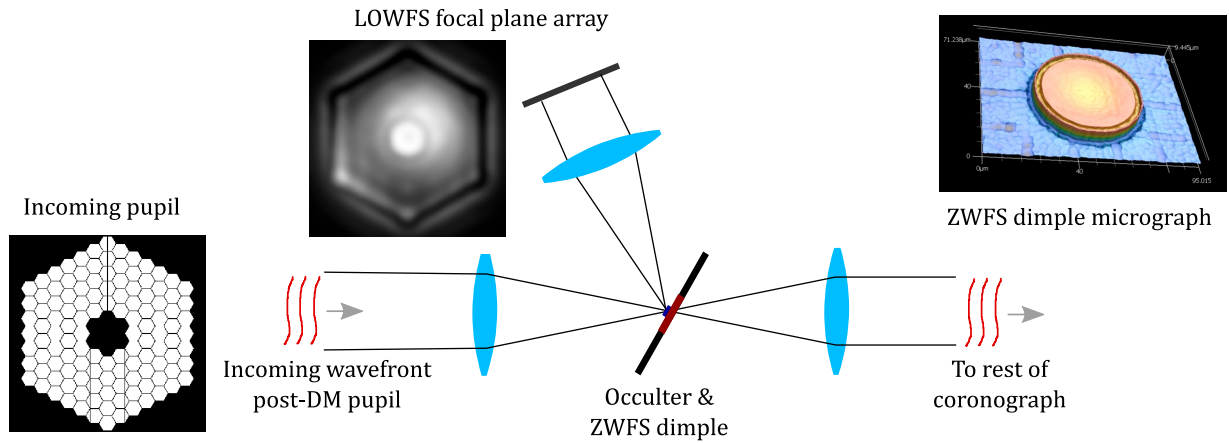

**Figure 11.5.** *Schematic of LOWFS integrated into coronagraph occulter assembly. LOWFS replaces the usual absorbing occulter spot with a reflecting spot (indicated in red) that has a small dimple (dark blue) yielding a small phase-only disturbance in the beam reflected to the LOWFS detector.*

**Figure 11.5**, the core of the starlight image is reflected off the backside of the occulting spot of the focal plane mask, which includes a small dimple at the center to create a spherical reference wavefront. This forms a Zernike wavefront sensor (ZWFS) (Wallace et al. 2011). The interference fringes that result provide a high-precision measure of the wavefront errors present in the beam. The spatial frequency of the measured wavefront is limited, since the occulting spot, which is only the width of the image core, functions as a low-pass spatial filter. LOWFS is not able to resolve higher spatial frequency effects, such as segment-to-segment phase errors, or deformations within segments.

Results on the WFIRST CGI testbed have demonstrated LOWFS line-of-sight sensitivity to 0.01 mas, consistent with a feedback closed-control bandwidth of ~10 Hz and feedforward control up to 100 Hz, and sensitivity to low order spatial modes (focus, astigmatism, coma, trefoil) down to ~10 pm at 0.05 Hz under the testbed source. These results are limited by testbed stability and photon noise. Low-light (V=2 mag for dark-hole creation and V=5 mag for LOWFS) experiments are on-going and the recent initial test results are very promising.

Another approach for LUVOIR is to sample the full beam from the star, so that all of the wavefront spatial frequencies that can be controlled are measured. This is done using a dichroic beam splitter to split the light, sending one band to the coronagraph and another to an Out-of-Band Wavefront Sensor (OBWFS): UV light for wavefront sensing, while observing in the visible, for instance. The OBWFS sensor uses the LOWFS architecture, including a ZWFS, but without the spatial filtering effect of the reflection from the occulting spot. It measures low, mid, and high spatial frequency wavefront errors, though the accuracy and speed with which each term can be measured depends on the magnitude of the guide star. **Figure 11.6** shows the spectral type and visual magnitude of the expected population of LUVOIR-A exoEarth target stars—note that the average is V~6 mag, and all targets are brighter than V~11 mag. Monte-Carlo simulation analysis shows that OBWFS is potentially capable of resolving segment-scale disturbances to the 1 pm level in 96 seconds with a V=5 mag guide star (**Table 11.3**). This assumes LUVOIR's idealized 15-m aperture with no throughput losses.

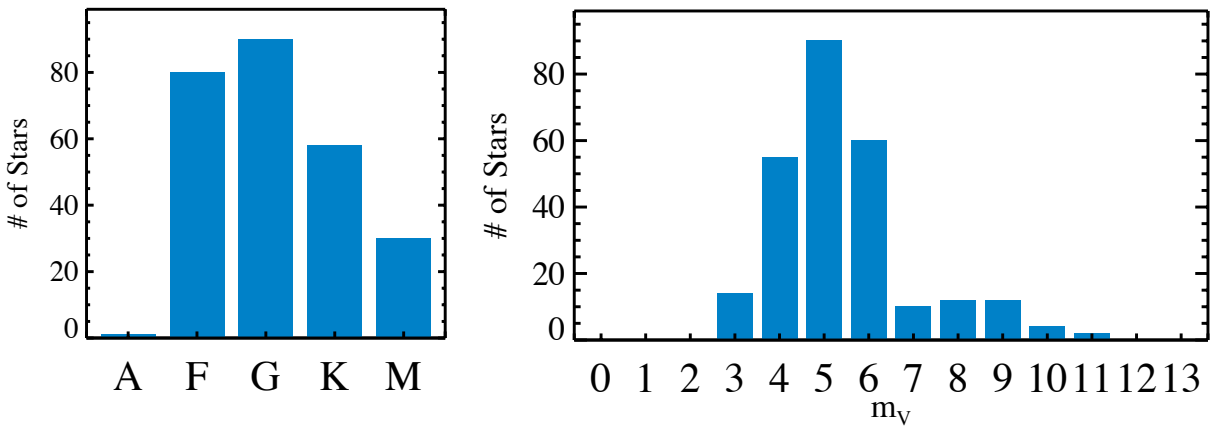

**Figure 11.6.** *Population of target stars for the LUVOIR-A exoEarth campaign. Most of the stars that will be used for low-order and out-of-band wavefront sensing are between V=4 mag and V=6 mag, and all target stars are brighter than V~11 mag.*

The LOWFS approach is attractive because it has a minimum of non-common path optical error since nearly all of the wavefront error (and wavefront error drift) measured in the LOWFS sensor is common to both the coronagraph and the sensor. The target low order wavefront will be near zero, consistent with the use of the classical linearized ZWFS control algorithms (Wallace et al. 2011).

For OBWFS, the target wavefront will be larger, in the tens of nanometers. Some of this is due to the non-common path, as the dichroic and other optics will be unique to the sensor, and as the wavefront sensor and coronagraph are operating in different wavelengths. Plus, the speckle-nulling control will use the DMs to generate grating-like features in the mid and high spatial frequency wavefront, features that help reject the effects of the shadowed aperture, and that compensate residual mid- and high-spatial frequency errors of the telescope—these features must be maintained. OBWFS will need to accurately measure picometer differential wavefront changes in the presence of tens of nanometers of fixed control wavefront error. This will tend to push operation of the ZWFS out of the linear regime, requiring use of higher dynamic-range, higher-performance nonlinear sensing algorithms. Nonlinear wavefront sensing algorithms that appear to meet OBWFS performance requirements are now being investigated.

The great advantage of OBWFS and LOWFS is that they can continuously measure the complete LUVOIR telescope and coronagraph wavefront: they directly observe all the effects that degrade contrast, while the coronagraph is observing. The greatest challenge they face is operating fast enough to keep up with these disturbances. A target star magnitude ≤ 5 allows segment control loop closure times of 96 seconds or less, and DM control loop closure times of 11 minutes or less (**Table 11.3**). For fainter stars, sensing times increase, eventually making it difficult to keep up with expected thermal and material creep effects. Fortunately, as **Figure 11.6** indicates, 90% of the LUVOIR-A stars are brighter than V=10. For dim science stars, a combination of LOWFS to measure line of sight and low order aberrations using rejected science light, and OBWFS tracking high order aberrations at a much longer exposure length may be more photon efficient.

One way to get around the dim guide star challenge is to provide an artificial guide star. A SmallSat or CubeSat equipped with a laser or bright beacon and placed near the line of sight between the telescope and the target star, would be able to provide a bright calibration source. The source could be made bright enough for OBWFS to measure even the > 64 × 64 actuator DM wavefront to picometer sensitivity in under a minute. Studies are now underway exploring the practical implementation issues of this method, including possible source brightness, formation flight requirements between the telescope and source spacecraft, and how best to implement the wavefront sensing on this source[3]. Preliminary indications are that a guide star satellite could be extremely beneficial, but at the moment it is not included in the baseline architecture.

**Table 11.3.** *OBWFS performance summary showing the estimated minimum integration times required to achieve target accuracy for particular spatial resolution levels for typical and near worst-case stellar magnitudes. Per* **Figure 11.6***, most of LUVOIR's target stars will be between V=4 mag and V=6 mag.*

| Mode | Spatial Resolution | Minimum Integration Time for V=5 mag | | Minimum Integration Time for V=10 mag | |
|------|--------------------|-----------------|----------------|-----------------|----------------|
| | | For 10 pm RMS error | For 1 pm RMS error | For 10 pm RMS error | For 1 pm RMS error |
| OBWFS | Segment Piston, Tip, and Tilt | 6 s | 96 s | 96 s | 158 min |
| OBWFS | DM actuator modes | 10 s | 11 min | 48 min | 80 hrs |

**Path to TRL 4:** LOWFS has been demonstrated by the WFIRST CGI project, establishing TRL 4 for LUVOIR. OBWFS should be demonstrated using lab testbeds, with results correlated to models.

**Path to TRL 5:** Testbed demonstration using a LUVOIR-traceable aperture, with realistic dynamic and thermal disturbances.

**Path to TRL 6:** Demonstration at a system level, with a subscale segmented telescope and a coronagraph instrument traceable to the LUVOIR ECLIPS design. The system level demonstration should incorporate other metrology methods, such as edge sensors and laser truss metrology, to show that separate control loops can be incorporated into a single control system.

### 11.3.3 Metrology of the telescope optics: Edge sensors and laser trusses

Another powerful tool for LUVOIR ultra-stability is high bandwidth metrology of the telescope optical alignments, coupled with precision rigid body actuation to stabilize the position and orientation of each optic or optical assembly in 6 degrees-of-freedom. Metrology does not require a guide star to sense and control the largest sources of instability, such as dephasing of the primary mirror segments or despace of the secondary mirror. As such it is complementary to OBWFS and LOWFS, providing much higher bandwidth, maintaining alignments during maneuvers, and ensuring telescope performance during general astrophysical observations, when the coronagraph is not operating.

Two metrology methods are especially well suited for LUVOIR. The first metrology method uses segment edge sensors to make high precision measurements of the position of each pair of adjacent segment edges (see **Figure 11.7**). Closed-loop control feeds the edge sensor measurements back to segment rigid body actuators to keep the segments aligned to each other. Further development of current edge sensor technologies is

3 See: https://www.nasa.gov/directorates/space-tech/esi/esi2016/Laser_Guide_Star

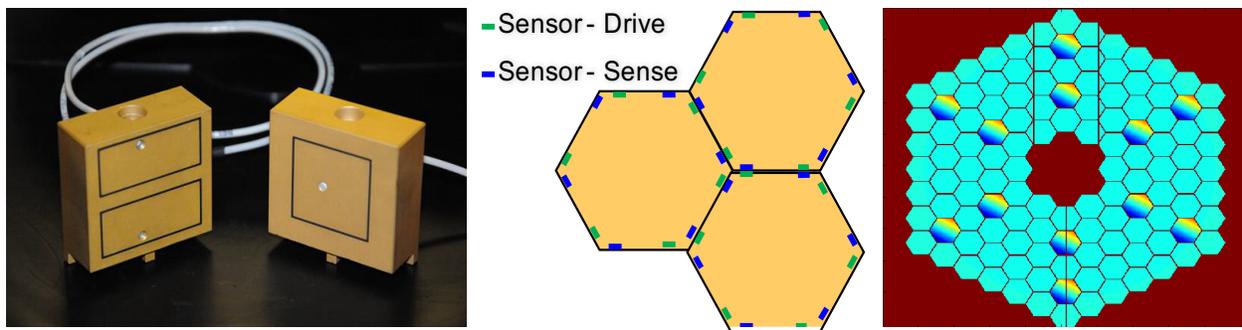

**Figure 11.7.** *Ground-based segmented telescopes commonly use non-contacting capacitive edge sensors, such as the Thirty Meter Telescope devices shown here. Configured with two sensors per adjacent segment edge, they are capable of measuring primary mirror segment-to-segment alignments. The hybrid edge sensor/laser truss metrology architecture sketched on the right uses laser metrology on the 12 segments shown tilted, and segment edge sensors on all segments. It offers better performance than either system alone, provided both have picometer precision in the most sensitive axes.*

needed to improve performance to the picometer-level sensitivity needed for space coronagraph applications.

While segment edge sensors are part of the LUVOIR-A baseline, the specific implementation of the edge sensor is still under study. Capacitive edge sensors have long been used on ground-based segmented telescopes, notably the Keck Telescopes on Mauna Kea, where they provide 1 nm noise and 3.2 nm/week drift in relative segment edge piston measurements. The Keck edge sensors use interleaved sensing elements to ensure high sensitivity and stability (Minor et al. 1990). Edge sensors planned for the upcoming Thirty Meter Telescope use face-on sensing plates, plus gap measurements—a non-interleaved approach better adapted for deployed-aperture space telescopes (Shelton et al. 2008). Capacitive gap measurement devices have demonstrated 15 pm precision for applications on the Laser Interferometer Space Antenna (LISA) mission, although for much smaller gaps than are planned between LUVOIR segments.

Challenges for capacitive edge sensors include the need for close and stable gaps between the segments, including the possible need to interleave sensor structures, and high voltage for high sensitivity. Other edge sensor designs, using inductive (Wasmeier et al. 2014) or optical measurements (Burt et al. 2012), are also reasonable candidates.

The second metrology method is laser truss metrology, using a network of laser distance gauges to measure the rigid body state of the segments relative to the secondary mirror, and measuring the secondary mirror relative to the instrument optics behind the primary mirror (see **Figure 11.8**). Laser metrology can measure alignment of all of the major optical elements relative to a common reference, such as a master optical bench, or reference point on the backplane support frame. Then the rigid body actuators on the segments and secondary mirror can effectively "rigidize" the entire telescope in inertial space.

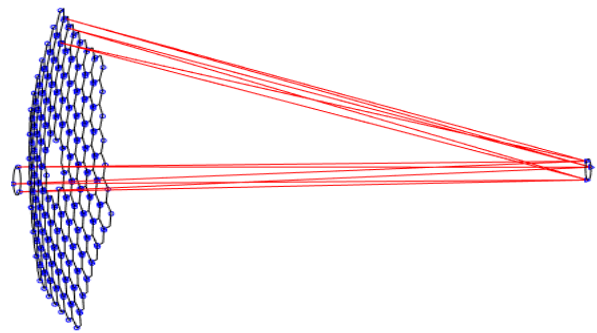

**Figure 11.8.** *Laser metrology truss. With 6 laser distance gauges between each segment and corner cubes attached to the secondary mirror, each segment can be independently aligned to the secondary. Another 6 gauges from the backend instrument bench to the secondary mirror ties the primary and secondary mirror to the instrument optics. These measurements are fed back to the segment and secondary mirror rigid body position actuators to keep the entire observatory in close alignment.*

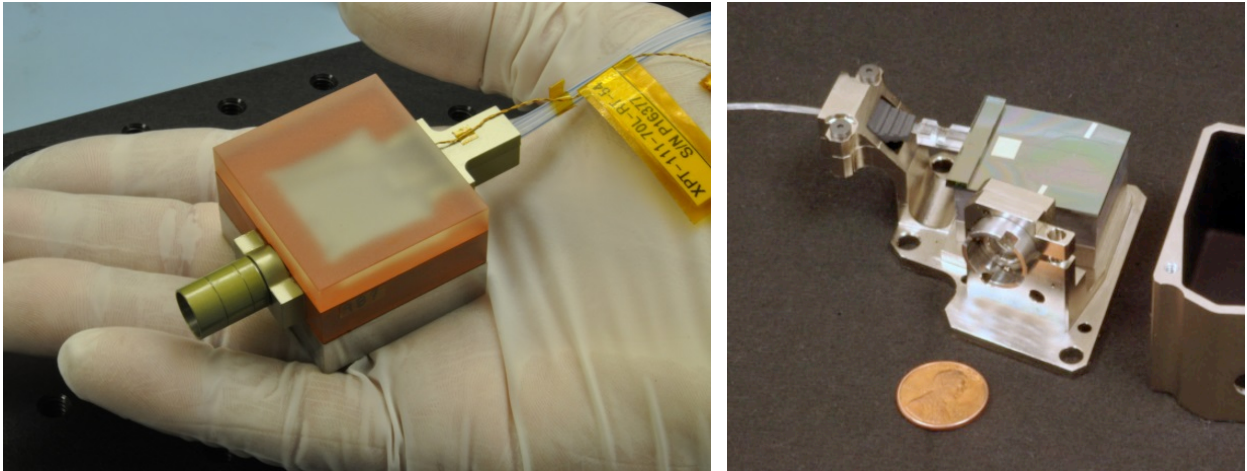

**Figure 11.9.** *Two examples of laser metrology beam launchers. Left: a beam launcher using integrated photonic devices for mixing outgoing and incoming beams. Credit: JPL. Right: a compact version of the SIM beam launcher devices. Credit: Lockheed Martin.*

Laser metrology draws its heritage ultimately from the Space Interferometry Mission (SIM) project, which demonstrated picometer precision on large ground-based testbeds, with large and heavy beam launchers and corner cubes. Subsequent development achieved much smaller and more compact laser metrology devices for attachment to lightweight mirror segments, targeting performance at the sub-nanometer level. The LISA Pathfinder mission has further refined laser metrology technology, flying laser gauges and electronics that achieve picometer accuracy. Combining elements of the SIM and LISA approaches could provide the picometer-precision alignment measurements that LUVOIR will need to preserve coronagraph high contrast performance during extended observations, and do it in a compact, unobtrusive package that will not interfere with image quality. **Figure 11.9** shows two example beam launcher devices from laser metrology systems.

Analysis shows that a hybrid metrology approach, illustrated in **Figure 11.7**, using segment edge sensors on every segment, and using a laser metrology truss on a few selected segments, can outperform either system on its own, provided that both systems have picometer-level sensitivity. This approach provides the end-to-end system measurement capabilities of a laser metrology truss, while the less complex segment edge sensor devices may be better positioned to measure lateral movements of the segments relative to each other. The optimal blend of measurements will depend on the ultimate precision of practical hardware for both systems and is a focus for upcoming technology studies.

**Path to TRL 4:** For edge sensors, develop one or more approaches (capacitive, inductive, optical) capable of picometer sensitivity, and demonstrate sensitivity in the lab. For laser metrology, develop improved beam launcher systems to reduce thermally-driven errors, and improved phasemeter electronics.

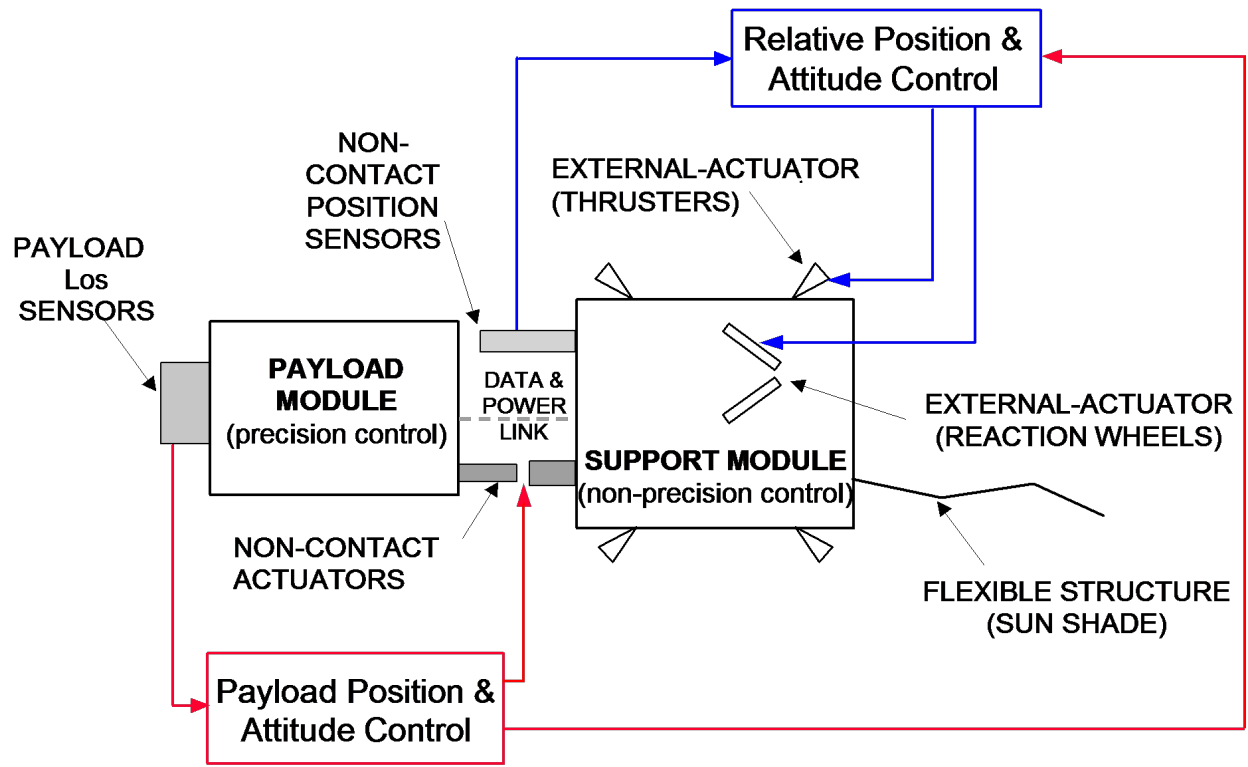

**Figure 11.10.** *Non-contacting vibration isolation "floats" the telescope across a narrow gap from the spacecraft. The payload controls the telescope line-of-sight by pushing against the spacecraft inertia using a set of six non-contact voicecoil actuators, while the spacecraft controls its inertial attitude such that interface stroke and gap are maintained. Requirements for spacecraft attitude control are no more stringent than those for conventional spacecraft, and do not derive from payload pointing requirements. Payload line-of-sight isolation from spacecraft disturbances is broadband, even down to low frequency, and is not affected by interface measurement noise. Power and signal are carried across the gap using very soft connections and may include the possibility of wireless transmission. Credit: L. Dewell (Lockheed Martin).*

**Path to TRL 5:** Demonstration of picometer sensitivity on a lab-scale testbed combining metrology and direct optical measurements. Depending on method, vacuum environment may be relevant.

**Path to TRL 6:** Demonstration at a system level, with a subscale segmented telescope and a coronagraph instrument traceable to the LUVOIR ECLIPS design. The system level demonstration should incorporate other wavefront sensing methods, such as LOWFS and OBWFS, to show that separate control loops can be incorporated into a single control system.

### 11.3.4 Non-contacting disturbance isolation

Even with the very best optics and optical controls, LUVOIR coronagraph performance could be limited by image and wavefront jitter due to spacecraft on-board vibrations. These come primarily from attitude control hardware. Reaction wheel assemblies, control moment gyros, and thrusters create vibrations that jitter the telescope optics and reduce image quality and contrast. The LUVOIR baseline design uses a non-contact isolation technology, illustrated in **Figure 11.10**, to prevent these vibrations from reaching the telescope (Dewell et al. 2017).

**Table 11.4.** *LUVOIR ultra-stable system technology status and path ahead. Green shaded options are in the LUVOIR-A baseline. Yellow options may provide alternative or potentially enhancing solutions given enough development.*

| Technology | Driving Requirements | Technical Challenges | Solution Paths | Current TRL | Path to TRL 4 | Path to TRL 5 | Path to TRL 6 | Note |
|---|---|---|---|---|---|---|---|---|
| Ultra-stable opto-mechanical systems architecture | **Contrast Stability:** < 1e-11 during coronagraph observations **Minimize** wavefront drift during maneuvers **Preserve** optical quality during general astrophysics observations | Changing thermal environment Vibration environment Sensor accuracy Actuator stability Material creep Risk, Cost, Complexity | Stable structures and materials Active thermal control Mix of wavefront sensing, metrology, and control | 2 | System modeling and simulation to TRL 3; Hardware subsystem demos to TRL 4 | Subsystem prototype demos with relevant vibration and thermal drifts. Validation by correlated models. | | Architecture will use wavefront sensing and control, metrology, active thermal control, and stable material technologies to find lowest risk and lowest cost solution that meets performance requirements. |
| Wavefront sensing and control concurrent with coronagraph observations | **Contrast Stability:** < 1e-11 during coronagraph observations **Bandwidth:** > 1 Hz (LOWFS) **Spatial Resolution:** > 12 cyc/ap (OBWFS) | Dim guide stars Non-common path wavefront errors High dynamic range On-board implementation | LOWFS | 4 | ✓ | Demo with corona-graph instrument and relevant vibration and thermal drifts; realistic source brightness. | System-level test with subscale segmented telescope and corona-graph instrument; Include relevant dynamic and thermal inputs with closed-loop control of system. Will require validated modes to fully verify performance on-orbit. | Limited in spatial resolution, but high bandwidth. WFIRST will demonstrate concept at TRL 9, though additional development for LUVOIR likely needed. |
| | | | OBWFS | 3 | | Lab test with corona-graph instrument | | Slow, but high spatial resolution observes control actions through entire system. |
| | | SmallSat/CubeSat flight technologies Precision stationkeeping Precision laser pointing | Laser guide star + OBWFS | 2 | | Smallsat demo | | Requires separate SmallSats or CubeSats to provide bright guide star within OBWFS field of view |
| Metrology systems | **Wavefront error stability:** < 10 pm RMS (within control spatial and temporal bands) | Picometer resolution Picometer stability Thermal sensitivity Size/Weight/Power of sensor elements High voltage (edge sensors) Cyclic error (laser truss) | Segment edge sensors | 3 | Demo measure-ment of relevant degrees of freedom | Demo metrology on subscale segments in presence of relevant vibration and thermal drifts. | | Need to develop sensor-head geometry capable of measuring desired degrees of freedom to picometer levels. |
| | | | Laser truss metrology | 3 | Improve beam-launcher size / stability, phase-meter electronics | | | Combine proven SIM and LISA-derived tech for compact, picometer-precision gauges |
| Picometer resolution rigid body actuators | **Stroke:** > 10 mm **Resolution:** < 10 pm **Creep:** < 1 pm / 10 min | Long performance lifetime Low size/weight/power Launch loads | Mechanical actuator for large stroke + piezo tip actuator for precision | 3 | Prototype design demo of stroke and resolution | Environ-mental testing; Lifetime and load tests | | Mech actuator heritage from JWST combined with commercial piezo actuators (e.g. Mad City Labs, Physik Instrumente) |
| Dynamic isolation systems | **Isolation:** 40 dB/decade **Slew Rate:** 3 ° / min | Power and signal cables bridging interface Large moments of inertia | Actively controlled non-contact interfaces | 4 | ✓ | Demo isolation with realistic disturbance inputs and with power and cable transmitted across the interface | | Build on demonstrated Disturbance Free Payload technology from Lockheed Martin, for example. |
| Microthruster attitude control to eliminate reaction wheel disturbances | **Thrust:** 1 N (coarse mode) 100 uN (fine mode) **Resolution:** 0.01 uN | Lifetime Redundancy Contamination | Electrospray thrusters for fine mode, electric propulsion for coarse mode | 5 | ✓ | ✓ | Demo thrusters with sufficient force and resolution for LUVOIR | Heritage from LISA Pathfinder System studies needed to design for LUVOIR |

**Path to TRL 4:** The non-contact isolation technology is at TRL 4, based on Lockheed Martin's Disturbance Free Payload concept and lab demonstrations, but system studies are required to optimize a LUVOIR implementation. Other approaches could be considered, and the system studies should include assessment of alternate methods.

**Path to TRL 5:** Demonstration of LUVOIR-traceable performance in a representative environment, including gravity offload and perhaps vacuum. Especially important will be demonstrating that cables providing power and signal can cross the gap without degrading the isolator performance.

**Path to TRL 6:** Demonstration at a system level, with a subscale segmented telescope. The system level demonstration should incorporate line-of-sight pointing measurements generated by a prototype fine guidance sensor traceable to HDI that are used to command a fast-steering mirror in addition to the VIPPS to show an integrated pointing control system.

### 11.3.5 Disturbance avoidance: Micro-thrusters for attitude control

Another method for eliminating vibrations from spinning-mass attitude control hardware is to eliminate such devices entirely, replacing them with extremely precise, micro-Newton to Newton level reaction control thrusters. This method has been flight proven by LISA Pathfinder, which used colloidal thrusters to achieve thrust of 30 micro-Newtons or less, with resolution down to 0.1 micro-Newtons. Other mature thruster technologies are available for pointing control when higher thrust levels are required, so that a LUVOIR sized telescope could be both rapidly repointed and precisely controlled without the vibrations inherent to reaction wheels or control moment gyros.

Micro-thrusters are a potential break-through alternative for attitude control, one that could reduce or eliminate the largest sources of dynamic disturbances for LUVOIR. Adoption of this approach needs to be considered at the system level to fully assess its advantages. These could include: use of lower stiffness structures and mirrors, for mass savings; precision pointing without the need for a fine steering mirror; and possibly improved image quality in all observing modes. At the same time, new contamination issues may arise; and the basic LUVOIR structural layout will need to be reconsidered.

### 11.3.6 Active control of the telescope wavefront

The telescope wavefront control actuation has a large trade space. Primary mirror segment and secondary mirror rigid body control is required to phase the segments and align the telescope following launch and maintain alignment during the life of the mission. Low-order segment figure actuation may be used to compensate for mirror fabrication errors, such as radius-of-curvature mismatch between segments, or residual gravity release effects. Higher-order segment figure actuation offers resilience to other possible wavefront errors, such as segment edge defects, and could even be used as part of the coronagraph speckle-nulling control. Wavefront control may also be applied at the telescope exit pupil using a deformable mirror external to the coronagraph. An active secondary mirror with high-order figure actuation, could also be used, though such corrections may not be effective over the entire telescope field of view due to pupil registration, distortion, and magnification limitations (McComas & Friedman 2002). Local segment control (i.e., sensing of the primary mirror segment rigid body motion and control of the segment motion) requires precision rigid body actuators, perhaps like those used on JWST, but with an additional ultra-fine actuation stage.

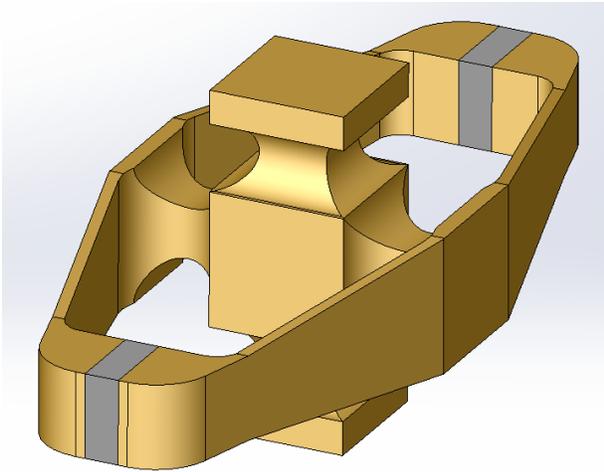

**Figure 11.11.** *Conceptual picometer actuator using PZT devices. The gray inserts on either end are the PZT devices. The gold structure is a flexure design that provides a mechanical reduction in step size between PZT actuation and physical motion at the actuator interfaces. The flexure also removes the PZT device from the load path, reducing forces on the device during launch. Credit: Ball Aerospace Technology Center.*

Control at the telescope is expected to be bandlimited in the spatial and temporal domains. In the spatial domain, the band is limited by the actuation spatial resolution, set by the segment geometry for rigid body control, and by actuator density if segment figure actuators or a deformable mirror is used. In the temporal domain the bands are constrained by the integrating action of observations and the chopping action of angular- or field-difference observing strategies. The exact temporal bandwidth is dependent on the science observation and architecture, but the control bandwidth is likely below 1 Hz, with passive stability needed at frequencies above the control bandwidth. The most stringent control capability within the 1 Hz bandwidth is expected to be segment piston at ~10 pm RMS of wavefront, followed by segment tilt at 25 pm RMS wavefront (Nemati et al. 2017).

This level of actuation for LUVOIR is TRL 3. An example is Mad City Labs, Inc.[4] closed-loop piezo actuators, which show 5 picometer step resolution and have been independently demonstrated to achieve <10 pm RMS on a 0.5 Hz sinusoid—good enough for LUVOIR. This actuator design was developed for a past flight application, but more development is needed for LUVOIR, including lightweighting and integration with a large-stroke actuator (such as the ones flown on JWST). The most important aspect of the Mad City Labs actuator is the demonstration of the small step resolution which provides confidence the LUVOIR requirements can be met. **Figure 11.11** shows an actuator concept that uses PZT devices in concert with a flexure design to achieve picometer-level step sizes while also isolating the PZT devices from launch loads.

**Path to TRL 4:** Develop flight prototype rigid body actuators, using one or more methods (PZT, mechanical, or a hybrid), and test in the lab.

**Path to TRL 5:** Demonstration of picometer-level actuation of a subscale segment with a prototype rigid body actuator in six-degrees of freedom. Metrology and direct optical measurements of segment displacements will verify performance. Vacuum testing of the actuator may be necessary.

**Path to TRL 6:** Demonstration at a system level, with a subscale segmented telescope. The system level demonstration should incorporate metrology methods, such as edge sensors and laser truss metrology, to show closed-loop control of the segment position(s). Vacuum testing may be necessary.

---

4 See http://www.madcitylabs.com/nanometz.html

### 11.3.7 Primary mirror segments

LUVOIR primary mirror segments and their supporting subsystems can draw on heritage from multiple sources, with some key options that will need to be traded to assure best performance and affordable cost, complexity, and risk. Here we note that the JWST mirror segments are made from beryllium, a material chosen for its high stability at cryogenic temperatures. Beryllium is not the best candidate for LUVOIR because at the operational temperature of 270 K, beryllium has a very high CTE. Beryllium mirrors are also time consuming and expensive to fabricate; it took about 8 years to complete a fully integrated JWST segment assembly.

Many space telescopes have used mirrors made of Ultra Low Expansion (ULE) glass, a material that is thermally very stable, and that can be figured and polished to UV requirements. ULE is commonly "lightweighted," fabricated with a honeycomb core between two thin facesheets, to provide needed high stiffness at a low mass. One facesheet is polished and coated, to provide the optically reflective surface (see **Figure 11.12**).

The LUVOIR-A design assumes the use of lightweighted ULE for each of its 120 primary mirror segments, building on technology first demonstrated by the Advanced Mirror Segment Demonstrator project, which built and tested prototype JWST mirror segments (Matthews et al. 2003). Each segment will be mounted via bonded invar pads that are then mounted to support flexures. The struts connect the mirror to six rigid body actuators, arranged as three separate bipods. All segments are then connected to a common backplane structure, a highly stable, stiff, composite structure that forms the backbone of the primary mirror assembly. This approach allows each segment to be independently positioned in six degrees-of-freedom, so they can be aligned and phased even if the backplane structure deforms. Another important part of each mirror segment will be an active thermal control system, with one or more heaters and temperature sensors acting to maintain the temperature of each segment at 270 K. Analysis indicates that thermal stability of 1 mK will provide ~1 pm mirror figure stability during operations (Eisenhower et al. 2015).

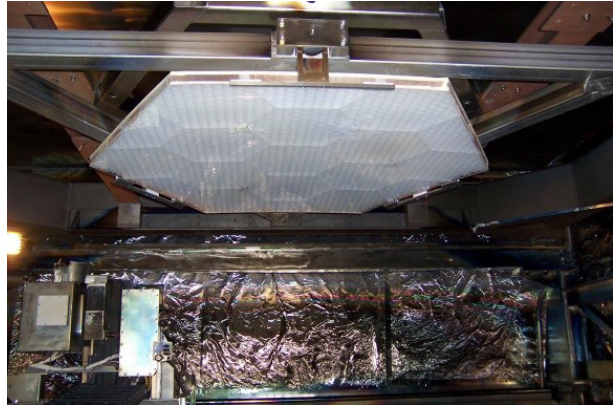

**Figure 11.12.** *A ULE mirror segment undergoing ion figuring for surface figure and radius of curvature correction. Capture Range Replication technology enables replication of mirrors to within capture range of final finishing processes for figure and surface finish, eliminating most of the grinding and polishing. This technology is especially useful when multiple mirrors with the same prescription are needed, as is the case with segmented optical systems. Credit: Harris Corporation.*

One consideration for LUVOIR is whether or not its segments should be equipped with surface figure actuators. To achieve diffraction-limited performance at 500 nm requires a surface figure error of ~10 nm RMS over the fully-phased 120-segment primary mirror surface. Mirrors are limited in

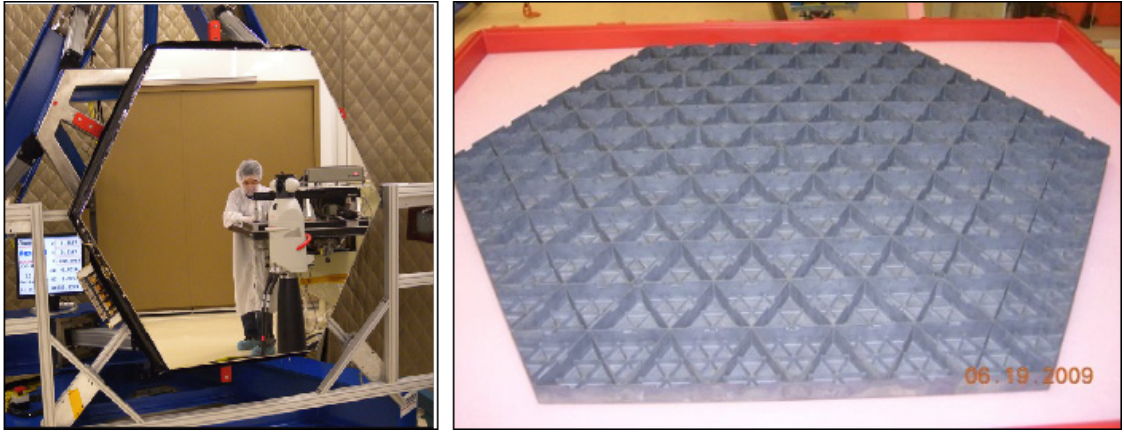

**Figure 11.13.** *SiC mirror segment technology. (Left) An AHM mirror being set up for test. (Right) A SiC mirror substrate showing its open back rib structure. Credit: JPL.*

their figure quality by a number of effects, including:

- Unpredicted/uncorrected gravity sag
- Mismatch of the radius of curvature between segments
- Print-through of mount points caused by thermal effects or adhesive and Invar creep.

Surface figure actuators enable on-orbit correction of these errors, insuring against contingencies, and relaxing difficult fabrication requirements. Surface figure actuators introduce weight and complexity, however. Lightweighted, surface-figure actuated ULE mirror segment systems have already been built and tested.

Another, highly actuated approach to primary mirror segments has been developed, using a Silicon Carbide (SiC) substrate technology equipped with many surface figure actuators. These "Actuated Hybrid Mirrors," shown in **Figure 11.13** repeatedly demonstrated 15 nm surface figure, after correction of fabrication and gravity sag errors of 1800 nm—a correction factor of >100×. Actuated Hybrid Mirrors use a replication-based fabrication process, with a nanolaminate foil face sheet bonded to the SiC substrate (Hickey et al. 2010).

Better performance is expected if the SiC substrate is clad with pure silicon instead of the nanolaminate, and then superpolished to meet as-built figure of <100 nm RMS (Wellman et al. 2012). Assuming the same correction factor as the nanolaminate version, operational surface figures of <10 nm would be achievable.

The advantage of highly actuated mirrors is the high level of figure control, enough to meet the required surface figure error performance at a wide range of temperatures, in 1-g as well as 0-g, even with relaxed polishing requirements. The high actuator density would even allow the use of the primary mirror as a third DM for shaping coronagraph wavefronts. A disadvantage is the relatively high coefficient of thermal expansion of SiC at LUVOIR's 270 K operating temperature. Analysis suggests that the high thermal controllability of SiC can be thermally stable at the needed level, though with 10× tighter temperature controls (~100 μK) than for the ULE. The passive stability of SiC improves as the operating temperature drops, approaching a coefficient of thermal expansion of 0 at around 100 K. Thus, SiC mirrors are well suited for cold or cryogenic applications. A final consideration is the simple fact that active SiC mirrors do not have the long heri-tage of ULE. Nonetheless, they offer important capabilities and provide an alternative for LUVOIR.

**Table 11.5.** *LUVOIR mirror technology status and path ahead. Green shaded options are in the LUVOIR-A baseline. Yellow options may provide alternative or potentially enhancing solutions given enough development. Red options would only be considered with substantial development and gains in system performance.*

| Technology | Driving Requirements | Technical Challenges | Solution Paths | Current TRL | Path to TRL 4 | Path to TRL 5 | Path to TRL6 | Note |
|---|---|---|---|---|---|---|---|---|
| Ultra-Low Expansion (ULE) glass mirror segments | **Figure Error:** < 10 nm RMS <br><br> **Figure Drift:** < 40 pm / 10 min <br><br> **Stiffness:** > 200 Hz first free mode <br><br> **Substrate Areal Density:** < 12 kg/m² | Radius of curvature matching Gravity sag prediction Print-through figure errors Epoxy and invar creep | No surface figure actuation | 5 | ✓ | ✓ | Segmented mirror system test with 2 or more segments and all baselined hardware (actuators, thermal control, edge sensors, laser metrology, etc.) | Heritage from Advanced Mirror System Demonstrator and Multiple Mirror System Demonstrator programs. |
| | | | Radius-of-curvature only actuation | 4 | ✓ | Environmental and optical performance testing of fully-integrated mirror segment | | |
| | | | Low-order surface figure actuators | 4 | ✓ | | | |
| Actuated Silicon Carbide mirror segments | | Silicon cladding Epoxy creep Actuator stability Electronic harnessing Mass | Polished, Si-clad SiC mirrors | 3 | Si cladding and polish | | | Heritage from Actuated Hybrid Mirrors with nanolaminate facesheets 3.5-m Herschel SiC mirror demonstrates flight heritage of base material |
| Zerodur glass-ceramic mirror segments | | Radius of curvature matching Gravity sag prediction Print-through figure errors Epoxy and invar creep | No surface figure actuation | 4 | ✓ | | | Open-back construction requires heavier, thicker mirrors for equal stiffness. Good thermal stability. |
| UV reflective coatings | **Average Reflectivity:** > 50% (103-115 nm) > 80% (115-200 nm) > 88% (200-850 nm) > 96% (850-2500 nm) | Protective coating Cleaning | Aluminum reflective coating with protective layers such as LiF, MgF, AlF3 | 3 | Demo coating on multiple runs to show repeatability | Life test in relevant radiation and I&T environments | Coating on multiple full-scale segments | Enhanced Al+LiF coatings demonstrate reflectivity performance Need to demonstrate thin capping layers of MgF₂ or AlF₃. |

**Path to TRL 6:** The path ahead for mirror segments will ultimately require building and testing full-scale segments at the subsystem level, including integrated thermal controls, rigid body actuators, and figure actuators if needed. Testing multiple segments together will be necessary to establish TRL 6, to verify radius of curvature matching and manufacturing repeatability, especially if figure actuation is not provided.

## 11.3.8  UV reflective coatings

General astrophysics and exoplanet science require high-throughput observations between 100 nm and 2.5 μm. The coating should achieve >50% reflectivity at 105 nm while not compromising performance at wavelengths > 200 nm compared to existing state-of-the-art. The coating process must be scalable to meter-class segments and repeatable to ensure uniform performance across an aperture comprised of >100 segments.

State-of-the-art enhanced Al+LiF coatings exhibit the desired performance with >85% reflectivity between 103 and 130 nm (Quijada et al. 2014). However, LiF is hygroscopic and will suffer from environmental degradation, impacting reflectivity. Atomic layer deposition of thin $MgF_2$ or $AlF_3$ overcoats will provide environmental stability while not affecting performance (Moore et al. 2014). Coating characterization is needed to perform polarization performance modeling and its impact on high-contrast imaging.

As to mirror cleaning, existing approaches (e.g., $CO_2$ snow, or electrostatic wands with AC excitation), work well for removing dust or water, but do not remove molecular contamination. Promising methods (e.g., electron or ion beams) could clean molecular layers as well as dust on substrates.

**Path to TRL 4:** Demonstrate the deposition of a protected Al+LiF+MgF$_2$ or Al+LiF+AlF$_3$ far-UV-optimized coating in separate coating runs to verify repeatability of the process. Characterize coating reflectivity, uniformity, and polarization performance to verify the coatings achieve LUVOIR requirements.

**Path to TRL 5:** Verify coating stability over time in space-simulated vacuum and ambient integration and test environments, as well as in a relevant radiation environment.

**Path to TRL 6:** Demonstrate coating process on multiple meter-class prototype segment assemblies.

## 11.4 Instrument technologies

The LUVOIR instrumentation has broad general astronomy and astrophysics capabilities as well as specialized coronagraph modes, as outlined in **Chapter 9**. These capabilities can be provided within current technologies for instrument elements such as mirrors, lenses, diffraction gratings, grisms, mechanisms, and structures. Except for the coronagraph elements described earlier, new instrument technology development is required only for detectors and for the LUMOS microshutter array, which is used for multi-object spectroscopy.

### 11.4.1 Detectors for LUVOIR instruments

The LUVOIR Study Team has baselined detector technologies with established heritage: CCDs, CMOS arrays, HgCdTe photodiode arrays, and microchannel plates. But as with any generational advance in instrumentation, the goal is to improve these technologies in areas of array size, sensitivity, noise performance, and radiation tolerance. These are the main areas of focus for LUVOIR detector technology development. At the same time, we note certain exciting, lower-TRL technologies that may provide high performance payoff in the future with adequate development; these will be monitored going forward.

### 11.4.1.1 Microchannel plates

In the far-UV, large-format, high dynamic range detectors are needed for multi-object spectroscopic observations at wavelengths as short as 100 nm. Microchannel plate technologies have been demonstrated in sub-orbital sounding rocket missions SISTINE, FORTIS and CHESS (Fleming et al. 2011; Hoadley et al. 2016). These suborbital efforts have matured independent aspects of microchannel plate detector technologies, including array size, photocathode sensitivity, and read-out electronics count rates, and establish TRL 4 for far-UV LUVOIR applications. Incorporating all aspects of recent demonstrations into a single, large format, high-count rate, high quantum efficiency detector array for a sounding rocket demonstration could raise the TRL to 6 for LUVOIR's applications. Micro-channel plates are baselined for the LUVOIR LUMOS instrument far-UV multi-object spectroscopy and imaging channels.

### 11.4.1.2 CCDs, electron multiplying CCDs, and hole multiplying CCDs

While CCD devices have a rich spaceflight heritage, many of LUVOIR's applications require key features not available in traditional devices. Single photon counting, large format detectors will enable key exoplanet observations, including multiplexed spectral observations of exoplanet systems. The electron-multiplying CCD (EMCCD) detectors baselined for the WFIRST CGI (Nemati 2014) provide the baseline solution for the LUVOIR ECLIPS visible channel, but their susceptibility to radiation damage might limit their useful life on orbit, at least for the most sensitive exoplanet observations (Nemati et al. 2016). Additional radiation testing and design work to further improve radiation tolerance is necessary to achieve long-life performance and will significantly enhance the exoplanet science yield potential for LUVOIR. Larger format arrays would also enhance the efficiency of high-resolution integral field spectrograph instrument designs. Hole-multiplying CCDs are promising, in that they have inherently higher radiation tolerance, but they will require further development to demonstrate similar photon counting capabilities and array sizes as EMCCDs.

EMCCDs and HMCCDs can also be δ-doped for enhanced sensitivity in the near-UV (200–400 nm). The δ-doping process is well demonstrated on CCD devices and has flight heritage. A δ-doped EMCCD device, shown in **Figure 11.14**, will fly on the next FIREBALL sounding rocket experiment (Nikzad et al. 2017).

### 11.4.1.3 CMOS and sCMOS detectors

CMOS devices are experiencing rapid development, driven by commercial applications. They generally offer larger formats and smaller pixels than CCD devices, but at the penalty of higher noise. Some CMOS devices, however, have demonstrated sub-electron read noise in some fraction of their pixels—so-called scientific or sCMOS devices. Improvement in this area could make them a possibly more robust alternative for LUVOIR exoplanet science. CMOS and sCMOS devices can also be δ-doped for enhanced UV performance. **Figure 11.15** shows a CMOS device prior to being δ-doped.

### 11.4.1.4 HgCdTe photodiode and avalanche photodiode arrays

In the NIR, Teledyne HgCdTe HAWAII 4RG ("H4RG") detectors developed for WFIRST already exhibit exceptionally good noise performance (single-digit read noise, $10^{-3}$ dark current). Further characterization of these large-format tile-able arrays should be pursued to understand their viability for NIR exoplanet science. Efforts to minimize read noise in the readout electronics would further improve science yield.

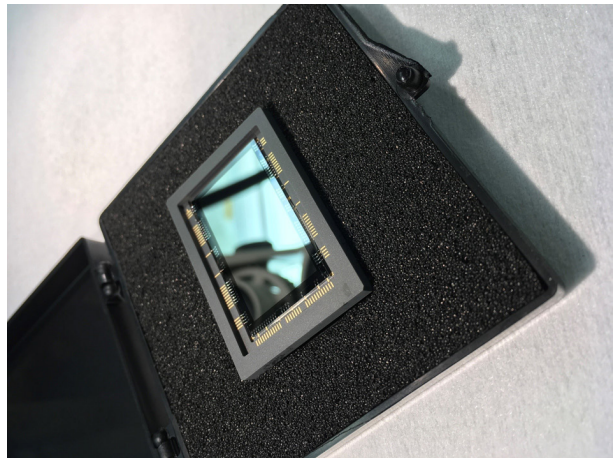

**Figure 11.14.** *A δ-doped, 2 Megapixel EMCCD device with visible-rejection filters to optimize for high in-band (120–150 nm) quantum efficiency, and high out-of-band rejection. Credit: S. Nikzad (JPL).*

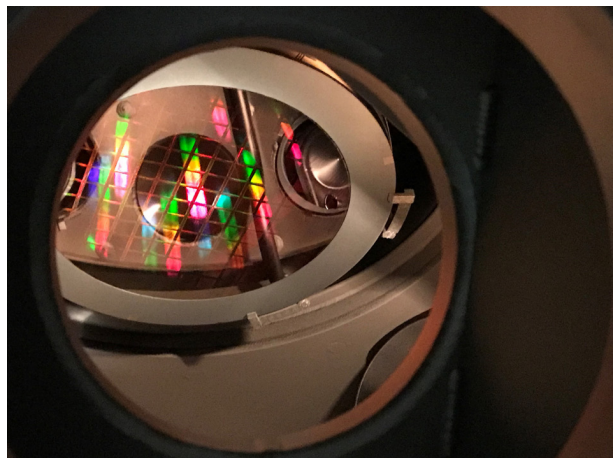

**Figure 11.15.** *An 8-inch wafer containing a monolithic CMOS sensor, prior to being δ-doped using molecular beam epitaxy. Credit: S. Nikzad (JPL).*

HgCdTe Avalanche Photodiode arrays offer a photon counting option in the NIR, similar to EMCCDs in the visible. Current devices, however, exhibit dark currents that

**Table 11.6.** *LUVOIR detector and microshutter technology status and path ahead. Green shaded options are in the LUVOIR-A baseline. Yellow options may provide alternative or potentially enhancing solutions given enough development. Red options would only be considered with substantial development and gains in system performance.*

| Instrument Channel | Driving Requirements | Technical Challenges | Solution Paths | Development status and path ahead | | | | Note |
|---|---|---|---|---|---|---|---|---|
| | | | | Current TRL | Path to TRL 4 | Path to TRL 5 | Path to TRL 6 | |
| LUMOS Far-UV Multi-object Spectrograph and LUMOS Far-UV Imager | Wavelength Range: 100 - 200 nm Array Size: 200 mm × 200 mm, Tileable to larger arrays Resol Size: ≤ 20 μm Read Noise: 0 e- Dark Current: ≤ 0.1 counts/cm²/s | Sensitivity Array sizes Dynamic range / bright object sensitivity | Micro-channel plates | 4 | ✓ | Sounding rocket test of 200 mm sized arrays; Demo of tiling capability | Full qualification of flight-packaged devices | Long development heritage through sounding rocket program, HST, etc. |
| LUMOS Near-UV Multi-object Spectrograph | Wavelength Range: 200 - 400 nm Array Size: ≥ 8k × 8k, Tileable to larger arrays Quantum Efficiency: ≥ 50% Pixel Size: ≤ 7 μm Read Noise: ≤ 5 e- Dark Current: ≤ 1e-3 e-/pix/s | Sensitivity Array sizes Radiation hardness | delta-doped CCD | 5 | ✓ | ✓ | Full qualification of flight prototype devices; Demo tiling of arrays | Sounding rocket heritage devices. Would need to show ability to tile multiple devices. Smaller pixels and better radiation tolerance would be desireable. |
| | | | delta-doped CMOS | 4 | ✓ | Sounding rocket demo; radiation testing | Full qualification of flight prototype devices; Demo tiling of arrays | |
| ECLIPS UV Channel | Wavelength Range: 200 - 525 nm Array Size: ≥ 1k × 1k Quantum Efficiency: ≥ 50% Read Noise: < 1 e- Dark Current: ≤ 1e-4 e-/pix/s | Radiation hardness High sensitivity in near-UV while maintaining performance through 500 nm | delta-doped EMCCD | 5 | ✓ | ✓ | Full qualification of flight prototype devices; Demo tiling of arrays | Sounding rocket heritage devices. Better radiation tolerance is desireable. |
| ECLIPS Visible Channel | Wavelength Range: 500 - 1000 nm Array Size: ≥ 4k × 4k Read Noise: < 1 e- Dark Current: ≤ 1e-4 e-/pix/s | Radiation hardness Sensitivity at 1000 nm for exoplanet characterization Large formats for use in integral field spectrographs | EMCCD | 5 | ✓ | ✓ | Demo improved radiation hardness and larger array size | Under development by WFIRST in 1k x 1k formats Radiation tolerance is low; lifetime of current devices is limited < 5 years Path to 4k x 4k devices is established |
| | | | HMCCD | 3 | Lab demo of device with similar capabilities to existing EMCCD devices | Flight package device with; Radiation testing | Full qualification of flight prototype device | Potential to deliver same performance as EMCCD, but with inherent radiation hardness. |
| ECLIPS NIR Channel | Wavelength Range: 1000 - 2000 nm Array Size: ≥ 4k × 4k Read Noise: < 3 e- Dark Current: ≤ 1e-3 e-/pix/s | Read noise Dark current | HgCdTe Photodiode Array | 6 | ✓ | ✓ | ✓ | Devices developed for WFIRST wide-field instrument Lower read noise and dark current desireable; potentially achievable through readout electronics optimization |
| | | | HgCdTe Avalanche Photodiode Array | 4 | ✓ | Reduce dark current to acceptable levels; Radiation testing | Full qualification of flight prototype device | Currently exhibit desired read noise, but need significant improvements in dark current Larger array sizes also needed |

| Instrument Channel | Driving Requirements | Technical Challenges | Solution Paths | Development status and path ahead | | | | Note |
|---|---|---|---|---|---|---|---|---|
| | | | | Current TRL | Path to TRL 4 | Path to TRL 5 | Path to TRL 6 | |
| HDI UVIS Channel | **Wavelength Range:** 200 - 1000 nm **Array Size:** ≥ 8k × 8k Tileable to larger arrays **Pixel Size:** < 7 μm **Read Noise:** < 5 e− **Dark Current:** ≤ 1e-3 e−/pix/s | Radiation hardness Array sizes Read noise High-speed region-of-interest readout for fine guiding | CCD/EMCCD | 5 | ✓ | ✓ | Full qualification of flight prototype device in large format, tiled array | CCD devices have high heritage and very good noise performance. Lack ability to define arbitrary region-of-interest for high-speed readout. Better radiation hardness is desireable. |
| | | | CMOS | 4 | ✓ | Demo large format devices with sufficiently low dark-current at operating temps.; Radiation testing | Full qualification of flight prototype device in large format, tiled array | Existing commercial devices have good performance, but need to be radiation tested and/or hardened |
| HDI NIR Channel | **Wavelength Range:** 1000 nm - 2500 nm **Array Size:** ≥ 4k × 4k Tileable to larger arrays **Pixel Size:** ≤ 10 μm **Read Noise:** ≤ 5 e- **Dark Current:** ≤ 1e-3 e-/pix/s | Read noise and dark current High-speed region-of-interest readout for fine guiding | HgCdTe Photodiode Array | 6 | ✓ | ✓ | ✓ | Devices developed for WFIRST wide-field instrument meet all requirements for LUVOIR HDI NIR Channel |
| Other Detector Devices | Any set of the above requirements | Low-TRL Cryocooling vibration compatibility with picometer stability requirements | Microwave Kinetic Inductance Detectors (MKID) and Transition Edge Sensors (TES) | 4 | ✓ | Demo large format devices; Radiation testing; Demo cryo-cooling operations do not interfere with picometer stability | Full qualification of flight prototype device in appropriate format for LUVOIR application | Broadband, zero read noise, zero dark current devices Built-in energy resolution provides on-chip low-resolution spectroscopy Operate at sub-Kelvin temperatures, requiring a cryocooler which may be incompatible with picometer stability requirements |
| | | | Quanta Image Sensor | 3 | Demo required quantum efficiency and basic performance specs | Demo flight-packaged device; radiation testing | Full qualification of flight prototype device in appropriate format for LUVOIR application | Room temperature, high-speed photon counting arrays Could replace CMOS or EMCCD devices in near-UV (with delta-doping) or visible |
| Microshutter Arrays | **Shutter Format:** 420 × 840 Tileable to larger arrays | Electrostrictive actuation Shutter yield and lifetime | Next-Gen Microshutter Arrays | 4 | ✓ | Environmental testing; Demo of tiling | Flight prototype qualification | |

are ~10,000× higher than is required for exoplanet imaging. It is believed this is largely due to the readout electronics and should be mitigated with better optimization (Rauscher et al. 2016). Larger array sizes would also need to be developed.

### 11.4.1.5 Other detector technologies

Newer, less mature technologies that could offer significant performance advantages in the long term include visible-band high temperature photon counting arrays, such as the Quanta Image Sensor (QIS) (Masoodian et al. 2017). With $\delta$-doping (applicable to virtually all silicon detectors), these could potentially provide photon counting detectors with high QE in the UV as well. Another potentially breakthrough technology that shall be monitored is the GaN material family-based photon counting UV avalanche diode arrays. They are intrinsically solar blind, with potential higher operating temperature and radiation tolerance. Cryogenic superconducting detectors such as microwave kinetic inductance detectors (MKIDs) and transition edge sensors (TES) are inherently noise free and provide native spectral resolution in the detector substrate, potentially enabling zero read noise exoplanet observations with built-in spectral resolution. However, the need for a cryocooler to achieve operating temperature may introduce unacceptable vibration into the system.

**Path to TRL 4, 5, and 6:** Specific recommendations for each detector technology are summarized in **Table 11.6**. We note, however, that recent detector development has been driven in large part by academic institutions. It is recommended that mission developers work closely with academic partners to achieve the biggest gains in advancement.

### 11.4.2 Microshutter arrays

The LUVOIR UV Multi-Object Spectrograph (LUMOS) instrument requires a 2 × 3 tiled-array of next-generation microshutters (Li et al. 2014) to enable the multi-object mode of observation. These microshutters leverage the heritage from JWST's Near Infrared Spectrograph (NIRSpec) instrument (Kutyrev et al. 2008) but add improvements to array size and shutter reliability.

The JWST NIRSpec uses a 2 × 2 tile of microshutter arrays (MSAs), with each tile having a format of 171 × 365 shutters. A permanent magnet is swept across the array to actuate the shutters. Next-generation microshutter arrays that eliminate the moving magnet and instead rely on low-voltage electrostrictive actuation have been demonstrated in the lab, with plans to fly on a sounding rocket in the near future. The demonstrated next-gen MSAs are directly applicable to LUVOIR, although they require engineering development to improve the format to 420 × 840 shutters and allow for array tiling.

**Path to TRL 5:** Environmental testing of the next-gen MSAs, including vibe, shock, acoustic, and thermal-vacuum testing, would constitute a component validation in a relevant environment.

**Path to TRL 6:** A tiled array of MSAs, each consisting of 420 x 840 shutters, needs to be demonstrated and subjected to the full range of flight qualifying environmental tests.

## 11.5 The technology path ahead

So, what is the right mix of LUVOIR technologies? Will an architecture with LOWFS and edge sensors alone be adequate? Or will OBWFS and laser truss metrology also be required? Will segment figure actuators be needed? Questions like these, when answered, will help to define the final performance, cost, and complexity of LUVOIR. They are being studied by the LUVOIR Study Team, following the methods in **Section 11.3.1** above.

The good news for now is that LUVOIR has many tools to work with, putting its scientific objectives within reach. Final determination of the best LUVOIR architecture will occur as the component technologies are demonstrated to meet the needed performance. It will happen through design trades, flowing out performance from components to exoEarth yield and other scientific metrics, following NASA system engineering practice, quantified by accurate, validated system-lev-el modeling and analysis, and grounded in rigorous testing.

As the LUVOIR Study Team continues to develop both architecture designs, it may be that additional technology needs will be identified, or that some of the technology gaps assessed here may be closed. The LUVOIR Study Team will continue to assess the impact and readiness of these technologies as the LUVOIR architectures evolve. It will conduct an inventory of technology development investments in each area, defining specific, actionable tasks that can be completed to achieve the objectives presented here, and estimating the cost and schedule required to execute those tasks. A complete technology development roadmap, with schedule and cost, will be included in the final report.

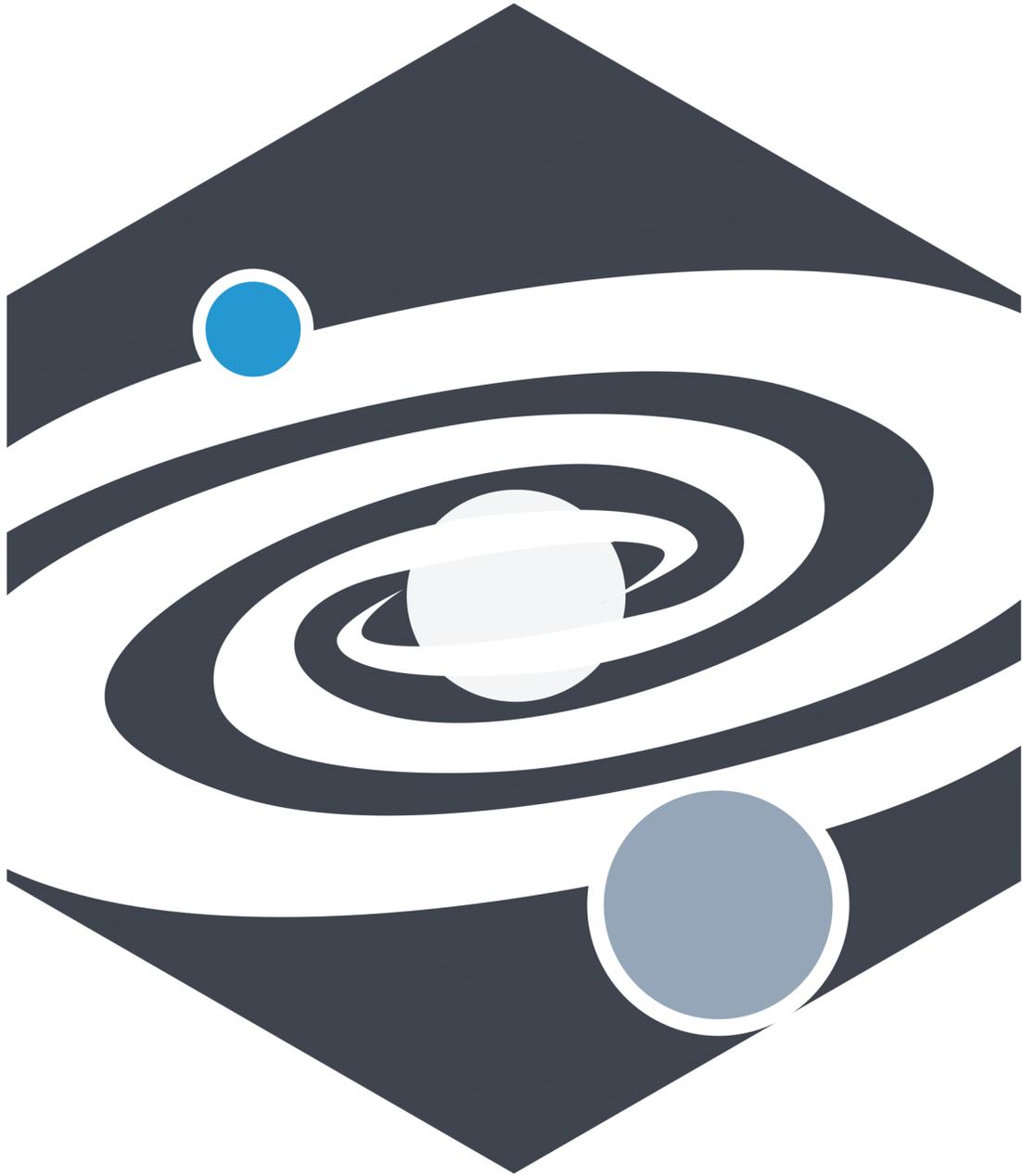

LUVOIR Cycle 1

# 12 LUVOIR Cycle 1

The signature science cases outlined in this document represent transformative leaps in their individual fields. In Hubble Space Telescope parlance, any one of these cases represents a Large or Treasury program. Here, we imagine how aspects of each signature science case can combine as part of a robust multi-scale, guest observer-driven portfolio to impact an enormous range of science in LUVOIR's first year of observations.

## 12.1  Doing the impossible

LUVOIR's combination of aperture, resolution, and sensitivity allow it to perform science that is otherwise impossible. For "impossible," consider two different operating definitions. First, impossible can be taken as its traditional meaning: science that fundamentally requires an observatory of LUVOIR's size and capability. The second definition speaks to science that *can* be done by observatories of lesser capability but *will not likely be done* due to the observations requiring prohibitively large amounts of time. For example, **Figure 12.1** shows the allocation of observing time on HST as a function of program size in orbits. As a general rule, small observations are executed frequently, Treasury-scale observations very rarely. For fully commissioned guest observer facilities, this trend is usually invariant with aperture. However, by virtue of its immense capability through a combination of aperture and instrumentation, one of LUVOIR's assets is its ability to move to the left programs that sit on the right of **Figure 12.1** for smaller observatories. For example, the iconic Hubble Ultra Deep Field (HUDF) which required over 11 days of observing with HST could be performed in approximately 1 hour with LUVOIR-A. Essentially, otherwise prohibitively large programs become routine, freeing up time for programs fitting our first definition of impossible.

## 12.2  A transformational first year of science

We do not know how a future time allocation committee will prioritize science programs, but we expect a healthy balance across exoplanet science, general astrophysics, and solar system science. We also note that in many cases parallel observations can be executed, facilitating exciting combinations of large program science. For example, a 25 hour-long ECLIPS spectroscopic observation of an exoEarth candidate can *simultaneously* return 5-band HDI imaging at a depth 2–3 magnitudes deeper than the HUDF. When we consider the power of parallel observations, and the multiple science cases addressed by deep field observations (let alone archival data uses), *any* description of a set of long LUVOIR observations for a single science case is likely to *significantly underestimate* the total return for that time investment.

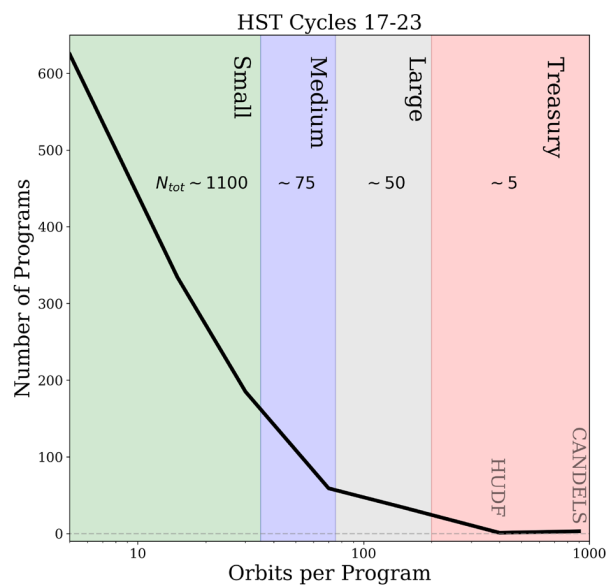

**Figure 12.1.** *Number of HST programs in Cycles 17-23 as a function of program execution time in orbits. Credit: STScI*

We illustrate LUVOIR's immense potential science return with a list of observing programs that require total exposure times of ~100 hours or fewer to execute, assuming the LUVOIR-A Architecture. These programs are drawn from the signature science cases described in this report. They do not, however, span the whole range of programs that the community may choose to execute as a guest observer-driven facility, nor are they intended to express any prioritization of science topics. They simply represent an example menu of large-scale program options that could be done in LUVOIR's first year, while leaving greater than three quarters of the time free for as-yet-to-be-determined programs at any scale, assuming a very conservative observatory efficiency of 50%. As neither the telescope+instrument sensitivities are finalized at this stage nor the exact targets known in many cases, *these example projects are approximate, preliminary, and purely illustrative*. It is important to emphasize that in nearly every case below, a parallel observation can be executed, usually with either HDI or LUMOS.

- **The exoEarth search begins:** ECLIPS makes ~ 25 visits to a variety of FGK stars (plus Proxima Centauri) within 7 parsecs. This provides ≥ 7-sigma photometric detections in three wavelength bands for any planets as bright or brighter than modern Earth within the ECLIPS field-of-view (FoV) at the time of each visit.

- **From darkness, light:** LUMOS observations of 10 fields provides observations of 10,000 galaxies between $0 < z < 1$, building the faint-end UV luminosity function for galaxies over 8 billion years of cosmic time and definitively constraining the amount of ionizing light they leak.

- **Water in motion:** Time-dependent, high spatial resolution, NIR and/or UV spectroscopic imaging of icy moons in the so-

lar system resolves geyser activity. Such observations could be done in coordination with an in-situ spacecraft, providing geyser location and timing information.

- **Reionization revealed:** Deep observations of a single HDI field reveals ~600 z=7 galaxies with enough statistical power to begin to reveal the effects (if any) of reionization suppression of early star formation.

- **Diverse worlds:** ECLIPS spectroscopy of ≥ 10 known planetary systems reveals the detailed atmospheric characteristics of a dozen or more planets including Venuses, mini-Neptunes, and Jupiters.

- **Feedback at the source:** Using its microshutter array, LUMOS maps the galactic outflows where they arise in star forming regions and follows the winds out to their interactions with the surrounding medium, directly resolving the complex physics of galactic feedback at unprecedented physical scales.

- **Habitable worlds:** For stars with promising planets from the exoEarth search, atmospheric characterization commences with ECLIPS spectroscopy in an integral field spectrograph band centered near 980 nm to search for water vapor absorption. A methane absorption feature also appears near this wavelength, allowing quick identification of any giant planets in the FoV.

- **The nurseries explored:** LUMOS obtains FUV-NUV observations of molecular diversity in the inner few AU of hundreds of young stellar and planetary systems.

The scientific impact of these example programs in the first year alone would be immense—and in many cases truly transformative. Moreover, this hypothetical sampling only represents a tiny fraction of what would be possible with a serviceable observatory intended to last *decades*. As such, LUVOIR is poised not just to continue in the tradition of the great observatories, but to re-define nearly every aspect of our knowledge of the cosmos.

# 13 Appendix A: Further LUVOIR science cases

To better capture the full range of LUVOIR's capabilities, this appendix includes short additional LUVOIR science programs contributed by the community and LUVOIR team members. The topics range from remote sensing of solar system bodies, to exoplanet observations, to a wide range of general astrophysics studies. In some instances, the authors have chosen to provide further details—or put their own perspective—on a case that is mentioned in the main science chapters. The cases are of varying length and appear in no particular order. This is a living document that will grow with time.

## 13.1    Observations of Venus with LUVOIR

Giada Arney (NASA GSFC), Valeria Cottini (NASA GSFC), Shawn Domagal-Goldman (NASA GSFC), Lori Glaze (NASA GSFC), Eric Lopez (NASA GSFC), Victoria Meadows (UW), Ravi Kopparapu (NASA GSFC), Roser Juanola Parramon (NASA GSFC)

### 13.1.1    Introduction

At its closest approach, Venus is the nearest planet to Earth, yet much still remains unknown about Earth's twisted sister. Many important Venus science questions can be addressed with sys-Earth telescopic observations. A sufficiently small solar elongation viewing angle for LUVOIR (≤ 45°) enables such observations of Venus. This, together with a darkening neutral density filter, opens the door to exciting and much needed data on the Venusian atmosphere. Three interesting case studies that could be investigated with LUVOIR are described briefly below as examples of the types of Venus science LUVOIR can make possible.

1. Observations from the JAXA Akatsuki orbiter have revealed an unusual stationary bow-shaped wave at the Venus cloud tops (65 km) observable at multiple wavelengths from the UV at 283 nm (corresponding to a $SO_2$ absorption band) to the longwave infrared at 8–12 µm (Fukuhara et al. 2017). Normal wind speeds at these high altitudes whip across the planet at roughly 100 m/s, but this UV-bright, bow-shaped feature remains stationary relative to the surface far below. The center of the bow-shaped feature (**Figure 13.1**) is located above the western slope of equatorial highland region Aphrodite Terra, and is interpreted to be a stationary atmospheric gravity wave associated with lower atmosphere wind flows over the terrain. However, the propagation of such waves to the cloud tops is difficult to reconcile with current understanding of convection in the Venus atmosphere (Seiff et al. 1985). Thus, the dynamics of the Venus atmosphere may be more complex than previously thought.

Monitoring the temporal evolution of features such as this could provide new insights into the physics of Venus atmosphere circulation.

2. Long temporal baseline (1970s–2012) monitoring of Venus across multiple Venus missions at λ = 215 and 283 nm has revealed quasi-periodic variations in high altitude (70 km) $SO_2$ abundance (Marcq et al. 2012) with reported variations from ~400 ppbv to less than 100 ppbv. This variability may be related to poorly understood oscillations in atmospheric circulation, and/or volcanic injections of $SO_2$ into the upper atmosphere. The amount of $SO_2$ currently in the atmosphere has been estimated to be in excess of equilibrated conditions by a factor of 100, implying a source (i.e., volcanism) within the past 20 million years (Bullock & Grinspoon 2001). Monitoring of $SO_2$ variations may therefore shed light on multiple processes related to atmospheric

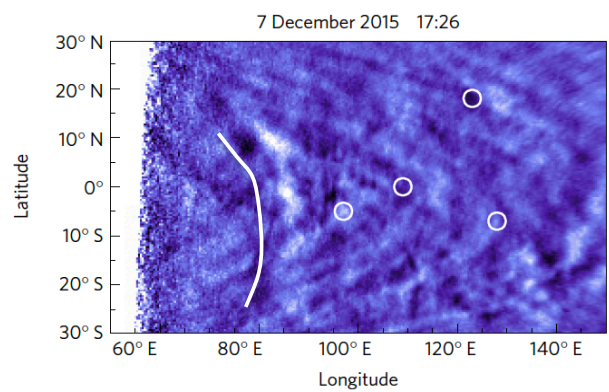

**Figure 13.1.** *A bow-shaped UV bright wave in the Venus upper atmosphere (highlighted with solid white line) seen with Akatsuki above the highland region Aphrodite Terra could be observed with LUVOIR and shed light on Venus atmospheric dynamics. Circles indicate displaced air parcels. Credit: Fukuhara et al. (2017)*

dynamics and/or volcanic processes. Such a monitoring campaign demands a UV-enabled space-based platform that can observe Venus over years or decades.

3. Besides $SO_2$, several other trace gases can be observed in the Venus spectrum, providing insights into chemical, dynamical, and photochemical processes that occur on Venus. For example, significant and surprising variability has been observed in the abundances and distributions of trace gases in the Venusian sub-cloud atmosphere (Arney et al. 2014). These variations hint at poorly understood chemical and physical mechanisms operating in the lower atmosphere. They also suggest the presence of $H_2SO_4$ virga events (rain that evaporates before reaching the surface). Between 1–2.5 μm, one can observe upwelling thermal radiation from the sub-cloud (0–45 km) atmosphere on the Venus nightside. This enables deep-atmosphere observations of HDO, $H_2O$, $SO_2$, HCl, CO, and OCS, which have been observed to vary both spatially and temporally on poorly constrained timescales.

Because exo-Venus planets may be one of the most common types of exoplanets (Kane et al. 2014), better understanding the planet next door will enable us to better interpret observations of exo-Venus worlds.

## 13.1.2    The role of LUVOIR

Currently, there is no planned NASA mission to Venus. Periodic monitoring of Venus over long time baselines can reveal important information about variations in atmospheric species that may constrain theories of dynamical, chemical, and geophysical processes occurring on Venus. LUVOIR-A is baselined to point to a minimum solar elongation angle of 33°, which is small enough to allow observations of Venus at even less than its maximum elongation.

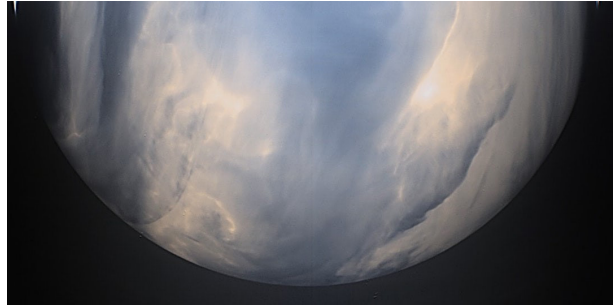

**Figure 13.2.** *A view of Venus from JAXA's Akatsuki. LUVOIR-A would obtain comparable spatial resolution. Credit: R. Juanola Parramon (NASA GSFC)/Damia Bouic/JAXA Akatsuki*

Observations towards crescent phase will enable observations of nightside thermal radiation upwelling from below the cloud deck. Certain UV observations (e.g., of high altitude $SO_2$ variations and of the "bow" shaped feature) require a space-based observatory, as they cannot be performed from the ground.

LUVOIR could achieve extremely good spatial resolution on Venus. **Figure 13.2** shows a view of Venus from the Akatsuki orbiter at a wavelength of ~2 μm showing variations in opacity of the lower cloud deck on the planet's nightside. LUVOIR could achieve a spatial resolution of 4 km/pixel at this wavelength (although note that the scattering footprint for radiation upwelling from the sub-cloud atmosphere is ~100 $km^2$).

## 13.1.3    The science program

Venus is significantly brighter than other sources LUVOIR will observe ($m_v$ = -4.4 on the dayside), necessitating neutral density filters to view this interesting target. To underscore this point, the brightest source the online LUMOS ETC includes has an AB magnitude of 15, and the brightest source included for HDI is magnitude 20. Venus is an extremely bright target for LUVOIR. However, LUVOIR is also considering observations of Jupiter, which has $m_v$ = -2.7, so Venus is comparable to other bright solar system targets under consideration.

The angular diameter of Venus is ~30" when the planet is close to maximum elongation. The LUVOIR-A LUMOS spectrometer currently has a wavelength range 100–400 nm, and it may be extended to 1000 nm. LUMOS has several available resolutions (R = 500, 16000, 63200, 100000), enabling moderate and very high-resolution spectroscopy. For comparison, previous observations probing isotopes in the Venus atmosphere have used $R = 10^5$ (Krasnopolsky et al. 2010), while measurement of gas species in the nightside thermal windows can be accomplished with R = 2000 (Arney et al. 2014), so the range of resolutions used by LUVOIR are useful for a variety of studies. The LUMOS field-of-view (FOV) is 2' x 2', so Venus fits comfortably within it.

LUVOIR-A HDI has a FOV of 2' x 3', which also allows for full views of Venus in a single frame. It has two channels: UVIS covering 200–950 nm and NIR covering 800–2500 nm. Nyquist sampled pixels at 400 nm (2.73 mas/pixel) and at 1200 nm (8.20 mas/pixel) mean that 10000 pixels will span Venus at 400 nm (~ 1.2 km/pixel) and 3600 pixels at 1200 nm (3.3 km/pixel). By comparison, the Venus Monitoring Camera aboard ESA's Venus Express orbiter could obtain 0.2–45 km/pixel depending on how far the spacecraft was from the planet (Markiewicz et al. 2007). LUVOIR allows for orbiter-quality monitoring of Venus.

LUVOIR-A ECLIPS will have a significantly smaller FOV than LUMOS and HDI, but it could still in principle observe Venus using open slots in the filter wheel. The ECLIPS imager has a FOV of 1.4" x 1.4" in the UV, 2.7" x 2.7" in the visible, and 5.6" x 5.6" in the NIR. Useful observations with ECLIPS could be obtained via a mosaic technique to cover the entire disk of Venus, or smaller sub-regions could be targeted.

---

### Program at a Glance

**Science goal:** Obtain high spatial and high spectral resolution data of Venus at UV-VIS-NIR wavelengths to monitor for atmospheric variability possibly related to atmospheric dynamics, chemistry, and/or volcanic activity.

**Program details:** Venus will have an angular diameter ~30" at maximum solar elongation, and its day-side magnitudes are: U = -2.79, V = -3.68, B = -4.38, R = -4.95, I = -5.08, Rc = -4.73, Ic = -5.04.

**Instrument(s) + configuration(s):** ECLIPS and LUMOS could be used to obtain medium to very high-resolution spectra. LUMOS can fit Venus in its FOV; ECLIPS would require mosaicing. HDI can obtain high spatial resolution images from the UV to NIR.

**Key observation requirements:** Observations of Venus will require neutral density filters and the ability to observe at a sufficiently small solar elongation angle (< 45°). This is compatible with the design of the notional LUVOIR-A.

## 13.2    Exo-cartography for terrestrial planets

Claire Marie Guimond (McGill University), Nicolas B. Cowan (McGill University)

### 13.2.1    Introduction

LUVOIR and its predecessors will have yielded us alien pale blue dots; the next step is to study their habitability. There are undoubtedly cases where understanding a planet requires understanding the diversity of its regions, especially since oceans and continents imply long-term habitability. Planets without a mix of water and land may not have a significant silicate weathering feedback, which regulates the $CO_2$ greenhouse on Earth and keeps her climate temperate (Cowan 2015).

With exo-cartography, we can infer the number, reflectance spectra, and longitudinal locations of major surface types (Fujii et al. 2017; Cowan and Strait 2013; Kawahara and Fujii 2011). This works because directly imaged planets show diurnal brightness variations as different surface and cloud features rotate into view (see Cowan and Fujii 2017). The lightcurve collected from the planet is the disk-integrated reflectance per exposure (**Figure 13.3**). In theory, we can invert lightcurves to piece out latitude-longitude albedo maps (Kawahara and Fujii 2010; **Figure 13.4**).

### 13.2.2    The role of LUVOIR

The very large aperture of LUVOIR will enable reflected light surface mapping and spin determination for terrestrial planets (Pallé et al. 2008; Oakley and Cash 2009; Cowan et al. 2009, 2011; Kawahara and Fujii 2010, 2011; Fujii and Kawahara 2012). Previous mapping papers have adopted the optimistic 1% photometric uncertainty (S/N of 100) for 1-hr integrations. For a 15-m telescope, this will only be possible for a super-Earth at 1 pc. However, Cowan et al. (2009) claimed they could do essentially the same science with 3% photometry in 1-hr integrations (24 data per rotation, each of S/N = 33).

As **Figure 13.5** shows, for an Earth twin at 10 pc, we can only expect an S/N of ~10 with one rotation, but for more slowly rotating planets and/or larger radii, this value can double or triple. Further, decreasing the time resolution (i.e., longitudinal sample rate) by a factor of 16 increases the per-integration S/N by a factor of 4—this would set the number of pixels in the final map. Stacking multiple epochs of observations can be problematic, as clouds strongly influence reflected light

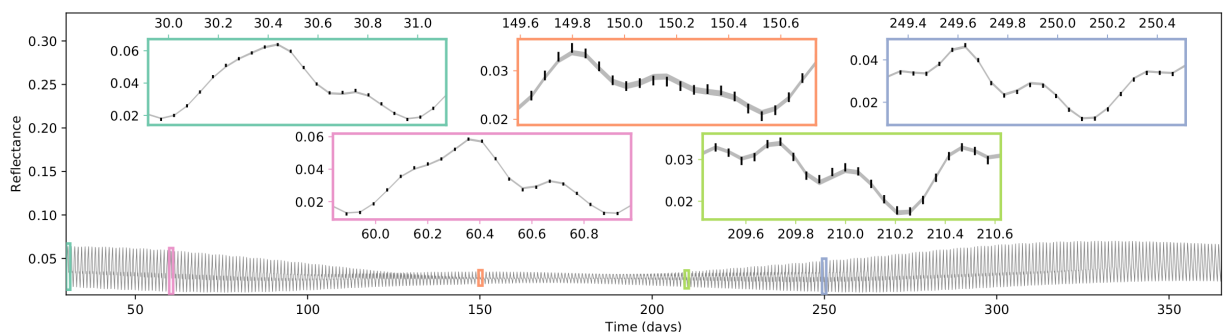

**Figure 13.3.** *Simulated light curve and resulting fit from an Earth analog with 0 degree (face-on) inclination and 90 degree obliquity. The simulated data and measurement errors are shown in the five insets, representing five epochs, each lasting one rotation, collected over the course of a year with a one hour cadence. The shaded region in each inset shows the central 90% posterior credible interval for the constructed light curve. From Farr et al. (2018, in prep.).*

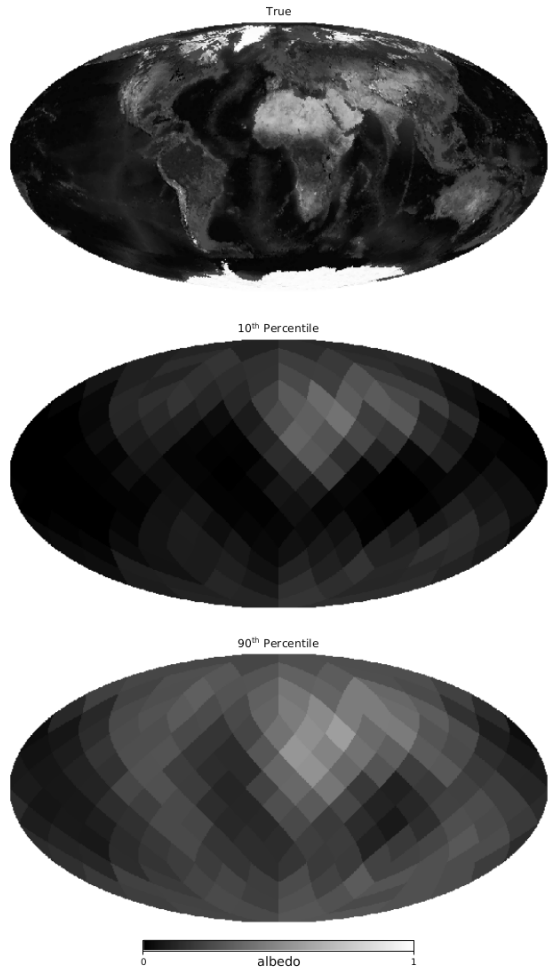

**Figure 13.4.** *Demonstration of albedo map retrieval using a simulated lightcurve of the Earth (top), with 10th percentile (middle) and 90th percentile (bottom) of the marginal posterior distributions for the albedo of each pixel. The 10th percentile map clearly shows a reflecting region (the Sahara Desert) while the 90th percentile clearly shows dark regions corresponding to the Pacific, Atlantic, and Indian Oceans. These maps therefore establish the presence of continents and oceans on the the planet. Adapted from Farr et al. (2018, in prep.).*

smallest targets, only spin orientation and low-resolution longitudinal maps will be retrievable.

### 13.2.3   The science program

This program will target the Earth-twin planets we expect to have been detected (Stark et al. 2014) in the habitable zone of G stars ($M_V$ ~ 4.8). We limit observations to one rotation period to better avoid confusion with diurnal cloud variation. Hence, the rotation period sets total integration time. We assume a sampling rate of four exposures per rotation—e.g., 6 hours of exposure for a 24-hour rotation—as the bare minimum to detect rotational variation.

We adopt the wavelength ranges 400–500 nm (UVIS channel) and 800–900 nm (UVIS/NIR channels). The inverted reflectivity difference across these bands has been fluxes, and these atmospheric features are prone to change between epochs (Oakley and Cash 2009).

Thus only with a 15-m class space telescope such as LUVOIR can we start to map super-Earth exoplanets. For the

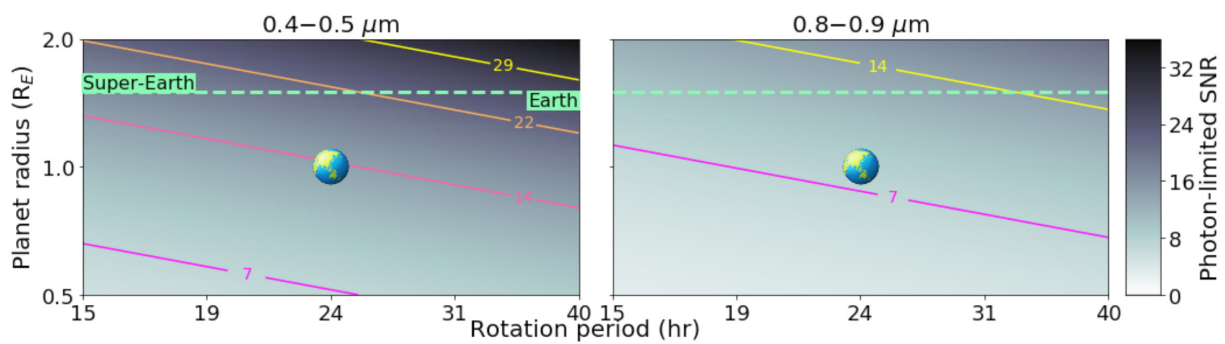

**Figure 13.5.** *Signal-to-noise ratio at the poisson limit for two bandpasses as a function of rotation period and planet radius, for a planet at 10 pc with semi-major axis of 1 AU, geometric albedo of 0.3, nominal telescope diameter of 15 m, and coronagraph throughput of 15%.*

**Program at a Glance**

**Science goal:** Construct rough longitudinal maps of Earth twins using rotational mapping techniques.

**Program details:** Observe brightness variations over one rotation period, using point source multi-band photometry, for the nearest handful of < 2 $R^E$ planets in the habitable zones of G stars.

**Instrument(s) + configuration(s):** Differential HDI photometry with ECLIPS.

**Key observation requirements:** Two imaging bands of 400–500 nm and 800–900 nm with signal-to-noise $\gtrsim$ 10, and planet-star contrast $\gtrsim$ $10^{-10}$.

shown to suppress the cloud signal and roughly recover the continental distribution (Kawahara and Fujii 2011).

The observations use LUVOIR's Extreme Coronagraph for Living Planetary Systems to suppress light from the host star, in conjunction with the High Definition Imager. The maximum angular separation between an Earth-twin on a 1-AU circular orbit 10 pc away is 100 mas, which would be within the outer working angle of the coronagraph. For the highest signal, planets will be imaged at the projected separation corresponding to the coronagraph's inner working angle. We will pursue targets with planet-star flux contrast $\geq 10^{-10}$, due to expected instrument limitations.

## 13.3 Extragalactic massive stars

Miriam Garcia (CAB, CSIC-INTA), Chris Evans (UKATC, STFC)

### 13.3.1 Introduction

Massive stars are cosmic engines and make valuable probes of the Universe. They are mighty sources of ionizing radiation and mechanical energy, giving rise to striking bubbles, giant HII regions, and galactic-scale outflows during their lives. In death they are the progenitors of supernovae and long-duration γ-ray bursts, with the latter so bright they can be detected in galaxies up to z = 9 (Robertson & Ellis, 2012). These dramatic explosions release chemical elements that are essential to life (e.g., oxygen) into the interstellar medium, and their remnants (neutron stars, pulsars, black holes) are the sites of extreme physics and sources of gravitational waves (Abbott et al. 2016).

A critical area of research in the field is characterizing the role of metallicity (abundance of elements heavier than H and He) on the lifecycle of massive stars. The Universe has become increasingly enriched in metals since the Big Bang so, to correctly interpret observations of distant galaxies, we need models of stellar physics and evolution that match their metal content. The ultimate goal is to investigate the paradigm of the first generation of (effectively metal-free) massive stars in the Universe (so-called Population III), a suggested source of reionisation at z > 6 (Haiman & Loeb, 1997).

A key observable of massive stars is their radiatively-driven stellar winds, via which solar masses of material can be lost throughout their lives, directly influencing their evolutionary sequence, feedback to the local medium, and pre-explosion core mass. The wind momentum depends strongly on metallicity (Kudritzki & Puls, 2000), implying a strong metallicity dependence to evolution and feedback. Only UV spectroscopy from space can provide the data required to assess the properties of such winds (e.g., Fullerton et al. 2006; Sundqvist et al. 2014). IUE, FUSE, and HST-STIS/COS have made unprecedented and long-lasting contributions to these studies.

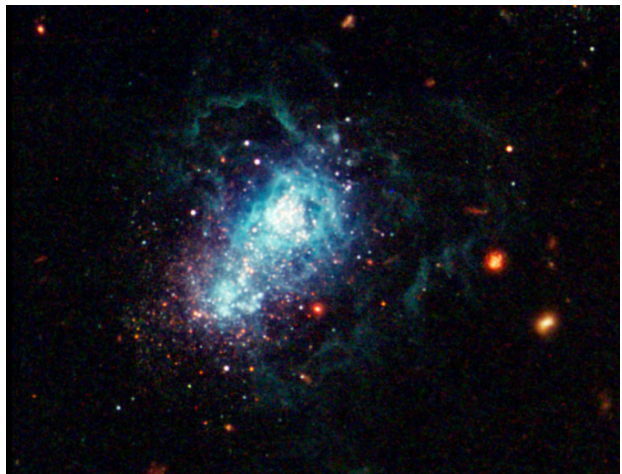
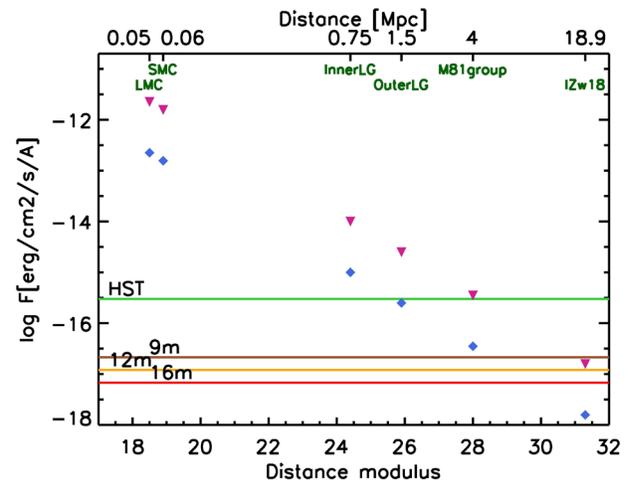

**Figure 13.6.** *Left: Resolving massive stars in the ultra-metal poor galaxy I Zw 18 at 18 Mpc is a long-standing goal for studies of stellar evolution. Right: UV fluxes (1500Å) for massive stars (triangles = B-type supergiants; diamonds = O-type stars) at increasing distances, compared to five orbits of HST spectroscopy (at R~2000). At least a 12m aperture is required for UV spectra of individual massive stars in I Zw 18.*

Studies in nearby galaxies such as the SMC (60kpc), IC1613 (750kpc), WLM, Sextans A and NGC3109 (~1.2Mpc) have enabled studies of the properties of massive stars with metallicities of 10–20% solar (e.g., Venn et al. 2003; Bresolin et al. 2007; Evans et al. 2007; Camacho et al. 2016) providing valuable empirical templates (e.g., Leitherer et al. 2001) and important tests of theory (e.g., Mokiem et al. 2007). However, significant uncertainties remain over their wind behavior at low metallicity, even when UV data are available (Tramper et al. 2011; Herrero et al. 2012; Garcia et al. 2014; Bouret et al. 2015) because of the sensitivity limits of current facilities. Moreover, we need to go at least an order-of-magnitude lower in metallicity to match conditions in distant absorption-line systems (e.g., Prochaska et al. 2003) and to move toward robust models of Pop. III stars, but these are simply out of reach of current facilities (HST, 8–10m ground-based telescopes).

### 13.3.2   The role of LUVOIR

The ambitious goal in this topic is studies of individual massive stars in I Zw 18, a star-forming galaxy at 18 Mpc (see **Figure 13.6**) with a metallicity of 2–3% solar (Vilchez & Iglesias-Paramo). The task will require outstanding spatial resolution and sensitivity. Optical/NIR spectroscopy, needed to determine some key physical properties (temperatures, luminosities) of individual stars, will be feasible with future ground-based, multi-object spectrographs aided by adaptive optics (e.g., TMT, ELT). However, UV data are crucial to build the complete description of the stars.

Such observations require the sensitivity and spatial resolution of *at least a 12-m aperture facility in space* (see **Figure 13.6**). High-quality UV spectroscopy over 1200–1800 Å of individual luminous stars will constrain their mass-loss rates, shocks, structure, and velocity fields of their winds. Only with access to UV spectra of these objects can we fully characterize their natures and test theoretical predictions for the properties of such stars in the early Universe.

### 13.3.3   The science program

LUMOS onboard LUVOIR will enable the first thorough characterization of the winds of metal-poor (from 10% down to 2–3% solar metallicity) massive stars with mid-resolution ($R = \lambda/\Delta\lambda \geq 5000$) with UV spectroscopy. The combination of the large aperture of LUVOIR and the multiplex of LUMOS will allow us to mine metal-poor galaxies out to the outer edges of the Local Group, other neighboring groups, and ultimately I Zw18 (see **Figure 13.6**). The data will provide definitive answers to pressing questions regarding the properties and evolution of high-mass stars including:

- The first observational parameterization of radiation-driven winds at very

---

**Program at a Glance**

**Science goal:** The physical properties of massive stars at very low metallicities, akin to those in the early Universe.

**Program details:** UV spectroscopy of individual massive stars in the galaxy I Zw18 at a distance of 18 Mpc.

**Instrument(s) + configuration(s):** Multi-object spectroscopy with LUMOS (G155L).

**Key observation requirements:** UV spectroscopy spanning 1000 to 2000 Å (essential), to 2500 Å (desirable).; R ≥ 5000; S/N > 20 per resolution element at 1500 Å.

---

low metallicities (<10% solar), with consequences for models of Pop. III stars and stellar evolution in the early Universe, and population-synthesis models used to interpret observations of high-z galaxies.

- Direct tests of 'chemically-homogeneous evolution' at such low metallicities, and a characterization of their winds. This evolutionary channel has been invoked to explain the binary system in the first LIGO detections (e.g., de Mink & Mandel, 2016) but lacks observational confirmation.

## 13.4 Ultraviolet haloes around edge-on galaxies

Benne W. Holwerda (University of Louisville) in collaboration with the SKIRT team (University of Gent)

### 13.4.1 Introduction

In the latest iterations of radiative transfer models of edge-on galaxies across from ultraviolet to the sub-millimeter regime (e.g., Popescu+ 2011, Holwerda+ 2012, Recently, De Geyter+ 2015, Mosenkov+ 2016, 2018), it has become clear that there is a dust-obscured ultraviolet component as well as a diffuse halo (see e.g., Seon+ 2014 and Seon & Draine, 2016, **Figure 13.7**).

Models of the ultraviolet radiative transfer are complicated by both the localized nature of the origin in galaxies and the scatter from the dusty ISM (see **Figure 13.8**). Similarly, the diffuse ultraviolet emission from inter-arm regions can either be from localized O-stars or scattered (Crocker+ 2015).

A way to test the nature of the diffuse UV emission in any of these galaxies that is completely independent from the panchromatic radiative transfer analysis is UV polarimetry, suggested by Baes & Viane (2016), Hodges-Kluck & Bregman (2014).

If the UV halo emission is dominated by scattered radiation, we would expect a strong linear polarisation signature, similar as seen in reflection nebulae. Unfortunately, there is currently no UV polarisation instrument available. The total emission at these wavelengths is probably dominated by direct light from the stellar halo, but it is possible that the signature of a very extended dust distribution might still be visible in polarised light. Detailed polarised radiative transfer simulations should be used to test this.

### 13.4.2 The role of LUVOIR

LUVOIR can image the diffuse ultraviolet haloes, obtain spectra to identify the predominant stellar population behind them, and measure polarization of the halo to determine how much of the light is scattered or not.

And because LUVOIR has the sensitivity and spatial resolution far surpassing existing UV observatories, it can map diffuse, extended ultraviolet haloes around galaxies at much earlier epochs in galaxy-formation, answering the question of whether dusty cirrus is common in the past.

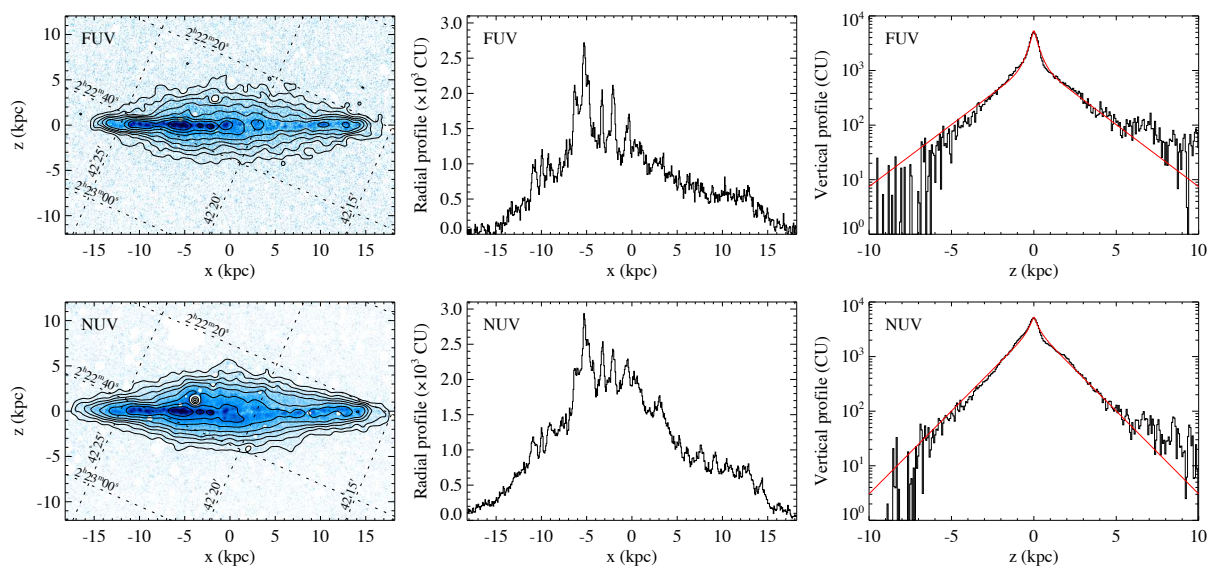

**Figure 13.7.** *The NGC 891 ultraviolet halo from Seon et al. (2014).*

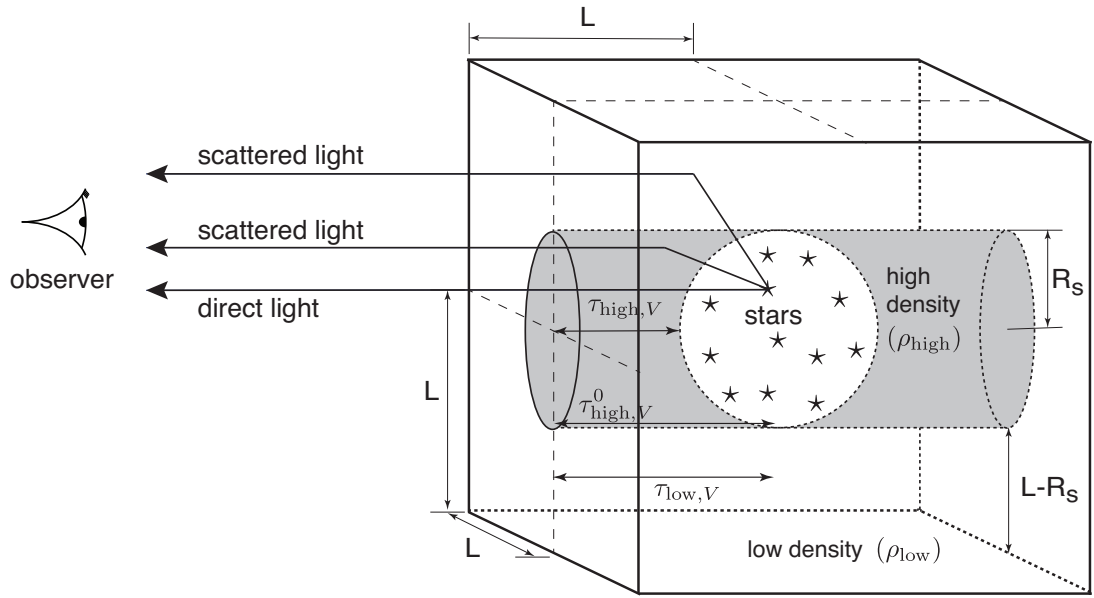

**Figure 13.8.** *A sketch of the model from Seon & Draine (2016). It shows ultraviolet rays directly or scattered into the line-of-sight towards the observer.*

### 13.4.3   The science program

The science program consists of deep imaging to detect ultraviolet haloes around galaxies. Ultraviolet spectroscopy of the haloes will determine the dominant stellar populations and polarization imaging will be of extreme importance as well. Exposure times depend on the surface brightnesses needed for each galaxy.

### References

Crocker et al. 2015, *AJ*, 808, 76

Baes & Viane, 2016, *A&A*, 587, 86

---

**Program at a Glance**

**Science goal:** To find and characterize the origin of the diffuse halo and inter-arm ultraviolet light in disk galaxies

**Program details:** To detect and characterize the ultraviolet halo, one targets edge-on spiral disk galaxies (e.g., NGC 891) and the inter-arm regions in well-resolved face-on galaxies.

**Instrument(s) + configuration(s):**
   LUMOS imaging – for detection and characterization of the extent.
   LUMOS spectroscopy – to determine the predominant stellar type responsible for the diffuse ultraviolet emission.
   POLLUX polarimetry – to determine how much of the diffuse light is generated *in situ* and how much scattered off cirrus clouds.

**Key observation requirements:** Key requirements are both FUV and NUV imaging to extremely low surface brightnesses ~32 mag/arcsec², R=3000 spectroscopy for stellar populations. A polarimeter to measure polarized light fraction.

---

## 13.5 Prospects for mapping terrestrial exoplanets with LUVOIR

Jacob Lustig-Yaeger (University of Washington), Victoria S. Meadows (University of Washington), Guadalupe Tovar (University of Washington), Edward Schwieterman (University of California, Riverside), Yuka Fujii (Tokyo Institute of Technology)

### 13.5.1 Introduction

Future direct imaging missions offer a means to characterize the surface habitability of terrestrial exoplanets, probing deeper atmospheric regions not accessible to transmission spectroscopy. Nulling the star with a coronagraph enables directly-imaged photometry and spectroscopy to capture disk-integrated light emitted and reflected from an exoplanet along an ensemble of relatively short atmospheric paths, compared to transmission spectroscopy which probes the upper regions of atmospheres along longer slanted paths (Fortney 2005). Photons on more direct paths through the atmosphere encounter lower optical depths, and therefore possess a greater sensitivity to the lower atmosphere and surface. Geometric arguments alone make direct imaging well suited for future habitability and biosignature assessments of terrestrial exoplanets.

To complement coronagraph spectroscopy of the disk-averaged surface and atmosphere, time-series observations offer a unique window into the two-dimensional surface heterogeneity of terrestrial exoplanets. Rotating planets that are spatially heterogeneous induce photometric variability in their observable light curves (Ford et al. 2001; Palle et al. 2008; Oakley et al. 2009). Numerous studies have explored time-series observations for exoplanet surface identification and mapping (for a recent review, see Cowan and Fujii 2017). Multi-wavelength, time-series observations of Earth have been used to construct longitudinal maps of land, ocean, and clouds (Cowan et al. 2009; Fujii et al. 2010, 2011). Further generalizing these observations and modeling methods have shown promise for uncovering the reflectance spectrum and longitudinal distribution of individual surfaces on exoplanets (Cowan et al. 2013), including the possibility of detecting extrasolar oceans (Fujii et al. 2017). Longitudinal mapping using multi-wavelength, time-series observations of terrestrial exoplanets offers a path towards identifying oceans and assessing habitability.

### 13.5.2 The role of LUVOIR

LUVOIR is critical for performing surface habitability studies of terrestrial exoplanets because it is a large aperture, space-based, coronagraph-equipped telescope. A space-based observatory is necessary for (a) achieving the coronagraph contrast ratio required to observe Earth-like exoplanets ($\sim 10^{-10}$), (b) escaping the diurnal cycle of Earth for continuous observations that may span multiple days, and (c) accessing wavelengths that are optically thick to space from the ground due to atmospheric opacity. In particular, the near-UV ($\sim 300$ nm) is sensitive to the cloud variability of Earth-like planets. The large aperture considered for LUVOIR ($\sim 12$–15m) will also allow both shorter cadence time-series ($\sim 1$ hour) – providing higher surface resolution to the inferred maps—and access to solar-analog systems out to a greater distance, which increases the yield for such investigations.

### 13.5.3 The science program

We propose a science program that enhances the science return from spectroscopic observations that motivate the design of the LUVOIR coronagraph. Approximately 100 hours will be required to detect the presence of $O_2$ in the atmosphere of an Earth-like

---

**Program at a Glance**

**Science goal:** Study the rotation rate, surface colors, geography, and habitability of directly imaged terrestrial exoplanets.

**Program details:** Construct time-series observations of habitable terrestrial exoplanets using the single spectrum exposures that will be required to build-up much longer integration times for spectroscopy.

**Instrument(s) + configuration(s):** ECLIPS coronagraph spectroscopy

**Key observation requirements:** Single coronagraph bandpass (0.7–0.8 µm), R = 100–200, S/N ≥ 0.5 per spectral element per exposure (S/N ≥ 2.5 per bandpass per exposure), coronagraph design contrast $10^{-10}$

---

exoplanet at 10 pc with the 15-m LUVOIR concept. Our program will leverage these long integrations to "mine" for rotationally induced exoplanet time variability using individual ~1 hour exposures from a longer composite spectrum. If these observations include data within the 0.7–0.8 µm LUVOIR bandpass (which contains the 0.76 µm $O_2$ A band), we can bin the spectrum into a single photometric point and construct a lightcurve of a terrestrial exoplanet over the total spectrum exposure time. Our simulations suggest that this dataset can be used to infer the rotation rate of Earth to within ~5% (Lustig-Yaeger et al. 2017). Binning the time-dependent spectra into two or more photometric points is an effective means of constructing simultaneous multi-wavelength lightcurves, which could potentially infer the crude longitudinal distribution and color of oceans on an Earth analog (Lustig-Yaeger et al. 2017). Other wavelengths can be used either instead of or in concert with the 0.7–0.8 µm LUVOIR bandpass, such as the 0.4–0.45 µm bandpass, where Rayleigh scattering masks surface features and enhances sensitivity to heterogeneous cloud coverage. However, the 0.7–0.8 µm bandpass is still preferred for the initial assessment because of its sensitivity to both atmospheric oxygen and the planetary surface.

This science program would allow a long spectral exposure, to yield two unique studies along both the spectral and temporal dimensions. Co-adding in time will give a long integration spectrum at the native instrument spectral resolution that can be used to study the atmospheric composition of the targeted exoplanet. Co-adding in wavelength will give a long baseline lightcurve with little to no spectral resolution that can be used to study the targeted exoplanet's rotation rate and surface map. Since the time-resolution of a lightcurve depends on the individual exposure times that comprise the time-series data, low noise or noiseless detectors that can support shorter exposure times will greatly improve the feasibility and scientific return of this program.

## 13.6   Spatially resolved maps of star-forming gas with LUVOIR

John M. O'Meara (Saint Michael's College)

### *13.6.1 Introduction*

Strong HI absorption, specifically the Damped Lyman alpha systems, have been used for decades to study the bulk of the neutral gas in the universe (Wolfe et al. 2005), and thus the reservoirs for star formation. Large samples of DLA now exist (e.g., Noterdaeme et al. 2012) and the metallicity for the DLA has been obtained for hundreds of systems (Rafelski et al. 2012). Despite their long history of study, the size of the DLA at high-z has gone largely unconstrained, as they are observed toward very small emitters on the sky (e.g., quasars or GRBs). If instead *spatially extended* objects such as galaxies are used as a background source, this would allow for a direct study of the spatial extent of any foreground intervening DLA gas, along with a map of its variation in metallicity.

This methodology can be applied to higher HI column density Lyman limit systems as well, provided their Lyman alpha line alone is sufficient to determine N(HI).

**Figure 13.9** demonstrates this idea showing variable absorption towards a strong gravitationally lensed galaxy at z~2.8 (O'Meara et al. 2018, in prep) with KCWI on Keck. Gravitationally lensed galaxies are unfortunately rare, limiting our ability to apply this technique en masse and to compare the results statistically to simulations. Furthermore, ground based observations of the DLA are limited to z>1.6, excluding the majority of the history of the universe, and with it much of the transition of galaxies from star-forming to passive, an epoch where a detailed understanding of the "where and when" of star-forming gas is critical.

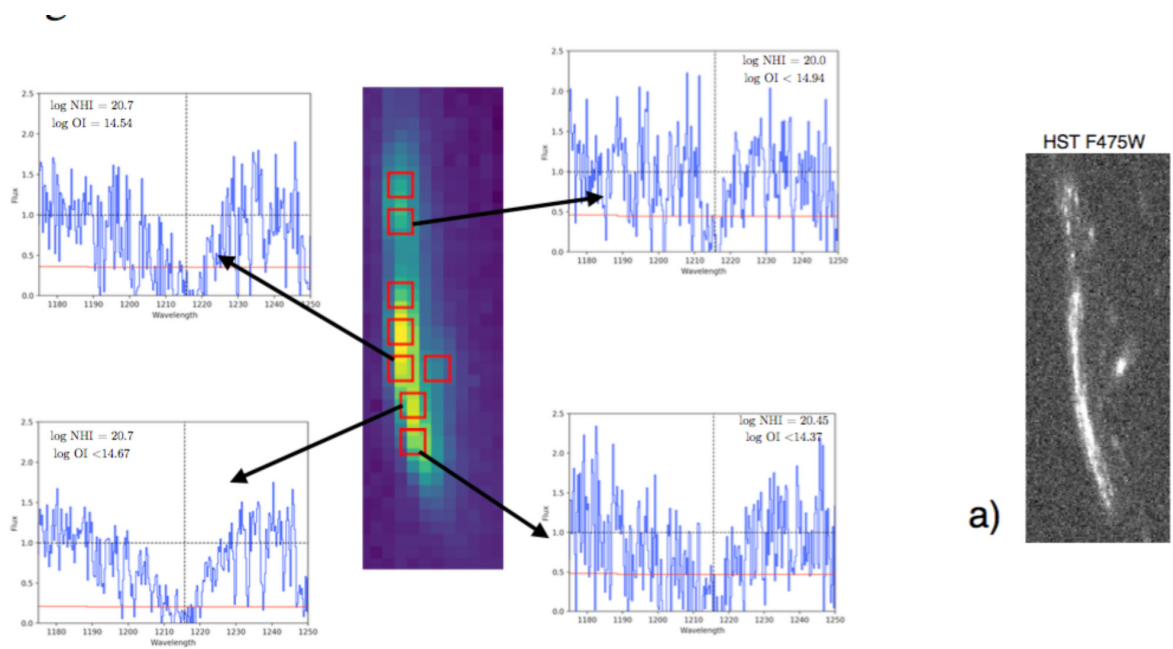

**Figure 13.9.** *Keck KCWI spectra of a z~2.8 lensed galaxy. Spectra from multiple positions along the galaxy show variations in HI column density in an intervening z~2.5 DLA. Metal lines are also observed for this system, allowing for constraints to be made on variations in gas metallicity on kpc scales, similar to scales in cosmological simulations.*

---

### Program at a Glance

**Science goal:** Spatially resolved maps of HI and metals in star-forming gas at $0 < z < 2$

**Program details:** LUMOS observations of $z < 2$ galaxies with intervening HI and metal absorption. Multiple LUMOS MOS shutters across each galaxy provide the spatial sampling.

**Instrument(s) + configuration(s):** LUMOS G120, G150, G180, G300M

**Key observation requirements:** S/N > 10 per spectrum. Wavelength coverage of HI Lyman-alpha and key metallicity tracers (OI, CII, SiII) all in the rest frame FUV.

---

### 13.6.2   The role of LUVOIR

LUVOIR provides two essential opportunities: First, the UV capabilities of LUMOS allow for the technique to be applied for HI Lyman alpha anywhere from $0 < z < 2.2$, essentially the last 11 billion years of cosmic history. At lower redshifts, the extent on the sky of the background galaxy increases, facilitating finer and finer spatial sampling. Second, LUVOIR's immense aperture allows for the technique to be applied to the more general population of galaxies. As a result, instead of relying on very rare gravitational lens scenarios, galaxies at FUV magnitudes of 20 (for LUMOS high resolution modes) or 23 (for low resolution modes still capable of detecting HI and metal absorption) are reachable in ~1 hr integration time.

### 13.6.3   The science program

LUMOS shutters will be placed across $z<2$ galaxies (identified either through ground based spectra or photometry), ideally in the M modes for higher resolution (FUV mag 20.5 or brighter). Selection of G120M, G150M, G180M, G300M will depend on the redshift of the intervening absorption, and will be made to cover HI Ly-a, along with key metal line diagnostics such as OI, SiII, and CII. Sources can be as faint as FUV mag 23, if the G145LL mode is employed. Observations will reach S/N = 10 or greater, which is possible for all M gratings for sources FUV mag 20.5 or brighter.

Each LUMOS microshutter spectrum will be analyzed for intervening HI and metal line absorption, and a spatial map of HI and metallicity variations will be made for each absorber.

Follow-up observations with either LUMOS or HDI may be warranted to observe the galaxies associated with the HI gas.

## 13.7   The quiescent UV spectra of cool dwarf stars

Allison Youngblood (NPP / NASA GSFC)

### 13.7.1   Introduction

The UV spectra of cool dwarf stars (F, G, K, & M) are dominated by emission lines originating from the magnetically heated upper layers of the stellar atmosphere (**Figure 13.10**); <10% of their photospheric/blackbody continuum is emitted in the UV ($\lambda$ < 3000 K). Chromospheric, transition region, and coronal emission lines from species like H I, Mg II, C IV, Si IV, and N V are ultimately powered by the star's magnetic dynamo, which also controls the rotational evolution of these stars via magnetized stellar winds

(Skumanich 1972). UV emission line fluxes and widths can diagnose temperature, pressure, density, and kinematics throughout the stellar atmosphere, even indicating $H_2$ fluorescence from the star's own UV photons (e.g., Kruczek et al. 2017, Jaeggli et al. 2018). When obtained for a wide range of masses and ages, UV spectra can elucidate the evolutionary processes of cool stars (e.g., Guinan et al. 2003).

Of special note is H I Lyman alpha (1215.67 Å), the brightest emission line in the FUV and NUV spectrum of a cool dwarf. A

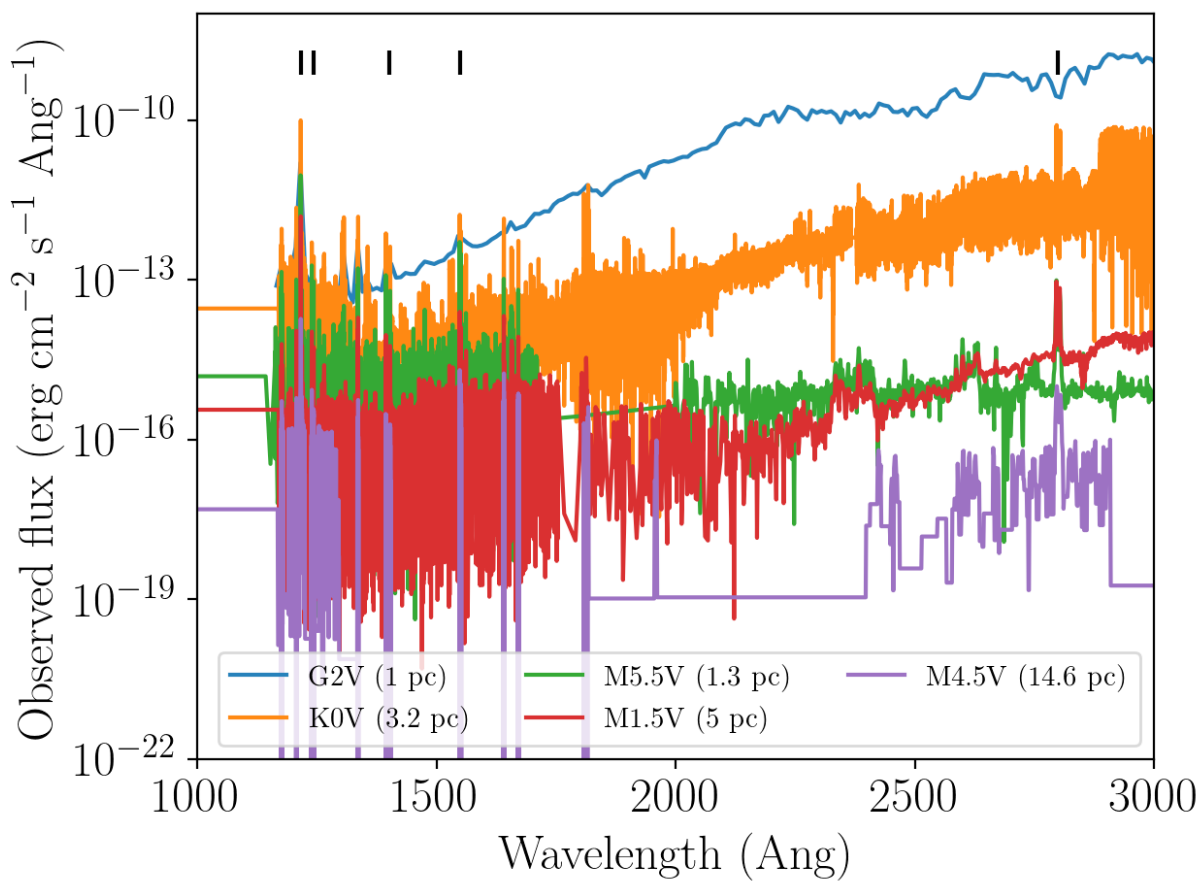

**Figure 13.10.** *The UV spectrum from 1000—3000 Å for a sample of cool dwarfs observed with Hubble and Solar Mesosphere Explorer. Lyman alpha, N V, Si IV, C IV, and Mg II are labeled with tick marks.*

---

### Program at a Glance

**Science goal:** Characterizing the quiescent UV spectra of cool dwarf stars (F, G, K, and M dwarfs) to better understand the physical conditions in their atmospheres and how they evolve in time.

**Program details:** F, G, K, and M dwarfs covering a wide parameter space of mass and age should be targeted.

**Instrument(s) + configuration(s):** LUMOS spectroscopy.

**Key observation requirements:** 1000-3000 Å, R=30,000, S/N > 10.

---

complete UV characterization of these stars cannot be obtained without it, but Lyman alpha observations are challenging. The Earth's extended atmosphere glows brightly in Lyman alpha, and interstellar H I attenuates the entire line core for even the closest stars. The core of the Lyman alpha line must be reconstructed from the observed line wings (e.g., Youngblood et al. 2016), which can prove difficult for faint targets.

### 13.7.2   The role of LUVOIR

LUVOIR's LUMOS instrument will have a sensitivity two orders of magnitude better than HST's STIS, enabling measurements of the chromospheric heating rate for even the most seemingly inactive cool stars (e.g., older M dwarfs; Guinan et al. 2016). LUMOS will provide access to more emission lines with a wider range of formation temperatures (probing from the lower chromosphere to the corona) not only because of its increased sensitivity but also its increased spectral range compared to STIS and COS (e.g., sensitive access to 1000–1200 Å). Its microshutter array can also create a long slit, which is essential for spatially separating the stellar Lyman alpha emission from the bright geocoronal Lyman alpha emission, and the sensitivity should be great enough to measure continuum emission.

### 13.7.3   The science program

To obtain a 1000 Å–3000 Å spectrum of a cool dwarf, a combination of at least two gratings must be employed. Using the medium resolution gratings (R~30,000) throughout will resolve line multiplets (e.g., C II, O IV) and the line widths themselves. Two lower resolution gratings (R~8,000 and 500) are also suitable alternatives for fainter targets.

After 5 hours of integration (representative of the integration time for the K and M dwarfs of HST's MUSCLES Treasury Survey; France et al. 2016), S/N = 10 for the faintest emission lines of an early M dwarf's FUV spectrum will be achieved for stars down to $FUV_{GALEX} = 23$ mag (9-m LUVOIR) or $FUV_{GALEX} = 25$ mag (15-m LUVOIR). For reference, the M2 V star GJ 832 at d = 5 pc is $FUV_{GALEX} = 21$ mag. LUMOS's microchannel plate detectors enable time-tagging photons as they arrive, so any flares that occur during integration will be able to be distinguished from quiescent emission, warranting their own analysis.

## 13.8    Exoplanet diversity in the LUVOIR era

Ravi Kopparapu (UMD/NASA GSFC), Eric Hebrard (University of Exeter), Rus Belikov (NASA Ames), Natalie M. Batalha (NASA Ames), Gijs D. Mulders (LPL), Chris Stark (STScI), Dillon Teal (NASA GSFC), Shawn Domagal-Goldman (NASA GSFC), Avi Mandell (NASA GSFC), Aki Roberge (NASA GSFC)

### 13.8.1   Introduction

In the search for exoEarth candidates with LUVOIR, we will undoubtedly detect a multitude of brighter planets. According to Stark et al. (2014), for an 8-m size telescope, the number of exoEarth candidates detected is ~ 20 (see Figure 4 in Stark et al. 2014), although this is strongly dependent on the value of $\eta_{Earth}$. At the same time, the number of stars that must be observed to detect these exo-Earth candidates numbers in the hundreds. If we assume that on average, every star has a planet of some size (Cassan et al. 2012; Suzuki et al. 2016), then there are hundreds of exoplanets of all sizes that can be observed. Not considering the ~20 exoEarth candidates, the bulk of the exoplanets will fall into "non-Earth" classifications, without any distinguishing features between them. This provides a motivation to devise a scheme for classifying exoplanets based on planetary size and corresponding comparative atmospheric characteristics of exoplanets.

With some exceptions of Venus-type exoplanets (Kane et al. 2012), there has not been an overarching way to classify planets beyond the habitable zone (HZ). Classifying different size planets based on the transition/condensation of different species at different stellar fluxes (i.e., orbital distances) provides a physical motivation for estimating exoplanet mission yields, separate from exoEarth candidate yields (**Figure 13.11**).

### 13.8.2   The role of LUVOIR

The histogram plot (**Figure 13.12**) visualizes the total scientific impact of the habitable planet candidate survey, along with several

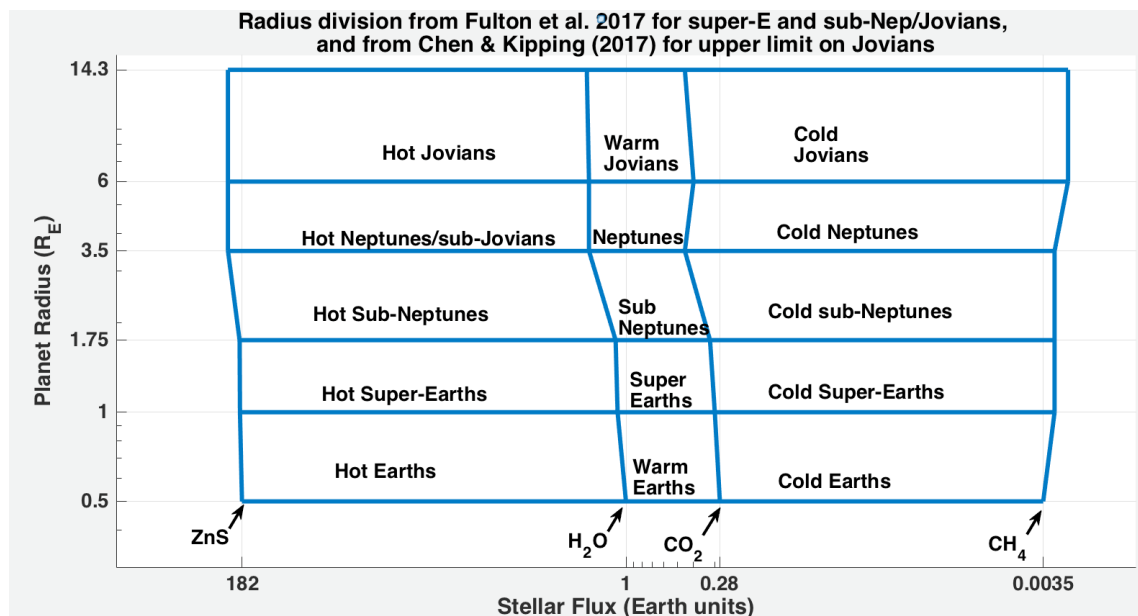

**Figure 13.11.** *The boundaries of the boxes represent where different chemical species are condensing in the atmosphere of a planet at a given stellar flux, according to equilibrium chemistry calculations Credit: Kopparapu et al. (2018)*

---

**Program at a Glance**

**Science goal:** Measure atmospheric compositions of a wide range of exoplanets with different sizes, orbits, and host stars.

**Program details:** IFS optical and near-IR spectra of a) all planets in the field during long integrations on stars with habitable zone planet candidates and b) a selected set of known planets around other stars.

**Instrument(s) + Configuration:** ECLIPS high-contrast, point-source spectroscopy

**Key observation requirements:** Spectral bandpass from 200 nm to 2000 nm, R ~ 100, Continuum SNR > 10

---

other classes of exoplanets, based on different telescope diameters. The y-axis gives the expected total numbers of exoplanets observed (yields), which are also given by the numbers above the bars. It is at the LUVOIR-type telescope sizes (~8-m and 15-m size) that one can truly see the diversity in exoplanet yields, and further characterize different classes of planets. We note that in general, larger apertures are less sensitive to changes in mission parameters than smaller apertures.

### 13.8.3   The science program

With a 4-m class mission, observations that are designed to maximize the yield of potential Earths will also yield the detection and characterization of all of the planet types discussed here, with the exception of hot Jupiters. Hot Jupiters are not observed by a 4-m class mission because the tight inner working angle, and because of the low abundances of hot Jupiters.

A 15-m telescope will bring the ability to not only observe planets, but to test the occurrence of different features within each of the planet types. It would observe dozens of each planet type, providing larger sample sizes which enables the study of each planet type as a population.

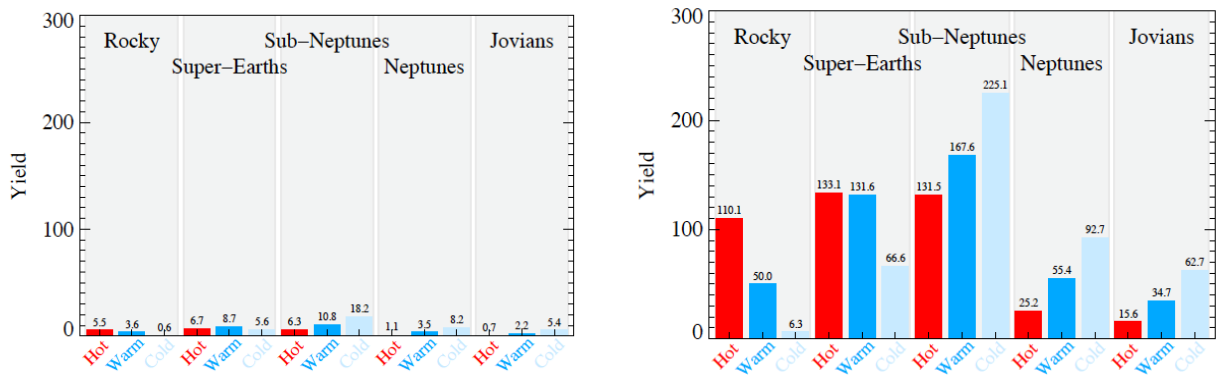

**Figure 13.12.** *Expected number of exoplanets observed for hot (red), warm (blue) and cold (ice-blue) incident stellar fluxes shown in* **Figure 13.11**. *The left panel is for a telescope size of 4-m and the right is for 15-m.*

## 13.9   Detecting exomoons with LUVOIR

Eric Agol (University of Washington, NASA Astrobiology Virtual Planetary Laboratory), Tiffany Jansen (Columbia University), Brianna Lacy (Princeton University), Tyler D. Robinson (Northern Arizona University), Victoria Meadows (University of Washington, NASA Astrobiology Virtual Planetary Laboratory)

### 13.9.1   Introduction

The detection of exomoons has proven elusive, and is becoming an increasingly prominent goal of exoplanet studies. LUVOIR can open the possibility of the detection of exomoons in the habitable zones of main sequence stars via spectroastrometry: the astrometric shift versus wavelength that occurs between wavelengths with flux dominated by the exoplanet vs. wavelengths dominated by the exomoon. The several requirements to reach this goal are: 1) a large telescope aperture; 2) astrometric calibration and stability; 3) ability to measure the centroid of the planet's light simultaneously over a broad range of wave-

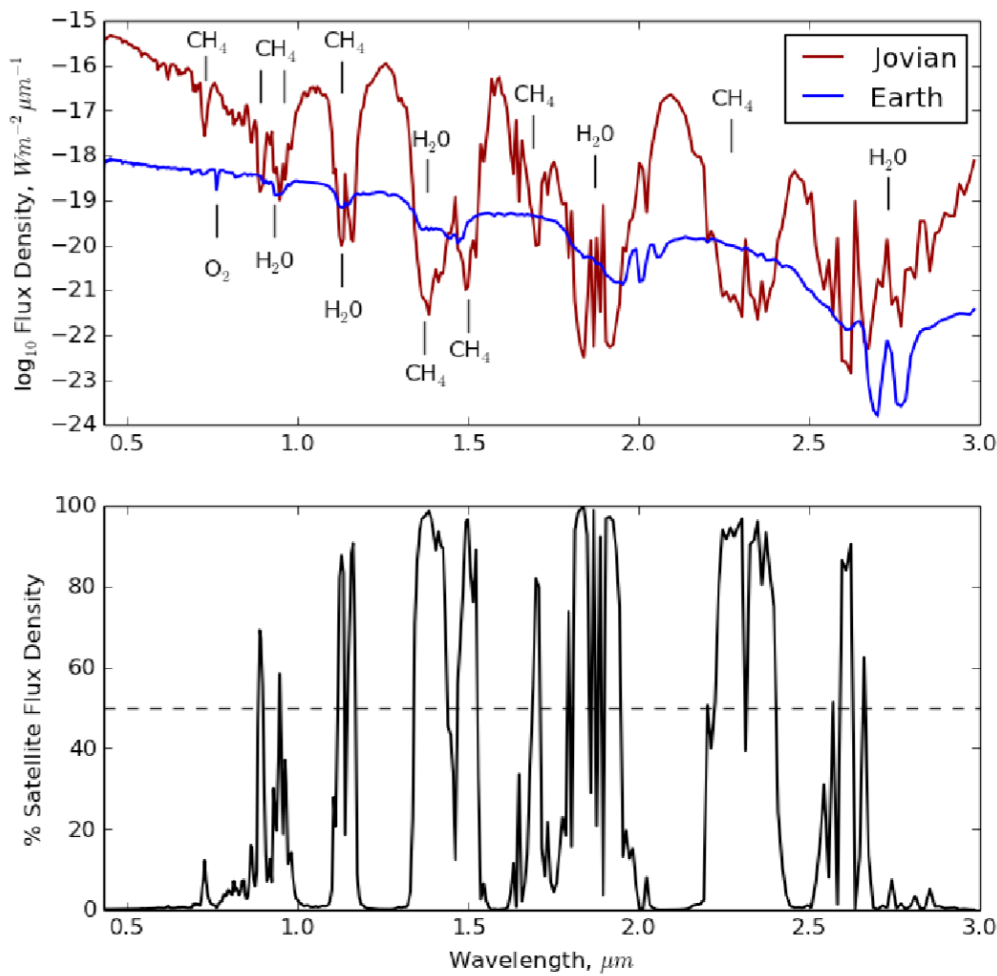

**Figure 13.13.** *The flux of the Earth-like moon and the warm Jupiter at quadrature phase angle as a function of wavelength in microns (top) and the contribution of flux due to the moon shown as a fraction of the total flux (bottom). The maximum fraction of the moon's flux for the Jovian-Earth system occurs at $\lambda$ = 1.83 $\mu$m, contributing 99.1% to the total flux.*

# Spectroastrometric detection of exomoons (Agol, Jansen, Lacy, Robinson & Meadows 2015)

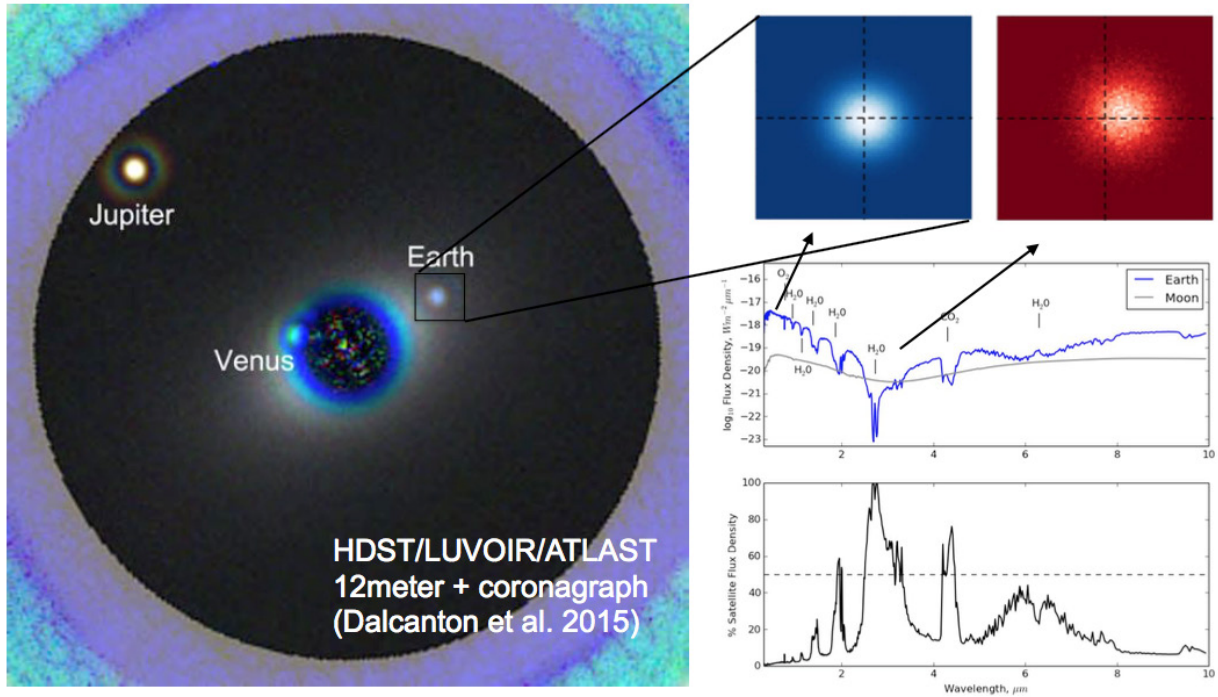

**Figure 13.14.** *Diagram indicating spectroastrometric signal.*

lengths; 4) ability to revisit the planet many times; and 4) extension to near-IR wavelengths. In addition to enabling the detection of exomoons, this technique may allow for the characterization of the planet via the orbit of the exomoon.

The Moon has likely played a critical role in the evolution of planet Earth, from influencing the geological and chemical composition (Smith 1977; Canup 2012), to modifying the mass and angular momentum of the Earth/Moon system (Canup & Asphaug 2001), to possibly stabilizing the obliquity of the Earth (Laskar et al. 1993). Lunar tides may have played an outsized role in influencing the evolution of life on Earth (Balbus 2014). Speculation that habitable exomoons could orbit giant planets makes an intriguing alternate niche for life (Williams et al. 1997;

Kaltenegger 2010). Caveats exist for each of these scenarios (Lissauer et al. 2012; Heller et al. 2014; Lammer et al. 2014), so a search for exomoons is necessary to ascertain the importance of moons on planetary physics and biology.

There are a range of proposed techniques for detection of exomoons, from transiting planet observations (Cabrera & Schneider 2007; Kipping 2009), to phase function measurements in direct imaging Moskovitz et al. (2009); Robinson (2011). Each of these techniques drawbacks: transiting exomoons may be swamped by stellar variability, while phase variations may be mimicked by variations in planetary atmospheres.

We recently proposed a new approach for the detection of exomoons that could result in an instantaneous detection of an exomoon,

and would allow for measurements of the exoplanet-moon system: spectroastrometry (Agol et al. 2015). This technique relies on the observation that a moon can outshine a planet at wavelengths where the planet is non-reflective, i.e., in strong molecular absorption bands (Williams & Knacke 2004; Moskovitz et al. 2009). The Earth is outshined by the Moon at $\approx 2.7$ µm, and the Earth would outshine a companion Jovian planet in several methane bands between 0.9–2.7 µm (**Figure 13.13**). The technique is summarized in **Figure 13.14**, which schematically shows how the centroid of the PSF may vary between wavelengths dominated by the planet versus wavelengths dominated by the moon.

The detection technique proceeds by simultaneous measurement of the centroid as a function of wavelength. The wavelengths at which the planet is dim would have a stronger contribution from the moon, and thus cause a shift of the centroid towards the position of the moon. Thus, even if the moon and planet are not resolved, the moon may still be detected via the spectroastrometric signal.

The astrometric signal, which is proportional to the angular separation of the planet and moon, scales as r/d, where r is the separation of the moon and planet on the sky, and d is the distance to the system. The astrometric noise scales as $(\lambda/D)(d/D)$, as the PSF width scales as $\lambda/D$, while the precision of the measurement of the centroid scales as d/D, where D is the diameter of the telescope. Thus, the signal-to-noise scales as $r(D/d)^2/\lambda$, and so the ability to search a large volume V for exomoons within the nearby Galaxy scales as $V \propto d^3 \propto D^3$. To increase the odds of a successful search, then, *requires the largest diameter telescope possible*.

This relation assumes that the Poisson-noise limit applies, and thus also requires: 1) high-contrast to avoid additional contribution to the noise from scattered light; 2) precise control of the PSF, the astrometric pointing, and the relative astrometric precision between wavelengths. Ideally this approach will be made more efficient with simultaneous astrometric measurements at a range of wavelengths, which might be accomplished with an IFU or an MKID device. One advantage of this technique is that only the *relative* position needs to be measured as a function of wavelength, not the absolute position. Another advantage is that the signal will vary with time as the moon orbits the planet, which makes it reproducible and difficult to confound with other systematic errors or sources of noise. This makes this technique more powerful than other proposed techniques for detecting exomoons, which may require observations at specific times or geometries (such as transit of the moon in front of the star, or transit of the moon in front of the planet), or may not have repeatability such as microlensing (Han & Han 2002). Spectroastrometry may also allow for the measurement of the mass of the planet+moon with Kepler's law, as well as a disentangling of the moon/planet spectrum (Agol et al. 2015).

### 13.9.2   The role of LUVOIR

Spectroastrometry of habitable-zone (HZ) Earths will not be feasible from ground-based telescopes as these may only probe the HZ of late M dwarf stars, for which the planets are close enough that they cannot harbor stable exomoons. A space telescope is required to achieve high contrast ($10^{-11}$–$10^{-10}$) to enable the detection of an exomoon, to avoid interference of the atmosphere with spectral features that show the sharpest variation in the spectroastrometric signal, to allow for repeated visits and long exposures, and to achieve stability of the PSF.

A large telescope will be necessary due to the strong scaling of the astrometric signal-to-noise with telescope aperture ($\propto D^2$), which will allow the search for this signal for a larger number of stars. The larger aperture allows for a sharper PSF and a larger number of photons, both of which improve the astrometric precision and allow for resolving out surface brightness due to (exo)zodi and speckles.

**Table 13.1.** *The moon and planetary parameters used in our model, where $D_{tele}$ is the telescope diameter, d is the distance from the observer, $t_{obs}$ is the duration of exposure, $\varepsilon$ is the telescope efficiency factor.*

| System | Moon radius (m) | Planet radius (m) | Planet-moon separation (m) | Orbital period (d) | $D_{tele}$ (m) | d (pc) | $t_{obs}$ (hr) | $\epsilon$ | Semi-major axis (AU) |
|---|---|---|---|---|---|---|---|---|---|
| Earth-Moon | $1.738 \times 10^6$ | $6.371 \times 10^6$ | $3.844 \times 10^8$ | 27.32 | 12 | 1.34 | 24 | 0.2 | 1.23 |
| Jovian-Earth | $6.371 \times 10^6$ | $6.991 \times 10^7$ | $3.064 \times 10^9$ | 34.60 | 12 | 10 | 24 | 0.2 | 1 |

Relative astrometric precision between bands of ≈0.1 mas should be sufficient for the expected ≈1 mas shift for the two cases examined here. For the Earth-Moon at Alpha Centauri example, near-infrared coverage to >1.5 μm will be required, while coverage to 2.7 μm would be preferred to obtain a strong Moon/Earth flux ratio. For the Earth-Jupiter case, coverage to 1 μm should allow for detection, while 2 μm will give a stronger Earth/Jupiter flux ratio. A broad range of wavelengths will need to be imaged simultaneously, which an IFU or MKID detector may allow for. The inner working angle and field of view may accommodate searches closer or further from stars, while for the two examples we explored we just examined the HZ, which is well covered by current LUVOIR specifications. A detailed technical and science case for spectroastrometry will require more realistic simulations, which are in progress.

### 13.9.3   The science program

It is impossible to forecast the properties of an ensemble of planet-moon systems, and so in lieu of this, here are two case studies that could drive technical requirements of LUVOIR: 1) an Earth-Moon twin orbiting Alpha Centauri A (1.34 pc); 2) an Earth-Jovian pair orbiting a G2V star at 10 pc with a separation of 30% of the Hill sphere. **Table 13.1** gives the properties of these two systems. One could imagine, of course, a broader range of possibilities with the Solar System as building blocks: e.g., a Mars-sized moon orbiting a Neptune-sized ice giant. However, our two case studies in some sense bracket a range of possibilities, although a true Jovian-satellite analog system would probably be challenging to detect with this technique. Spectroastrometry may be used to search for rings of planets since Saturn is dark in

---

**Program at a Glance**

**Science goal:** Detection of moons orbiting exoplanets.

**Program details:** Precisely measure the positions of directly imaged exoplanets at different wavelengths and times to look for photocenter shifts (spectroastrometry).

**Instrument(s) + configuration(s):** ECLIPS IFS spectral imaging

**Key observation requirements:** Contrast > $10^{-9}$ for Jupiter orbited by Earth-like moon; Contrast > $10^{-11}$ for Earth orbited by Moon; NIR wavelengths (> 1.0 μm for Earth-Jupiter case, > 1.5 μm for Earth-Moon case); Relative astrometric precision between wavelength bands of ≈ 0.1 mas

---

methane bands, while its ring system is still reflective. The pattern of illumination of the rings will impart a centroid offset relative to the centroid of Saturn.

The Moon-Earth analog centroid offset is ≈0.4 mas at 1.4 µm, 0.9 mas at 1.9 µm, and 2 mas at 2.7 µm. Thus, near-infrared capability is required to apply this technique to an Earth-Moon analog system. *Relative astrometric stability between wavebands at better than 80 µas would be required to make a detection exceeding 5–σ*. The planet-star contrast at these wavelengths is ≈$10^{-11}$, and so is going to be affected significantly by speckles if a contrast of $10^{-10}$ is achieved. *Thus, pushing towards a contrast of $10^{-11}$ may be necessary for detection of a Moon-like exomoon*.

The Earth-Jupiter analog (with Jupiter moved into 1 AU) has a centroid offset of 1.3 mas at 0.86 µm. A 12-meter telescope with R=80 could detect this offset at S/N=13 with a 24-hour exposure, assuming the Poisson-noise limit. The planet/star contrast at this wavelength is ≈$10^{-9}$, and so will not be as affected by speckles for a telescope design that approaches $10^{-10}$ contrast.

Our investigation has the drawback that we neglected exozodiacal light and speckles, as well as other sources of instrumental noise, so further work to create realistic simulations is needed to estimate the impact of these on the spectroastrometric signal (Jansen et al. in prep.).

## 13.10  LUVOIR for stars, stellar evolution, and the local universe

Bruce Elmegreen (IBM T.J. Watson Research Center)

### 13.10.1  Introduction

LUVOIR is a ~12-meter class UV-Vis-NIR space telescope for the 2030s and beyond. Because of the large aperture and short observing wavelength, it will have very high angular resolution, corresponding to an Airy disk of 5 x $10^{-8}$ radians for optical light (10 milliarcsec). It will also have broad wavelength coverage from 100 nm to 2.5 microns, allowing a wide range of science investigations. The field of view is proposed to be very wide, making it ideal for surveys: the High Definition Imager would have a field of view of 2' x 3' with 2.73 mas/pix at UVIS and 8.2 mas/pix at NIR. This means the image sizes will be 2.9 Gpix and 0.32 Gpix, respectively, and the file sizes 1–10 GByte per image. It will also have a multi-object spectrograph from 100 nm–850 nm, as well as other instruments. With these specifications, taken from Dalcanton et al. (2015) and Elmegreen et al. (2017), LUVOIR will revolutionize the study of stars, stellar evolution, and the local universe.

### 13.10.2  The role of LUVOIR

The accelerating progress of our view of the sky over the last two millennia—the time since systematic astronomical studies and star catalogs began—is shown schematically in **Figure 13.15**, which plots the number of distinguishable pixels in the sky versus time. The first major jump from 1 arcmin resolution with the eye to 1 arcsec resolution with a telescope started in early 1600 with a long gap before the Hubble Space Telescope improved the resolution further by factor of 10. The next jump to LUVOIR comes rather quickly on this scale, showing another factor of 10 gain with a 2.6-year doubling time in the most recent era. This pace is comparable to the advance in general technology, reflecting a combination of what's possible and what gains we can expect. From this long-term viewpoint, LUVOIR will improve our resolution of the sky by the same factor over HST as HST improved it over Galileo's first telescope. Some of the science gains from this improvement will be discussed in the next section.

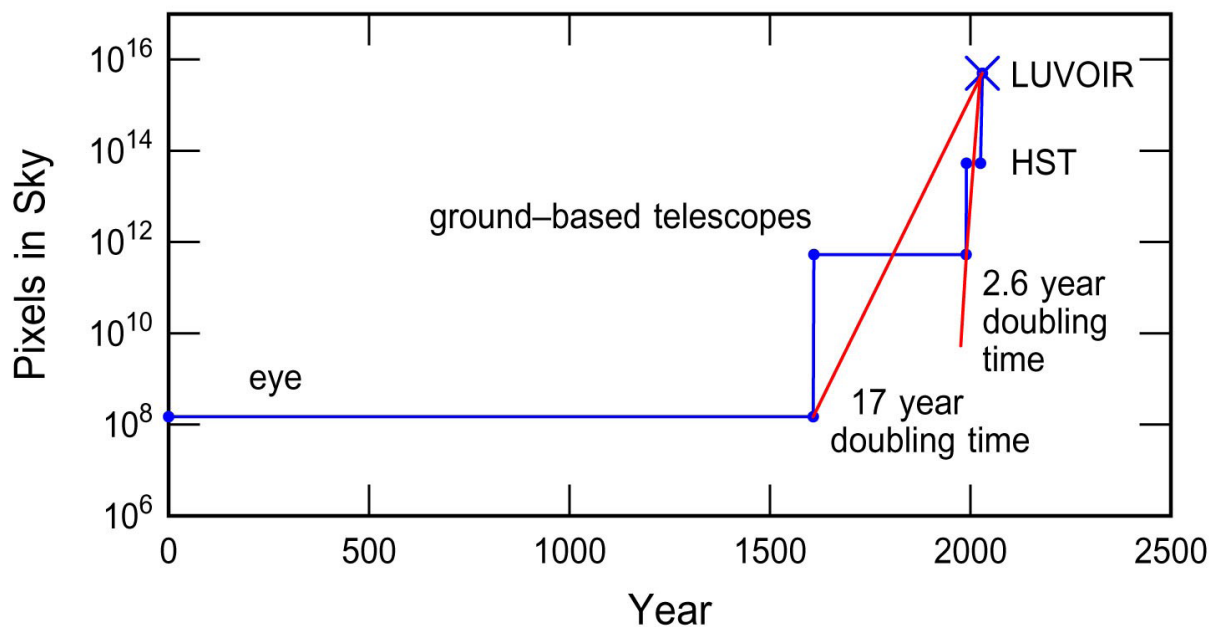

**Figure 13.15.** *The total number of pixels in the sky, calculated as the ratio of the sky solid angle to the angular resolution of the instrument, is plotted versus the year in history. Three major technology jumps have occurred, the invention of telescopes, the placement of telescopes in space, and the ability to have large telescopes in space. The first two each revolutionized astronomy, and the third is likely to as well.*

### 13.10.3 The science program

**Local star formation**

Star formation in the solar neighborhood will be resolved at 1–10 AU scales, showing dusty accretion to protostars, planet-forming regions of young disks, central jets, and close binary stars in mutual formation. The Orion protoplanetary disks discovered with HST—the so-called proplyds (O'dell et al. 1993)—will be observable at 5 AU resolution where their interaction with the hot radiation and winds from nearby massive stars will be visible. Interactions like these could have introduced radioisotopes from supernovae into the early Solar Nebula, giving our meteorites their Al-Mg anomalies (MacPherson & Boss 2011). The dusty structures of the Taurus filaments, which are regions of low-mass star formation, will be observed at 1.4 AU resolution, which is $10^{-4}$ times their size. This will easily resolve the transition from supersonic to subsonic turbulence and reveal for the first time the thermal, turbulent and gravitational processes that form low mass stars and their disks.

**Imaging stellar surfaces**

LUVOIR will be able to resolve and image some dozen nearby stars (Ochsenbein & Halbwachs 1982; van Belle et al. 1996) to look for star spots, limb darkening, rotation, flares and other possible features in optical and UV light. For example, Betelgeuse has been resolved with optical interferometry

(Haubois 2010), Antares was resolved with the VLT1 interferometer (Ohnaka et al. 2013) and Mira was resolved with HST (Karovska 1997). The LUVOIR resolution of 10 mas corresponds to a solar radius at a distance of 0.45 pc. Larger stars like supergiants will be resolvable to proportionally larger distances.

**Local group galaxies**

The Local Group of galaxies, including M31, M33, the LMC, SMC, and many smaller galaxies, will be resolved at 0.05 pc, allowing us to study hundreds of supernova remnants and planetary nebula in great detail. From the structures and line emission of these nebulae, we will learn about the final stages in stellar lives, nucleosynthetic element dispersal in the surrounding gas, and dust formation in dense remnant winds.

The high resolution of LUVOIR will allow us to distinguish between individual stars as well. According to the calculation of integration time versus point source absolute magnitude on page 49 of Dalcanton et al. (2015), in 1 hour we can see a solar type star at a distance of 2.8 Mpc using the F555W filter at an apparent AB magnitude of 32. Within 2.8 Mpc there are 9 large galaxies, i.e., with absolute magnitude less than -16, and about 60 small galaxies. At the stellar surface density in the solar neighborhood, which is about 60 solar masses per $pc^2$, there are 150 stars per $pc^2$ larger than 0.01 solar masses and 10 stars $pc^{-2}$ larger than 1 solar mass (using the Kroupa 2001 initial stellar mass function). The mean projected separations between these stars are 6 mas and 24 mas at 2.8 Mpc. The second number is larger than the angular resolution of LUVOIR, so we will be able to separate solar mass stars in galaxy disks within 2.8 Mpc, i.e., for dozens of galaxies of various types and for the highly diverse conditions in these galaxies. By counting these stars and the more massive stars, which are separated by even larger distances, we can determine the field star mass function and approximate star formation history in all regions with fairly low extinction. Variations in this field star mass function for a given history would suggest comparable variations in the stellar initial mass function. This will be the first time that the stellar IMF can be directly compared to environmental conditions and star formation rates. Such comparisons should give us greater understanding of the origin of the IMF.

## Nuclear regions of nearby galaxies

LUVOIR will resolve nearby nuclear star clusters, disks and black hole activity with 1 pc resolution or better out to 20 Mpc, which includes thousands of galaxies. NGC 300, for example, has a nuclear star cluster with a 3-pc radius (van der Marel et al. 2007) that corresponds to 300 mas—easily resolved by LUVOIR. In M31, the nuclear region has a 108 solar mass black hole with intriguing red and blue structures nearby, as observed by HST (Lauer et al. 2012). The active galactic nucleus in NGC 4261 has an optical disk 1.7 arcseconds across (Ferrarese et al. 1996), which is 170 resolution elements. The implications are enormous for studies of black hole accretion, nuclear disk storms, small-scale flaring activity, nuclear spirals and torques.

## Proper motions with sub-pixel accuracy

Proper motions with HST could reach 0.02 pixel accuracy (Anderson & King 2000, 2003) with simultaneous fitting of the average stellar point spread function and the stellar positions. 0.02 pixel for LUVOIR equals $10^{-9}$ radians or 0.21 mas. This means that with a 5-year baseline, the proper motion that can be measured, in km/s, equals 0.2 times the distance in kpc. With this we can observe a wide variety of interesting and important motions for the first time: the random motions of dust features in local star forming regions at the sonic speed, 0.1 km/s, the expansion of supernova remnants, planetary nebulae and "pillars of creation" at 1 km/s, the internal motions of Milky Way globular clusters to within 2 km/s, the rotation of the LMC and SMC to within 10 km/s, and the rotation of M31 within 140 km/s. For example, proper motions have already been measured in the LMC (Kallivayalil et al. 2013) although with less precision. This is a completely new capability to study the dynamics of nearby dust and emission line structures, and of stars and stellar systems, especially in regions where the Gaia satellite will not be able to distinguish individual stars.

## Extensions to high redshifts

As noted in the LUVOIR technical summaries (Dalcanton et al. 2015), the angular resolution will be about 100 pc or better for all redshifts. This is equivalent to ground-based resolution (1 arcsec) for all galaxies within 20 Mpc, which is essentially the Hubble Atlas of Galaxies (Sandage 1984). Deep images at high redshift will cover the same co-moving volume as the Sloan Digital Sky Survey with the same or better resolution (Dalcanton et al. 2015). As a result, we will be able to see for the first time at high redshift the substructure of star formation clumps in young galaxies, we will resolve young bars and bulges, see multiple nuclei from recent mergers, map spiral arms, and discover dwarf galaxy companions. The radial profiles will be resolved too, allowing some understanding of the origin of the exponential radial structure in disks. Decades of research on local galaxies observed from the ground will suddenly have a counterpart in the epoch of galaxy formation.

## Multi-object spectroscopy

LUVOIR will observe spectral line emission from low-lying states of atoms, which are primarily in the UV and unobservable from the ground. Such observations were a primary driver of going into space in the first place as mentioned by Spitzer (1946) in his "Report to Project Rand: Astronomical Advantages of an Extra-Terrestrial Observatory." Most atoms are in these ground states and both absorption and emission lines are in the UV or FUV. Dalcanton et al. (2015) have a diagram of the observable wavelengths for many important lines as a function of lookback time in the Universe. Most low-level ion transitions cannot be observed from the ground after a lookback time of some 10 Gyr, which means these lines are invisible for most of the Universe unless they are observed from space.

With FUV emission lines, we can see interstellar and active galactic nuclei emission from $10^6$ K gas at redshifts greater than 0.3 and compare it to the observable x-ray gas at the same temperature. We can observe circumgalactic (CGM) emission in Lyman alpha, OVI, and CIV faster than on the ground by factors of 10 to 100 because of the lower levels of background light in space (Dalcanton et al. 2015). An example of a Lyman alpha blob in space is shown in Cai et al. (2017).

LUVOIR will also observe absorption lines from ground states in the FUV. We expect 100 times as many background galaxies in a deep field at redshifts less than 1 to 2 for observations of foreground CGM absorption, and 100 times as many QSOs (Dalcanton et al. 2015). This will allow us to map the CGM around galaxies, trace element formation, outflows and inflows and see galaxies building up and quenching by outflows for the first time.

Extensive observations of the CGM in both emission and absorption will revolutionize our understanding of galaxy formation and evolution, just as most of the other new capabilities mentioned here will be revolutionary in their own fields. Not only will we see more and smaller galaxies at high redshift, but we will also start to fill in the space between the galaxies.

**Figure 13.16** shows the impact of this change on another plot with a historical perspective. This is the sky-covering fraction of known astronomical objects as a function of year. For all time before the invention of

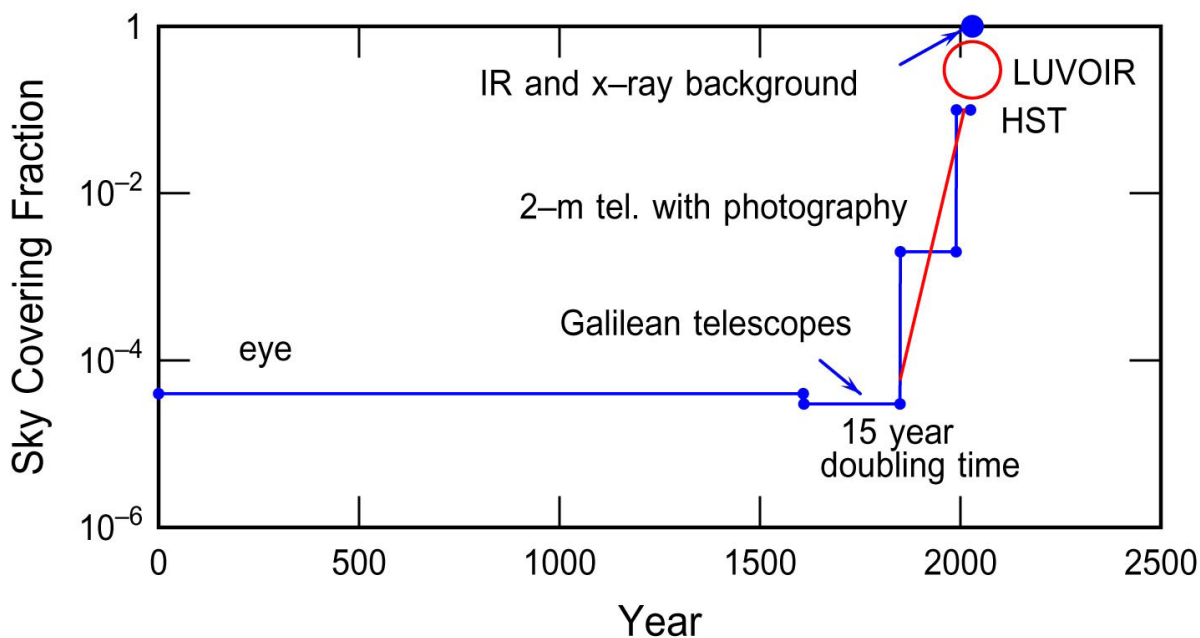

**Figure 13.16.** *The fraction of the sky covered with known astronomical objects is plotted versus the year in history. As telescopes reach fainter object and deeper fields, more and more objects become known. LUVOIR will add to this progression by mapping the circumgalactic gas.*

---

### Program at a Glance

**Science goal:** Study local star formation, the stellar initial mass function, and nuclear regions in local galaxies, image stellar surfaces to explore activity, probe dynamics of nearby dust structures, determine properties of young galaxies, and map the circumgalactic medium.

**Program details:** High-resolution imaging and multi-object spectroscopy of local stars, dust and galaxies out to high redshift, young galaxies.

**Instrument(s) + configuration(s):** HDI imaging, multi-object spectroscopy

**Key observation requirements:** Resolution < 10 mas

---

telescopes, humans could only see some 5000 stars at 1 arcmin resolution. The fraction of the sky covered by these light sources was very small, letting us think that most of the Universe was empty space. When the telescope was invented, the sky covering fraction got even smaller because, although many more stars could be seen, each was 60 times smaller at 1 arcsecond resolution, so more blank space could be seen as well. However, with larger telescopes to collect fainter light (in William Herschel's era), and with photographic images that allowed collection of this faint light over long periods of time (thousands of times longer than the eye could integrate), the sky started to be seen as covered with faint galaxies and nebulae, filling in the blank space and increasing the covering fraction by a factor of 100. HST increased it by another factor of 100 by finding faint galaxies at high redshift between all the nearby galaxies and stars, covering deep fields at nearly the 10% level. LUVOIR will make the next jump with its ability to see gas between the galaxies, and its possible first-time resolution of unknown objects that have appeared up to now to be only a smooth background. LUVOIR will literally fill the sky with discoveries!

My thanks to Aki Roberge, Daniela Calzetti, and Bradley Peterson for their invitation to speak on this topic at the AAS Winter Meeting 2018. Many of the numbers and ideas presented here are from the comprehensive summaries by Dalcanton et al. (2015) and Elmegreen et al. (2017).

## 13.11 UV characterization of exoplanet host stars: keys to atmospheric photochemistry and evolution

Kevin France (University of Colorado at Boulder)

### 13.11.1 Introduction

The planetary effective surface temperature alone is insufficient to accurately interpret biosignature gases when they are observed with LUVOIR, particularly for planets orbiting low-mass stars (K and M dwarfs). The UV stellar spectrum drives and regulates the upper atmospheric heating and chemistry on Earth-like planets, and it is critical to the definition and interpretation of oxygen species and other biosignature gases (e.g., Seager et al. 2013). As discussed in **Chapter 3**, the specifics of the stellar spectrum may produce false-positives in our search for biologic activity (Hu et al. 2012; Tian et al. 2014; Domagal-Goldman et al. 2014; Harman et al. 2015).

### 13.11.2 The role of LUVOIR

The chemistry of important potential biomarker molecules in the atmosphere of an Earth-like planet depends sensitively on the strength and shape of the host star's UV spectrum. $H_2O$, $CH_4$, and $CO_2$ are sensitive to far-UV radiation (FUV; 100–175 nm), while the atmospheric oxygen chemistry is driven by a combination of FUV and near-UV (NUV; 175–320 nm) radiation (**Figure 13.17**). Additionally, the temporal variability in UV emission lines can be an indicator of strong charged particle release (e.g., Youngblood et al. 2017). Energetic particle deposition into the atmosphere of an Earth-like planet during a large M dwarf flare can lead to significant atmospheric $O_3$ depletions (> 90% for large flares; Segura et al. 2010). This alters the atmospheric chemistry and increases the penetration depth of UV photons that could potentially sterilize (or catalyze) surface life. Given that particle fluxes are not (typically) directly measured for stars other than the Sun, UV observations offer the best estimates of these important particle environments.

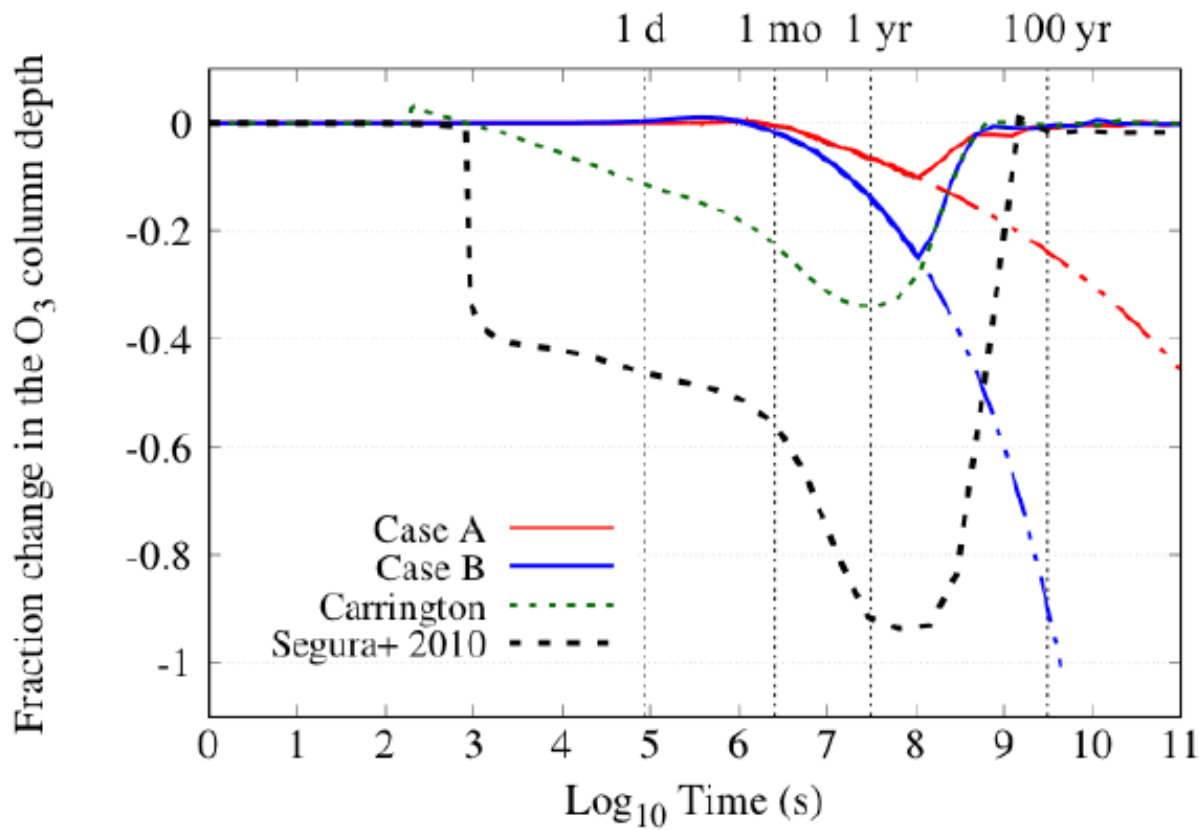

**Figure 13.17.** *Impact of FUV-derived particle impact on long-term ozone depletion (Youngblood et al. 2017, Tilley et al. 2018.) Red and blue curves are particle fluences constrained by the FUV data.*

The high sensitivity and temporal resolution of the LUVOIR-LUMOS spectrograph will enable a thorough characterization of the UV host star spectrum for every potentially inhabited star observed by LUVOIR. At present, there are no other UV-capable facilities projected to be available to the astronomical community in the 2030s, making LUVOIR's contribution essential for this important host star characterization. Furthermore, the high sensitivity enabled by the large aperture and high efficiency of LUMOS (France et al. 2017) mean that regular monitoring of these systems will be possible during exoplanet characterization campaigns with minimum impact to the primary exoplanet observations.

### 13.11.3 The science program

This study requires the acquisition of a UV spectrum for each LUVOIR target with a habitable zone planet showing atmospheric spectral signatures. The UV spectrum is

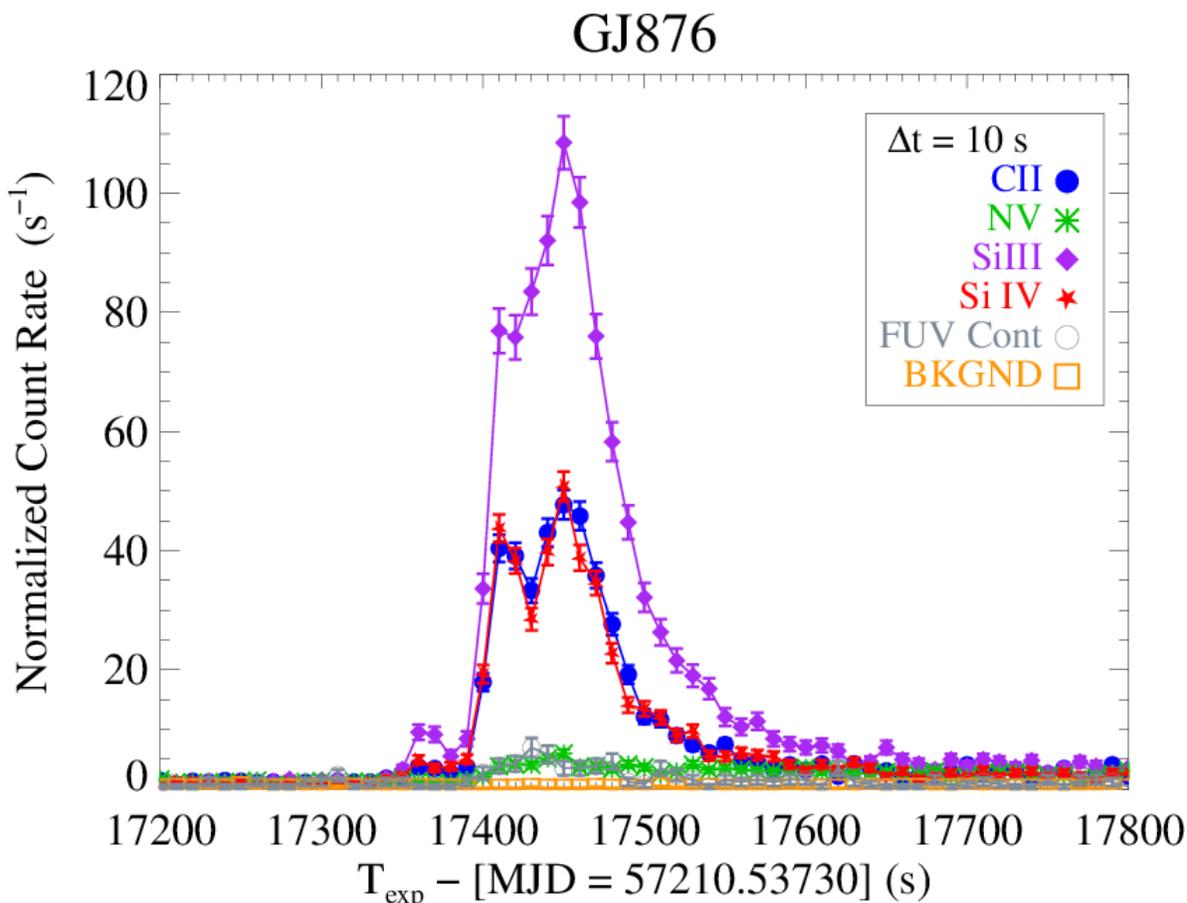

**Figure 13.18.** *FUV flare from the local M dwarf GJ 876 with an effective HZ strength equivalent of an X38-class solar flare (France et al. 2016; Youngblood et al. 2017).*

---

### Program at a Glance

**Science goal:** Characterization of the exoplanet host star's high-energy irradiance enabling detailed photochemical and atmospheric evolution modeling.

**Program details:** FUV and NUV time-resolved spectroscopy of the host star of every exoplanetary system targeted by LUVOIR.

**Instrument(s) + configuration(s):** LUMOS medium-resolution (R = 30,000), single-object spectroscopy

**Key observation requirements:** S/N of 20 per spectral resolution element

---

then essential to accurately model and interpret those features. This is particularly true for M and K dwarf host stars, where stellar atmosphere models still require a comprehensive panchromatic data set to optimize the temperature-pressure profile of the star's chromosphere, transition region, and corona (e.g., Fontenla et al. 2016). In practice, a comprehensive study of low-mass stars is required to understand the full parameter-space of age, mass, and metallicity of the system. This requires a sample of ~400 stars, including the LUVOIR Earth-like planet candidates, and a range of stellar ages that will likely require going beyond LUVOIR's direct imaging survey volume.

As most of these objects are well distributed on the sky, this will be primarily an object-by-object survey of F-M stars. We will require spectral coverage across the major UV and optical diagnostics for chromospheric (~$10^4$ K), transition region (~$10^5$ K), and coronal (~$10^6$ K) gas. This includes lines of O VI (103.2 nm; **Figure 13.17**), Fe XIX (111.9 nm), Lyman-alpha (121.6 nm), Fe XII (124.2 nm), NV (123.8 nm), O I (130.4 nm), C II (133.5 nm), Fe XXI (135.4 nm), Si IV & O IV]

(140 nm), C IV (154.8 nm), He II (164.0nm), C I (165.7 nm), Fe II (240 and 260 nm), Mg II (280 nm), Ca II (394 nm), and H-alpha (656.3 nm).

Time-resolution of 1 second is required because the characteristic time scale for strong stellar/solar flares is ~1–5 minutes, and 1 second allows us to resolve flare lightcurves – important for inferring properties of CMEs and to understand the duration of energy deposition into the orbiting planet's atmosphere. This requires a photon-counting detector.

---

## 13.12 White dwarfs as probes of fundamental astrophysics

Martin Barstow (University of Leicester) & Boris Gänsicke (University of Warwick)

### 13.12.1 Introduction

White dwarfs (WDs) are the remnants of all stars with initial masses less than 8 $M_\odot$, and they provide important laboratories for the study of stellar evolutionary processes and the behavior of matter at extremes of temperature and density. As some of the oldest objects in the Galaxy they are useful cosmological clocks, placing strong limits on the ages of globular clusters and disk populations. They are implicated in the production of Type Ia supernovae, on which the cosmological distance scales and the existence of dark energy are predicated, even though the precise mechanism(s) remain unresolved.

In the strong gravitational fields associated with white dwarfs, it is predicted that their atmospheres should be pure H or He (depending on the prior evolution), devoid of heavier elements, which sink out of the surface layers. However, many studies

(e.g., Barstow et al. 1993, 2003) have demonstrated that white dwarf atmospheres containing metals are ubiquitous (**Figure 13.19**). While the presence of this material was initially attributed to the effect of radiative levitation, this mechanism was unable to explain the detailed abundances or the presence of metals in cool white dwarfs, where the radiative effects are negligible. It is now evident that many white dwarfs are accreting material from extrasolar planetary debris (e.g., Jura et al. 2009, 2012; Gänsicke et al. 2012; Barstow et al. 2014). Consequently, the study of white dwarf atmospheres provides a unique opportunity to determine the composition of these bodies.

It has been shown in theories of quantum gravity that fundamental constants, such as the fine structure constant ($\alpha$) and the electron/proton mass ratio ($\mu$) can vary in the presence of a strong gravitational field.

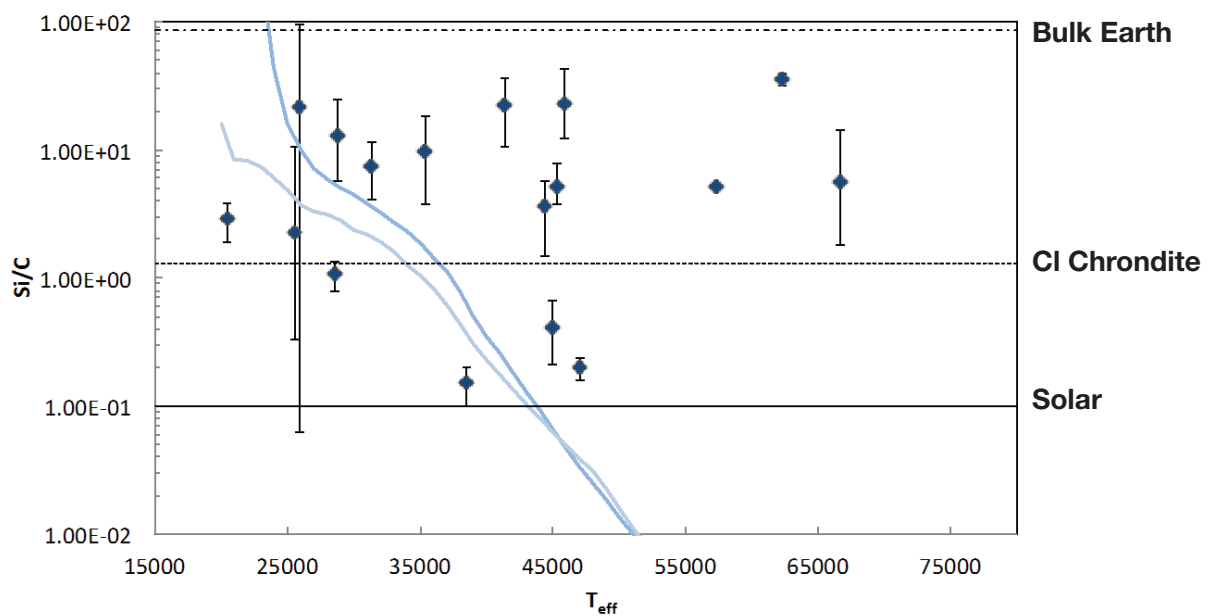

**Figure 13.19.** *Si to C ratio for a sample of 17 white dwarfs observed in the far-UV by FUSE. Most stars show evidence for accretion of rocky material.*

Such variations are expected to manifest as small shifts in the wavelengths of atomic and molecular transitions. With UV spectra containing the absorption lines of many such transitions, white dwarfs have been used to study the potential effects of these variations (Berengut et al. 2013; Bagdonaite et al. 2014) but the work is limited to a few of the very brightest white dwarfs.

### 13.12.2 The role of LUVOIR

White dwarfs have been studied in the UV for around 40 years, initially by IUE and then HST. While this work has yielded many important and exciting scientific results, it has generally been limited to a small and heavily biased sample. White dwarfs are Earth-sized, and hence intrinsically faint, and even with the improved throughput of COS, high-resolution ultraviolet spectroscopy can only be obtained for either very nearby (<20 pc) or young (<100 Myr) and therefore hot white dwarfs, severely limiting our understanding of the underlying physics, and the wider diagnostic power of these stellar remnants. LUVOIR will provide a unique opportunity to address these shortcomings. The larger aperture will enable observations of fainter objects, increasing the accessible sample, but also yield reduced data collection times, making the collection of suitable stellar samples more efficient. In addition, the dramatic improvement in the diffraction limited resolution enabled by the larger aperture, coupled with coronagraphic capability for the most extreme luminosity ratios, will open up the possibility of resolving binary systems with smaller separations and/or at greater distances.

### 13.12.3 The science program

Knowledge of the WD ages is important in measuring the ages of stellar populations. However, such results depend on a thorough understanding of the evolution of WDs themselves and on predictions of the cooling rates. In turn, the masses, radii, and photospheric compositions affect these rates. Studies of WDs in binary systems potentially represent a direct test of the evolutionary models and the mass-radius relation, since the WD mass can be independently determined from the orbital and physical elements of the system. In practice, however, few such systems have been available to be studied in sufficient detail to make these comparisons. An important science goal is to build up statistically significant samples of binary systems where orbits can be determined and follow-up spectroscopy carried out.

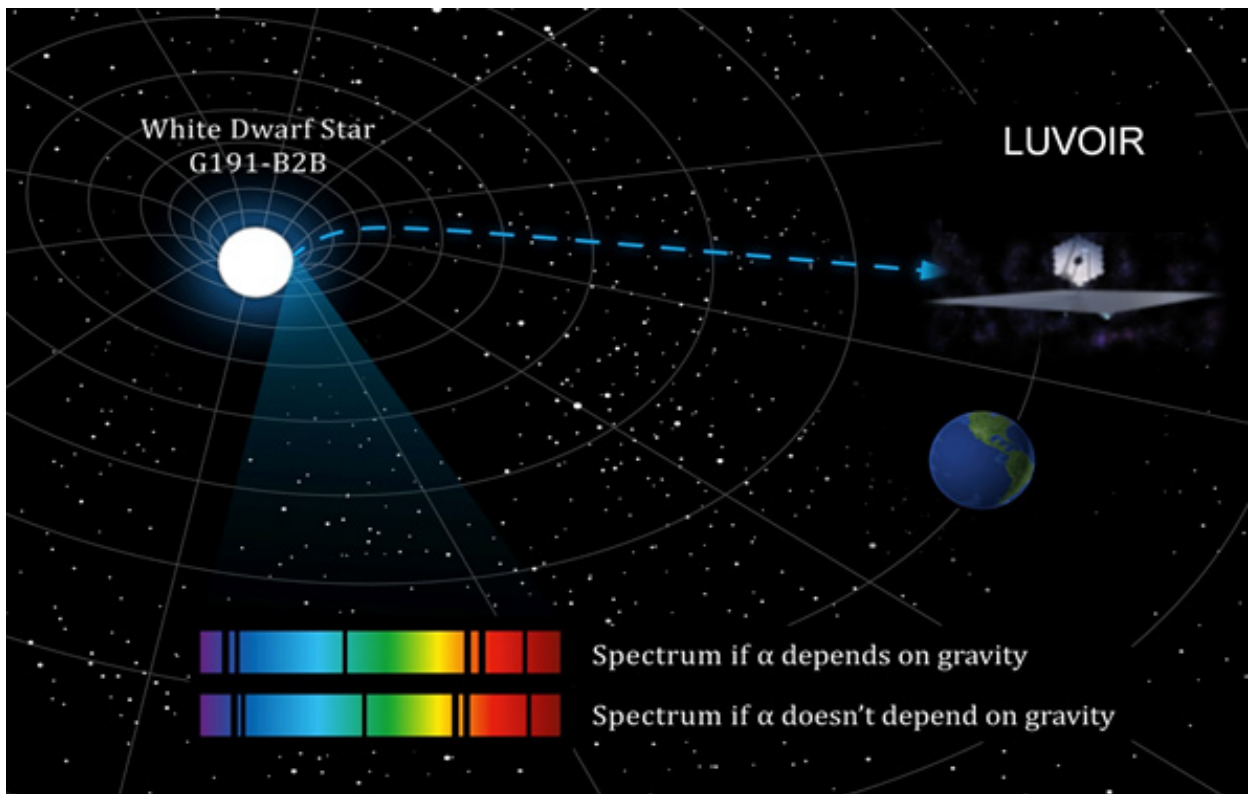

**Figure 13.20.** *Schematic illustration of the effect of a dependence of the fine structure constant on gravity (image courtesy of Julian Berengut).*

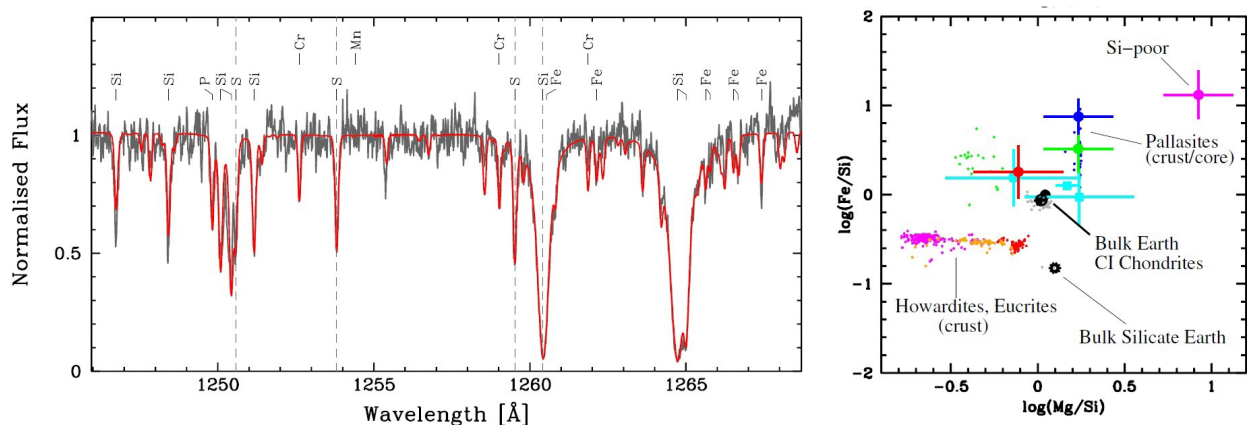

**Figure 13.21.** *The bulk abundances of disrupted exo-planetary bodies measured from far-ultraviolet spectroscopy of white dwarfs (left) are consistent with rocky parent bodies (right: large dots = white dwarfs, small dots = solar system meteorites).*

Priorities in exoplanet research are rapidly moving from finding planets to characterizing their physical properties. Of key importance is their chemical composition, which feeds back into our understanding of planet formation. Mass and radius measurements of transiting planets yield bulk densities, from which interior structures and compositions can be deduced (Valencia et al. 2010). However, those results are model-dependent and subject to degeneracies (Rogers & Seager 2010; Dorn et al. 2015). Transmission spectroscopy can provide insight into the atmospheric compositions (Sing et al. 2013;

Deming et al. 2013), though cloud decks detected on a number of super earths systematically limit the use of this method (Kreidberg et al. 2014). For the foreseeable future, far-ultraviolet spectroscopy of white dwarfs accreting planetary debris remains the only way to directly and accurately measure the bulk abundances of exoplanetary bodies. Significant progress will be made through the acquisition of a large sample of high-resolution UV spectra to provide these measurements.

Observing potential variations in the fine structure constant in white dwarf spectra is very challenging, requiring extremely high S/N and deep understanding of systematic wavelength calibration effects. Statistically, there is also benefit in observing a significant sample of objects to compare results between them. Any observed effect should be reproduced in stars of similar gravity. Furthermore, extending the sample to the extreme range of white dwarf gravities allows exploration of the dependence of $\alpha$ & $\mu$ on gravity, or at least places important limits, which can constrain the possible range of theories.

A large UVOIR telescope in the 10 to 16-m aperture range will enable high S/N observations of several thousand white

<div style="border:1px solid">

**Program at a Glance**

**Science goal:** Understanding the physical properties of white dwarfs, their evolution and their composition—leading to new insights into fundamental physics and the composition of rocky extra-solar planets.

**Program details:** High resolution UV spectra, some with polarimetry; High resolution imaging; coronagraphic imaging for a subset of systems

**Instrument(s) + configuration(s):** LUMOS/HDI/ONIRS/POLLUX

**Key observation requirements:** 25:1 in spectra for most surveys, but 100:1 + for fine structure constant studies; diffraction limited imaging at v ~ 20

</div>

dwarfs, increasing potential sample sizes for the above programs by 1 to 2 orders of magnitude. The length of typical exposures (approx. 0.1 hour) will likely be small compared to observational overheads. Therefore, attention will need to be paid to minimizing target acquisition and readout times and optimizing the pattern of slews between targets to achieve a high observing efficiency so that projects requiring large numbers of short exposure are not too costly.

## 13.13 Small bodies of the inner solar system

Andrew S. Rivkin (Johns Hopkins University Applied Physics Laboratory), Geronimo L. Villanueva (NASA Goddard Space Flight Center)

### 13.13.1 Introduction

The term "asteroid" covers a wide range of compositions from metal to rock to primordial mixtures of ice, organics, and silicates, and sizes ranging from planetary embryos to objects that can fit comfortably inside a grad student office. The range of research addressing the asteroids is similarly broad, and LUVOIR particularly promises to allow major advances due to its high spatial resolution, high-contrast imaging, and sensitivity.

1) Dawn's visits to Ceres and Vesta demonstrate that regional and local variations occur on large asteroids. There are over 200 asteroids with diameters >100 km, and roughly 30 with diameters >200 km.

We want to understand the homogeneity/heterogeneity of large asteroid surfaces.

2) Ongoing activity on Ceres is a matter of ongoing debate. Main belt comets are known, but we have no data sensitive enough to detect ice/water/sublimation.

3) Binary systems are commonly found on NEOs with radar, but they are difficult to characterize. The Ida/Dactyl system was discovered by Galileo flyby, but unobservable from current Earth-based systems (or JWST). Study of these systems is important to understand collisional evolution.

4) Positional measurements of very high precision are of potential use for certain objects like hazardous NEOs, or small asteroids that are going to make close

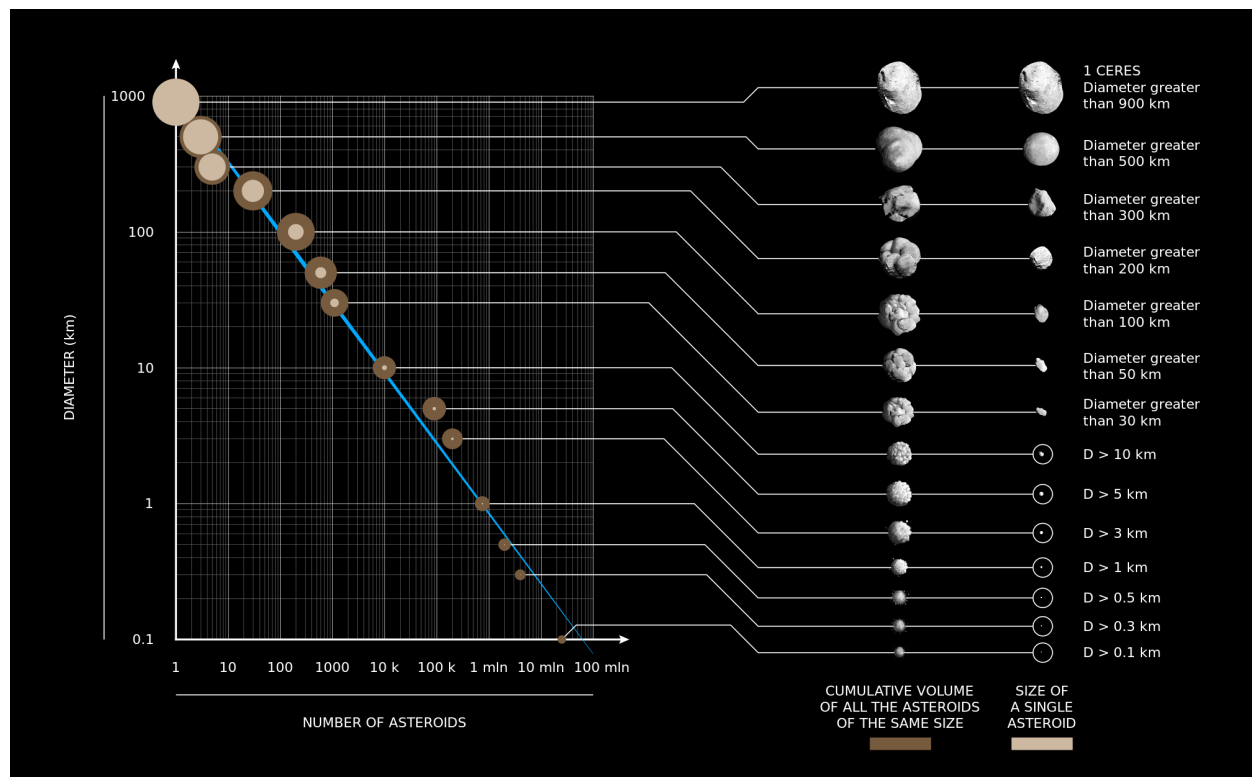

**Figure 13.22.** *The asteroids of the Solar System, categorized by size and number. Credit: M. Colombo (DensityDesign Research Lab).*

passes to large asteroids (so mass can be determined).

### 13.13.2 The role of LUVOIR

LUVOIR will permit orders of magnitude greater characterization of asteroids than what is currently done, with the detection capability up to 50 meters bodies (millions of asteroids, see **Figure 13.22**). Specifically, LUVOIR can go ~4 times deeper in ~10% of the time that HST requires, and it will be much more sensitive than any of the ELTs for this kind of observation (mainly due to reduced background brightness).

### 13.13.3 The science program

**Imaging**

- FOV: Primarily useful for recovery of lost objects or discovery surveys, neither of which is likely to be a driver for LUVOIR observations. Observations of impact ejecta or active asteroid comae/tails likely to need similar or smaller field-of-view as cometary observations.

- Filters/Wavelength coverage: Filters used during the Eight Color Asteroid Survey (ECAS) of the 1980s covered wavelengths from 300–1100 nm and are still in use on some spacecraft today. Nevertheless, their exact placement could be revisited. Longer wavelength regions include important absorptions due to silicates (~2000 nm) and clays (~2200–2300 nm). In addition, the ability to use a neutral density filter could be useful—several targets of interest have been unobservable by earlier space observatories because they are too bright. Objects of potential interest range can be as bright as V~7.5. The faint limit for discovered and cataloged objects is currently at V~21–22, but may be pushed fainter with new surveys.

- Spatial Resolution: A spatial resolution of 0.01" would allow ~10 km resolution in the middle of the asteroid belt and ~15 km resolution at its far edge. This resolution would resolve the largest 200 asteroids to have ~100 pixels or more, and allow the satellites of Mars and the very largest Trojan asteroids to be similarly resolved. For binary system studies, Ida/Dactyl

---

### Program at a Glance

**Science goal:** Characterization of the smallest bodies in the asteroid belt (> 50 m), ensuring the early detection of potentially hazardous objects (planetary defense) and opening a new window into the evolution of our planet and the inner solar system.

**Program details:** Observations of asteroids may include targeted imaging of large objects to study surface features, targeted spectroscopy of objects to obtain measurements not obtainable from Earth due to wavelength or SNR requirements, targeted astrometric measurements of near-Earth objects, or parallel observations of objects in the field during measurements of astrophysical targets.

**Instrument(s) + configuration(s):** HDI and LUMOS appear likely to be the instruments of main utility for asteroid studies, although the multi-object capability of LUMOS is not likely to be used for asteroids.

**Key observation requirements:** Absorption bands on asteroids typically are 10% or less in band depth in the wavelengths in question. Objects of possible interest range from V < 10 to V > 20, potentially reaching arbitrary faintness (for instance, a small NEO observed near aphelion). Tracking rates of objects of interest will range from near-sidereal to > 100 arcsec per hour.

---

can be used as an example: They are separated by roughly 70 milliarcseconds, and have a magnitude difference of 6.7 magnitudes.

- Depth: Albedo variations on objects are typically a few percent. Detection of sublimation from active asteroids will have requirements similar to comets at a similar solar distance.

- Comments: Irregular satellites may also be lumped in with asteroids. The giant planets have dozens of irregular satellites that are thought to be captured asteroids. For many of them no physical properties are available.

**Spectroscopy**

- Wavelength coverage: Major asteroid types typically have absorptions near 1 and 2 μm or else a shallow absorption centered near 700 nm when considering wavelengths shortward of 2.5 μm. Very little asteroid data exists shortward of 400 nm or so, but HST and spacecraft observations of Ceres and Lutetia show evidence of features at wavelengths as short as 110–120 nm.

- Resolution: Spectral resolutions are typically low for asteroid observations, as absorptions are typically very broad. A low spectral resolution mode (R ~ 200 to 250?) will be of use.

- Multiplexing/IFU: possible, but difficult to co-locate several dynamical objects in a single FOV.

- Depth: Band depths in the 1–2 μm region are typically a few percent.

## 13.14 Resolved photometry of young super star clusters

Søren S. Larsen (Department of Astrophysics/IMAPP, Radboud University, Nijmegen, The Netherlands)

### 13.14.1 Introduction

Star clusters were traditionally viewed as excellent examples of "simple stellar populations"—consisting of stars with a single age and a single chemical composition. While this still appears to hold true for low-mass *open* clusters, it is now clear that massive *globular* clusters that inhabit the Galactic halo are far more complex systems. Color-magnitude diagrams (CMDs) from the Hubble Space Telescope show *multiple populations* that reveal themselves through parallel main sequences, split red giant branches, and other features not reproduced by standard models for stellar evolution (Gratton et al. 2012; Bastian & Lardo 2017).

Two key capabilities of HST have been crucial in uncovering the variety of this phenomenon in old GCs: 1) from space, it is possible to achieve exquisite photometric accuracy that is very hard to reach from the ground, especially in crowded environments; 2) space-based observations provide access to important spectral features in the UV (CH, CN, and NH molecular bands) that are sensitive to the light-element abundance variations that trace the multiple populations.

The Magellanic Clouds provide the closest examples of *young* star clusters with masses approaching those of ancient GCs. HST imaging has revealed a surprising complexity of the CMDs in these clusters, too, with extended main sequence turn-offs, parallel young main sequences, and other puzzling features (Milone et al. 2016). It is unclear, however, to what extent these phenomena are related to the multiple populations observed in old GCs.

The stellar populations in ancient GCs tend to become increasingly complex with increasing mass, and the same may well be the case for their younger counterparts. However, small number statistics are a crucial limitation in young clusters, especially for post-main sequence stars, as even a $10^5$ $M_\odot$ cluster only contains 20–30 post-MS stars (Larsen et al. 2011). Young clusters with masses well above $10^5$ $M_\odot$, also known as *Super Star Clusters* (SSCs), are rare and tend to be located beyond the Local Group.

**HST ACS/WFC**      **6.5 m LUVOIR / HDI**      **15 m LUVOIR / HDI**

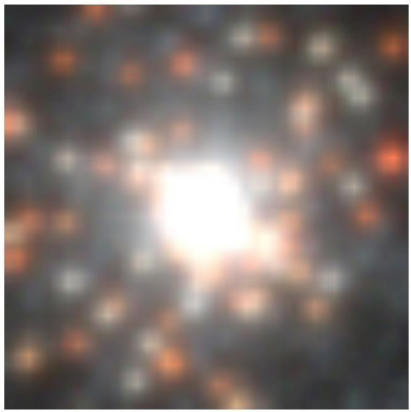 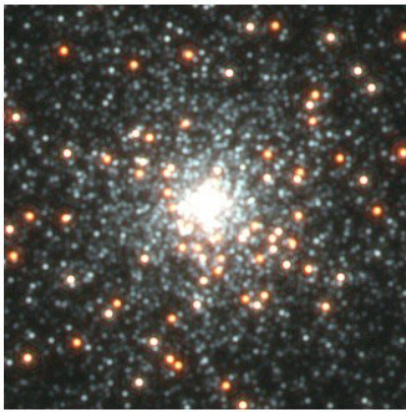 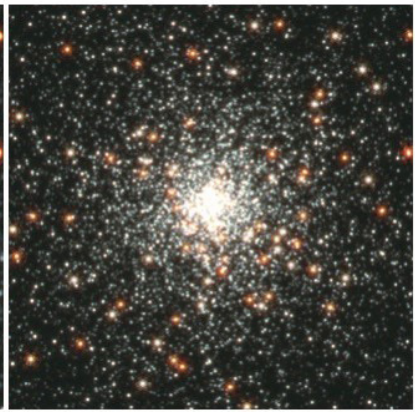

**Figure 13.23.** *Simulated 1800 s exposures (2"x2") of a 50 Myr old star cluster in the galaxy NGC 1313 (Larsen et al. 2011), at a distance of 4 Mpc. This cluster has a mass of about 200,000 $M_\odot$.*

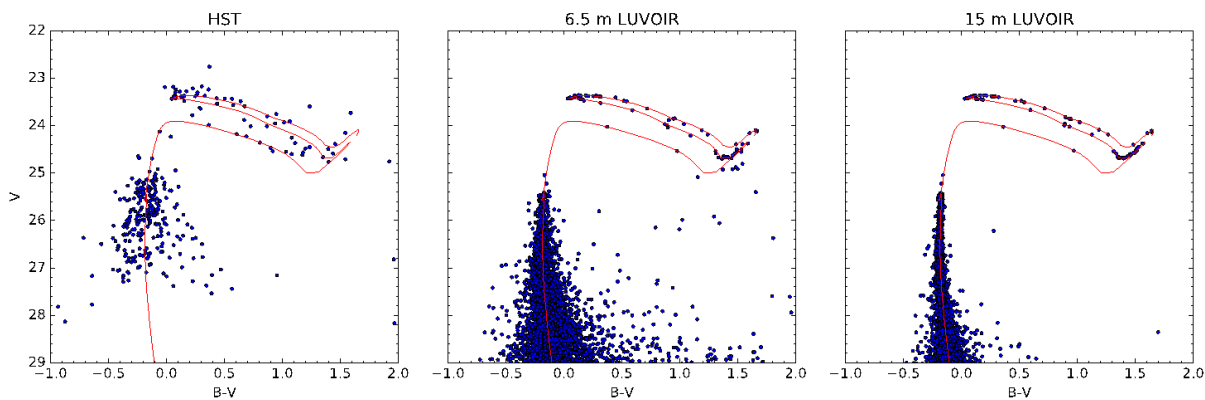

**Figure 13.24.** *Color-magnitude diagrams for the simulated images in* **Figure 13.23***.*

HST has played a crucial role in finding these objects, but the next step—characterising their stellar contents—will require a much larger space-based telescope. Nevertheless, it is already evident that the CMDs of some young SSCs are not well reproduced by standard models for stellar evolution (Larsen et al. 2011). A number of effects, including stellar rotation, binary evolution, age spreads, and chemical abundance anomalies may all play a role. Current work is already pushing HST to the limit of its capabilities, and it is clear that there is still a lot to learn.

### 13.14.2 The role of LUVOIR

Imaging with a 10-m class LUVOIR will make it possible to obtain exquisite photometry for individual stars in SSCs well beyond the Local Group. In evolved stars, the wavelength range 250–450 nm provides access to strong molecular features that are sensitive to light element abundances (CN, CH, NH). For hot stars, observations at longer wavelengths (e.g., ground-based AO-assisted imaging) lack the $T_{eff}$ sensitivity necessary to properly characterize (sub-)populations within the SSCs.

The difference between the current state-of-the-art and LUVOIR is illustrated in **Figure 13.23**, which shows simulated images of an actual cluster in the galaxy NGC1313, at a distance of 4 Mpc (Larsen et al. 2011). HST can only resolve the supergiants and the brightest main sequence stars in this 50 Myr old cluster and provide a crude CMD, while LUVOIR can provide accurate photometry reaching far down the main sequence (**Figure 13.24**). Blue/UV photometry with LUVOIR

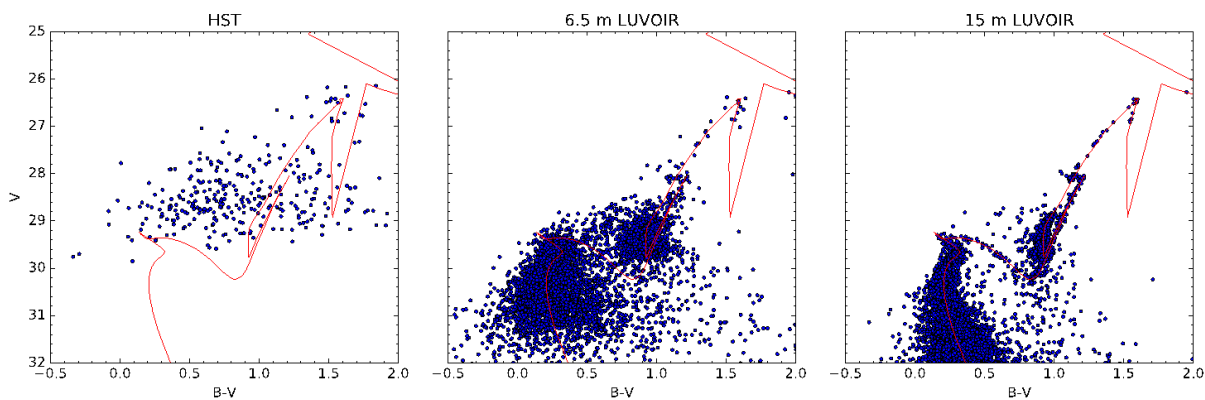

**Figure 13.25.** *Color-magnitude diagrams for the same cluster, but for an age of 1 Gyr and an exposure time of 5 hours per filter.*

---

**Program at a Glance**

**Science goal:** Characterize the stellar contents of massive star clusters in nearby (~5 Mpc) galaxies via high quality color-magnitude diagrams.

**Program details:** Imaging in blue/UV filters (F275W/F336W/F343N/F439W equivalent), exposure times from ~30 min to several hours per filter.

**Instrument(s) + configuration(s):** HDI

**Key observation requirements:** SNR should be ~100 per filter (errors ~0.015 mag) to clearly detect C, N abundance variations in cool stars.

---

would establish whether variations in light-element abundances (C, N) are present in this and other young clusters, and would also put tight constraints on any age spreads. In **Figure 13.25**, the same cluster as in **Figure 13.23** has been simulated for an age of 1 Gyr. Even the brightest stars are now completely beyond reach of HST, while LUVOIR will be able to reach well below the main sequence turn-off in a few hours of exposure time per bandpass. The difference between a 6.5-m and 15-m aperture is also evident. If past experience is any guide, LUVOIR CMDs of such clusters will likely reveal surprising new features not yet imagined.

### 13.14.3 The science program

We envision a LUVOIR survey of SSCs in a sample of relatively nearby star-forming galaxies, out to distances of ~5 Mpc. The targets would cover a variety of galaxy types, metallicities, environments, etc., ranging from dwarf starburst galaxies to large spirals. The survey could also include nuclear star clusters. The global characteristics of the SSC populations in most nearby galaxies are already well known, and will be studied in more detail by missions such as Euclid and WFIRST prior to LUVOIR. Somewhat further away, the rich cluster population in the Antennae merger would offer interesting, but challenging targets.

## 13.15  The detection of terrestrial planets in the habitable zones of A-stars

Ramses Ramirez (Earth-Life Science Institute, Tokyo Institute of Technology), Aki Roberge (NASA GSFC)

### 13.15.1  Introduction

Most habitable zone (HZ) studies focus on terrestrial planets orbiting F,G,K,M main-sequence stars. This is largely because an arbitrarily long timescale of 1 Gyr, corresponding to the minimum thought for life to arise, is also as long as the entire main-sequence lifetime of an early F-star. However, life on Earth had already appeared by ~3.8 Gyr ago, if not sooner. Indeed, life may have evolved even earlier had our planet not been beset by ongoing impacts during the late heavy bombardment. Evidence from zircons also suggests habitable conditions on Earth by 4.3 Gyr ago, only ~300 Myr after formation of the planet. Thus, in order to explore the range of conditions under which life can arise, we should include A-stars in the search for habitable planets.

A-star systems for which exoplanets have already been directly imaged include: Fomalhaut (7.7 parsecs), Beta Pictoris (19.4 pc), HR 8799 (39 pc), HD 95086 (90 pc), and HD 15082 (118 pc). Direct imaging is currently the best way to discover exoplanets around A-stars, as indirect methods like radial velocity and transits have proven ineffective. The known A-stars planets are massive (e.g., Jovian or larger) and located far beyond their host stars' respective HZs. Previous A-stars surveys have not found smaller terrestrial planets, since the required ultra-high contrast is not yet available. Should terrestrial planets exist within the HZs of these stars, however, we may be able to detect them using the ECLIPS coronagraph to block out extraneous starlight.

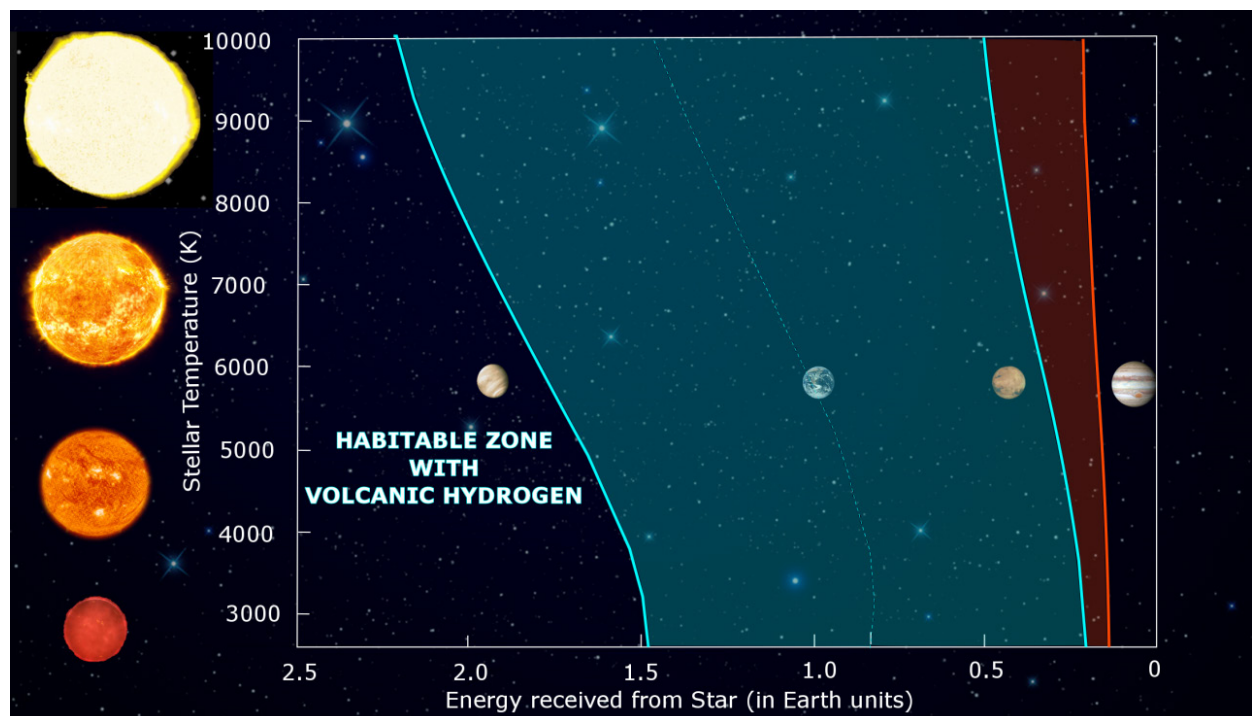

**Figure 13.26.** *The classical $CO_2$-$H_2O$ habitable zone (blue) with volcanic hydrogen extension (red). Adapted from Ramirez and Kaltenegger (2017).*

---

**Program at a Glance**

**Science goal:** Discover terrestrial exoplanets in the habitable zones of A-type stars and probe their atmospheres to investigate their habitability.

**Program details:** Direct imaging and low-resolution Vis/NIR spectroscopy of super-Earth exoplanets in the habitable zones of early-type stars with spectral types ~A7V and later, out to ~76 parsecs.

**Instrument(s) + configuration(s):** ECLIPS imaging and IFS spectroscopy

**Key observation requirements:** Contrast floor < 3 × 10$^{-11}$, IWA < 3.5 λ/D

---

### 13.15.2  The role of LUVOIR

The extreme high contrast needed to directly observe Earth-like planets around Sun-like stars (~ 1 × 10$^{-10}$) is only possible with a space-based telescope and ultra-high performance starlight suppression instrument (like LUVOIR). The situation is harder for terrestrial planets around early-type stars, as the planet-to-star flux ratio is even smaller (see below).

### 13.15.3  The science program

The effective stellar flux of a planet near the conservative inner and outer edges of the classical HZ of Fomalhaut (A4V; $T_{eff}$ = 8600K; L = ~16.63 $L_{sun}$) is ~1.25 and 0.5 times that received by the Earth, respectively (**Figure 13.26**), which corresponds to orbital distances of ~ 3.65 and 5.8 AU. The inner working angle (IWA) of ECLIPS is 3.5 λ/D = 48 milliarcsec at 1 micron. So HZ inner edge is exterior to the IWA for A4V stars out to 76 parsecs, at wavelengths shorter than 1 micron. This shows how the large separation of the HZ for A-type stars is advantageous for high-contrast imaging and spectroscopy.

However, the planet-to-star flux ratio is smaller than it is for the Earth around the Sun. A planet at the equivalent insolation distance has about the same absolute bolometric magnitude, no matter the star. Equation 15 in Turnbull et al. (2012) gives the approximate planet-to-star flux ratio for an Earth-twin planet at the inner edge of the HZ. Adapting that equation for a 1.4 $R_{Earth}$ super-Earth exoplanet gives

$$\frac{F_{super\text{-}Earth}}{F_{star}} \sim \left(1.4^2\right) * 2 \times 10^{-10} \big/ L_{star}$$

Setting the planet-to-star flux ratio to the ECLIPS contrast limit of ~ 3 × 10$^{-11}$, we can estimate the most luminous star for which an Earth-like super-Earth at the inner edge of the HZ can be detected. The limiting stellar luminosity is < 13.1 $L_{Sun}$. Therefore, such exoplanets can be detected around stars with spectral types of ~A7V and later, albeit with very long exposure times. We will calculate the required exposure times at a later date.

---

## 13.16  Quest for the first quasars

Yoshiki Matsuoka (Ehime University)

### 13.16.1  Introduction

One of the greatest achievements of modern optical/infrared astronomy is the discovery of ubiquitous supermassive black holes (SMBHs) throughout the universe (e.g,, Kormendy & Ho 2013). Almost all massive galaxies in the local universe harbor a central SMBH, as evidenced by stellar and gaseous kinematic measurements. In the more distant universe, we know the presence of numerous quasars and active galactic nuclei (AGNs), whose energetic radiation is believed to originate from mass accretion onto a SMBH. As such, SMBHs are recognized as a major constituent of the baryonic universe. Furthermore, they may well control the fate of the host galaxies, as implied from the correlation between SMBH mass and galaxy bulge mass, and powerful gas outflows observed in quasar host galaxies, among other observational facts.

On the other hand, the origin of SMBHs is not yet known. The discoveries of SMBHs exceeding a billion solar masses ($M_{sun}$) at $z > 6$, where the universe is less than a billion years old, have provoked controversy on their seeding mechanism; SMBHs may be born as (i) remnants of first stars with $10^2$–$10^3$ $M_{sun}$, which must be followed by extremely efficient mass accretion over a long time, (ii) products of runaway collapse of primordial star clusters, resulting in $10^3$–$10^4$ $M_{sun}$ seeds, or (iii) massive seeds with $10^4$–$10^5$ $M_{sun}$, resulting from direct collapse of primordial gas (e.g., Volonteri 2012). Observations of "first quasars," representing the earliest stage of SMBH assembly from one or more types of the above seeds, are undoubtedly a key to disentangling the origin of SMBHs (see **Figure 13.27**).

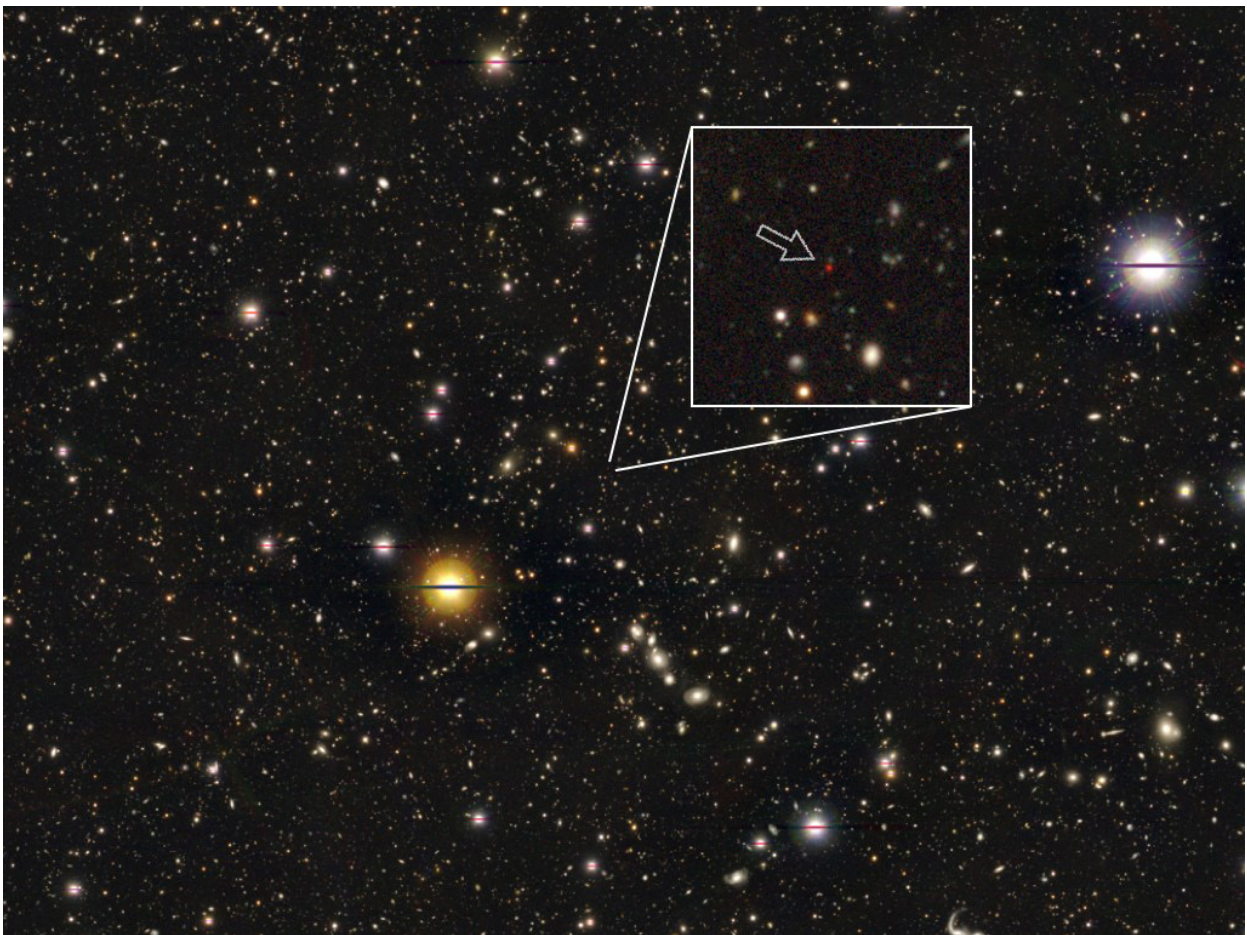

**Figure 13.27.** *A distant quasar discovered at z ~ 6, where the universe is less than a billion years old (Matsuoka et al. 2016, ApJ, 828:26). LUVOIR will probe the even more distant universe, where "first quasars" are being formed.*

### 13.16.2 The role of LUVOIR

Quasars are expected to be extremely faint at the early stage of SMBH mass assembly, which took place in the very distant universe ($z > 10$). For example, a $10^5$ $M_{sun}$ SMBH at $z = 10^{-20}$, radiating at the Eddington limit, will have a rest-frame UV magnitude of 30–32 AB mag, which will be observed in near-IR wavelengths at $\lambda < 3$ µm. LUVOIR will be the first telescope with sufficient photometric sensitivity to detect such faint and distant signals.

### 13.16.3 The science program

The High Definition Imager (HDI) will be used for a square-degree scale imaging survey through a few near-IR bands. The survey field should overlap other multi-wavelength data sets, obtained by other space and ground-based instruments, in order to maximize various scientific potential when combined with the LUVOIR data. We will select $z > 10$ sources with the dropout technique; any source at $z > 10$ will be completely dark at $\lambda < 1.3$ µm due to IGM absorption, which gives rise to a strong spectral break identifiable with near-IR multi-band photometry. Candidate first quasars will be searched for with broadband colors and inferred luminosity. Unfortunately, the emergent spectrum from a first quasar has been poorly studied so far. It may have an extremely red optical/IR color, if surrounding hydrogen gas heavily obscures the SMBH (Pacucci et al. 2015). On the other hand, if there are escape paths of photons from the vicinity of the SMBH, then they would make first quasars appear much bluer. Alternatively, most luminous dropout sources at the relevant cosmic epoch may simply be first quasars if a coeval (proto-) galaxy cannot produce a similar amount of radiation energy; this is indeed the case at the current observational frontier of $z = 6$–7, where virtually all sources with high luminosity (rest-frame UV absolute magnitude $M_{1450} < -24$ mag; Matsuoka et al. 2016) are quasars. The candidate first quasars thus identified will be followed up with ground-based extremely large telescopes, which will confirm the nature of the candidates through high-resolution near-IR spectroscopy. The grism mode of HDI may also be used to narrow down the photometric candidates. Ultimately, the spectroscopic information and statistical properties, such as the number density as a function of SMBH mass and redshift, will

---

### Program at a Glance

**Science goal:** We aim to discover "first quasars," which represent the early evolutionary stage of supermassive black holes, with HDI extremely deep imaging.

**Program details:** a square-degree scale imaging survey in multi near-IR bands

**Instrument(s) + configuration(s):** HDI, imaging (+ grism spectroscopy)

**Key observation requirements:** 10σ depth of ~32 AB mag

---

be compared with theoretical predictions to disentangle the seeding mechanisms of SMBHs.

## 13.17 Protostellar outflows/jets

Christian Schneider (Hamburger Sternwarte) & Gregory Herczeg (Kavli Institute for Astronomy and Astrophysics)

### 13.17.1 Introduction

Stars grow by accreting matter from their surrounding protoplanetary disk, which requires the efficient redistribution of angular momentum—a process still highly uncertain. Simulations of disks with non-ideal magneto-hydrodynamics suggest a magnetically driven disk wind may extract angular momentum from the disk, thereby leading to accretion (see **Figure 13.28** below and reviews by Hartmann et al. 2016 and Bai & Stone 2013). The accretion onto the star is readily observed (e.g., Balmer jump, H-alpha emission), but observational constraints on the physics of mass transport, with implications for planet growth and migration as well as the dispersal of the natal envelope, need new instrumentation.

### 13.17.2 The role of LUVOIR

By the time of LUVOIR, sophisticated simulations of protoplanetary disks will exist, which will require observational tests of angular momentum transport from the mass loss rates and wind velocities in the inner AU around the central star. Current observations of disk winds are limited to unresolved line emission or line-of-sight absorption, with large uncertainties (e.g., Edwards et al. 2006; Rigliaco et al. 2013). Similar observational challenges limit the interpretation of possible jet rotation signatures (Bacciotti et al. 2002, Coffey et al. 2012) and jet collimation shocks (Schneider et al. 2013). Currently, the major instrumental limitations are a lack of spatial resolution compared with insufficient sensitivity, e.g., the FUV emission lines require several orbits with HST for a single long-slit spectrum and several slit positions are required to sufficiently sample the spatial structure of the jet. LUVOIR will overcome these issues and is needed even in the era of ALMA, JWST or 30-m telescopes as none of these are capable of imaging disk winds or accretion flows close to the star, nor do they cover the relevant temperature regime. The primary disk wind diagnostics are at UV to optical wavelengths, where high spectral and spatial resolution observations (preferably with an IFU) are not feasible, even with AO systems. Further, disk images in FUV CO and $H_2$ emission will reveal accretion flows within the disk and onto the star or protoplanets in exquisite detail inaccessible to ground-based instruments.

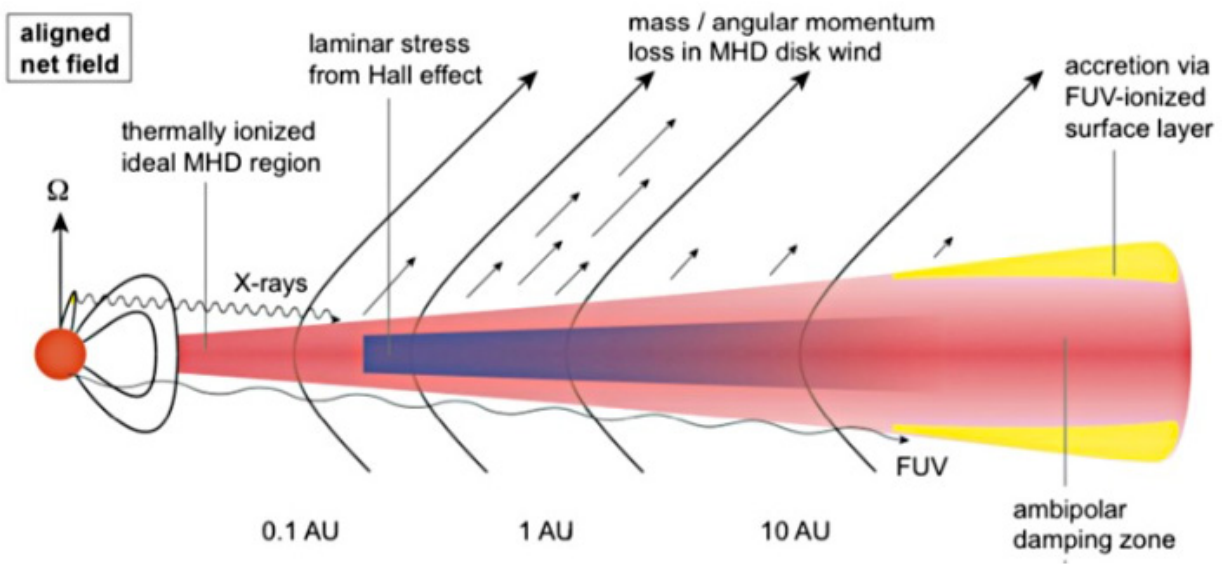

**Figure 13.28.** *Sketch of a protoplanetary disk and angular momentum transport mechanisms. Credit: Simon et al. 2015, MNRAS 454 .*

### 13.17.3 The science program

**Measure the launching and mass flux in disk winds.** The launching of MHD winds from the disk may drive the accretion flow. These winds are thought to be launched from a large range of disk radii resulting in streamlines with different velocities and temperatures. Therefore, a large repertoire of diagnostics is needed to sample the total mass flux. Measuring the mass flux and velocity for objects of different evolutionary stages and masses with will allow us to measure the angular momentum extracted by the disk wind. This mass loss combined with accurate accretion rates will provide rigorous tests of models for launching disk winds and to understand the energy balance of jets and accretion.

**Measure jet collimation and rotation properties.** Jets are collimated by a magnetic field anchored in the protoplanetary disk. Depending on the anchor/launch point in the disk, different streamlines possess different collimation properties. Sampling the collimation properties and jet rotation for different outflow components is needed to study jet lunching, the relevance for envelope dispersal, and will provide valuable information on the disk's magnetic field.

**Measure the mass flows in the inner disk.** ALMA and scattered light observations revealed that disks have large radial dust traps and azimuthal asymmetries likely induced by disk physics, chemistry, or planet-disk interactions. While the disk should be Keplerian, recent ALMA observations suggest radial mass flows. At the highest spatial resolution, LUVOIR will be able to reveal any non-Keplerian flows in the inner AU and onto protoplanets in the disk.

### Description of Observations

To measure mass flux rates, we need to sample relevant temperature regimes. Examples of important emission lines are $H_2$, [O I], [S II], [N II], [O III], and C IV with line fluxes in the range of $10^{-14}$ erg/s/cm$^{-2}$ for the nearby jets and contrast ratios of 1:100 at 0.1 arcsec from the star. Typical outflow velocities range from a few 10 to several 100 km/s.

With LUMOS, one can efficiently measure the collimation and rotation signatures of the jet close to the launching region on scales of ~10 AU (~100 mas, nearby star forming region) with high spectral resolution to trace

---

**Program at a Glance**

**Science goal:** Understand mass-flows around forming stars

**Program details:** Spatially and spectrally resolved observations of young stellar objects, forming planets, and protostellar jets

**Instrument(s) + configuration(s):** LUMOS high spectral-resolution spatially resolved spectroscopy, other instruments?

**Key observation requirements:** R > 10,000, spatial resolution better than 0.1 arcsec, contrast of 100:1 at 0.1 arcsec.

---

different flow components using the micro-shutter array. Roughly speaking, the flow velocity close to the jet base relates to the lunching radius. Sampling the collimation properties in several velocity channels in the range from <10 km/s up to several 100 km/s will be critical for this. With expected rotation signatures in the km/s range at scales of 0.1 arcsec (nearby SF region), LUMOS is ideally suited to perform high S/N, high spectral resolution (R>10,000) observations. Also, the region of interest, a few 100 AU translating to an apparent size of a few arcsec for the nearby star formation regions, is ideally matched to LUMOS.

## 13.18 Exploring the high energy processes in microquasars as an exemplar case for high time resolution astrophysics

Warren Skidmore (Thirty Meter Telescope) and members of the TMT Time Domain ISDT

### 13.18.1 Introduction

High time resolution (between a few seconds to a few milliseconds) spectroscopic observing capabilities spanning the UV into the optical have applications across many areas of astrophysics, from the solar system to the high redshift universe. Here we describe one exemplar science case that requires observing capabilities that can be applied to many other cases.

Black hole X-ray binaries have a stellar mass black hole (a few to ~20 $M_\odot$) around which an accretion disk exists that is fed from a mass losing secondary star (**Figure 13.29**). In many systems, high-speed jets of material are expelled in a situation very similar to that in AGN, and this earns these systems the name of microquasars. Microquasars provide an opportunity to study the poorly understood jet acceleration mechanism and the mysterious process that harnesses the energy and angular momentum of accretion disk material to drive the jets. In AGN the region in which these processes are taking place is hidden by dust and gas but in microquasars they are directly visible. However, the timescales involved demand special capabilities from astronomical instruments.

In microquasars there are multiple processes occurring simultaneously: shocks at the base of the jets, flares from the accretion disk, reprocessing of X-rays to lower energies, synchrotron and cyclotron processes involving optically thick and optically thin emitting materials. Each process has different temporal and spectral characteristics. To disentangle the observed signals from each of the processes it is necessary to examine both the temporal and spectral characteristics together using time resolved spectroscopy that includes the UV and has the largest coverage to longer wavelengths.

### 13.18.2 The role of LUVOIR

A typical brightness of these targets is R~12, B~13.6, but much of the variability may be constrained to emission lines. A large space

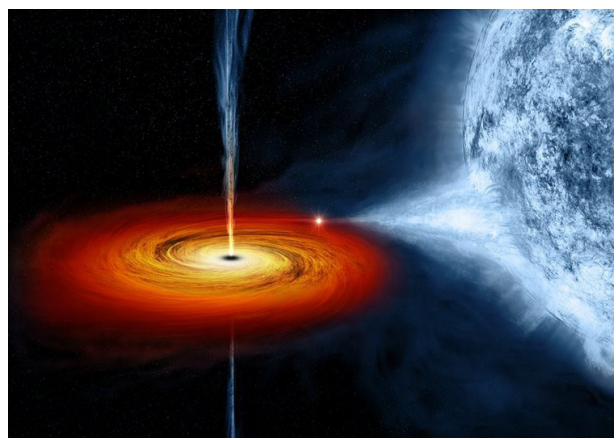

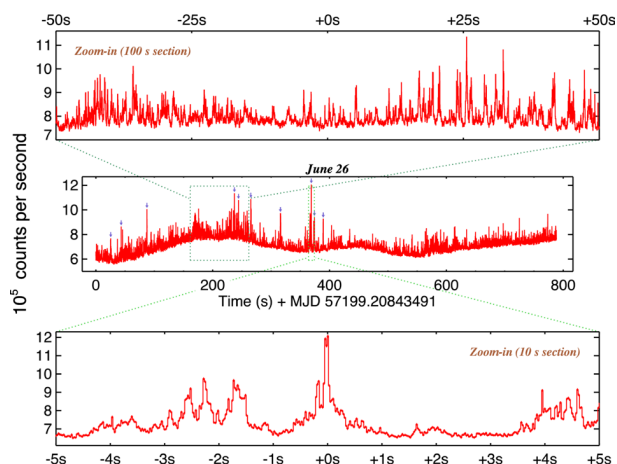

**Figure 13.29.** *Artist's impression of the micro-quasar Cygnus X-1, showing material flowing from the mass-losing star and forming an accretion disk and the jets being driven out from the vicinity of the black hole. Right - Rapid optical variations in V404 Cyg observed with ULTRACAM on the 4.2-m WHT. Credit: Gandhi et al. (2016)*

---

**Program at a Glance**

**Science goal:** Identify the observable signals of the stages of the process of converting accretion energy into power to drive the jets in microquasars and characterize each of those stages.

**Program details:** 100 to 400 nm, R = 500 LUMOS observations of a science target and 2 to 4 local field stars with $T_{exp}$=0.033 s (100 to 400 nm) and 0.01 s (200 to 1000 nm)

**Instrument(s) + configuration(s):** 1. LUMOS multi-object (up to 5 targets) with fast readout modes and G145 LL grating; 2. HDI grism multi-object (up to 5 targets) with fast readout modes

**Key observation requirements:** Wide wavelength coverage, low spectral resolution, short exposure and fast readout modes.

---

based telescope with a moderate resolution UV/optical spectrograph is necessary in order to have the necessary wavelength coverage and acceptable S/N at the required sampling frequencies. These include:

- Far UV to blue (100 to 400 nm): Sampling frequency of ~30Hz with spectral resolution of R~500.
- Near UV to near IR (200 to 1000 nm): Sampling frequency of ~100 Hz with spectral resolution of R~500.

### 13.18.3 The science program

1. We would separately observe 4 microquasar targets (target brightness R~12, B~13.6, plus 2 to 4 local field stars to act as local photometric standards) using the LUMOS spectrograph with the G145 LL grating in a low multiplexing, wide wavelength coverage mode (100 to 400 nm). Each target would be observed for a duration of 2 hours with integration times of 0.033 sec giving S/N~20.

2. As above (4 separate targets and local standards), except using the R~500 GRISM mode on HDI and integration times of 0.01 s giving S/N~10 to 20.

Further advances could be made using:

- Time resolved spectropolarimetric observations, as some of the variable emission is likely to be highly polarized.
- Coordinated simultaneous observations between X-ray to look at reprocessed signals and map the system (cross correlating to accuracies of ~1 millisecond)

However, these observations are not described here except to note the importance of high accuracy time stamps (better than ~10 μs).

## References

Gandhi, P., et al. 2016, *MNRAS*, 459, 554

---

## 13.19 Comets and minor planets: the importance of the small things

Noemi Pinilla-Alonso (Florida Space Institute, University of Central Florida, Orlando, USA),
Alvaro Alvarez-Candal (Observatorio Nacional, Rio de Janeiro, Brazil)

### 13.19.1 Introduction

Minor planets and comets are rocky and/ or icy objects, usually ranging in size from a few meters to a few hundreds of kilometers. They comprise near-Earth and main belt asteroids, Trojans (of Jupiter and other giant planets), trans-Neptunian objects, Centaurs, comets, and a recently discovered category called the transitional objects (de Leon et al. 2018).

The study of minor planets over the last decades has led to dramatic changes in our understanding of the process of planet formation and evolution, and the relationship between our Solar System and other planetary systems. Small bodies also serve as large populations of "test particles" recording the dynamical history of the giant planets, revealing the nature of the Solar System impactor population over time, and illustrating the size distributions of planetesimals, which were the building blocks of planets. The number of discoveries regarding exoplanets and debris disks is continuously increasing, and therefore it is crucial to first understand our own solar system's provenance and evolution in order to better interpret what is going on in newly discovered planetary systems

Telescope observations from the ground or in Earth orbit telescopes have increased in the last decade and form the basis for understanding these small bodies. Also, detailed information from some particular targets is available through missions such as Rosetta, Dawn and New Horizons. However, when compared with the number of small bodies estimated to be part of the Solar System, or even with the ones already detected (~700,000 asteroids, ~ 3,000 trans-

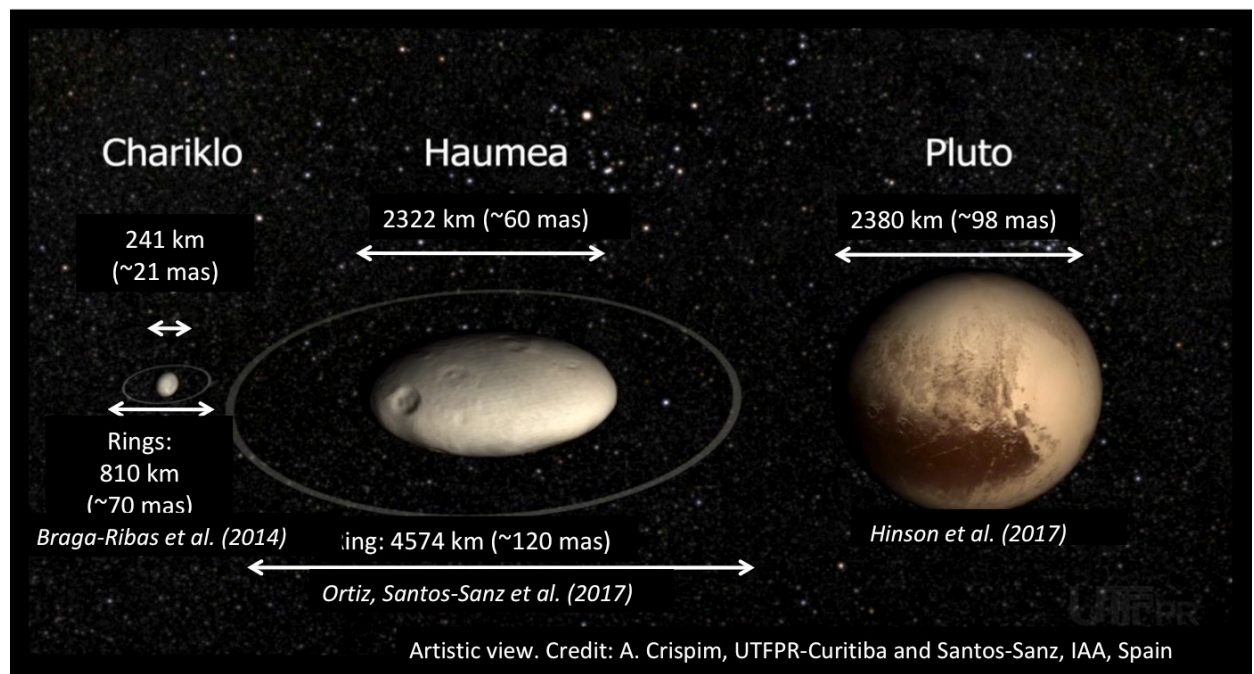

**Figure 13.30.** *The Chariklo and Haumea rings systems, to scale, detected by occultations. Pluto is shown for comparison.*

Neptunian objects, ~7,000 Jupiter Trojans, and ~5,500 comets), actual knowledge of the small bodies population is sparse.

A key program for a space telescope such as LUVOIR dedicated to observations of comets and minor planets, from the ultraviolet to the infrared wavelengths, would revolutionize the actual knowledge of the Solar System in a lot of different areas. Below I detail some of them

## 13.19.2 The role of LUVOIR

**Size/Shape/Ring detection and characterization:** In the last years rings have been detected around three minor bodies (**Figure 13.30**) (Braga-Ribas et al. 2014, Ortiz et al. 2015, Ortiz et al. 2017). The viewing geometries of these rings change secularly due to the movement of these bodies around the Sun. High angular and spatial resolution would allow direct confirmation of the presence of these rings and possibly detection of other rings or small moons. Spectroscopy of the systems under different viewing circumstances would allow compositional studies of these rings (Duffard et al. 2014).

Direct spatial resolution would also enable studies of the shape of minor planets. The most effective tool in that regard is radar imaging, but this only works for targets that pass very close to the Earth. Very recently, a survey using Adaptive Optics for the ~100 brightest asteroids is aiming to determine the shapes of these bodies, VLT/SHEPRE (Garufi et al. 2017). For the rest of the populations, most asteroids, Trojans, Centaur and TNOs, only inversion techniques applied to lightcurves provide an approach to measuring the shapes of these bodies.

**Surface Composition:** Some solid ices ( e.g., Cruz-Diaz et al. 2014a,b), and mineral spectral features (Cloutis et al. 2008) occur at wavelengths accessible with LUVOIR but not from the ground (see **Figure 13.31** and **Figure 13.32**). The ultraviolet end of the spectrum is particularly interesting. Wavelengths below 400 nm have been largely unexplored. These wavelengths are particularly sensitive to water (either in the form of water ice or in the form of hydrated materials), ammonia compounds, $CO_2$, complex organic materials and Fe-rich materials, which result as a product of space weathering. They are also more sensitive to hydrated minerals than the NIR-wavelegths (~3 μm). In particular, the study of small bodies at the UV wavelengths with LUVOIR would provide new insights in tracing the water ice and organics, the seeds of life on Earth, in the Solar System. Also it would be key in studying the nature of ices at different penetration depths than VNIR spectroscopy.

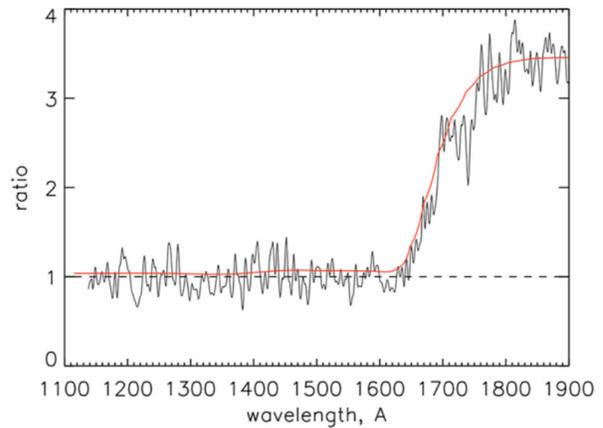

**Figure 13.31.** *Spectral ratio of bright terrain to average dark terrain on Iapetus. The ratio is very similar to pure water ice suggesting that the major difference between two terrains is the amount of water ice present (Hendrix et al. 2010).*

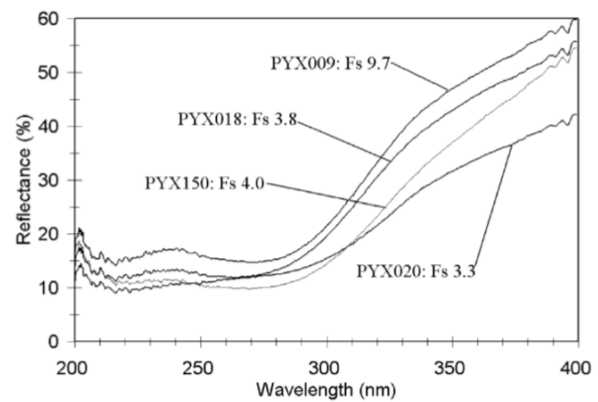

**Figure 13.32.** *Reflectance spectra of three different sets of high-calcium pyroxenes with different ratios of $F_2+$ to $Fe_3+$ (Cloutis et al 2008).*

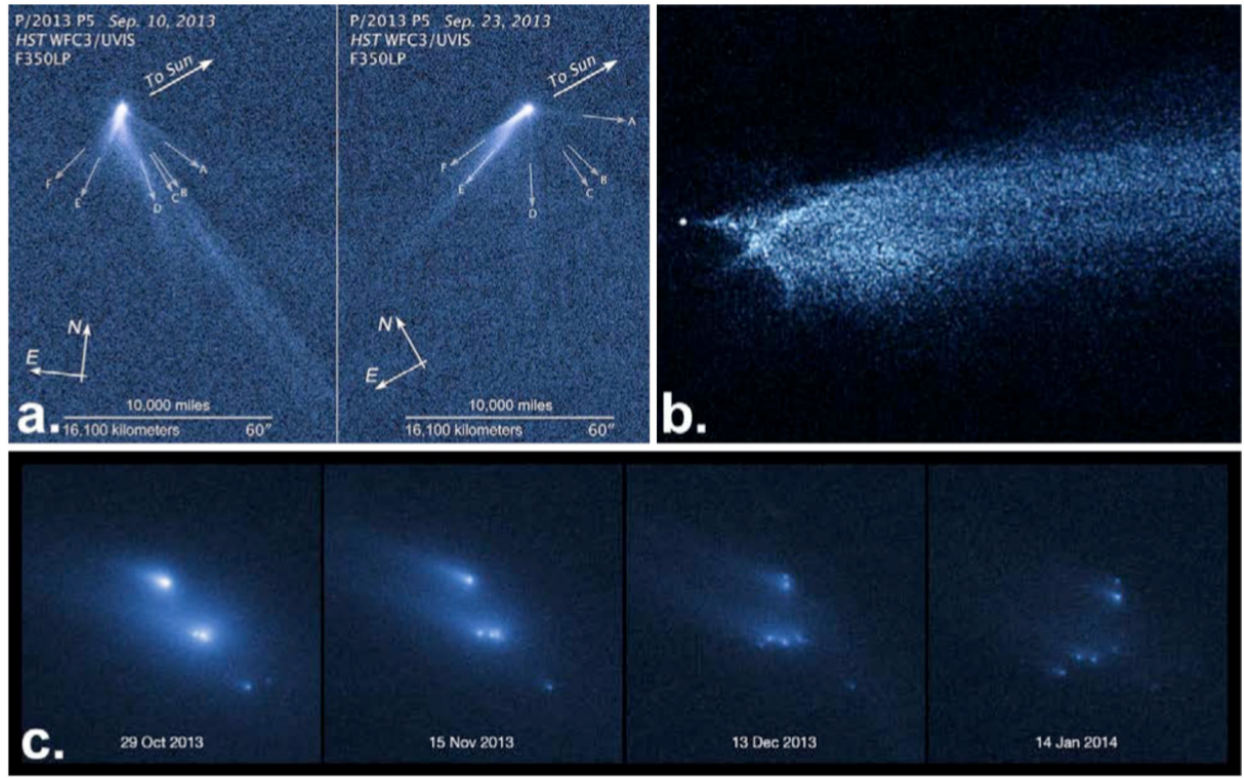

**Figure 13.33.** *Images of active asteroids a) P/2013 P5 multiple tails; b)P/2010 A2(LINEAR) tail and nucleus; c) P/2013 R3 (Catalina-PANSTARRS) breaking up (Jewitt et al. 2015).*

**Binary detections and characterization:** The number of detected and well-characterized binary systems among the TNO population is as low as some tens. The great LUVOIR improvement in angular resolution would open a new field that is dormant now because of the lack of adequate technology. The determination of the density of minor bodies is a key question in the determination of density and mass that are crucial to better understand the structure of the interior of these bodies and the physical characteristics of the superficial materials (rubble-pile vs. monolithic, porosity etc.).

At wavelengths below 2 μm JWST will not provide any power in excess of HST in terms of ability to resolve tightly bound binary TNO systems, and even at 2 μm it will be similar to WFC3 UVIS (Benecchi et al. 2009). At mid-IR wavelengths, JWST's spatial resolution (as low as ~1" at the longest wavelengths) is insufficient to resolve all but the rare, wide binary TNO systems (Parker et al. 2016).

Additionally, one key distinguishing feature of TNO binary systems is the common optical colors of the components that would give indication of formation of the system (Benecchi et al. 2009), in this regards multi color imaging of binary systems coupled with better angular resolution would enable us to explore this subject.

**Activity (comets and active asteroids):** The traditional separation between the rocky inactive asteroids and the icy active comets has disappeared during the last two decades due to the discovery of several objects with the dynamics of asteroids displaying activity as comets (the active asteroids, AAs Jewitt et al 2015). These asteroids display activity in the form of particle loss at rates ranging from $10^{-2}$ to 4 kg/s (Jewitt, et al. 2015). For the currently known active asteroids, several driver processes have been identified, including solar radiation sweeping of particles, electrostatic effects, ice sublimation, and thermal disintegration of surface regolith. Understanding the mechanisms that lead to the activity on these bodies will have implications on both our understanding of the origins of the Solar System and the future of space exploration, and in particular planetary defense and mitigation.

The activity of these targets can last from weeks to months. In some cases it has led to the disappearance of the target after a complete break-up, while in others the activity has been observed recurrently (see **Figure 13.33**). For the cases of a total break up, observation of the fragments is crucial and has to happen soon after the discovery. Using HST limits to the size of fragments has been placed to Hv~23.5 and a size of ~35 m (Moreno et al. 2017).

### 13.19.3 The science program

**Size/Shape/Ring detection and characterization:** This program requires high spatial and angular resolution for direct observations of rings and an observing mode allowing multiple visits or a chain of observations in order to cover a whole rotation for the shape studies (HDI). The superb angular resolution provided by LUVOIR in the UVIS mode (where there happens to be the best balance between collected light and spatial resolution) of 2.73 mas per pixel will allow detection of the ring system of Chariklo with a spread of 25 pixels (its major projected axis). See **Figure 13.30** for the size of the ring systems in comparison with dwarf-planet Pluto.

**Surface Composition:** For this purpose of compositional studies we could use

LUMOS and HDI. We would need low and medium spectroscopy (<30000) of small bodies. According to the LUMOS-ETC provided by STScI we can obtain S/N > 10 in 1 h for all targets with AB mag < 22. Regular detailed observations on some targets are desirable as this may mean rotational coverage, i.e., multiple visits with time constrain.

Using the HDI we could characterize almost all known Centaurs and trans-Neptunian objects, which are typically fainter and more difficult to characterize (as listed in the Minor Planets Center). Using HDI ETC we find we could get spectrophotometry with high S/N in about 1 hr. of exposure per filter. Survey mode will be highly desirable.

Solar System science would extraordinarily benefit from Large Programs (to systematically collect VNIR data from targets in particular populations not reachable with other telescopes) and Fillers (to increase the knowledge on a population without any special requirement in the number of targets observed) in VNIR but also in the UV, where no data exist up to date and even a small number of targets would provide groundbreaking information.

**Binary detections and characterization:** This program requires HDI spectro-photometry. The minimum angular distance in TNO binaries observed in Benecchi et al. 2009 is comparable to the angular size of Chariklo's ring system (which we have showed if would have an excellent coverage by LUVOIR). Using the UVIS mode, we could easily observe binary systems with half that angular size and study the properties of tight binary systems. Observations should be carried out as large programs, with multiple visits to determine orbits.

**Activity (comets and active asteroids):** Exceptional imaging capabilities would allow detection of the tiniest fragments. Multi-object spectroscopy would be desirable to be able to study the different fragments as

---

<div style="border:1px solid #000; padding:10px;">

### Program at a Glance

**Science goal:** Detection and characterization of minor bodies, their rings, and their binarity. Measurements of water ice and organics in Centaurs and TNOs. Activity in comets and asteroids.

**Program details:** High spatial and angular resolution imaging. Low and medium spectroscopy (<30000). Multi-object spectroscopy.

**Instrument(s) + configuration(s):** HDI (UVIS mode), LUMOS.

**Key observation requirements:** S/N > 10 in 1 h for all targets with AB mag < 22. Telescope tracking the source.

</div>

well as the coma. A target-of-opportunity program would be desirable, with the possibility of activating it in the first 24 hours after the discovery of the activity, and subsequent follow-up that could be concentrated in one week or over several months.

Last but not least, most of these observations have to be done with the telescope tracking the source, so this capability is extremely important.

## 13.20 Detecting liquid surface water on exoplanets

Tyler D. Robinson (Northern Arizona University & Virtual Planetary Laboratory) and Mark S. Marley (NASA Ames Research Center)

### 13.20.1 Introduction

Planetary habitability requires liquid water stability at the surface of a terrestrial planet (e.g., Kasting et al. 1993). The remote characterization of habitability for planets detected in reflected light will require either (1) information about the near-surface atmospheric state for a confirmed terrestrial world, or (2) the direct detection of surface liquid water via ocean glint and/or polarization measurements. This brief report explores what telescope and instrument requirements would be required to make either of these two observations.

Surface liquid water stability depends on both temperature and pressure from the overlying atmosphere. Unfortunately, we know of no studies that investigate how well atmospheric pressure and temperature can be constrained from visible and near-infrared spectral observations of terrestrial planets—more work here is clearly required. A Rayleigh scattering slope, or detection of dimer or pressure-induced absorption features (Misra et al. 2014b), could indicate pressure. Detection of Rayleigh scattering requires observations at blue-visible wavelengths, and would not place strict requirements on spectral resolution. Misra et al. (2014b) emphasize dimer detection at 1.06 μm at a resolution (λ/Δλ) of 100.

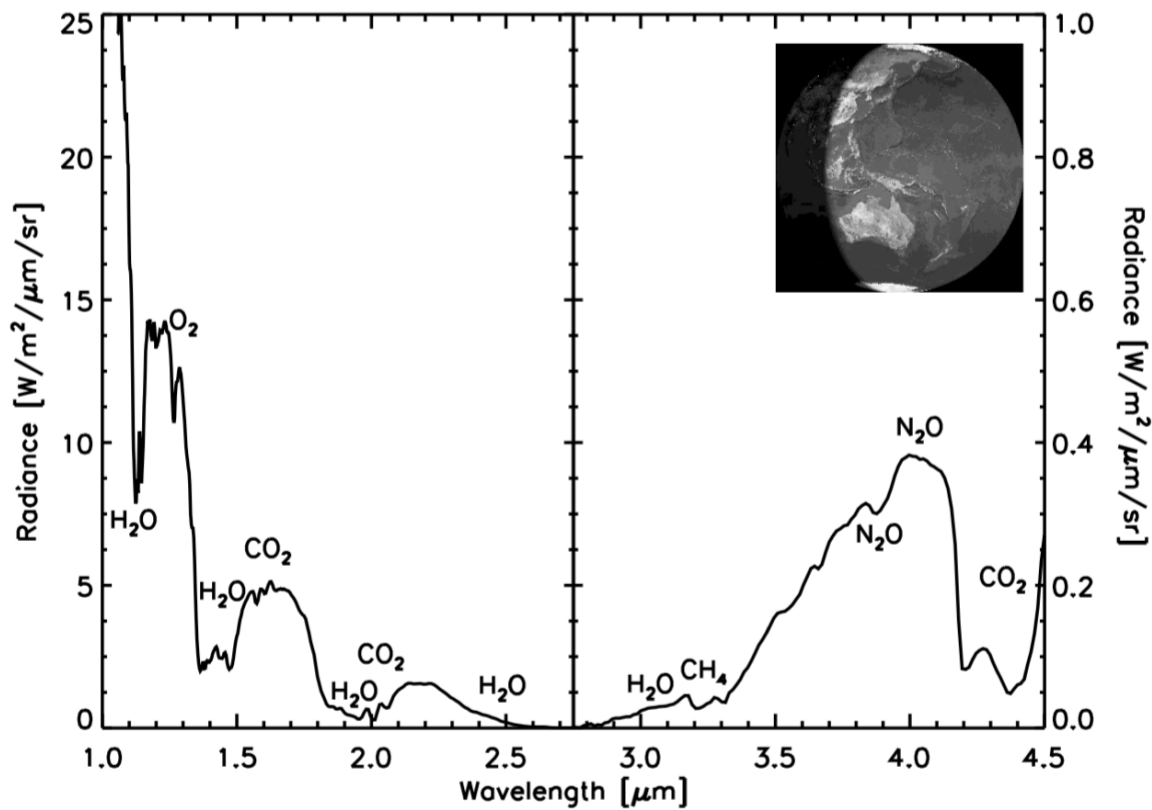

**Figure 13.34.** *A moderate resolution, near-infrared spectrum of Earth, from the validated models of Robinson et al. (2011). Key features are labeled. Note the separate y-axis for wavelengths beyond 2.75 um, which helps show the rise in thermal emission beyond this wavelength.*

Temperatures will be difficult to constrain in the absence of a detection of planetary thermal radiation. However, it may be possible to constrain temperature, and also habitability, by detecting a water vapor profile consistent with condensation (i.e., a condensation curve). Detection of clouds and weather (e.g., from variability) would also indicate condensation, but not necessarily at/near the surface. **Figure 13.35** shows pressure at the τ=1 level for a clear sky Earth atmosphere, and this roughly indicates the pressure levels sensed by a spectrum. Comparison of high- (λ/Δλ=1,000) and moderate- (λ/Δλ=100) resolution spectra shows that, for water vapor, not much vertical resolution is gained by pushing beyond a spectral resolution of roughly 100. Notably, while the 0.94 μm water band probes the near-surface environment, pushing to the 1.4 μm and 1.9 μm water bands is necessary for probing Earth's middle and upper troposphere. Thus, these bands may be essential to constraining water vapor profiles for potentially habitable exoplanets.

Direct detection of surface liquid water may prove a challenging endeavor, but would be complimentary to any atmospheric spectral retrieval analysis (like that outlined above). The near- infrared is best suited to polarization or glint detection, owing to the lack of Rayleigh scattering here (Robinson et al. 2010; Zugger et al. 2011), and detection could be accomplished with photometric observations at continuum wavelengths (i.e., between strong atmospheric absorption bands). Performing observations in the near-infrared would also minimize glint false positive signatures from polar ices (Cowan et al. 2012), whose reflectivity is diminished at these wavelengths. For an Earth-twin, clouds and hazes would likely obscure any polarization signature from surface oceans (Zugger et al. 2011). Thus, glint is the best option for the direct detection of surface liquid water, and, at least for Earth, is most pronounced near a star-planet-observer (phase) angle of 150°, where a glinting planet would be nearly twice as bright as a non-glinting planet. The ability to measure planetary phase functions to such close planet-star separations will depend on the inner-working angle (IWA) of the high-contrast field-of-view, and cannot be achieved for planetary systems with inclinations below about 60 degrees. For an Earth-Sun twin system at 10 parsecs, the required IWA would be 1.8 λ/D (at the 1.33 μm continuum and for a 10-meter diameter aperture). As another example, taking an Earth-like planet near the inner edge of the habitable zone for a mid-K dwarf, the requirement shrinks to 0.9 λ/D at 10 parsecs. Thus, an IWA of 2 λ/D could allow glint detection out to almost 10 pc for habitable planets around G dwarfs, and almost 5 pc for K dwarfs. These distances decrease to 6 pc and 3 pc, respectively, for a 3 λ/D IWA.

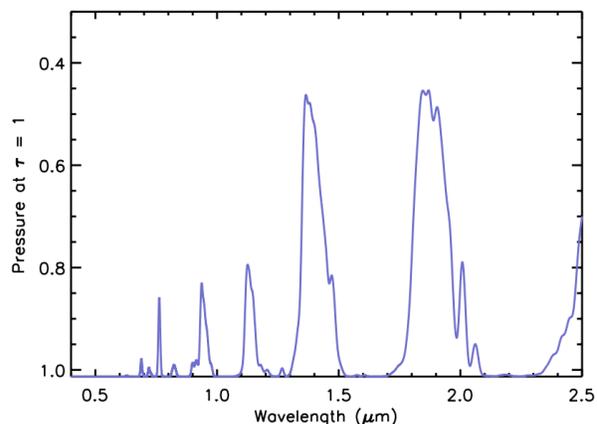

**Figure 13.35.** *The pressure of the τ=1 level for a clearsky Earth atmosphere at λ/Δλ=100. Note the strong water vapor bands at 0.94, 1.1, 1.4, and 1.9 μm.*

We briefly note that the James Webb Space Telescope (JWST) will not be capable of probing the near-surface environment

**Table 13.2.** *Observation requirements for detecting key atmospheric features that constrain the presence of liquid surface water on rocky exoplanets.*

## Optical/Near-IR Direct Spectroscopy of $CO_2$ Features

| Observation Requirement | Major Progress | Substantial Progress | Incremental Progress |
|---|---|---|---|
| Wavelengths | 0.4-2.2 µm | 0.4-1.7 µm | n/a |
| Spatial resolution | | | |
| Spectral resolution | 200 | 100 | n/a |
| Field-of-view | | | |
| Contrast | $10^{-10}$ | $10^{-10}$ | n/a |
| Telescope aperture | 12 | 8 | n/a |
| SNR | 50 | 50 | n/a |

## Optical/Near-IR Direct Spectroscopy of $H_2O$ Features

| Observation Requirement | Major Progress | Substantial Progress | Incremental Progress |
|---|---|---|---|
| Wavelengths | 0.4-1.6 µm | 0.4-1.6 µm | 0.4-1.0 µm |
| Spatial resolution | | | |
| Spectral resolution | 70 | 70 | 70 |
| Field-of-view | | | |
| Contrast | $10^{-10}$ | $10^{-10}$ | $10^{-10}$ |
| Telescope aperture | 12 | 8 | 4 |
| SNR | 50 | 20 | 10 |

of habitable zone planets orbiting stellar types earlier than M (Misra et al. 2014a). Depending on systematics, JWST may be able to characterize potentially habitable worlds around nearby M dwarf stars, but these worlds possess their own unique challenges to habitability (e.g., Luger & Barnes 2015). Finally, while not discussed at length here, the capability of measuring profiles of atmospheric water vapor through an atmosphere will also be useful for studies of gaseous worlds interior to and throughout the habitable zone, as water should be a key radiatively active species (and greenhouse gas) for these planets (Cahoy et al. 2010).

In summary, R=70 (or greater) spectra of Earth-like planets would be suitable for constraining water vapor abundance profiles.

Future work is necessary to determine the minimum signal-to-noise ratios required for such an analysis. Broadband detection of glint in the phase function of a habitable zone exoplanet is possible in the near-infrared with a 2 $\lambda$/D IWA and becomes restricted for a 3 $\lambda$/D IWA. The tables above contain additional information on requirements for detecting water vapor and carbon dioxide features (for climate modeling constraints).

**Note on near-infrared $CO_2$ features:** Carbon dioxide is a key greenhouse gas for Earth, whose greenhouse forcing is essential to maintaining Earth's habitability (Hansen et al. 2013). Thus, inferring atmospheric $CO_2$ concentrations from near-infrared spectra of Earth-like planets would be essential to running predictive/forward climate models for these worlds (thereby helping to inform our understanding of the potential habitability of these planets). The strongest $CO_2$ features in Earth's near-infrared, reflected-light spectrum are a pair of double features at 1.59 $\mu$m and 2.03 $\mu$m. Both would require a spectral resolution of $\lambda$/$\Delta\lambda$=100 to place a single spectral resolution element across the bands, and double this resolution could enable better detections. The features at the shorter wavelength are shallower, requiring a SNR of order 40 to distinguish. The longer wavelength features are deeper, requiring an SNR only of order 10, but may be more difficult to detect due to the overall decrease in stellar brightness at these longer wavelengths. These details are represented in **Table 13.2**.

### 13.20.2 The role of LUVOIR

While the above observations are enabled by a direct imaging mission, they are difficult and require either long observations to increase the signal-to-noise on a spectral feature, or multiple visits to a target to observe it at multiple phases. Additionally, some of these phases will be at small inner working angles. All of these problems benefit from an observatory with a larger aperture, which increases the signal from the planet, decreases the amount of time for an observation (which in turn enables multiple visits), and decreases the inner working angle of the observatory. Finally, the wavelength range of LUVOIR encompasses multiple water vapor features, from 0.94–1.9 $\mu$m, will allow crude determination of the water vapor profile of an exoplanet's troposphere.

### 13.20.3 The science program

Detection of glint from liquid surface water on rocky exoplanets will require multiple visits of ECLIPS coronagraphic imaging to the same target. These multiple visits will conduct phase-dependent imaging in polarized light, and (if time allows) direct spectroscopy to measure $H_2O$ vapor concentrations.

---

**Program at a Glance**

**Science goal:** Definitively show the presence of liquid water on the surface of rocky exoplanets.

**Program details:** Coronagraphic imaging at phase angles as close to 150° as possible to maximize glint signal. More generally, glint can be observed between gibbous and crescent phase. Direct spectroscopy to measure atmospheric water vapor concentrations. Direct spectroscopy to measure NIR $CO_2$ features and constrain greenhouse effect.

**Instrument(s) + configuration(s):** ECLIPS imaging and IFS

**Key observation requirements:** Contrast < $10^{-10}$; Multiple visits; Water vapor features at 0.94, 1.1, 1.4, and 1.9 $\mu$m, R ~ 100; $CO_2$ feature at 1.59 $\mu$m, R > 100.

---

## 13.21 Geology and surface processes in the solar system

Noah Petro (NASA Goddard Space Flight Center)

### 13.21.1 Introduction

All solid objects in the Solar System undergo regular variations due to internal and external forces. However, we know now that these variations can cause measurable changes to the surface and exosphere of small, airless, bodies. The large moons of Jupiter and Saturn, for example, experience tidal heating that can trigger geysers or possibly volcanic eruptions. However, in order to measure these events requires nearly constant monitoring of an object over multiple hours or even days, so that the full range of variation in surface properties (surface temperature changes, albedo and/or compositional changes) can be measured. Spectral measurements of the erupted material (either volcanic or via geyser, see **Figure 13.36**) and comparison to the surface will provide important constraints on compositional variations of the source regions of these moons. Additionally, watching an entire eruption and being able to characterize any compositional changes that occur will provide insights into the mechanism by which these eruptions occur.

### 13.21.2 The role of LUVOIR

A highly capable observatory in space provides an excellent opportunity to make high-resolution (spatial, spectral, and temporal) observations of planetary bodies.

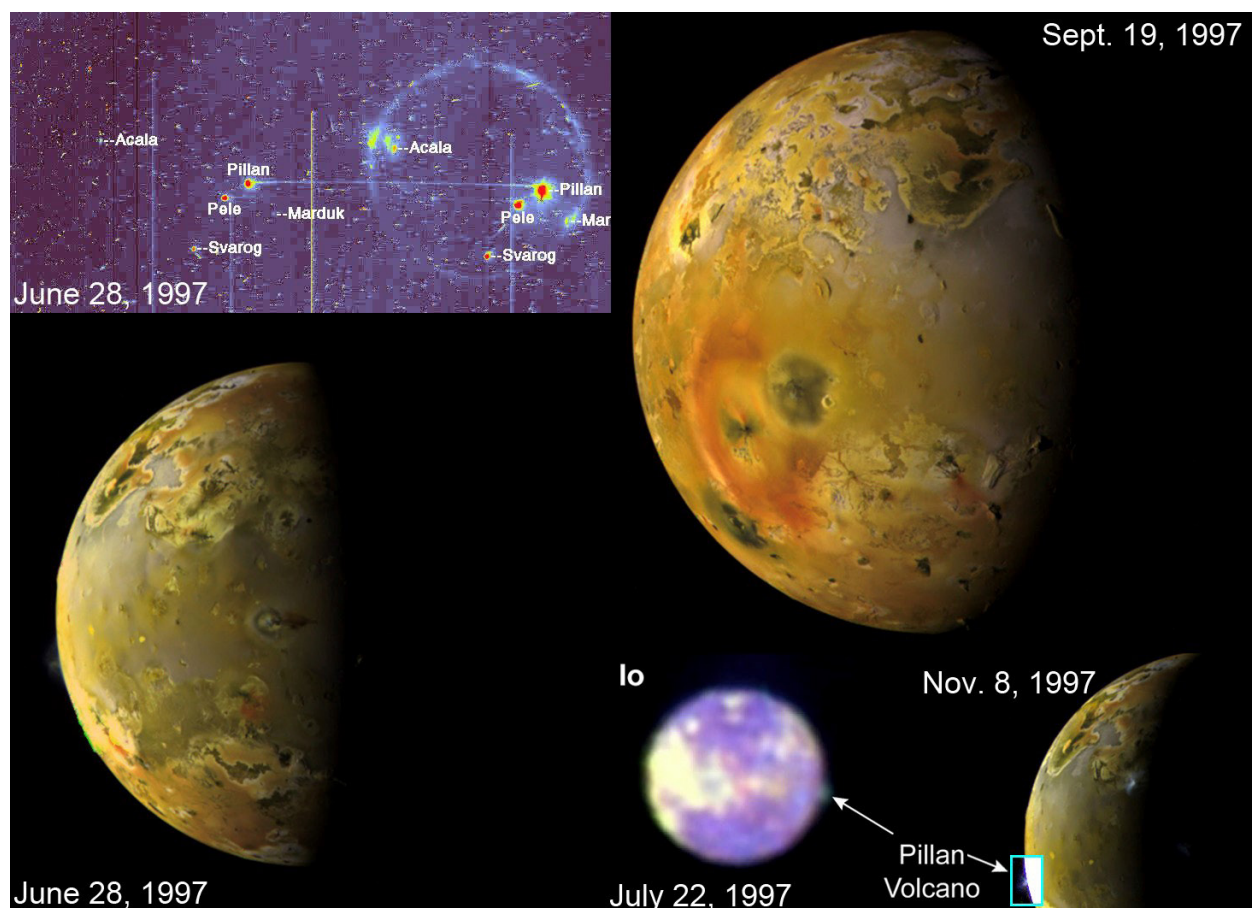

**Figure 13.36.** *Galileo (June/28/1997: daylight and eclipse, Nov/8/1997, Sep/19/1997) and HST (July/22/1997) observations of the eruptions on Io. Montage by Jason Perry.*

---

### Program at a Glance

**Science goal:** Planetary observations to identify surface and environmental changes on the moons of Jupiter and Saturn.

**Program details:** Observing silicate and icy bodies extensively (over days and possibly weeks) allows for the identification of changes such as volcanic or eruptive activity on outer Solar System moons. Measurements into thermal bands will support observations of Io's volcanic activity.

**Instrument(s) + configuration(s):** HDI

**Key observation requirements:** Imaging at 0.5–3.5 µm; several meter-km spatial resolution. Observations of the moons of Jupiter and Saturn require tracking and high-precision pointing.

---

Any of the observations listed above could be the focus of one or more dedicated missions to those bodies. LUVOIR could fulfill, complement, or supplement any mission objectives for a mission to those bodies.

### 13.21.3 The science program

Observations of the moons of Saturn and Jupiter would require planning for optimal viewing of their transits about their planet. Imaging with HDI would be useful for characterizing the composition of these moons and to identify activity there. Ideally, wavelengths should go beyond 3 µm for water observations (longer for thermal observations).

## 13.22 Transit spectroscopy of Earth-sized planets around M-dwarfs

Avi Mandell (NASA GSFC) and Eric Lopez (NASA GSFC)

### 13.22.1 Introduction

Studying transiting planets is highly complementary to studies of directly imaged planets: (1) we can readily measure the mass and radius of transiting planets, linking atmospheric properties to bulk composition and formation, (2) many transiting planets are strongly irradiated resulting in novel atmospheric physics, and (3) the most common temperate terrestrial planets orbit close to red dwarf stars (M-dwarfs) and are difficult to image directly, but are comparatively likely to transit at high signal to noise. The Transiting Exoplanet Survey Satellite (TESS) will discover transiting planets orbiting the brightest stars and should discover a small number of temperate terrestrial planets transiting nearby early-to-mid M-dwarfs. Furthermore, ground-based surveys of very-late M-dwarfs and sub-stellar primaries may yield additional targets—in fact, one of the best Earth-sized HZ targets to date is TRAPPIST-1e (Gillon et al. 2017). Follow-up of these discoveries should provide the first opportunity to place constraints on the atmospheres and habitability of temperate terrestrial planets.

JWST will be a fantastic platform for examining larger and brighter planets, resulting in a revolution in our understanding of hot planets orbiting close to their parent stars. However, characterizing the smaller, cooler worlds will be incredibly time-intensive: JWST will need months of integration time to provide tantalizing constraints on the presence of an atmosphere. The amplitude of spectral features for a temperate terrestrial planet transiting in front of a nearby M-dwarf is comparable to the single-transit photon-counting precision with JWST; therefore, in the absence of a systematic noise floor, 100 transits of such a planet could yield

10-$\sigma$ detections of greenhouse gases—and this neglects the effects of cloud opacity in damping the signal of spectral absorption (Cowan et al. 2014). Spending a total of one month of JWST time to characterize the atmosphere of a potentially habitable world is compelling, but the observations would have to be spread out over nearly a decade for a planet in a month-long orbit (this scheduling problem is somewhat alleviated for planets in the habitable zones of later M-dwarfs and sub-stellar companions, which have shorter orbital periods).

### 13.22.2 The role of LUVOIR

JWST will most likely make important inroads into the exploration of temperate Earth-sized planets around M-dwarfs, but it is entirely possible that little will be known about the atmospheres of these planets by the time JWST ends its mission—and further (and deeper) study will be left for a future flagship mission with equal or greater photon-gathering power.

LUVOIR will be a capable successor to JWST in this regard. LUVOIR Architecture A will have a collecting area 178 m$^2$, a factor of 7x larger than JWST, and observations will reach the same SNR with 2.7x less integration time. The primary instrument for transit spectroscopy with LUVOIR will be the High Definition Imager (HDI), due to the broad simultaneous wavelength coverage (200 nm–2.5 µm) and the ability to spatially scan the spectra of bright stars across the large focal plane detectors. HDI will have sets of grisms and will operate similarly to the Wide Field Camera 3 (WFC3) instrument currently on HST. It will have the capability of full-throughput observations in either short (200–900 nm) or long (800 nm–2.5 µm) wavelength channels, or simultaneous

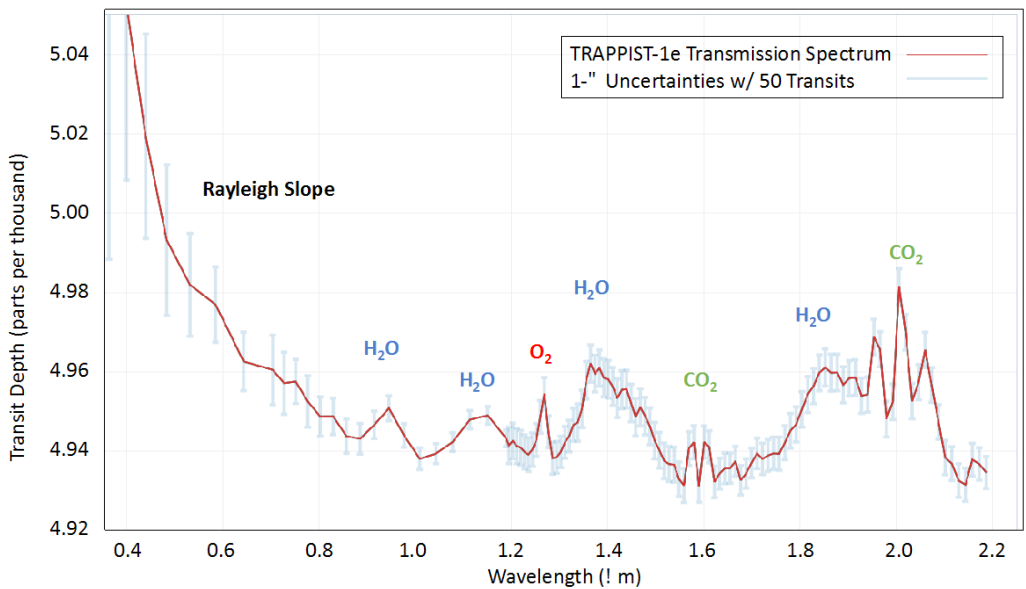

**Figure 13.37.** *Simulated transit spectrum of TRAPPIST-1e with LUVOIR-A (15-m), assuming 50 transits combined. The effective resolution of the spectrum is R=150 at λ > 1.2 μm, R = 30 at 700 nm < λ < 1.2 μm, and R = 10 at λ < 700 nm. Spectral bands of multiple key atmospheric species are visible by eye, enabling constraints on habitability. Credit:* Planetary Spectrum Generator Tool

observations with both bands but at half the throughput. Spectral resolution will be R~500, enabling full characterization of spectral bands of molecular as well as atomic species.

LUVOIR will improve on JWST measurements between 0.8 and 2.5 μm, a band which covers molecular features of $H_2O$ and $CH_4$ and therefore provides constraints on the water vapor content and oxidation state of the atmosphere. At shorter wavelengths (0.2–0.8 μm), LUVOIR's capabilities will be unique. In particular, measurements of

Rayleigh scattering and possibly $O_3$ at 200–300 nm will be the first searches for these key biomarkers of photosynthetic life.

### 13.22.3 The science program

The TRAPPIST-1 system will be a high priority for any future transit spectroscopy science program. **Figure 13.37** illustrates what could be accomplished with 50 transits on the potentially habitable Earth-sized planet TRAPPIST-1e. A number of molecules are visible even by eye, and it is clear that strong constraints on the atmospheric chemistry

---

**Program at a Glance**

**Science goal:** Characterizing the atmospheres of potentially habitable planets to constrain the chemical compositions and search for signs of life.

**Program details:** Transit spectroscopy of Earth-sized planets in the habitable zones of nearby M-dwarf and brown dwarf primaries.

**Instrument(s) + configuration(s):** HDI grism spectroscopy

**Key observation requirements:** 0.4 nm–2.5 μm; R~150; ultra-high precision spectroscopy of bright sources (photometric uncertainty < 3 ppm).

and even on biomarker species such as $O_2$ will be possible.

Sullivan et al. (2015) modeled the yield of planets discovered by the TESS mission, determining that the mission would discover between 2 and 7 temperate Earth-sized planets orbiting M-dwarfs with K < 9. These stellar hosts will be larger than TRAPPIST-1, but will also be brighter, and may therefore provide a similar SNR with the same amount of observing time.

## 13.23 The transmission spectra of rock atmospheres on magma worlds

Eric Lopez (NASA GSFC) and Avi Mandell (NASA GSFC)

### 13.23.1 Introduction

One of the most exciting revelations from planetary transit surveys has been the discovery of a new population of extremely irradiated rocky planets on ultra-short-period (USP) orbits (e.g., Charbonneau et al. 2009; Léger et al. 2009; Batalha et al. 2010; Sanchis-Ojeda et al. 2014). The irradiation on these planets is so intense, with dayside temperatures reaching over 2500 K, that they are expected to have dayside surfaces that are partially or completely molten (Kite et al. 2017). Although completely uninhabitable, these planets have the potential to teach us a great deal about geophysical processes under extreme conditions. Indeed, models predict that these planets should partially vaporize their rocky mantles and outgas significant but highly refractory atmospheres dominated by silicates and heavy metals, potentially allowing us to directly probe their bulk mantle compositions (Miguel et al. 2011). Moreover, models also predict that alkali metals in this outgassed material may be escaping at significant rates, producing highly extended metallic exospheres possibly reaching all the way to the Roche lobe (Kite et al. 2017). However, these exospheres would be highly time-variable, driving the need to obtain higher S/N over a single transit. Indeed, recently there was a claimed detection of significant thermal variability in the dayside emission from the USP 55 Cancri e (Demory et al. 2016) along with the possible detection of highly extended exospheric Na and Ca+ absorption (Ridden-Harper et al. 2016).

Given their short periods and extremely high temperatures, JWST is extremely well suited for characterizing the overall thermal emission from these planets (Samuel et al. 2014) and to detect emission from SiO at 4 and 10 microns (Ito et al. 2015). However, due to its wavelength coverage JWST will be limited in its ability characterize the atmospheric compositions of USP planets with transmission or emission spectroscopy since many of the primary lines for the relevant metal species aside from SiO are found in the optical and near UV between 300 and 800 nm (e.g., Na, K, Ca, Mg, Fe, TiO, VO), and therefore are inaccessible with JWST. At the same time, the intrinsic rarity of these USP planets means that even after the NASA's TESS mission concludes, most of the known USPs will be around relatively faint FGK stars (V-mag >11), making them difficult to characterize with HST due to its smaller aperture.

### 13.23.2 The role of LUVOIR

With its combination of large aperture and broad UV and optical wavelength coverage, LUVOIR using will be uniquely suited to probe the compositions of vaporized atmospheres on USP planets. Between 300 and 800 nm, a range covered by the UVIS channel on HDI, are a series major spectral features for the key metal species predicted by outgassing models. In particular, features like sodium D at 590 nm, calcium H & K at 390 nm, potassium at 770nm, and the TiO from 600 to 800 nm, should all be detectable in either transmission or emission spectroscopy along with a wide range of iron and magnesium lines between 300 and 600 nm. For example, modeling the emission spectra of several known USPs, Ito et al. (2015) predicted that both the 590 nm Na and 770 nm K features should be observable with moderate spectral resolution (R~500) in emission at ~9 ppm for a 2500 K USP planet and ~36 ppm for a 3000 K planet.

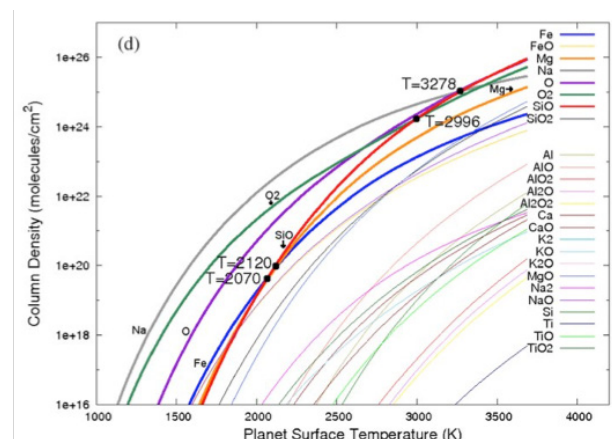

**Figure 13.38.** *Predicted compositions of outgassed atmospheres for a Bulk Silicate Earth composition from Miguel et al. (2011), black points show transitions where the dominant species changes.*

Additionally, if these planets do intend possess extended exospheres then these should be detectable in transmission at S/N in single transit, which is important given the expected variability in both the planetary atmospheric escape rates and the stellar spectrum. The possible detection of exospheric Na & Ca+ on 55 Cancri e by Ridden-Harper et al. (2016) claimed a 7% transit in the 590 nm sodium line extending all the way to the planet's Roche lobe at ~5 Earth radii.

Given the extremely short orbital periods, typical transit durations are just 1.5–2 hours. However, the 15-m LUVOIR Architecture A should be able to reach a precision of ~16 ppm at a R~500 for a 12th magnitude G-star at ~150 pc in a single 1.5-hour integration. This will allow the detection of any exospheric metals at high S/N in single transit, while for a planet with dayside temperature of 3000 K, sodium should be detectable in emission from the lower atmosphere at ~5σ with just 5 transits or ~10 hours of integration.

### 13.23.3 The science program

Given the relatively short integration times needed, it should be possible to obtain multi-epoch observations for multiple spectral features on multiple exoplanets, especially in transmission. Currently there are already five well-studied USP planets (55 Cancri e, CoRoT-7b, Kepler-10b, Kepler-78b, and K2-141b), which would be well suited to these observations, with the likelihood that at least a few more will be found by upcoming surveys like TESS and PLATO. With just ~50 hours of observation time per planet it will be possible to obtain ~10 transits and 5–20 eclipses per planet, which will allow us to search for extended exospheres in multiple species, including all the common alkalis, characterize those species in emission close to the planet's surface, and study their time variability.

---

**Program at a Glance**

**Science goal:** Characterizing the compositions of vaporized silicate and metal atmospheres from extremely hot molten rocky planets.

**Program details:** ~250 hours for multi-wavelength and multi-epoch NUV and optical transmission and emission spectroscopy for the best ~5 ultra-short-period exoplanets.

**Instrument(s) + configuration(s):** HDI UVIS spectroscopy

**Key observation requirements:** 300–800 nm; R=500; ultra-high precision spectroscopy of bright sources (photometric uncertainty ~ 2–7 ppm)

---

## 13.24 A statistical search for global habitability and biospheres beyond Earth

Shawn Domagal-Goldman (NASA GSFC), Jacob Bean (University of Chicago), Eliza Kempton (Grinnell College, University of Maryland), Sara Walker (Arizona State University)

### 13.24.1 Introduction

Here, we propose a statistical search for habitability and life. Such a search would provide tests for our understanding of the top-level factors controlling planetary habitability and raise the overall confidence that we are not alone. This search would be conducted by searching for signs of habitability and of life on multiple planets, both in and beyond the habitable zone.

The statistical search for habitability would turn the concept of the habitable zone into a testable hypothesis. According to the habitable zone hypothesis, there is a region around the star for which feedback processes in the climate system work to stabilize surface temperatures to allow for sustained, global reservoirs of liquid surface water. Beyond this zone, the climate would stabilize in water-poor regimes, with vastly different surface temperatures. Because of all the feedbacks in the climate system, multiple aspects of a planet's spectral signatures should also change beyond the habitable zone: $CO_2$ concentrations (and their features) should increase on either side of the habitable zone (due to lack of a $CO_2$ weathering feedback); water-soluble gases and aerosols should increase in concentration; and water vapor and water cloud features themselves should disappear. For a more thorough discussion of this concept, and how it can begin with near-term observations, see Bean, Abbot, & Kempton (2017).

A statistical search for life should raise the overall confidence in our detection of life beyond Earth compared to what is achievable by the search for life on a single exoplanet (Walker et al., 2018). For example, consider an exoplanet for which LUVOIR detects the presence of water ($H_2O$), molecular oxygen ($O_2$) and ozone ($O_3$), but not methane ($CH_4$). This is a plausible scenario, given the long integration times required to detect $CH_4$ in LUVOIR's wavelength range. The combination of these measurements, as well as constraints on UV fluxes from observations of the host star, would produce a scenario for which the most likely explanation is life on that planet. However, the lack of $CH_4$ on that planet may be enough to prevent a conclusion as bold as "We are not alone." As an alternative, consider a scenario where LUVOIR detects this same set of features multiple times on worlds in different planetary systems. This would suggest that either our understanding of photochemical processes is woefully incomplete and under-predicting some $O_2$-producing mechanism, or that life is present on at least some subset of these worlds.

This example would also apply to other biosignatures. An organic haze in the presence of a $CO_2$-rich atmosphere has been recently proposed as a biosignature (Arney et al., 2018). While consideration to false positives has been given, a haze-biosignature has not undergone the same amount of thorough scrutiny as $O_2/O_3$. Thus, one might place a lower amount of confidence in this biosignature than the detection of $O_2/O_3$. To quantify this example, if we are to assume that ~50% of $CO_2$-rich, haze-bearing habitable zone worlds have life, the detection of multiple such planets would increase the likelihood that *at least one* of these planets has a biosphere (whereas a search focused on a single target would not be able to draw such a strong conclusion).

This search does not necessarily need to focus on the same signature being present on multiple worlds. It would also apply to any combination of potential biosignature across multiple planets, and thereby allow the mission to conduct both a broad survey for easy-to-detect biosignatures on multiple planets and a detailed assessment of the biosignatures on LUVOIR's best targets. The approach could therefore by optimized so that detailed assessments (such as the measurement of longitudinal-dependent colors/spectra) are only conducted on relatively close targets, for which such assessments are capable of being conducted with reasonable investments of telescope time.

A statistical search will also be useful in the case of no (or very poor) signs of life being detected by LUVOIR. Instead of just a full null result, the process of quantifying the likelihood of life on each planet, with a statistical combination of those likelihoods, will yield an overall confidence level that sets an upper-bound on the frequency of life. Such a conclusion is not really possible unless we conduct a statistical survey of multiple worlds rather than a deep dive on a single planet. Further, the more planets we assess, the stronger our conclusions will be. Detecting no biosignatures on 55 habitable-zone planets would give a 2-$\sigma$ upper limit on the frequency of life of 6%. Thus, our uncertainty on "eta$_{life}$" would decrease by an order of magnitude, from a range of 0–1.0 to a range of 0–0.06.

Again, it is useful to consider an example: if LUVOIR finds multiple Earth-sized worlds in the habitable zone of other stars and determines that many of them have "anti-biosignatures" (such as high atmospheric concentrations of $H_2$ and either CO or $CO_2$), the confidence level that life is rare might be high.

There are two requirements to such an approach to the search for life:

1) The science community must be able to ascribe a quantitative estimate a given data set resulted from a biosphere on any particular planet; and

2) We must be able to collect spectroscopic data on many worlds.

The former requirement demands development in the research tools utilized by the astrobiology and exoplanet communities. Options for quantifying our assessment that a given planet has a biosphere are discussed and reviewed by Walker et al. (2018, in press), and leverage similar efforts to quantify signs of life in our own Solar System. The latter requirement inevitably leads to a LUVOIR-sized telescope.

### 13.24.2 The role of LUVOIR

A statistical approach to habitability and the habitable zone does not require LUVOIR, as it could begin with ground- and space-based surveys for $H_2O$ and $CO_2$ features via transit spectroscopy (Bean et al., 2017). LUVOIR would extend this statistical approach to new targets, with new observational techniques. Specifically, LUVOIR would expand the search to planets around Sun-like (F, G, K) stars. It would also provide new information on clouds and aerosols on planets around Sun-like and M-type stars, via a wavelength range that is complementary to currently-planned transit spectroscopy observatories.

While other missions might attempt to search for signs of life, and perhaps detect them on one world, a statistical approach requires that this search be conducted on a large number of worlds. This, in turn, demands a large-aperture telescope, to drive up the yield of planets that the mission can discover and characterize. LUVOIR will be able to detect over 50 such planets. This is a large enough sample size to conduct a statistical

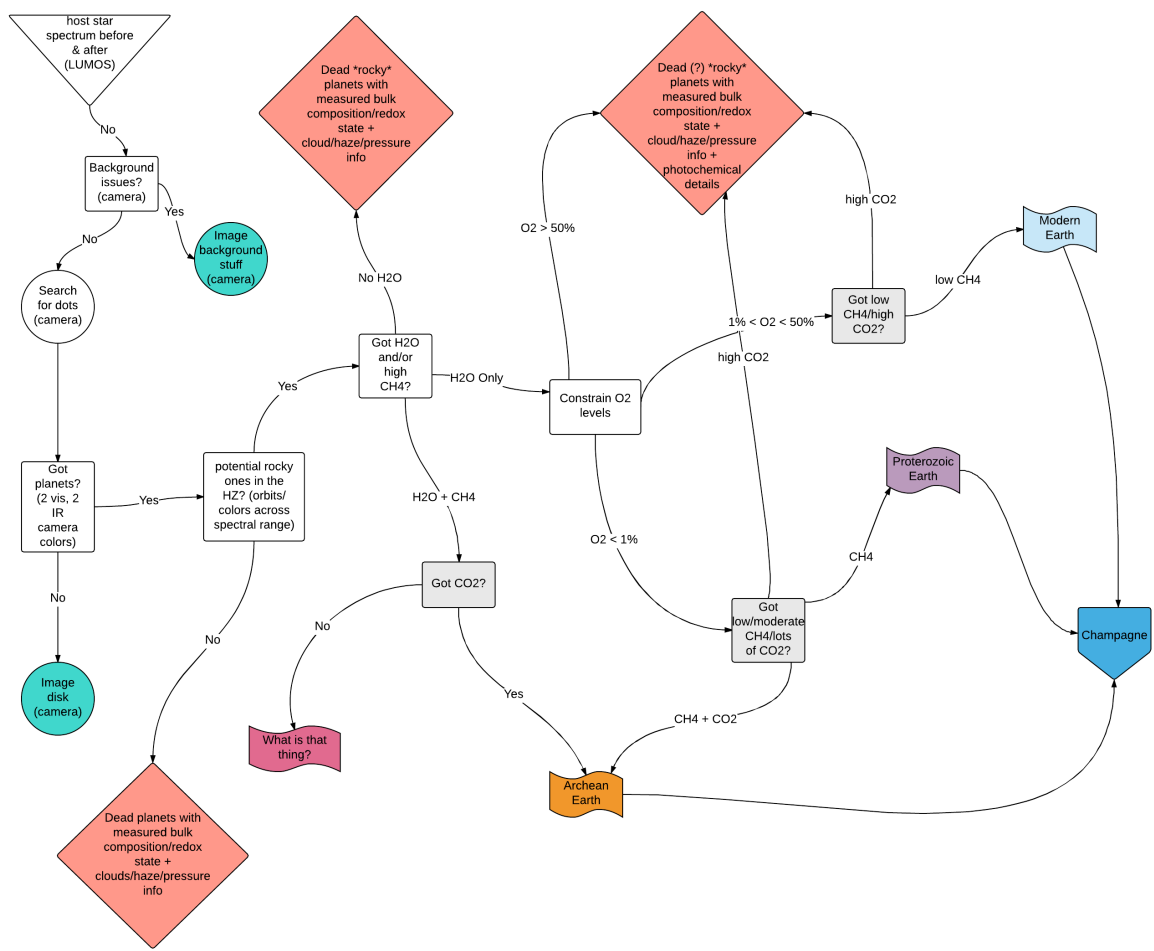

**Figure 13.39.** *An example of the observing strategy for a statistical approach to the search for life on biosignatures. Credit: S. Domagal-Goldman (NASA GSFC)*

search for signs of life on exoplanets. If detectable biospheres exist on at least 10% of potentially habitable worlds, LUVOIR will find at least one such biosphere. But if life is relatively common, LUVOIR has the potential to find signs of life on many worlds. If it is rare, we will be able to set upper bounds on the frequency of life.

### 13.24.3 The science program

A statistical search for habitability and life would be consistent with the main exoplanet observing program outlined in **Chapter 3**. It would be conducted almost exclusively via ECLIPS observations of exoplanets in the habitable zones of their host stars, and LUMOS observations would obtain the UV spectral energy distribution of those host stars. It would use different bands within the ECLIPS wavelength range, depending on the distance to the target and the spectral feature in question, and likely would run ECLIPS to the point where integration on individual bands becomes time-limited. Details of this approach are below.

The main difference of the approach outlined here to the main exoplanet characterization program is in how optimized the biosignature search becomes, and how the data from that search are processed.

---

**Program at a Glance**

**Science goal:** To find signs of life—or conclusively rule out life in our Sun's local neighborhood.

**Program details:** This requires an optimized search for biosignatures

**Instrument(s) + configuration(s):** This would require all of the bandpasses and spectroscopy modes of ECLIPS.

**Key observation requirements:** The main requirement here is a broad wavelength range for ECLIPS, running from ~0.2 μm to ~2 μm, so that different signatures with different integration times can be detected. This also requires a large-aperture telescope, in order to detect and characterize dozens of potentially habitable exoplanets).

---

This approach would likely *require* an optimized search for biosignatures, spending as little time as possible confirming/rejecting biosignatures on each planet. This search would likely focus on planets with atmospheric $H_2O$, which would be found during LUVOIR's planet detection phase. In addition to weeding out dry planets, this initial phase would also include a search for $H_2O$ beyond the habitable zone. This would help test the habitable zone concept by mapping the presence of $H_2O$ water vapor as a function of stellar irradiation (Bean et al. 2017).

Then, the search would turn towards easy-to-detect, but low-confidence biosignatures, such as $O_3$ and atmospheric organic haze, both of which can be detected in the UV channel of ECLIPS. This initial biosignature search would include some planets near—but not in—the habitable zone. This would provide a basis for comparison of presumed "dead" planets, which could lead to more robust conclusions about the detection of life and help inform the science community if something about our understanding of photochemical and geological processes is dramatically incomplete.

For planets with $O_3$ or an organic haze, LUVOIR would proceed to the detection of $O_2$ (for $O_3$-bearing worlds) and $CO_2$ (for haze-bearing worlds). However, it would only do so if the detection of this second feature was permissible in a reasonable integration time. (The definition of "reasonable" would be up to a future time allocation committee, but history suggests $\lesssim$ 100 hours.) If $O_2$ or $CO_2$ is detected, LUVOIR would then proceed to further spectral measurements to rule out false positives, search for secondary biosignature gases (such as $CH_4$) or attempt to characterize the surface. Again, this would be limited on a target-by-target bases based on required integration time. A full decision tree for such a search is shown in **Figure 13.39**.

The totality of these observations, at varying degrees of detail depending on the target, would maximize LUVOIR's ability to test our ideas of planetary habitability and search for life across the set of exoplanets it characterizes.

---

# 14 Appendix B: Science calculation and simulation details

## 14.1  Estimating LUVOIR's exoplanet yield

Christopher C. Stark (STScI)

### 14.1.1  Methodology

To estimate the yield of directly imaged planets, we assume that LUVOIR must conduct a blind survey to search for and characterize potentially Earth-like exoplanets. While the efficiency of the LUVOIR exoplanet survey and the quality of its data products would benefit from a precursor survey identifying potentially Earth-like planets, we conservatively assume such a survey does not exist at the time of launch. Thus, the yield of such a blind survey is a probabilistic quantity, which depends on LUVOIR's coronagraphic capabilities, the occurrence rate of planets of various types, their detectability, and the unknown distribution of planets around individual nearby stars.

To calculate expected exoplanet yields, we used the Altruistic Yield Optimization (AYO) yield code of Stark et al. (2014), which employs the completeness techniques introduced by Brown (2005). Briefly, for each star in our master target list, we randomly distribute a large number of synthetic planets of a given type, forming a "cloud" of synthetic planets around each star, as shown in **Figure 14.1**. Planet types are defined by a range of radii, albedo, and orbital elements. We *calculate the flux from each synthetic planet in reflected light given its properties, orbit, and phase*, and then determine the exposure time required to detect it at SNR=7. Based on these detection times and the exposure time of a given observation, we can calculate the fraction of the synthetic planets that are detectable, i.e., the completeness, as a function of exposure time. The completeness simply expresses the probability of detecting that planet type, if such a planet exists. The average yield of an observation is the product of the completeness and the occurrence rate of a given planet type. We repeat this process for every observation until the total program lifetime is exceeded, arriving at an average total mission yield. In reality, yields may vary from this average due to the random distribution of planets around individual stars; we incorporate this source of uncertainty in our yield calculations by accounting for the Poisson probability distribution of planets for each star.

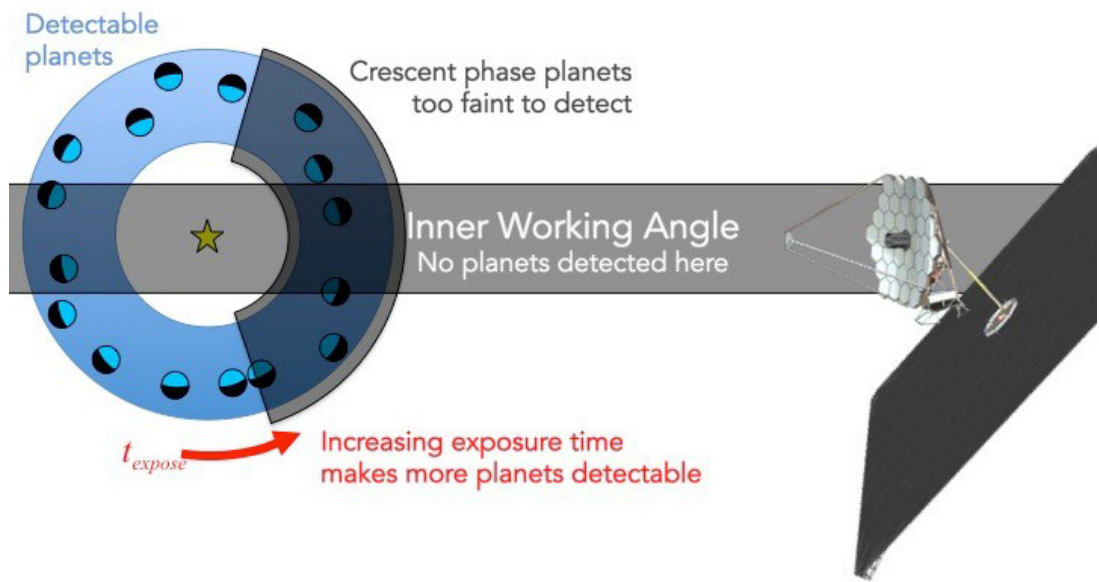

**Figure 14.1.** *The completeness of an observation is the fraction of detectable planets to total planets and is a function of the exposure time. The yield of an observation is the product of completeness and the probability that such a planet actually exists (the occurrence rate).*

We employ the techniques of Stark et al. (2015) and Stark et al. (2016), which optimize the observation plan to maximize the yield of potentially Earth-like planets. For a coronagraph-based search, this involves optimizing the targets selected for observation, the exposure time of each observation, the delay time between each observation of a given star, the number of observations of each star, and the planet phase for spectral characterization (Stark et al. 2015). We do not explicitly schedule the observations. We expect the ability to schedule the observations will have a negligible impact on the exoplanet yield given LUVOIR's extremely large field of regard and rapid retargeting capabilities.

### 14.1.2    Inputs and assumptions

Yield estimates require simulating the execution of a mission at a high level. They are therefore dependent on a large number of assumptions about the target stars, the planetary systems they host, and the capabilities of the mission. Given the inherent uncertainties in many of these assumptions, consistency between yield analyses is of primary importance. We adopt inputs and assumptions that are consistent with the choices made by the Exoplanets Standard Definitions and Analysis Team and those made by the HabEx STDT. We now review and justify our fiducial assumptions about the parameters that affect the yield.

#### 14.1.2.1  Astrophysical assumptions

### 14.1.2.1.1 Planet types & occurrence rates

We followed the planet categorization scheme of Kopparapu et al. (2018), which consists of a 3 by 5 grid of planets binned by temperature (hot, warm, and cold) and planet radius ("rocky," "super-Earths," "sub-Neptunes," "Neptunes," and "Jupiters"), as shown in **Figure 14.2**. Each planet was assigned a single albedo defined by its radius (listed in **Figure 14.2**), a Lambertian phase function, and all planets were assumed to be on circular orbits. We assume that the semi-major axis boundaries that define the temperature bins of each planet type scale with the bolometric stellar insolation, such that they scale with the square root of the bolometric stellar luminosity.

For exoEarth candidates we adopted the green outlined region in **Figure 14.2**. By this definition, exoEarth candidates are on circular orbits and reside within the conservative HZ, spanning 0.95–1.67 AU for a solar-twin star (Kopparapu et al. 2013). We only include planets with radii smaller than 1.4 Earth radii and radii larger than or equal to $0.8a^{-0.5}$, where $a$ is semi-major axis for a solar-twin star. The lower limit on our definition of the radius of exoEarth candidate is derived from an empirical atmospheric loss relationship derived from solar system bodies (Zahnle & Catling 2017). The upper limit on planet radius is a conservative interpretation of an empirically measured transition between rocky and gaseous planets at smaller semi-major axes (Rogers 2015). All exoEarth candidates were assigned Earth's geometric albedo of 0.2, assumed to be valid at all wavelengths of interest.

We adopted the exoplanet occurrence rate values from the NASA ExoPAG SAG13 meta-analysis (Kopparapu et al. 2018), given by

$$\frac{d^2 N(R,P)}{d \ln R \, d \ln P} = \Gamma R^\alpha P^\beta,$$

where N(R,P) is the number of planets per star in a bin centered on radius R and period P, R is in Earth radii and P is in years, and $[\Gamma, \alpha, \beta] = [0.38, -0.19, 0.26]$ for $R < 3.4$ $R_{Earth}$ and $[\Gamma, \alpha, \beta] = [0.73, -1.18, 0.59]$ for $R \geq 3.4$ $R_{Earth}$. **Figure 14.2** lists the occurrence rates when integrating over the boundaries of each planet type. Within each planet type, we adopted the SAG13 radius and period distribution above. With this distribution, within a given planet temperature bin, small planets outnumber large planets.

The SAG13 meta-analysis is a crowd-sourced average of published and unpublished occurrence rates, averaged over FGK spectral types. Uncertainties on the SAG13 occurrence rates are not well understood and are simply set to the standard deviation of the crowd-sourced values. Because of the large uncertainties in the SAG13 occurrence rates, we have weak constraints on how occurrence rates change with spectral type. Thus, we simply assume that the occurrence rates for each planet type bin are independent of spectral type. **Table 14.1** summarizes the key astrophysical assumptions underlying our exoEarth candidate yield calculations.

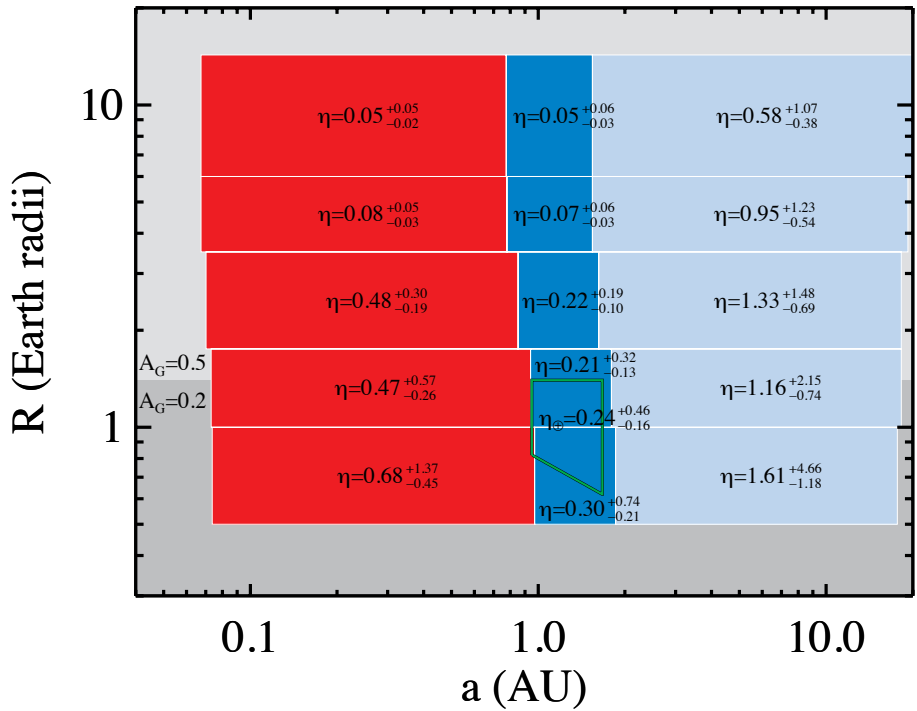

**Figure 14.2.** *Planet classifications for a solar twin used for yield modeling, including bin-integrated occurrence rates ($\eta$) and geometric albedos ($A_G$). Planets are binned into hot (red), warm (blue), and cold (ice blue) temperature bins and rocky, super-Earth, sub-Neptune, Neptune, and Jupiter size bins. The green outline indicates the boundaries of exoEarth candidates. The semi-major axis boundaries shown are for a solar-twin star; for other types of stars, semi-major axis boundaries are scaled to maintain a constant bolometric stellar insolation.*

**Table 14.1.** *Summary of astrophysical assumptions.*

| Parameter | Value | Description |
|---|---|---|
| $\eta_\oplus$ | 0.24 | Fraction of Sun-like stars with an exoEarth candidate |
| $R_{\rm p}$ | $[0.6, 1.4]\ R_\oplus$ | Planet radius[a] |
| $a$ | $[0.95, 1.67]$ AU | Semi-major axis[b] |
| $e$ | 0 | Eccentricity (circular orbits) |
| $\cos i$ | $[-1, 1]$ | Cosine of inclination (uniform distribution) |
| $\omega$ | $[0, 2\pi]$ | Argument of pericenter (uniform distribution) |
| $M$ | $[0, 2\pi]$ | Mean anomaly (uniform distribution) |
| $\Phi$ | Lambertian | Phase function |
| $A_G$ | 0.2 | Geometric albedo of planet from 0.55–1 $\mu m$ |
| $z_c$ | 23 mag arcsec$^{-2}$ | Average V band surface brightness of zodiacal light for coronagraph observations[c] |
| $z_s$ | 22 mag arcsec$^{-2}$ | Average V band surface brightness of zodiacal light for starshade observations[c] |
| $x$ | 22 mag arcsec$^{-2}$ | V band surface brightness of 1 zodi of exozodiacal dust[d] |
| $n$ | 3 | Number of zodis for all stars |

[a]Distribution is a function of $a$ according to the SAG13 occurrence rates.

[b]$a$ given for a solar twin. The habitable zone is scaled by $\sqrt{L_\star/L_\odot}$.

[c]Local zodi calculated based on ecliptic pointing of telescope. On average, starshade observes into brighter zodiacal light.

[d]For Solar twin. Varies with spectral type, as zodi definition fixes optical depth.

## 14.1.2.1.2 Exozodiacal & zodiacal dust

Exozodiacal dust adds background noise, thereby reducing the SNR of a planet detection relative to the case of no exozodiacal dust. Unfortunately, we currently have only weak constraints on the amount of zodiacal dust around our target stars. Upcoming Large Binocular Telescope Interferometer (LBTI) measurements will better inform this current working assumption in the near term, but at the time of this report are unavailable.

We therefore simply adopted a baseline exozodi level of 3 zodis. Our definition of 1 zodi is a uniform (optically-thin) optical depth producing a V band surface brightness of 22 mag arcsec$^{-2}$ at a projected separation of 1 AU around a solar twin. Thus, the exozodi surface brightness drops off as the inverse square of the projected separation (Stark et al. 2014). Because the HZ boundaries scale by the *bolometric* stellar insolation, the V band surface brightness of 1 zodi of exozodi varies with spectral type (Stark et al. 2014).

The solar system's zodiacal brightness varies with ecliptic latitude and longitude; the closer one observes toward the Sun, the brighter the zodiacal cloud will appear. We calculated the zodiacal brightness for each target star by making simple assumptions about typical telescope pointing (Leinert et al. 1998). We assume the coronagraph can observe near where the local zodi is minimized and adopted a solar longitude of 135 degrees for all targets.

## 14.1.2.1.3 Target catalog

Our input star catalog was formed using the methods of Stark et al. (in prep). Briefly, the target list is equivalent to the union of the original Hipparcos catalog and the GAIA TGAS catalog. For each star, we adopted the most recent measured parallax value from the original Hipparcos, updated Hipparcos, and GAIA TGAS catalogs, then down-selected to stars within 50 pc. BVI photometry and spectral types were obtained from the Hipparcos catalog. Additional bands and missing spectral types were supplemented using SIMBAD. We filtered out all stars identified as luminosity class I–III, leaving only main sequence stars, sub-giants, and unclassified luminosity classes.

We note that while the accuracy of any individual star's parameters may be important when planning actual observations, yield estimates can be very robust to these inaccuracies, as their effects average out when considering a large target sample. Given LUVOIR's extremely large sample size of hundreds of stars, inaccuracies in the target catalog should have a negligible impact on yield.

## 14.1.2.2  Binary stars

Detecting exoplanets in binary star systems presents additional challenges. Light from companion stars outside of the coronagraph's field of view, but within the field of view of the telescope, will reflect off the primary and secondary mirrors. Due to high-frequency surface figure errors and contamination, some of this light is scattered into the coronagraph's field of view. For some binary systems, this stray light can become brighter than an exoEarth.

We directly calculate the stray light from binary stars in the final image plane. We adopt the numerical stray light models of Sirbu et al. (2018, in prep). These models predict the power in the wings of the PSF at large separations assuming a λ/20 RMS surface roughness and an f⁻³ envelope, where f is the spatial frequency of optical aberrations. We assume that the stray light can be measured or modeled and include it simply as an additional source of background noise. We make no artificial cuts to the target list based on binarity and allow the benefit-to-cost optimization in the AYO yield code to determine whether or not stray light noise makes a target unobservable. In practice, the AYO prioritization does reject a number of binary systems with contrast ratios close to unity and/or close separations. We note that including the full amount of light scattered by the companion is actually conservative, as the companion scattered starlight could be actively reduced with specialized observation methods. For example, LUVOIR could use multi-star or super-Nyquist wavefront control coronagraphic techniques (Thomas et al. 2015; Sirbu et al. 2017).

### 14.1.2.3  Mission parameters

We adopted a total of 2 years for the exoEarth survey time (including overheads). Coronagraph observations were assessed a 5 hour overhead on all exposures for wavefront control, a reasonable assumption based on the WFIRST CGI operations concept. Total exposure time and overheads were required to fit within the exoplanet science time budget.

For planet detections, we required an SNR=7 evaluated over the full bandpass of the detection instrument, where both signal and noise are evaluated in a simple photometric aperture of 0.7 $\lambda$/D in radius (where in this case, D is the diameter of the outer edge of the Lyot stop). The SNR was evaluated according to Eq. 7 in Stark et al. (2014), which includes a conservative factor of 2 on the background Poisson noise to account for a simple background subtraction. We also included a background term for detector noise, discussed in **Section 14.1.2.4.2**. For spectral characterizations, we required a spectrum with R=70 and SNR=5 per spectral channel which we evaluated at a wavelength of 1 micron, to search for water vapor in all detected exoEarth candidates.

### 14.1.2.4  Instrument performance assumptions

### 14.1.2.4.1  Coronagraph assumptions

Coronagraph performance was estimated via a wave propagation model, assuming an idealized optical system and perfect wavefront control. We adopted two coronagraphs designed for the LUVOIR Architecture A segmented, on-axis primary: an apodized pupil Lyot coronagraph (APLC) and a charge 6 vortex coronagraph (VC). The APLC coronagraph consisted of 3 masks: a small-IWA mask (10% bandwidth, 3.8–12 $\lambda/D$ working angle), a medium-IWA mask (15% bandwidth, 6–20 $\lambda/D$ working angle), and a large-IWA mask (11–33 $\lambda D$ working angle). To each star we independently assigned either one of the APLC masks or the VC coronagraph, based on the wavelength of observation and a cost-to-benefit ratio. We simulated the leaked starlight as a function of stellar diameter and the off-axis PSFs as a function of angular separation, providing inputs to the yield calculations according to the standards of Stark & Krist (2017).

The wave propagation model does not include some known systematic noise sources, such as residual spatial speckle noise caused by dynamic wavefront errors. Realistic estimates would require full end-to-end simulations of a well-defined telescope, instrument, and observing procedure. These effects will impact the coronagraph noise floor, the properties and frequency of false positives, and the final yield.

The APLC coronagraph design consists of a binary apodizer mask in the entrance pupil, followed by an image plane coronagraph mask, followed by a Lyot stop (N'Diaye et al. 2016). The apodizer is optimized to maximize coronagraph throughput while achieving $10^{-10}$ raw contrast over a desired range of working angles (e.g., Zimmerman et al. 2016). In the end, we designed three separate masks with overlapping working angles, allowing us to maximize the throughput for any size habitable zone. The resulting APLC designs are robust to stellar diameter and jitter (tolerance ~1 mas stellar diameter and ~0.5 mas jitter respectively for the LUVOIR Architecture A primary at 600 nm).

The apodized vortex coronagraph design (Mawet et al. 2013) consists of a grayscale apodizer mask in the entrance pupil, followed by a vortex phase mask in the focal plane (Mawet et al. 2009), and an annular Lyot stop. We chose a charge 6 vortex mask to balance sensitivity to low order aberrations and throughput at small angular separations (Ruane et al. 2017). We used the Auxiliary

**Table 14.2.** *Summary of adopted coronagraph performance. Listed contrast is for a theoretical point source; contrasts used in simulations included the effects of finite stellar diameter. While only the spatially averaged raw contrast and coronagraph throughput are indicated, AYO simulations used their actual values at the planet angular separation.*

| Parameter | $APLC_1$ | $APLC_2$ | $APLC_3$ | VC | Description |
|---|---|---|---|---|---|
| $\zeta$ | $4 \times 10^{-11}$ | $5 \times 10^{-11}$ | $5 \times 10^{-11}$ | $3 \times 10^{-10}$ | Raw contrast[a] |
| $\Delta mag_{floor}$ | 26.5 | 26.5 | 26.5 | 26.5 | Systematic noise floor (faintest detectable point source) |
| $T_{core}$ | 0.19 | 0.26 | 0.26 | 0.17 | Coronagraphic core throughput[a] |
| $T$ | 0.23 | 0.23 | 0.23 | 0.23 | End-to-end VIS channel detection throughput (including QE, excluding core throughput) |
| IWA | 3.8 | 6.3 | 11.4 | 3.7 | Inner working angle ($\lambda/D$) |
| OWA | 11.5 | 19.5 | 33 | 10 | Outer working angle ($\lambda/D$)[b] |
| $\Delta\lambda$ | 10% | 15% | 15% | 20% | Bandwidth |

[a]Average value between the IWA and OWA.

[b]Separation at which core throughput reaches half the maximum value.

Field Optimization algorithm (Jewell et al. 2017) to optimize the grayscale apodizer and reduce the diffracted starlight due to the central obscuration, the secondary mirror support structures, and gap between mirror segments in an annular region about the star.

**Table 14.2** summarizes the coronagraph performance that we adopted. We note that although these metrics may provide a useful high-level understanding of coronagraph performance, some metrics should be interpreted with caution. For example, the inner working angle (IWA) estimates where the planet's throughput reaches 50% of the maximum value, but this does not mean that there is no planet signal interior to the IWA. On the contrary, the vortex coronagraph provides useful (albeit lower) throughput down to ~2 $\lambda$/D, such that bright, short-period planets may be detectable interior to the quoted 3.7 $\lambda$/D IWA.

The bandpass of the APLC designs is limited by the design of the apodized masks and was chosen based on a throughput-bandwidth tradeoff analysis. To work with a segmented on-axis telescope, the vortex coronagraph design also uses an apodized mask, which was designed assuming 20% bandwidth. All of these bandwidths are less than or equal to the expected simultaneous bandwidth of the wavefront control system. High Contrast Imaging Testbed results indicate that surpassing a bandwidth of $\Delta\lambda/\lambda=0.2$ is challenging with a conventional dual DM coronagraph layout, thereby justifying our adopted maximum bandwidth of 20%.

The total throughput of the system in **Table 14.2** is evaluated at visible wavelengths and includes the reflectivity of all optical surfaces, the detector quantum efficiency (QE), detector readout inefficiencies in photon-counting mode, IFS throughput, and a 5% contamination budget. Detector parameters are discussed below. This throughput metric does not include the core throughput of the coronagraph, which was taken into account separately via the off-axis PSF simulations discussed above.

### 14.1.2.4.2 Detector & other performance assumptions

**Table 14.3** lists the detector noise parameters that we adopted for yield calculations. We calculated the total detector noise count rate in the photometric aperture as

$$\mathrm{CR_{b,detector}} \approx n_{\mathrm{pix}}\left(\xi + \mathrm{RN}^2/\tau_{\mathrm{expose}} + 6.73 f\mathrm{CIC}\right),$$

where f is the photon counting rate and $n_{\mathrm{pix}}$ is the number of pixels contributing to the signal and noise. We tuned f to each individual target, such that our photon-counting detector time-resolves photons from sources 10 times a bright as an Earth-twin at quadrature.

**Table 14.3.** *Photon-counting CCD noise parameters adopted for yield modeling.*

| Parameter | Value | Description |
|-----------|-------|-------------|
| $\xi$ | $3 \times 10^{-5}$ counts pix$^{-1}$ sec$^{-1}$ | Dark current |
| RN | 0 counts pix$^{-1}$ read$^{-1}$ | Read noise (N/A) |
| $\tau_{\mathrm{read}}$ | 1000 s | Read time (N/A) |
| CIC | $1.3 \times 10^{-3}$ counts pix$^{-1}$ clock$^{-1}$ | Clock induced charge |

We assumed the IFS splits the core of the PSF into 4 lenslets at the shortest wavelength, each of which are dispersed into 6 pixels per spectral channel for a total of 24 pixels per spectral channel at the shortest wavelength. For broadband coronagraphic detections using the imager, we adopted 4 pixels for the core of the planet. We note that the assumed detector noise is sufficiently low that small changes to the number of pixels have a negligible impact on yield.

### 14.1.3 Operations concepts

Yield is commonly thought of as the number of planets detected and/or characterized. As shown by Stark et al. (2016), the yield of a mission is very sensitive to precisely what measurements are required for "characterization," and how the mission goes about making those measurements. Thus, the yield depends on the science products desired and how the mission conducts the observations.

#### 14.1.3.1 Desired science products

LUVOIR is designed to be capable of obtaining many data products on exoplanets. For the exoplanet yield analysis, we considered three primary data products on planets identified as exoEarth candidates:

1. Photometry: to detect planets and measure brightness and color
2. Spectra: to assess chemical composition of atmospheres
3. Orbit measurement: to determine if planet resides in HZ and measure spectro-photometric phase variations

In the following sections, we describe how LUVOIR will obtain these data products in an efficient manner to maximize the yield of the mission.

#### 14.1.3.2 Dealing with confusion

Upon initial detection of a possible companion, the nature of the source may be unclear. We will have only photometry, possibly one color, and a stellocentric separation to determine the nature of the object. Color, brightness, and the fact the source is unresolved may allow us to discriminate between many background galaxies and exoplanets. However, recent work has shown that other planets can mimic the color of exoEarth candidates (e.g., Krissansen-Totton et al. 2016). Further, planets that most easily mimic Earth are small, hot terrestrial planets, which have even higher occurrence rates than exoEarth candidates (van Gorkom & Stark, in prep); planet-planet confusion may be common. However, performing costly characterizations on all planets mimicking an Earth could decrease the efficiency of the exoplanet survey and reduce the yield of exoEarth candidates; we may need to disambiguate point sources to identify high priority planets. LUVOIR is capable of dealing with these expected sources of confusion without significantly impacting the yield.

#### 14.1.3.3 Order of operations

The order in which observations are conducted will impact the final yield of the mission. For example, taking spectra of every object consistent with an exoEarth candidate immediately upon discovery would be costly, as spectral characterization times can be long and we are likely to find many other planets at coincidental phases that mimic exoEarths. A more efficient order of operations would play to the strengths of the coronagraph, e.g., by first following up initial detections with orbit measurements, followed by spectra of interesting systems when planets are known to be at advantageous phases.

Ultimately these decisions will depend on uncertain quantities, like $\eta_{Earth}$ for nearby FGK stars and the rate of confusion with background objects. A precise operations concept will require further detailed study and will surely be adapted "on the fly" during mission operations.

### 14.1.3.4 Simulating operations concepts

To simulate a given operations concept, we would need to generate a fictitious universe and simulate the execution of the mission one observation at a time, adapting to the detections, non-detections, and false positives as we go using decision-making logic. While current yield codes are capable of doing most of this (Savransky & Garrett 2016), the critical decision-making logic step is extremely complex and in its infancy. Realistic decision-making processes require simulating the precision of multiple types of measurements, estimating the likelihood that a planet is an exoEarth versus a background object or another planet, and determining the optimum criteria for decision-making.

In light of these complexities, we relied on the findings of Stark et al. (2016), wherein the impact of different operations concepts on yield was estimated by adopting general rules that define the observation plan. For example, to include orbit determination, Stark et al. (2016) required each planetary system be observed at least six times to a depth consistent with detecting an exoEarth. Using these methods, Stark et al. (2016) found that for coronagraph-based missions like LUVOIR, orbit determination is not particularly costly while spectral characterization can be very costly.

Therefore, we adopt the following operations scenario:

1. Detect planets using two bands within the UV and VIS coronagraph channels simultaneously (~450–500 and 500–550 nm), likely providing color information for the majority of detections.

2. Revisit all systems as necessary with the coronagraph until the orbits of high-priority planets are sufficiently constrained (likely more than 6 times each over the course of months to years)

3. Based on the color, orbit, brightness, and phase variations, identify high-priority targets for spectral characterization.

4. Schedule and conduct spectral characterization observations on each exoEarth candidate at an optimized orbital phase to search for the presence of water vapor in the planet's atmosphere.

This operations scenario is both realistic and robust to error. By requiring orbit measurement regardless of what is detected, the operations concept is straightforward, does not rely on any confusion mitigation immediately after a detection, and proper motion will be established for free for all detected planets. Because the LUVOIR coronagraph's field of regard is greater than a hemisphere at any given time, we expect that the revisit schedule for each star can easily be optimized to maximize detections, constrain orbits, and minimize characterization time without detailed consideration of whether or not the targets are inaccessible.

### 14.1.4 Exoplanet yield-informed trades

Exoplanet yield was considered continuously during the LUVOIR Architecture A design study. Four major design trades were made based at least in part on yield studies:

### 14.1.4.1 Optimizing the primary geometry

Concurrent with the LUVOIR Architecture A study, the Segmented Coronagraph Design Analysis (SCDA) study concluded that two major factors greatly impacted the yield of coronagraphs for segmented apertures: the obscuration ratio of the secondary mirror and the inscribed diameter of the primary.

The LUVOIR design team and the SCDA team at STScI (PI: R. Soummer) studied 10 possible apertures for LUVOIR, ranging from engineering-optimized designs to coronagraph-optimized designs. The studied apertures produced a range of yields that varied by a factor of two. The final aperture was selected based on its high yield and scalability.

### 14.1.4.2 IFS vs fiber-fed spectrograph

Two proposed options were considered to obtain spectra of exoplanets: an integral field spectrograph (IFS) and a fiber-fed spectrograph. Both of these instruments have design strengths and weaknesses. The former has heritage from the WFIRST CGI, but packaging considerations and detector size place limitations on the field of view and maximum resolution. The latter is compact and provides potentially higher resolution, but only provides spectra of one source at a time and requires a potentially complex operations concept. In the end, a yield study revealed that the throughput reduction required for the fiber-red spectrograph significantly reduced exoplanet yield to an unacceptable level.

### 14.1.4.3 NUV coronagraph channel

In addition to introducing additional engineering challenges, the presence of a NUV coronagraph channel reduces the throughput of all coronagraph channels by ~30%, as it requires all optical surfaces after the telescope and prior to the dichroics to be aluminum. In addition, the end-to-end throughput of the NUV channel coronagraph is roughly half that of the visible channel. We considered replacing the NUV coronagraph channel with a second visible channel and co-adding the coronagraph channels to double the detection bandwidth. We found that this trade increased yield by ~15%. The STDT deter-mined that NUV coronagraphic capability was a higher priority than a 15% increase in exoplanet yield.

### 14.1.4.4 TMA vs Cassegrain design

The current LUVOIR Architecture A is a TMA design with 4 telescope optics (primary, secondary, tertiary, and FSM). By switching to a Cassegrain design, potentially two aluminum surfaces could be removed, increasing the throughput of all instruments by ~18%. We determined that an 18% increase in throughput would increase yield by ~6%. The STDT and design team determined that a 6% increase in yield did not warrant the increased complexity of the Cassegrain design given the timeline of the study.

## 14.2   The LUVOIR science simulation tools

The LUVOIR team has produced a set of online software tools to enable quick, accurate performance modeling and to facilitate team and community input to the LUVOIR concept study. These tools were developed by the STDT's Simulations Working Group, led by Jason Tumlinson (STScI), with substantial contributions from Giada Arney (NASA GSFC), Tyler Robinson (NAU), and Graham Kanarek (STScI).

The online tools are hosted by STScI (at http://luvoir.stsci.edu) and are accessible though the LUVOIR website at https://asd.gsfc.nasa.gov/luvoir/. For all tools, the aim has been to keep their usage and results simple and easy to grasp for first-time users. Each tool includes an "Info" tab that describes its usage and assumptions. The user may save plots and other results. These tools will continue to evolve during the LUVOIR study and additional new tools may be added.

The tools are written in python and make extensive use of the bokeh python library (bokeh.pydata.org). The underlying code is all open-source and available at github.com/tumlinson/luvoir_simtools. The github repository includes instructions for how to use the code in local mode and its dependencies, as well as an ipython notebook with basic usage of the tools that can be adapted to many purposes.

**Figure 14.3.** *Front page for online LUVOIR Science Simulation Tools at* https://asd.gsfc.nasa.gov/luvoir/tools/

### 14.2.1    Multiplanet yields visualization

This tool visualizes multi-planet yield calculations based on the work in Stark et al. (2014, 2015, 2016). These yields assume one year of science exposure time (1.5 to 2 years total time including overheads) and the planet mass/orbit bins shown under the "Planets" tab. The bar chart at left shows the expected numbers of different kinds of planets observed.

In the main panel, the planets colored in gold are shaded according to the fractional "yield" for that star (aka. completeness) in the altruistic yield calculation. These estimated (probabilistic) yields are then sampled with random draws to highlight a hypothetical sample of detected "warm rocky" planets, which are marked in light blue. This sample is randomly drawn again according to the "probability of life" slider, which specifies the fraction of warm rocky planets with remotely detectable biosignature gases, which are marked in flashing green. This tool demonstrates why dozens of habitable planet candidates are needed to perform a survey for habitable conditions and life in the nearby galaxy that can provide a statistically meaningful answer even if it is a null result.

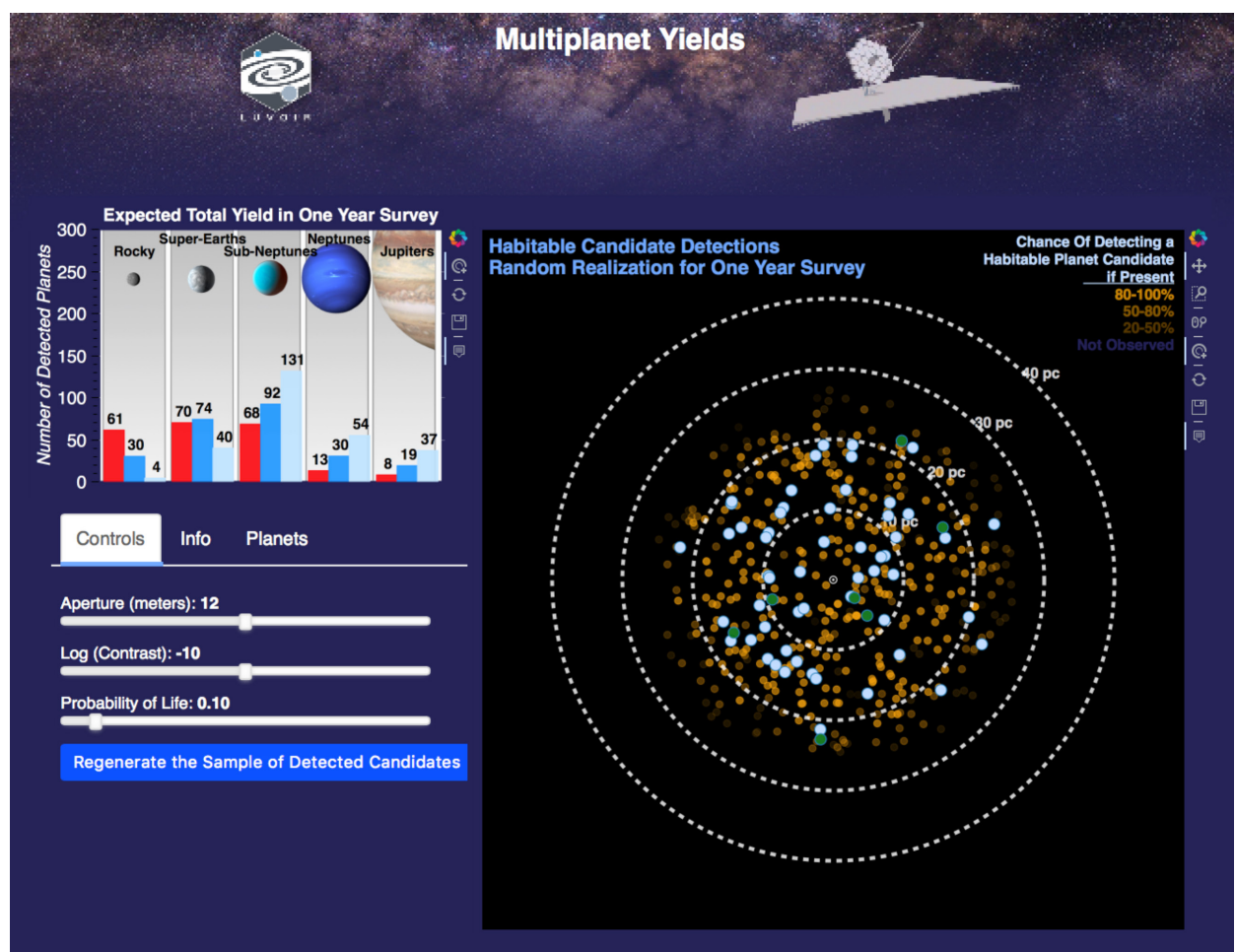

**Figure 14.4.** *Multiplanet yields visualization tool.*

### 14.2.2    Coronagraphic spectra of varied planets

This richly featured tool is based upon the coronagraph noise model from Robinson et al. (2016) and calculates realistic, noisy spectra of many kinds of planets for an ECLIPS-like coronagraph. Many model planet spectra are available, and both LUVOIR and ground-based spectra can be simulated. The ground-based mode includes thermal radiation from the atmosphere and the wavelength-dependent Earth atmosphere transmissivity. The library of planetary spectra users can select includes a range of planet types such as Earth-like worlds at different periods of Earth history, other solar system planets, false positive biosignature planets, mini-Neptunes, and warm Jupiters.

Noise terms include thermal radiation from the telescope, dark current, read noise, zodiacal light, exozodiacal light, and stellar light leakage. Users can adjust these noise terms and can also adjust basic observational parameters such as the size of the planet, the observer-system separation distance, the telescope diameter, the inner and outer working angles, the exposure time, the telescope temperature, throughput terms, and spectral resolution. Users may also specify a desired signal-to-noise ratio (SNR), which the tool uses to calculate the exposure time required as a function of wavelength to achieve this SNR.

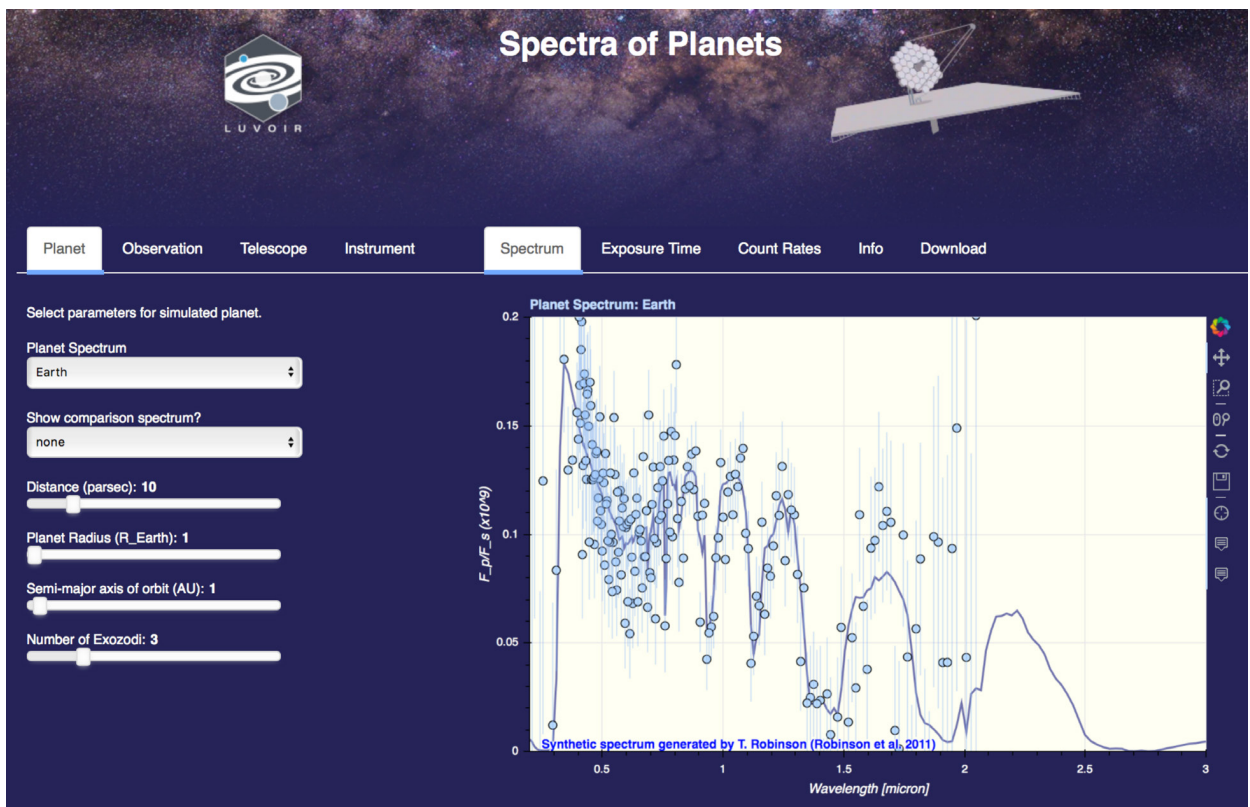

**Figure 14.5.** *Simulated coronagraphic planet spectra. The "Spectrum" tab shows the spectrum of the selected planet with added noise. The "Exposure Time" tab shows the wavelength-dependent exposure time required to obtain a user-specified SNR. The "Count Rates" tab shows the wavelength-dependent planet flux and noise terms in counts/second.*

### 14.2.3   The High-Definition Imager exposure time calculator

This tool computes S/N limits and exposure times for imaging with HDI and LUVOIR-A-like telescopes. A range of input spectral energy distributions are available. This is one of the tools that permits the user to save and restore their specific calculations. By supplying a unique string, the user names a file that will be stored on the STScI server.

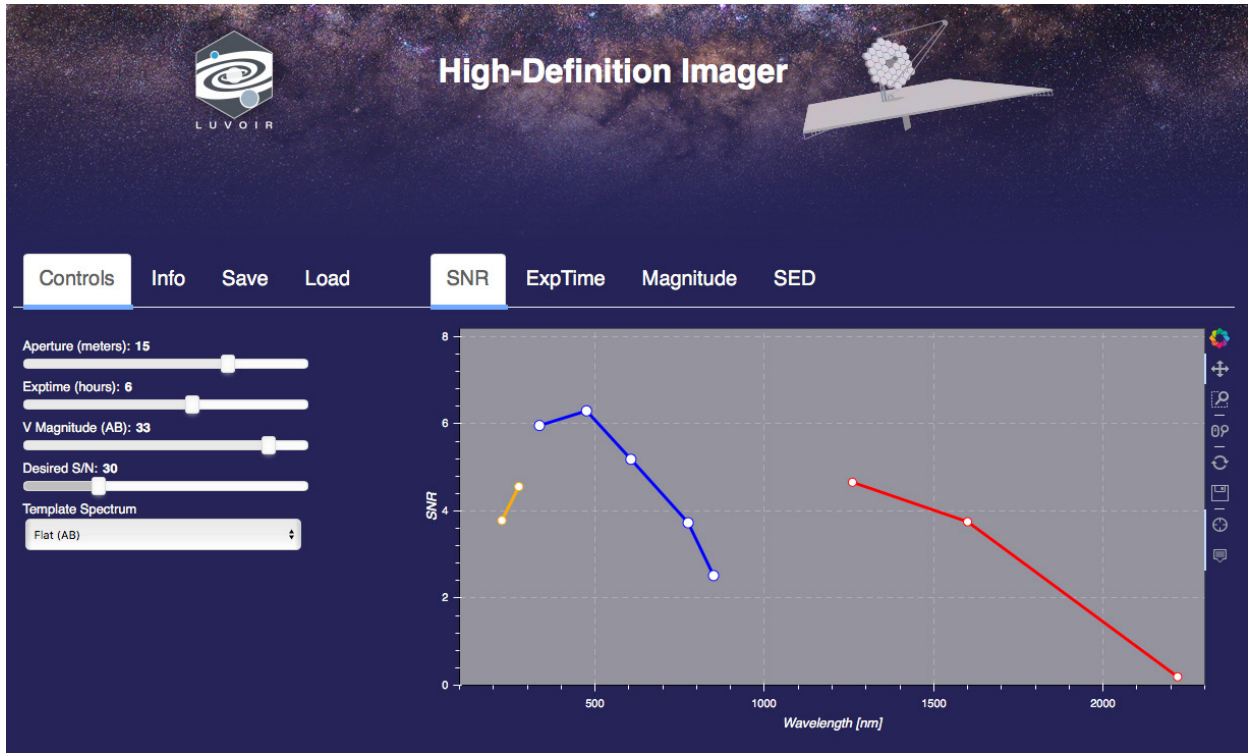

**Figure 14.6.** *HDI exposure time calculator results. For a flat continuum source normalized to AB = 33 mag, the S/N returned in the V band is S/N = 5. The "ExpTime" and "Magnitude" tabs show alternate results.*

### 14.2.4   The LUMOS exposure time calculator

This tool uses UV-band input spectra provided by the STScI pysynphot package to calculate S/N values for the FUV modes of LUMOS paired with LUVOIR-A-like telescopes. Various pysynphot spectral templates are available, and the normalization is done in the GALEX FUV band. S/N is always given per resolution element.

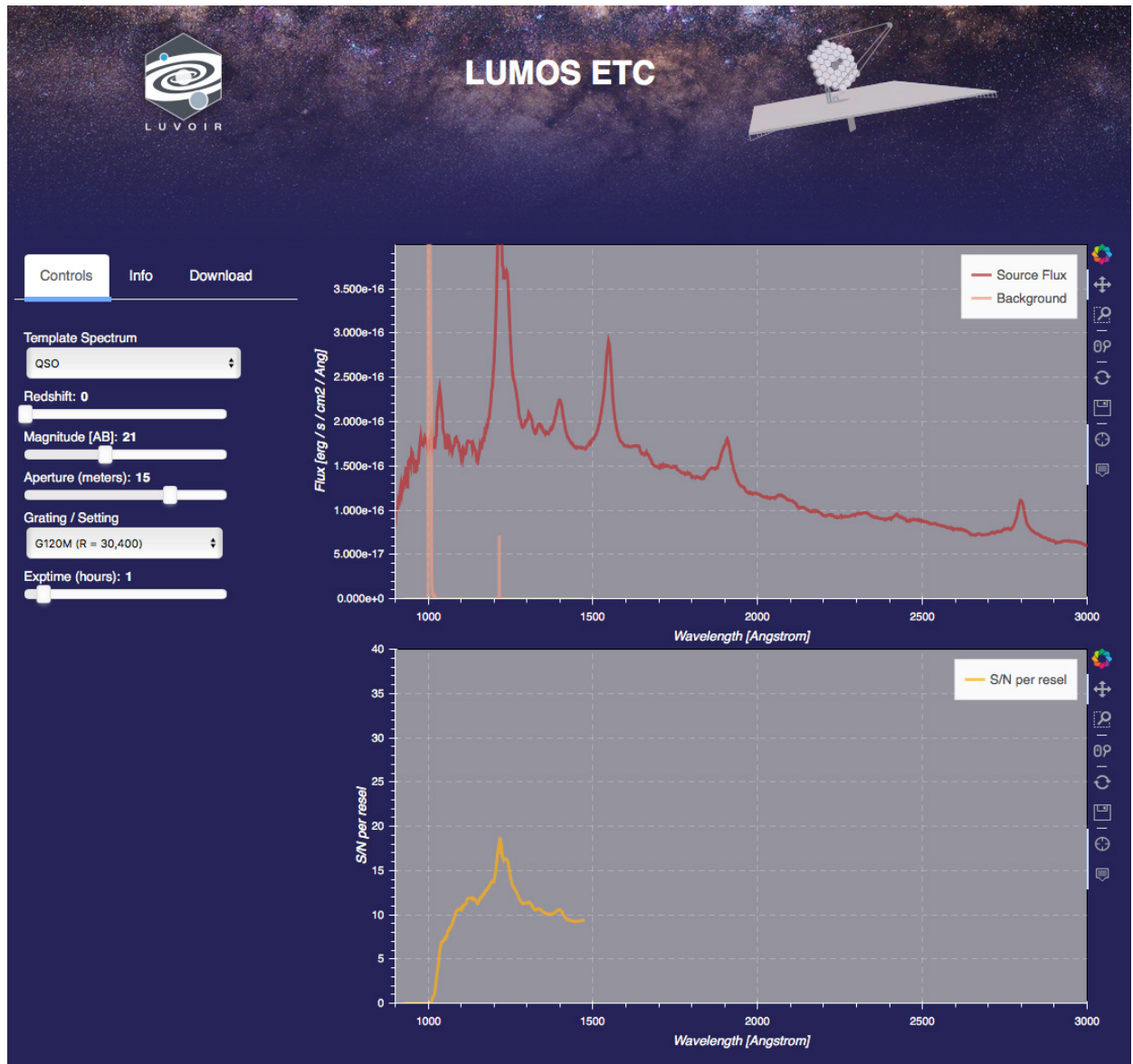

**Figure 14.7.** *LUMOS exposure time calculator.*

### 14.2.5   LUVOIR imaging comparisons

This basic tool compares images at Hubble and LUVOIR spatial resolutions to illustrate the gains in image quality going from a 2.4 to a 12-m UV/optical telescope.

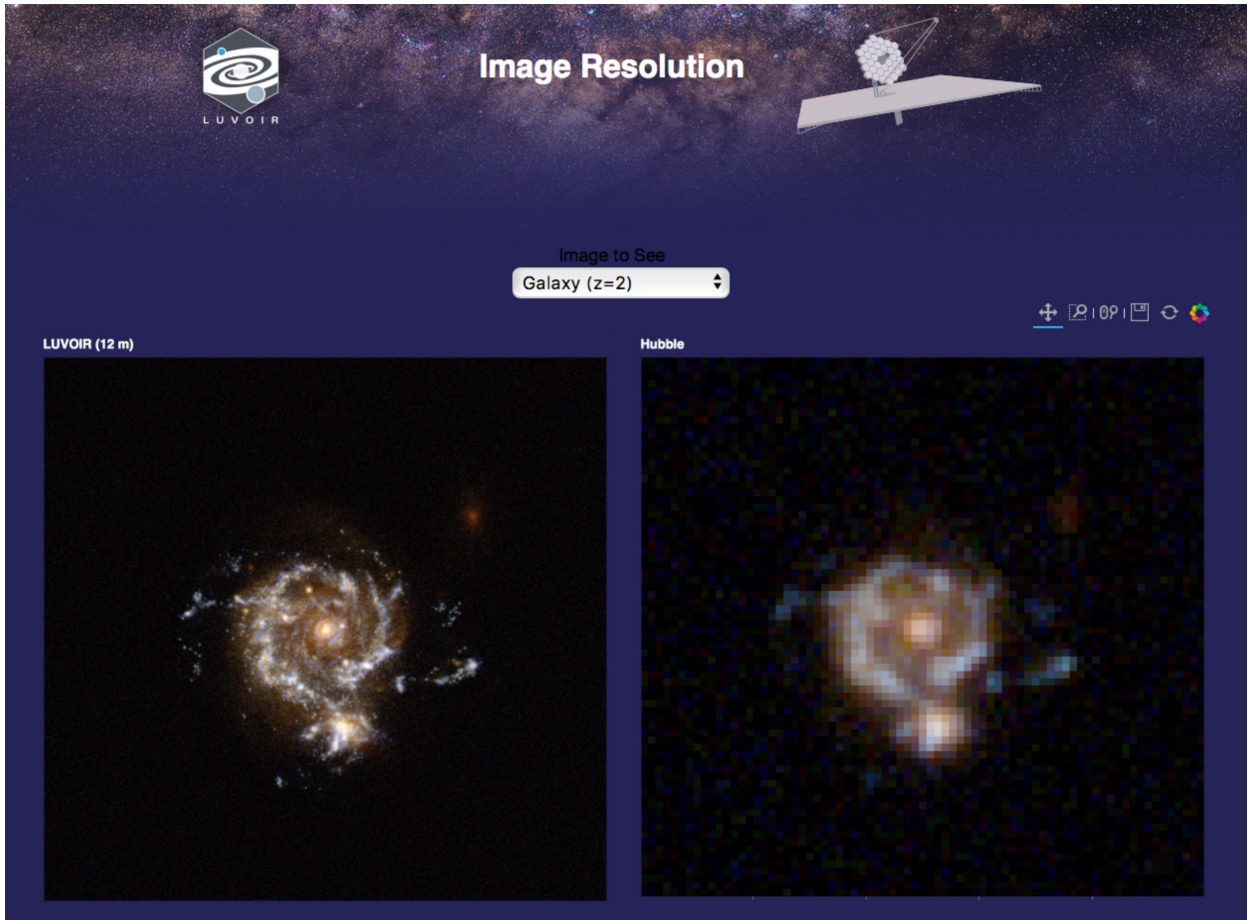

**Figure 14.8.** *Gallery of imaging simulations to visualize LUVOIR's spatial resolution.*

## 15  Appendix C: Complete science traceability matrix for LUVOIR-A

### Science

| | Is there life elsewhere? Chapter 3 | | | | | | | | | | | | |
|---|---|---|---|---|---|---|---|---|---|---|---|---|---|
| **Objectives** | Frequency of habitable exoplanet candidates | | | Characterize exoplanet host stars | Confirmation of exoplanet habitable conditions | | | | Search for signs of life on exoplanets | | | | Investigate habitability of Solar System ocean moons | | |
| **Measurement** | Survey for dozens of rocky planets around F,G,K,M stars | Establish habitable zone orbits of rocky planets | Search for water vapor in habitable zone rocky planet atmospheres | Measure UV spectrum and activity of host stars | Constrain abundances of major atmospheric constituents (e.g. H2O, CO2, CH4) | Detect surface liquid water and constrain atmospheric water vapor profile | Determine atmospheric properties (Rayleigh scattering, haze) | | Constrain abundances of atmospheric biosignature gases (e.g. O2, O3, CH4), and false positive indicators (e.g. high-contrast CO2, O4) | | | | Observe changes in surface morphology | Monitor plume activity and study plume morphology and dynamics | Determine moon atmospheric properties |

### Requirements / Implementation

ECLIPS — UV / VIS / NIR

| | R/T/G | Parameter | | | | | | | | | | | | | | |
|---|---|---|---|---|---|---|---|---|---|---|---|---|---|---|---|---|
| UV | R | Coronagraph contrast 1E-10 | ✓ | ✓ | | | | ✓ | | ✓ | | | | | | |
| | R | Coronagraph instantaneous bandpass 10% | ✓ | ✓ | | | | ✓ | | ✓ | | | | | | |
| | R | Coronagraph UV bandpass 200 nm - 525 nm | ✓ | ✓ | | | | ✓ | | ✓ | | | | | | |
| | R | Coronagraph Inner Working Angle in NUV (λ/D) 4 | ✓ | ✓ | | | | ✓ | | ✓ | | | | | | |
| VIS | R | Coronagraph contrast 1E-10 | ✓ | ✓ | ✓ | ✓ | | | | ✓ | ✓ | | | | | |
| | R | Coronagraph instantaneous bandpass 10% | ✓ | ✓ | ✓ | ✓ | | | | ✓ | ✓ | | | | | |
| | R | Coronagraph VIS bandpass 515 nm - 1.03 µm | ✓ | ✓ | ✓ | ✓ | | | | ✓ | ✓ | | | | | |
| | R | Coronagraph Inner Working Angle in visible (λ/D) 4 | ✓ | ✓ | ✓ | ✓ | | | | ✓ | ✓ | | | | | |
| | R | Coronagraph visible spatially-resolved spectroscopic resolution 140 | | | ✓ | ✓ | | | | ✓ | | | | | | |
| NIR | R | Coronagraph contrast 1E-10 | | | ✓ | ✓ | | | | ✓ | ✓ | | | | | ✓ |
| | R | Coronagraph instantaneous bandpass 10% | | | ✓ | ✓ | | | | ✓ | ✓ | | | | | ✓ |
| | R | Coronagraph NIR bandpass 1.0 µm - 2.0 µm | | | ✓ | ✓ | | | | ✓ | ✓ | | | | | ✓ |
| | R | Coronagraph Inner Working Angle in NIR (λ/D) 2 | | | ✓ | ✓ | | | | ✓ | ✓ | | | | | |
| | R | Coronagraph NIR spatially-resolved spectroscopic resolution 70 or 200 | | | ✓ | ✓ | | | | ✓ | ✓ | | | | | ✓ |

### Requirements / Implementation

LUMOS — Far-UV MOS / Near-UV MOS / Far-UV Imager

| | R/T/G | Parameter | | | | | | | | | | | | | | |
|---|---|---|---|---|---|---|---|---|---|---|---|---|---|---|---|---|
| Far-UV MOS | R | Spectrograph FUV Bandpass 1000 Å - 2000 Å | | | | ✓ | | | | | | | | | ✓ | |
| | R | Spectrograph Low Resolution Resolving Power 300 | | | | | | | | | | | | | ✓ | |
| | R | Spectrograph High Resolution Resolving Power 30,000 | | | | ✓ | | | | | | | | | ✓ | |
| | R | Spectrograph MOS shutter size 100 µm x 200 µm | | | | | | | | | | | | | ✓ | |
| | R | Spectrograph MOS FOV 2' x 2' | | | | | | | | | | | | | ✓ | |
| Near-UV MOS | R | Spectrograph NUV Bandpass 2000 Å - 4000 Å | | | | ✓ | | | | | | | | | | |
| | R | Spectrograph High Resolution Resolving Power 30,000 | | | | ✓ | | | | | | | | | | |
| | R | Spectrograph MOS shutter size 100 µm x 200 µm | | | | | | | | | | | | | | |
| | R | Spectrograph MOS FOV 2' x 2' | | | | | | | | | | | | | | |
| Far-UV Imager | R | FUV imaging bandpass 1150 Å - 2000 Å | | | | | | | | | | | | | | |
| | R | FUV imaging angular resolution 15 mas | | | | | | | | | | | | | | |

### Requirements / Implementation

HDI — UVIS / NIR Channel

| | R/T/G | Parameter | | | | | | | | | | | | | | |
|---|---|---|---|---|---|---|---|---|---|---|---|---|---|---|---|---|
| UVIS | R | UVIS imaging wavelength coverage 220 nm - 950 nm | | | | | | | | | | | | | ✓ | |
| | R | UVIS imaging angular resolution Nyquist sampled at 400 nm | | | | | | | | | | | | | ✓ | |
| | R | UVIS imaging astrometric precision Detect relative shifts of 0.0002 pixel within 96 hours of geometry calibration | | | | | | | | | | | | | | |
| NIR Channel | R | NIR imaging wavelength coverage 950 nm - 1.8 µm | | | | | | | | | | | | | ✓ | |
| | R | NIR imaging angular resolution Nyquist sampled at 1.2 µm | | | | | | | | | | | | | ✓ | |

### Requirements / Implementation

Observatory / OTE

| | R/T/G | Parameter | | | | | | | | | | | | | | |
|---|---|---|---|---|---|---|---|---|---|---|---|---|---|---|---|---|
| | R | Total collecting area 135 m² | ✓ | ✓ | ✓ | ✓ | ✓ | ✓ | ✓ | ✓ | ✓ | ✓ | | | ✓ | ✓ |
| | R | Aperture Diameter 15 m | ✓ | ✓ | ✓ | ✓ | ✓ | ✓ | ✓ | ✓ | ✓ | ✓ | | ✓ | ✓ | ✓ |
| | R | Obscuration Ratio (SM circumscribed diameter / PM inscribed diameter) 15% | ✓ | ✓ | ✓ | ✓ | ✓ | ✓ | ✓ | ✓ | ✓ | ✓ | | ✓ | | |
| | R | Diffraction-limited imaging at the specified wavelength 500 nm | ✓ | ✓ | ✓ | ✓ | ✓ | ✓ | ✓ | ✓ | ✓ | ✓ | | | | |
| | R | Solar Elongation Viewing Angle 45° | | ✓ | | ✓ | | | | | | ✓ | | ✓ | ✓ | ✓ |
| | R | Pointing stability (1σ per axis over an observation) 0.30 mas | ✓ | ✓ | ✓ | ✓ | ✓ | ✓ | ✓ | ✓ | ✓ | ✓ | | ✓ | | |
| | R | Repoint anywhere in anti-sun hemisphere within timeframe 45 min | | ✓ | | | ✓ | | | | | | | ✓ | ✓ | ✓ |
| | R | Fast object tracking 60 mas/s | | | | | | | | | | | | ✓ | ✓ | ✓ |
| | R | Capability for operating instruments in parallel. | | | | | | | | | | | | | | |

# Science — Observatory Traceability Matrix

**Observation column legend**

- O1 = High-contrast exoplanet spectroscopy between 500 nm - 2.0 μm (SNR>10, R=70-200)
- O2 = High-contrast photometry between 200 nm - 1 μm (SNR=5 in ≥10% bands)
- O3 = Multi-epoch astrometric imaging (~0.1 μas precision)
- O4 = High-contrast spectroscopy between 500 nm - 2.0 μm (SNR=10, R=70-200)
- O5 = High-contrast imaging between 500 nm - 1.0 μm
- O6 = High resolution spectroscopy from 1000 Å - 4000 Å (SNR=10, R≥60,000)
- O7 = Multi-epoch spectroscopy 500 nm - 2.5 μm
- O8 = Spatially resolved spectroscopy between 1.0 μm - 2.0 μm (R=70,200) and between 1000 Å - 2000 Å (R=30,000)
- O9 = Multi-object spectroscopy from 1000 Å - 4000 Å (SNR=30, R=30,000)
- O10 = Multi-object spectroscopy from 1000 Å - 3000 Å (SNR=20, R=30,000)
- O11 = Multi-object spectroscopy from 1000 Å - 3000 Å (SNR=10, R=30,000)
- O12 = Emission line maps from 1000 Å - 3000 Å (SNR=5, R=30,000)
- O13 = Multi-object spectroscopy 1000 Å - 4000 Å (SNR=30, R=30,000)
- O14 = g- and r-band imaging (SNR>5)

## Science header

| Level | How do we fit in? (Chapter 4) | | | | | | | How do galaxies evolve? (Chapter 5) | | | | |
|---|---|---|---|---|---|---|---|---|---|---|---|---|
| **Objectives** | Constrain exoplanet atmospheric composition | Detect and characterize atmospheric hazes and clouds | Determine planetary system architectures | Determine morphology and architecture of debris disks | Determine bulk composition of extrasolar planetesimals | Characterize full population of Solar System minor bodies | Constrain the physics of atmospheric escape | Measuring the baryons over cosmic time from 10-10^7 K | High-definition exploration of the CGM | Characterize nearby galaxy inflows and outflows in detail | Explore star formation histories across the Hubble sequence | Color-Magnitude diagrams for stars in galaxies out to tens of Mpc across multiple galaxy types |
| **Measurement** | Constrain abundances of gases (e.g. H2O, CO2, CH4, O2) | Constrain haze and cloud abundances | Determine planet mass and orbital parameters | Constrain atmospheric composition | Spatially map dust distribution | Measure abundances of multiple atomic and molecular species in debris disk gas | Measure size, distribution, orbital parameters | Characterize surface composition and volatile outgassing | Detect, map, and characterize outflows | Absorption line measures: Column density, velocity width, redshift for multiple ions | Absorption line measures: Column density, velocity width, redshift for multiple ions | Emission line maps of multiple ions / Spatially map inflowing/outflowing gas in absorption |

## Requirements / Implementation matrix

### ECLIPS

| Band | Parameter | Threshold (T) | Goal (G) | Observations |
|---|---|---|---|---|
| UV | Coronagraph contrast | 1E-10 | | ✓ O2 |
| UV | Coronagraph instantaneous bandpass | 10% | 15% | ✓ O2 |
| UV | Coronagraph UV bandpass | 200 nm - 525 nm | 200 nm - 525 nm | ✓ O2 |
| UV | Coronagraph Inner Working Angle in NUV (λ/D) | 4 | 3 | ✓ O2 |
| VIS | Coronagraph contrast | 1E-10 | | ✓ O1, O2, O4, O5 |
| VIS | Coronagraph instantaneous bandpass | 10% | 15% | ✓ O1, O4 |
| VIS | Coronagraph VIS bandpass | 515 nm - 1.03 μm | 515 nm - 1.03 μm | ✓ O1, O4 |
| VIS | Coronagraph Inner Working Angle in visible (λ/D) | 4 | 3 | ✓ O1, O4 |
| VIS | Coronagraph visible spatially-resolved spectroscopic resolution | 140 | 140 | ✓ O1, O4 |
| NIR | Coronagraph contrast | 1E-10 | | ✓ O1, O4 |
| NIR | Coronagraph instantaneous bandpass | 10% | 15% | ✓ O1, O7 |
| NIR | Coronagraph NIR bandpass | 1.0 μm - 2.0 μm | 1.0 μm - 2.0 μm | ✓ O1, O4 |
| NIR | Coronagraph Inner Working Angle in NIR (λ/D) | 2 | 2 | ✓ O1, O4 |
| NIR | Coronagraph NIR spatially-resolved spectroscopic resolution | 70 or 200 | 70 or 200 | ✓ O4, O7 |

### LUMOS

| Module | Parameter | Threshold (T) | Goal (G) | Observations |
|---|---|---|---|---|
| Far-UV MOS | Spectrograph FUV Bandpass | 1000 Å - 2000 Å | 1000 Å - 2000 Å | ✓ O6, O9, O10, O11, O12, O13 |
| Far-UV MOS | Spectrograph Low Resolution Resolving Power | 300 | 500 | ✓ O11, O12 |
| Far-UV MOS | Spectrograph High Resolution Resolving Power | 30,000 | 60,000 | ✓ O6, O9, O10, O11, O12, O13 |
| Far-UV MOS | Spectrograph MOS shutter size | 100 μm x 200 μm | 100 μm x 200 μm | ✓ O9, O10, O11, O12, O13 |
| Far-UV MOS | Spectrograph MOS FOV | 2' x 2' | 3' x 3' | ✓ O9, O10, O11, O12, O13 |
| Near-UV MOS | Spectrograph NUV Bandpass | 2000 Å - 4000 Å | 2000 Å - 4000 Å | ✓ O9, O13 |
| Near-UV MOS | Spectrograph High Resolution Resolving Power | 30,000 | | ✓ O9, O13 |
| Near-UV MOS | Spectrograph MOS shutter size | 100 μm x 200 μm | 100 μm x 200 μm | ✓ O9, O10, O11, O12, O13 |
| Near-UV MOS | Spectrograph MOS FOV | 2' x 2' | 3' x 3' | ✓ O9, O13 |
| Far-UV Imager | FUV imaging bandpass | 1150 Å - 2000 Å | 1000 Å - 2000 Å | ✓ O13 |
| Far-UV Imager | FUV imaging angular resolution | 25 mas | 15 mas | ✓ O13 |

### HDI

| Channel | Parameter | Threshold (T) | Goal (G) | Observations |
|---|---|---|---|---|
| UVIS | UVIS imaging wavelength coverage | 220 nm - 900 nm | 200 nm - 950 nm | ✓ O3, O7, O14 |
| UVIS | UVIS imaging angular resolution | Nyquist sampled at 400 nm | Nyquist sampled at 400 nm | ✓ O3, O7, O14 |
| UVIS | UVIS imaging astrometric precision | Detect relative shifts of 0.0002 pixel within 96 hours of geometry calibration | Detect relative shifts of 0.0001 pixel within 48 hours of geometry calibration | ✓ O3 |
| NIR Channel | NIR imaging wavelength coverage | 950 nm - 1.8 μm | 850 nm - 2.5 μm | ✓ O7 |
| NIR Channel | NIR imaging angular resolution | Nyquist sampled at 1.2 μm | Nyquist sampled at 1.2 μm | ✓ O7 |

### Observatory / OTE

| Parameter | Threshold (T) | Goal (G) | Observations |
|---|---|---|---|
| Total collecting area | 135 m² | 155 m² | ✓ O1, O4, O6, O7, O9, O10, O11, O12, O13, O14 |
| Aperture Diameter | 15 m | | ✓ O3, O7, O14 |
| Obscuration Ratio (SM circumscribed diameter / PM inscribed diameter) | 15% | 10% | ✓ O1, O2, O4, O5 |
| Diffraction-limited imaging at the specified wavelength | 500 nm | | ✓ O1, O2, O4, O5, O14 |
| Solar Elongation Viewing Angle | 45° | 90° | ✓ O3, O7 |
| Pointing stability (1σ per axis over an observation) | 0.30 mas | 0.25 mas | ✓ O1, O4, O14 |
| Repoint anywhere in anti-sun hemisphere within timeframe | 45 min | 30 min | ✓ O3, O7 |
| Fast object tracking | 60 mas/s | | ✓ O7 |
| Capability for operating instruments in parallel | - | - | ✓ O11, O12, O13 |

## Science

| Question / Report Chapter | What are the building blocks of cosmic structure? Chapter 6 | | | | How do stars and planets form? Chapter 7 | | | | |
|---|---|---|---|---|---|---|---|---|---|
| **Objectives** | Determine the turnover in the high-redshift galaxy luminosity function | Constrain the nature of dark matter | Determine the escape fraction of ionizing radiation out to r~1 | | Constrain massive star formation and determine the stellar IMF | | Characterize the composition of planet forming material | Follow the rise of the periodic table | |
| **Measurement** | Incidence frequency of z* galaxies at faint magnitudes | Matter power spectrum on small scales via observations of dwarf galaxies near Milky Way analogues | Stellar proper motions in Local Group galaxies | Amount of flux below the Lyman break for N>5000 galaxies from 0.2 < z < 1.2 | Imaging of individual stars and star clusters | Spectroscopy of individual stars and star clusters | Spectroscopy of atoms and molecules in protoplanetary disks | Spectroscopy of extremely metal-poor stars in the Milky Way halo | Spectroscopy of late-type stars in the Milky Way halo |
| **Observations** | i-, z-, and H-band imaging (SNR>5 to z>33.5) | V- and R-band imaging (SNR>5 to V>31) | Multi-epoch astrometric imaging (~0.5 μas precision) | Multi-object spectroscopy between 1000 Å - 2000 Å (SNR>5, R>500) | U-, B-, V-, R-, I- band imaging (SNR 30-300) | Multi-object spectroscopy from 1000 Å - 4000 Å (R=30,000) | Multi-object spectroscopy from 1000 Å - 4000 Å (R=30,000) | Multi-object spectroscopy from 1000 Å - 4000 Å (SNR>30, R=30,000) | Multi-object spectroscopy from 1000 Å - 4000 Å (SNR>80, R=30,000) |

## Requirements / Implementation

Instruments: ECLIPS (UV / VIS / NIR), LUMOS (Far-UV MOS / Near-UV MOS / Far-UV Imager), HDI (UVIS / NIR Channel), Observatory / OTE. Each parameter lists Threshold (T) and Goal (G) values.

| Instrument | Parameter | Threshold | Goal | C1 | C2 | C3 | C4 | C5 | C6 | C7 | C8 | C9 |
|---|---|---|---|---|---|---|---|---|---|---|---|---|
| ECLIPS UV | Coronagraph contrast | 1E-10 | | | | | | | | | | |
| | Coronagraph instantaneous bandpass | 10% | 15% | | | | | | | | | |
| | Coronagraph UV bandpass | 200 nm - 525 nm | 200 nm - 525 nm | | | | | | | | | |
| | Coronagraph Inner Working Angle in NUV (λ/D) | 4 | 3 | | | | | | | | | |
| ECLIPS VIS | Coronagraph contrast | 1E-10 | | | | | | | | | | |
| | Coronagraph instantaneous bandpass | 10% | 15% | | | | | | | | | |
| | Coronagraph VIS bandpass | 515 nm - 1.03 μm | 515 nm - 1.03 μm | | | | | | | | | |
| | Coronagraph Inner Working Angle in visible (λ/D) | 4 | 3 | | | | | | | | | |
| | Coronagraph visible spatially-resolved spectroscopic resolution | 140 | 140 | | | | | | | | | |
| ECLIPS NIR | Coronagraph contrast | 1E-10 | | | | | | | | | | |
| | Coronagraph instantaneous bandpass | 10% | 15% | | | | | | | | | |
| | Coronagraph NIR bandpass | 1.0 μm - 2.0 μm | 1.0 μm - 2.0 μm | | | | | | | | | |
| | Coronagraph Inner Working Angle in NIR (λ/D) | 2 | 2 | | | | | | | | | |
| | Coronagraph NIR spatially-resolved spectroscopic resolution | 70 or 200 | 70 or 200 | | | | | | | | | |
| LUMOS Far-UV MOS | Spectrograph FUV Bandpass | 1000 Å - 2000 Å | 1000 Å - 2000 Å | | | | ✓ | | ✓ | ✓ | ✓ | ✓ |
| | Spectrograph Low Resolution Resolving Power | 300 | 500 | | | | ✓ | | | | | |
| | Spectrograph High Resolution Resolving Power | 30,000 | 60,000 | | | | | | ✓ | ✓ | ✓ | ✓ |
| | Spectrograph MOS shutter size | 100 μm x 200 μm | 100 μm x 200 μm | | | | ✓ | | ✓ | ✓ | ✓ | ✓ |
| | Spectrograph MOS FOV | 2' x 2' | 3' x 3' | | | | ✓ | | ✓ | ✓ | ✓ | ✓ |
| LUMOS Near-UV MOS | Spectrograph NUV Bandpass | 2000 Å - 4000 Å | 2000 Å - 4000 Å | | | | | | ✓ | ✓ | ✓ | ✓ |
| | Spectrograph High Resolution Resolving Power | 30,000 | 40,000 | | | | | | ✓ | ✓ | ✓ | ✓ |
| | Spectrograph MOS shutter size | 100 μm x 200 μm | 100 μm x 200 μm | | | | | | ✓ | ✓ | ✓ | ✓ |
| | Spectrograph MOS FOV | 2' x 2' | 3' x 3' | | | | | | ✓ | ✓ | ✓ | ✓ |
| LUMOS Far-UV Imager | FUV imaging bandpass | 1150 Å - 2000 Å | 1000 Å - 2000 Å | | | | | | | | | |
| | FUV imaging angular resolution | 25 mas | 15 mas | | | | | | | | | |
| HDI UVIS | UVIS imaging wavelength coverage | 220 nm - 900 nm | 200 nm - 950 nm | ✓ | ✓ | ✓ | | ✓ | | | | |
| | UVIS imaging angular resolution | Nyquist sampled at 400 nm | Nyquist sampled at 400 nm | ✓ | ✓ | ✓ | | ✓ | | | | |
| | UVIS imaging astrometric precision | Detect relative shifts of 0.0002 pixel within 96 hours of geometry calibration | Detect relative shifts of 0.0001 pixel within 48 hours of geometry calibration | | | ✓ | | | | | | |
| HDI NIR Channel | NIR imaging wavelength coverage | 950 nm - 1.8 μm | 850 nm - 2.5 μm | ✓ | | | | ✓ | | | | |
| | NIR imaging angular resolution | Nyquist sampled at 1.2μm | Nyquist sampled at 1.2μm | | | | | ✓ | | | | |
| Observatory / OTE | Total collecting area | 155 m² | 135 m² | ✓ | ✓ | ✓ | ✓ | ✓ | ✓ | ✓ | ✓ | ✓ |
| | Aperture Diameter | 15 m | 15 m | ✓ | ✓ | ✓ | | ✓ | ✓ | ✓ | ✓ | ✓ |
| | Obscuration Ratio (SM circumscribed diameter / PM inscribed diameter) | 15% | 10% | | | | | | | | | |
| | Diffraction-limited imaging at the specified wavelength | 500 nm | 500 nm | | ✓ | ✓ | | ✓ | | | | |
| | Solar Elongation Viewing Angle | 45° | 30° | | | ✓ | | | | | | |
| | Pointing stability (1σ per axis over an observation) | 0.30 mas | 0.25 mas | ✓ | ✓ | ✓ | | | | | | |
| | Repoint anywhere in anti-sun hemisphere within timeframe | 45 min | 30 min | | | | ✓ | | | | | |
| | Fast object tracking | 60 mas/s | 60 mas/s | | | | | | | | | |
| | Capability to operate instruments in parallel | - | - | ✓ | | | ✓ | ✓ | | | | |

# 16 Appendix D: Further technical information

In this appendix, we present whitepapers with additional information on key technological issues that drove LUVOIR design trades.

## 16.1   LUVOIR telescope temperature considerations

Lee Feinberg (NASA GSFC)

### 16.1.1   Introduction

At first glance, an obvious goal for an observatory such as LUVOIR that operates in the near infrared is that it should be as cold as possible to reduce the thermal background emission. However, there are number of technical and programmatic considerations that compete with this goal. This whitepaper discusses the issues associated with operating LUVOIR at temperatures below a nominal room temperature of 293 K.

The difficulties of designing and validating a large space telescope that operates at temperatures below room temperature have been explored in great detail through the development of JWST, and primarily relate to three engineering challenges:

1. The need to fabricate and assemble a highly stable system that is guaranteed to function properly at cold temperatures.

2. The need to demonstrate the full end-to-end optical and mechanical performance of the telescope before launch.

3. The need to eliminate contamination prior to and during operations in order to achieve peak performance.

All of these challenges have engineering solutions, but the question is whether the gain in science capability justifies the cost and difficulty associated with those solutions.

### 16.1.2   Temperature vs. thermal background radiation

The first question is what temperature a telescope needs to operate at to enable the desired long-wavelength science. To assess this, one must consider the temperature of everything in the optical chain (primary mirror, secondary mirror and struts, instrument optics and components, etc.). For simplicity, consider the radiance vs. wavelength of the telescope, including coatings, for varying temperatures as shown in **Figure 16.1**. This analysis assumes Aluminum-coated mirrors, which has worse emissivity in the NIR, but is the necessary coating to enable observations with LUVOIR in the UV.

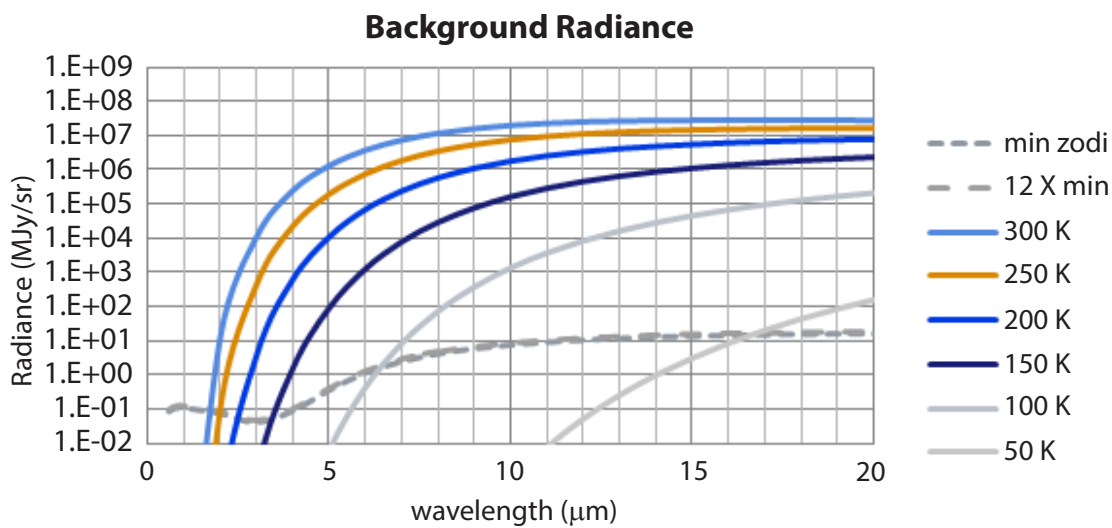

**Figure 16.1.** *Telescope background radiance as a function of temperature. Credit: P. Lightsey / Ball Aerospace Technology Center.*

| Issue Under Consideration | 293 K (Room Temperature) | 260 K (Molecular Deposition Begins) | 200 K (Allows 2.5 µm Zodi-limited Observing) | 140 K (Water Condenses; Stable Cryo Materials) | 50 K (JWST Heritage) |
|---|---|---|---|---|---|
| Glass (ULE, Zerodur) Stability | green | green | green | green | red |
| SiC Stability | green | green | green | green | green |
| Beryllium Stability | red | red | red | yellow | green |
| Composite Stability | green | yellow | yellow | yellow | green |
| Composite Complexity | green | yellow | yellow | yellow | red |
| Moisture Release / Coefficient of Moisture Expansion | red | red | red | green | green |
| Molecular Contamination for UV Throughput | green | yellow | yellow | yellow | red |
| Cryo/Cold Figuring and Polishing Complexity | green | yellow | yellow | red | red |
| System Alignment at Room Temperature | green | yellow | yellow | yellow | red |
| System Optical Testing and Verification | green | yellow | yellow | yellow | red |
| Environmental Cryo/Thermal Testing | green | yellow | yellow | yellow | red |
| Damping of Dynamic Disturbances | green | green | green | red | red |
| Material Stress and Strength Properties | green | yellow | yellow | red | red |
| Shock | green | green | green | yellow | yellow |
| Use of Lubricants | green | yellow | yellow | yellow | yellow |
| Epoxy and Bonds | green | yellow | yellow | yellow | yellow |
| Coronagraph Instrument Technologies | green | yellow | yellow | yellow | red |
| Heater Power / Thermal Management | green | yellow | yellow | green | yellow |

*Figure 16.2. Material and system issue risks by temperature region. Red is considered high risk, yellow is moderate risk, green is low risk.*

**Figure 16.1** plots the thermal emission at different telescope temperatures compared the limiting Solar System zodiacal light background ("1 Zodi"). Also shown for comparison is 1.2× Zodi. **Figure 16.1** shows that the telescope temperature must be 200 K to achieve faint-source observations that are Zodi background-limited at 2.5 µm. The figure also shows that reducing the telescope temperature only 50 K from room temperature increases the background limit by only ~0.3 µm. However, achieving a Zodi-limited background flux is not necessary for bright-target applications such as stellar, planetary, or exoplanetary applications, and reductions in the thermal background at long wavelengths will still lead to improvements in sensitivity. We therefore address the impacts of decreasing the telescope temperature on the overall telescope engineering design.

### 16.1.3   Impacts

There are a number of issues associated with operating a telescope at cold temperatures and the impacts vary depending on temperature and system architecture. A summary of these issues and the level of complexity they add at different temperature regimes is shown in **Figure 16.2**. The temperatures chosen were natural break points, which derive from both basic physics and extensive experience. For example, we know that quartz crystal microbalance (QCM) monitors during JWST testing start to show large depositions at 260 K (a point where molecular deposition starts to happen), 200 K is above where water begins to condense on optical surfaces (roughly 140 K to 170 K), 140 K is where some materials have become CTE-stable (coefficient of thermal expansion), and 50 K is the JWST heritage where virtually all contaminants are frozen and Beryllium and JWST M55J laminates are stable.

The remainder of the white paper summarizes of each of the considerations from **Figure 16.2**. The topics are grouped by whether they are a significant factor in the trade space or not.

## 16.1.3.1  Significant factors

**Molecular Sticking vs. UV Sensitivity**— UV systems are extremely sensitive to even monolayers of absorbing molecules. As contaminants build up, the shortest wavelengths will first see reductions in throughput for layers as thin as a few angstroms. First order effects are due to absorption, specific to the type of molecular contamination. Quarter-wave effects will also come into play for thicker layers of contamination.

To deal with molecular absorption, we must carefully select, bake, and certify all materials to minimize molecular adherence. There is a long heritage of doing this at room temperature where the sticking coefficients in vacuum are low. But even with extreme care, epoxies, plasticizers used in cables, hydrocarbons from lubricants, and many other small residuals will exist and will stick to mirrors at colder temperatures. Water also desorbs and begins to stick to surfaces in the 140–170 K range. For cryogenic optics or components, it is very possible that frequent bakeouts would be needed where baked off products are carefully managed (venting of cold fingers). There is not a lot of heritage of doing this for UV systems because of the complexity.

Given these challenges, is any reduction in the system operating temperature acceptable? During JWST instrument testing, it was routinely found that at 243 K the QCM sensor would show significant depositions, while at 258 K it did not. This suggests that 258 K is an approximate breakpoint where molecules begin depositing (how many are UV absorbing will depend on the species). An advantage of operating at a temperature of 260 K is that mirror bakeouts of contaminants would not be as time consuming or technically challenging (the power required to bake out mirrors would be a small addition compared to what would already be needed to maintain nominal operations, and the time to achieve a mirror bakeout would be reduced).

One possible strategy with cold systems is to operate above where water condenses in vacuum (170K) and is frozen in structures by operating at 200K. However, other molecular species will have already deposited. One could also use a warm window to isolate an instrument, but at shorter near-UV wavelengths the availability of wide-band, highly transmissive windows is limited

Take-home Message: While it is not physically impossible to engineer solutions to achieve high UV sensitivities at cryogenic temperatures, it would add a high degree of complexity with many unknowns and with no true heritage, especially below 260 K where molecular deposition begins to significantly increase.

**Mirror Stability vs. Temperature**— Mirror stability is driven by the architecture of the mirror subsystem, which combines the material stability of the mirror substrate, with local thermal controls designed to keep the temperature of the mirrors constant in the operational environment. However, mirror materials have different stability characteristics at room temperature compared with lower temperatures, and the engineering solution for mirror segment control will be dominated by what materials are used and the thermal conditions of the segments. Exoplanet coronagraphy with LUVOIR requires individual mirror segment stability of a few picometers RMS, which is a very significant technological challenge.

Material options for LUVOIR include ULE®, Zerodur®, and Silicon Carbide (SiC) for room temperature operation, and the same materials plus beryllium and silicon for colder temperatures. The current LUVOIR architecture is based on the high material stability of ULE® or Zerodur® with low thermal and optical control authority, operating near room temperature at 270 K. To date, ULE®

CTE and modeling has been demonstrated to have the best thermal stability performance at room temperature to meet the challenging stability requirements. ULE® at room temperature has heritage for high stability studies dating back to the Terrestrial Planet Finder-Coronagraph (TPF-C) and Exo-C Probe studies. A study conducted by the Smithsonian Astrophysical Observatory (Eisenhower et al. 2015), performed detailed modeling of ULE® segment at room temperature with CTE values measured from a real representative mirror boule. Eisenhower et al. shows wavefront stability changes as small as 0.5 pm for a 1.2 meter flat-to-flat hexagonal mirror (approximately LUVOIR size) controlled with a 1 mK-stable heater plate (deemed feasible but challenging). This is a key result for demonstrating the feasibility of LUVOIR.

Another consideration for LUVOIR mirrors is the degree of optical figure actuation that will be needed to meet stringent wavefront performance goals. ULE® mirrors can be equipped with mechanical actuators, to permit a certain degree of correctability. It remains to be shown how this will impact the thermal stability of the segment subsystem, however.

Other approaches are possible, with substrates with high thermal controllability and very precise thermal sensing and control, allowing use of stronger materials such as SiC and operation at a wider range of temperatures. SiC mirrors have been demonstrated with high levels of correctability by incorporating electrostrictive or piezoelectric actuators into the substrate structure. This gives SiC active mirrors the ability to operate at the required optical performance level both at 1-g and 0-g. It provides correctability for wavefront errors that can easily arise (and have often arisen in the past) during fabrication, test, assembly or launch. SiC mirrors have lower passive thermal stability at room temperature, but higher thermal controllability than ULE®. SiC at room temperature is a viable approach for LUVOIR mirrors provided that very precise thermal controls (<< 0.1 mK) are used. If temperatures below 150 K are desired, SiC stability improves to be better than conventional ULE®.

While additional details on mirror CTE and stability are controlled by International Trafficking and Arms Regulations (ITAR), the key point is that a ULE®-based, room-temperature approach appears to be feasible, while offering the least departure from traditional practice. While ULE® and Zerodur® can be tailored to cryogenic temperatures, picometer stability performance has not been assessed. At a minimum, going away from room temperature would risk the best possible stability performance that has been demonstrated and that builds on a large database and history including heritage back to the TPF-C design. While it is true other materials like SiC could have advantages for optical control, thermal control, mass efficiency, or dynamics stability, the need for these advantages is yet to be determined.

Other materials like SiC, silicon, and beryllium have high CTE at room temperature but are thermally conductive and may offer stable solutions at colder temperatures (e.g., at 150 K). While these solutions could offer stable solutions at cryogenic temperatures, more work would need to be done with substrate CTE measurements and modeling equivalent to the Eisenhower analysis. Note that while some mirror manufacturers will consider thermal conductivity when assessing stability, the LUVOIR architecture is not driven by thermal conductivity but rather is driven primarily by CTE performance and thermal inertia.

UV and high contrast systems require very tight controls on both mid-frequency wavefront and surface roughness errors. To this end, glass mirrors like ULE®, Fused Silica, and Zerodur® have been used on most UV and EUV telescopes including Hubble and Fuse. Polishing of Si-clad or chemical-vapor deposition (CVD)-clad SiC can also achieve good surface roughness, although CVD SiC is harder to polish. Another solution for SiC is to embed actuators for higher control authority.

Take-home Message: The mirror materials for ultra-precise mirror stability with the best demonstrated performance and highest heritage are ULE® and Zerodur®, but both of these materials are designed to operate at or close to room temperature. Operating at a different temperature will necessitate a significant mirror technology demonstration effort.

**Damping of Dynamic Disturbances—** Thermal damping affects dynamic stability and the change in damping is as much as 10× from room temperature to 50 K. In general, warmer is better for dynamic damping, regardless of material. Less damping would impact wavefront error and line-of-sight dynamic stability of the system.

**Coronagraph Instrument Technologies vs. Temperature—** To achieve longer wavelength performance for coronagraphy, not only does the telescope need to be cold, so does the entire corona-graph instrument, including the deformable mirrors, Lyot stops, occulting mask, and op-tical train. All of these coronagraph technolo-gies have an extensive technological history for picometer stability and high contrast at room temperature. While some actuators can work at colder temperatures, a whole technology development program would be needed including the picometer stability performance of such systems at colder tem-peratures. This essentially would restart the coronagraph technology effort and would prevent using WFIRST heritage directly. Of course, this is only applicable for corona-graphic measurements, and would not apply to other infrared instruments.

**Cryo/Cold Figuring and Polishing Complexity—** The wavefront budget for LUVOIR is tight for a 500 nm diffraction-limited wavefront performance. A single primary mirror segment needs to have <10 nm RMS wavefront (5 nm RMS surface), which is roughly 4× better than what was done on JWST. This includes gravity backout and metrology uncertainties and is already considered a very challenging requirement at room temperature (gravity effects and the need to match radius of curvature are the dominating terms). Due to CTE effects, mirrors distort as they go cold by many nanometers. Cold optics will need to be tested at temperature and go through a cryo-figuring/cryo-polishing iteration, adding metrology uncertainty, and requiring more time. A single cryogenic test for a JWST mirror would take anywhere from 3–6 months when all of the logistics and pre- and post-integration and testing are considered. Testing many segments could add many years to the critical path of the telescope.

**Material Stress and Strength Properties —** Any material mismatch (mirror to mounts, mounts to flexures, etc.) will have temperature-induced stresses. This can impact strength margins and every material mismatch will need to be analyzed and likely tested at temperature via pull tests. Also, other material properties like stiffness can vary with temperature and material testing may be needed. These issues required complex flexures to be included on JWST that required a significant amount of time and testing in the design phase and additional time in the production phase. Just the additional time to design the very complex flexures for JWST likely added more than a year to the design phase of the mirror segments.

**System Alignment at Room Temperature—** It is highly desirable to be able to align the system at room temperature and know that it will still be aligned at operating temperature. Otherwise accurate models will be needed to predict alignment changes at temperature. Those models must be verified with tests at temperature. Alternatively, cryo-actuators can be used to compensate for misalignments, as was done on JWST segments. This will introduce additional requirements and complexities. Operating at or near room temperature (as cold as 260 K where alignment shifts are small) would allow for alignment and even some optical testing in a cleanroom environment, which is much more efficient and lower cost.

**Epoxy and Bonds—**Acrylics used to hold multi-layer insulation (MLI) and epoxies used to bond interfaces have glass transition temperatures that can impact their strength or cause other problems like contamination release. These transition temperatures are typically in the 240–220 K range. To deal with this, cryo-strength and contamination testing is needed. In addition, every epoxy bond and joint will not only need to be analyzed for room temperature launch loads, but also for cryo-strength margins. Likely this means considerable testing for cryo-material strength at the proposed temperature, which was a cause of cost and complexity on JWST.

**Heater Power and Thermal Management—**The biggest engineering advantage for cold operation of the telescope is reducing the required heater power. A well-insulated mirror would need about 20–30 Watts to maintain room temperature at a Sun-Earth L2 orbit. Total power just for the mirrors on a 12-meter telescope will be approximately 1.5 kW and the backplane could require as much or more, though efficient use of insulation can help reduce the total power requirement. While this is a large power requirement, it is within engineering capabilities. In addition, strides are being made in solar array efficiency and it may be that the mass and cost of the arrays in this timeframe are no larger than other large observatories.

**Shock—**At cold temperatures, shock is not absorbed as well. For JWST, shock has played a key role in driving the launch restraint mechanism selection and drove extensive cryo-shock testing. Leveraging this experience from JWST, the impact of the issue could be reduced but will still be a complexity driver depending on how cold an operating temperature is chosen.

## 16.1.3.2 Less significant factors

**Composite Stability and Complexity—**For totally passive systems, backplane CTE for stability is a similar issue to that for mirrors. However, one not only needs to consider stability but also cryo-stress and cryo-distortion of the backplane itself. On JWST, cryogenic temperatures drove the backplane schedule in a large way. Every joint type and bond had to be assessed for cool-down stresses (in addition to launch loads). Both stability and thermal distortion had to be modeled in the design and then verified through cryogenic testing. The CTE of every tube needed to be measured and statistical studies done to assess backplane stability. A whole technology was needed to develop the 50 K-stable backplane material. There is a huge database of experience with room temperature composites. Going cold, especially to a temperature that is not 50 K, could require extensive material testing and possibly even new laminate design.

One way to mitigate backplane stability (and stability only) is to use edge sensors or laser metrology in a loop with a segmented DM to mitigate segment jitter. Segment motions due to thermal changes can be controlled with feedback to segment actuators. If a metrology solution is developed, thermal stability will prove less important than dynamic stability, which would favor a SiC backplane solution. The likely edge sensor technologies also have a room temperature heritage (capacitive and laser metrology technologies) so going cold might complicate this problem.

**Environmental Cryo/Thermal Testing—**Everything will need to be tested at its nominal operating temperature, which for cryogenic systems will take extra time and add complexity. This includes thermal balance testing and verification of all electrical connections where impedances and phase vary with temperature.

**System Optical Testing and Verification—**It is important to test an optical system at operating temperature in order to verify system wavefront error, alignment, wavefront sensing and control, and other optical performance metrics. If the telescope is significantly colder than room temperature, then room temperature optical testing will likely not be sufficient. The system testing of JWST was a major cost and complexity driver and just cooling down and warming up can take months and require costs to plan and practice.

**Use of Lubricants—**At colder temperatures, typical warm lubricants will need to be replaced. This can also mean mechanism bearings may need to be redesigned. While cryo-lubricants do exist, cryo-actuators are more complex and generally more expensive.

## 16.2   Getting to orbit: launch vehicles

Norman Rioux & Matthew R. Bolcar (NASA GSFC)

The LUVOIR mission has the potential to enable revolutionary scientific breakthroughs with the largest telescope aperture ever deployed in space. While in the more distant future, telescopes might be assembled in space, the most economical and immediate path forward is to put a large aperture in space with a single launch. In the 2030s when LUVOIR is expected to fly, the launch vehicle industry will not be exactly the same as it is now. Here we summarize our understanding of the current and future launch capabilities.

The LUVOIR observatory will be in orbit at the Sun-Earth L2 (SEL2) point, which provides a stable thermal environment and excellent field of regard. Constraints on the size of the telescope aperture and the instrument suite include the mass-to-orbit and the size of the fairing that the launcher can provide. A mission with the ambitious goals of LUVOIR will surely need a heavy lift launch vehicle; however, as with all technology, the near-term landscape for launch vehicle options includes some mature options with well-known characteristics and some less-mature options whose characteristics are still somewhat in flux and therefore are more difficult to evaluate with certainty. In order to mitigate the risk (and its associated costs) inherent in each different launch vehicle option, it is prudent to adopt telescope designs with the flexibility to use a variety of different launch vehicles and fairings.

Below we outline the current options for heavy-lift vehicles and describe their various attributes. **Figure 16.3** depicts representative values of lift capabilities for a variety of launch vehicle configurations to achieve Sun-Earth L2 orbit. **Figure 16.4** depicts the outer diameters of the associated fairings, and **Figure 16.5** depicts the height of the associated fairings. Not all of this space will be available to the payload; margins of roughly 0.5 to 1 m around the edges are needed to allow for payload motions during launch. These launch vehicles range in maturity from existing vehicles with proven flight records to vehicles that are undergoing development. Below we highlight a few of the launch providers most relevant for LUVOIR.

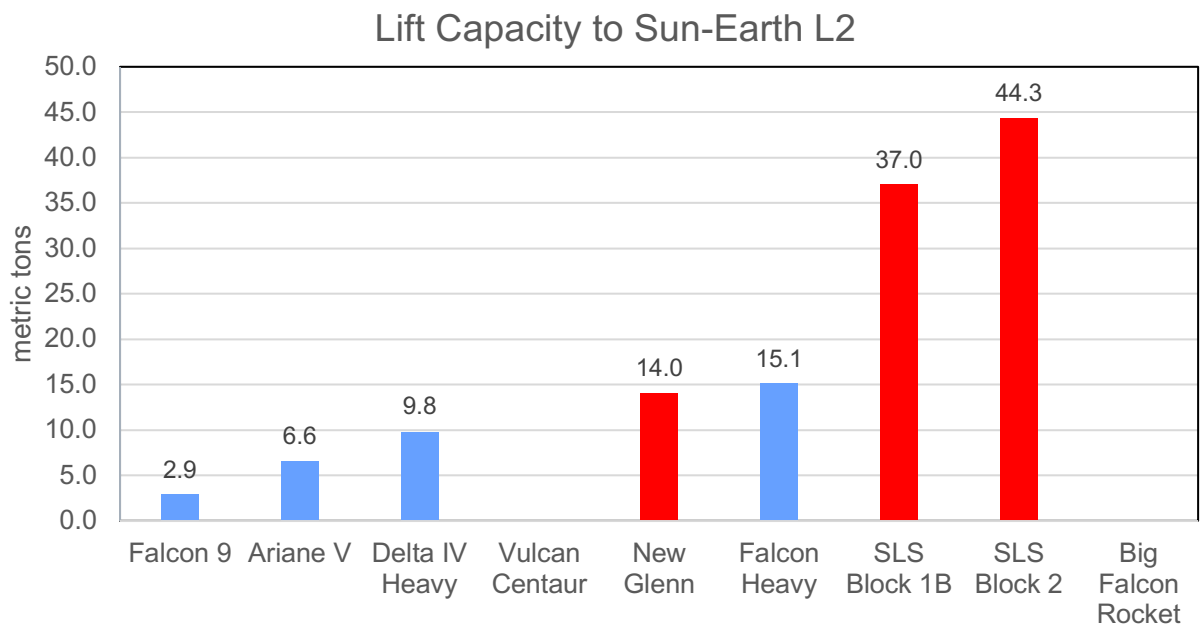

**Figure 16.3.** *Launch mass to Sun-Earth L2 orbit for a variety of current (blue) and future (red) launch vehicles. These are representative values subject to refinement of designs in development and evolution of existing vehicles. Estimates for the Vulcan Centaur and Big Falcon Rocket are not available.*

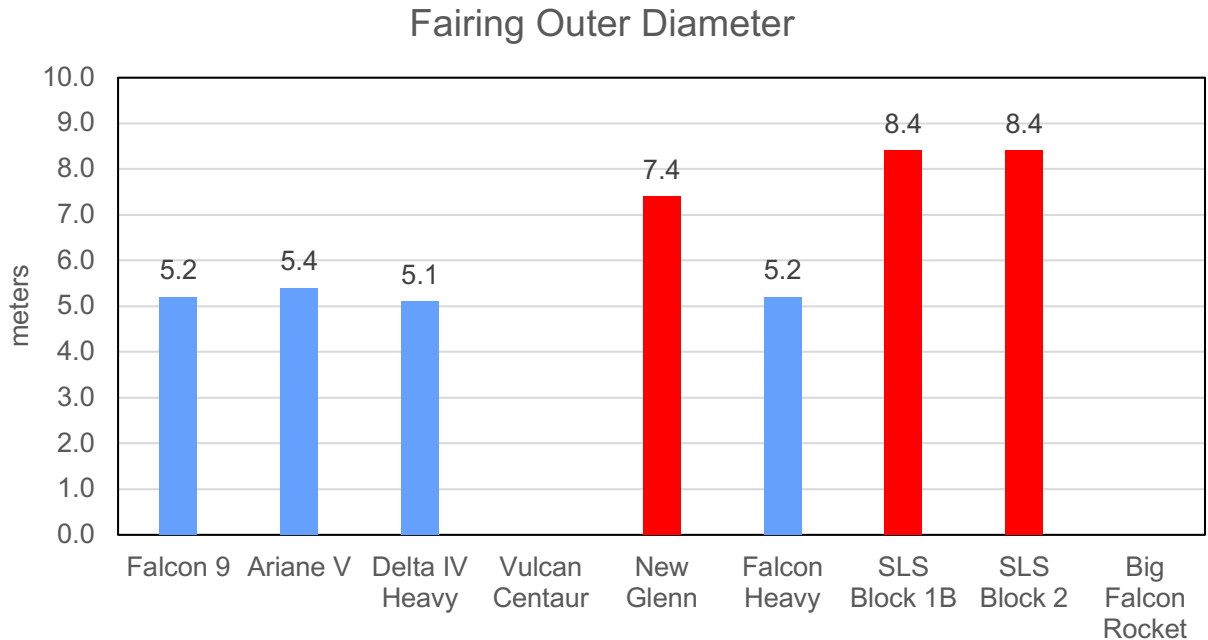

**Figure 16.4.** *Fairing outer diameters for a variety of current (blue) and future (red) launch vehicles. These outer diameters correspond to the exterior physical extent of the fairing. Inner diameters are developed through coupled loads analyses with particular payloads and are roughly 0.5 m to 1 m smaller.*

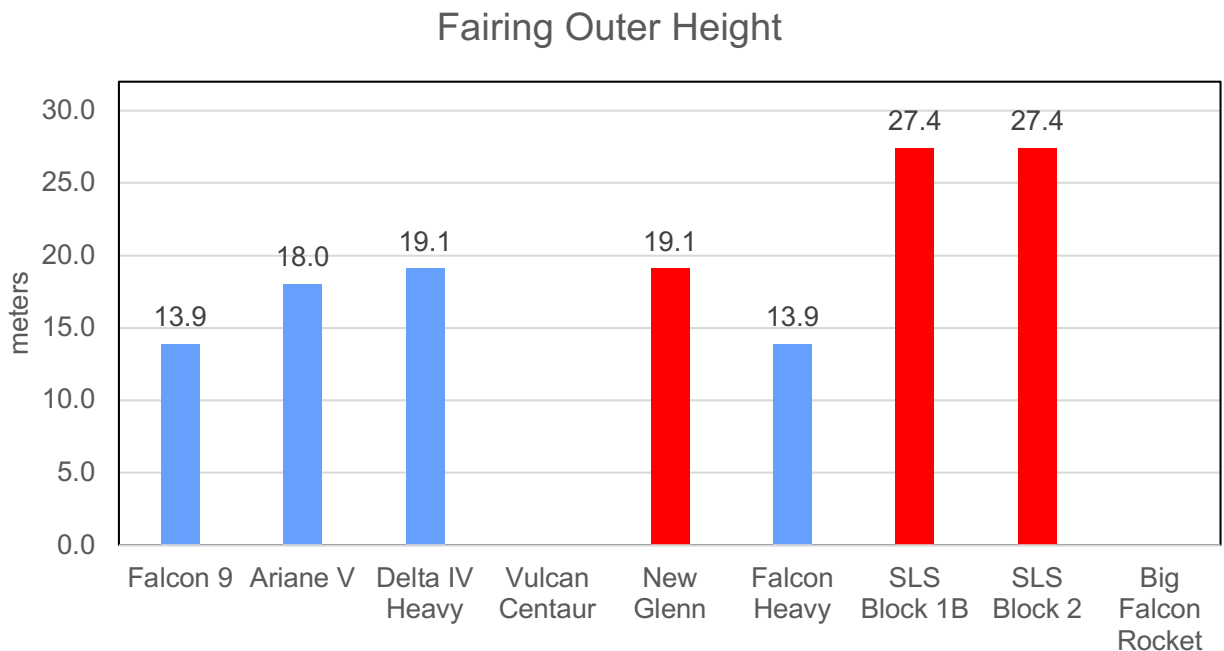

**Figure 16.5.** *Fairing heights for a variety of current (blue) and future (red) launch vehicles. These outer heights correspond to the exterior physical extent of the fairing. Useable inner heights vary depending on the fairing shape and dynamic envelopes.*

### United Launch Alliance (ULA)

**ULA's Delta IV Heavy** is an existing heavy lift launch vehicle with a proven track record. It supports a fairing with a 5-m outer diameter and a 4.6-m inner diameter, and ~10,000 kg lift capacity to SEL2. United Launch Alliance (ULA) has announced contracts for Delta IV (both "Medium" and "Heavy" variants) launches through 2023. ULA is also developing the Vulcan Centaur as their next generation medium-to-heavy lift vehicle. Little detail is currently available on the Vulcan performance, other than that it will be of a similar class to other next-gen vehicles such as the Falcon Heavy and New Glenn.

### SpaceX

**In Feb. 2018, SpaceX successfully launched of the first Falcon Heavy** vehicle. It offers lift capability in excess of the Delta IV Heavy, ~15,000 kg to SEL2. Currently, SpaceX plans to use the same 5.2 m-diameter, 13.9 m-tall fairing for both Falcon 9 and Falcon Heavy variants; there are no plans to develop a larger fairing for the Falcon Heavy. SpaceX has also announced plans to develop the Big Falcon Rocket (BFR), with capabilities exceeding that of the Saturn V. Space X has announced an "aspirational goal" of a first BFR cargo mission to Mars in 2022.

### Blue Origin

Similar to SpaceX, Blue Origin is also developing a series of reuseable launch vehicles intended to increase the cadence of commercial spaceflight launches while reducing costs. Blue Origin's heavy-lift variant, the New Glenn, is expected to be able to lift ~14,000 kg to SEL2, but with a 7.4 m-diameter, 18.5 m-tall fairing.

### NASA

**NASA is also currently developing the Space Launch System** (SLS) in three stages (Block 1, 1B, and 2) with increasing lift and fairing capacities. Options are under study for fairings with 5 m, 8.4 m, and 10 m outer diameters. The first SLS Block 1 launch is expected by 2020 with Block 1B variant coming online by 2022. A Block 2 launch is not expected until ~2030.

### Acknowledgements

Launch vehicle and fairing capacity data presented here were largely provided via private communications with representatives from SpaceX, Blue Origin, and NASA's Launch Services Program.

## 16.3   Starlight suppression with coronagraphs

Stuart Shaklan (JPL)

Stellar coronagraphs are instruments designed to suppress the veiling glare of starlight so that faint planets can be seen adjacent to their parent stars. The glare is caused by both diffraction (sidelobes) from the aperture boundaries including the outer limit of the pupil, the secondary obscuration, secondary support structures, mirror segment gaps, and scatter from the imperfect optical surfaces in the telescope and instrument. Generally speaking, coronagraphs are easier to design for filled, off-axis apertures than for partially obscured, segmented, on-axis apertures.

The diffraction problem can be addressed in three ways:

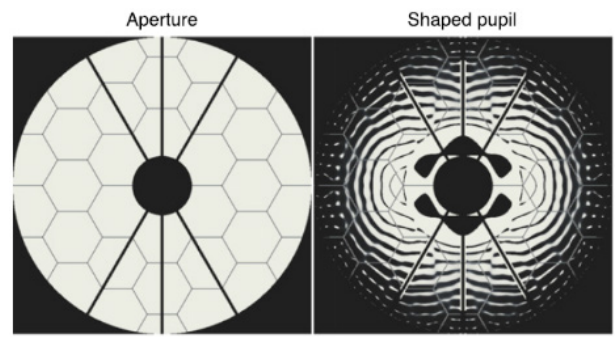

**Figure 16.6.** *Pupil apodization on a segmented aperture. Credit: N'Diaye et al. (2016). ApJ, 818, 163*

- Pupil Apodization: The first solution is to apodize the pupil. Apodization, like a low-pass electronic filter, reduces sidelobes by cancelling or removing diffracted light from the beam and allowing for the use of a small central obscuration to block the undiffracted starlight. This is accomplished either by placing a specially designed mask at a pupil image, or by using optics to concentrate the beam in an advantageous way. **Figure 16.6** shows a grayscale apodization mask function for a segmented aperture telescope, while **Figure 16.7 2** shows an optical remapping solution. Examples of this strategy include the phase-induced amplitude apodization (PIAA) or the apodized pupil lyot coronagraph (APLC).

- Lyot Mask: The second approach is to use masks in the image plane designed to diffract most of the on-axis light to the outer edges of the beam. This is then followed by another mask at the reimaged pupil, called the Lyot plane, to block the outer edges of the beam and remove the diffracted starlight (**Figure 16.8**). Off-axis light from a nearby star passes around the first mask and then through the gap in the Lyot stop. Examples of this strategy include the hybrid lyot coronagraph (HLC) and vector vortex coronagraph (VVC).

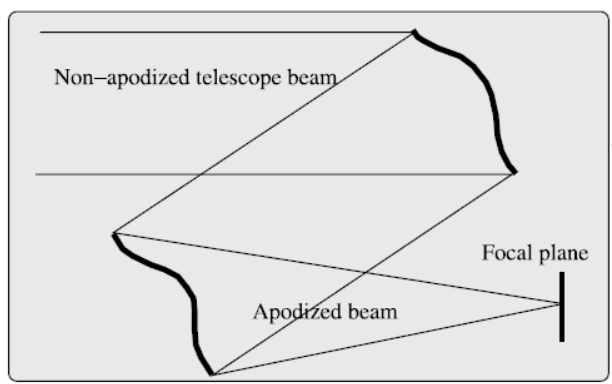
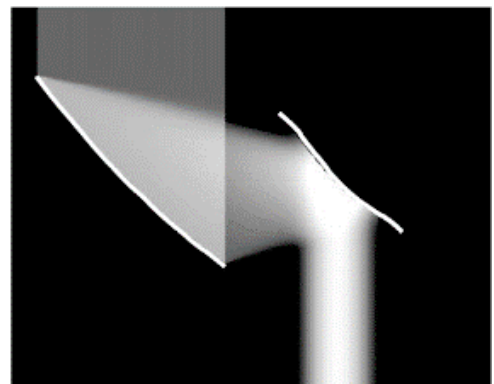

**Figure 16.7.** *Pupil remapping. Credit: Guyon et al. (2005). ApJ, 622, 744*

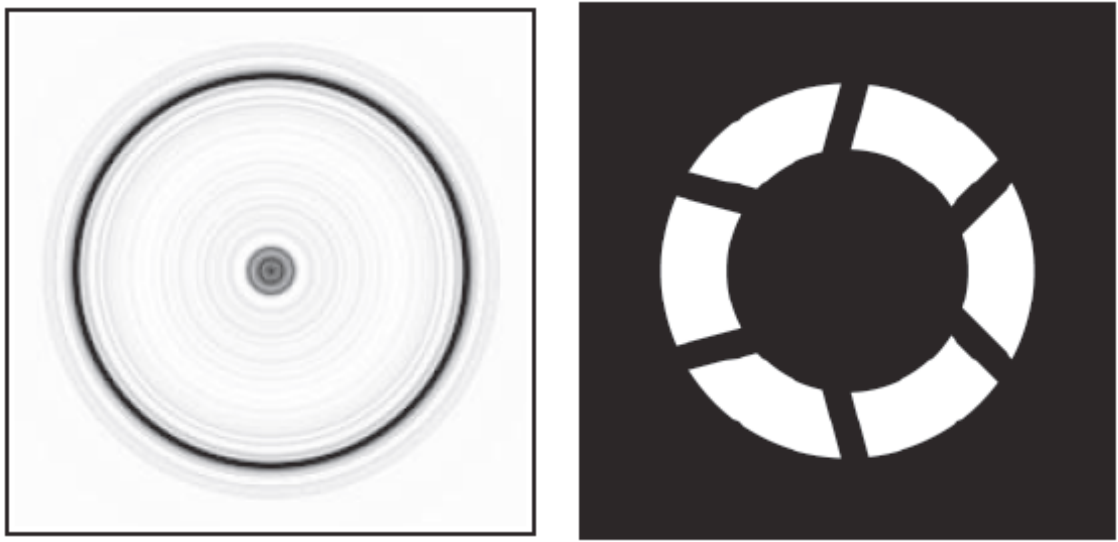

**Figure 16.8.** *Hybrid Lyot image plane mask (left) and Lyot mask (right) for the WFIRST coronagraph. Credit: Trauger et al. (2013). SPIE, 8864, 886412*

• Nulling Interferometry: The third solution, the Visible Nulling Coronagraph (VNC), is to split the pupil light into two beams and then recombine them using a phase shift and beam shear to cancel the on-axis starlight but not off-axis source flux (**Figure 16.9**).

Coronagraph designers for space-based corongraphy are gravitating toward hybrid approaches combining pupil apodization and image plane masking to deal with segmented apertures, but other solutions are still being explored.

With diffracted light eliminated from the system, scattered light originating from aberrations, coating defects, and contamination remains and must be removed. This is achieved by flattening the wavefront using one or more deformable mirrors (DMs), with typically >1000 actuators within the pupil (**Figure 16.10**). To control the wavefront, it must first be sensed. This is done by adjusting the DM surface several times while recording the change in image plane illumination. An algorithm then determines the required wavefront

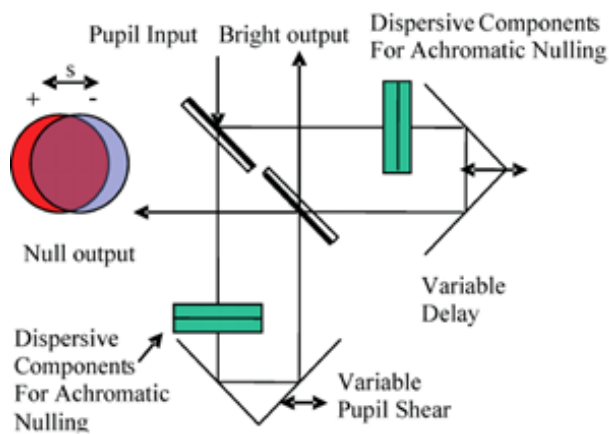

**Figure 16.9.** *Nulling coronagraph. Credit: Shao et al. (2006). SPIE, 6265, 626517*

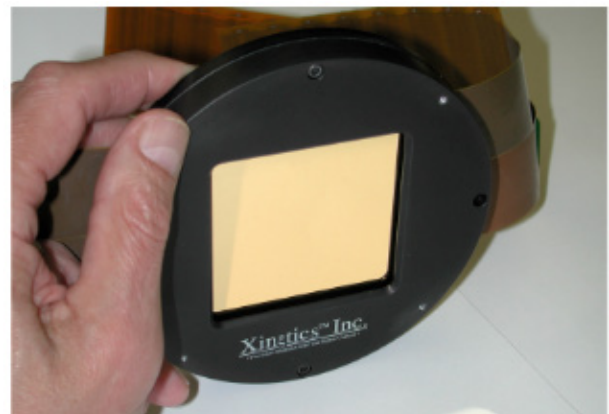

**Figure 16.10.** *64 x 64 element DM with a fused silica facesheet. Credit: Trauger et al. (2011). SPIE, 8151, 81510G*

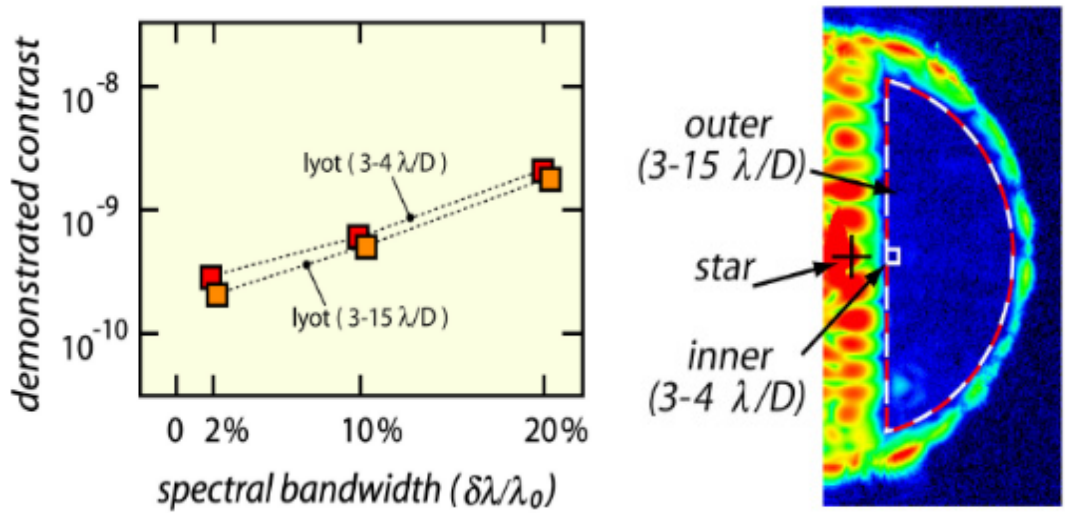

**Figure 16.11.** *Dark hole in broadband light, demonstrated at the JPL High Contrast Imaging Testbed. Credit: Trauger et al. (2011). SPIE, 8151, 81510G*

correction and commands the DMs to form new surface shapes. This is repeated until a "dark hole" is formed with a level of glare low enough to expose a planet. **Figure 16.11** shows a dark hole achieved in the laboratory in a 10% wavelength bandpass.

The desired level of suppression is $10^{-10}$ for imaging of an Earth-like planet around a Sun-like star; that is, the residual scatter in the image plane after diffraction and wavefront control is 10 billion times below the level of the incident starlight. Amazingly, this can be achieved using optics of equivalent quality to those currently used on the Hubble Space Telescope, and is limited mainly by the ability to accurately set the DM and to hold the system stable. The stability issue is perhaps the most challenging, with sub-nanometer requirements imposed on the wavefront. This is particularly challenging when it comes to low-order aberrations such as pointing, focus, coma, and astigmatism. To measure and control these terms, low-order wavefront sensors with fast response times using the starlight rejected by the coronagraphic masks are being developed.

The effectiveness of a coronagraph is evaluated based on several performance metrics: the inner working angle (IWA; i.e., how close to the star you can observe), wavelength bandwidth over which it can operate effectively, the throughput of the system, and the raw contrast (i.e., the level of glare suppression). Typically, achieving high performance at small IWA leads to either lower system throughput or increased sensitivity to aberrations and the finite stellar diameter. The challenge of suppression increases with bandwidth; the broader the band (more signal photons), the more background appears, and the more challenging the wavefront sensing and control becomes. Perhaps most importantly, the IWA scales proportionally with wavelength ($\propto \lambda/D$, where D is the telescope diameter); the same coronagraph that works at 50 milliarcsec (mas) at a wavelength of 500 nm will be limited to about 100 mas at 1 μm. This is an important factor when characterizing exoplanet spectra in the near-infrared.

## 16.4   Impacts of optical coatings on polarization and coronagraphy

Matthew R. Bolcar (NASA GSFC)

The LUVOIR mission's exoplanet science objectives require high-contrast imaging with an internal coronagraph. A coronagraph's wavefront control system uses deformable mirrors to correct amplitude and wavefront aberrations in the optical system, allowing the diffracted starlight to be suppressed so that orbiting exoplanets can be directly observed. Metallic coatings that are typically used on mirror surfaces present a challenge for a coronagraph's wavefront control system. Specifically, the orthogonal polarization states of light (often denoted by "*s*" and "*p*") are affected differently by the coating, as a function of wavelength, angle of incidence, and coating properties.

There are two effects of concern. The first effect is polarization aberration, where each polarization state sees a different amplitude (*diattenuation*) and phase change (*retardance*) upon reflecting from a metallic surface (see **Figure 16.12**). The second effect is cross-polarization leakage, where some portion of a polarization state is converted into the orthogonal state upon reflection. The result of these two effects is that when unpolarized light (such as starlight) reflects from a metallic surface, four independent (or incoherent) electric fields are created: *s*-incident light reflected in the *s*-state ("*ss*"), *s*-incident light reflected in the *p*-state ("*sp*"), *p*-incident light reflected in the *p*-state ("*pp*"), and *p*-incident light reflected in the *s*-state ("*ps*"). Since these fields are incoherent (i.e., they have uncorrelated amplitude and spatial variations), a coronagraph cannot sense and correct each one individually. Instead, the coronagraph wavefront control system will sense and correct the average of the four fields. The remaining uncorrected portion will contribute to leaked starlight that could potentially obscure an exoplanet. It is important to note that the cross-polarization terms (*sp* and *ps*) are orders of magnitude

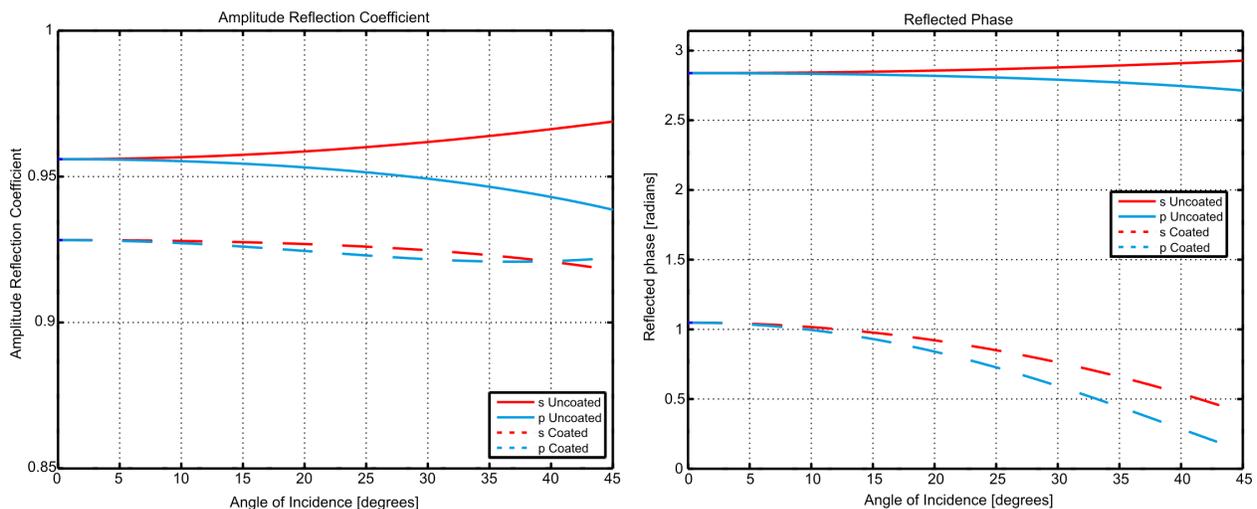

**Figure 16.12.** *Left: Amplitude reflection coefficient for each polarization state as a function of angle of incidence for a bare aluminum mirror (solid lines) and an aluminum mirror coated with a quarter-wave of MgF$_2$ (dashed lines) at a wavelength of 550 nm. Right: The reflected phase shift for the same two cases. In each case, the orthogonal polarization states experience a different amplitude and phase change upon reflection. Angles of incidence at the primary and secondary mirrors for an on-axis, ~12-m-class telescope would typically be less than 15 degrees.*

smaller than the primary terms (ss and pp) (Balasubramanian et al. 2005).

There are several ways to address the polarization issues:

**Telescope Design Considerations:** Both polarization effects are strongly dependent on the angle-of-incidence (AOI) of the light at the optical surface. Slower (high-F/#) optics will have lower AOIs across the surface of the mirrors, thus minimizing the polarization effects. Flat mirrors that are used to fold the optical path at large angles should also be avoided or used in crossed-pairs such that the effects are cancelled out. Slower optics, however, can lead to longer optical systems with smaller fields-of-view. The impacts of polarization aberration and cross-polarization leakage must therefore be traded against volume constraints and science objectives.

**Coating Properties:** The polarization effects are also dependent on the coating properties, specifically the index of refraction; choosing appropriate materials can help minimize the effects. To enable the LUVOIR mission's ultraviolet (UV) science objectives requires a protected aluminum coating on at least the primary and secondary mirrors. It is expected that there is little that can be done from a coating perspective to further reduce the polarization effects, aside from ensuring that the protective overcoat material does not significantly increase the effects over the base aluminum layer (Balasubramanian et al. 2011).

**Coronagraph Architecture:** Another effective way to deal with polarization aberrations is to split the light at the coronagraph and only observe one polarization at a time. This can be done serially by a single instrument: a polarized filter would select a single polarization state for which the aberrations would be sensed and observed, allowing for exoplanet detection and correction in that polarization state; the orthogonal polariza-tion state could then be selected and the observation repeated. The obvious drawback is requiring twice the amount of time to capture all of the exoplanet photons. Alternatively, the polarization states can be split with a polar-izing beam-splitter, with each being sent to a separate coronagraph instrument. This al-lows for simultaneous observation of all of the exoplanet photons, at the expense of requir-ing two coronagraph instruments, each with its own focal plane, filter wheels, deformable mirrors, and associated electronics. An addi-tional avenue of research includes introduc-ing polarization-controlling diffractive optical elements to correct the effects.

Regardless of the approach taken, there are two questions that remain to be answered. The first question is: how effective are polarizing filters or polarizing beam-splitters at separating the orthogonal states? Achieving $10^{-10}$ raw contrast may require polarizing optical components that are beyond the state-of-the-art. The second question regards the cross-polarization leakage term. When a single polarization state is selected (say, $s$), both the $ss$ and $ps$ components are transmitted. If the coronagraph wavefront control system senses and corrects the $ss$ component, then the $ps$ component will contribute a low-level static speckle background that may obscure an exoplanet. Modeling is being performed on the LUVOIR architectures to fully understand the magnitude of these polarization effects and determine if they are significant enough to be of concern.

It is important to note that coating-induced polarization aberration and cross-polarization leakage will be generated by *any* metallic mirror coating. **Figure 16.13** shows the diattenuation and retardance for both a bare aluminum-coated and bare silver-coated mirror. Both aluminum and silver have similar order-of-magnitude effects. It is therefore a false assumption to believe that high-contrast coronagraphy performance can be significantly improved by switching from an aluminum coating to a silver coating.

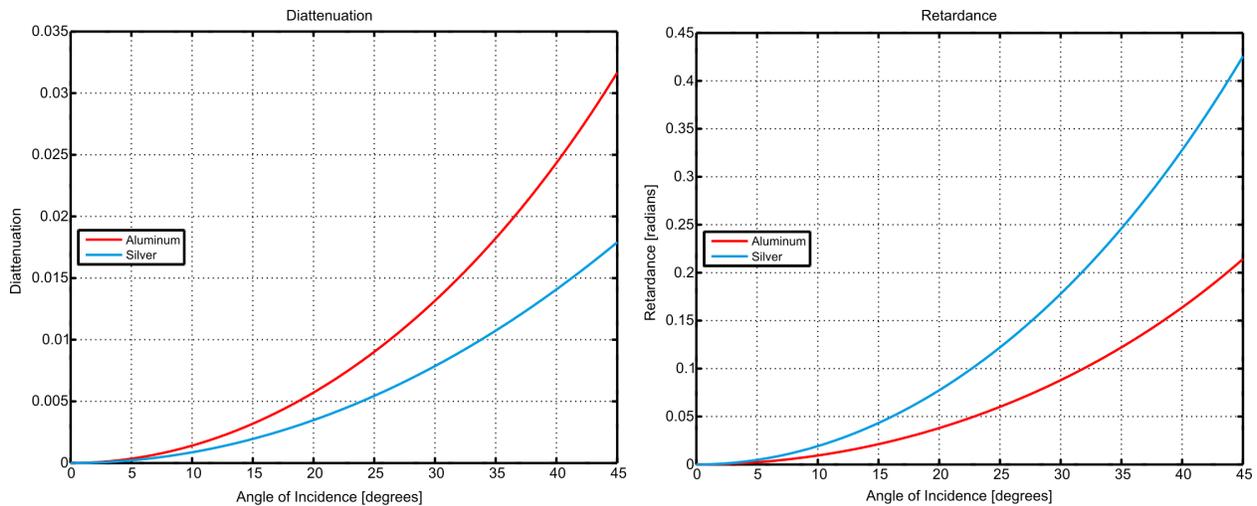

**Figure 16.13.** *Left: The diattenuation (normalized reflected amplitude difference between polarization states) as a function of angle of incidence for bare aluminum and bare silver coatings at a wavelength of 550 nm. Right: The retardance (absolute reflected phase difference between polarization states) as a function of angle of incidence for the same two metals. Angles of incidence at the primary and secondary mirrors for an on-axis, 12-m-class telescope would typically be less than 15 degrees.*

## 16.5 UV coatings and short-wavelength cutoff

Matthew R. Bolcar (NASA GSFC)

Enabling the LUVOIR mission's compelling ultraviolet (UV) science goals, while also maintaining broadband capabilities to support the exoplanet science mission and other general astrophysics observations, will require a high-performance broadband reflective coating. Of the common metallic mirror coatings (gold, silver, aluminum), only aluminum is capable of providing high reflectivity into the UV at wavelengths as short as 100 nm. However, almost immediately upon exposure to air (even at very low pressures), an oxidation layer forms on the surface of aluminum-coated mirrors that dramatically reduces its UV performance (Balasubramanian et al. 2017). Protective overcoats (usually a fluoride such as LiF, $MgF_2$, or $AlF_3$) are used to arrest the oxidation once it has begun and protect the aluminum layer from further oxidation. These overcoat layers themselves can also impact the reflectivity of the mirror at the shortest wavelengths. Current technology development efforts are focused on improving deposition processes to maximize protected aluminum coating reflectivity at wavelengths below 150 nm, while maintaining high reflectivity at longer wavelengths through the visible and near-infrared. **Figure 16.14** shows a number of reflectivity curves for demonstrated and theoretical protected aluminum coatings (Hennessy et al. 2016).

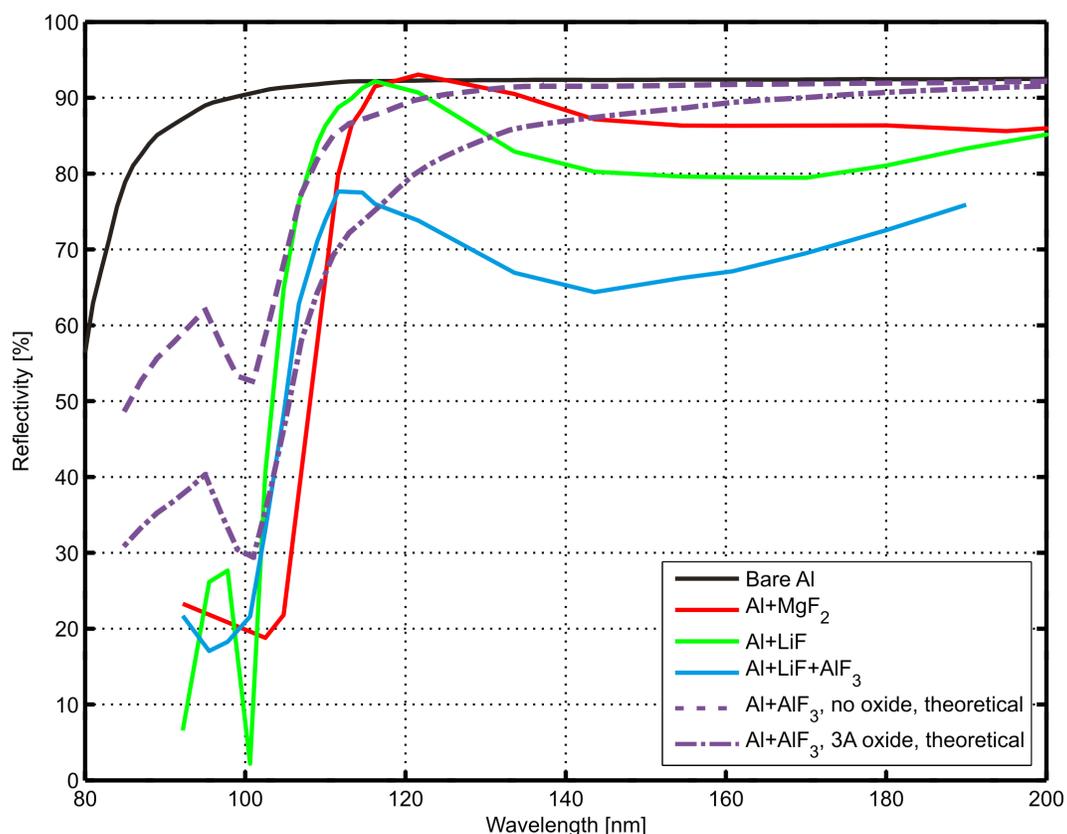

**Figure 16.14.** *Theoretical and demonstrated performance for various aluminum coatings. The theoretical, un-oxidized bare Al performance is shown in black. Demonstrated performance for $MgF_2$ and LiF overcoats are shown in red and green, respectively. A LiF overcoat protected by a thin layer of $AlF_3$ is shown in blue. Finally, theoretical performance for an $AlF_3$ overcoat is shown in purple for two scenarios: without an interstitial oxide layer (dashed) and with a 3 Å interstitial oxide layer (dot-dashed). This figure was adapted with permission from Hennessy et al. (2016), with additional contributions from M. Quijada (GSFC) and K. Balasubramanian (JPL).*

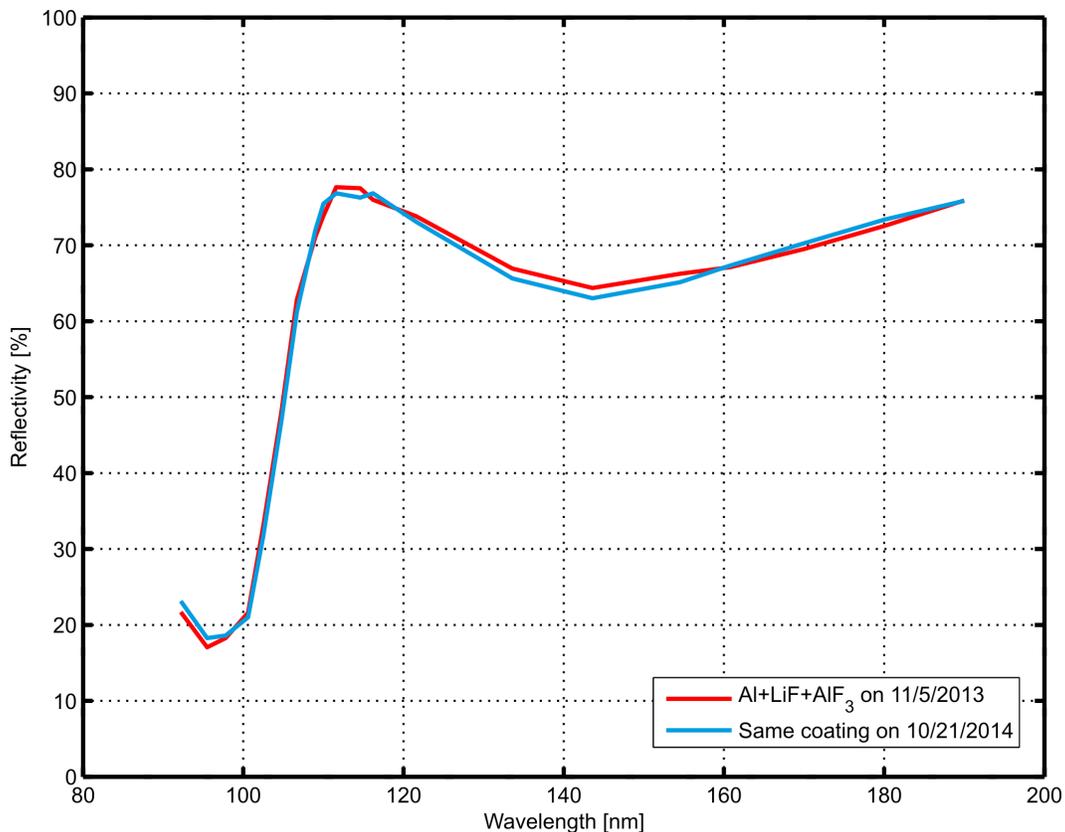

**Figure 16.15.** *Demonstrated stability for an aluminum coating protected by LiF, and further overcoated with a thin layer of AlF$_3$. The AlF$_3$ helps prevent the LiF coating from absorbing water vapor from the ambient environment, thus preserving the as-deposited performance of the coating. Data provided courtesy K. Balasubramanian (JPL).*

Typical protected aluminum coatings use a single, thin layer of MgF$_2$ as the protectant. This is the coating used on the Hubble Space Telescope and provides excellent reflectivity (>85%) at wavelengths longer than ~120 nm. Shortward of 120 nm, the reflectivity sharply drops to about 20%. Current research efforts are developing Atomic Layer Deposition (ALD) processes for depositing a very thin (~few nanometers) layer of MgF$_2$, which may provide protection from oxidation while maintaining the high-reflectivity of bare aluminum.

Aluminum protected by LiF provides reflectivity greater than 80% at wavelengths as short as 110 nm and depends critically on the coating process employed. Its cutoff wavelength is around 100 nm. It is important to also note that LiF is a hygroscopic material that deteriorates when exposed to water vapor. Mirrors coated with LiF would need to be held under a constant dry purge during the entire integration and test phase, as well as launch. This will be extremely challenging for a system as large and complex as LUVOIR. Alternatively, protecting the LiF layer with a thin layer of AlF$_3$ or MgF$_2$ can prevent the absorption of water vapor and stabilize the performance of the coating over long periods of time (see **Figure 16.15**).

**Figure 16.14** also shows two theoretical curves for an aluminum mirror protected by $AlF_3$ only (purple curves). The first curve is a theoretical best case in which no oxide layer has formed on the Al undercoat. The second curve shows the theoretical performance assuming a 3 Å layer of oxide has formed between the Al and the $AlF_3$ overcoat. Layer thicknesses and process conditions affect the resulting performance and may allow further optimization.

Several technology development efforts are currently underway to optimize these coatings and the deposition processes, including both physical vapor deposition (PVD) and atomic layer deposition (ALD). New techniques that allow the overcoat to be deposited immediately after the deposition of the base Al layer will reduce the thickness of the oxide layer that forms or prevent it all together. Process improvements will also help increase the reflectivity, as well as its uniformity across the mirror surface.

## 16.6   Ultraviolet detectors for cosmic origins and exoplanet science with LUVOIR

Kevin France (University of Colorado – Boulder) & David Schiminovich (Columbia University)

The LUVOIR Surveyor is envisioned as the `Space Observatory of the 21st Century'. With up to 40 times the geometric collecting area of the Hubble Space Telescope and highly sensitive, multiplexed instruments, LUVOIR is poised to provide transformative scientific measurements of a broad range of astrophysical objects. Ultraviolet imaging and spectroscopic capabilities are central to the majority of the key scientific goals of LUVOIR, from quantifying the flows of matter between galaxies and the intergalactic medium, to understanding how host star's UV radiation regulates the atmospheric photochemistry on habitable planets.

Carrying out high-precision ultraviolet astronomy across such a wide range of sources requires detector systems working below the atmospheric cut-off (100–400 nm) with low-noise and/or photon-counting capability, high quantum detection efficiency, large format size, and high temporal resolution. Ideally, all of these characteristics would be encompassed in a single detector, but multiple technologies may be required to accomplish LUVOIR's suite of science investigations. We present an overview of UV detector technologies so the LUVOIR STDT and NASA can make informed recommendations for directed technology investment to support first generation UV instrumentation for LUVOIR. We note that with the goal of serviceability, some technologies that are less mature today may be optimized by the time second generation LUVOIR instruments are proposed, and as such, a long-term but adaptable technology maturation plan would be desirable for UV detectors.

**Microchannel Plate Devices:** Microchannel plates (MCPs) with "solar-blind" photocathodes, low dark noise rates, and zero read noise have a rich flight heritage on astronomy, heliophysics, and planetary science missions.

Advantages: MCP-based detectors are inherently photon-counting, operate at room temperature, can be ruggedized for 10+ year lifetimes in space, are scalable to large formats, and offer relatively high quantum efficiency at short UV wavelengths ($\lambda < 130$ nm).

Disadvantages: MCPs have limited dynamic range for bright objects that require instrument safety protocols, do not regularly support signal-to-noise (S/N) > 100 observations owing to fixed pattern noise, and experience issues with long-term "gain-sag" (burn-in at locations of prolonged high illumination).

**CCD and sCMOS:** Charge coupled devices (CCDs) have significant heritage in space and on the ground, can be $\delta$-doped to improve UV performance, and anti-reflection coatings can be further optimized to offer higher QE over a selected bandpass. Research is ongoing to extend the range of QE enhancements.

Advantages: UV-optimized CCDs have flight-heritage on astrophysics suborbital missions and on solar missions at shorter UV wavelengths. The large dynamic range and flat-field characteristics of these CCDs are well suited for high S/N observations. CCDs are moving to larger formats, and advancements in electron-multiplying CCDs (EMCCDs) offer the prospect of sub-electron read noise photon-counting operation. Other silicon-based technologies such as low-noise scientific complementary metal-oxide-semiconductor (sCMOS) are now being optimized for the UV, using similar processing techniques to those being used for CCDs.

**Table 16.1.** *UV detector targets for LUVOIR and a comparison with the state-of-the-art in laboratory, flight and commercial technology (adapted from Bolcar et al. 2016).*

| Parameter | Goal | State-of-the-Art | |
|---|---|---|---|
| Operational Bandpass | 90 nm–400 nm | < 90 nm–300 nm | MCP |
| | | < 90 nm–400 nm | EMCCD |
| | | ~200 nm–400 nm | sCMOS |
| Read Noise | 0 | 0 | MCP |
| | | N/A for multi. mode | EMCCD |
| | | 0.8–1.0 e$^-$ | sCMOS |
| Dark Current and/or Spurious Count Rate | ≤0.1–1 counts/cm$^2$/s ≤1–10$^{-3}$ e$^-$/resol/hr | 0.05–0.5 counts/cm$^2$/s | MCP |
| | | > 0.005 e$^-$/resol/hr | EMCCD |
| | | > 0.005 e$^-$/resol/hr | sCMOS |
| Quantum Efficiency (Peak) | 75% (Far UV–Near UV) | 45–20% FUV–NUV | MCP |
| | | 30–50% FUV–NUV | EMCCD |
| | | 20–50% NUV | sCMOS |
| Resol Size | ≤ 10 μm | 20 μm | MCP |
| | | 8–13 μm | EMCCD |
| | | ~5–11 μm | sCMOS |
| Dynamic Range (Max. Count Rate) | ≥ 10$^4$ Hz / resol (as needed) | 2kHz / resol 5 MHz global | MCP |
| | | Readout dependent | EMCCD |
| | | 10$^5$ Hz / resol | sCMOS |
| Time Resolution | ≤ 1 ms (as needed) | << 1 ms | MCP |
| | | < 10 ms | EMCCD |
| | | < 10 ms | sCMOS |
| Format | ≥ 8–16k pixels per side with high fill factor | 8k × 8k | MCP |
| | | ~1k × 2k (4k × 4k proto) | EMCCD |
| | | ~4k × 4k | sCMOS |
| Radiation Tolerance | Good | Good | MCP |
| | | TBD | EMCCD |
| | | Good | sCMOS |

Disadvantages: CCDs are not inherently "solar-blind," due to the low band-gap of silicon, so care must be taken to mitigate sensitivity to scattered visible-band radiation in UV applications. Testing and optimization of these devices for radiation hardness in an L2-like environment is also an active area of research. Initial results indicate that EMCCDs in particular suffer degrade charge transfer efficiency due to extended periods of radiation exposure. Hole-multiplying CCDs (HMCCD), a p-type version of EMCCDs, are currently under development and would be inherently radiation hard. sCMOS devices require similar investigation into radiation hardness and low-temperature operation.

**Advanced Concepts:** Less mature UV detector technologies, such as microwave kinetic inductance detectors (MKIDs), offer the possibility of energy resolution at the pixel level. Uncertainties in the scalability to significant pixel/spaxel counts and cryogenic operation currently limit the utility of these devices for LUVOIR, but these issues may be quantified and possibly overcome with additional technology investment.

**Challenges unique to UV:** Key technology challenges that remain unique to the UV involve boosting efficiency and reducing noise, along with several other issues linked to these goals:

Quantum Efficiency: UV sensitive detectors have quantum efficiencies at or below 20–50% within the band, leading to the possibility that future technologies may provide factors of up to 2–5× improvement in overall UV sensitivity. Some UV sensor technologies have a comparable or higher QE in the visible, raising potential "red leak" issues that may require additional filtering. A related challenge is the low transmittance of UV bandpass filters, which limits options for efficient red-blocking or band selection. Multi-layer filter advancements offer a promising avenue of research.

Noise: The sky background is several magnitudes fainter in the UV than in the visible, which presents both an opportunity and a challenge. The opportunity is the exploitation of a low-background window for the study of faint objects. For example, for broadband FUV observations, the sky background is 28.5 mag/sq. arcsec, or fewer than ~1 photon per resol for 10–30 min exposure times on a 10-m class telescope, even less for narrowband observations or spectroscopy. The challenge is that taking full advantage of this low background requires detectors that do not themselves limit faint observations, motivating low noise or photon-counting technologies, with read noise and/or dark current typically lower than required for standard broadband visible observations. Borosilicate MCPs with anti-coincidence shielding should mitigate the higher particle backgrounds inherent to operation at L2.

Dynamic Range: These same photon-counting detectors may also have limited dynamic range, particularly in the large formats required for LUVOIR, and this is a challenge that needs to be addressed by all technologies being considered. A detailed comparison of the dynamic range performance for each of the leading detector technologies, given realistic instrumentation and observing programs, would be a valuable exercise for the LUVOIR study.

Improving performance in these areas can significantly impact science return. The required exposure time to reach a given S/N for a particular target in a sky-background-limited observation scales as $t \sim QE$ and increases to $t \sim QE^2$ in the detector dark current-limited regime, motivating efficiency gains. Similarly, the required exposure times decrease linearly with improvement in dark current and/or read noise, down to the very low sky-background limit. No single technology leads performance in all three areas (efficiency, read noise, and dark current). At this point, the optimal UV detector is likely to be application or instrument-specific.

# 17 Appendix E: Additional LUVOIR instrument concepts

Here we present information on some possible alternate and/or second-generation instrument concepts for LUVOIR.

## 17.1   Starshades for LUVOIR

Eric Cady (JPL) & Aki Roberge (NASA GSFC)

### 17.1.1   Introduction

Starshades are a relatively newer idea for providing the extreme high-contrast needed for exoplanet direct observations. They have strengths and weaknesses that are complimentary to those of coronagraphs. A starshade is a independent spacecraft flying in formation with the telescope (**Figure 17.1**). The goal is to keep the telescope in the shadow cast by the starshade, and keep both spacecraft aligned with the target star. The larger the telescope, the larger the starshade needs to be. The edges of a starshade have a very particular shape to control diffraction and deepen the shadow at the location of the telescope. An example of a small starshade mission concept can be seen in this video: https://exoplanets.nasa.gov/resources/1015/. A lecture on the Theory and Development of Starshades given at the 2014 Sagan Summer Workshop is available here: https://www.youtube.com/watch?feature=player_detailpage&v=h5w6z0jow1Q#t=0.

A starshade blocks the unwanted bright light from an exoplanet host star before it enters the telescope, while allowing light from nearby planets to pass nearly unattenuated. Therefore, internally scattered light reaching the detector is minimized. Telescope segments and obstructions do not need to be masked out and the wavefront does not need to be corrected with deformable mirrors. A starshade can provide a very small inner working angle (IWA), approximately given by the starshade radius (R) divided by the starshade-telescope separation (z), down to around the diffraction limit of the telescope ($\sim$ 1.2 $\lambda$/D). For planets in habitable zones of nearby sun-like stars, the starshade-telescope separations are tens to hundreds of thousands of km.

Starshades have no intrinsic outer working angle (OWA). They can be designed to operate over large bandpasses (a few times larger than today's coronagraph bandpasses) and to provide small IWAs at virtually any wavelength. For a fixed IWA and contrast level, the required starshade size increases with wavelength. Since the starshade is not part of the optics train, internal reflections are kept to a minimum and total throughput is high. This makes them excellent for deep spectroscopy, especially in the NIR where coronagraphs struggle to provide small IWAs.

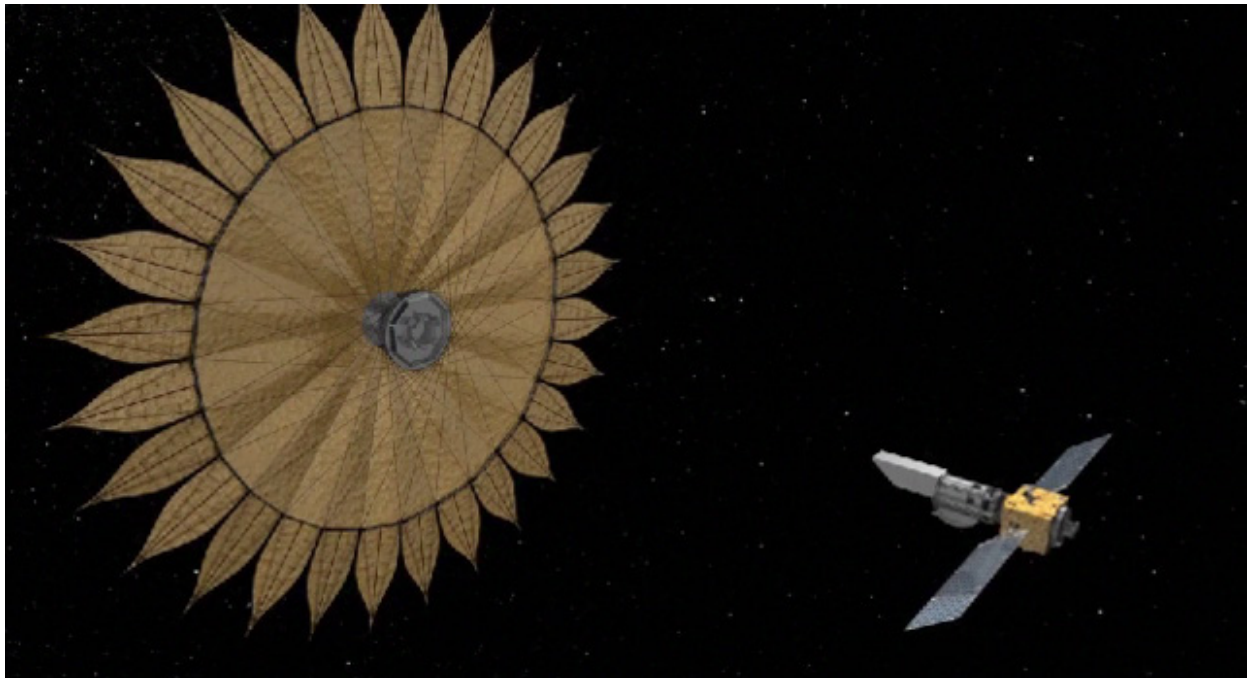

**Figure 17.1.** *Artist's conception of the Exo-S mission concept, a starshade paired with a small telescope. The image captures a moment just after starshade deployment. Credit: NASA / JPL / Caltech.*

On the weaknesses side, the need to slew the starshade over huge arcs to realign it with different target stars means there are long intervals (days to weeks) between the high-contrast observations and the total number of observations is fuel-limited. The telescope can do other kinds of astronomical observations in the intervals, but starshades are relatively inefficient for high-contrast imaging surveys. The starshade and telescope must be precisely aligned during observations to maintain high contrast. Keeping the telescope in the darkest part of the starshade shadow generally translates to lateral position precision of about a meter (the separation precision is much less stringent).

Full-scale end-to-end system tests on the ground are not possible, although sub-scale tests are being done in the lab and in the field (e.g., Cady et al. 2010; Smith et al. 2016). The large sizes of starshades means that they must be folded up for launch and deployed in space. The exact shape of the optical edge must be accurate (on the order of 100 μm tolerance for contrast in the $10^{-10}$ range) after deployment. Further, the thin edge of the starshade must be engineered to minimize sunlight scattering back into the telescope. More information on starshade technology deleveopment may be found at https://exoplanets.nasa.gov/exep/technology/technology-overview/

## 17.1.2   Scope of LUVOIR starshade study

Starshades have several attractive features, and their strengths complement those of coronagraphs. Therefore, a series of starshade designs for LUVOIR were investigated over the May/June 2017 period. Here we discuss the design parameters and constraints, the optimization methodology, and the results.

**Performance Constraints:** Starshade designs were created for 4 architectures with the following constraints:

Case 1.   9-m LUVOIR aperture, long wavelength cutoff at 1000 nm

Case 2.   15-m LUVOIR aperture, long wavelength cutoff at 1000 nm

Case 3.   9-m LUVOIR aperture, long wavelength cutoff at 2500 nm

Case 4.   15-m LUVOIR aperture, long wavelength cutoff at 2500 nm

After consultation with the LUVOIR team, the short wavelength cutoff for the 4 architectures was set to a goal of 250 nm. The starshade inner working angle (IWA) for each architecture was chosen to match that of a coronagraph on the same architecture (3.5 λ/D at 500 nm). For 9-m designs, the IWA was 40 milliarcsec (mas); for 15-m designs, the IWA was 24 mas.

For the LUVOIR designs, we set a requirement that the mean contrast in a 1.0 λ/D annulus centered at the IWA in the final telescope focal plane be no greater than $1 \times 10^{-12}$. The same metric was used for HabEx starshade designs in separate work and is consistent with producing $10^{-10}$ contrast after running through an error budgeting process with achievable tolerances.

**Engineering Constraints:** Each starshade was optimized for a diameter 2-m larger than the telescope aperture, to allow ~1-m formation-flying offset between starshade and telescope in all directions. Petal tip sizes and inter-petal gaps were bounded at ≥1 mm, and the number of petals was fixed at 28, which had been previously preferred in other contexts for mechanical reasons. Starshade diameter and petal length were permitted to float as part of the overall optimization.

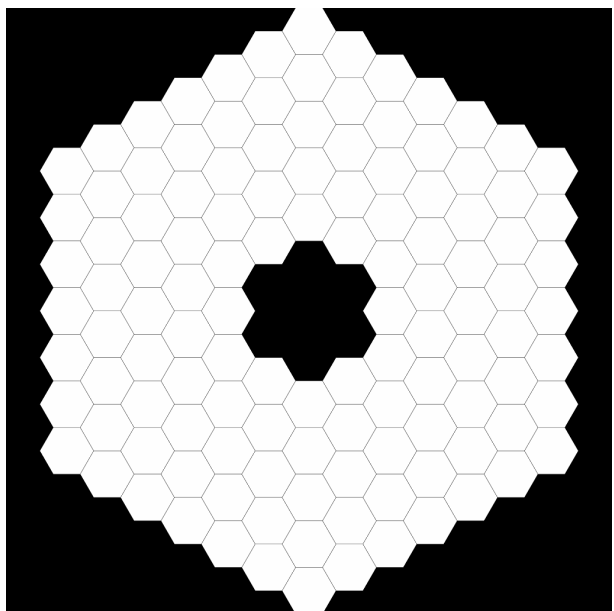

**Figure 17.2.** *Preliminary LUVOIR 15-m pupil used to verify performance of 15-m designs. A version scaled by 9/15 was used for the 9-m designs.*

**Other:** Candidate designs were checked with a representation of the preliminary 15-m LUVOIR aperture provided by M. Bolcar. In the absence of a finalized LUVOIR-B design at the time of this analysis, a 9/15 scale version of the 15-m design was used instead for evaluating the 9-m cases. This preliminary aperture is shown in **Figure 17.2**. The resulting starshade designs are not sensitive to the exact telescope aperture geometry.

### 17.1.3   Starshade size and shape optimization

The optimization procedure used for the LUVOIR starshades has been covered in other works (e.g., Cady et al. 2009). Very briefly, a smoothly varying radial apodization function, A(r), is approximated with a binary shape that has transmission of 1 or 0 everywhere (**Figure 17.3**). To select an appropriate A(r) for a LUVOIR starshade, we choose scientific constraints (IWA, bounds of science bandpass, telescope diameter) and engineering constraints (total starshade diameter, petal length, number of petals, minimum feature sizes, edge smoothness, formation-flying tolerance). These parameters are sufficient to populate a constrained linear optimization with a min-max cost function. We select the science IWA as the geometric IWA, the radius physically subtended by the starshade on the sky. The geometric IWA is defined as $\theta = \arctan(R/z) \approx R/z$, where R is the starshade radius and z is the starshade-telescope separation. We choose $\theta$ based on science goals for a particular starshade, and let the distance z float as $R/\theta$.

Other choices of science IWA are possible, such as angular radius that permits some fractional exoplanet throughput, but there are two strong advantages of using the geometric IWA. First, the parameter is independent of A(r), and so is known prior to beginning the optimization. Second, since the starshade has no structure beyond the IWA, errors in starshade manufacturing will only appear as sources from inside the angular extent of the starshade, and so using the geometric IWA as the science IWA during design means the ultimate performance will be much less sensitive to manufacturing errors.

Implicit in these designs is an assumption that the apodization is solid in the center and tapers monotonically outward, with that taper occurring over a sub-region of the radial extent. Leaving a solid central disk is a practical choice, as the spacecraft bus can be located there without interfering with the apodization profile. Manufacturing considerations require that A(r) cannot go smoothly to 0 at R or 1 at the radius of the central disk, as the petal tips or gaps between petals would become impossibly narrow. We prevent this by constraining feature sizes directly in the optimization.

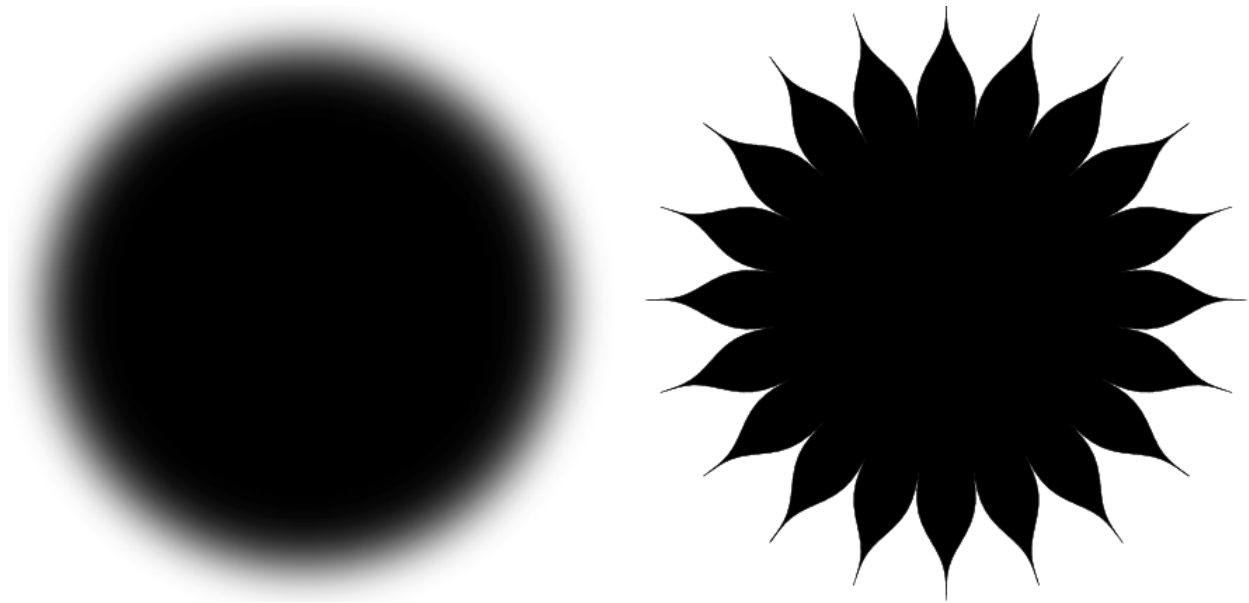

**Figure 17.3.** *An apodizer is converted into a binary starshade by replacing the tapering outer annulus with a series of N binary structures ("petals'").*

Linear optimizations provide globally optimal solutions for their constraints, so in practice we iterate on the scientific and engineering parameters and rerun linear optimizations until an apodization function consistent with the science case is produced. Once a promising apodization is identified, it is propagated through the target telescope to the focal plane using a propagator that runs on the exact locus of edge points, rather than the azimuthally symmetric approximation used for the optimization. Final evaluation of the starshade performance is done in the focal plane.

## 17.1.4  Results

In the first round of design, results were found for the two 250–1000 nm cases:

Case 1.  Starshade is a 40 mas IWA design for the 250–1000 nm band with the 9-m telescope (labeled OH22). It has a diameter of 95 m with 27.3 m petals. The telescope-starshade separation distance is 244,939 km.

Case 2.  Starshade is a 24 mas IWA design for the 250–1000 nm band with the 15-m telescope (labeled OM11). It is 160 m in diameter with 46 m petals. The telescope-starshade separation distance is 687,549 km.

The two shades, with the telescope apertures for scale, are shown in **Figure 17.4**. For the 250–2500 nm bandpass (Cases 3 and 4), no designs that satisfied the science constraints were found for starshade diameters < 200 m.

Given the large sizes of these designs, a second round of investigation was done to look at the effect on IWA of fixing the Case 1 diameter to 70 m; this raised the geometric IWA to 57 mas. From this 70 m case, loosening the science constraints only produced weak improvement in IWA; raising the permitted contrast by 10x only decreased the IWA to 53 mas, while separately raising the lower wavelength cutoff to 500 nm decreased the IWA to 54 mas.

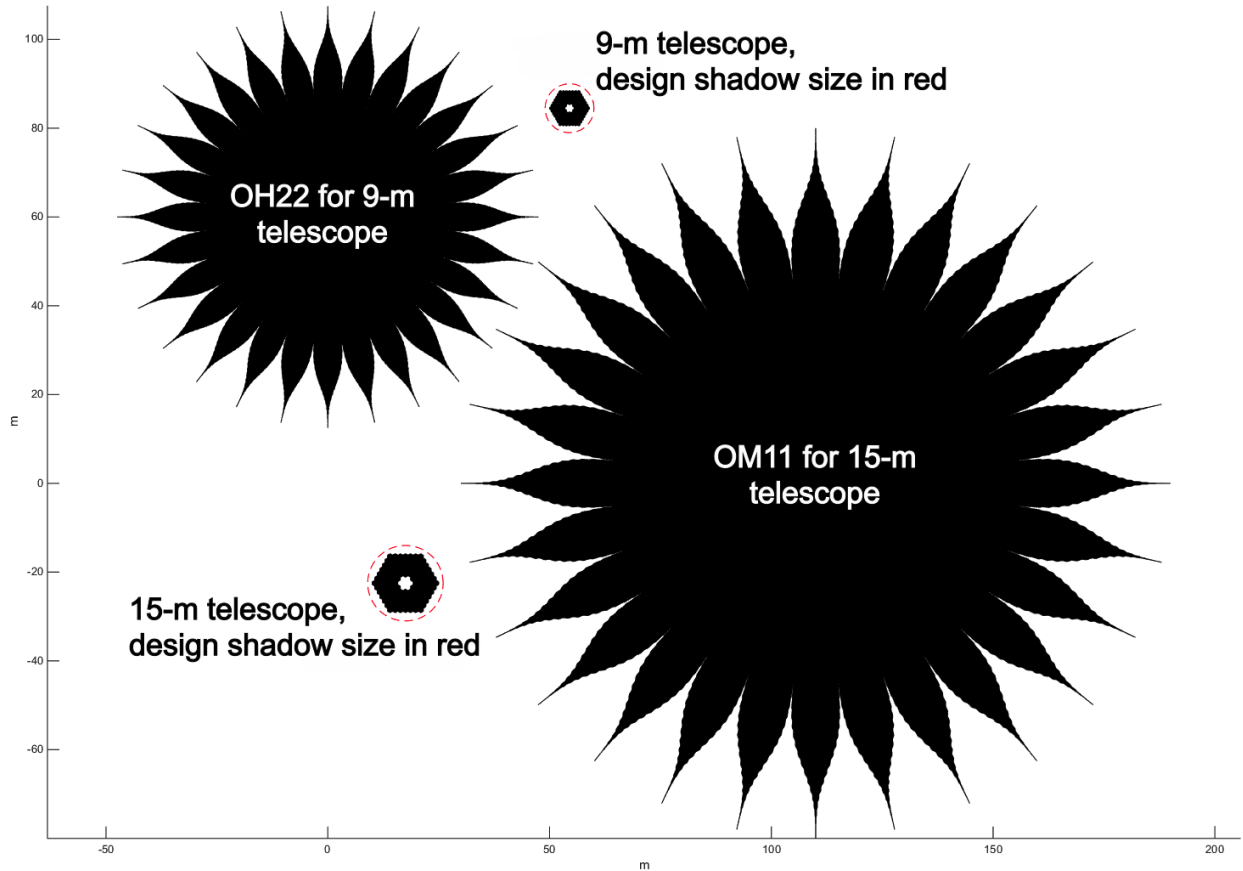

**Figure 17.4.** *The two starshades for a 250–1000 nm bandpass, to scale, along with the telescope apertures and shadows they were designed to.*

We conclude that a NUV/visible starshade for a 9-m telescope may be a viable starlight suppression option, albeit one that surpasses the current "nominal" maximum starshade size of ~70 m. However, a high performance starshade paired with a 15-m telescope is likely unfeasible.

## 17.2   Optical/Near-Infrared Spectrograph (ONIRS)

### 17.2.1   Overview

There are a variety of science cases that would be enhanced or enabled by an Optical/Near-Infrared Spectrograph (ONIRS) with different capabilities that are already offered within the ECLIPS and HDI instruments. Some of those capabilities are:

1. Optical spectroscopy at resolutions greater than R~140 (ECLIPS IFS) or R~600 (HDI grism).

2. NIR spectroscopy at resolutions greater than R~70 (ECLIPS IFS), R~200 (ECLIPS point-source spectroscopy), or R~600 (HDI grism).

3. NIR bandpass longer than 2.0 μm (ECLIPS) or 2.5 μm (HDI), for use on bright targets where the telescope thermal background does not limit performance.

These additional capabilities would enhance a wide range of scientific investigations, including probes of the chemical evolution of high-redshift galaxies, studies of the dynamics and composition of the circumgalactic medium, and measurements of black hole dynamical masses. Closer to home, ONIRS would also improve our understanding of star formation and the initial mass function, protostellar outflows and jets, winds from protoplanetary disks, the composition of exoplanet atmospheres, and the surfaces of solar system bodies. Very high spectral resolution in the optical/NIR could also enable exoplanet mass measurements via the radial velocity technique.

These observing capabilities could be achieved in multiple ways, and likely are not all compatible with a single instrument. For example, solar system surface studies will demand spatially resolved spectroscopy, while the ultra-high spectral resolution

(R~10⁵) and stable wavelength calibration needed for radial velocity work are more easily achieved with a fiber-fed point-source spectrograph. High resolution optical spectroscopy (Capability #1) could likely be added to the LUMOS instrument, a design option we will be considering in the coming months. An instrument concept that has Capability #3 appears in **Section 17.3**.

The LUVOIR STDT considered ONIRS for a first-generation instrument study. However, the STDT chose to defer detailed study of ONIRS, due to the limited number of instrument bays, some uncertainty about the performance of optical/NIR instruments planned for the future ground-based Extremely Large Telescopes (ELTs), and time/resource constraints within the present mission concept study.

### 17.2.2   Preliminary design drivers

We note that work on ONIRS did not progress to the point of a clear and detailed set of science goals and observation needs, or an actual instrument concept. However, we here present some information on particular science applications that would drive requirements on spectral resolution and wavelength range.

*Exoplanet masses with radial velocity.* Space-based radial velocity observations may present some advantages over ground-based ones, thanks to higher PSF stability and the absence of telluric features. A discussion of ultra-precise exoplanet RV from space appears in Plavchan et al. (2018). To summarize, measuring the masses of exoplanets down to Earth-mass planets around Sun-like stars (signal ~ 9 cm/s), radial velocity precision on the order of 1 cm/s will be required. Achieving this will demand very high spectral resolution to combat stellar noise (R~300,000), extremely precise wavelength stability and calibration, and sufficient sensitivity to keep individual exposure times short at such high spectral resolution. A laser frequency comb is a likely candidate for a high-precision wavelength calibration source.

*Exoplanet spectroscopy.* Particular enhancements in exoplanet transit spectroscopy could be achieved by extending the bandpass for low-resolution point-source spectroscopy beyond ~2 μm. A much fuller discussion of this science case appears in **Section 17.3**.

New high-dispersion spectroscopy methods are being used to probe exoplanet atmospheres with ground-based telescopes. This technique employs R~100,000 spectroscopy to resolve the individual lines within molecular absorption bands and cross-correlates many spectra taken over the orbital period of the planet against a highly accurate molecular absorption template. The technique is well-suited to measuring the spin of giant planets (e.g., Snellen et al. 2014) and detecting molecules in non-transiting giant planets (e.g., Birkby et al. 2017). Coupled with a high-contrast instrument, it may allow detection of molecules in the atmospheres of Earth-size planets around M dwarf stars at relaxed contrast levels (e.g., Snellen et al. 2015). Such high-dispersion + high-contrast instruments are under consideration for the future ground-based ELTs. However, there is some uncertainty about the optimal spectral resolution for using this technique on a space-based platform and the limits of its performance (e.g., Wang et al. 2017). Furthermore, it is not clear how to obtain absolute molecular abundances with this technique, as the planet's continuum flux level is not measured, and systematic uncertainties and degeneracies have not been thoroughly studied. With experience gained from employing this technique on the ELTs, its value on a space-based platform will become clearer and may lead to a desire for a second-generation LUVOIR instrument.

*Composition of solar system bodies.* Important spectral features for characterization of solar system atmospheres and surfaces lie at wavelengths longward of ~2 μm, for example, $N_2$ ices at 2.15 μm and organics like ethane ($C_2H_6$) at 3.4 μm. Such observations will demand spatially resolved spectroscopy, either with an IFS or a linear-variable filter camera. An example of the latter is the Ralph instrument on the New Horizons spacecraft. As many solar system bodies are bright, it may be possible to usefully observe them beyond ~2 μm even in the presence of the thermal background from a warm telescope like LUVOIR. Furthermore, a large space-based telescope like LUVOIR will provide unprecedented spatial resolution for such observations, as extreme adaptive optics systems on the ELTs may struggle to provide high spatial resolution over large fields of view.

## 17.3  Extension of LUVOIR's capabilities to 5 μm or beyond

Michael Werner (Caltech-JPL)

We present the scientific rationale and technical implications for the use of LUVOIR out to at least 5 μm. We consider primarily the benefits of such an extension to the study of exoplanets, which is a pillar of the LUVOIR scientific case. We discuss two possible realizations of LUVOIR (9.2-m, T = 293 K) and (12-m, T = 273 K) and also present a simple instrument concept that could achieve our scientific objectives with limited impact on the rest of the facility. This white paper is based on Werner et al. (2016) (hereafter W16), which provides considerably greater detail. The examples below are based on a 9.2-m LUVOIR and thus underestimate the power of the 15-m version under study.

### 17.3.1   Introduction

The rationale for operating LUVOIR out to at least 5 μm may be summarized as follows. One can be confident that, whenever LUVOIR launches, the study of exoplanet atmospheres will be an exciting and crucial scientific discipline. Numerous molecular species likely to be important probes of exoplanet atmospheres have bands in the infrared beyond ~2 μm including $H_2O$, CO, $CO_2$, $CH_4$, $NH_3$, and $O_3$. Many of these same species are abundant in the Earth's atmosphere and are difficult to access from the ground but are easily studied by a space telescope such as LUVOIR. In addition, recent developments suggest that atmospheric hazes or clouds may be common in exoplanet atmospheres and that they can interfere with our ability to determine molecular abundances using optical and near-infrared measurements only (e.g., Sing et al. 2016). Observations further into the infrared have the potential to improve constraints on molecular abundances by either punching through the haze or better defining its properties and extent. As a general principle, the study of exoplanet atmospheres will benefit greatly from inclusion of the widest range of molecules over the widest wavelength range. LUVOIR's infrared capabilities would inform our overall understanding of exoplanet atmospheres and, in ways which may be difficult to foresee at present, support our quest to identify habitable or potentially habitable planets through atmospheric studies. Of course, an infrared-capable LUVOIR could also carry out exciting studies of targets other than exoplanets.

### 17.3.2   Use of a warm telescope for infrared exoplanet studies from space

Sensitivity is not the only rationale for an infrared space telescope. Here we consider instead the benefits of a 273–300K space telescope, in comparison to a comparable- or even larger-sized telescope at a mountaintop observatory. For present purposes, these benefits come in four categories:

Access to the entire infrared spectrum. From outside the Earth's atmosphere, we can access the entire infrared spectrum and all infrared-active molecular species. This may include important species with critical diagnostic features longward of 5 μm, as well as the prominent species with spectral features in the 2–5 μm band discussed here.

Clear skies and long observations. Many exoplanet observations—for example, transits, eclipses, and phase curves—will be of long duration and have timing constraints. A space telescope, operating always under cloudless skies and easily able to observe a particular target for days rather than hours, provides great practical advantages for this type of study.

Stability. The study of exoplanet spectra using combined light spectroscopy (i.e.,

transit spectroscopy) requires achieving a measurement precision on the order of parts per million (ppm) for the most difficult cases. Achieving such photometric precision on a large, warm telescope in space would certainly be easier than on a similar telescope on the ground, looking through the warm and turbulent atmosphere.

Sensitivity. A large space telescope should achieve higher sensitivity than a comparably sized telescope on the ground because of the absence of atmospheric absorption and emission and the routine achievement of diffraction-limited performance.

The scientific, technical, and functional considerations outlined above motivated us to explore extending the wavelength range of LUVOIR to 5 µm, without imposing a constraint on the telescope temperature, but by adding an instrument. We begin by exploring the scientific benefits of this extended wavelength coverage. We also present a strawman design for an infrared spectrometer, covering the 1–5 µm wavelength range with resolving power R = 200, which could execute the science described below with minimal system impact. The 5-micron cutoff reflects current detector technology. A more thorough study should explore longer wavelengths, noting, for example, that the 5–8 µm spectral interval is largely inaccessible from the ground. It could also include exploration of eclipse spectroscopy; here we consider only transit spectroscopy.

### 17.3.3  Exoplanet science: atmospheric characterization with transit spectroscopy

#### 17.3.3.1  Telescope size can trump background noise

Transit spectroscopy thrives on observations of the brightest stars, for which the main noise source, even with a warm telescope,

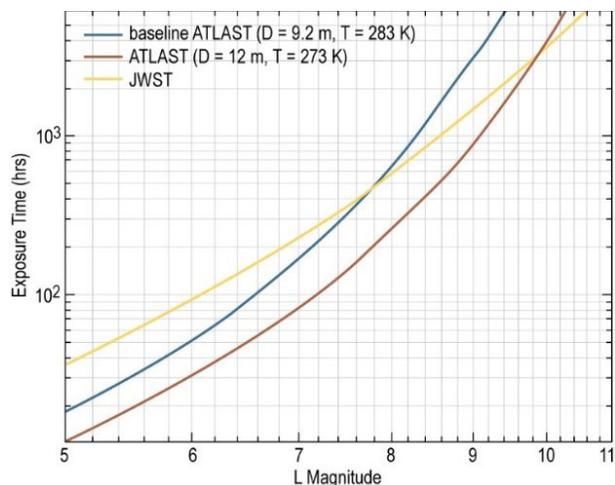

**Figure 17.5.** *Exposure time (or inverse sensitivity) as a function of stellar magnitude for detection of a small and narrow feature (10 ppm at R=200) at 4 µm due to a transiting exoplanet. Three telescope configurations are shown: 9.2-m and 12-m LUVOIRs [called ATLAST in W16] and JWST. LUVOIR is more sensitive than JWST for bright host stars, with the 12-m LUVOIR telescope taking ~4× less time to make the same observation, as is to be expected in the stellar photon limited case. For the LUVOIR configurations, a total photon throughput of 0.35 and a (single mode) background with emissivity 0.2 is assumed; these parameters are commensurate with those of the instrument described below. For JWST, the actual expected performance of the NIRCAM grism (Greene et al. 2007) is adopted. The specific example adopted here is quite stressing in terms of photometric precision requirement, but the gain of the warm telescope would be preserved in less demanding observations as well.*

can be the photon noise of the star itself. For these observations of bright stars there is nothing gained, in principle, by cooling the telescope. In **Figure 17.5**, we consider the time taken (or inverse sensitivity) for three telescope configurations to detect a similar small (10 ppm at R=200) and narrow feature at 4 µm, as a function of the stellar magnitude at that wavelength. The configurations are two versions of LUVOIR and JWST (6.5-

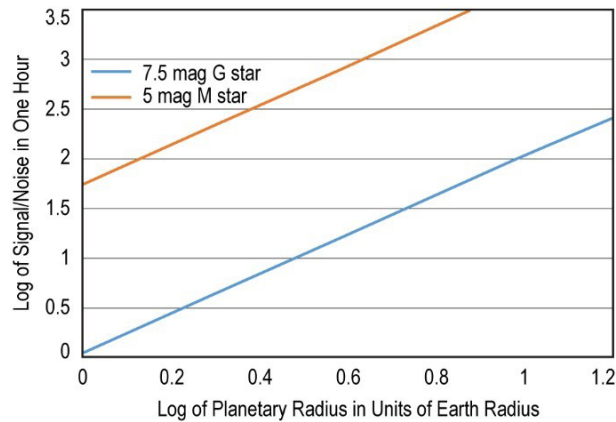

**Figure 17.6.** *Signal/noise on the transit depth for each resolution element in one hour for planetary transits at 4 µm. A 9.2-m, 293 K telescope and the R=200 fiber-fed spectrograph described below were adopted for the calculation.*

m, cryogenic telescope). LUVOIR is more sensitive than JWST for bright host stars, with the 12-m LUVOIR telescope taking ~4× less time to make the same observation for stars brighter than [L]~9, as is to be expected in the stellar photon limited case. The 15-m LUVOIR would be about 40% faster than the 12-m version. For transit spectroscopy, a large warm telescope becomes more powerful than a smaller but cooler one.

## 17.3.3.2  There will be targets

Exoplanet direct imaging is a major driver of the LUVOIR concept, which includes coronagraphy to obtain direct images and low-resolution spectra of exoplanets. The infrared functionality discussed here would provide a powerful complement to this direct imaging work through the use of combined light studies, greatly increasing the scope and wavelength coverage of LUVOIR exoplanet studies. Observations of transiting exoplanets with LUVOIR can yield an enormous scientific return due to the combination of a large aperture and thousands or tens of thousands of targets, each presenting a different scientific puzzle.

In addition to Kepler, K2, and TESS [scheduled for launch in 2018], the PLATO mission, to be launched by ESA in 2028 for a six+ year mission, will yield planets for study by LUVOIR. Based on information provided by David Brown of the PLATO team, PLATO should find over 3400 transiting planets around stars brighter than 11th magnitude in the visible. The predicted S/N per hour vs. planet size for the some of the most readily studied exoplanets which PLATO might discover is shown in **Figure 17.6**. We take S/N > 100 as our target, because only a portion of the total transit signal is contributed by the atmosphere. This is readily achieved in a single transit for Jovian planets with R~10 $R_{Earth}$. A handful of smaller planets will orbit stars of ~5th mag, and their spectra can be readily obtained. Co-adding transits will reach still smaller planets. The 5th magnitude star can be observed with LUVOIR and the R=200 spectrometer described below without saturating the detectors, provided that frame rates faster than 1 Hz are achievable.

## 17.3.3.3  Abundance constraints: an example

A critical performance metric for characterizing exoplanet atmospheres via spectroscopy is the degree of constraint the measurements provide on the abundance, also termed the atmospheric mixing ratio. W16 evaluated the effects of the LUVOIR wavelength range on the constraints that LUVOIR transit spectroscopy might provide. We note a recent paper by Greene et al. (2016), which, for the specific case of JWST, shows the importance of observations longward of 2.5 µm for characterizing exoplanet atmospheres. As an illustrative example, W16 considered the case of a 2.6 $R_{Earth}$, 6.1 $M_{Earth}$ transitional super-Earth/sub-Neptune planet with a 600 K atmosphere orbiting a K star, assuming solar metallicity, thermal equilibrium chemistry, and a clear atmosphere, observed with the 9.2-m LUVOIR. A number of independent observations were simulated and the results analyzed with a multi-parameter MCMC retrieval using the CHIMERA code.

**Table 17.1.** *Limiting magnitude for determination of molecular abundances with an overall uncertainty of less than one order of magnitude for a 2.6 $R_{Earth}$ super-Earth/sub-Neptune (T~600 K) orbiting a K star, assuming 24 hours of observation time spread over a number of transits.*

|  | Molecule | 0.4–2.5 μm | 0.4–5 μm |
|---|---|---|---|
| Mean mol. weight 2.3 | $H_2O$ | $H_{mag} < 10.3$ | $H_{mag} < 11.3$ |
|  | $CH_4$ | $H_{mag} < 10.3$ | $H_{mag} < 12.3$ |
| Mean mol. weight 28 | $H_2O$ | $H_{mag} < 6.3$ | $H_{mag} < 8.3$ |
|  | $CH_4$ | $H_{mag} < 6.3$ | $H_{mag} < 8.3$ |

The results of these experiments are presented in **Table 17.1**, which shows that with 0.4–5.0 μm spectral coverage, LUVOIR has the potential to determine the abundance of water and methane in many TESS/K2/PLATO targets. If the spectral coverage were limited to 2.5 μm, brighter stars are required, so substantially fewer targets would become available for study. This demonstrates the benefits of the enhanced wavelength range we are proposing, particularly for the high mean molecular weight atmosphere, which approximately corresponds to an Earth-like atmosphere. In this case, a very bright star ([H] < 6.3) is required for abundance determinations with less than an order of magnitude uncertainty; stars this bright that harbor transiting planets may be rare statistically. As expected, the lower mean molecular weight atmosphere, with its larger scale height, would be much more readily observed. Several recent papers (Misra, Meadows, & Crisp 2014; Betremieux & Kaltennenger 2014) have highlighted the effects of refraction in the lower part of an exoplanet's atmosphere on our ability to study abundances using transit measurements. The importance of this effect depends on both the size of the star and the star-planet distance, and it will have to be taken into consideration in more sophisticated predictions.

### 17.3.3.4 Studying a true Earth analog

A major thrust of exoplanet studies is, of course, the push to identify and study potentially habitable Earth-like planets. Recent work has shown that simply identifying $O_2$ and $O_3$ in a particular atmosphere is not adequate. Other molecular species, notably $CH_4$ but also $N_2$, CO, and $CO_2$, may have to be studied in order to rule out or confirm the potential habitability of a particular planet which shows $O_2$ and $O_3$. Understanding the prevalence of these species (all of which have strong spectral features longward of 2.5 μm) in exoplanet atmospheres, thus takes on special interest, even if it may not in general be possible to search for all of these molecules in the atmosphere of the same planet.

The detection of biomarkers in the atmosphere of a true Earth analog presents major challenges. This is a difficult detection, so we present it as a limiting case; studying larger planets with more extended atmospheres should be *much* easier. The average methane volume mixing ratio in the Earth's atmosphere is quite low ($1.7 \times 10^{-6}$ mol/mol), a factor of 200 less than $CO_2$, but it is the major source of departure from thermodynamic equilibrium if the atmosphere is considered as a closed system (without oceans and rocks). The strong methane 3.4 μm feature will be blended with a water absorption at low R, but nonetheless contributes up to 10 km (a little more than the 8 km scale height) of extra absorption in the transmission spectroscopy of an Earth. The weaker methane features at shorter wavelengths are yet more heavily blended. In atmospheres with high cloud coverage, the strong IR molecular features absorb at altitudes above the cloud deck, which is a major advantage of long wavelength spectroscopy.

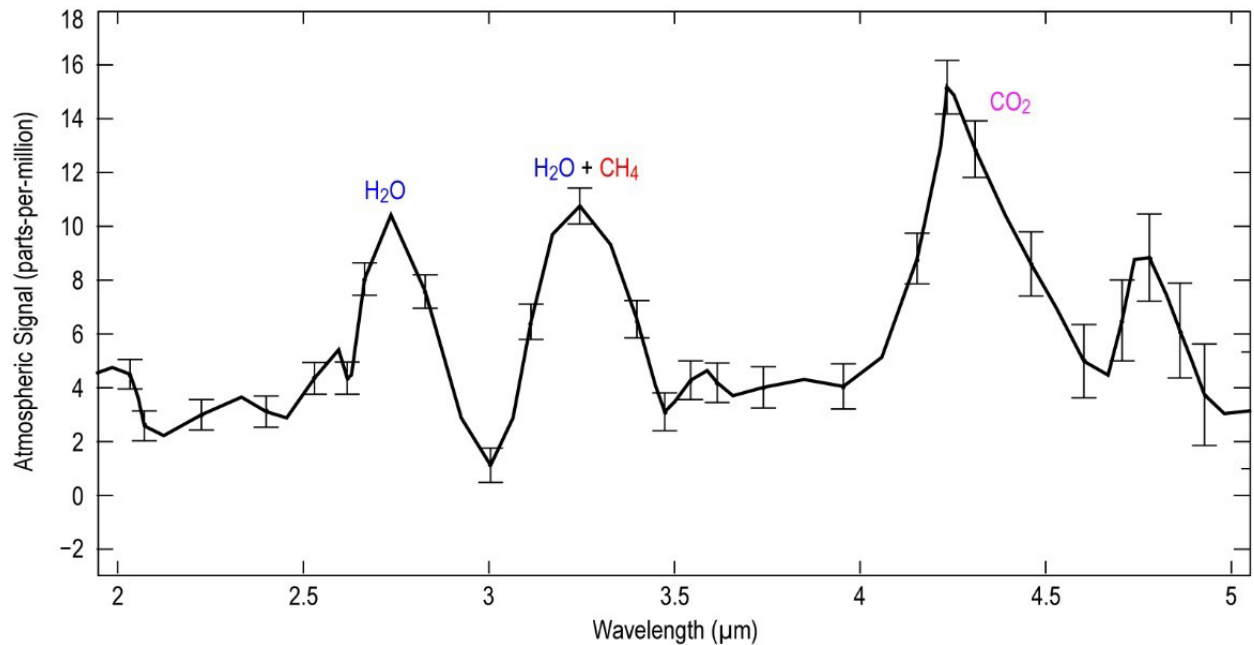

**Figure 17.7.** *A 2–5 μm R~20 model transmission spectrum of a 1 $R_{Earth}$ habitable zone planet with an Earth-like oxidizing atmosphere transiting an M4V star ($R_{star}$ = 0.32 $R_{Sun}$, $T_{star}$ = 3100 K) which is 15 pc from Earth. The 1-sigma errors per wavelength bin, obtained with the smaller LUVOIR design (D=9.2 m, T=273 K), are plotted along with the model. The total integration time is 120 hours in transit, corresponding to 60 transits of about 2 hours each. Features of $H_2O$, $CH_4$, and $CO_2$ are detected. $CH_4$ is blended with $H_2O$ at ~3.3 μm and determining its independent presence and total column would require full atmospheric retrieval; in the Earth's atmosphere, $CH_4$ and $H_2O$ contribute roughly equally to the 3.3 μm feature. A total throughput of 0.35 and an emissivity of 0.2 is assumed here. The y-axis is labelled in parts per million (ppm) relative to the total signal produced by the star. The overall depth of the transit for this case would be about 800 ppm; the features shown in the spectrum would be manifest as changes in the transit depth with wavelength. The precision of a few ppm called for here would be a challenging target for IR detector technology.*

We have computed the transit spectrum of an Earth analog orbiting a late type star, exploiting the advantages conferred by M stars for transit measurements; in addition, the refraction effect mentioned above is much less of a concern for M stars than for solar type stars. Conservatively assuming that the occurrence of habitable zone rocky planets is 0.2 around late type stars, the nearest M-dwarf with a favorable transit will be about ~15 pc away.

In **Figure 17.7** we show a simulated LUVOIR R=20 transit spectrum of an Earth transiting an M4V star (distance 15 pc, $L_{mag}$ = 7.7) in the 2–5 μm region; the underlying model simply uses an Earth transit spectrum and the instrument prescription described in this paper, for which the spectral resolution can be degraded from 200 to 20 with a √10 increase in S/N per resolution element.

While the total time required for the spectrum shown in **Figure 17.7** (~60 transits or 120 hours) is large, it is not outrageously so (two exoplanets have had similar amounts of Hubble time awarded). The orbital period of a habitable zone planet around our M4 star is 16 days, so about 2.5 years would be required to obtain spectra similar to that shown in **Figure 17.7**.

### 17.3.4   Summary: an infrared capability for LUVOIR

We have demonstrated that a ~10-m diameter telescope operating in space at room temperature would be able to make unique and important scientific measurements in the infrared out to at least 5 µm. By using the technique of transit spectroscopy, where for bright stars the noise is set by the stellar photons rather than the thermal background, such a telescope could outperform smaller but lower temperature telescopes for studies which could prove critical for the study of exoplanet atmospheres. It would also be capable of obtaining spectra of millions of infrared sources from the WISE catalog and provide new information about the formation of solar systems, which would link directly to the exoplanet observations. We feel that such a capability should be carefully considered for incorporation into LUVOIR, now or in the future, particularly if further studies of instrument concepts such as that presented below show that they can be easily incorporated into the payload.

### 17.3.5   A fiber-fed infrared spectrometer for LUVOIR

W16 describe a fiber-fed IR spectrometer for LUVOIR. The spectral resolving power of 200 is chosen to permit disambiguation of molecular features in exoplanet atmospheres and also to permit study of the profiles of solid state absorption features in galactic protostellar and protoplanetary disks. We emphasize that other spectrograph designs and means of feeding the spectrograph could provide equivalent performance, and that coupling the starlight into and out of the fiber could stress the LUVOIR pointing system and perhaps other design parameters. An alternative approach using mirrors instead of fibers to feed the spectrograph ought certainly to be considered, for example.

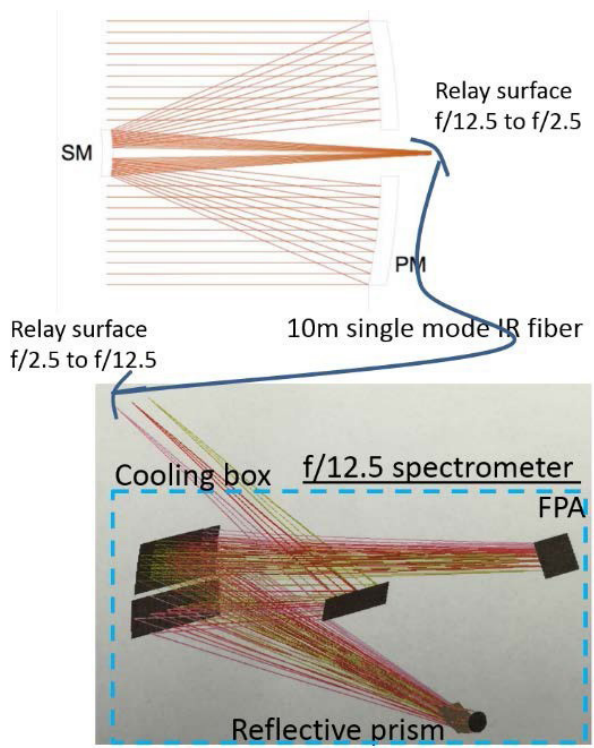

**Figure 17.8.** *Fiber-fed, prism-based IR spectrometer configuration.*

This particular approach was chosen because, on the surface at least, it appears to minimize the system impact of the thermal infrared capability. **Figure 17.8** is a concept sketch of the system. A pick-off mirror will be inserted into the LUVOIR telescope focal plane to relay the star signal into a prism-based spectrometer system (f/12.5), featuring a MWIR FPA (5 µm-cutoff H2RG from Teledyne). The reflective prism is made of calcium fluoride. A 10 m IR fiber (Fluoride Fiber ZBLAN SM from Le Verre Fluore) is used to carry the light from the telescope to the spectrometer, which is mounted on the exterior structure of LUVOIR so that it can cool radiatively. To get optimal fiber coupling, 2 relay mirrors with power (f/12.5 to f/2.5) are utilized. To minimize the spectrometer background noise, a cold shield box covers the full spectrometer system, and the second relay surface illuminates the spectrometer slit, which is mounted on the side of the otherwise sealed cold shield box. The performance of this system is detailed in W16 and used in the calculations presented above.

The design of the spectrometer described here was supported by a grant from the JPL Research and Technology Development fund. Portions of the work described in this paper were carried out at the Jet Propulsion Laboratory, California Institute Technology, under a contract with the National Aeronautics and Space Administration.